\documentclass[english]{mythesis}

\usepackage{my}
\usepackage{mydate}
\usepackage{myref}
\usepackage{enumitem}
\usepackage{datetime}
\usepackage{mypdf}
\usepackage{mysource}
\usepackage[numbers]{natbib}
\usepackage{longtable}
\usepackage{lscape}

\usepackage{pdflscape}
\usepackage{afterpage}
\newcolumntype{P}[1]{>{\hspace{0pt}}p{#1}}

\makeatletter
\newcommand\notsotiny{\@setfontsize\notsotiny{6.81415}{7.6828}}

\makeatother

\usepackage[colour,noprint]{myhyper}
\usepackage[acronym,nonumberlist]{glossaries}

\usepackage{setspace}
\usepackage{forest}

\usepackage{etoolbox}
\RequirePackage{savesym}
\savesymbol{quote}
\RequirePackage{csquotes}
\restoresymbol{CS}{quote}

\savesymbol{clipbox}
\RequirePackage{adjustbox}
\restoresymbol{AB}{clipbox}

\usepackage{bm}

\usepackage{amsmath}

\usepackage{amsfonts}

\usepackage{url}

\usepackage{booktabs}

\usepackage{xcolor}

\usepackage{multirow}

\usepackage{graphicx}

\usepackage{nicefrac}

\usepackage{mathtools}

\usepackage{chngpage}

\usepackage{animate}

\newcolumntype{M}[1]{>{\begin{varwidth}[t]{#1}}l<{\end{varwidth}}}

\makenoidxglossaries

\newglossaryentry{latex}{
        name=latex,
        description={help}}

\newacronym{OCT}{OCT}{Optical Coherence Tomography}

\newacronym{SLO}{SLO}{Scanning Laser Ophthalmoscopy}

\newacronym{EDI}{EDI}{Enhanced Depth Imaging}

\newacronym{PCA}{PCA}{Posterior Ciliary Arteries}

\newacronym{AMD}{AMD}{Age-Related Macular Degeneration}

\newacronym{RPE}{RPE}{Retinal Pigment Epithelium}

\newacronym{CNN}{CNN}{Convolutional Neural Network}

\newacronym{GPET}{GPET}{Gaussian Process Edge Tracing}

\newacronym{MMCQ}{MMCQ}{Multi-scale Median Cut Quantisation}

\newacronym{OCT-A}{OCT-A}{Optical Coherence Tomography - Angiography}

\newacronym{SD-OCT}{SD-OCT}{Spectral Domain Optical Coherence Tomography}

\newacronym{EDI-OCT}{EDI-OCT}{Enhanced Depth Imaging Optical Coherence Tomography}

\newacronym{TD-OCT}{TD-OCT}{Time Domain Optical Coherence Tomography}

\newacronym{SS-OCT}{SS-OCT}{Swept Source Optical Coherence Tomography}

\newacronym{CVI}{CVI}{Choroid Vascular Index}

\newacronym{SD}{SD}{Standard Deviation}

\newacronym{HRA}{HRA}{Heidelberg Retina Angiography}

\newacronym{ART}{ART}{Automatic Real Time}

\newacronym{PPD}{PPD}{Posterior Predictive Distribution}

\newacronym{ETDRS}{ETDRS}{Early Treatment Diabetic Retinopathy Study}

\newacronym{GCU}{GCU}{Glasgow Caledonian University}

\newacronym{MAE}{MAE}{Mean Absolute Error}

\newacronym{MD}{MD}{Mean Difference}

\newacronym{FOV}{FOV}{Field of View}

\newacronym{CI}{CI}{Confidence Interval}

\newacronym{RBF}{RBF}{Radial Basis Function}

\newacronym{CLAHE}{CLAHE}{Contrast Limited Adaptive Histogram Equalisation}

\newacronym{CKD}{CKD}{Chronic Kidney Disease}

\newacronym{eGFR}{eGFR}{estimated Glomerular Filtration Rate}

\newacronym{CT}{CT}{Choroid Thickness}

\newacronym{D-RISC}{D-RISC}{Direct Retinal Imaging for Shock Resuscitation in Critically Ill Adults}

\newacronym{OR}{OR}{Odds Ratio}

\newacronym{GPU}{GPU}{Graphics Processing Unit}

\newacronym{ITU}{ITU}{Intensive Therapy Unit}

\newacronym{HDU}{HDU}{High Dependency Unit}

\newacronym{OIT}{OIT}{Oblique Illumination Test}

\newacronym{OCTANE}{OCTANE}{Optical Coherence Tomography And NEphropathy}

\newacronym{FH}{FH}{Family History}

\newacronym{APOE4}{APOE4}{Apolipoprotein E4}

\newacronym{CA}{CA}{Choroid Area}

\newacronym{ROI}{ROI}{Region Of Interest}
                      
\newacronym{EOI}{EOI}{Edge Of Interest}

\newacronym{AUC}{AUC}{Area Under the Receiver Operating Characteristic Curve}

\newacronym{DVCKD}{DVCKD}{Diurnal Variation for Chronic Kidney Disease}

\newacronym{SNR}{SNR}{Signal-to-Noise Ratio}

\newacronym{SFCT}{SFCT}{Subfoveal Choroid Thickness}

\newacronym{CFP}{CFP}{Colour Fundus Photography}

\newacronym{SOTA}{SOTA}{State-Of-The-Art}

\usepackage{algorithm}
\usepackage{algorithmicx}
\usepackage{algpseudocode}
\usepackage{tabularx}
\makeatletter
\newcommand{\multiline}[1]{%
  \begin{tabularx}{\dimexpr\linewidth-\ALG@thistlm}[t]{@{}X@{}}
    #1
  \end{tabularx}
}
\makeatother

\DeclarePairedDelimiter{\ceil}{\lceil}{\rceil} %
\DeclareMathOperator*{\argmax}{\text{argmax}}  %
\DeclareMathOperator*{\argmin}{\text{argmin}}  %
\DeclareMathOperator*{\trans}{\text{T}}  %

\usepackage{times}
\sffamily

\logoleft{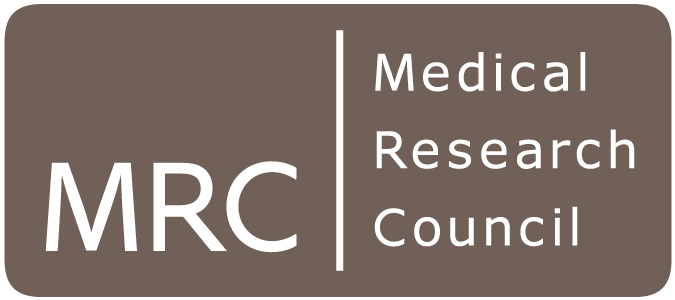}
\logoright{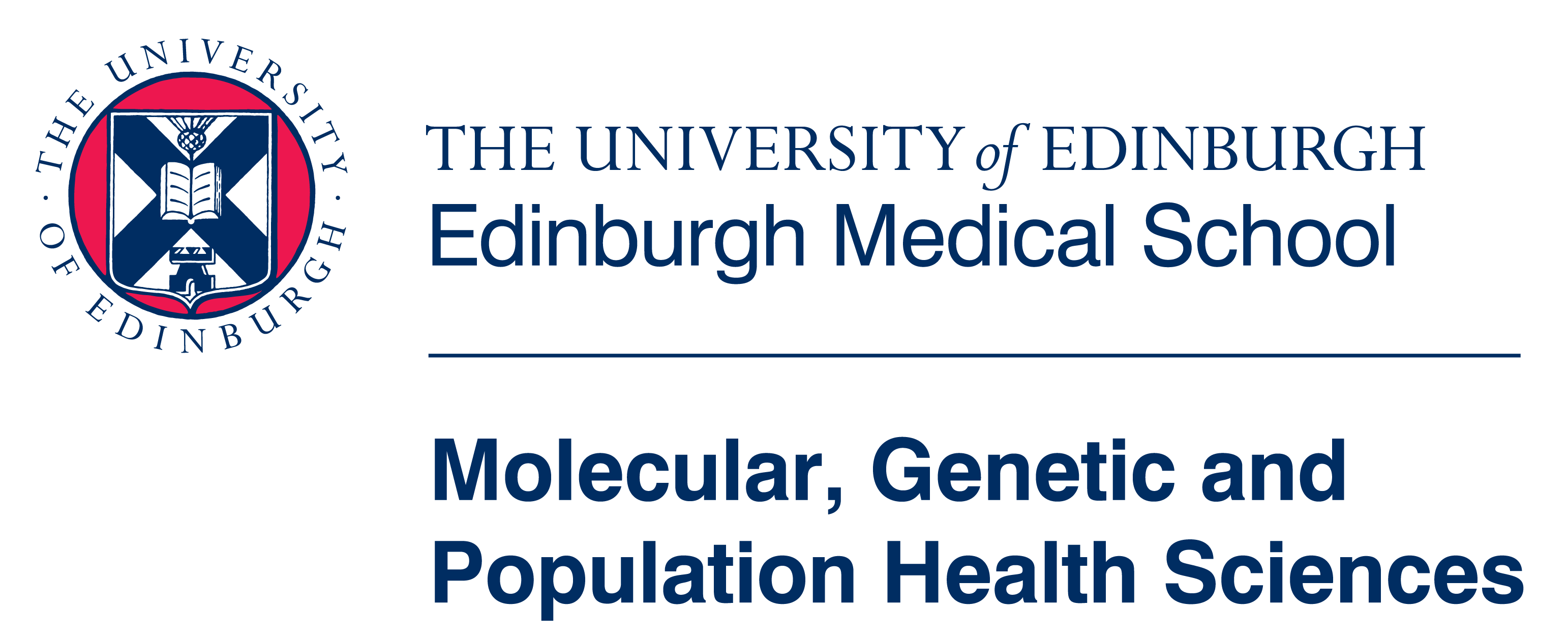}

\hypersetup{%
	pdftitle={Choroidal Image Analysis for  OCT Image Sequences with Applications in Systemic Health},%
	pdfauthor={Jamie Burke},%
	pdfkeywords={computer vision, optical coherence tomography, choroid, systemic health},%
	pdflang={EN-gb},%
	pdfsubject={Thesis}%
}

\title{Choroidal Image Analysis for  OCT Image Sequences with Applications in Systemic Health}
\author{\myname{Jamie Burke}}
\university{The University of Edinburgh}

\presentationyear{2024}
\grade{Doctorate of Philosophy in Precision Medicine}

\let\cleardoublepage\clearpage

\begin{document}

\pagestyle{empty}

\cleardoublepage
\pagestyle{fancy}

\frontmatter

\maketitle

\begin{myacknowledgements}
{\setstretch{1.0}
First and foremost, I must extend my thanks to my supervisory team: Dr. Stuart King, Dr. Ian J.C. MacCormick, and Prof. Kenny Baillie. Your insights across both technical and clinical domains have been invaluable to the work presented for this degree and in this thesis. In particular, I want to thank Stuart for his unwavering technical and personal support throughout the COVID-19 lockdown(s), without which this doctorate degree would have certainly lacked direction. I also wish to express special gratitude to Ian for his seemingly infinite patience with my naive questions, manuscript reviews, expert ophthalmology input, and for helping to frame much of the research toward a clinical audience, always offering a long-term, forward-thinking perspective.

The first half of my PhD was completed in the School of Mathematics, and I’d like to thank the \textit{Usual Suspects}, a group of excellent individuals with whom I had the pleasure of spending much of my first research period. You always managed to brighten my day, even in the deep and dark halls of JCMB.

Breaking up my PhD studies were two valuable internships hosted by Public Health Scotland and NHS England. Although unrelated to my PhD research, these were valuable experiences outside academia that helped broaden my understanding of both healthcare and data science. I’d like to acknowledge Theresa Ryan and Dr. Dan Schofield for their supervision at these institutions, both of whom made my time there worthwhile and enjoyable. I also want to give an honourable mention to Dr. Jim Rafferty, who helped supervise my work with NHS England, and who, alongside Dan, is fully convinced that everything, everywhere is a graph model, in some strange and peculiar way.

Upon returning to my PhD studies, I was fortunate enough to be invited to the Robert O Curle (ROC) Ophthalmology Suite at the Institute for Regeneration and Repair, thanks to the invitation of Dr. Tom MacGillivray. Consequently, I'd like to acknowledge him as an effective fourth supervisor, who helped me navigate and publish in the field of retinal imaging. It was at the ROC lab where I felt most at home, surrounded by like-minded computer vision researchers in ophthalmology, and where I navigated successful and collaborative projects. In particular, I’d like to acknowledge Adam Threlfall, Borja Marin, Darwon Rashid, Fabian Yii, Miracle Ozzoude, Sam Gibbon, and Ylenia Giarratano for their collaboration, company and banter.

I'd also like to acknowledge Justin Engelmann for the fantastic work we've done together. Without your wizard-like skills in deep learning, this thesis would only be half of what it is today. It’s been a pleasure working with someone as passionate, hard-working, and successful as you. I also want to give an honourable mention to Charlene Hamid. Without your expertise in OCT imaging and acquisition, much of the data in this thesis wouldn't have been available, and my time acquiring and extracting data was made significantly easier with your help.

It would be remiss not to mention the wonderful people in my life outside academia who have kept me afloat over the last five years. To my good friends Thomas Rodger and Conan Phimister, your company at the flat was incredible and always appreciated. Thank you for listening to me ramble on about my work and for letting me comfort-feed you some very (deliciously) gluttonous pesto pasta. To my family --- Katherine, Jim, Ollie, Rachel, Nathan, and Katharine --- thank you for putting up with my spells of absence and general grumpiness toward life and research over the past five years. I promise to get a real job now. You are all excellent individuals, and I’m very lucky to call you my family. I must also give a special acknowledgement to my brother Ollie, who always believed in me and encouraged me along this academic path. I therefore extend my endless gratitude (and occasional frustration) to you, good sir.

Finally, there is one very important individual I couldn't imagine not mentioning --- my partner, Chloe Campbell. You have been an exceptional constant in my life these past five years and have somehow managed to put up with all my idiosyncrasies. I am forever grateful for your unwavering encouragement and emotional support, even though you still don’t quite know what the choroid is. You, and the gang, have played a crucial role in keeping me motivated and goal-oriented, and you've been an indispensable distraction, allowing me to live my happiest life outside of work. You are simply fantastic, and I couldn't have asked for a better collaborator to navigate life with.
\vfill
}

\end{myacknowledgements}

\begin{mydeclaration}
{\setstretch{1.0}
    \textit{I declare that this thesis has been composed by myself and that the work has not been submitted for any other degree or professional qualification. I confirm that the work submitted is my own, except where work which has formed part of jointly-authored publications has been included. My contribution and those of the other authors to this work is indicated below. I confirm that appropriate credit has been given within this thesis where reference has been made to the work of others.}
    
    I outline below whether the content of any research chapter has been published, and highlight the author of declaration in bold type, with lead authors underlined and supervisors in italic type.
    \begin{itemize}\setlength\itemsep{0em}
        \item I was the lead researcher for the work presented in chapter \ref{chp:chapter-GPET}. The methodology was conceived by myself and Stuart King, and I wrote the software and carried out the method's evaluation. This work was previously published in \cite{burke2021edge, burke2023evaluation}:
        \begin{itemize}\setlength\itemsep{0em}
            \item ``Edge Tracing using Gaussian Process Regression'' by (in publication authorship order): \underline{\textbf{Jamie Burke}} and \textit{Stuart King};
            \item ``Evaluation of an Automated Choroid Segmentation Algorithm in a Longitudinal Kidney Donor and Recipient Cohort'' by (in publication authorship order): \underline{\textbf{Jamie Burke}}, Dan Pugh, Tariq Farrah, Charlene Hamid, Emily Godden, Thomas J. MacGillivray, Neeraj Dhaun, \textit{J. Kenneth Baillie},  \textit{Stuart King} and  \textit{Ian J.C. MacCormick}.
        \end{itemize}
    
        \item I was the lead researcher for the work presented in chapter \ref{chp:chapter-mmcq}.  The methodology was conceived by myself and Stuart King, and I wrote the software and carried out the method's evaluation and reproducibility. This work was has not been published.
    
        \item There is no lead author for the research presented in chapter \ref{chp:chapter-deepgpet}. The methodology was conceived by myself and Justin Engelmann, under shared co-authorship. I helped collect the data, generate the ground-truth labels for modelling and carried out the method's reproducibility. Justin Engelmann trained the deep learning model, and we both co-wrote the software and performed the method's evaluation. This work was previously published in \cite{burke2023open}:
        \begin{itemize}\setlength\itemsep{0em}
            \item ``An Open-Source Deep Learning Algorithm for Efficient and Fully Automatic Analysis of the Choroid in Optical Coherence Tomography'' by (in publication authorship order): \underline{\textbf{Jamie Burke}}, \underline{Justin Engelmann}, Charlene Hamid, Megan Reid-Schachter, Tom Pearson, Dan Pugh, Neeraj Dhaun, Amos Storkey, \textit{Stuart King}, Thomas J. MacGillivray, Miguel O. Bernabeu and \textit{Ian J.C. MacCormick}. 
        \end{itemize}
    
        \item There is no lead author for the research presented in chapter \ref{chp:chapter-choroidalyzer}. The methodology was conceived by myself and Justin Engelmann, under shared co-authorship. I helped collect the data, generate the ground-truth labels for modelling and carried out the method's reproducibility. Justin Engelmann trained the deep learning model, and we both co-wrote the software and performed the method's evaluation. This work was previously published in \cite{engelmann2024choroidalyzer}:
        \begin{itemize}\setlength\itemsep{0em}
            \item ``Choroidalyzer: An Open-Source, End-to-End Pipeline for Choroidal Analysis in Optical Coherence Tomography'' by (in publication authorship order): \underline{Justin Engelmann}, \underline{\textbf{Jamie Burke}}, Charlene Hamid, Megan Reid-Schachter, Dan Pugh, Neeraj Dhaun, Diana Moukaddem, Lyle Gray, Niall Strang, Paul McGraw, Amos Storkey, Paul J. Steptoe, \textit{Stuart King}, Thomas J. MacGillivray, Miguel O. Bernabeu and \textit{Ian J.C. MacCormick}. 
        \end{itemize}
    
        \item I was the lead researcher who carried out the reproducibility analysis of the methods presented in chapters \ref{chp:chapter-deepgpet} and \ref{chp:chapter-choroidalyzer}. The methodology was conceived by myself. I helped collect the data, wrote the software and carried out the reproducibility analysis. This work has been submitted and is in peer-review as \cite{burke2024octolyzer}:
         \begin{itemize}\setlength\itemsep{0em}
            \item ``OCTolyzer: Fully automatic analysis toolkit for segmentation and feature extracting in optical coherence tomography (OCT) data'' by (in publication authorship order): \underline{\textbf{Jamie Burke}}, Justin Engelmann, Samuel Gibbon, Charlene Hamid, Niall Strang, Neeraj Dhaun, Thomas J. MacGillivray, \textit{Stuart King} and \textit{Ian J.C. MacCormick}.
        \end{itemize}
    
        \item I was the lead researcher who carried out the data analysis in section \ref{sec:ch_app_sec_ckd}. I conceived the idea with help from senior researchers Neeraj Dhaun, Stuart King, J. Kenneth Baillie, Ian J.C. MacCormick and Thomas J. MacGillivray. I carried out the image analysis, data analysis and the manuscript write-up. This work was previously published in \cite{burke2023evaluation}:
         \begin{itemize}\setlength\itemsep{0em}
            \item ``Evaluation of an Automated Choroid Segmentation Algorithm in a Longitudinal Kidney Donor and Recipient Cohort'' by (in publication authorship order): \underline{\textbf{Jamie Burke}}, Dan Pugh, Tariq Farrah, Charlene Hamid, Emily Godden, Thomas J. MacGillivray, Neeraj Dhaun, \textit{J. Kenneth Baillie},  \textit{Stuart King} and  \textit{Ian J.C. MacCormick}.
        \end{itemize}
    
        \item There is no lead author for the research carried out in section \ref{sec:ch_app_sec_prevent}. Samuel Gibbon and I conceived the idea, under shared co-authorship. I carried out the image analysis, and Samuel Gibbon and I both carried out the data analysis and the manuscript write-up. This work was previously published in \cite{burke2024prevent}:
         \begin{itemize}\setlength\itemsep{0em}
            \item ``Association between choroidal microvasculature in the eye and Alzheimer's disease risk in cognitively healthy midlife adults'' by (in publication authorship order): \underline{\textbf{Jamie Burke}}, \underline{Samuel Gibbon}, Audrey Low, Charlene Hamid, Megan Reid-Schachter, Graciela Muniz-Terrera, Craig W. Ritchie, Baljean Dhillon, John T O’Brien, \textit{Stuart King}, \textit{Ian J.C. MacCormick}, Thomas J. MacGillivray. 
        \end{itemize}
    
        \item There is no lead author for the research carried out in section \ref{sec:ch_app_sec_shock}. J. Kenneth Baillie conceived the idea, with help from Ian J.C. MacCormick, David Griffiths and George Cooper on carrying out project administration and ethical approval. I carried out the data collection and the image analysis, and George Cooper and I both carried out the data analysis and manuscript write-up. This work has been submitted and in peer-review as \cite{burke2024drisc}:
         \begin{itemize}\setlength\itemsep{0em}
            \item ``Direct Retinal Imaging for Shock Resuscitation in Critical Ill Adults II (D-RISC II)'' by (in publication authorship order): \underline{George Cooper}, \underline{\textbf{Jamie Burke}}, Charlene Hamid, Emily Godden, Neeraj Dhaun, \textit{Stuart King}, Thomas J. MacGillivray, \textit{J. Kenneth Baillie}, David M. Griffith and \textit{Ian J.C. MacCormick}.
        \end{itemize}
    \end{itemize}

    Using TexCount, there are $\approx$74,000 words in this thesis. Finally, the use of ``we'', ``our'' and ``the author'' in the main text is purely stylistic.
}
\end{mydeclaration}

\nopagebreak

\begin{myabstract}[english]
    The retina is a light-sensitive tissue at the back of the eye and is responsible for vision. Light-sensitive photoreceptor cells in the outer retina detect light and, through a series of neuronal and vascular layers, process it into signals for the brain. The photoreceptors are perfused and maintained indirectly by the choroid and choriocapillaris, a highly vascularised layer posterior to the retina. The choroid is an extension of the central nervous system and has parallels with the renal cortex, but choroidal blood flow is four-fold higher per unit mass than the kidney and ten-fold higher than the brain. Thus, there has been growing interest in the structure and function of the choroidal circulation reflecting physiological status of systemic disease in the kidney and brain. The choroid can be imaged using optical coherence tomography (OCT), a non-invasive imaging technique which uses interferometry to capture three-dimensional, cross-sectional visualisations of ocular tissue at micron resolution. Advancements in OCT technology now permit deeper penetration and improved visualisation of the choroidal vessels. However, conventional methods of characterising and quantifying this vascular space have not kept pace with the improvements in OCT technology which visualise it, resulting in non-standardised manual or semi-automatic approaches as commonplace methods for choroidal measurement. The ability to measure anatomy consistently at micron-scale both intra- and inter-patient is paramount to capturing the inherent biological change or signal being studied --- a signal which can be corrupted when exposed to human subjectivity. 
    
    In this thesis, I develop and evaluate several novel methods to analyse the choroid in OCT image sequences, with each successive method markedly improving on its predecessors. In the first instance, I develop two semi-automatic approaches for choroid region (Gaussian Process Edge Tracing, \textit{GPET}) and vessel (Multi-scale Median Cut Quantisation, \textit{MMCQ}) analysis, which improve on manual approaches but ultimately are biased by the end-user's biological interpretation and technical experience. As a first step to fully automatic choroid segmentation, I develop \textit{DeepGPET} as a deep learning-based method for choroid region segmentation, which significantly improves on semi-automatic approaches in terms of time, reproducibility, and end-user accessibility. However, DeepGPET lacks choroidal vessel quantification and still requires manual input for generating standardised, choroid-derived measurements. Improving on this, I developed \textit{Choroidalyzer}, a fully automatic, deep learning-based, end-to-end pipeline which fully characterises the choroidal space and vessels, and automatically generates clinically meaningful and reproducible choroid-derived metrics. I provide rigorous evaluation of these four approaches, and consider their use-case and potential clinical value in three distinct applications into systemic health: OCTANE: evaluating longitudinal choroidal change and its association with renal function in transplant recipients and donors; PREVENT: investigating associations between the choroid and risk factors for developing later-life Alzheimer's disease in a mid-life cohort; D-RISCii: assessing choroidal variation and feasibility of OCT imaging in critical care. This thesis has contributed several new approaches to the research community which are all open-source and freely available, enabling consistent and reproducible measurement of the choroid. This thesis also highlights the potential role the choroid may play in reflecting pathophysiology in the kidney, brain and wider systemic health from iatrogenic shock, thus helping accelerate the nascent field of choroidal analysis in OCT image sequences.

    \begin{mykeywords}
    \mykeyword choroid
    \mykeyword optical coherence tomography
    \mykeyword image analysis
    \mykeyword deep learning
    \mykeyword systemic health
     \end{mykeywords}

\end{myabstract}

\begin{mylaysummary}

    The retina is an essential part of the eye which is responsible for vision through detecting light and converting it into signals for the brain. The light-sensitive photoreceptors in the retina play a crucial role in this process and are supported and maintained by a dense network of blood vessels in the choroid, a predominantly vascular layer located behind the retina. The choroid is part of the central nervous system, shares similarities with organs like the kidney, and has a much higher blood flow rate per unit mass compared to the kidney or brain, in part to support the energy demands of the photoreceptor cells. Consequently, there is growing interest from the research community for studying the choroid in the context of systemic diseases that affect these organs. The choroid can be visualised using optical coherence tomography (OCT), a non-invasive imaging technique that provides detailed, three-dimensional visualisations of the back of the eye at micro-metre resolution. Recent advancements in OCT technology have significantly improved visualisation of the choroidal vessels, thus permitting a unique assessment into the structure and function of this microvascular network. However, traditional, manual or semi-automatic approaches to measure the choroid have not kept pace with the improving OCT technology that visualise it, leading to inconsistent results which have the potential to be impacted by human subjectivity.

    In this thesis, I develop and evaluate new methods for analysing the choroid in OCT image sequences. I first create two semi-automatic approaches to analyse the choroidal space (Gaussian Process Edge Tracing, \textit{GPET}) and vessels (Multi-scale Median Cut Quantisation, \textit{MMCQ}), which improve on manual methods but still depend on user expertise. I then develop a fully automatic deep learning method, \textit{DeepGPET}, for isolating the choroidal space in  OCT images. This method is faster and more consistent, though still requires manual input to measure the choroid. As a final improvement for analysing the choroid in OCT image sequences, I develop \textit{Choroidalyzer}, a fully automatic, deep learning-based toolkit that comprehensively quantifies the choroid and its blood vessels. Choroidalyzer is a significant improvement on previous approaches and provides reliable and clinically meaningful measurements of the choroid without human bias. I thoroughly test these new approaches to choroidal image analysis, and explore their potential for characterising the choroid in the context of systemic disease using three distinct applications: OCTANE: evaluating longitudinal choroidal change and its association with renal function in transplant recipients and donors; PREVENT: investigating associations between the choroid and risk factors for developing later-life Alzheimer's disease in a mid-life cohort; D-RISCii: assessing choroidal variation and feasibility of OCT imaging in critical care. This thesis contributes several new, freely available tools for quicker, accurate, and more consistent measurement of the choroid compared with previous manual or semi-automatic approaches. This thesis also emphasises the potential role of the choroid in reflecting overall systemic health, ultimately helping accelerate the nascent field of choroidal analysis in OCT image sequences.

\end{mylaysummary}

\nopagebreak

{\setstretch{1.0}
\mytoc

\addcontentsline{toc}{chapter}{Acronyms}
\printnoidxglossary
\printnoidxglossary[type=\acronymtype]

\addcontentsline{toc}{chapter}{List of Figures}
\listoffigures
\addcontentsline{toc}{chapter}{List of Tables}
\listoftables
\addcontentsline{toc}{chapter}{List of Algorithms}
\listofalgorithms
}
\mainmatter

\begin{mychapter}[chp:chapter-Intro]{Introduction}\label{chp:chapter-introduction}

    The choroid is a crucial vascular layer at the back of the eye which sits posterior to the retina, effectively maintaining the visual system through nourishment of the photoreceptors \cite{nickla2010multifunctional}. The retina has often been the primary anatomical layer of interest in ophthalmology, while the choroid has seen neglect by virtue of the challenges faced by historical imaging methods. Previous methods used to image the retina relying on light reflection (laser doppler flowmetry), sound (ultrasonography) or fluorescence (fluorescein angiography) are hindered from imaging the choroid at high resolution by the pigment in the outer retinal pigment epithelium (\acrshort{RPE}) and choroid \cite{mrejen2013optical}. Additional to the light absorption and back-scattering of the anterior retina, some of these approaches remain a challenge to collect repeated measurements of the choroid \cite{mrejen2013optical}. 
    
    Optical coherence tomography (\acrshort{OCT}) \cite{huang1991optical} has provided non-invasive access to cross-sectional retinochoroidal structures at the posterior pole since 1991. However, traditional time-domain (\acrshort{TD-OCT}) and conventional spectral domain \acrshort{OCT} (\acrshort{SD-OCT}) are both affected by melanin and the scattering properties of blood vessels in the retina and choroid \cite{tan2024techniques}. Improved visualisation of the choroid was introduced in the late 2000s with enhanced depth imaging \acrshort{OCT} (\acrshort{EDI-OCT}) \cite{spaide2008enhanced} and later in 2012 with swept-source \acrshort{OCT} (\acrshort{SS-OCT}) which both permitted deeper penetration of ocular tissue over their traditional counterparts.
    
    This enhanced appearance of the choroid has allowed many to study the deeper vascular tissue of the eye in both retinal and systemic disease, conducting biomarker investigation with representative measurements like choroidal thickness \cite{tan2024techniques}. However, approaches to measuring choroidal thickness vary significantly, both in terms of methodology \cite{mazzaferri2017open, cheong2018novel, ma2022longitudinal, xu2022automatic, farrah2023choroidal, xuan2023deep} and in terms of protocol \cite{yiu2014characterization, sezer2016choroid, singh2019choroidal, aksoy2023choroidal}. 
    
    This has lead to a deficit in the research community in accurately and reliably measuring the choroid between and within individuals \cite{boonarpha2015standardization, xie2021evaluation}. The ability to measure anatomy consistently at micron-scale is paramount to capturing the inherent biological change or signal being studied \cite{breher2019metrological, brehar2020comparison} --- a signal which can be corrupted by human subjectivity. 
    
    While the advent of Fourier domain \acrshort{EDI-OCT} and \acrshort{SS-OCT} has enabled improved visualisation of the choroid, current approaches to measure it have not kept pace with these advancements. This has lead to the research community continuing to adopt traditional, semi-automatic image processing methods or manual grading in their studies \cite{mazzaferri2017open, ss2019octtools, sonoda2014choroidal, agrawal2020exploring}. This is by virtue of their accessibility through mediums such as Fiji (ImageJ) \cite{ma2022longitudinal, schindelin2012fiji, rishi2022analysis} or availability on open-source platforms like GitHub \cite{mazzaferri2017open, ss2019octtools}. 

    However, current methods are inadequate because of human subjectivity and improper usage and reporting of approaches to measuring the choroid. This has led to a concern around their overall reproducibility and accuracy. Additionally, more accurate and domain-specific approaches based on deep learning are often not openly available to the research community, leading to institutions using their own bespoke methods. Together, these contribute to a lack of standardisation within choroidal image analysis in \acrshort{OCT} image sequences.
    
    This absence of accurate and reliable methods for measuring the choroid drives a real need for more accurate approaches to choroid analysis which are open-source, easily accessible and rigorously validated. This would ultimately equip researchers with the ability to extract reproducible and clinically meaningful measurements of the choroid, thus improving standardisation in the field of choroidal image analysis in \acrshort{OCT} image sequences.
    
    In this thesis, we develop and evaluate several new approaches to measure the choroid from \acrshort{OCT} image sequences. We first consider the applicability of traditional image processing techniques, which begins to address problems around reproducibility and human subjectivity. We then consider the utility of more advanced and domain-specific computer vision algorithms based on deep learning. In the latter, these approaches permit fully automatic, end-to-end choroidal image analysis in \acrshort{OCT} image sequences, completely removing manual subjectivity from the measurement procedure. We show their ability to improve standardisation of choroidal image analysis by assessing their measurement error in detail.
    
    Additionally, we release these tools to the wider research community as open-source to facilitate end-user accessibility. These tools are designed to not require proprietary software to use, nor require any domain-specific knowledge in image processing or computer science. We anticipate the accessibility, accuracy and reliability of these tools to help improve the standardisation of choroidal measurements in \acrshort{OCT} image sequences. In addition to the development of these novel algorithms, we demonstrate their use-case on three novel applications of the choroid in systemic disease to show their wider applicability to the nascent field of choroidal image analysis in \acrshort{OCT} image sequences.

    \begin{mysection}[]{Guiding principles}
        This thesis hopes to advance the field of choroidal image analysis in \acrshort{OCT} through following a set of guiding principles:

        \begin{enumerate}
            \item \textbf{Measurement protocol}: Follow strict measurement protocol which accounts for pixel resolution, skew and curvature of intra-ocular structures, enabling improved measurement of the choroid in \acrshort{OCT} image sequences between and within patients.

            \item \textbf{Reproducibility}: Assess each method's measurement variability to help future end-users differentiate true biological change from measurement error in their studies. This will promote their reliability to produce clinically meaningful measurements of the choroid in \acrshort{OCT} image sequences.

            \item \textbf{Rigorous validation}: Conduct thorough validation of the new approaches developed in this thesis, both quantitatively and qualitatively, to transparently identify their strengths and limitations. This will communicate each method's applicability to choroidal image analysis in \acrshort{OCT} image sequences.

            \item \textbf{End-user accessibility}: Release these tools as open-source to ensure their accessibility to the broader research community. This will facilitate the adoption of these methods across varying levels of expertise, encouraging wider usage and standardisation of choroidal measurements in \acrshort{OCT} image sequences.

            \item \textbf{Application to systemic disease}: Demonstrate the potential of choroidal analysis in reflecting microvascular function, injury and response to systemic disease. The thesis will explore the emerging relevance of the choroid in this field and hope to encourage further research in this nascent area.
            
        \end{enumerate}

        This thesis ultimately aims to contribute to the field of choroidal image analysis in \acrshort{OCT} image sequences and its potential role in systemic disease. Through providing the research community with accessible and rigorously validated tools for extracting reproducible and clinically meaningful measurements of the choroid, the author hopes to help improve the standardisation of these measurements in the research community. 
        
        Finally, the author hopes this thesis will contribute to the nascent field of `oculomics' by describing the potential for the eye to inform about systemic health. This will be accomplished through applying these novel methods in the context of renal transplant in chronic kidney disease, Alzheimer's disease risk and fluid resuscitation in multi-organ failure.

    \end{mysection}

    \begin{mysection}[]{The structure of this thesis}
            
        \begin{figure}[tb]
            \centering
            \includegraphics[width=0.9\textwidth]{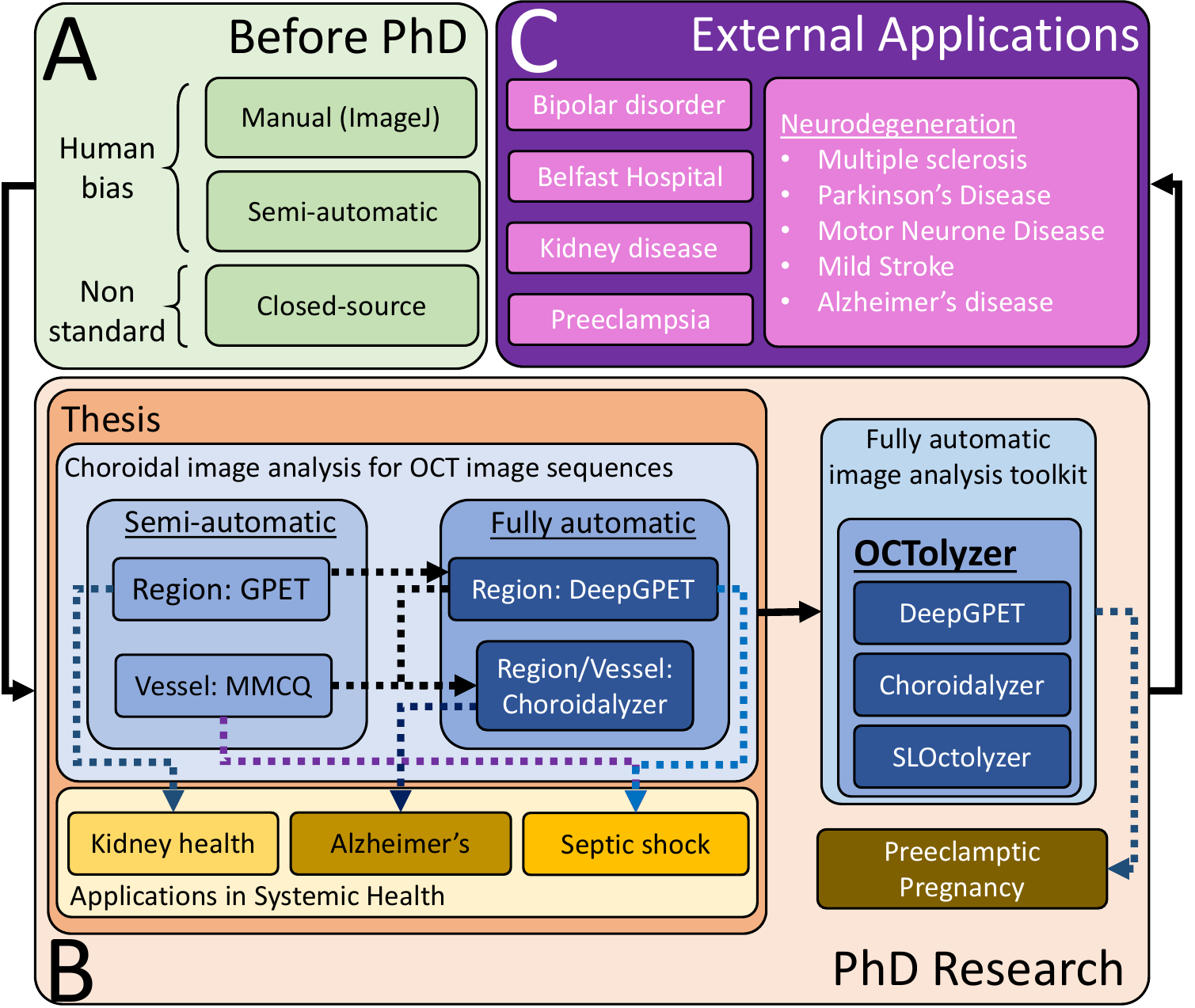}
            \caption[Schematic of work conducted during research period.]{Schematic of the PhD research journey. (A) Current approaches to choroidal image analysis. (B) Work conducted during research period, with boxes and arrows identifying analysis methods from applications and their relationships. (C) Past, current and planned future applications of the methods developed from this thesis.}
            \label{fig:INTRO_PhD_schematic}
        \end{figure}
    
        This thesis is composed of an introductory chapter, five research chapters and a concluding discussions chapter. Chapter \ref{chp:chapter-introduction} introduces the choroidal microvasculature, it's anatomical composition and relationship with systemic disease. We also discuss how the choroid is imaged with \acrshort{OCT} and how it is measured, with a brief overview on the current approaches used. 
        
        Research chapters \ref{chp:chapter-GPET} -- \ref{chp:chapter-choroidalyzer} discuss, define and rigorously evaluate the accuracy and reproducibility of the automatic segmentation tools developed over the course of the research period. These research chapters introduce and evaluate Gaussian Process Edge Tracing (\acrshort{GPET}, chapter \ref{chp:chapter-GPET}) for semi-automatic choroid region segmentation; Multi-scale Median Cut Quantisation (\acrshort{MMCQ}, chapter \ref{chp:chapter-mmcq}) for semi-automatic choroid vessel segmentation; DeepGPET (chapter \ref{chp:chapter-deepgpet}), for fully automatic, deep learning based choroid region segmentation; and Choroidalyzer (chapter \ref{chp:chapter-choroidalyzer}), for end-to-end, deep learning based choroidal region and vessel segmentation, alongside automatic fovea detection and choroid measurement. 
        
        The final research chapter, chapter \ref{chp:chapter-applications}, presents three applications of choroidal image analysis in \acrshort{OCT} image sequences in systemic disease using the aforementioned tools. These include investigating choroidal changes in a longitudinal kidney donor and recipient cohort, estimating associations between choroidal microvasculature and Alzheimer's disease risk, and evaluating the feasibility of \acrshort{OCT} imaging and choroidal variation for shock resuscitation in critically ill adults.

        Finally, chapter \ref{chp:chapter-discussion} will discuss the key takeaways from this thesis and the core contributions this thesis provides to the research community. Moreover, this chapter will outline the current limitations and challenges faced by the work in this thesis. Lastly, this chapter will outline avenues of further work for choroidal image analysis in \acrshort{OCT} image sequences. Figure \ref{fig:INTRO_PhD_schematic} visualises the work contained in this thesis and over the PhD research period.

    \end{mysection}

    \begin{mysection}[]{PhD outputs} 

        \begin{mysubsection}[]{Publication portfolio} 
        
            Below lists publication outputs which were published and peer-reviewed by February 2025. Author of thesis in bold type, with lead author(s) underlined.
            
            \begin{itemize}
            \setlength\itemsep{0.1em}
            
                \item \underline{\textbf{Burke, Jamie}}, and Stuart King. ``\textit{Edge tracing using Gaussian process regression.}'' IEEE Transactions on Image Processing 31 (2021): 138-148.
    
                \item \underline{\textbf{Burke, Jamie}}, Dan Pugh, Tariq Farrah, Charlene Hamid, Emily Godden, Thomas J. MacGillivray, Neeraj Dhaun, J. Kenneth Baillie, Stuart King, and Ian J.C. MacCormick. ``\textit{Evaluation of an automated choroid segmentation algorithm in a longitudinal kidney donor and recipient cohort.}'' Translational Vision Science \& Technology 12, no. 11 (2023): 19-19.
                
                \item \underline{\textbf{Burke, Jamie}}, Neeraj Dhaun, Baljean Dhillon, Kyle J. Wilson, Nicholas A.V. Beare, and Ian J.C. MacCormick. ``\textit{The retinal contribution to the kidney–brain axis in severe malaria.}'' Trends in parasitology 39, no. 6 (2023): 410-411.
                
                \item \underline{\textbf{Burke, Jamie}}, \underline{Justin Engelmann}, Charlene Hamid, Megan Reid-Schachter, Tom Pearson, Dan Pugh, Neeraj Dhaun, Amos Storkey, Stuart King, Thomas J. MacGillivray, Miguel O. Bernabeu, and Ian J.C. MacCormick. ``\textit{An Open-Source Deep Learning Algorithm for Efficient and Fully Automatic Analysis of the Choroid in Optical Coherence Tomography.}'' Translational Vision Science \& Technology 12, no. 11 (2023): 27-27.
            
                \item \underline{Engelmann, Justin}, \underline{\textbf{Jamie Burke}}, Charlene Hamid, Megan Reid-Schachter, Dan Pugh, Neeraj Dhaun, Diana Moukaddem, Lyle Gray, Paul McGraw, Amos Storkey, Niall Strang, Paul J. Steptoe, Stuart King, Thomas J. MacGillivray, Miguel O. Bernabeu, and Ian J.C. MacCormick. ``\textit{Choroidalyzer: An open-source, end-to-end pipeline for choroidal analysis in optical coherence tomography.}'' Investigative Ophthalmology \& Visual Science 65, no. 6 (2024): 6-6. (Mathematical and Computational Ophthalmology Special Issue).

                \item \underline{\textbf{Burke, Jamie}}, Samuel Gibbon, Justin Engelmann, Adam Threlfall, Ylenia Giarratano, Charlene Hamid, Stuart King, Ian J.C. MacCormick, and Thomas J. MacGillivray. ``\textit{SLOctolyzer: Fully automatic analysis toolkit for segmentation and feature extracting in scanning laser ophthalmoscopy images.}'' Translational Vision Science \& Technology 13, no. 11 (2024): 7-7.

                \item \underline{\textbf{Burke, Jamie}}, \underline{Samuel Gibbon}, Audrey Low, Charlene Hamid, Megan Reid‐Schachter, Graciela Muniz‐Terrera, Craig W. Ritchie et al. ``\textit{Association between choroidal microvasculature in the eye and Alzheimer's disease risk in cognitively healthy mid‐life adults: A pilot study.}'' Alzheimer's \& Dementia: Diagnosis, Assessment \& Disease Monitoring 17, no. 1 (2025): e70075.

            \end{itemize}
                
            Below lists publication outputs which are either in peer-review, or prepared but currently not submitted by December 2024. Author of thesis in bold type, with shared lead authorship in underline type.
            
            \begin{itemize}
            \setlength\itemsep{0.1em}
                \item \underline{\textbf{Burke, Jamie}}, Ashley Akbari, Rowena Bailey, Kevin Fasusi, Ronan A. Lyons, Jonathan Pearson, James Rafferty, and Daniel Schofield. ``\textit{Representing multimorbid disease progressions using directed hypergraphs.}'' medRxiv (2023): 2023-08. Submitted and in peer-review at IEEE Journal of Biomedical and Health Informatics as of February 2025.
            
                \item \underline{\textbf{Burke, Jamie}}, Justin Engelmann, Charlene Hamid, Diana Moukaddem, Dan Pugh, Neeraj Dhaun, Amos Storkey, Niall Strang, Stuart King, Thomas J. MacGillivray, Miguel O. Bernabeu, and Ian J.C. MacCormick. ``\textit{Domain-specific augmentations with resolution agnostic self-attention mechanism improves choroid segmentation in optical coherence tomography images.}'' arXiv preprint arXiv:2405.14453 (2024). Manuscript prepared but currently not submitted as of February 2025.
    
                \item \underline{\textbf{Burke, Jamie}}, Justin Engelmann, Charlene Hamid, Diana Moukaddem, Dan Pugh, Niall Strang, Neerah Dhaun, Thomas J. MacGillivray, Stuart King, Ian J.C. MacCormick ``\textit{OCTolyzer: Fully automatic toolkit for segmentation and feature extraction in optical coherence tomography and scanning laser ophthalmoscopy data.}'' arXiv preprint arXiv:2407.14128 (2024). Submitted and in peer-review at Elsevier Computers in Biology and Medicine as of February 2025.

                \item \underline{George Cooper}, \underline{\textbf{Burke, Jamie}}, Charlene Hamid, Emily Godden, Neeraj Dhaun, Stuart King, Tom MacGillivray, J. Kenneth Baillie, David M. Griffith and Ian J.C. MacCormick. ``\textit{D-RISC II: Direct Retinal Imaging for Shock Resuscitation in Critical Ill Adults II.}'' Submitted and in peer-review at BMC Critical Care as of February 2025. 
                
            \end{itemize}
        
        \end{mysubsection}
        \vfill
        \begin{mysubsection}[]{Technical portfolio} 
        
        \begin{itemize}
        \setlength\itemsep{0.1em}
        
            \item Primary owner / maintainer:
            
             \begin{itemize}
             \setlength\itemsep{0.1em}
             
                \item Semi-automatic choroid region segmentation using Gaussian process regression, \newline \verb|GPET|: \url{https://github.com/jaburke166/gaussian_process_edge_trace}
                \item Semi-automatic choroid vessel segmentation using multi-scale median cut quantisation, \newline \verb|MMCQ|: \url{https://github.com/jaburke166/mmcq}
                \item Fully automatic choroid region segmentation in optical coherence tomography data, \newline \verb|DeepGPET|: \url{https://github.com/jaburke166/deepgpet}
                \item Fully automatic pipeline for segmentation and feature extraction in scanning laser ophthalmoscopy images, \newline \verb|SLOctolyzer|: \url{https://github.com/jaburke166/SLOctolyzer}
                \item Fully automatic pipeline for segmentation and feature extraction in optical coherence tomography and scanning laser ophthalmoscopy data, \newline \verb|OCTolyzer|: \url{https://github.com/jaburke166/OCTolyzer}
                \end{itemize}
                
            \item Contributed significantly to:
            
            \begin{itemize}
            
                \item Fully automatic, end-to-end choroid region and vessel analysis in optical coherence tomography data, \newline \verb|Choroidalyzer|: \url{https://github.com/justinengelmann/choroidalyzer}
                \item Directed hypergraphs for longitudinal multimorbidity data, \newline \verb|Hypergraphs-mm|: \url{https://github.com/nhsx/hypergraph-mm}
                
            \end{itemize}
            
        \end{itemize}
        
        \end{mysubsection}
        \vfill
    
    \end{mysection}

    \begin{mysection}[]{The choroidal microvasculature}\label{sec:INTRO_choroid}

            The choroid is a highly vascularised structure of the eye located between the retina and the sclera \cite{nickla2010multifunctional}, and plays a crucial role in the physiology of the eye. It is one of the most vascularised tissues in the body \cite{reiner2018neural}, citing the highest blood flow per unit tissue weight over any other \cite{nickla2010multifunctional}. 
            
            The uveal circulation branches from the ophthalmic artery and enters the eye external to the optic nerve, receiving 95\% of the blood flow directed to the eye \cite{mrejen2013optical}, splitting into anterior and posterior segments. The anterior segment reaches the iris and ciliary body, while the posterior segment, or posterior uveal tract, supplies the choroid and accounts for more than 70\% of the entire uveal blood flow \cite{parver1980choroidal}.

            Embryologically, the eye first forms as an outpocketing of the forebrain, called the optic vesicle \cite{purves2001neuroscience}. This undergoes invagination to form the retina and the \acrshort{RPE}, each of which have their own dedicated circulations \cite{mrejen2013optical}, the latter being supplied by the choroid eventually. The choroid begins to form around six weeks of gestation as the choriocapillaris \cite{mrejen2013optical}, located posterior to the RPE. By eight weeks, the medial and lateral short posterior ciliary arteries (\acrshort{PCA}s), which stem from the posterior uveal tract \cite{hayreh2023uveal}, enter the choroid and branch into smaller arterioles toward the choriocapillaris \cite{mrejen2013optical}. The choriocapillaris are drained exclusively by vortex veins, in the absence of venous equivalents to the posterior ciliary arteries \cite{ruskell1997peripapillary}. 
            
            Anatomically, the choroid is anchored posterior to the retina facilitating subretinal fluid exchange between the \acrshort{RPE} and choriocapillaris and is connected laterally to the optic nerve. The choroid is attached anteriorly to the sclera by loose connective tissue which creates a potential space, the suprachoroid \cite{mrejen2013optical}. This potential space is traversed by vessels and nerves, and ranges in thickness from 10 to 35 microns \cite{saidkasimova_suprachoroidal_2021}. However, this is often not visible using current imaging methods, with a thin suprachoroid reported to be visible in up to 44\% of healthy people (74 eyes) aged 55 -- 85 in \acrshort{EDI-OCT} \cite{yiu2014characterization} and a similar proportion (from 200 eyes) in \acrshort{SS-OCT} \cite{chandrasekera2018posterior}. The inconsistency in the visibility of the suprachoroid in \acrshort{OCT} when measuring the choroid can lead to poor agreement \cite{yiu2014characterization, chandrasekera2018posterior}, and so for the purposes of standardisation we consider the suprachoroid to be a separate entity to the choroidal vascular layer.
            
            The choroid, bound between the suprachoroid and Bruch's membrane, is structurally organised into three vascular layers commonly based on vessel size: the outer layer, known as Haller's layer, consists of large blood vessels which are sourced from the posterior uveal tract. These larger vessels branch into smaller vessels in the middle segment, known as Sattler's layer which then form capillary networks at Bruch's membrane, known as the choriocapillaris \cite{mrejen2013optical}. 
            
            The choriocapillaris is a dense, anastomosed network of fenestrated vessels organised in a honeycomb-like, lobular pattern \cite{lutty2010development}, forming a thin sheet at Bruch's membrane and is critical for the exchange of nutrients and waste with the \acrshort{RPE} and outer retinal layers. The choriocapillaris layer has the highest capillary density in the body \cite{lejoyeux2022choriocapillaris} and is most dense and thickest under the fovea, with an approximate depth of 10$\mu$m \cite{nickla2010multifunctional}. Individual capillaries are densely packed together with broad and flat lumens of approximately 20 -- 50 $\mu$m in diameter \cite{lutty2010development}. This means the choroidal capillaries have low vascular resistance, thus facilitating high choroidal blood flow \cite{delaey2000regulatory}, to enable sufficient exchange with the \acrshort{RPE} \cite{tan2024techniques}. 
            
            The choroid is the major source of oxygen to the photoreceptors, acting as the only source for the fovea --- where inner retinal vessels are absent \cite{nickla2010multifunctional}. The nature of choroidal blood flow is exceptional, with a blood flow rate approximately 20 to 30 times greater than that of the retina \cite{alm1973ocular}. This immense flow rate is more than sufficient --- but not necessary --- to meet the energy demands of the photoreceptors, which are heavily reliant on the choroid for oxygen and nutrient supply at all times \cite{reiner2018neural}. Any significant reduction in choroidal blood flow can lead to rapid photoreceptor degeneration, contributing to conditions like Age-Related Macular Degeneration (\acrshort{AMD}) \cite{nickla2010multifunctional}.
            
            Moreover, The structural and functional integrity of the retinal and choroidal vasculatures is essential for normal retinal function \cite{schmetterer2012ocular}, and deficits in choroidal perfusion cannot be maintained by the retina and vice versa \cite{schmetterer2012ocular}. Unlike the highly permeable and freely anastomosing choriocapillaris, the choroid has a segmental blood distribution network \cite{cheung2021watersheds}, delivering oxygen to localised, avascular zones of the retina \cite{hayreh1975segmental}. This leaves strips known as watershed zones between the various \acrshort{PCA}s, the short \acrshort{PCA}s, the choroidal arteries, the arterioles, and the vortex veins \cite{hayreh1990vivo}. This can lead to localised areas of filling defects and lower vascular perfusion which is thought to contribute to \acrshort{AMD} or choroidal neovascularisation \cite{cheung2021watersheds}. 

            Relative to the choroidal circulation, the retinal circulation is characterised by low flow and high oxygen extraction (approximate arteriovenous difference in partial pressure) of approximately 40\% \cite{alm1973ocular}. Conversely, choroidal blood flow has markedly higher flow rate with relatively low oxygen extraction of 3 -- 4\% \cite{delaey2000regulatory}. The high blood flow is thus more than sufficient but not necessary to nourish the high metabolic demands of the photoreceptors \cite{delaey2000regulatory}, which have the highest rate of oxygen use per unit tissue of any tissue in the body \cite{wangsa2003retinal}. This suggests that alongside support for the photoreceptor metabolic activity and continuous outer segment renewal \cite{reiner2018neural}, the excessive flow characteristics of the choroidal vasculature have other functions \cite{delaey2000regulatory}, including potentially thermoregulation to cool the retina \cite{parver1980choroidal} and compensation for lens focusing during accommodation \cite{nickla2010multifunctional}. However, some believe the high blood flow is necessary to promote diffusion of oxygen and nutrients through the outer blood-retina barrier \cite{mrejen2013optical, reiner2018neural}.
            
            The retinal vasculature is thought to have efficient autoregulatory functions \cite{schmetterer2012ocular}, an intrinsic property of many vascular tissues in the body which maintains perfusion pressure (constant blood flow) despite changes in systemic blood pressure and/or intracranial or intraocular pressure \cite{harris1998regulation}. For the retina, blood flow is maintained across fluctuations in ocular perfusion pressure (the difference between mean arterial pressure and intra-ocular pressure) through various myogenic and metabolic functions \cite{schmetterer2012ocular}. This compensation constitutes to changes in retinal vessel calibre (vasodilation/vasoconstriction), which maintains the oxygen partial pressure at a low and normal physiological level \cite{harris1998regulation}.

            Interestingly, choroidal blood flow is not entirely autoregulated using traditional myogenic responses like the retina \cite{nickla2010multifunctional}. Instead, blood pressure management of choroidal vessels and response to retinal activity in the choriocapillaris are directly and indirectly under control by intrinsic choroidal neurons, respectively \cite{reiner2018neural}. The choroid is well endowed with several choroidal nerve fibers (neurons), mostly situated at the central retina \cite{nickla2010multifunctional}. These neurons innervate (regulate) choroidal blood flow through vasodilatory (from parasympathetic and sensory nerve fibers) and vasoconstrictive influence (from sympathetic nerve fibers) \cite{reiner2018neural}. Although, compensation for changes in ocular perfusion pressure is considered to be influenced by local, autoregulatory myogenic mechanisms \cite{mrejen2013optical}.
        
            \begin{mysubsection}[]{Systemic disease} 

                Examination of the choroid has become increasingly popular with the technological advancements in non-invasive \acrshort{OCT} imaging devices. Conventional \acrshort{SD-OCT} led to the primary focus of \acrshort{OCT} research on the cross-sectional retinal layers, with little attention paid to the choroid due to limited visibility. However, the advent of \acrshort{EDI-OCT} \cite{spaide2008enhanced} and \acrshort{SS-OCT} \cite{he1999synthesized} have significantly improved imaging of choroidal structures. Improved, non-invasive imaging of the choroid has in turn contributed to the nascent field of `oculomics' \cite{wagner2020insights}, where the role of the choroid as a potential biomarker in numerous systemic diseases has been investigated \cite{tan2016state}.

                The choroid's poor autoregulation results in choroidal fluctuations corresponding to changes in intra-ocular pressure, which itself fluctuates in response to systemic blood pressure due to diurnal variation \cite{iwase2015diurnal}. The extent of diurnal variation in the choroid has been investigated previously \cite{chakraborty_diurnal_2011, tan2012diurnal, usui_circadian_2012, kinoshita_diurnal_2017, singh2019diurnal, ostrin_imi-dynamic_2023}, reporting choroidal fluctuations of approximately 30 $\mu$m over the course of the day, with the majority of research studies reporting larger choroid thickness measurements in the morning, compared to the afternoon and evening \cite{iwase2015diurnal}. 

                Additionally, the choroid's poor autoregulation makes it more susceptible to systemic changes in blood pressure. This has helped motivate investigations into conditions affecting systemic blood pressure, such as hypertension \cite{papathanasiou2022choroidal} and chronic kidney disease \cite{balmforth2016chorioretinal}. Additionally, the eye and the kidney show a striking resemblance in anatomy, physiology and response to disease \cite{wong2014kidney}, and in particular the choroidal microcirculation seemingly reflects the renal microcirculation \cite{farrah2020eye}. The endothelial cells of the choriocapillaris and renal cortex have similarly sized fenestrated vessel walls allowing subretinal fluid exchange in the choroid and blood filtration in the kidneys \cite{farrah2020eye}. Moreover, the circulatory systems in the choroid and renal cortex have similar proportions of blood flow in contrast to their retinal and renal medullary counterparts \cite{nickla2010multifunctional}, with choroidal blood flow four-fold higher per unit mass than the kidney \cite{farrah2020eye}.

                Consequently, there has been interest and recent evidence of the choroid reflecting renal function. Choroidal changes have been linked to chronic kidney disease (\acrshort{CKD}), with patients often exhibiting choroidal thinning correlating with severity of kidney dysfunction \cite{farrah2023choroidal, choi2020strong}. This relationship highlights the potential of choroidal imaging as a non-invasive method to monitor microvascular injury in \acrshort{CKD}, as shown by Balmforth, et al. \cite{balmforth2016chorioretinal}, who suggested that an overactive sympathetic nervous system and systemic endothelial dysfunction were potential causes to choroidal thinning in pre-dialysis \acrshort{CKD}. Additionally, in patients undergoing haemodialysis, the choroid can show significant fluctuations in thickness \cite{cho2019evaluation, shoshtari2021impact}, possibly due to changes in fluid balance \cite{canaud2019fluid}, offering insights into the impact of dialysis on systemic and ocular health.

                Emerging evidence suggests potential links between the eye and neurodegeneration, such as in Alzheimer's disease \cite{ashraf2023retinal}, Parkinson's disease \cite{wagner_retinal_2023}, multiple sclerosis \cite{suh2024oculomics} and cerebral small vessel disease \cite{kwa2002retinal}. For instance, Alzheimer's disease patients or those with mild cognitive impairment often exhibit thinning of the choroid \cite{ma2022longitudinal, gharbiya2014choroidal}, which could be linked to hypoperfusion and atrophy. However, the research community acknowledges the need for more methods to measure the choroid \cite{xie2021evaluation, suh2024oculomics}, as paradoxical results have also been found in similar research contexts \cite{asanad2019retinal, robbins_choroidal_2021}. Although, the choroid's response in the early prodromal stages of Alzheimer's disease has not yet been elucidated, and this will be discussed in section \ref{sec:ch_app_sec_prevent}. 

                Relative to invasive imaging of the cerebrovascular network \cite{wang2024insight}, which is limited by resolution \cite{cabrera2019identification}, non-invasive \acrshort{OCT} imaging of the intra-ocular structures enables consistent and reproducible sampling of a microvascular site which is part of the central nervous system and may mirror cerebral hypoperfusion and microvascular dysfunction \cite{courtie_retinal_2020}. This has contributed to retinal biomarker investigation in systemic inflammatory conditions like septic shock \cite{courtie_retinal_2020, courtie2022optical}, which affects microvascular flow and permeability. Recent pre-clinical evidence \cite{park2021visualization} in a septic rat model reported reduced choroidal blood flow during shock, and other related studies have demonstrated feasibility of retinal imaging and measurement in \acrshort{OCT} and \acrshort{OCT}-Angiography (\acrshort{OCT-A}) in a critical care environment \cite{liu_optical_2019, courtie2021stability}.
                
                The choroid's utility in systemic disease will continue to grow with advancements in the technology and accessibility of \acrshort{OCT} \cite{chopra_optical_2021, song_review_2021}. Ultimately, with sufficient imaging technology and standardised analysis tools, the choroid has the potential to provide a valuable window into systemic microvascular health and the health of vital organs like the kidney and the brain. 
            
            \end{mysubsection}
        
    \end{mysection}	
        
    \begin{mysection}[]{Optical coherence tomography}

        \begin{figure}[tb]
            \centering
            \includegraphics[width=\linewidth]{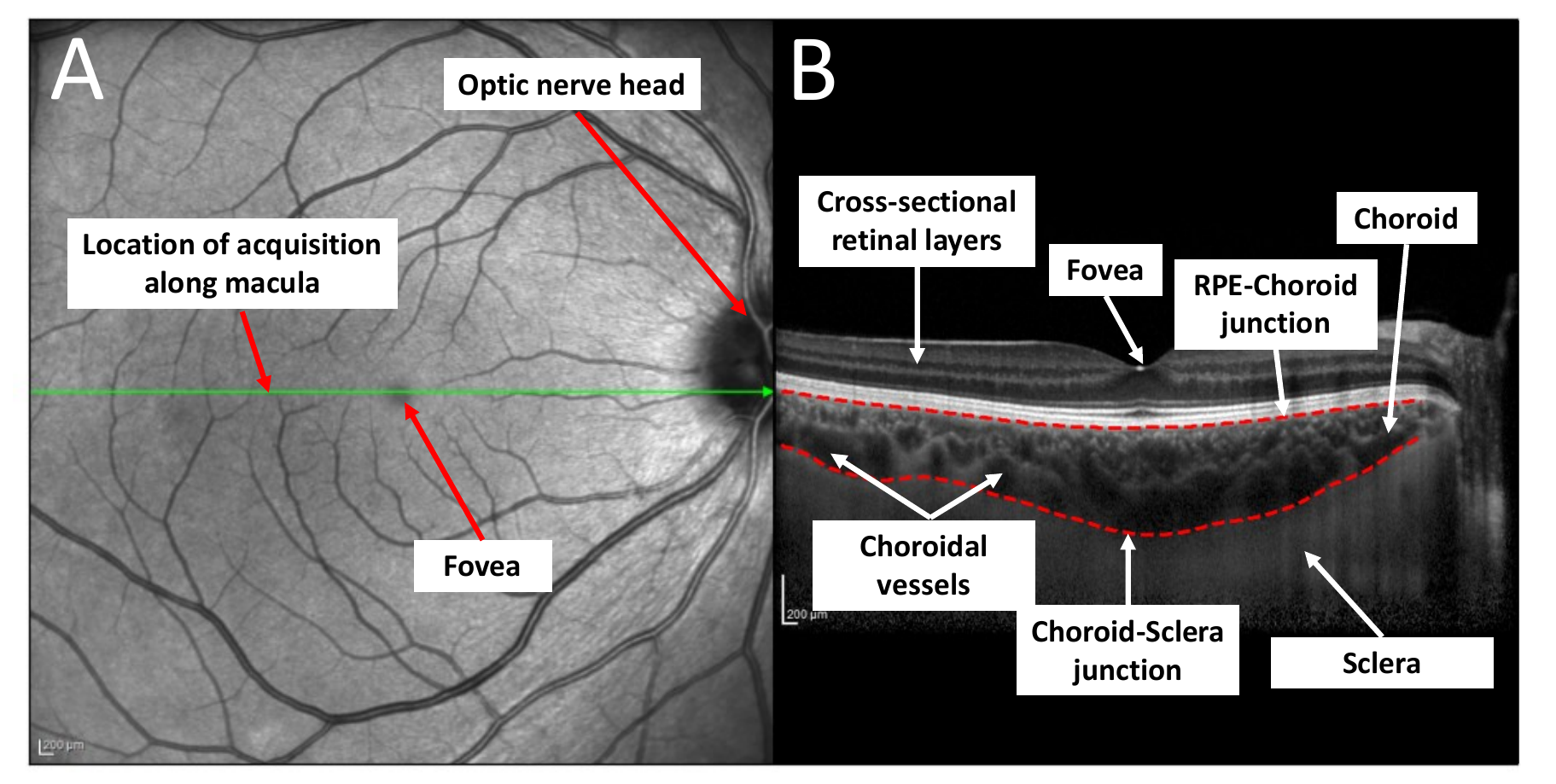}
            \caption[Exemplar, annotated \acrshort{SLO} and \acrshort{OCT} scan.]{Typical horizontal-line, OCT scan from a Heidelberg Engineering OCT imaging device. En face localiser SLO image (A) and corresponding OCT B-scan (B), with annotations overlaid of major anatomical landmarks.}
            \label{fig:INTRO_annotated_OCT_SLO}
        \end{figure}

        Previous methods for imaging/measuring the choroid, which sits posterior to the hyperreflective RPE, have used reflectance, sound and fluorescence. Light reflectance based on laser Doppler flowmetry \cite{riva1994choroidal} suffers from it's limited representation, through only measuring localised choroidal blood flow at the choriocapillaris, typically underneath the fovea \cite{mrejen2013optical}, which is not representative across the posterior pole \cite{hayreh1975segmental, kim2021choroidal}. Ultrasonography use sound waves and vibrations to detect ocular tissue based on sound reflectance \cite{hewick2004comparison}, but typically have very low resolution and poor localisation which can result in less precise choroidal measurements \cite{mrejen2013optical}. 
        
        Indocyanine green angiography is more accurate for en face choroidal perfusion measurements in the early angiographic sequence (dye fill time) \cite{flower1973clinical}, fluorescing over reasonably long wavelengths which can penetrate the choroid (790 -- 805 nm) compared to fluorescein angiography \cite{mrejen2013optical}. However, this method is limited by it's invasive methodology \cite{tan2024techniques}, light absorption and scattering by the pigment in the \acrshort{RPE} and choroid, and blood flow in the choroid \cite{mrejen2013optical}. Other methods have been considered such as ultrasonic colour Doppler imaging (combining ultrasonography with Doppler flowmetry) \cite{stalmans2011use} and laser interferometry (measuring perfusion based on the distance between the cornea and retina during the cardiac cycle) \cite{schmetterer2000comparison}. However, these methods also face the aforementioned challenges around localisation and reproducibility \cite{tan2024techniques}.

        It has been established that optical imaging techniques help improve the resolution of imaging the choroid \cite{mrejen2013optical}, and optical coherence tomography, or \acrshort{OCT}, has become the gold standard for clinical evaluation and imaging of the chorioretinal structures \cite{tan2024techniques, fujimoto2001optical}. \acrshort{OCT} was first introduced in 1991 \cite{huang1991optical} and uses low coherence interferometry to construct a three-dimensional map of the intra-ocular structures at the posterior pole. 
        
        The general principle is that a tightly focused light beam is scanned across the posterior pole and measurements on back-scattered light which reflects from ocular tissue at various depths can then be processed. Interferometry leverages the phase difference between different light beams to infer reflectivity of the ocular tissue, otherwise known as \textit{interference}. This is done first in the axial direction (A-scan) at a single lateral position, and iterated over in the transverse direction to create a two-dimensional cross-sectional image of the chorioretinal structures (B-scan). Doing this across the macula along the frontal plane (the final perpendicular axis to the transverse and axial direction) will enable imaging of three-dimensional intra-ocular structures \cite{fujimoto2001optical}.

        Figure \ref{fig:INTRO_annotated_OCT_SLO} shows a standard \acrshort{OCT} B-scan capture in panel (B), with the corresponding \textit{localiser} en face image in panel (A). Landmarks are annotated on both images, and in particular the green horizontal line in panel (A) corresponds to the cross-sectional location of the B-scan on the macula. The localiser image is an en face scan of the retinal vasculature which can vary depending on the \acrshort{OCT} system. For most datasets used in this thesis, we adopt the Heidelberg Engineering \acrshort{OCT} system which leverages confocal imaging methods to capture an infra-red scanning laser ophthalmoscopy (\acrshort{SLO}) image. The primary purpose of this image is to fixate the \acrshort{OCT} beam at the correct location in the posterior pole (the green horizontal line in figure \ref{fig:INTRO_annotated_OCT_SLO}(A)).

        \begin{mysubsection}[]{Time-domain OCT}

            \begin{figure}[tb]
                \centering
                \includegraphics[width=\linewidth]{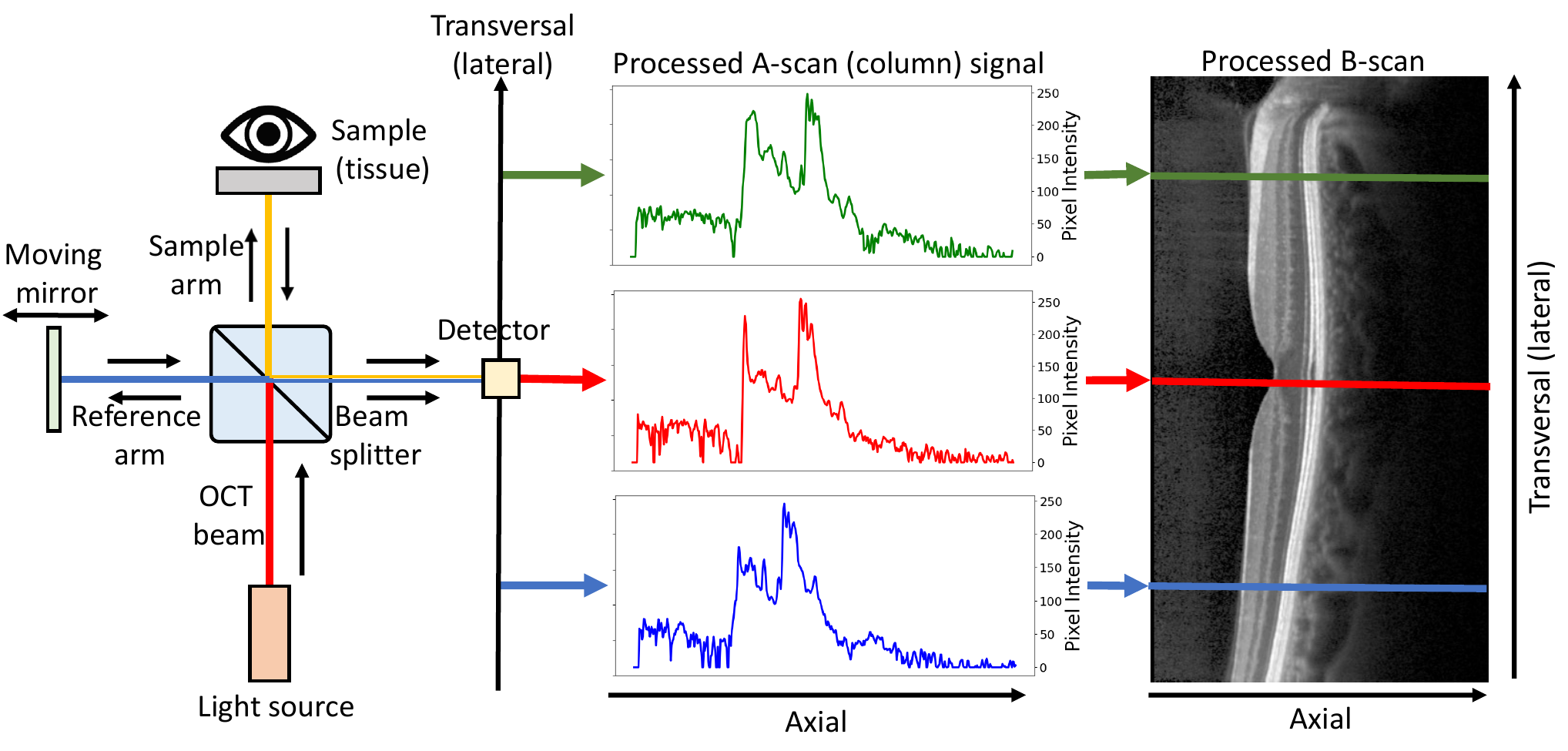}
                \caption[Simple schematic of an \acrshort{OCT} imaging system.]{Schematic of how \acrshort{TD-OCT} works to generate an \acrshort{OCT} B-scan. Note that this B-scan is from an \acrshort{SD-OCT} device and is purely demonstrative of how the system works in general.}
                \label{fig:INTRO_OCT_diagram}
            \end{figure}
        
            Figure \ref{fig:INTRO_OCT_diagram} shows a schematic of the conceptually simplest approach to \acrshort{OCT}, time-domain \acrshort{OCT} (\acrshort{TD-OCT}). \acrshort{TD-OCT} works using low coherence light, i.e. light waves which have a short coherence length and can only interfere with each other over very short distances \cite{STIFTER2010324}. This interference is leveraged in \acrshort{OCT} to create precise depth measurement with high axial resolution. 
            
            In the \acrshort{OCT} setup in figure \ref{fig:INTRO_OCT_diagram}, a single light source (superluminescent diode) emits a beam of broadband light with a short coherence length which is split into identical beams along two separate arms using a beam splitter. One beam penetrates the ocular tissue along the sample arm and back-scatters features of depth and reflectivity to the beam splitter, while the other beam moves along the reference arm and reflects off a mirror which is moving in the axial direction.
            
            Because we use low coherence light, each position of the moving mirror generates a reference beam which interferes with one point in the sample beam to produce a signal. This location along the sample beam is equal to the optical length of the reference arm and corresponds to an axial depth in the ocular tissue. It is this optical path length and resulting interference between the reference and sample beams which creates the axial position and reflectance (pixel intensity) for a single pixel in an \acrshort{OCT} B-scan \cite{drexler2014optical}. 
            
            Varying the reference arm in length, i.e. the axial position of the moving mirror, has the effect of generating a reflectivity depth profile for a specific transverse (lateral) position. This is known as an A-scan, and represents a whole column (in the axial direction) of an \acrshort{OCT} B-scan. By scanning a reflectivity depth profile (A-scan) along the transverse axis for the same sample, a two-dimensional \acrshort{OCT} B-scan is generated by stitching the corresponding A-scans together. Finally, by repeating this process along the sagittal direction (vertically), three-dimensional \acrshort{OCT} volume data (sequential and parallel \acrshort{OCT} B-scans) is collected of the intra-ocular structures at the posterior pole.

            While \acrshort{TD-OCT} is conceptually simple and easy to manufacture \cite{mrejen2013optical}, it has some limitations. To create a good quality image requires significant sampling time as the \acrshort{OCT} beam goes through the full depth during the sampling of each axial location per lateral position. However, this is limited due to safety standards, thus lowering the overall signal-to-noise ratio (\acrshort{SNR}) \cite{mrejen2013optical}. The slower acquisition speed (400 A-scans/second) \cite{sezer2016choroid}, light absorption from the \acrshort{RPE} and limited axial resolution ($\approx$ 10 $\mu$m) \cite{gabriele2011optical} prevent sufficient choroid visualisation.
            
        \end{mysubsection}

        \begin{mysubsection}[]{Spectral-domain OCT}
        
            \begin{figure}[tb]
                \centering
                \includegraphics[width=\linewidth]{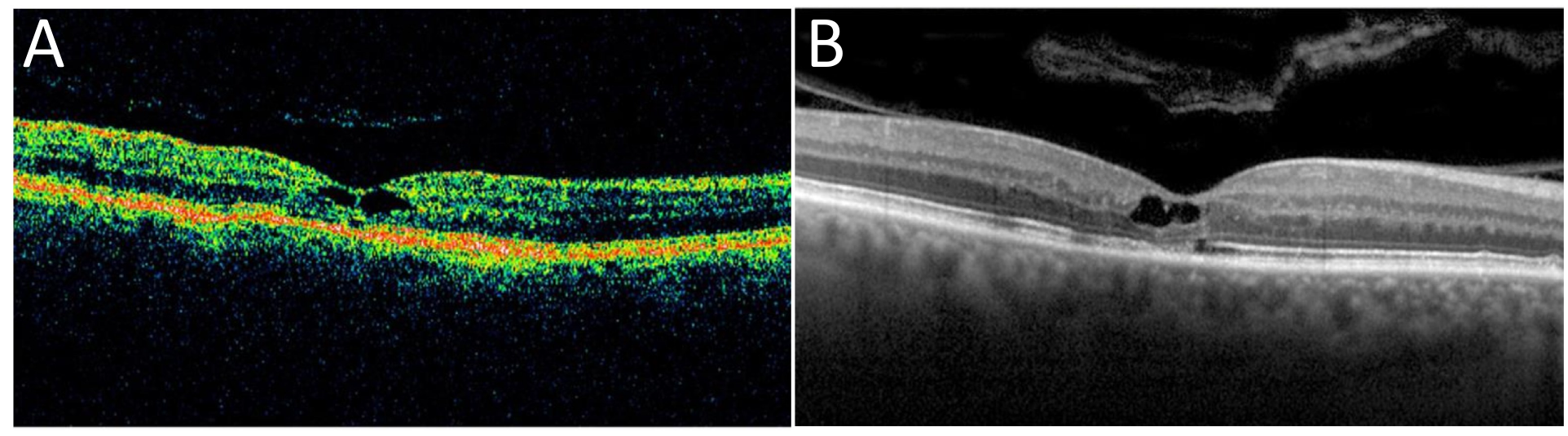}
                \caption[Comparison of \acrshort{TD-OCT} and conventional \acrshort{SD-OCT}.]{Comparison of \acrshort{TD-OCT} and \acrshort{SD-OCT} for a single \acrshort{OCT} B-scan of the same eye and approximately the same anatomical location. Reprinted with permission from Christopher Mody \cite{mody2016past}.}
                \label{fig:INTRO_TD_vs_SD_diagram}
            \end{figure}
            
            A significant advance in \acrshort{OCT} technology was removing the need for a moving reference arm and being able to collect reflectivity depth profiles (A-scans) in a single pass by working in the Fourier domain. This is called spectral domain \acrshort{OCT} (\acrshort{SD-OCT}) \cite{wojtkowski2002vivo}. Low coherence light is split along the reference and sample arms, but the reference arm is fixed in \acrshort{SD-OCT} image acquisition (the mirror in figure \ref{fig:INTRO_OCT_diagram} does not move anymore). For each A-scan, the reference and sample beams are combined at the beam-splitter to create an interference pattern, as is typical in \acrshort{TD-OCT}. However, in \acrshort{SD-OCT} this pattern is passed through a diffraction grating which disperses the signal into individual frequency components \cite{Paschottaoptical_coherence_tomography} (the detector in figure \ref{fig:INTRO_OCT_diagram} is now a spectrometer with a diffraction grating). While \acrshort{TD-OCT} encodes the interference pattern (and depth of ocular tissue) as a function of the reference arm distance, \acrshort{SD-OCT} encodes the interference pattern (and corresponding depth) as a function of frequency (or wavelength).
            
            This spectral data encodes depth and reflectivity information of the ocular tissue. In particular, each frequency (or wavelength) in the spectra corresponds to a particular depth of the ocular tissue back-scattered from the sample beam, and is associated with a specific strength (or power). A charge-coupled device (spectrometer) decodes this spectral data using a Fourier transform, resolving and reconstructing a depth reflectivity profile, which is the A-scan. Once A-scans have been collected along the transversal axis at multiple lateral positions, the B-scan can be generated to create a two-dimensional \acrshort{OCT} B-scan. Figure \ref{fig:INTRO_TD_vs_SD_diagram} shows the difference in image quality and signal between the same eye using \acrshort{TD-OCT} in panel (A) and using \acrshort{SD-OCT} in panel (B) at the same anatomical location.

            Advantageously, \acrshort{SD-OCT} generates an A-scan in a single pass, rather than requiring multiple sampling points like in \acrshort{TD-OCT}. Therefore, \acrshort{SD-OCT} is much more efficient and has a significantly higher scanning rate between 20,000 -- 52,000 A-scans/second with a higher axial resolution of around 5 -- 7 $\mu$m \cite{adhi2013optical}. However, while \acrshort{SD-OCT} utilise infra-red wavelengths up to 850nm for greater penetration of the retinal and subretinal space \cite{mrejen2013optical}, light scattering from the \acrshort{RPE} and the vascular nature of the choroid prevent penetration to the Choroid-Sclera boundary often \cite{spaide2008enhanced}. Thus, conventional \acrshort{SD-OCT} struggles to visualise the choroid and Choroid-Sclera boundary well, as shown in figure \ref{fig:INTRO_TD_vs_SD_diagram}(B).
            
            Additionally, \acrshort{SD-OCT} suffers from the sensitivity `roll-off' \cite{tan2024techniques}, where deeper ocular structures tend to have poorer \acrshort{SNR}. This sensitivity roll-off is in part due to the sampling of the spectrometer once it detects the interferometric signal, after it's been passed through the grating \cite{spaide2008enhanced}. The sampling is non-linear in the axial direction (as each A-scan forms), resulting in lower resolution (and signal) with greater depth. Additionally, increasing depth is associated with increasing frequency in the spectral data, and the limitations of the spectrometer prevents sampling of the highest frequency responses, resulting in a loss of measured signal intensity with greater depth \cite{spaide2008enhanced}. 
            
        \end{mysubsection}

        \begin{mysubsection}[]{Enhanced depth imaging OCT}\label{sec:INTRO_OCT_EDI}
        
            \begin{figure}[tb]
                \centering
                \includegraphics[width=\linewidth]{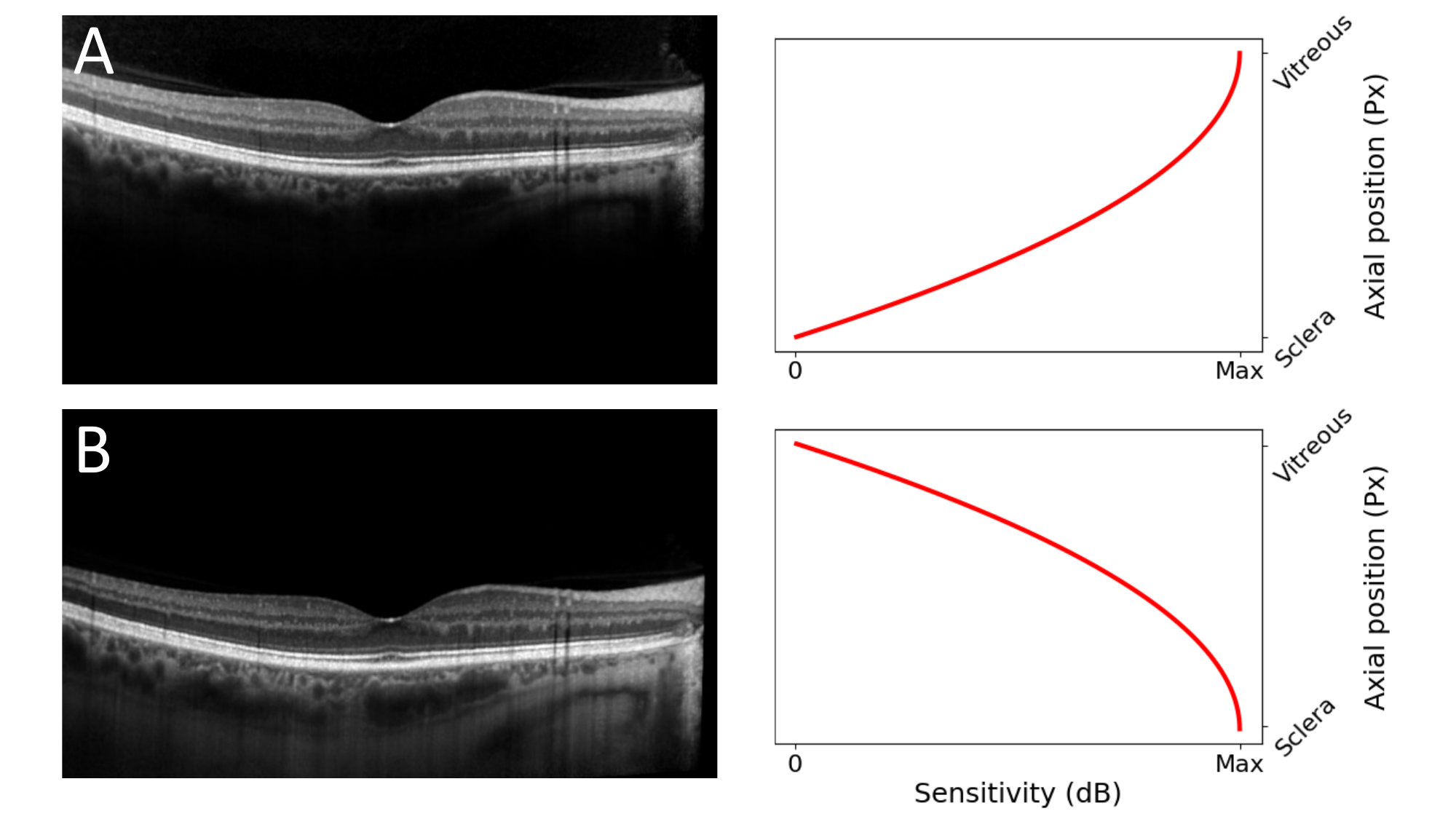}
                \caption[Comparison of conventional \acrshort{SD-OCT} and \acrshort{EDI-OCT}.]{Comparison of an \acrshort{SD-OCT} B-scan without \acrshort{EDI} mode (A) and with \acrshort{EDI} mode (B). Beside each B-scan plots the approximate signal sensitivity as a function of axial distance (from sclera to vitreous).}
                \label{fig:INTRO_nEDI_vs_EDI_diagram}
            \end{figure}

            \begin{figure}[tb]
                \centering
                \includegraphics[width=\linewidth]{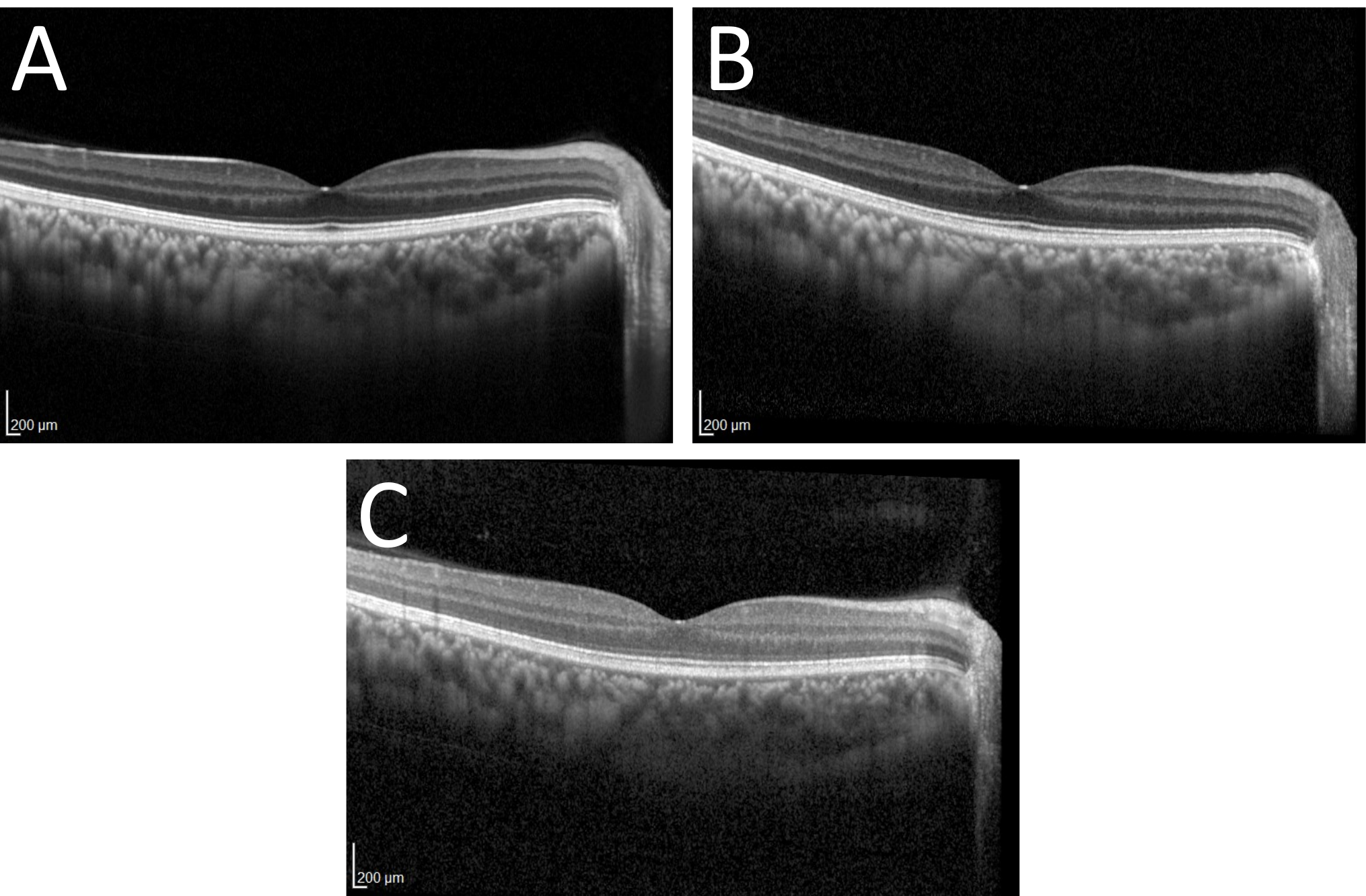}
                \caption[Comparison of B-scans with different \acrshort{ART} values with \acrshort{EDI} mode toggled on and off.]{Different \acrshort{OCT} B-scans of the same eye using the Heidelberg Engineering Spectralis OCT1 module. (A) \acrshort{EDI} mode with an \acrshort{ART} of 100. (B) \acrshort{EDI} mode with an \acrshort{ART} of 25. Conventional SD-\acrshort{OCT}, known as non-\acrshort{EDI} \acrshort{OCT} in this thesis, with an \acrshort{ART} of 9.}
                \label{fig:INTRO_ART_OCT}
            \end{figure}

            The axial position (ocular tissue depth) which has the highest sensitivity (or signal) is called the `zero-delay line', and corresponds to the ocular tissue depth which equals the length of the reference arm \cite{mrejen2013optical}. The sensitivity, and thus signal, decreases as a function of distance away from this depth, as these represent higher frequencies which are harder to sample. In conventional \acrshort{SD-OCT} the zero-delay line is at the vitreoretinal interface, which is sensible since the vitreous is almost transparent and requires high sensitivity for good imaging \cite{mrejen2013optical}. The trade-off is that this produces poor quality choroid visualisation, as exemplified in figure \ref{fig:INTRO_TD_vs_SD_diagram}(B).

            To mitigate for the sensitivity roll-off at the depth of the choroid, \acrshort{SD-OCT} can leverage the conjugate property of the Fourier transform and alter the position of this zero-delay line. These adaptations to conventional \acrshort{SD-OCT} give rise to enhanced depth imaging (\acrshort{EDI-OCT}) \cite{spaide2008enhanced}. 
            
            A consequence of taking the Fourier transform of the interferometric signal is that it gives rise to the real and conjugate A-scan, which is referred to as the real and inverted A-scan \cite{spaide2008enhanced} (or B-scan after stitching the A-scans along the transverse axis). In practice, the real B-scan is used with peak sensitivity at the vitreoretinal interface, which corresponds to a sensitivity profile which decreases with greater depth. If the camera head is positioned nearer to the eye, this shifts the zero-delay line toward the inner sclera. Here, the inverted B-scan has an inverted sensitivity profile to that of the real B-scan, such that signal strength \textit{increases} with greater depth (toward the zero-delay line, i.e. inner sclera) \cite{mrejen2013optical}. 
            
            This has the effect of enhancing deeper structures such as the choroid, as shown in figure \ref{fig:INTRO_nEDI_vs_EDI_diagram}. In panel (A), conventional \acrshort{SD-OCT} captured a B-scan (using the `real' A-scan signals) whose sensitivity profile is shown to the right, and decreases non-linearly with increasing depth through the eye. This, as well as the light scattering properties of the choroid, prevent sufficient visualisation. In panel (B) we show an \acrshort{EDI-OCT} B-scan. For this acquisition, the camera head is moved closer, and the `inverted' B-scan is used as the output, which shows an inverted sensitivity profile to the right. 

            Heidelberg Engineering imaging devices now have \acrshort{EDI-OCT} as an explicit option during image capture, by selecting the `\acrshort{EDI}' button. In addition to \acrshort{EDI} as an option, Heidelberg Engineering also have active eye tracking (TruTrack, Heidelberg Engineering, Heidelberg, Germany) of the pupil during image capture \cite{spaide2008enhanced}. This can allow for multiple (up to 100) B-scans to be captured during the same acquisition and averaged over to reduce speckle noise, which is called Automatic Real Time (\acrshort{ART}) averaging. Figure \ref{fig:INTRO_ART_OCT} shows an \acrshort{OCT} B-scan of the same eye with decreasing image quality, where \acrshort{EDI} mode with an \acrshort{ART} of 100 provides the most enhanced visualisation of the choroidal vessels and Choroid-Sclera boundary.
            
        \end{mysubsection}

        \begin{mysubsection}[]{Swept-source OCT}

            A further advancement in Fourier domain \acrshort{OCT} was the introduction of swept-source \acrshort{OCT} (\acrshort{SS-OCT}), which utilise longer wavelengths for greater depth penetration. Additionally, through the use of a wavelength-swept laser and photodetector hardware \cite{he1999synthesized}, \acrshort{SS-OCT} is able to mitigate for the sensitivity roll-off from which \acrshort{SD-OCT} suffers \cite{tan2024techniques}. In figure \ref{fig:INTRO_OCT_diagram}, the \acrshort{OCT} beam source emits a sweeping light source across a range of wavelengths and the detector is now a simple photodiode like in \acrshort{TD-OCT}, rather than a broadband light source and spectrometer as in \acrshort{SD-OCT}.

            In \acrshort{SS-OCT} image capture, a tunable laser is swept over the interference signal between the reference and sample arms across a range of wavelengths and in an ordinal fashion. Through the use of a simple photodiode, the laser's output has a much higher throughput, processing A-scan reflectivity depths much more efficiently \cite{mrejen2013optical} and without the need for a grating or spectrometer \cite{kishi2016impact}. This resulted in improved acquisition speed of up to 4x when it was introduced in 2012, relative to commercial \acrshort{SD-OCT} systems at the time \cite{teussink2019state}. 
            
            In today's commercial market, Topcon's commercial \acrshort{SS-OCT} DRI Triton reports a a superior A-scan rate of 100,000 A-scans/second \cite{topcon_triton} over Heidelberg Engineering's Spectralis OCT2 module's 85,000 A-scans/second \cite{heidelberg_oct2_module}. Nevertheless, some commercial \acrshort{SD-OCT} systems are still rivalling \acrshort{SS-OCT} systems, with Optopol's commercially available NX Revo-130 reporting scan rates of 130,000 A-scans/second \cite{optopol_sdoct}. However, it is thought that \acrshort{SS-OCT} systems have potential for enabling \acrshort{OCT} imaging speeds beyond 400,000 A-scans/second and even reaching speeds up to 1,000,000 A-scans/second \cite{drexler2014optical}. 
            
            While acquisition rate is important for overcoming barriers to patient focus and concentration, especially in the context of disease, it is not the only metric on which to evaluate \acrshort{OCT} systems. In the particular application of choroidal image analysis, deep, subretinal tissue penetration and signal strength are key. Interestingly, there exists a direct trade-off between acquisition speed and signal strength \cite{bille2019high}. Signal strength is positively associated with \textit{integration time} \cite{teussink2019state}, the length of time it takes to capture an A-scan, and is thus inversely associated with acquisition rate. However, modern systems can counteract higher acquisition rate (and thus low signal strength) using post-acquisition image processing methods such as B-scan averaging (\acrshort{ART}) \cite{spaide2008enhanced}.
            
            \acrshort{SS-OCT} can provide greater depth into the ocular tissue due to the longer coherence length of the wavelength-swept laser, relative to the shorter coherence length of the broadband light source of \acrshort{SD-OCT} systems. This has a direct impact on the sensitivity roll-off, which \acrshort{SD-OCT} suffers from more so than \acrshort{SS-OCT} \cite{mrejen2013optical}. Additionally, the use of a wavelength-swept laser enables a broader range of wavelengths up to 1050nm (with an approximate bandwidth of 100nm \cite{kishi2016impact}) which allow for both the vitreous and choroid to be imaged simultaneously, rather than having to prioritise one as in \acrshort{SD-OCT} and \acrshort{EDI-OCT} \cite{mrejen2013optical}. Ultimately, this allows for greater axial depth penetration in scattering tissues relative to shorter wavelength systems such as \acrshort{SD-OCT} \cite{teussink2019state}. 

            \begin{figure}
                \centering
                \includegraphics[width=0.8\linewidth]{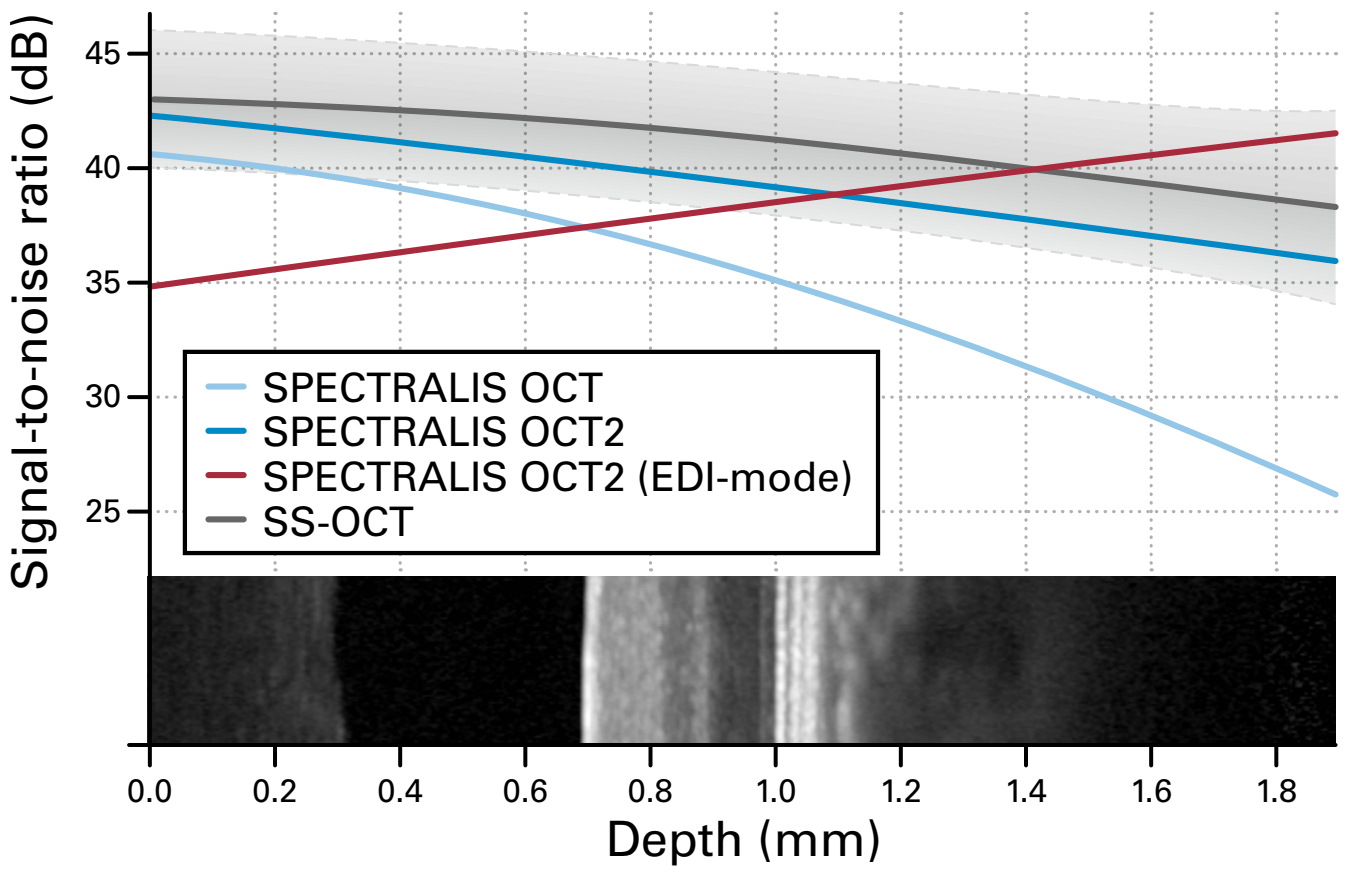}
                \caption[\acrshort{SNR} sensitivity roll-off curves for Spectralis \acrshort{SD-OCT} and hypothetical \acrshort{SS-OCT}.]{\acrshort{SNR} (sensitivity) roll-off curves (against ocular tissue depth) for \acrshort{SD-OCT} Spectralis \acrshort{OCT}, Spectralis \acrshort{OCT}2 and Spectralis \acrshort{OCT}2 with \acrshort{EDI}-mode, compared to a hypothetical \acrshort{SS-OCT} device. Reprinted from Teussink, et al. \cite{teussink2019state} with permission from Heidelberg Engineering GmbH.}
                \label{fig:INTRO_SD_SS_SNR}
            \end{figure}

            \begin{figure}
                \centering
                \includegraphics[width=0.8\linewidth]{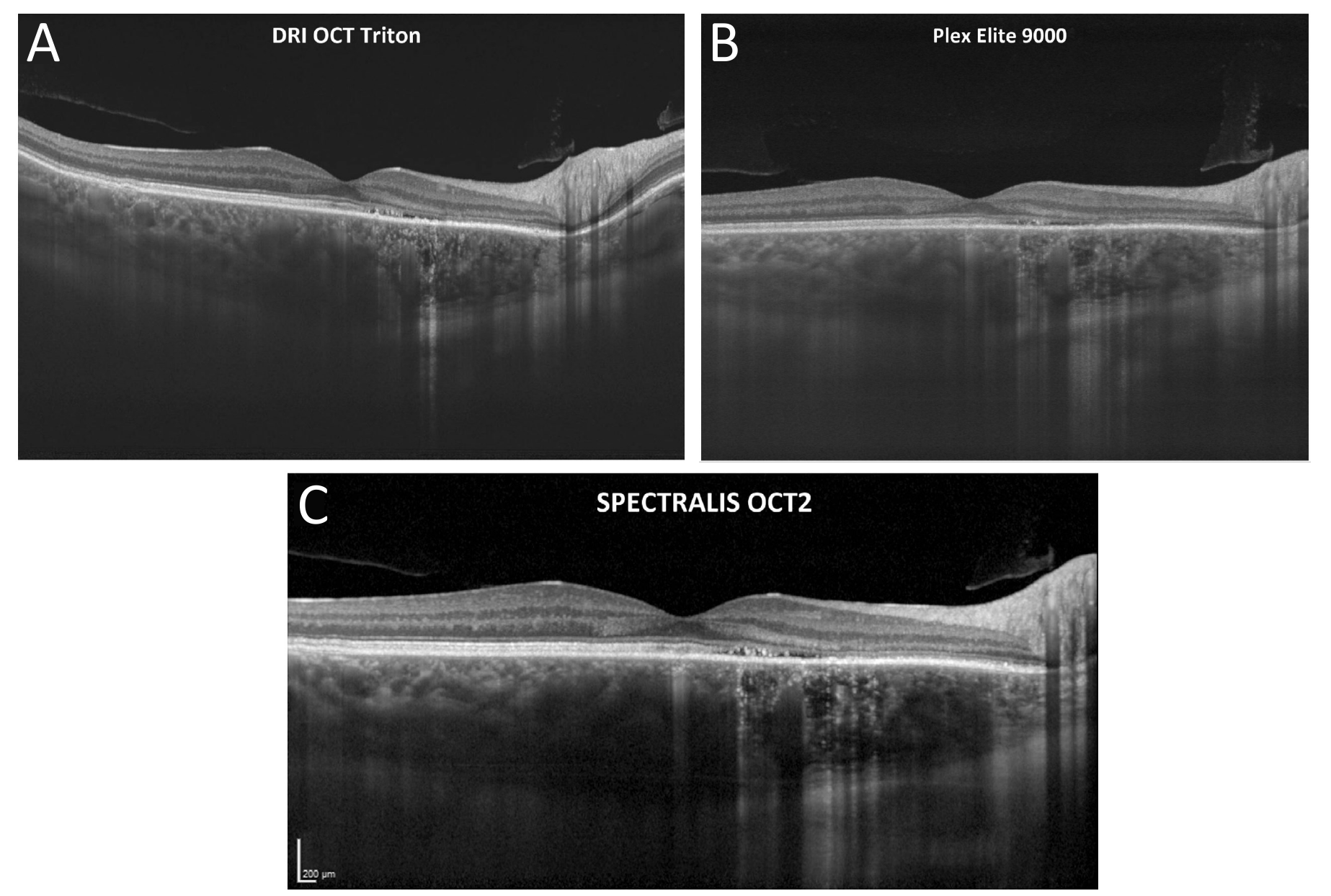}
                \caption[Comparison of \acrshort{SS-OCT} and \acrshort{EDI-OCT} from different \acrshort{OCT} devices.]{Comparison of visibility of the Choroid-Sclera boundary in a patient with central serous chorioretinopathy, acquired using the Topcon DRI Triton (A), Zeis Plex Elite 9000 (B) and Heidelberg Engineering \acrshort{OCT}2 Module with \acrshort{EDI} mode (C). Note that panels (B) and (C) use an \acrshort{ART} of 100. Reprinted from Teussink, et al. \cite{teussink2019state} with permission from Heidelberg Engineering GmbH.}
                \label{fig:INTRO_SD_SS_example}
            \end{figure}
            
            However, the longer coherence length of the swept-source laser negatively impacts axial resolution, which is partially compensated for by the higher bandwidth of 100nm (compared to \acrshort{SD-OCT} systems which usually have a bandwidth of 60nm around the central 850nm wavelength). Nevertheless, this leads to commercial \acrshort{SD-OCT} systems having better optical in-depth resolution (7 $\mu$m for Heidelberg Engineering's Spectralis standard \acrshort{OCT} module \cite{engeineering_hardware_instructions} compared with 8 $\mu$m for Topcon's DRI Triton \cite{topcon_triton}). Additionally, although longer wavelengths penetrate farther, they also suffer from greater water absorption which can lead to light attenuation \cite{teussink2019state}, as the beam has to penetrate through the vitreous which is primarily water \cite{mrejen2013optical}. This ultimately has an impact on the signal strength of \acrshort{SS-OCT} systems.
            
            Signal strength is often measured as the sensitivity of the \acrshort{OCT} system or \acrshort{SNR}. Teussink, et al. \cite{teussink2019state} state that ``\acrshort{SNR} is determined by the signal of a specific sample, how much light reaches the sample, how much light is reflected, and how much this light is attenuated before reaching the detector''. In their work, they estimate that longer wavelengths of 1050nm (with a bandwidth of 100nm) only lead to a factor increase in \acrshort{SNR} of approximately 1.7, relative to an 850nm wavelength \acrshort{SD-OCT} system (with a bandwidth of 60nm). This was observed even with a hypothetically higher input power (at the maximum safety limits) of the wavelength-swept laser source (meaning that more light would be reaching the sample) \cite{teussink2019state}.

            For choroidal image analysis in \acrshort{OCT} image sequences, maximal \acrshort{SNR} at greater depth is optimal. As the sensitivity roll-off is improved in \acrshort{SS-OCT} capture relative to conventional \acrshort{SD-OCT}, this system is more optimal for choroidal visualisation. However, for spectral domain \acrshort{EDI-OCT}, this sensitivity roll-off curve is inverted (figure \ref{fig:INTRO_nEDI_vs_EDI_diagram}). Thus, regardless of higher wavelength and lower sensitivity roll-off, \acrshort{EDI-OCT} still obtains higher \acrshort{SNR} values for deeper ocular structures \cite{teussink2019state}, such as the larger choroidal vessels nearer the Choroid-Sclera boundary. Figure \ref{fig:INTRO_SD_SS_SNR} shows the sensitivity roll-off curves for the Heidelberg Engineering Spectralis OCT1, OCT2 and \acrshort{EDI}-OCT2 systems, and comparison to a hypothetical \acrshort{SS-OCT} device. 

            Figure \ref{fig:INTRO_SD_SS_example} compares the visibility of the Choroid-Sclera boundary for a patient with an enlarged choroidal space due to pachychoroid disease (central serous chorioretinopathy). The same location was scanned using Topcon's \acrshort{SS-OCT} DRI Triton, Zeiss' \acrshort{SS-OCT} Plex Elite 9000 and Heidelberg Engineering's \acrshort{SD-OCT} Spectralis \acrshort{OCT}2 module with \acrshort{EDI} mode activated. Note that for Zeiss and Heidelberg Engineering, these B-scans were captured with an \acrshort{ART} of 100. While the Choroid-Sclera boundary is visible across all three devices, the \acrshort{SD-OCT} capture in panel (C) has greater contrast of the choroidal space.
                        
        \end{mysubsection}

    \end{mysection}

    \begin{mysection}[]{Segmenting the choroidal microvasculature} \label{subsec:intro_litreview}

         \begin{figure}[tb]
            \centering
            \includegraphics[width=\linewidth]{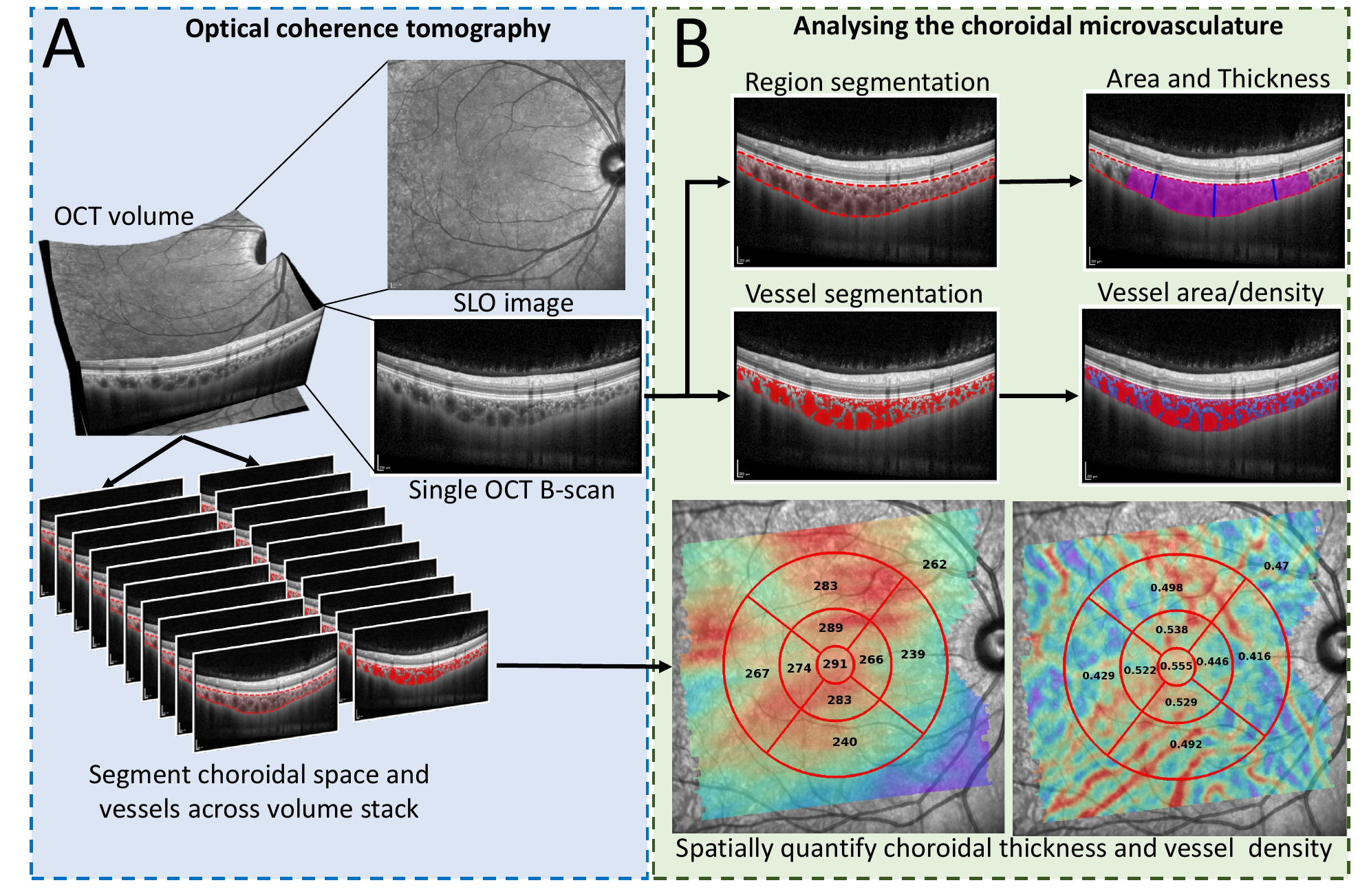}
            \caption[Choroid segmentation tasks and measurements of interest.]{\acrshort{EDI-OCT} volume scan acquisition (A), with segmentation tasks and measurements sought for choroidal image analysis (B).}
            \label{fig:INTRO_measure_choroid_diagram}
        \end{figure}

        Figure \ref{fig:INTRO_measure_choroid_diagram} shows a diagram with the output from a typical \acrshort{OCT} volume scan acquisition in panel (A), showing the \acrshort{OCT} volume block with the sequential B-scans and corresponding localiser \acrshort{SLO} image. In panel (B), we show the segmentation tasks necessary for choroidal measurement, region and vessel segmentation. We also show the typical measurements of interest that are commonly sought after in choroidal image analysis including, for a single B-scan, thickness and area of the choroidal space and choroidal vessel density. This can also be extrapolated for an \acrshort{OCT} volume scan to allow full choroidal mapping across the macula, measuring local choroid thickness and vessel density at different spatial locations of the macula.

        Measurement of the choroid in \acrshort{OCT} images require segmentation of the choroidal space and vessels, which is generally a harder problem than retinal layer segmentation. This is primarily due to the hyperreflective Bruch's complex which can influence the \acrshort{OCT} beam's optical signal as it penetrates deep to the choroid. This can result in poor Choroid-Sclera boundary visualisation. While \acrshort{EDI-OCT} and \acrshort{SS-OCT} improve choroid visualisation, natural speckle noise still plays a significant role in poor Choroid-Sclera boundary visualisation \cite{teussink2019state}. Furthermore, the Choroid-Sclera boundary is typically identified from locating the posterior-most vessels appearing as dark blobs suspended in brighter-appearing connective tissue \cite{sohrab2012pilot, branchini2013analysis}. However, this is not always the case and the homogeneous appearance of the sclera and choroidal interstitial space can often obscure the true boundary location.
            
        The choroidal vessels are highly heterogeneous in terms of size, shape, and edge contrast which make them hard to discern exactly. The choroid is also often plagued by low \acrshort{SNR} which corresponds to high speckle and low contrast between the interstitial and vascular compartments. A final challenge is the oblique presentation of the vessels from cross-sectional visualisations. The choroidal vascular network is a dense mesh of blood vessels which makes it challenging to visualise individual vessels at perfect, circular cross-sections. Instead, oblique viewpoints of the choroidal vessels are seen on each \acrshort{OCT} B-scan. Additionally, certain A-scans of the \acrshort{OCT} B-scan are corrupted by shadows cast by superficial retinal vessels which branch out of the optic nerve head lying along the inner retinal layers. During acquisition, the \acrshort{OCT} beam may penetrate through the retinal vasculature lying along the inner surface of the retina and sitting perpendicular to the incident laser light. This darkens the appearance of the interstitial and vascular compartments, which can often confuse automatic segmentation approaches \cite{mao2019deep}. 
            
        These reasons all contribute to the challenging tasks of choroid segmentation. While manual segmentation of the choroidal space and vessels is often done using Fiji (ImageJ) \cite{schindelin2012fiji} or on the imaging device, this is a complex and time-consuming task, and prohibitively so for choroid vessel segmentation. Moreover, manual choroid segmentation suffers from high intra- and inter-rater variability \cite{rahman2011repeatability, shao2013reproducibility, sim2013repeatability}. Ultimately, manual segmentation is subject to the expertise of the grader, which may vary depending on experience as different graders may have different definitions of the Choroid-Sclera boundary --- especially if the suprachoroidal space is present \cite{yiu2014characterization}. The subjective nature of manual segmentation prevents standardisation of choroidal measurements in the research community.

        \begin{mysubsection}[]{Region segmentation}\label{subsubsec:intro_litreview_region}

            \begin{mysubsubsection}[]{Manual measurement}

                The lack of open-source and easily accessible computational approaches to characterise the choroid in \acrshort{OCT} often lead to the research community leveraging manual methods \cite{cheong2018novel, yiu2014characterization, balmforth2016chorioretinal, rahman2011repeatability}, and this has also continued as a standard approach even in recent years \cite{ma2022longitudinal, farrah2023choroidal, aksoy2023choroidal, shoshtari2021impact, robbins_choroidal_2021}. Due to the time it would take a manual grader to segment the choroid manually across the macula, representative thickness measurements are taken, usually underneath the fovea, i.e. subfoveal choroid thickness (\acrshort{SFCT}). However, these often suffer from wide variability among graders \cite{rahman2011repeatability, shao2013reproducibility, sim2013repeatability} --- especially when detecting the Choroid-Sclera boundary --- due to poor visualisation \cite{boonarpha2015standardization} and also inconsistent definition \cite{chandrasekera2018posterior}. 
            
                Rahman, et al. \cite{rahman2011repeatability} showed that intra- and inter-observer reproducibility had an average difference of 23 $\mu$m and 32 $\mu$m, respectively, while Shao, et al. \cite{shao2013reproducibility} showed good agreement between graders of around 3 $\mu$m on average. Boonarpha, et al. \cite{boonarpha2015standardization} showed that the repeatability of manual choroid thickness measurements vary according to image quality, ranging from a mean difference between raters of 4 $\mu$m and 27 $\mu$m across the spectrum of Choroid-Sclera boundary visualisation. Sim, et al. \cite{sim2013repeatability} found that inter-rater variability in thickness from two graders improved from 6 -- 40 $\mu$m on average to 4 -- 21 $\mu$m after introducing a strict manual measurement protocol for individual Sattler, Haller and choroid layer segmentation.
                
                Yiu, et al. \cite{yiu2014characterization} found that varying the definition of the Choroid-Sclera boundary to include/exclude the suprachoroidal space lead to error in manual \acrshort{SFCT} ranging from 17 $\mu$m to 70 $\mu$m. Chandrasekera, et al. \cite{chandrasekera2018posterior} showed that agreement between automatic and manual detection of the Choroid-Sclera boundary worsened when the suprachoroidal space was visible (increase of mean difference in thickness of 17 $\mu$m to 23 $\mu$m). Conversely, Vuong, et al. \cite{vuong2016repeatability} found that only 15\% of 91 eyes had a visible suprachoroidal space in a population with average age 59.5, and that it's appearance did not impact manual measurement error, which ranged between 1 -- 37 $\mu$m on average when considering different Choroid-Sclera boundary definitions.
                
                Given that effect size can vary across disease, with some being as low as 20 -- 30 $\mu$m change (such as when tracking myopia progression \cite{breher2019metrological, flores2013relationship}), the wide variability in reproducibility from manual grading from poor measurement protocol and human bias \cite{boonarpha2015standardization, xie2021evaluation} poses a problem for standardised choroidal image analysis in \acrshort{OCT} image sequences. 
            
                Additionally, measuring choroidal thickness on \acrshort{OCT} has also been poorly standardised in the research community, as the choroid can appear skewed on an \acrshort{OCT} B-scan and some institutions account for this skew \cite{sezer2016choroid, boonarpha2015standardization, balmforth2016chorioretinal, chen2012topographic}, while others do not \cite{cheong2018novel, yiu2014characterization, aksoy2023choroidal, shoshtari2021impact, rahman2011repeatability, cho2014influence}. We will show in section \ref{subsec:INTRO_MEASURE_ROI} that accounting for this choroidal curvature can have an inordinate impact on measuring choroidal thickness because of the pixel-to-micron scaling imposed by the \acrshort{OCT} B-scan. We will also show in chapter \ref{chp:chapter-GPET} that manual graders are a poor reference standard for measuring choroid thickness consistently within patients when accounting for this choroidal curvature.
                
            \end{mysubsubsection}

            \begin{mysubsubsection}[]{Semi- and Fully-automatic}

                The inconsistencies introduced from manual grading has been a driving force for more objective approaches based on computer science and mathematics, with traditional image processing methods initially playing a significant role. These methods often required manual intervention or pre-defined parameters, making them semi-automatic and less efficient compared to more modern, fully automatic methods which use deep learning. Table \ref{tab:INTRO_region_methods} gives a brief overview of the range of semi-automatic and fully-automatic approaches developed for choroid region segmentation over the last 12 years. Performance metrics have explicitly not been included due to lack of publicly available datasets to standardise and benchmark performance with.

                \afterpage{%
                \begin{landscape}
                    \centering                    \scriptsize

{\begin{longtable}{p{1.5cm}p{1.5cm}p{2cm}p{5cm}p{3cm}p{3cm}P{1cm}p{1cm}}
\toprule
Study & Method & Data & Description & Advantages & Limitations & Reproducibilty & Open-source \\ \hline
\endfirsthead
\endhead
\midrule
\endfoot
\endlastfoot
Kajic, et al. \cite{kajic2012automated} & Semi-automatic & N=12 (871), HD-OCT**, Manual & Two-stage statistical model based on texture, shape and Dijkstra's algorithm \cite{dijkstra2022note}. & Reliable performance on low SNR B-scans. & Many tuning parameters requiring domain-knowledge. & No & No \\

Torzicky, et al. \cite{torzicky2012automated} & Semi-automatic & N=5 (5), PS-OCT$^+$, Manual & threshold technique based on the degree of polarisation uniformity image. & Leverages phase retardation to identify choroid. & Limited evaluation data, not generalisable outside PS-OCT. & Yes & No \\

Duan, et al. \cite{duan2012automated} & Semi-automatic & N=6 (384), PS-OCT$^+$, Manual & Gradient and regression lines on phase retardation tomography from SS-OCT & Leverages phase retardation to identify choroid. & Jagged boundary, requires error correction, not generalisable outside PS-OCT. & Yes & No \\

Alonso-Caneiro, et al. \cite{alonso2013automatic} & Semi-automatic & N=119 (1173), EDI-OCT, Manual* & Two-stage use of Dijkstra's algorithm \cite{dijkstra2022note} after several brightness-adjusted enhancement with gradient-based edge weights. & Large evaluation set. & Extensive and interdependent pre-processing, slow (45 seconds/B-scan). & No & No \\

Tian, et al. \cite{tian2013automatic} & Semi-automatic & N=45 (45), EDI-OCT, Manual & Gradient-based Dijkstra's algorithm \cite{dijkstra2022note}, posed as dynamic programming. & Fast (1.25 seconds/B-scan). & Non-smooth Choroid-Boundary segmentations from local optimisation. & No & No \\

Lu, et al. \cite{lu2013automated} & Semi-automatic & N=30 (30), EDI-OCT, Manual & Two-stage contour model and gradient-based graph search, posed as dynamic programming. & Easy-to-use. &  Human-interactive, small evaluation set. & No & No \\

Hu, et al. \cite{hu2013semiautomated} & Semi-automatic & N=30 (<1100), SD-OCT, Manual* & 3D graph-search method for OCT volumes. & Allows whole-volume segmentation. & Evaluated on poor quality data, tuned parameters specific to pathology. & No & No \\

Danesh, et al. \cite{danesh2014segmentation} & Semi-automatic & N=10 (100), EDI-OCT, Manual & Two-stage tracing using Gaussian mixture models with wavelet features and graph-cut. & Multiresolution pipeline obtaining coarse and fine details. & Requires manual labelling for training for inference, slow (13 -- 21s/B-scan). & No & No \\

Vupparaboina, et al. \cite{vupparaboina2015automated} & Semi-automatic & N=5 (485), EDI-OCT, Manual & Tracing using tensor voting, Hessian analysis and structural similarity index. & Promotes smoothness and connectivity. & Computationally expensive and thus slow. & Yes & No \\

Chen, et al. \cite{chen2015automated} & Semi-automatic & N=66 (212), HD-OCT**, Manual & Graph min-cut-max-flow and the energy minimisation. & Suitable for HD-OCT segmentation. & Sensitive pre-processing, threshold-based upper bound detection not robust, slow (9s/B-scan). & Yes & No \\

Masood, et al. \cite{masood2018automatic} & Semi-automatic & N=21 (525), EDI-OCT, Manual & Graph-based energy minimisation with normalised cuts. & More precise than manual methods, better than conventional image processing alone. & Requires manual parameter tuning, slow (40s/B-scan). & No & No \\

Salafani, et al. \cite{salafian2018automatic} & Semi-automatic & N=11 (32), EDI-OCT, Manual & Neutrosophic space Dijkstra's algorithm \cite{dijkstra2022note} extension for boundary detection. & Handles uncertainty well to improve over standard graph methods. & Sensitise pre-processing and post-manual correction. & No & No \\

Chen, et al. \cite{chen2018automated} & Semi-automatic & N=42 (978), EDI-OCT, Manual* & 3D choroid segmentation on SD-OCT using EDI-OCT and registration with graph methods. & Enables low-quality SD-OCT segmentation, slow (4.5min/volume). & Restricted to volume segmentation, dependent on high quality EDI-OCT. & No & No \\

Wang, et al. \cite{wang2017automatic} & Semi-automatic & N=30 (7680), SS-OCT, Manual & Combined Markov Random Fields (MRF) with level sets. & MRF and Regularisation to promote smoothness and noise robustness. & Multiple parameter tuning, questionable ground-truth comparison. & No & No \\

Mazzaferri, et al. \cite{mazzaferri2017open} & Semi-automatic & N=30 (30), EDI-OCT, Manual* & Dijkstra's algorithm \cite{dijkstra2022note} with several pre-processing steps, including B-scan flattening, Gabor transform edge detection. & Encoded uncertainty in edge-map detection. & Sensitive and multi-stage pre-processing, requires domain-specific knowledge to use. & No & Yes\textsuperscript{\textdagger} \\

George, et al. \cite{george2019two} & Semi-automatic & N=19 (589), EDI-OCT, Manual & Two-stage contour evolution model RPE-Choroid and Choroid-Sclera boundaries & General purpose for choroid and cornea segmentation. & Outputs bumpy contours, sensitive to posterior choroidal vessels. & No & No \\

Chen, et al. \cite{chen2017automated} & Deep learning & N=31 (62), EDI-OCT, Manual & SegNet \cite{badrinarayanan2017segnet} with graph-based post-processing. & Local graph-method, global CNN feature learning. Evaluated on pathology. & Dependent on graph-based post-processing, jagged boundaries. & No & No \\

Sui, et al. \cite{sui2017choroid} & Deep learning & N=73 (912), EDI-OCT, Manual & Learned graph-edge weights from CNN with Dijkstra's algorithm \cite{dijkstra2022note}. & Smooth segmentations from global/local features from CNN/graph theory. Leverages off-line augmentation to increase dataset & Data leakage in evaluation. & No & No \\

Al-Bander, et al. \cite{al2017novel} & Deep learning & N=169 (169), EDI-OCT, Manual & Patch-based CNN based on super-pixel segmentation. & Leverages global CNN representation learning. & Unsmooth boundary segmentations from super-pixel CNN classification. & No & No \\

Masood, et al. \cite{masood2019automatic} & Deep learning & N=21 (525), EDI-OCT, Manual & Patch-based CNN classifier into ``choroid'', ``not-choroid''. & Removes handcrafted features from pipeline. & Semi-automatic RPE-Choroid tracing, not pixel-level segmentation. & No & No \\

He, et al. \cite{he2021choroid} & Deep learning & N=108 (146), EDI-OCT, Manual & Patch-based CNN classifier into ``choroid boundary'', ``not-choroid boundary''. & Regularisation ensures smoothness and removes handcrafted features from pipeline. & Semi-automatic RPE-Choroid tracing and RANSAC \cite{fischler1981random} for patch-stitching. & No & No \\

Hsia, et al. \cite{hsia2021automatic} & Deep learning & N=30 (750), EDI-OCT, Manual & Mask R-CNN for feature extraction, region proposal and boundary detection. & Leverages fine-tuning and ensemble modelling on small dataset. & Slow execution ($\approx$6s/B-scan on GPU), required post-processing to remove dents. & No & No \\

Chen, et al. \cite{chen2022application} & Deep learning & N=123 (3075), EDI-OCT, Manual & Combines Mask R-CNN to extract square choroid ROI and use fully convolutional network \cite{long2015fully} for segmentation. & Leverages fine-tuning. & Over complex, required post-processing to remove dents. & No & No \\

Tsuji, et al. \cite{tsuji2020semantic} & Deep learning & N=34 (290), SS-OCT, Manual & SegNet \cite{badrinarayanan2017segnet} for pixel-level segmentation with class weight balancing. & No requirement of pre-processing, less sensitive to noise than graph methods & Tunable, post-processing to smooth Choroid-Sclera boundary. & Yes & No \\

Mao, et al. \cite{mao2020deep} & Deep learning & N=20 (5120), SS-OCT, Manual & Modified UNet \cite{ronneberger2015u} with learnable skip connections and global pooling. & Leverages augmentation on-the-fly, outperforms standard UNet. & Post-processing necessary in non-signal regions, data leakage potential, no details on evaluation data. & No & No \\

Devalla, et al. \cite{devalla2018drunet} & Deep learning & N=100 (100), EDI-OCT, Manual & UNet \cite{ronneberger2015u} with dilated convolutional blocks for multi-layer retinal and choroid segmentation. & Comprehensive intra-ocular segmentation. & Evaluation focus not on the choroid. & No & No \\

Chai, et al. \cite{chai2020perceptual} & Deep learning & Unknown (790), SS-OCT, Manual & Adversarial learning to minimise domain-shift discrepancy from Topcon to Nidek. & No requirement of ground truths in target domain (Nidek). & Low-quality perceptual loss, No detail on OCT mode of devices, small evaluation. & No & No \\

Zhang, et al. \cite{zhang2020automatic} & Deep learning & N=20 (1280), SS-OCT, Manual & Bio-Net: Several UNet's \cite{ronneberger2015u}, regularised to match choroid thickness. & Leveraged to find shadow-free en face choroid visualisations. & High complexity from multiple models. & No & No \\

Cahyo, et al. \cite{cahyo2021multi} & Deep learning & N=99 (42496), SS-OCT, Manual & SA-Net: UNet \cite{ronneberger2015u} with spatial feature reconstruction using consecutive B-scans. & Large dataset, leverages features from adjacent B-scans. & High complexity from multiple models, unfeasible on edge-devices. & No & No \\

Xu, et al. \cite{xu2022automatic} & Deep learning & N=35 (1400), SS-OCT, Manual & Enhanced UNet \cite{ronneberger2015u} by adding dense connection per block \cite{huang2017densely}. & Dense feature extraction, evaluated on high myopia. & Poor segmentation on raw B-scans, sensitive pre-processing required. & No & No \\

Wang, et al. \cite{wang2024pgkd} & Deep learning & Unknown (1062), EDI-OCT, Manual  & PGKD-Net: Transformer-assisted cascade network & Evaluated on several tasks, and high myopia. & Requires prior mask to reproduce, high complexity implies scalable only with GPU resources. & No & No \\ 
\bottomrule

\caption[Previous semi- and fully-automatic approaches for choroid region segmentation in \acrshort{OCT}.]{Semi-automatic and fully-automatic approaches to choroid region segmentation over the last 12 years. The data column is structured to describe the number of eyes (total B-scans) used for training and/or evaluation, the type of OCT data and whether model development/training used manual labels. * : Manual labelling by two or more graders; ** : High-penetration \acrshort{OCT}; $^+$ : Polarization Sensitive \acrshort{OCT}; \textsuperscript{\textdagger}: Requires proprietary software MatLab.} \label{tab:INTRO_region_methods} \\
\end{longtable}}

                \end{landscape}
                }

                Briefly, many semi-automatic approaches utilised graph-based methods such as Dijkstra's algorithm \cite{dijkstra2022note} for individual edge tracing of the RPE-Choroid and Choroid-Sclera boundaries. These methods represent the image as a graph, where each pixel is a node, and the segmentation task is modelled as a graph partitioning problem \cite{mazzaferri2017open, lu2013automated, masood2018automatic, salafian2018automatic, kajic2012automated, alonso2013automatic, hussain2018automated, chen2018automated, hu2013semiautomated, danesh2014segmentation}. Other traditional, semi-automatic approaches leveraged statistical models based on level sets and different enhancement strategies to simplify the edge tracing problem \cite{torzicky2012automated, duan2012automated, chen2015automated, wang2017automatic, vupparaboina2015automated, george2019two}. While these semi-automatic approaches improve on just manual labelling alone, they typically rely on handcrafted features and pre-defined criteria which lack generalisation. Additionally, many of these approaches take a local perspective on optimisation or require many tuning parameters and thus domain-specific knowledge to operate successfully. Moreover, these methods are multi-stage and heavily rely on many pre-processing steps which, if sub-optimal, can affect the output segmentation.

                As datasets grew and computational resources became more available, fully automatic approaches based on convolutional neural networks (\acrshort{CNN}s) become preferable over semi-automatic ones and could provide faster and more generalisable inference \cite{o2020deep}. The earlier approaches were a combination of feature extracting \acrshort{CNN}s which took a patch-based binary classification approach to region segmentation, utilising semi-automatic algorithms for pre-/post-processing and model fitting  \cite{chen2017automated, tsuji2020semantic, sui2017choroid, al2017novel, masood2019automatic, he2021choroid, chen2022application, hsia2021automatic}. More recently, there has been a push to develop fully automatic methods using fully convolutional networks (FCNs), such as the popularised UNet deep learning architecture \cite{ronneberger2015u} with residual \cite{he2016deep} or dense learning \cite{huang2017densely}. This permits pixel-level segmentation and Kugelman, et al. \cite{kugelman2019automatic} found that this type of encoder--decoder deep learning architecture outperformed patch-based methods. Thus, the UNet model has become the reference standard for deep learning approaches toward choroid region segmentation \cite{xu2022automatic, mao2020deep, devalla2018drunet, chai2020perceptual, cahyo2021multi, zhang2020automatic, wang2024pgkd}.

            \end{mysubsubsection}

        \end{mysubsection}

        \begin{mysubsection}[]{Vessel segmentation}\label{subsubsec:intro_litreview_vessel}
                
            While manual segmentation of the choroidal region requires identification of a single landmark in the OCT B-scan, segmenting the choroidal vessels require multiple, interdependent and heterogeneous blobs to be identified which is prohibitively time consuming and subject to human bias. Thus, approaches toward choroid vessel segmentation leveraged semi-automatic and fully-automatic approaches from the get-go. Table \ref{tab:INTRO_vessel_methods} gives a brief overview of the range of semi-automatic and fully-automatic approaches developed for choroid vessel segmentation over the last 12 years. Performance metrics have explicitly not been included due to lack of publicly available datasets to standardise and benchmark performance with.

            \afterpage{%
            \begin{landscape}
                \centering                \notsotiny

{\begin{longtable}{p{1.5cm}p{1.5cm}p{1.75cm}p{4cm}p{4cm}p{4cm}P{0.75cm}p{0.75cm}}
\toprule
Study & Method & Data & Description & Advantages & Limitations & Reproducibility & Open-source \\ \hline
\endfirsthead
\endhead
\midrule
\endfoot
\endlastfoot
Zhang, et al. \cite{zhang2012automated} & Semi-automatic & N=43, SD-OCT, None & Multiscale Hessian matrix analysis. & Addresses size-based heterogeneity of vessels. & Assumed circularity of vessel, large data loss (N=19) from poor choroid visualisation. & Yes & No \\

Kajic, et al. \cite{kajic2012automated_3d, esmaeelpour2014choroidal} & Semi-automatic & N=12 (871), EDI-OCT, Manual & 3D multi-scale edge filtering. & Addresses size-based heterogeneity of vessels. Pathological evaluation. & Vessel estimation assumed circularity. & No & No \\

Mahajan, et al. \cite{mahajan2013automated} & Semi-automatic & N=12 (12), SD-OCT, Manual & Multiple global thresholding with Frangi's filter \cite{frangi1998multiscale}. & Domain-specific method to address size/contrast heterogeneity of vessels. & Inter-dependent pipeline on global thresholding, limited evaluation. & No & No \\

Branchini, et al. \cite{branchini2013analysis} & Semi-automatic & N=42 (42), SD-OCT, None & Otsu's global thresholding after manually selecting choroidal area. & Simple application. & Sensitive to noise, limited to large vessels, requires manual selection. & No & No \\

Sonoda, et al. \cite{sonoda2014choroidal} & Semi-automatic & N=20 (20), EDI-OCT, Manual & Niblack’s local thresholding after manual selection of vessel brightness. & Open-source protocol, commonly used, assessed in pathology. & Did not specify Niblack parameters, requires manual brightness adjustment, limited to high quality choroids and strict grading protocol, slow ($\approx$ 5 min/B-scan). & Yes & Yes** \\

Vupparaboina, et al. \cite{vupparaboina2016optical} & Semi-automatic & N=2 (388), SD-OCT, None & Multi-stage enhancement with histogram equalisation and global thresholding. & Less human involvement than Sonoda, et al. \cite{sonoda2014choroidal}, less noise sensitive. & Limited and only qualitative evaluation, global thresholding not adept to multi-scale vessels. & No & No \\

Agrawal, et al. \cite{agrawal2016choroidal} & Semi-automatic & N=345 (345), EDI-OCT, Manual*  & Modified Niblack's method with morphology. & More automated than Sonoda, et al. \cite{sonoda2014choroidal} Commonly used across studies. & Lack of standardisation of Niblack’s parameters across studies, slow ($\approx$ 1 min/B-scan). & Yes & Yes\textsuperscript{\textdagger} \\

Liu, et al. \cite{liu2019robust} & Deep learning & N=10 (40), SS-OCT, Manual* & UNet based architecture with a large ResNet encoder. & No pre-processing required, assessed inter-rater variability. & High computational demand, manual vessel labelling. & No & No \\ \bottomrule

\caption[Previous semi- and fully-automatic approaches for choroid vessel segmentation in \acrshort{OCT}.]{Semi-automatic and fully-automatic approaches to choroid vessel segmentation over the last 12 years. The data column is structured to describe the number of eyes (total B-scans) used for training and/or evaluation, the type of OCT data and whether model development/training used manual labels. *: Manual labelling by 2 or more graders; ** : Measurement protocol provided in \cite{sonoda2014choroidal}; \textsuperscript{\textdagger} : \href{https://www.ocularimaging.net/land}{Website} available but membership, permission and co-author requirements \cite{betzler2022choroidal}.} \label{tab:INTRO_vessel_methods}
\end{longtable}
}

            \end{landscape}
            }
            
            Some of the earliest methods were oriented around binarisation and thresholding, whereby choroidal pixels are classified based on pixel intensity values into choroid vessel and surrounding interstitial tissue (choroid stroma) \cite{branchini2013analysis, vupparaboina2016optical}. Other methods were more domain-specific to address the heterogeneity of the choroidal vessel size and contrast \cite{zhang2012automated, kajic2012automated_3d, esmaeelpour2014choroidal, mahajan2013automated} but were heavily inter-dependent on a priori information such vessel shape. 

            Local thresholding methods were preferred over global ones as an attempt to address the heterogeneity in size of the choroidal vasculature. Namely, the Niblack autolocal thresholding algorithm \cite{niblack1985introduction} became commonly used in the research community by virtue of two research groups  \cite{sonoda2014choroidal, agrawal2016choroidal} releasing measurement protocols for Fiji (ImageJ) \cite{schindelin2012fiji} or membership-only image analysis platforms \cite{betzler2022choroidal} which use modified versions of Niblack. These semi-automatic approaches suffer from manual intervention as part of the protocol, thus making large-scale volumetric analysis unfeasible. Therefore, some attempts have been made to replace semi-automatic algorithms using deep learning \cite{liu2019robust}, with UNet architectures having superior performance over SegNet \cite{badrinarayanan2017segnet} and DeepLab \cite{chen2017rethinking} models, according to ground-truth labels generated by Niblack's method, as reported by Muller, et al. \cite{muller2022application}.

        \end{mysubsection}
        
        \begin{mysubsection}[]{Synthesising region and vessel segmentation}

            All of the previous approaches have considered the tasks of region and vessel segmentations as separate ones. However, it is advantageous to consider the inter-dependence of the choroidal space and vessels during the modelling process, and recent advancements have focused on synergistically segmenting both the choroidal region and its vasculature to provide more comprehensive analyses of the choroid. Table \ref{tab:INTRO_region_vessel_methods} gives a brief overview of the range of semi-automatic and fully-automatic approaches developed for choroid region and vessel segmentation. Performance metrics have explicitly not been included due to lack of publicly available datasets to standardise and benchmark performance with.

            \afterpage{%
            \begin{landscape}
                \centering                \notsotiny

{\begin{longtable}{p{1cm}p{1cm}p{2cm}p{4.5cm}p{3.75cm}p{4cm}P{0.75cm}p{0.75cm}}
\toprule
Study & Method & Data & Description & Advantages & Limitations & Reproducibility & Open-source \\ \hline
\endfirsthead
\endhead
\midrule
\endfoot
\endlastfoot
Srinath, et al. \cite{srinath2014automated} & Semi-automatic & Unknown, EDI-OCT, None & Hessian analysis for boundary with structural similarity index, level sets for vessels. & Domain-specific algorithm based on hetergeneity of intensities in choroid. & Sensitive pre-processing, jagged segmentation boundaries, fails to segment small vessels. & No & No \\

Hussain, et al. \cite{hussain2018automated} & Semi-automatic & N=10 (190), EDI-OCT, Manual* & Global Otsu thresholding for vessels, leverages vessel features to obtain edge weights for Dijkstra's algorithm. & Novel depth-intensity normalisation for choroid enhancement. & Requires several tuning parameters, domain-specific knowledge, noise-sensitive. & No & No \\

Li, et al. \cite{li2021automated} & Deep learning & N=92 (92), EDI-OCT, Manual* & Two separate UNet's \cite{ronneberger2015u} for vessels and (Haller-Sattler, Choroid-Sclera) multi-layer segmentation. & Comprehensive choroid segmentation. & Challenging ground truth validation. & No & No \\

Zheng et al. \cite{zheng2021deep} & Deep learning & N=23 (667), SS-OCT, Auto/Manual & UNet \cite{ronneberger2015u} with residual connections for boundary, Niblack method for vessels. & Improved feature learning and map reconstruction. & Semi-automatic Niblack vessel segmentation, unnecessary loss of resolution, healthy cohort. & Yes & No \\

Khaing, et al. \cite{khaing2021choroidnet} & Deep learning & N=160 (160), SD-OCT, Manual & ChoroidNET: Two dilated UNet \cite{ronneberger2015u} models for sequential choroid layer and vessel segmentation. & Utilises inter-dependency of choroidal space and vessels, tested on pathology. & Small training set, manual vessel labelling. & No & No \\

Zhu, et al. \cite{zhu2022synergistically} & Deep learning & N=92 (2852), EDI-OCT, Manual* & CUNet: UNet \cite{ronneberger2015u} style with single feature extraction encoder, two task-specific decoders for region and vessels. & Adaptive multi-task loss as regularisation, end-to-end, evaluated on myopia. &  Potential data leakage. Manual vessel annotation. & No & No \\

Wang, et al. \cite{wang2023choroidal} & Deep learning & N=65 (396), SS-OCT, Manual*  & CVI-Net: UNet \cite{ronneberger2015u} and pyramid pooling layers for multi-scale feature learning. & Addresses heterogeneity of choroidal vessel size. & Small dataset, extensive pre-processing required, manual vessel labels. & No & No \\

Xuan, et al. \cite{xuan2023deep} & Deep learning & N=162 (2162), SS-OCT, Manual & Medical transformer \cite{valanarasu2021medical} for region, UNet \cite{ronneberger2015u}, Niblack for vessels. & Permits standardised choroid measurements, attempted automation of Niblack. & Small reproducibility dataset (20 B-scans), Use of semi-automatic Niblack method. & Yes & Yes** \\

Arian, et al. \cite{arian2023automatic} & Deep learning & N=156 (537), EDI-OCT, Manual & Modified-loss UNet \cite{ronneberger2015u} for retinal pathology. & Specified Niblack training parameters. & Semi-automatic Niblack method. Limited to fovea-centred evaluation. & No & Yes\textsuperscript{\textdagger} \\

Wen, et al. \cite{wen2024transformer} & Deep learning & N=10 (120), SS-OCT, Manual & TACLNet: Transformer-assisted UNet \cite{ronneberger2015u}. & Multi-scale feature learning, generalisable to other tasks, outperforms previous methods. & Small dataset, manual vessel labelling & No & No \\ 

\bottomrule
\caption[Previous semi- and fully-automatic approaches for choroid region and vessel segmentation in \acrshort{OCT}.]{Approaches to choroid region and vessel segmentation. The data column is structured to describe the number of eyes (total B-scans) used for training and/or evaluation, the type of OCT data and whether model development/training used manual labels. * : Manual labelling by two or more graders. ** : \href{https://choroid-ai.com/}{Website} inactive as of December 2024; \textsuperscript{\textdagger} : \href{https://shorturl.at/lzoG1}{Codebase} not plug-and-play, only fully-automatic region segmentation, released \href{https://shorturl.at/ZPfHM}{dataset}.}\label{tab:INTRO_region_vessel_methods}
\end{longtable}
}

            \end{landscape}
            }

            Traditional methods have been proposed for choroid layer and vessel segmentation \cite{hussain2018automated, srinath2014automated}, but were complex, multi-stage and heavily parameter dependent. Attempts have been made at deep learning-based segmentation of Haller's and Sattler's layers \cite{li2021automated}, despite the oblique presentation of choroidal vessels on OCT and no reliable ground truth or standardised protocol to differentiate the two vascular layers \cite{mrejen2013optical}. Other deep learning methods segmented the region and vessels but still considered their problem formulations as separate entities \cite{xuan2023deep, zheng2021deep, arian2023automatic}, applying semi-automatic Niblack's method for vessel segmentation. Although, Xuan, et al. \cite{xuan2023deep} did acknowledge the need for standardisation of choroidal measurements and included additional \acrshort{OCT} B-scan foveal pit detection in their model DCAP. 
            
            Finally, more comprehensive approaches combined the tasks of region and vessel segmentation into single deep learning-based models, with the creation of models based on UNet architectures \cite{ronneberger2015u} like ChoroidNet \cite{khaing2021choroidnet} CUNet \cite{zhu2022synergistically} and CVI-Net \cite{wang2023choroidal}. A final advancement was the development of transformer-assisted TACLNet, proposed by Wen, et al. \cite{wen2024transformer} which showed superiority performance across CUNet \cite{zhu2022synergistically}, ChoroidNet \cite{khaing2021choroidnet}, CVI-Net \cite{wang2023choroidal} and the regular UNet \cite{ronneberger2015u}.

        \end{mysubsection}

        \begin{mysubsection}[]{Current challenges}

            \begin{mysubsubsection}[]{Vessel segmentation} \label{sec:INTRO_litreview_challenge_vessels}

                Regarding choroid vessel segmentation, the research community largely adopt the semi-automatic vessel segmentation protocols outlined by Sonoda, et al. \cite{sonoda2014choroidal} and Agrawal, et al. \cite{agrawal2016choroidal} which both use the semi-automatic Niblack thresholding algorithm. However, despite the widespread use of Niblack's algorithm for choroid vessel segmentation, there is concern surrounding its improper usage. A crucial element missing from the established software and measurement protocols \cite{sonoda2014choroidal, betzler2022choroidal} is how to use the Niblack method and, specifically, how to tune the internal parameters of the local thresholding tool. The platforms available to conduct image analysis, such as Fiji (ImageJ) \cite{schindelin2012fiji}, have further compounded this issue, as they often default to specific parameter settings that may not be reported by researchers. To the best of our knowledge, only two studies have reported specific Niblack parameters \cite{muller2022application, arian2023automatic}. This lack of transparency makes it difficult to compare results across different studies or to replicate findings, thus hampering efforts to standardise the reporting of choroidal measurements.
                
                There have also been several other reports hindering the standardisation of Niblack's autolocal thresholding algorithm \cite{muller2022application, wei2018comparison, elbay2022comparison, rosa2024impact, ma_validation_2024}. Wei, et al. \cite{wei2018comparison} found a reported difference of around 3\% in choroid vessel density (and an intra-class correlation of 0.35) measured following the protocols outlined by Sonoda, et al. \cite{sonoda2014choroidal} and Agrawal, et al. \cite{agrawal2016choroidal}. Elbay, et al. \cite{elbay2022comparison} observed the modified Niblack method proposed by Agrawal, et al. \cite{agrawal2016choroidal} to be sensitive to the type of scaling used to view the \acrshort{OCT} B-scan (see section \ref{subsec:INTRO_MEASURE_VIEW}), reporting errors of up to 4\% in choroid vessel density. Muller, et al. \cite{muller2022application} found that tuning of only one of the internal Niblack parameters could result in differences in choroid vessel density between 2\% and 20\%. Rosa, et al. \cite{rosa2024impact} found a significant difference of around 7.5\% in choroid vessel density when using the Niblack method proposed by Agrawal, et al. \cite{agrawal2016choroidal} and altering the image brightness of the \acrshort{OCT} B-scan. Finally, Ma, et al. \cite{ma_validation_2024} reported that the mean difference between  \acrshort{SD-OCT} and \acrshort{SS-OCT} data in three dimensional vessel density was 13\% using the protocol outlined by Sonoda, et al. \cite{sonoda2014choroidal}. 
            
                These are crucial studies which accompany the concern around the Niblack method's potential lack of reproducibility and measurement error. Agrawal, et al. \cite{agrawal2020exploring} published a major review on choroid vessel density in retinal and systemic disease reporting average changes between healthy and diseased eyes between 2\% and 6\%. This suggests that improper usage of these current methods and their lack of standardisation could result in measurement error exceeding observed effect sizes in disease, which could lead to paradoxical results inherently skewed by measurement variability. This was posited by Rosa, et al. \cite{rosa2024impact} who, in their discussion, found contradictory results among the literature when measuring the choroid vessel density in Vogt-Koyanagi-Harada disease \cite{liu2018choroidal, kawano2016relative, agrawal2016Koyanagi, jaisankar2019choroidal} and Non-Arteritic Anterior Ischemic Optic Neuropathy \cite{pellegrini2019choroidal, guduru2019choroidal}, using the methods outlined by Sonoda, et al. \cite{sonoda2014choroidal} and Agrawal, et al. \cite{agrawal2016choroidal}.
                
                Conversely to Ma, et al. \cite{ma_validation_2024}, Agrawal, et al. \cite{agrawal2019choroidal} reported a mean difference in choroid vessel density of only 3\% between two-dimensional \acrshort{SD-OCT} and \acrshort{SS-OCT} data. While not perfectly comparable to three-dimensional vessel density, recent evidence suggest that choroid vessel density does not change across different macular scanning locations \cite{agrawal2017influence, goud2019new, kim2022influence} (but does between macular and peripheral locations \cite{singh2018wide}). Thus, it is possible that the contradictory findings between Ma, et al. \cite{ma_validation_2024} and Agrawal, et al. \cite{agrawal2019choroidal} may be linked to improper use of Niblack's method, such that it may be sufficient to measure the choroidal vessels only with suitable domain knowledge and expertise of the algorithm. Given that Agrawal, et al. \cite{agrawal2019choroidal} developed and released the modified Niblack method, it's likely they know how best to use it. Unfortunately, there lacks a standardised protocol for using and specifying internal parameters for semi-automatic methods like Niblack which is so heavily adopted in the community \cite{agrawal2020exploring}. It would be pertinent for one to exist to educate researchers on the pitfalls of its parameter specification, similar to Chu, et al. \cite{chu2020quantification} who do so for using the Phansalkar thresholding algorithm for the choriocapillaris in \acrshort{OCT-A}.
                
            \end{mysubsubsection}

            \begin{mysubsubsection}[]{Bias, reproducibility and accessibility} 

                A problem in the literature is the generation of high quality ground truth data for deep learning model development, particularly for choroid vessel segmentation. The majority of the sophisticated deep learning-based are developed and evaluated against manual ground truth labels which are subject to human bias (tables \ref{tab:INTRO_region_methods} -- \ref{tab:INTRO_region_vessel_methods}, Data column). Few consider the value of developing models based on ground truth labels generated using semi-automatic or fully automatic approaches, which explicitly encode a degree of determinism and consistency into the ground truth generation process. Maloca, et al. \cite{maloca2023human} argues that such approaches to generating ground truth labels are preferable as they reduce subjectivity and thus bias and inconsistency, especially for the choroid.

                Furthermore, there is a distinct lack of reporting the reproducibility of automatic methods in medical imaging research and deep learning-based segmentation (tables \ref{tab:INTRO_region_methods} -- \ref{tab:INTRO_region_vessel_methods}, Reproducibility column, only 9 (18\%) out of all 49 reviewed papers conducted reproducibility analysis). For those studies which do report reproducibility, they typically adopt population-level, dimensionless metrics like the intra-class correlation or coefficient of variation/repeatability. Even unit-dependent metrics like average signed differences can be subject to cancellation effects and asymmetric error distributions. Without an \textit{interpretable} and \textit{contextual} assessment of reproducibility, there is no way to differentiate measurement variability from the true biological effect being measured. This is particularly pertinent for interpreting study findings where effect sizes are likely to be small, such as in choroid vessel density between healthy and disease eyes (2\% -- 6\%) \cite{agrawal2020exploring} or choroid thickness in myopia progression (20 -- 30 $\mu$m) \cite{breher2019metrological, flores2013relationship}.
                
                There are currently no accessible, open-source algorithms for large-scale, fully-automatic choroid region and vessel segmentation (tables \ref{tab:INTRO_region_methods} -- \ref{tab:INTRO_region_vessel_methods}, Open-source column, only 5 (10\%) out of all 49 reviewed papers released their codebase). All available algorithms fall into one of three categories: First, manual or semi-automatic methods \cite{ss2019octtools, sonoda2014choroidal, schindelin2012fiji, agrawal2016choroidal, brandt2018octmarker} that require human supervision (introducing subjectivity) and/or membership-access only. Second, fully-automatic deep learning-based methods that are not openly accessible, either only providing the code but not the trained model necessary to use the method \cite{kugelman2019automatic, wang2024pgkd} or not providing any access at the time of writing \cite{xuan2023deep} (December 2024). Third, releasing the model as fully-automatic but comprising of many steps, requiring underlying domain-specific knowledge or a license for proprietary software \cite{mazzaferri2017open, arian2023automatic}. 

             \end{mysubsubsection}

        \end{mysubsection}

        \begin{mysubsection}[]{Summary}

            The field of choroid region and vessel segmentation has seen significant evolution. Beginning from traditional image processing like graph-based \cite{mazzaferri2017open, kajic2012automated, hussain2018automated} and thresholding methods \cite{sonoda2014choroidal, agrawal2020exploring, agrawal2016choroidal}, which suffer from domain-specific expertise, free-parameter tuning, inter-dependent pipelines and limitations on generalisability to complex edge-cases, and ending at advanced, fully-automatic deep learning-based methods, which offered superior accuracy and large-scale automation \cite{o2020deep}, marking a significant advancement in the field of ophthalmic image analysis. While initially these methods treated the two tasks separately \cite{xu2022automatic, zhang2020automatic, wang2024pgkd, liu2019robust}, researchers leveraged deep learning methods for synergistic segmentation of the region and vessels \cite{xuan2023deep, khaing2021choroidnet, zhu2022synergistically, wang2023choroidal, wen2024transformer}. 
                    
            However, these methods are still limited. Their ground truth label generation is still subject to human subjectivity when produced manually, and there is a lack of interpretable reproducibility analysis to help the research community differentiate measurement variability from true biological change. Additionally, there is a significant absence of open-source, deep learning methods available to the wider research community. This has resulted in much of the research community still dependent on manual or semi-automatic methods which are made available by image analysis platforms like Fiji (ImageJ) \cite{schindelin2012fiji}. This can have unfortunate consequences regarding the standardisation of choroidal measurements in \acrshort{OCT} image sequences, particularly if different definitions of the Choroid-Sclera boundary are assumed \cite{yiu2014characterization, boonarpha2015standardization, xie2021evaluation, chandrasekera2018posterior}, or improper usage of Niblack's method continues \cite{muller2022application, wei2018comparison, elbay2022comparison, rosa2024impact, ma_validation_2024}.
            
        \end{mysubsection}
        
    \end{mysection}	
            
    \begin{mysection}[]{Measuring the choroidal microvasculature} \label{subsec:ch1_INTRO_measure_choroid}

        \begin{figure}[tb]
            \centering
            \includegraphics[width=0.8\linewidth]{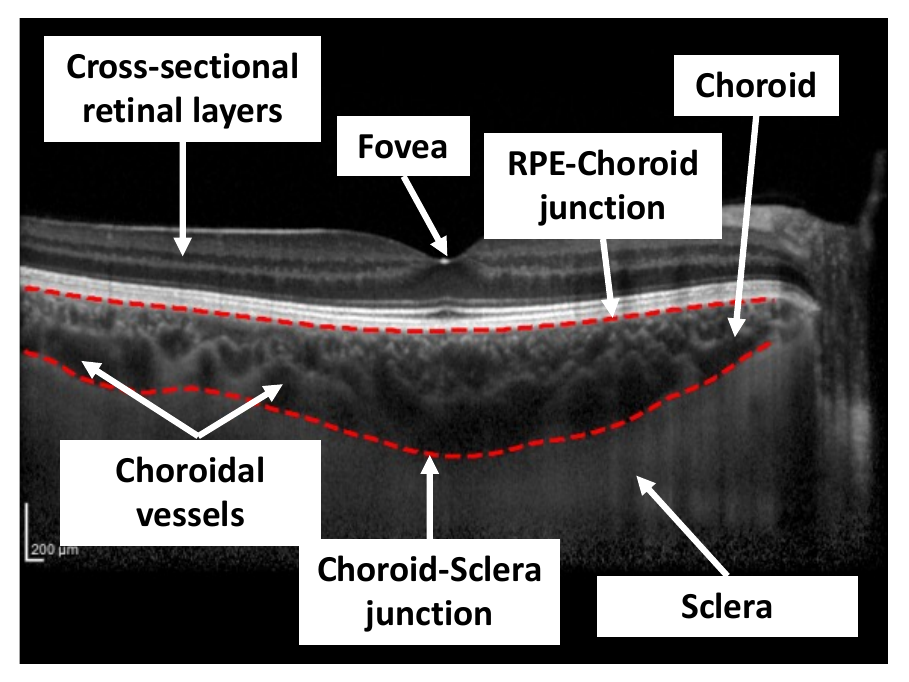}
            \caption[Exemplar, annotated \acrshort{OCT} B-scan.]{\acrshort{EDI-OCT} B-scan showing the choroid beneath the retina, posterior to the hyperreflective outer retinal layers and fixed to the \acrshort{RPE}. Red-dashed lines mark the \acrshort{RPE}-Choroid and Choroid-Sclera boundaries, with annotated landmarks. Due to the limited resolution of current \acrshort{OCT} devices (about 3.87 $\mu$m-per-pixel for Heidelberg Engineering Spectralis \acrshort{OCT}1 devices), the choriocapillaris layer is barely visible. The choroidal space appears as bright regions (stroma) with darker areas representing vasculature, a pattern empirically observed \cite{sohrab2012pilot, branchini2013analysis} and accepted in the research community \cite{agrawal2020exploring}.}
            \label{fig:INTRO_octbscan}
        \end{figure}

        \begin{figure}[tb]
            \centering
            \includegraphics[width=\linewidth]{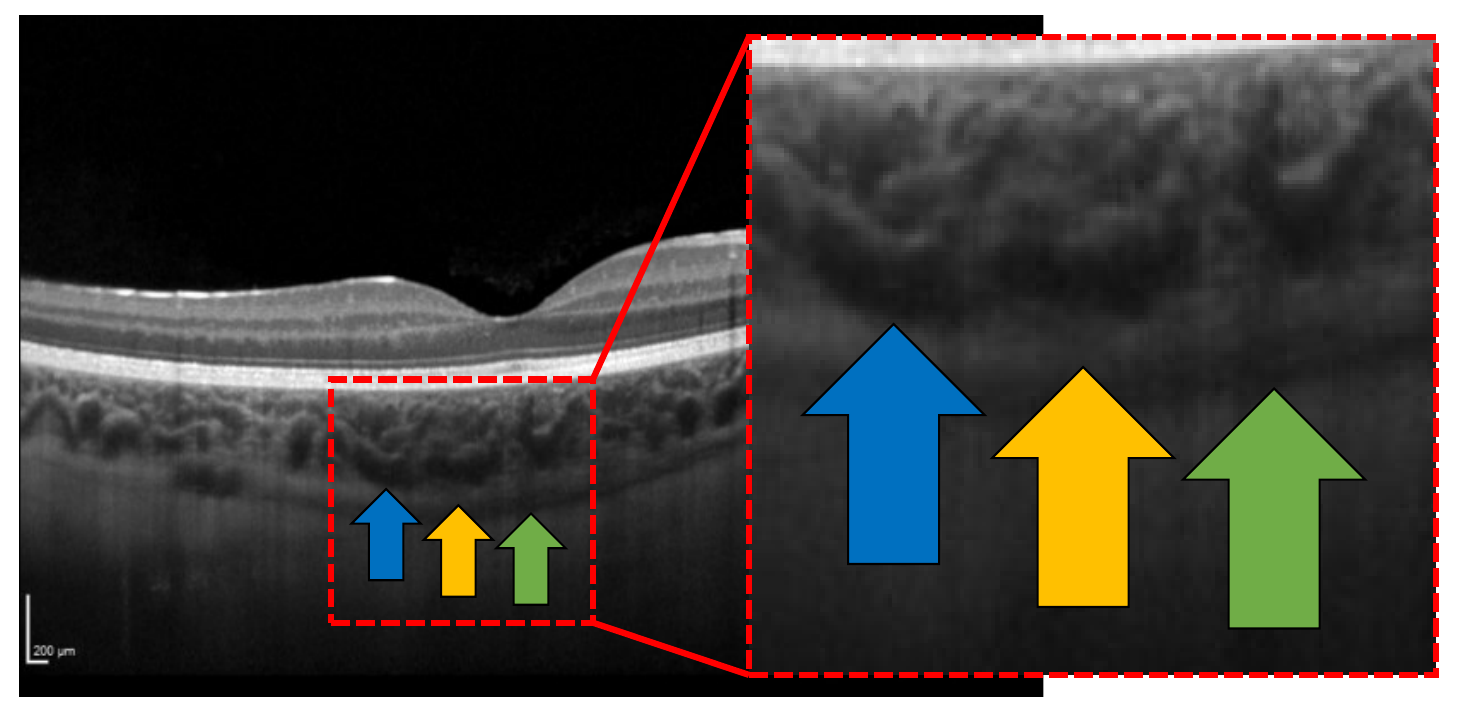}
            \caption[Different definitions of the Choroid-Sclera boundary.]{OCT B-scan with visible suprachoroidal space, with coloured arrows corresponding to the three distinct definitions of the Choroid-Sclera boundary.}
            \label{fig:INTRO_CS_definition}
        \end{figure}

        Figure \ref{fig:INTRO_octbscan} shows an exemplar \acrshort{EDI-OCT} B-scan with the RPE-Choroid and Choroid-Sclera boundaries delineating the choroidal space, with important biological landmarks commonly captured in the B-scan also visually defined. Conceptually, there there is debate around the exact location of the Choroid-Sclera boundary, and figure \ref{fig:INTRO_CS_definition} demonstrates this. Depending on the appearance of the choroid (and the visibility of the suprachoroidal layer) in an \acrshort{OCT} B-scan, this may either be \cite{yiu2014characterization}:
        \begin{enumerate}\setlength\itemsep{0em}
            \item the hyperreflective line at the posterior most location of the largest choroidal vessels in Haller's layer, i.e. inner choroid junction (figure \ref{fig:INTRO_CS_definition}, blue arrow);
            \item the hyporeflective line separating the posterior of Haller's layer from the suprachoroid, i.e. inner suprachoroid junction (figure \ref{fig:INTRO_CS_definition}, yellow arrow);
            \item the hyperreflective line separating the suprachoroidal layer from the sclera, i.e. inner scleral junction (figure \ref{fig:INTRO_CS_definition}, green arrow).
        \end{enumerate}
        The disparity in these definitions can have an inordinate impact on choroidal measurements --- Yiu, et al. \cite{yiu2014characterization} found that error in manual \acrshort{SFCT} can vary from 17 $\mu$m to 70 $\mu$m when different definitions of the Choroid-Sclera boundary were used, especially when the suprachoroidal layer was observed.
            
        The original authors who proposed \acrshort{EDI-OCT} \cite{spaide2008enhanced} define the choroid to include the suprachoroidal layer \cite{mrejen2013optical}, and thus define the Choroid-Sclera boundary at the point of the inner sclera \cite{margolis2009pilot}, i.e. definition (3) (green arrow in figure \ref{fig:INTRO_CS_definition}). However, even that definition is still debated and can be hard to discern in practice, as the suprachoroidal space is not always present due to \acrshort{OCT} device limitations, age, health and other factors \cite{yiu2014characterization}. Inconsistency arises when certain choroids exhibit a suprachoroidal space and others do not, under the assumption that this space has no relationship to the research objective \cite{yiu2014characterization, chandrasekera2018posterior}.

        Chandrasekera, et al. \cite{chandrasekera2018posterior} showed that only 38\% of 200 eyes showed a visible suprachoroid in \acrshort{SS-OCT} data, and Yiu, et al. \cite{yiu2014characterization} showed a similar proportion (45\%) in 74 eyes using \acrshort{EDI-OCT} data, noting that visibility of the suprachoroid was higher in older eyes. Additionally, both studies found that choroidal thickness was significantly different when considering the different definitions of the choroid outlined above. Moreover, Chandrasekera, et al. \cite{chandrasekera2018posterior} additionally found that agreement between semi-automatic and manual detection of the Choroid-Sclera boundary worsened when the suprachoroidal space was visible (increased mean difference in thickness of 17 $\mu$m to 23 $\mu$m), reporting that automatic methods would prefer definition (1) while manual measurement would opt for definition (3). 

        Thus, to remain objective in our analyses throughout this thesis --- given the inconsistent visibility of the suprachoroidal layer, and the fact that it is an avascular potential space \cite{saidkasimova_anatomy_2021} unlike the highly vascularised choroidal layers \cite{nickla2010multifunctional} --- we opted to follow definition (1) (blue arrow in figure \ref{fig:INTRO_CS_definition}). 
        
        Therefore, the choroidal space in an \acrshort{OCT} B-scan for this thesis is defined as the space posterior to the boundary delineating the \acrshort{RPE} layer and Bruch's membrane complex (\acrshort{RPE}-Choroid) and superior to the boundary delineating the sclera or suprachoroid, if visible, from the posterior most point of Haller's layer (Choroid-Sclera).

        \begin{mysubsection}[]{OCT B-scan} \label{subsec:ch1_INTRO_measure_bsan}

            \begin{figure}[tb]
                \centering
                \includegraphics[width=\linewidth]{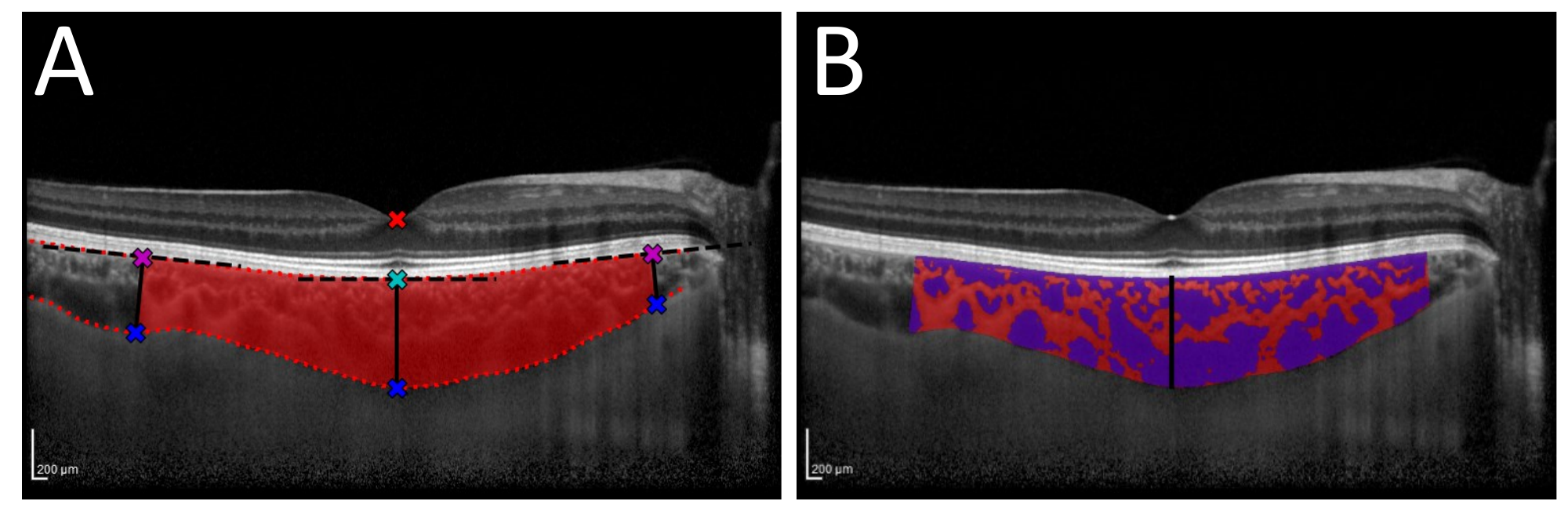}
                \caption[Demonstration of \acrshort{ROI} constructing and thickness measurement.]{(A) \acrshort{OCT} B-scan with measurement landmarks overlaid to describe construction of the fovea-centred \acrshort{ROI}. (B) \acrshort{OCT} B-scan with prescribed \acrshort{ROI} in red, \acrshort{SFCT} in black, and vessels in purple.}
                \label{fig:INTRO_choroid_ROI}
            \end{figure}

            Measuring the choroid in the macula is usually done with reference to a particular region of interest (\acrshort{ROI}). This is to ensure comparisons between and within patients can be done accurately and reliably. For the work in this thesis, we consider measuring the macular choroid in a fovea-centred \acrshort{ROI}, with some micron distance either side of the fovea. The specific location of the fovea is its centre, at the foveal pit. The foveal pit in an \acrshort{OCT} B-scan is defined in this thesis as a single pixel coordinate at the centre of the foveola centralis. The foveola is a circular zone of approximately 350 micron width which is avascular, creating a depression at the centre of the macula which pushes the inner retinal layers laterally \cite{FORRESTER20161}. The single pixel coordinate is one which appears most brightly illuminated on the B-scan in this depression, often aligning with a ridge formed at the photoreceptor layer.

            Figure \ref{fig:INTRO_choroid_ROI} demonstrates the construction of a two-dimensional \acrshort{ROI} in an \acrshort{OCT} B-scan (A), with measurements visualised in (B). In panel (A), we show the landmarks used to define a two-dimensional \acrshort{ROI} (red), given a 3000 $\mu$m distance either side of the fovea and describe its construction below:
            \begin{enumerate}\setlength\itemsep{0em}
                \item Given the location of the fovea (red cross in \ref{fig:INTRO_choroid_ROI}(A)), find the nearest location along the \acrshort{RPE}-Choroid boundary, which defines the centre of the \acrshort{ROI} (cyan cross in \ref{fig:INTRO_choroid_ROI}(A)).
                \item Traversing the \acrshort{RPE}-Choroid boundary some micron distance (3000 $\mu$m in this example) either side of this central pixel defines the endpoints of the \acrshort{ROI} (purple crosses in \ref{fig:INTRO_choroid_ROI}(A)) along this boundary.
                \item Tangential lines at each \acrshort{RPE}-Choroid reference point are drawn (black dashed lines in \ref{fig:INTRO_choroid_ROI}(A)), using an offset of 15 pixels in each direction to the reference point to estimate tangential gradient around reference points.
                \item Tangential lines are rotated 90 degrees (black solid lines in \ref{fig:INTRO_choroid_ROI}(A)) and extended until they intersect the Choroid-Sclera boundary, detecting the reference points along this boundary (blue crosses in \ref{fig:INTRO_choroid_ROI}(A)). 
            \end{enumerate}
            The \acrshort{ROI} is defined as the choroidal space (red shaded region in \ref{fig:INTRO_choroid_ROI}(A)) which is bound between these solid black lines which extend \textit{locally perpendicular} from the \acrshort{RPE}-Choroid boundary. Importantly, the prescribed, fovea-centred \acrshort{ROI} accounts for any choroidal curvature from poor acquisition, or degree of myopia, and ensures a standardised measurement protocol for measuring the choroid.

            In figure \ref{fig:INTRO_choroid_ROI}(B), we show the 3000 $\mu$m, fovea-centred \acrshort{ROI} in red, with segmented vessels in this \acrshort{ROI} in blue, and a black solid line representing the one-dimensional, subfoveal choroid thickness  (\acrshort{SFCT}) for this B-scan. Within this \acrshort{ROI}, there are four metrics of interest to measure the choroid: \acrshort{SFCT}, choroid area, choroid vascular area and choroid vascular index (\acrshort{CVI}). \acrshort{SFCT} measures the thickness of the choroid directly underneath the fovea, and locally perpendicular to the \acrshort{RPE}-Choroid boundary (black solid line in figure \ref{fig:INTRO_choroid_ROI}(B)). Using the pixel length-scales in the axial and lateral directions, the pixel length of this black line is converted into microns. It is most common to measure the choroid through a one-dimensional measurement like \acrshort{SFCT} \cite{tan2024techniques}. However, because of the segmental nature of the choroidal vascular bed \cite{hayreh1975segmental}, this is misguided \cite{kim2021choroidal}. Thus, it is common to take several thickness measurements across the \acrshort{ROI} and average them \cite{xie2021evaluation}.
            
            Instead of one-dimensional thickness, two-dimensional measurements like area are more representative and robust measurements of the choroid \cite{xie2021evaluation}. Choroid area counts the number of pixels in the total \acrshort{ROI} (red and purple pixels in figure \ref{fig:INTRO_choroid_ROI}(B)), and this count is converted into mm$^2$ by a conversion factor equal to the total mm$^2$ area of a single pixel, dictated by the axial and lateral pixel length-scales. Choroid vessel area is computed much the same but only for the vessel pixels (purple pixels in \ref{fig:INTRO_choroid_ROI}(B)). \acrshort{CVI} is a dimensionless ratio of choroid vascular area to total choroid area within the prescribed \acrshort{ROI} (ratio of purple pixels to red and purple pixels in figure \ref{fig:INTRO_choroid_ROI}(B)).
    
            \begin{mysubsubsection}[]{Pixel length-scales}\label{subsec:INTRO_MEASURE_SCALES}

                \begin{figure}[tb]
                    \centering
                    \includegraphics[width=\linewidth]{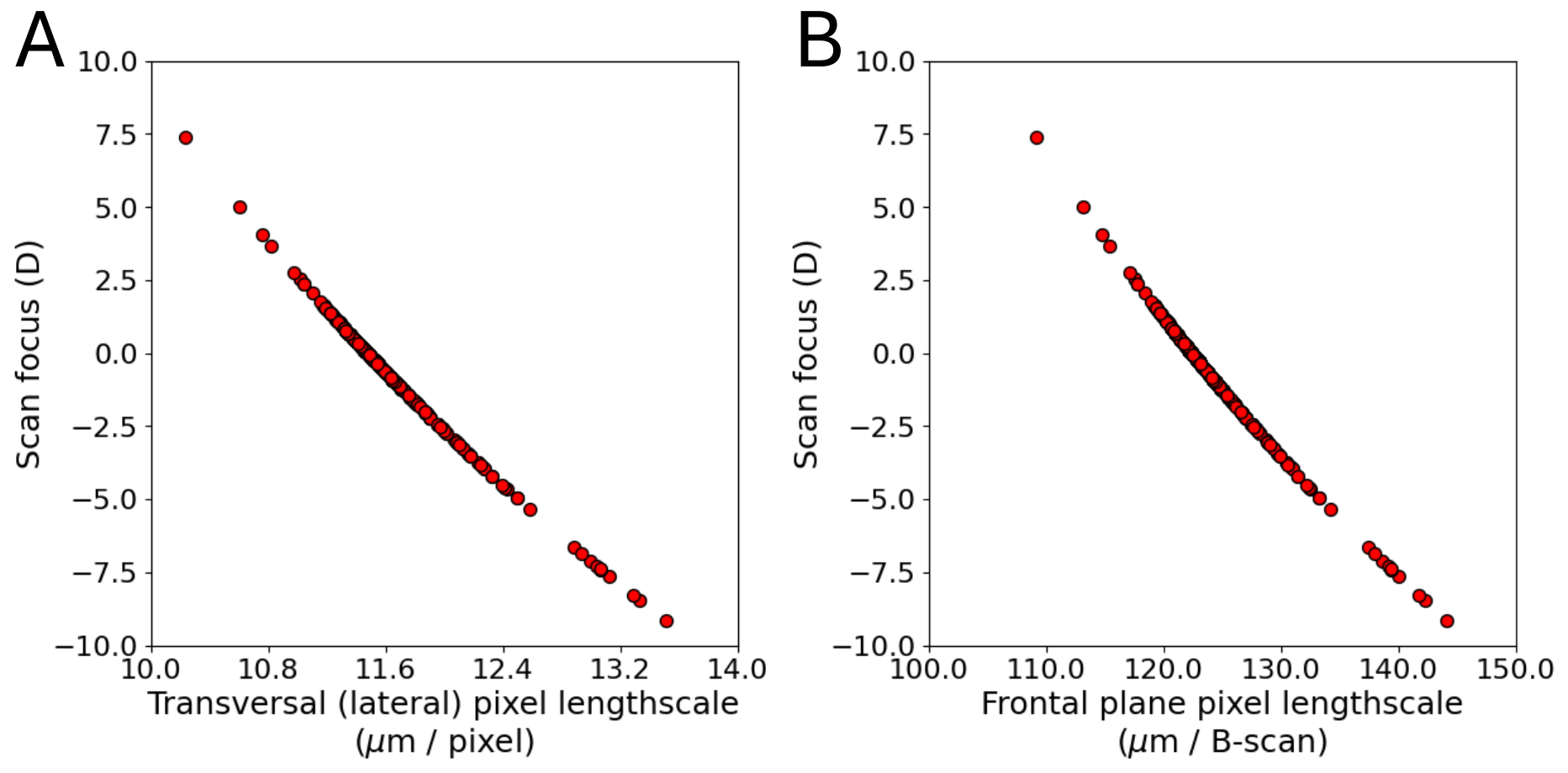}
                    \caption[Relationship between scan focus and lateral pixel length-scale on the Heidelberg Engineering \acrshort{OCT} system.]{Relationship between the scan focus value and the lateral (A) and frontal (B) pixel length-scales for Heidelberg Engineering \acrshort{OCT} systems.}
                    \label{fig:INTRO_Bscan_scales}
                \end{figure}

                The axial and lateral pixel length-scales in an \acrshort{OCT} B-scan are typically not equal, which is known as an \textit{anisotropic} scaling. For \acrshort{OCT} imaging devices, the ratio of axial to lateral pixel length-scale is approximately 1:3. Heidelberg Engineering's Spectralis \acrshort{OCT}1 module has a fixed axial pixel length-scale of 3.87 $\mu$m-per-pixel and, for a perfectly emmetropic model eye, a lateral pixel length-scale of 11.47 $\mu$m-per-pixel for pixel resolution 768 $\times$ 768. Topcon's DRI Triton Plus has a fixed axial pixel length-scale of 2.61 $\mu$m-per-pixel and a fixed lateral pixel length-scale of 8.79 $\mu$m-per-pixel for pixel resolution 992 $\times$ 1024.
                
                For some \acrshort{OCT} imaging devices (Heidelberg Engineering's \acrshort{OCT} Spectralis, for example), the lateral pixel length-scale has the ability to change across different examinations depending on the corneal surface properties and refraction of the patient's eye, as well as the operators use of the machine. While the former can be approximately encoded into the device, this is not always known and thus a phantom eye model is used instead. In the latter, during acquisition the operator can adjust various settings to compensate for the individual refractive surfaces in each patient's eye \cite{engeineering_hardware_instructions}. 

                An important tuning parameter during acquisition for the technician is the scan focus knob on the imaging device. This compensates for the refractive error of the patient's eye, and thus acts as an approximation for the patient's refractive error (measured in dioptres, D). While suitable focus on the chorioretinal structures is possible with the scan focus knob set to `0' --- perfectly emmetropic ---  it can be adjusted freely and is particularly useful in cases of high myopia or hyperopia \cite{engeineering_hardware_instructions}. Adjusting the focus alters the lateral pixel length-scale in an \acrshort{OCT} B-scan, as well as the distance between sequential B-scans in an \acrshort{OCT} volume scan (the frontal plane length-scale). This relationship can be observed in figure \ref{fig:INTRO_Bscan_scales}, which is approximately linear in the emmetropic range, between -1D and 1D. This is a crucial detail for producing standardised choroidal measurements, and can in particular have significant consequences for measurements which are computed using different \acrshort{ROI} definitions.
                
            \end{mysubsubsection}

            \begin{mysubsubsection}[]{Different viewing modes} \label{subsec:INTRO_MEASURE_VIEW}

                \begin{figure}[!t]
                    \centering
                    \includegraphics[width=\linewidth]{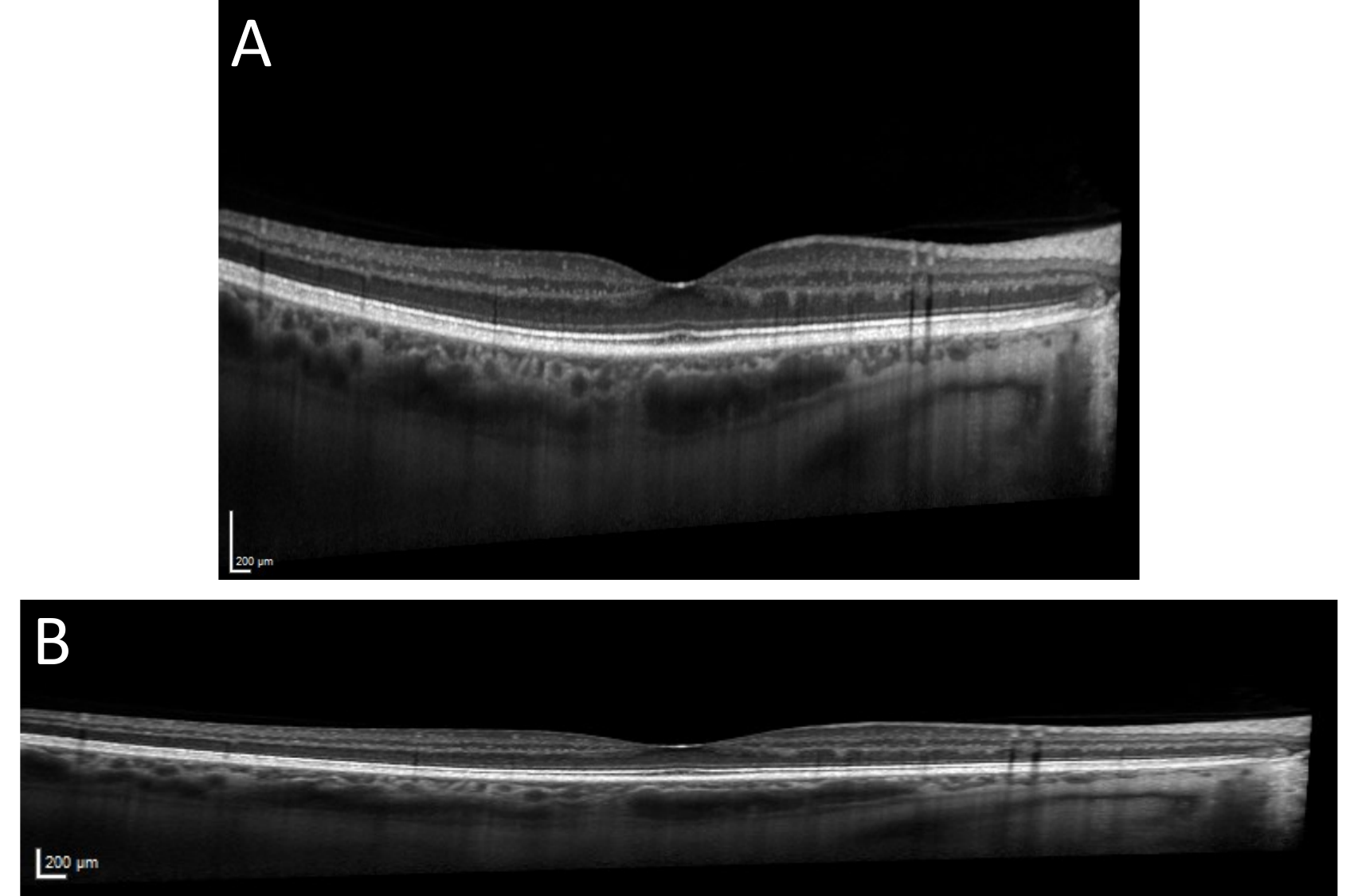}
                    \caption[Viewing modes on the Heidelberg Engineering \acrshort{OCT} system.]{An \acrshort{EDI-OCT} B-scan presented in 1:1 pixel mode (A) and 1:1 micron mode (B).}
                    \label{fig:INTRO_nEDI_vs_EDI_diagram}
                \end{figure}
    
                A way to circumvent anisotropic scaling is to present the chorioretinal structures in a B-scan whose axial and lateral pixel length-scales are equal. Default scaling used by most in the research community presents the \acrshort{OCT} B-scan in `1:1 pixel mode' (figure \ref{fig:INTRO_nEDI_vs_EDI_diagram}(A)), allowing chorioretinal structures to be visualised more easily for the end-user by appearing more bloated in the axial direction which aids the interpretation and readability of the B-scan. Another option available to present the chorioretinal structures is `1:1 micron mode' (figure \ref{fig:INTRO_nEDI_vs_EDI_diagram}(B)) which, although squashes the chorioretinal structures axially, is advised by device manufacturers to help mitigate for potential manual measurement error \cite{engineering_perfect_unknown}. 

                Indeed, Lawali, et al. \cite{lawali2023measurement} found that different scale types produced significantly different subfoveal retinal thickness measurements when measured manually. However, with the introduction of automatic methods which can segment and subsequently measure the chorioretinal anatomy in a programmatic way, presenting the B-scan using the default `1:1 pixel mode' is sufficient, and is used for the remainder of this thesis. However, it is for images with anisotropic scaling that the anatomic curvature is compensated for at the point of measurement.
    
            \end{mysubsubsection}

            \begin{mysubsubsection}[]{Regions of interest}\label{subsec:INTRO_MEASURE_ROI}

                \begin{figure}[tb]
                    \centering
                    \includegraphics[width=\linewidth]{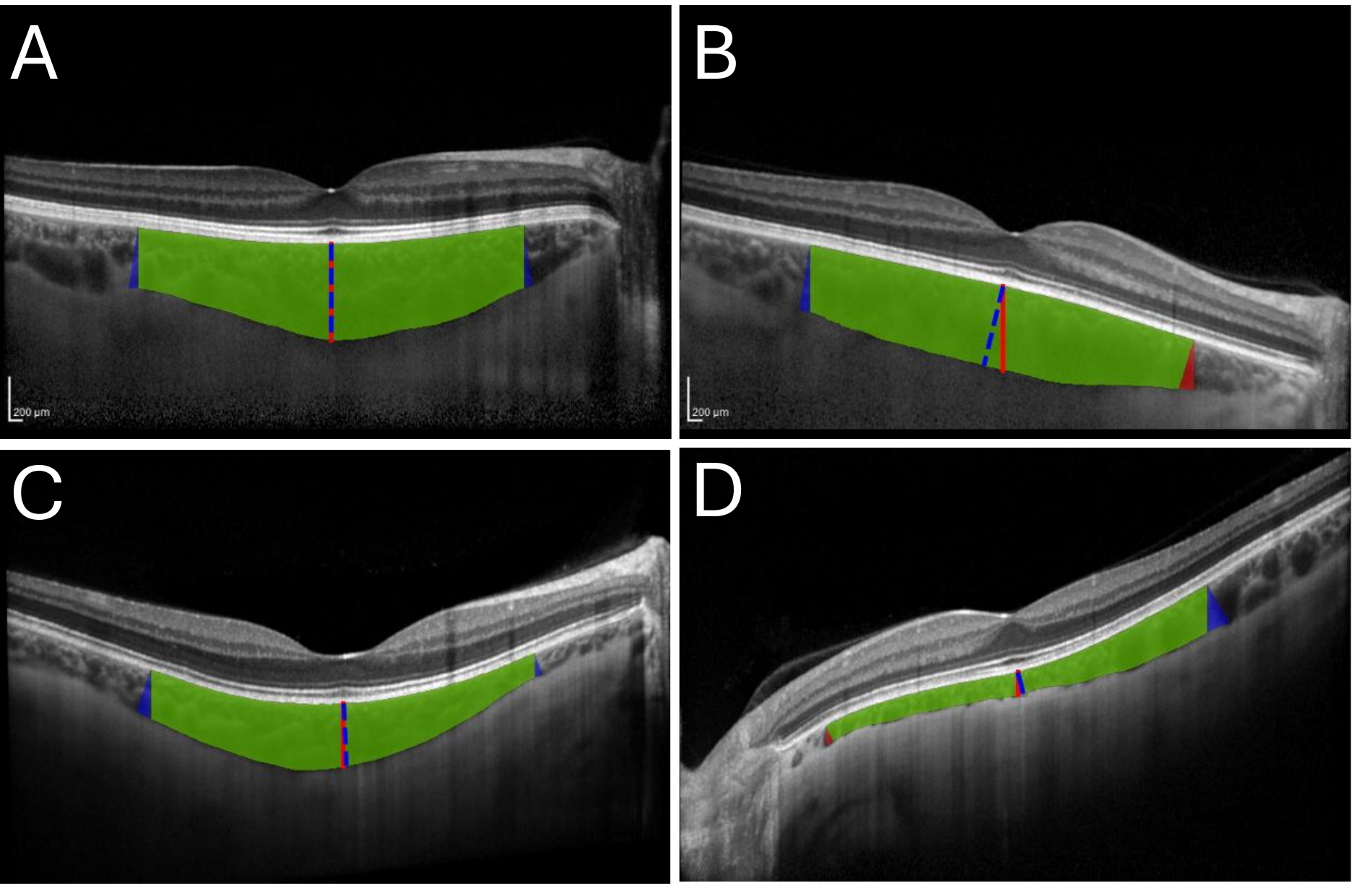}
                    \caption[\acrshort{ROI}s aligned to the image- and choroid-axis for \acrshort{OCT} B-scans with skew and/or curvature.]{\acrshort{OCT} B-scans with \acrshort{SFCT} and \acrshort{ROI} overlaid for a flat (A), skewed (B), curved (C) and skewed-and-curved (D) choroids. Lines drawn in solid red and dashed blue for image- and choroid-aligned \acrshort{SFCT}. Green pixels represent overlap between the two image- and choroid-aligned \acrshort{ROI}s, with red and blue pixels representing regions exclusive to the image- and choroid-aligned \acrshort{ROI}s.}
                    \label{fig:INTRO_ROI_impact}
                \end{figure}

                \begin{table}[tb]\footnotesize
                    \begin{adjustwidth}{-1in}{-1in}  
                    \centering
                    \scalebox{0.8}{
\begin{tabular}{llllllllllllll}
\toprule
\multirow{3}{2cm}{Choroid curvature} & \multirow{3}{*}{Method of measurement} & \multicolumn{12}{c}{Choroid measurement} \\
 &  & \multicolumn{3}{c}{\acrshort{SFCT} {[}$\mu$m{]}} & \multicolumn{3}{c}{Average thickness {[}$\mu$m{]}} & \multicolumn{3}{c}{Area {[}mm$^2${]}} & \multicolumn{3}{c}{\acrshort{CVI}} \\
 \cmidrule(l){3-5} \cmidrule(l){6-8} \cmidrule(l){9-11} \cmidrule(l){12-14} 
 &  & Choroid & Image & $\Delta_r$ & Choroid & Image & $\Delta_r$ & Choroid & Image & $\Delta_r$ & Choroid & Image & $\Delta$ \\
 \midrule
 
\multirow{3}{*}{Flat} & Hyperope & 424 & 423 & 0\% & 335 & 332 & 1\% & 1.915 & 1.902 & 1\% & 0.483 & 0.482 & 0.001 \\
 & Emmetrope & 424 & 423 & 0\% & 342 & 337 & 1\% & 2.071 & 2.054 & 1\% & 0.484 & 0.482 & 0.002 \\
 & Myope & 424 & 423 & 0\% & 361 & 351 & 3\% & 2.173 & 2.138 & 2\% & 0.49 & 0.496 & 0.006 \\
\midrule
 
\multirow{3}{*}{Skew} & Hyperope & 409 & 369 & 11\% & 347 & 305 & 14\% & 1.847 & 1.838 & 0\% & 0.543 & 0.544 & 0.001 \\
 & Emmetrope & 424 & 369 & 15\% & 374 & 315 & 19\% & 1.914 & 1.902 & 1\% & 0.554 & 0.555 & 0.001 \\
 & Myope & 448 & 369 & 21\% & 413 & 327 & 26\% & 1.981 & 1.982 & 0\% & 0.573 & 0.581 & 0.008 \\
\midrule
 
\multirow{3}{*}{Curved} & Hyperope & 277 & 278 & 0\% & 222 & 204 & 9\% & 1.259 & 1.246 & 1\% & 0.535 & 0.537 & 0.002 \\
 & Emmetrope & 278 & 278 & 0\% & 239 & 213 & 12\% & 1.325 & 1.305 & 2\% & 0.54 & 0.546 & 0.006 \\
 & Myope & 278 & 278 & 0\% & 261 & 225 & 16\% & 1.406 & 1.375 & 2\% & 0.566 & 0.567 & 0.001 \\
\midrule
 
\multirow{3}{2cm}{Skew   \& Curved} & Hyperope & 111 & 100 & 11\% & 168 & 127 & 32\% & 0.722 & 0.695 & 4\% & 0.345 & 0.329 & 0.016 \\
 & Emmetrope & 116 & 100 & 16\% & 179 & 127 & 41\% & 0.792 & 0.762 & 4\% & 0.379 & 0.363 & 0.016 \\
 & Myope & 123 & 100 & 23\% & 189 & 123 & 54\% & 0.776 & 0.745 & 4\% & 0.349 & 0.338 & 0.009 \\
 \bottomrule
\end{tabular}
}

                    \end{adjustwidth}
                    \caption[Choroid measurement error between different \acrshort{ROI} definitions.]{Choroidal measurements for the flat, skewed, curved and skewed-and-curved choroids from figure \ref{fig:INTRO_ROI_impact}, while simulating different refractive errors. Highly myopic: -9.17D, emmetropic: 0D, highly hyperopic: 7.37D. $\Delta_r$ is absolute, relative difference measured as $\nicefrac{|c-i|}{i}$ where $c$ and $i$ are the choroid- and image-aligned values. $\Delta$ is absolute difference.}
                    \label{tab:INTRO_ROI_metrics}
                \end{table}
                
                Historically, retinochoroidal metrics are measured according to the image-axis \cite{cheong2018novel, yiu2014characterization, aksoy2023choroidal, shoshtari2021impact, rahman2011repeatability, cho2014influence}. That is, the \acrshort{ROI} and subsequent measurements are defined and computed per A-scan (column-wise) which most imaging devices also adopt. However, this assumes there is little to no curvature in the retinochoroidal tissue in the B-scan. The shape of the patient's eye (degree of myopia) and quality of image capture both contribute to the appearance of the retina and choroid on the \acrshort{OCT} B-scan. This is why in this thesis we ensure the \acrshort{ROI}, and subsequent measurements, are choroid-aligned, i.e. locally perpendicular to the \acrshort{RPE}-Choroid boundary. This \acrshort{ROI} is aligned with the predominant axis which the choroid's \acrshort{RPE}-Choroid boundary lies along, which can deviate from the lateral image-axis. This deviation is known as \textit{skew} or \textit{curvature}, defined as consistent or inconsistent deviation in the angle of elevation of the choroid from the lateral image-axis, respectively.

                Figure \ref{fig:INTRO_ROI_impact} shows \acrshort{OCT} B-scans whose intra-ocular structures exhibit different types of skew or curvature. The choroid in panel (A) appears predominantly flat, while the skewed choroid in panel (B) shows consistent deviation away from the horizontal image-axis. The choroid in panel (C) is curved and thus shows inconsistent deviation from the horizontal image-axis. In panel (D) we see both skew and curvature.
                
                In figure \ref{fig:INTRO_ROI_impact}, we overlay the image-aligned and choroid-aligned \acrshort{ROI}s and \acrshort{SFCT}. While the majority of the \acrshort{ROI}s overlap (green), there are distinct regions which are mutually exclusive between the \acrshort{ROI} definitions which, depending on the extent of skew and/or curvature, can have an impact on downstream choroidal measurements. For example, the skew exhibited by the choroids in panels (B, D) underneath and fovea result in deviation between the lines drawn measuring \acrshort{SFCT} when considering different \acrshort{ROI}s (red solid line compared to blue dashed-line). This is in contrast to the choroids in panels (A, C), which are locally flat underneath the fovea, resulting in identical lines drawn to measure \acrshort{SFCT} regardless of the \acrshort{ROI} definition.

                The extent of deviation in choroidal measurements was investigated for these four examples by considering a fovea-centred \acrshort{ROI} covering a 6000 $\mu$m distance and measuring \acrshort{SFCT}, average thickness across the \acrshort{ROI}, total choroid area and \acrshort{CVI}. For these four eyes, we consider three cases where the patient's eye is highly myopic, emmetropic and highly hyperopic, simulating different lateral pixel length-scales (for a Heidelberg Engineering \acrshort{OCT} imaging device). We define emmetropic as when the scan focus is 0, which corresponds to a lateral scaling of 11.47 $\mu$m-per-pixel. We define highly myopic and hyperopic as the extremal observations from figure \ref{fig:INTRO_Bscan_scales}, i.e. refractive errors (approximated by the scan focus knob in dioptres, D) -9.17D and 7.37D, respectively. These correspond to lateral pixel length-scales of 13.51 $\mu$m-per-pixel and 10.24 $\mu$m-per-pixel, respectively.

                Table \ref{tab:INTRO_ROI_metrics} shows the extent of deviation between the image- and choroid-aligned choroidal measurements. As expected, measurements were stable for the flat choroid, but significant deviations in \acrshort{SFCT} and average thickness was observed across the board for the remaining choroids (the curved choroid did not deviate in \acrshort{SFCT} because it is locally flat under the fovea as in figure \ref{fig:INTRO_ROI_impact}(C)). Interestingly, deviation in all choroidal measurements (but primarily in \acrshort{SFCT} and average thickness) increased as the eye became more myopic. This is due to the anisotropic scaling which becomes more disproportionate with increasing degree of myopia (figure \ref{fig:INTRO_Bscan_scales}). This suggests not only the importance of ensuring a consistent \acrshort{ROI} definition, but also that these pixel length-scales are taken into account when reporting physical measurements of populations which could have a wide variety of refractive errors.
                
                For the skewed choroids, which were not aligned with the image axis underneath the fovea, the average \acrshort{SFCT} deviation was 16\% (37 $\mu$m in absolute value), and 31\% (58 $\mu$m in absolute value) for average thickness across the \acrshort{ROI}. Across all choroids, there was a deviation in average thickness of 19\% (37 $\mu$m in absolute value) and the largest deviation observed was the skewed choroid for a hypothetically myopic eye, which saw a deviation of 23\% and 54\% (79 $\mu$m and 86 $\mu$m in absolute value) for \acrshort{SFCT} and average thickness, respectively. 

                Two-dimensional measurements like area and \acrshort{CVI} appear more robust to these different \acrshort{ROI} definitions, with an average absolute difference in \acrshort{CVI} of 0.006 (0.6\%) across all choroids. For area, choroidal curvature did cause some minor deviation of 2\% (0.026 mm$^2$ in absolute value) for the curved choroids. 
                
                While the deviations in two dimensional measurements are unlikely to be clinically significant (Agrawal, et al. \cite{agrawal2020exploring} reported differences in \acrshort{CVI} of between 2\% -- 6\% between healthy and diseased eyes), the reported deviations in thickness are. For example, reported effect sizes in choroidal thickness can be as low as 20 -- 30 $\mu$m when considering myopia progression \cite{flores2013relationship, breher2019metrological}, while changes in \acrshort{SFCT} is reported to decrease between 5\% -- 12\% during haemodialysis, as reported by Farrah, et al. \cite{farrah2020eye}, and choroidal thinning by approximately 25\% in end-stage \acrshort{CKD}, as reported by Balmforth, et al. \cite{balmforth2016chorioretinal}.

                \begin{figure}[tb]
                    \begin{adjustwidth}{-1in}{-1in}
                    \centering
                    \includegraphics[width=\textwidth]{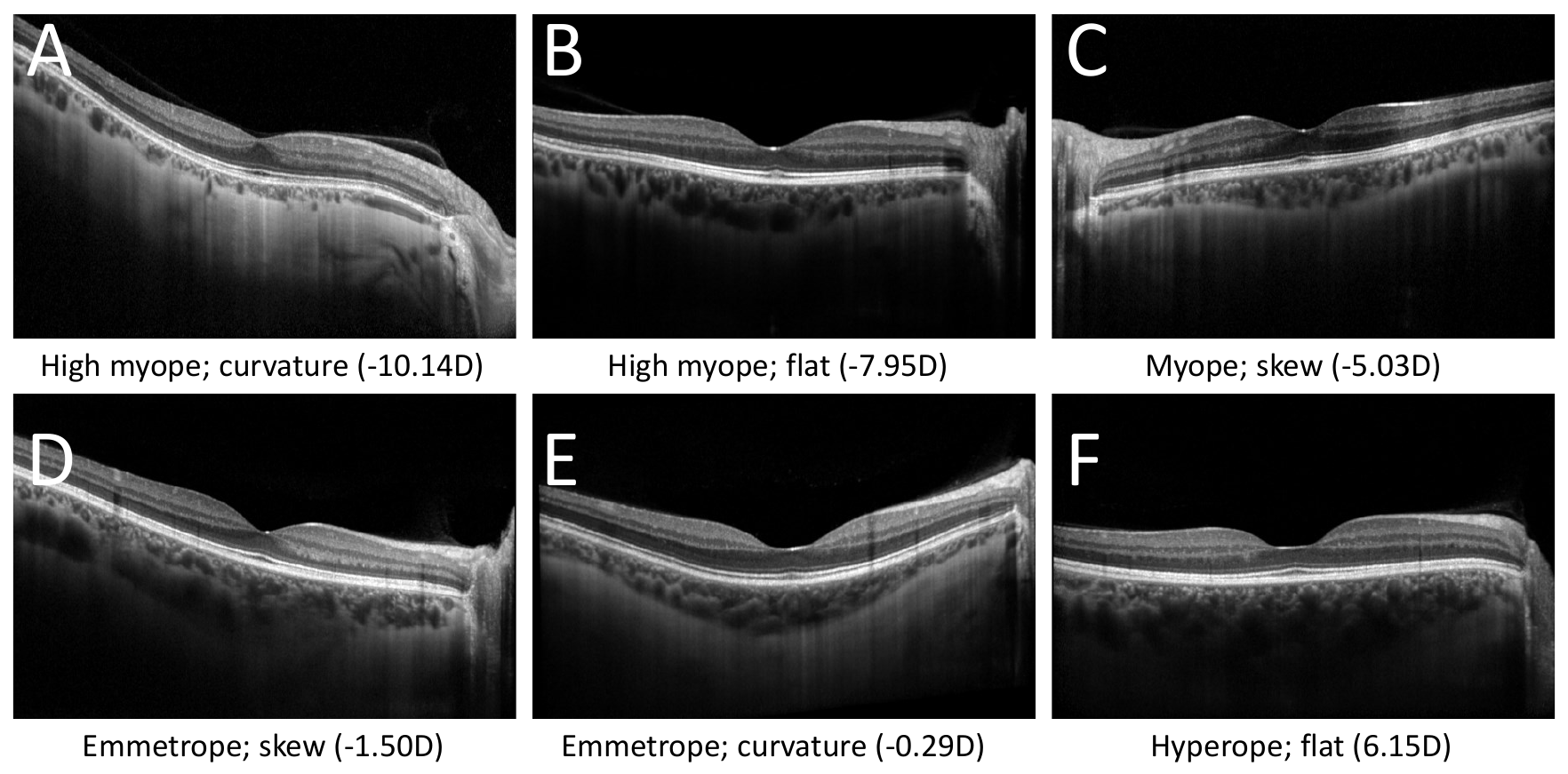}
                    \end{adjustwidth}
                    \caption[Different examples of choroid skew and curvature for varying refractive errors.]{A selection of \acrshort{OCT} B-scans which show varying curvature and skew, with their refractive error annotated.}
                    \label{fig:INTRO_ROI_refr_skew_curve}
                \end{figure}

                \acrshort{SFCT} is a one-dimensional straight line distance measure which is highly susceptible to changes in anisotropic pixel length-scales. As will be discussed in chapter \ref{chp:chapter-GPET}, this anisotropic scaling and difficulty to follow standardised measurement protocol can have an inordinate impact on comparing measurements between and within patients. Two-dimensional measurements showed less bias because of a potential cancellation effect between the different \acrshort{ROI}s. For example, in figure \ref{fig:INTRO_ROI_impact}(B), the red and blue shaded regions form two distinct triangles either side of the overlapped portion of each \acrshort{ROI} (shaded green region), and thus contribute similar amounts to their respective image- and choroid-aligned area value, respectively. However, in cases where there is choroidal curvature, this cancellation effect is lost and can cause measurement error like the choroid in figure \ref{fig:INTRO_ROI_impact}(C) where area deviated most (table \ref{tab:INTRO_ROI_metrics}). 

                Figure \ref{fig:INTRO_ROI_refr_skew_curve} shows six \acrshort{OCT} B-scans from eyes with varying refractive error --- measured using the Oculus Myopia Master (Oculus, Wetzlar, Germany) --- visualising the different kinds of curvature and skew that can occur across the spectrum of refractive error. Skewedness of the intra-ocular structures is primarily due to poor acquisition, and curvature from high myopia \cite{arrigo2023clinical}. However, modern day \acrshort{OCT} image capture has added functionality to circumvent choroidal curvature and high myopia \cite{engineering_perfect_unknown}. This can be seen in panel (B), whose eye had a refractive error of approximately 8D, but showed no signs of curvature. This is in contrast to the degree of curvature and skew observed in panel (A) for the eye whose refractive error was -10D. The intra-ocular structures are skewed in panels (C) and (D) from poor acquisition, while in panel (E) we see significant curvature for an emmetropic eye (-0.29D). Finally, in panel (F) we see that for this hyperopic eye (6.15D), the choroid is large and the intra-ocular structures are flattened due to the size of the eye. These examples suggest that the extent of curvature and skewedness can occur across the range of refractive errors.
 
            \end{mysubsubsection}

        \end{mysubsection}
            
        \begin{mysubsection}[]{OCT volume scan}\label{subsubsec:ch1_INTRO_oct_volume}

            The introduction of automatic segmentation methods has permitted near real-time segmentation of the choroid across macular \acrshort{OCT} volume scans. Measuring the macular choroid is a more representative quantification than subfoveal choroid thickness \cite{xie2021evaluation}, allowing choroid volume and thickness to be estimated at different locations in the macula.

            An \acrshort{OCT} volume is made up of several, parallel and linear \acrshort{OCT} B-scans spaced evenly across the macula. Given an \acrshort{OCT} volume whose choroid has been fully segmented, measurements across the macular choroid can be estimated by interpolating choroid-derived thickness and volume (and vessel density) across the B-scans. This constitutes generating a two-dimensional thickness map which encodes the depth of the choroid at each location on the macula and en face localiser. 

            \begin{figure}[tb]
                \centering
                \includegraphics[width=\textwidth]{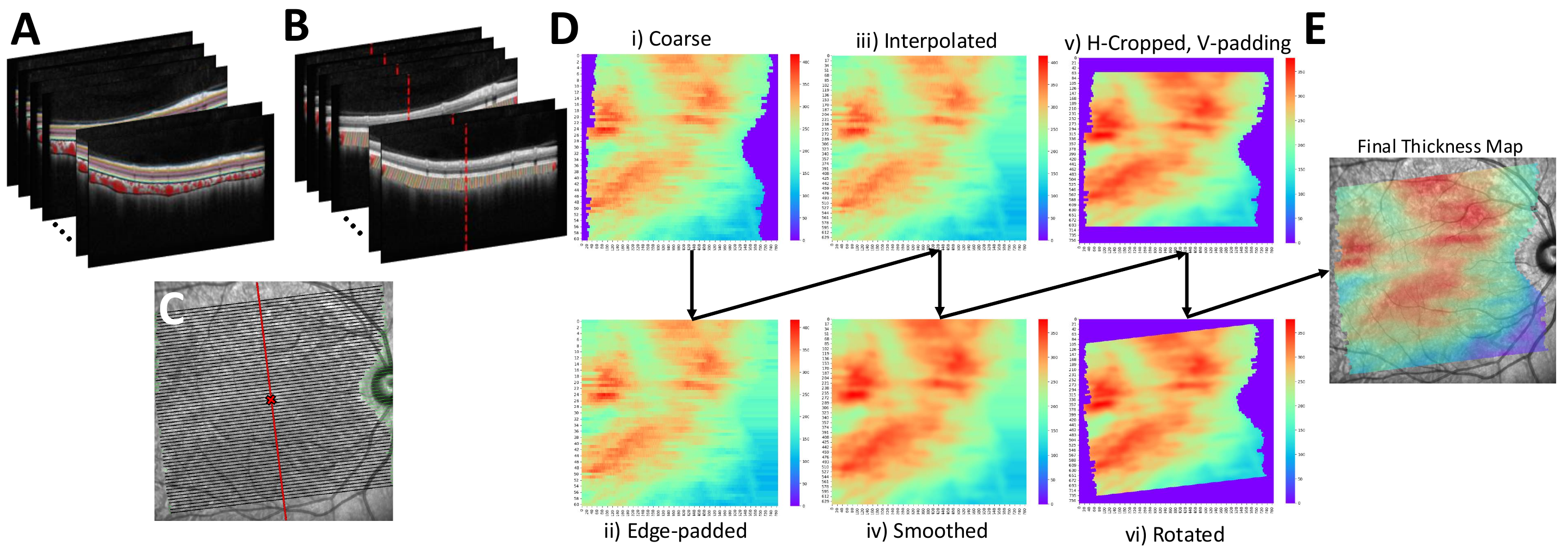}
                \caption[Procedure to generate spatial en face choroid maps from \acrshort{OCT}.]{Procedure to generate a choroid thickness map. (A) \acrshort{OCT} volume scan with segmentations overlaid. (B) Valid thicknesses measured for the choroid across the volume scan. (C) Thickness array alignment using fovea. (D) Step-wise process to generate thickness map. (E) Final thickness (heat)map overlaid onto en face localiser, with redder colours implying higher depth/density.}
                \label{fig:INTRO_thickness_map_diagram}
            \end{figure}

            The procedure generating the thickness map is shown in figure \ref{fig:INTRO_thickness_map_diagram}(D) and described below. the procedure for measuring choroid vessel maps is analogous.
            \begin{enumerate}\setlength\itemsep{0em}
                \item For every B-scan, thickness measurements are taken at every available A-scan, measuring the micron distance between the \acrshort{RPE}-Choroid and Choroid-Sclera boundary, locally perpendicular to the \acrshort{RPE}-Choroid boundary (figure \ref{fig:INTRO_thickness_map_diagram}(B));
                \item Each B-scans' thickness array is aligned (figure \ref{fig:INTRO_thickness_map_diagram}(C)) with the lateral position of the fovea detected on the en face localiser (red dotted and solid lines in figures \ref{fig:INTRO_thickness_map_diagram}(B) and \ref{fig:INTRO_thickness_map_diagram}(C), respectively). Thickness array alignment per B-scan is required since the anatomical layer segmentations (black diagonal lines in figure \ref{fig:INTRO_thickness_map_diagram}(C)) do not necessarily cover the full lateral width of the acquisition region of interest (green diagonal lines in figure \ref{fig:INTRO_thickness_map_diagram}(C));

                \item The aligned thickness arrays are stacked along the frontal plane to create a coarse, two-dimensional map of thickness values (\ref{fig:INTRO_thickness_map_diagram}(Di));
                \item This is padded horizontally by duplicating the edge values to prevent edge artefacts from interpolation (\ref{fig:INTRO_thickness_map_diagram}(Dii));
                \item This is then interpolated to the same pixel resolution as the SLO localiser using bi-linear interpolation (\ref{fig:INTRO_thickness_map_diagram}(Diii));
                \item Since the raw thickness arrays only sample the depth of the choroid at discrete locations, a Gaussian filter whose standard deviation equals the frontal-plane pixel resolution --- the pixel distance between sequential B-scans --- is applied for smoothing (\ref{fig:INTRO_thickness_map_diagram}(Div));
                \item The smoothed map is vertically padded to centre the map onto the fovea of the en face localiser SLO, and the padded edges from step (2) are removed (\ref{fig:INTRO_thickness_map_diagram}(Dv));
                \item Finally, this map is rotated to the angle of elevation of the acquisition \acrshort{ROI} (\ref{fig:INTRO_thickness_map_diagram}(Dvi)), as shown by the green and black diagonal lines in figure \ref{fig:INTRO_thickness_map_diagram}(C).
                
            \end{enumerate}

            \begin{figure}[tb]
                \centering
                \includegraphics[width=\textwidth]{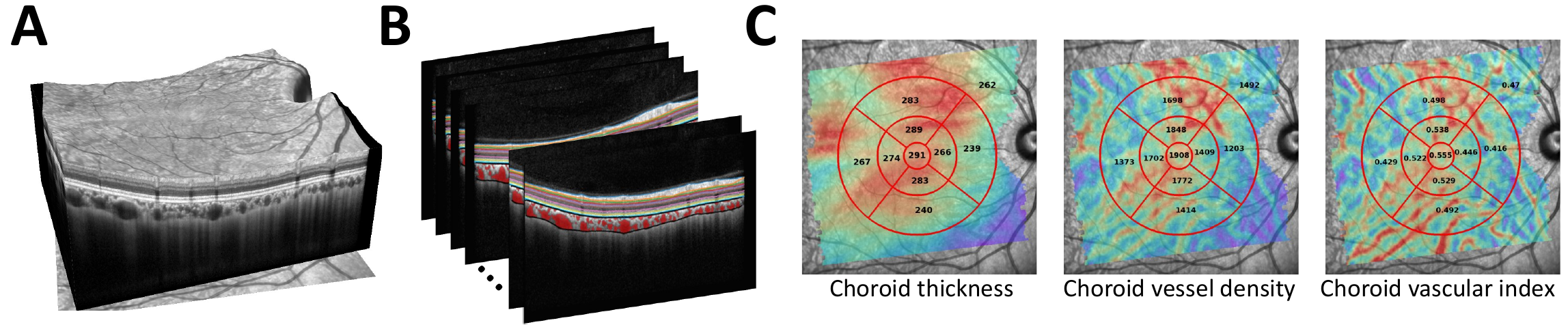}
                \caption[Diagram of generating spatial choroid measurements across the macula from \acrshort{OCT} volume scans.]{Diagram of how measurements are computed for a macular \acrshort{OCT} volume scan. (A) 3D render of an \acrshort{OCT} volume scan. (B) Sequential B-scans with segmentations overlaid. (C) Choroid thickness, vessel density and \acrshort{CVI} (heat)maps, with average measurements along the \acrshort{ETDRS} grid overlaid (redder colours imply higher depth/density).}
                \label{fig:INTRO_volume_diagram}
            \end{figure}

            The thickness map (in microns) can then be interpolated to estimate choroidal volume (in mm$^3$), or can be averaged for more localised measurements. Choroid vessels maps of interest include vessel density in micron$^2$, volume in mm$^3$ and \acrshort{CVI} (dimensionless). 
            
            A standard macular grid was developed over 30 years ago from a landmark, multi-centre clinical trial the Early Treatment Diabetic Retinopathy Study (\acrshort{ETDRS}) \cite{early1991grading}. The \acrshort{ETDRS} grid centres three concentric circles with 1mm, 3mm and 6mm diameters at the foveal pit as seen on the en face localiser, and further splits the 3mm and 6mm circles into four, equally sized quadrants. The quadrants are aligned with the angle of elevation of the \acrshort{OCT} volume's acquisition \acrshort{ROI}. Average thickness, vessel density, \acrshort{CVI} and interpolated volumes are measured in these 9 sub-fields. Figure \ref{fig:INTRO_volume_diagram} shows representative choroid thickness, vessel density and \acrshort{CVI} maps across the macula, generated from an OCT volume scan, superimposed onto the en face localiser with measurements in each sub-field of the \acrshort{ETDRS} grid. 
            
        \end{mysubsection}

        \begin{mysubsection}[]{OCT peripapillary scan} \label{subsec:ch1_INTRO_measure_peri}
        
            \begin{figure}[tb]
                \centering
                \includegraphics[width=\textwidth]{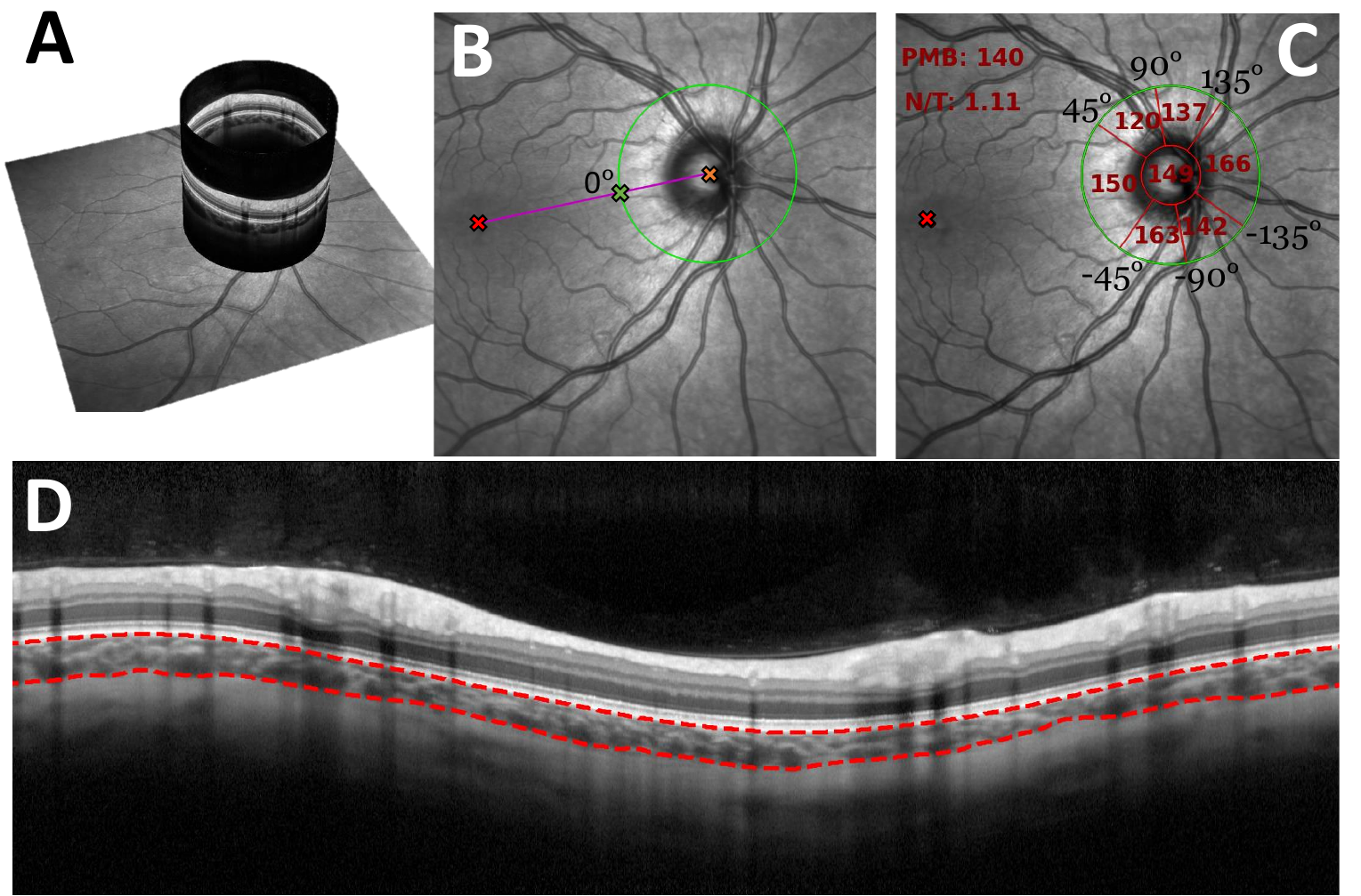}
                \caption[Diagram of generating spatial choroid measurements in \acrshort{OCT} peripapillary B-scans.]{Diagram demonstrating how peripapillary thickness measurements for an \acrshort{OCT} peripapillary B-scan are made. (A) Circular \acrshort{OCT} B-scan overlaid onto the en face localiser. (B) Detection of the centre of the temporal peripapillary grid. (C) Peripapillary grid definition on the en face localiser, with average choroid thickness values overlaid per sub-field. (D) Circular \acrshort{OCT} B-scan with choroid layer segmentation overlaid.}
                \label{fig:INTRO_peri_diagram}
            \end{figure}

            A peripapillary B-scan is a circular scan centred at the optic disc and is continuous along the frontal plane (the lateral ends of the B-scan are continuous). Figure \ref{fig:INTRO_peri_diagram} shows the procedure to obtain average choroid thickness in each of the peripapillary sub-fields, often used for assessing glaucoma \cite{engineering_glaucoma}. These sub-fields measure chorioretinal thickness spatially around the optic disc, as the circular B-scan traverses the disc's circumference at an offset. Panel (A) shows the en face localiser with the circular B-scan overlaid. 
            
            Upon choroid layer segmentation of the choroid in a peripapillary \acrshort{OCT} B-scan (figure \ref{fig:INTRO_peri_diagram}(D)), we first generate a thickness array. This measures a micron distance for every A-scan (column). We do not measure locally perpendicular to the RPE-Choroid boundary because the sinuous nature of the peripapillary B-scan is a result of the circular-shaped acquisition line, and not natural curvature from the anatomy. This is in contrast to the linear, macular line-scans previously discussed.

            To create the peripapillary grid, we first define the centre (0 degrees) of the temporal peripapillary sub-field (figure \ref{fig:INTRO_peri_diagram}(B)). This is done by cross-referencing the point along the circular acquisition region (on the en face localiser) which is co-linear with the optic disc centre and fovea centre. This position has a corresponding lateral position along the circular B-scan. If the en face localiser is unavailable, then the centre is taken as the lateral centre of the B-scan. However, without the en face localiser, the peripapillary grid will likely be off-centre which could have significant clinical consequences \cite{tuncer2018does}.
        
            Using this temporal centre, the thickness array can then be divided into temporal (-45 -- 45 degrees), supero-temporal (45 -- 90 degrees), supero-nasal (90 -- 135 degrees), nasal (135 -- -135 degrees), infero-nasal (-135 -- -90 degrees) and infero-temporal (-90 -- -45 degrees) sub-fields. Average thickness values are taken per sub-field, as well as the nasal-to-temporal (N/T) ratio and papillomacular bundle (PMB) (-30 -- 30 degrees), with a final average over the whole B-scan. Figure \ref{fig:INTRO_peri_diagram}(C) shows these values overlaid onto the peripapillary grid, with the N/T and PMB values shown in the top-left. 
            
        \end{mysubsection}

        \begin{mysubsection}[]{Comparing measurements}\label{subsec:INTRO_metrics}

            Table \ref{tab:INTRO_metrics} defines and describes the metrics used throughout this thesis to compare agreement between two sets of measurements, either choroid-derived measurements or segmentation maps outputted from a segmentation model. This section serves as a brief description of these metrics for the reader. For choroid-derived metrics, consider sets $x$ and $y$ to be of length $n$. For segmentation metrics, we define the $2 \times 2$ confusion matrix $C$ which compares the agreement between two binary segmentation maps as
            \begin{equation}
                C = \left(\begin{array}{cc}
                   TP & FP \\
                   FN & TN \\
                   \end{array}\right).
            \end{equation}
            Upon comparing a predicted, binary segmentation map $S_p$ to the true segmentation map $S_t$, we define $TP$ and $TN$ as the number of true positives and negatives, i.e. the sum of all positive (1) and negative (0) matches between maps $S_p$ and $S_t$, respectively. $FP$ is the number of false positives, i.e. where $S_p$ is equal to 1 but $S_t$ is equal to 0, and vice versa for $FN$, the number of false negatives.

            \begin{table}[!b]
                \begin{adjustwidth}{-0.5in}{-0.5in}
                \centering
                \scalebox{0.9}{\begin{tabular}{p{3.5cm}lllp{3cm}}
\toprule
\multicolumn{1}{l}{Name} & Range & Formula & Optimum & Description \\
\midrule
\multicolumn{1}{l}{Choroid-derived metrics} &  &  &  &  \\
\cmidrule(l){1-1}
Pearson ($r$) & {[}0, 1{]} & $\nicefrac{\text{Cov}(x, y)}{\sigma(x)\sigma(y)}$ & High & Linear correlation \\
Spearman ($s$) & {[}0, 1{]} & $1 - \nicefrac{6}{n(n^2-1)}\sum_{i=1}^{n}d_i^2$ & High & Monotonic correlation \\
Mean Difference (MD) & Any & $\nicefrac{1}{n}\sum_{i=1}^{n}(x_i - y_i)$ & Low & Prediction error \\
Mean Absolute Error (MAE) & $\geq$ 0 & $\nicefrac{1}{n}\sum_{i=1}^{n}|x_i - y_i|$ & Low & Unsigned prediction error \\
Element-wise measurement noise ($\lambda_i$) & $\geq$ 0 & $\nicefrac{\sigma([x_i, y_i])}{\sigma(x)}$ & Low & Proportional measurement variability \\
\multicolumn{1}{l}{} &  &  &  &  \\
\multicolumn{1}{l}{Segmentation metrics} &  &  &  &  \\
\cmidrule(l){1-1}
Precision & {[}0, 1{]} & $\nicefrac{\text{TP}}{\text{TP} + \text{FP}}$ & High & Positive predictive value \\
Recall & {[}0, 1{]} & $\nicefrac{\text{TP}}{\text{TP} + \text{FN}}$ & High & True positive rate (sensitivity) \\
Dice coefficient & {[}0, 1{]} & $\nicefrac{2(\text{Precision} \times \text{Recall})}{\text{Precision} + \text{Recall}}$ & High & Harmonic mean of precision and recall \\
Area-under-curve (AUC) & {[}0, 1{]} & NA & High & True positive to false positive trade-off \\
\bottomrule
\end{tabular}}

                \end{adjustwidth}
                \caption[Description of choroid-derived and segmentation metrics used in this thesis.]{Definition and brief description of metrics used throughout this thesis for choroid-derived measurements and for segmentation maps.}
                \label{tab:INTRO_metrics}
            \end{table}

            \begin{mysubsubsection}[]{Choroid-derived metrics}

                For comparing choroid-derived measurements, we use Pearson $r$ and Spearman $s$ correlation coefficients, mean difference, mean absolute error, and element-wise measurement noise $\lambda$. Pearson $r$ measures the linear relationship between measurements in $x$ and $y$, while Spearman $s$ measures their monotonic relationship. In table $\ref{tab:INTRO_metrics}$, $\text{Cov}(x,y)$ is the covariance between measurement sets $x$ and $y$, $\sigma(x)$ measures the standard deviation of the values in $x$ and $d_i$ measures the difference in rank between $x_i$ and $y_i$, after ordering sets $x$ and $y$. Pearson's $r$ assumes a linear relationship between $x$ and $y$, while Spearman does not. On the assumption that $x$ and $y$ are measuring the same thing, then there should by definition hold a linear relationship. Anything other than a perfect linear relationship would imply disagreement between the approaches which computed the measurements in $x$ and $y$. In the case where a perfect linear relationship exists, Pearson and Spearman correlation coefficients are equal.

                While $r$ and $s$ provides a dimensionless representation of agreement between $x$ and $y$, this sometimes lacks clear interpretation on how close $x$ and $y$ match each other. Mean difference (\acrshort{MD}) and mean absolute error (\acrshort{MAE}) are more interpretable measures of agreement between $x$ and $y$. This is because they are measured in the same units as the original data. However, \acrshort{MAE} is a more robust metric to assess agreement than \acrshort{MD}. This is because errors are averaged using their absolute value, so they will contribute positively toward the error regardless of their sign (positive/negative). Conversely, \acrshort{MD} has the potential to cancel out error if their magnitude is equal but with opposite signs, and this cancellation effect also has the effect of shrinking the systematic bias estimate of asymmetric error distributions. Thus, \acrshort{MAE} provides a more accurate and reliable reflection of the magnitude of error compared to \acrshort{MD}, making it a more informative metric.

                The aforementioned metrics are a population-level description of agreement between $x$ and $y$. However, it is of particular interest to measure an individual-level description of agreement, which is what $\lambda$ achieves, introduced by Engelmann, et al. \cite{engelmann2024applicability} in his study on assessing the repeatability of retinal fractal dimension in oculomics. Say that each element of $x$ and $y$ represents a measurement of an individual eye. Then
                \begin{equation}
                    \lambda = \frac{\text{SD of within-eye measurements}}{\text{SD of between-eye measurements}}.
                \end{equation}
                Specifically, from the equation in table \ref{tab:INTRO_metrics}, for the $i^{\text{th}}$ eye the measurement noise is
                \begin{align}
                    \lambda_i &= \frac{\sigma([x_i, y_i])}{\sigma(x)} \\
                              &= \frac{\nicefrac{1}{\sqrt{2}}|x_i - y_i|}{\sigma(x)},
                \end{align}
                where the numerator, $\sigma([x_i, y_i])$, is the within-eye standard deviation between values $x_i$ and $y_i$, which is simply the range between $x_i$ and $y_i$. The denominator is the standard deviation across all eye measurements in $x$, arbitrarily chosen between $x$ and $y$. This ratio expresses the size of the error for the $i^{\text{th}}$ eye in units of the overall population variability, i.e. how many times the size of the error for the $i^{\text{th}}$ eye is contained within the overall population variability. This is particularly helpful to assess the effect size of the measurement error at the subject level.

                Lower values of $\lambda_i$ are desirable, indicating better reproducibility and less noise in individual measurements. $\lambda_i$=0 is the optimal value, representing perfect precision for the $i^{\text{th}}$ eye's measurements in $x$ and $y$. Values where $\lambda_i < 1$ would mean that the within-eye variability is less than the between-eye variability, indicating greater measurement reliability and consistency at the eye-level. A smaller $\lambda_i$ suggests the measurement process between $x$ and $y$ is stable such that any observed variability is more attributable to differences between eyes rather than the measurement error within the $i^{\text{th}}$ eye. Computing $\lambda_i$ for all $i=1,\dots,n$ outputs a distribution of eye-level measurement error, which can be represented with high-level summary statistics.

                Note that this metric is particularly helpful for differentiating between measurement error and the overall measured effect in a study population. Say $x$ and $y$ represent the same measurements produced by the same method, thus they are \textit{repeated} measurements. For a given population (with repeated data to generate $x$ and $y$), the average measurement noise $\overline{\lambda} = \nicefrac{1}{n}\sum^n_i \lambda_i$ represents the average individual-level error relative to the population variability. Using the same population for a particular research objective, such as measuring the effect of a drug or treatment, if the measured effect relative to the population variability is greater than the estimated $\overline{\lambda}$, you can be confident that measurement error does not fully explain the observed signal, and thus what remains is true biological change. This kind of interpretation is crucial for ensuring confidence in an observed effect in research studies.

                A final consideration is when there are repeated measurements per eye to take into account (as will be the case in this thesis). During each examination, multiple measurements $x^{(j)}_i$, $y^{(j)}_i$ for $j=1,\dots,m$ may be taken for each eye $i=1,\dots,n$. For example, an \acrshort{OCT} volume scan may be measured in the 9 sub-fields of the \acrshort{ETDRS} grid \cite{early1991grading} ($m = 9$). In this case, we would average over the $m$ measurements to create a representative value for the $i^{\text{th}}$ eye for both sets of measurements, $\overline{x}_i$, $\overline{y}_i$, where
                \begin{equation}
                    \overline{x}_i = \nicefrac{1}{m}\sum_j^m x^{(j)}_i, 
                \end{equation}
                for $i=1,\dots,n$ and similarly for $\overline{y}_i$. Then we arrive at our setup from before, measuring the eye-level measurement noise for all $n$ eyes using a single comparative pair of values in $\overline{x}$ and $\overline{y}$.
                
            \end{mysubsubsection}

            \begin{mysubsubsection}[]{Segmentation metrics}
            
                Standard segmentation metrics of interest to compare a predicted segmentation map and ground truth segmentation map are \textit{precision}, \textit{recall}, the \textit{Dice} similarity coefficient and the area under the receiver operating characteristic (AUCROC) curve (area-under-curve, \textit{\acrshort{AUC}}).
                
                Given a predicted binary segmentation map, compared to the true segmentation mask, precision measures the positive predictive value of your segmentation task. Thus, precision measures the proportion of correctly labelled pixels against all pixels positively labelled in the predicted map. For choroidal space/vessel segmentation, precision measures the extent of over-segmentation in the model. Recall is the sensitivity of the segmentation model and is the true positive rate, measuring how accurate the model is at predicting positive instances overall. For choroidal space/vessel segmentation, recall measures the extent of under-segmentation in the model \cite{monteiro2006performance}. 
                
                In machine learning, there is an inverse relationship between precision and recall, commonly referred to as the precision-recall trade-off. Increasing precision typically reduces recall and vice versa, because optimising for one often comes at the expense of another. High precision is achieved by making more conservative positive predictions, which reduces the $FP$ count but may increase the $FN$ count, lowering recall. Conversely, taking a liberal approach to positive predictions (decreasing $FN$) increases recall, but may lower precision (increasing $FP$) \cite{franti2023soft}.
                
                Instead of reporting precision and recall separately, the Dice similarity coefficient combines precision and recall to give an overall representation of predictive performance for a segmentation model. In particular, the Dice coefficient will measure the extent of over- and under-segmentation of a predicted binary segmentation map relative to to its ground-truth segmentation map. It is referred to as the \textit{harmonic} mean between precision and recall, as it assumes an equal weighting between precision and recall \cite{muller2022towards}.

                So far, we have considered segmentation metrics for a predicted binary segmentation map. However, segmentation models often output a probabilistic segmentation map, where elements take on values in the range [0, 1]. This map represents the degree of confidence the model has in predicting a positive label for each pixel. The map can be binarised through selecting a threshold value $T$ between [0,1], where precision, recall and Dice coefficient can then be computed. Often we wish to measure the overall model performance using these probability values, rather than the thresholded binary map, and the \acrshort{AUC} does exactly this.
                
                \acrshort{AUC} measures the trade-off between the true positive rate (recall) and false positive rate ($\nicefrac{FP}{FP+TN}$) across different thresholds $T$ in [0, 1]. For choroidal space/vessel segmentation, \acrshort{AUC} provides a single value representing how well the segmentation model's raw predictions are at differentiating between the choroid/vessel and the rest of the \acrshort{OCT} B-scan. Note that all aforementioned segmentation metrics here are bound between 0 and 1, with higher values indicating better performance.

            \end{mysubsubsection}
        
        \end{mysubsection}
        
    \end{mysection}	

    \begin{mysection}[]{Executive summary}

        In this chapter, we introduced the choroid, a crucial vascular layer of the eye which has historically been neglected due to poor imaging quality. However, recent advancements in \acrshort{OCT}, particularly with \acrshort{EDI-OCT} and \acrshort{SS-OCT}, have improved the ability to visualise and measure this microvascular bed. This has contributed to new investigations into its role in systemic disease given emerging evidence of its ability to reflect microvascular dysfunction in hypertensive conditions, renal function, neurodegeneration and systemic inflammation.

        We also observed that the choroid's complex vascular structure presents significant challenges in measurement due to imaging issues like speckle noise, low signal-to-noise ratio, retinal vessel shadowing and ambiguous vessel wall definition. Current approaches to measure the choroid often rely on manual measurement or improper usage of open-source semi-automatic tools, which introduces variability across and within patients, limiting reproducibility and accuracy. Other, more accurate methods remain closed-source, preventing standardisation of choroidal measurements. This emphasises the need for open-source, reproducible methods to enable consistent and accurate measurement of the choroid.

        We also introduced choroidal measurements of interest and found that measurement error can arise from differences in definitions of the Choroid-Sclera boundary, measurement protocol, or image characteristics such as skew and curvature. Therefore, a strict protocol accounting for factors like the presentation of intra-ocular structures on the B-scan is essential for consistent and standardised measurements of the choroid. This is especially pertinent when pixel length-scales are subject to change across individuals. Otherwise, this has the potential to exacerbate potential measurement error in one-dimensional metrics such as choroid thickness.

    \end{mysection}          
    
\end{mychapter}

\begin{mychapter}[]{GPET: Semi-automatic choroid region segmentation using Gaussian process regression} \label{chp:chapter-GPET}
    
    \begin{mysection}[]{Introduction}

        In this chapter, we have a particular interest in segmenting the choroidal space from an \acrshort{OCT} B-scan. Here, we introduce a method which takes a traditional approach to semantic image segmentation using an edge-based model based on Bayesian machine learning. That is, we formulate the problem of segmenting the choroidal space in an \acrshort{OCT} B-scan as identifying the upper and lower boundaries delineating the choroid, i.e. the \acrshort{RPE}-Choroid and Choroid-Sclera boundaries. Advantageously, this method leverages uncertainty quantification and non-parametric learning using a Bayesian modelling approach known as Gaussian process regression.

        Semantic image segmentation is an important tool for deriving information from an image, and in medical imaging plays a crucial role in many applications, facilitating higher-level image analysis such as feature measurement \cite{zhou2022automorph} or disease detection \cite{maccormick2019accurate}. The goal of any image segmentation algorithm is to partition an image into meaningful sub-regions. Edge-based segmentation algorithms do this by identifying boundaries, or edges between those regions. Human visual systems are able to interpret and recognise scenes and objects just based on the basic structure of their outlines \cite{walther2011simple}, making edge-based approaches useful in mimicking some of the mechanisms of our own vision. 

        The most popular frameworks to employ semantic segmentation in medical imaging utilise deep learning and \acrshort{CNN}s. However, this can be a challenge when high quality annotated datasets are limited, which was the case at the time of developing this algorithm. Nevertheless, even if data was indeed available then the ground labels generated for model development would have been done manually. This has the potential to inject human bias into the model development and segmentation procedure, particularly for manual detection of the Choroid-Sclera boundary \cite{maloca2023human}.

        Two common types of models suited for tracing individual edge structures in an image like an OCT B-scan are those which consider the pixels in an image as nodes of a graph, and those which adapt a contour to an edge structure using splines or level sets. Note that both of these methods have been used previously for semi-automatic choroid region segmentation (table \ref{tab:INTRO_region_methods}). However, and in particular to the Choroid-Sclera boundary, these approaches are typically unable to capture uncertainty or incorporate prior knowledge into the segmentation procedure.

        The use of Bayesian frameworks in computer vision has been popularised due to their non-parametric learning and uncertainty quantification. Some image processing applications involve denoising \cite{liu2007using}, object categorisation \cite{kapoor2010gaussian}, detection of pneumonia in chest X-rays \cite{frank2020gaussian} or super-resolution of images \cite{he2011single}. However, there has not been an extensive amount of work on applying Gaussian processes to image segmentation. Freytag, et al. \cite{freytag2012efficient} performed semantic segmentation by using histogram intersection kernels for fast and exact Gaussian process classification. Tu and Zhu \cite{tu2002image} proposed a Bayesian statistical framework to image segmentation by learning object appearance models for different region types, placing prior distributions over region size, region number and boundary smoothness. Simek and Barnard \cite{simek2015gaussian} used a two-dimensional Gaussian process regression model to segment Arabidopsis leaves by modelling the blade and petiole as two random functions, joining them at their boundaries using a smoothing constraint. 
        
        The method introduced in this chapter, Gaussian Process Edge Tracing (GPET), provides a tunable, robust and probabilistic approach to edge-based image segmentation of the choroid in OCT B-scans. \acrshort{GPET} models an edge of interest (\acrshort{EOI}) in an image as a Gaussian process in a regression setting. Using a suitable estimate of the \acrshort{EOI} using gradient-based pre-processing steps and tunable initialisation, the model iteratively searches for edge pixels which it uses to update it's belief on the location of the true \acrshort{EOI}. Moreover, the internal parameters of the model allows the user to tune the fitting procedure to the properties of the \acrshort{EOI}. 

        \begin{mysubsection}[]{Gaussian process regression} 

            Gaussian processes are a Bayesian machine learning technique used to perform regression on a set of observations. Bayesian methods are able to treat the learning process incrementally by updating their beliefs in regions near these observations once they have been incorporated into the model. While providing precision in regions near observations, Gaussian processes also provide a way of capturing the uncertainty in regions farther away from these observations, where the model has little information to learn from. The modelling of this uncertainty induces a distribution of functions that plausibly fits the set of observations. 

            For Gaussian processes to be used in a regression setting, we typically have some noisy observations $y$ which correspond to some underlying function $f$ which we want to model. A Gaussian process regressor models this underlying function $f$ as a Gaussian process and assumes the following relationship
            \begin{equation}\label{eq:gpr_assume_model}
                y(x) = f(x) + \epsilon,
            \end{equation}
            where $\epsilon$ represents the underlying noise --- assumed to be Gaussian --- perturbing the true function values $f(x)$ (which we \textit{don't} have access to during modelling) to the noisy observations $y$ (which we \textit{do} have access to during modelling). Formally, the identically, independently distributed Gaussian noise $\epsilon \sim \mathcal{N}\big(0, \sigma_y^2\big)$ is characterised by the observation noise variance $\sigma_y^2$ which governs the uncertainty that the model assumes in the set of observations $y(x)$.

            We can think of any continuous, real-valued function $f$ as being an infinitely long list, extending to the far ends of the number line (of real values), i.e. a list $\{f(x) : x \in (-\infty, \infty)\}$ defined at input values $x$ within the open interval $(-\infty, \infty)$. However, to prevent dealing with the mathematics of infinity, we can consider the function $f$ evaluated at inputs $x$ which comes from a large, finite discretisation of this infinite input space, $X$, yielding a very long list of function values $\{f(x) : x \in X\}$.

            Although, while we still can't deal mathematically with the extortionate size of this list, we can however place a multivariate Gaussian distribution on them. This distribution exerts a probabilistic belief on what these function values look like and how pairwise function values relate (or co-vary) with each other. This induces a \textit{Gaussian process distribution} which is analogous to a stochastic process --- a generalisation of a joint probability distribution of random variables to a probabilistic distribution of functions \cite{williams2006gaussian}. Simply put, a Gaussian process is a multivariate Gaussian distribution over function values in the very long list $\{f(x) : x \in X\}$. 

            By modelling $f$ as a multivariate Gaussian distribution, the function values $f(x)$ become random values, and thus $f$ becomes a random function which can be fully specified by a mean function $m(x)$ and covariance function $k(x,x')$, defined by
            \begin{align}
                \mu(x) &= \mathbb{E}\left[f(x)\right];\\
                k(x, x') &= \mathbb{E}\left[(f(x)-\mu(x))(f(x')-\mu(x'))\right],
            \end{align}
            where $x, x'$ are inputs, $\mu(x)$ is the expected (average) value of the random function $f$ at input $x$, and covariance $k(x, x')$ measures the correlation value between the random function values at inputs $x$ and $x'$. 
            
            With these ingredients, the problem formulation can be formalised as a random function $f(x)$ modelled by a Gaussian process such that
           \begin{equation}\label{eq:gp_defn}
                f(x) \sim \mathcal{GP}\big(\mu(x), k(x, x')\big).
           \end{equation}
            In relation to segmenting the \acrshort{RPE}-Choroid (upper) and Choroid-Sclera (lower) boundaries in an \acrshort{OCT} B-scan, we consider the input values $x$ to be the column indexes (the horizontal axis) of the B-scan, $f$ to be the true row locations of the upper (or lower) boundary at inputs $x$, and $y$ to be some subset of row locations which we use to model the Gaussian process as closely to $f$ as possible. Thus, since we are working with one-dimensional column and row indexes, we can assume for the following high-level description that inputs $x, x'$ and outputs $f(x), y(x)$ are all one-dimensional values. Consider chapter 2.2 of Rasmussen and William's book for more generalisable details \cite{williams2006gaussian}.

           \begin{mysubsubsection}[]{Prior distributions and covariance functions}

                Using Bayesian statistics to fit functions requires constructing a \textit{prior distribution} for the function values. This prior distribution encodes our initial belief of the properties exhibited by the function $f$ that we wish to fit to the data $y$. The choice of prior reduces to choosing an initial mean function $\mu(x)$ and covariance function $k(x, x')$. However, we can assume without any loss of generality that $\mu(x) = 0$ for all inputs $x$, because it is the covariance function which has the largest impact on model specification and fitting. 
                 
                The covariance function is used to build the covariance matrix between random function values and typically depends on the input observations of these function values. The choice of covariance function determines the properties of the random functions drawn from the Gaussian process distribution. Considering that this approach is being used to model boundaries separating biological tissue, these are typically continuous (or approximately smooth). This is an essential property we want to encode in our covariance function.
                 
                Therefore, because we want to model continuous functions, function values of nearby inputs should be of similar value and hence have large correlation. Thus, a covariance function is chosen which measures the similarity between function values based on the distance between their input locations. These are known as isotropic stationary covariance functions, which are translation invariant and only depend on the distance between inputs $x$ and $x'$. 
                
                The stationary kernels which are of interest in this chapter are the squared exponential and Matern covariance functions. The squared exponential, also known as the radial basis covariance function (\acrshort{RBF}), produce smooth and continuous functions. For input locations $x, x'$, the squared exponential kernel is defined as
                \begin{equation}\label{eq:GP_cov_SE}
                    k(x, x') = \sigma_f^2\exp{\big(-\frac{1}{2\sigma_l^2}r^2\big)},
                \end{equation}
                where $r = |x - x'|$, i.e. the absolute difference between inputs $x$ and $x'$. Importantly, this function value is largest when $x$ is closest to $x'$, i.e. when the input locations of the function values $f(x)$ and $f(x')$ are closest to each other. There are two parameters $\sigma_f$ and $\sigma_l$ which act as the typical deviation (amplitude) from the mean function and sinuosity (curvature) of the random functions drawn from the Gaussian process, respectively. Formally, $\sigma_f$ and $\sigma_l$ are the function variance and length-scale hyperparameters of the covariance function, respectively. $\sigma_f$ dictates the typical amplitude of the prior functions, where approximately 66\% of the values making up each random function lie within $\pm \sigma_f$. $\sigma_l$ is the length-scale which roughly describes the typical distance between turning points of the random function.
                
                \acrshort{RBF}s produce very smooth curves when sampling from the Gaussian process (because this function is infinitely differentiable), but its strong assumption on smoothness and continuity is one of its pitfalls when modelling real-life physical processes. For example, the Choroid-Sclera boundary is not always perfectly smooth, especially in the presence of retinal pathology. A collection of more flexible covariance functions is the Matern class, of which for our purposes we consider only two types, defined as 
                \begin{align}
                    k_{\nu=\nicefrac{3}{2}}(x, x') &= \sigma_f^2\bigg(1 + \frac{\sqrt{3}r}{\sigma_l}\bigg)\exp\bigg(-\frac{\sqrt{3}r}{\sigma_l}\bigg)\\
                    k_{\nu=\nicefrac{5}{2}}(x, x') &= \sigma_f^2\bigg(1 + \frac{\sqrt{3}r}{\sigma_l} + \frac{5r^2}{3\sigma_l^2}\bigg)\exp\bigg(-\frac{\sqrt{5}r}{\sigma_l}\bigg).
                \end{align}
                Here, $\nu$ is a positive parameter that dictates the differentiability, and therefore continuity, of the random functions. Rasmussen and Williams consider $\nu=\nicefrac{3}{2}$ and $\nu=\nicefrac{5}{2}$ to be particularly useful for modelling machine learning problems \cite{williams2006gaussian}. The generalised formula for the Matern class, as well as many other examples of covariance functions, can be found in chapter 4.2 of Rasmussen and William's book \cite{williams2006gaussian}. 

                The use of $r$ in these formulas means that Gaussian processes which use these covariance functions will generate function values with higher covariance when their input locations lie close to each other. This property (and the differentiability of the covariance functions) defines how smooth the random functions drawn from the Gaussian process are.
    
                Once a covariance function has been chosen, assuming a trivial mean function, then the initial prior distribution of the Gaussian process is fully defined and we can draw \textit{realisations} from it, which are just random samples drawn from a multivariate Gaussian distribution. However, we interpret these random samples as random functions. Define the prior distribution of the Gaussian process, $f(x)$, such that
                \begin{equation}\label{eq:prior}
                    f(x) \sim \mathcal{GP}(0, k(x, x')).
                \end{equation}             
                Samples can then be drawn from this prior distribution by defining input points $X_* = \{x_i : i = 1,\dots,n\}$ and drawing a random list of function values $\{f_*(x_i) : i = 1,\dots,n\}$, or random function $f_*$, from a multivariate Gaussian distribution with an $n \times n$ covariance matrix $K(X_*,X_*)$ such that 
                \begin{equation}
                    f_* \sim \mathcal{N}(0, K(X_*,X_*)),
                \end{equation}
                where $K(X_*, X_*)$ is a square $n \times n$ matrix whose $(i, j)^{\text{th}}$ element is the covariance $k(x_i, x_j)$ between function values $f(x_i)$ and $f(x_j)$, dependent on inputs $x_i, x_j$.

                \begin{figure}[tb]
                    \centering
                    \includegraphics[width=\textwidth]{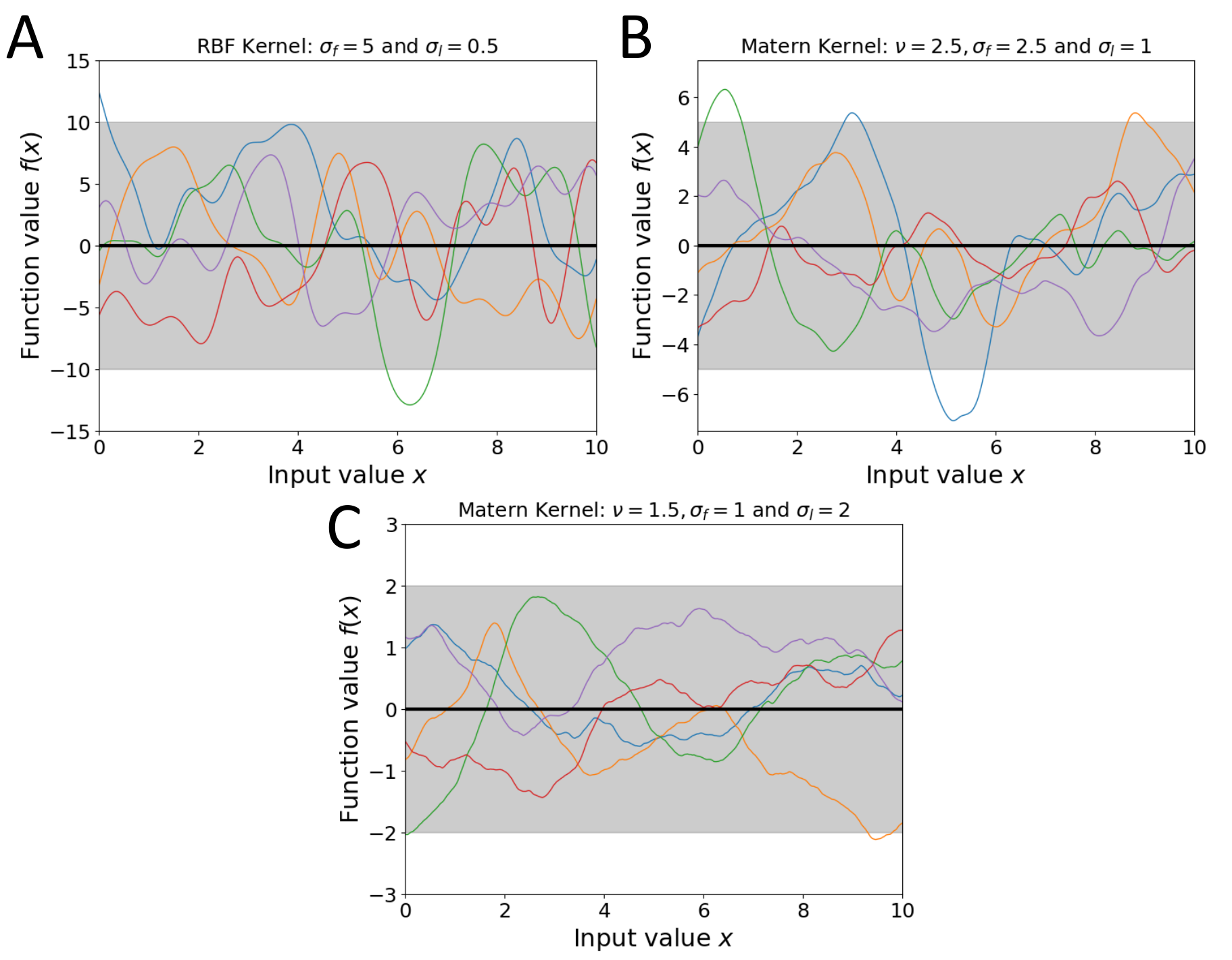}
                    \caption[Example Gaussian process prior distributions with different covariance functions.]{Five random functions drawn from the prior distribution of three different Gaussian processes, each with a different kernel and hyperparameters.}
                    \label{fig:GPET_prior}
                \end{figure}
                
                Figure \ref{fig:GPET_prior} shows 3 example Gaussian processes using the \acrshort{RBF} (A) and Matern (B -- C) covariance functions, with varying settings of $(\sigma_f, \sigma_l)$ (and $\nu$ for the Matern covariance function). Here, we've defined $X_*$ to be the interval $[0, 10]$ and drawn five random functions from their respective prior distributions. The shaded regions represent the point-wise mean-function value for each input location $\pm 2$ standard deviations, corresponding to each prior distributions' 95\% credible region\footnote{This is the preferred terminology for a Bayesian model's 95\% confidence interval.} --- note that the mean function (solid black line) is 0 by design as $\mu(x)=0$ for all $x$. In these examples, the 95\% credible region of each Gaussian process is a rectangular area with its length spanning the input range since there there are no observations fitted to the model (as is the case for a prior distribution). Note that different settings of $(\sigma_f, \sigma_l)$ (and $\nu$) yield different curvatures and amplitudes of the random functions drawn from each Gaussian process, while the different covariance function changes the smoothness of the random functions.
           
           \end{mysubsubsection}

            \begin{mysubsubsection}[]{Posterior distributions}

                In the aim of model-fitting its now in our interest to update our Gaussian process by incorporating a set of observations $\{(x_i, f(x_i)) : i=1, \dots, n\}$ representing the true function $f$. However, the observations are typically corrupted by an element of noise in real-world problems. Denote these noisy observations as $\mathcal{D} = \{(x_i, y(x_i)) : i=1,\dots,n\}$ where $y(x_i) \sim \mathcal{N}(f(x_i), \sigma_y^2)$, where we assume $\sigma_y^2$ is fixed and known and that $y(x_i)$ deviates around the underlying true function value $f(x_i)$ for all inputs $x_i$. By integrating these noisy observations into our Gaussian process, we update our probabilistic belief about the true underlying function which induces a \textit{posterior distribution}. 
                
                This posterior distribution allows us to make predictions, or sample random functions as we did in the previous section, using the properties imposed by the covariance function \textit{and} after taking into account the noisy function values, $y$, we've observed. In order to do this, we first form the joint distribution between $y$ and the function values we want to know (or predict/sample at), $f_*$. 
                
                Define vector $\bm{y} = \{y(x_i) : i=1,\dots,n\}$ as the list of observed function values at locations $X = \{x_i : i=1, \dots, n\}$. Similarly define $\bm{f_*} = \{f_*(x_j) : j=1,\dots,n_*\}$ as the function values we want to predict at locations $X_* = \{x_j : j=1, \dots, n_*\}$. This joint distribution is defined as
                \begin{equation}\label{eq:gp_joint_dist}
                    p(\bm{y}, \bm{f_*}) = \mathcal{N}\Bigg(\begin{bmatrix} \bm{y} \\ \bm{f}_*\end{bmatrix}; \begin{bmatrix} \bm{0} \\ \bm{0}\end{bmatrix}, \begin{bmatrix} K(X,X) + \sigma_y^2\mathbb{I}_n & K(X, X_*) \\ K(X_*,X) & K(X_*, X_*)\end{bmatrix}\Bigg),
                \end{equation}
                where $\mathbb{I}_n$ is the $n \times n$ identity matrix. The covariance matrix of this joint distribution is a block matrix comprising of 4 matrices computed using the chosen covariance function and the input locations of the observations, $X$ and input locations we want to predict/sample at, $X_*$. 
                 
                The posterior distribution of the Gaussian process is then induced by conditioning on the known, but noisy, observations. We restrict the samples we can draw from the joint distribution to only describe functions which accommodate these noisy observations. The posterior distribution of this Gaussian process, after observing our noisy observations, is defined as
                \begin{align}\label{eq:gp_posterior}
                    p(\bm{f}_* : \bm{y}) &= \mathcal{N}(\bm{f}_*; \mathbb{E}[\bm{f}_*], \text{Cov}[\bm{f}_*]); \\\label{eq:gp_updated_mean}
                    \mathbb{E}[\bm{f}_*] &= K(X_*, X)[K(X,X)+ \sigma^2_y\mathbb{I}_n]^{-1}\bm{y};\\ \label{eq:gp_updated_cov}
                    \text{Cov}[\bm{f}_*] &= K(X_*, X_*) - K(X_*, X)[K(X,X)+ \sigma^2_y\mathbb{I}_n]^{-1}K(X, X_*).
                \end{align}
                The additive diagonal matrix $\sigma_y^2\mathbb{I}_n$ is used to perturb the posterior functions away from the observations (and also for numerical stability during sampling). This conditioning updates the mean function and covariance matrix shown by equations \eqref{eq:gp_updated_mean} and \eqref{eq:gp_updated_cov}, respectively. This new distribution extracts an update on our belief on the mean and covariance of the unobserved function values $\bm{f}_*$ of the Gaussian process. The mathematical details of constructing this posterior distribution is shown in appendix A.2 of Rasmussen and William's book \cite{williams2006gaussian}. The second term in $\text{Cov}[\bm{f}_*]$ implies that with the more data observed, the more certain we are of predictions --- especially for function values whose input locations are near the observations' input locations. 

                Typically in Bayesian machine learning, the hyperparameters of the covariance function $\sigma_f, \sigma_l$ and observation noise $\sigma_y$ are also updated to accommodate the new observations in the posterior distribution\footnote{However, this is not always necessary and they can remain fixed, if desired.}. Hyperparameter optimisation collates all kernel hyperparameters and observation noise into $\bm{\theta}$, and these are optimised through maximising the log marginal likelihood of the observation set,
                \begin{align}
                    \hspace*{-0.25cm}\bm{\theta} &= \argmin_{\bm{\theta}}\Big\{\text{log}\hspace{3pt} p(\bm{y} \hspace{3pt} | \hspace{3pt} X, \bm{\theta})\Big\};\label{eq:optim_theta}\\ \label{eq:marginal_likelihood}
                    \hspace*{-0.25cm}\text{log} \hspace{3pt} p(\bm{y} \hspace{3pt} | \hspace{3pt} X, \bm{\theta}) &= -\frac{1}{2}\bm{y}^{\trans}A^{-1}\bm{y} - \frac{1}{2}\text{log}\hspace{1pt}\big|A\big| - \frac{n}{2}\text{log}\hspace{1pt}2\pi,
                \end{align}
                with $n = |\mathcal{D}|$, i.e. the sample size, and $A = K(X,X)+\sigma_y^2\mathbb{I}$. For the equations above and for the remainder of this chapter, we implicitly assume conditioning on $\bm{\theta}$ and test input locations $X_*$ for readability.

                \begin{figure}[tb]
                    \centering
                    \includegraphics[width=\textwidth]{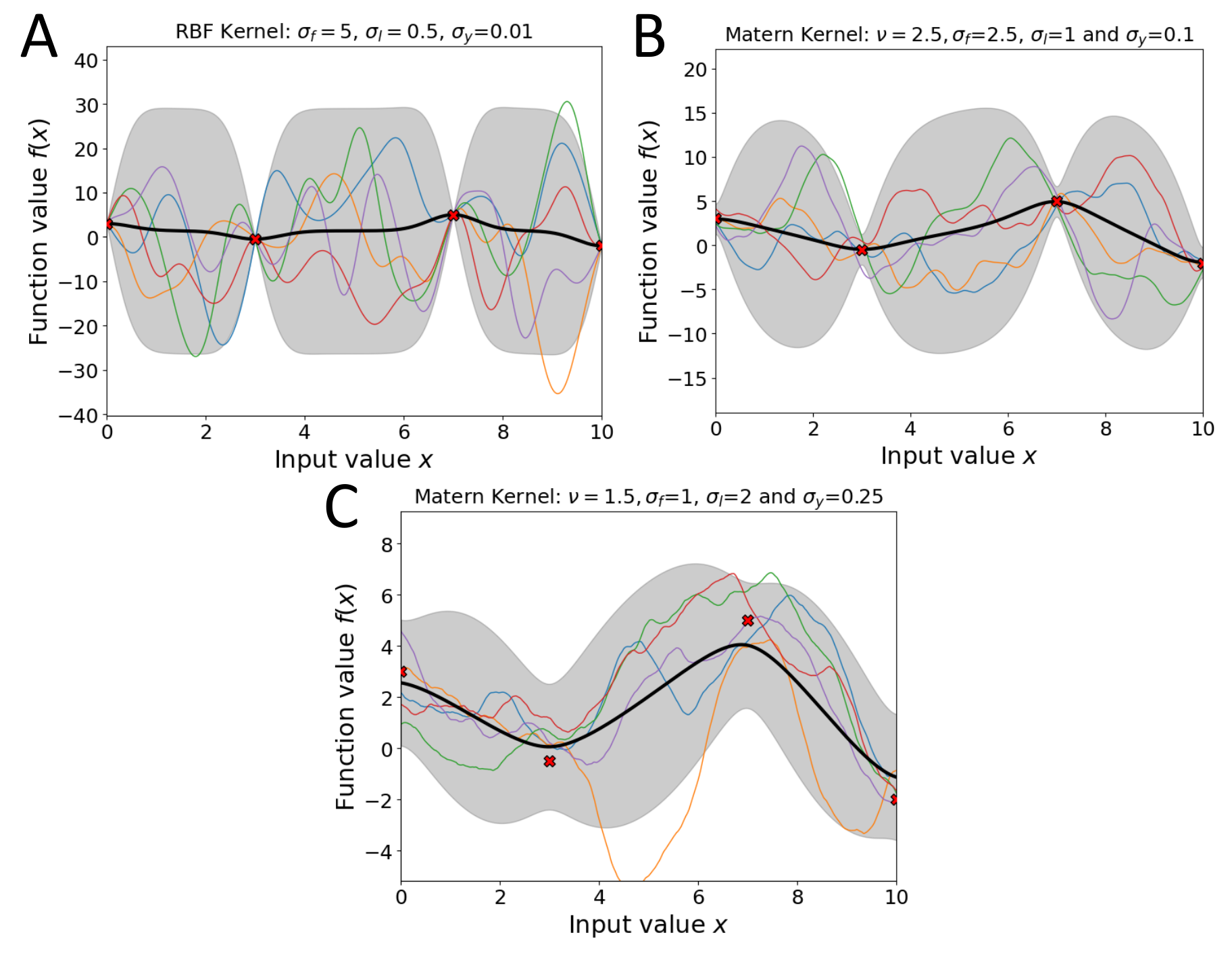}
                    \caption[Example Gaussian process posterior distributions with different covariance functions.]{Five random functions drawn from the posterior distribution of three different Gaussian processes, each with a different covariance function, hyperparameters, and different observation noise $\sigma_y$.}
                    \label{fig:GPET_post}
                \end{figure}
                
                Equipped with this posterior distribution we can now draw functions $f_*$ from it in a similar fashion to the prior distribution. Therefore, we draw a list of function values $\{f_*(x_i) : i = 1,\dots,n\}$ for a random function $f_*$ from the posterior distribution, which is simply a multivariate Gaussian distribution such that
                \begin{equation}
                    f_* \sim \mathcal{N}(\mathbb{E}[\bm{f}_*], \text{Cov}[\bm{f}_*]),
                \end{equation}
                where $\mathbb{E}[\bm{f}_*]$ and $\text{Cov}[\bm{f}_*]$ are given by equations \eqref{eq:gp_updated_mean} and \eqref{eq:gp_updated_cov}. Figure \ref{fig:GPET_post} shows the Gaussian processes from figure \ref{fig:GPET_prior}, whose prior distributions have now incorporated new observations (red crosses) to generate posterior distributions. Note that hyperparameter optimisation has not happened here because the edge-fitting procedure to be described below does not leverage hyperparameter optimisation until convergence, contrary to typical application of Bayesian machine learning methods.
                
                For each Gaussian process, a different level of noise $\sigma_y^2$ has been chosen to demonstrate how the value of $\sigma_y$ influences the posterior distribution, and the randomly sampled functions, as they near the observations. The change in mean function and 95\% credible regions represent how each of the Gaussian process models have changed their belief in what the underlying function looks like after fitting the data. Note how the credible regions reduce in area near observations, and especially so when $\sigma_y$ is very small. It's important to note that the choice of covariance function, and its hyperparameters ($\sigma_f$, $\sigma_l$), must be chosen according to the data being fitted. Furthermore, we must also have an idea of how much noise, $\sigma_y$, exists in the observations so that the model doesn't become too confident (figure \ref{fig:GPET_prior}(A)), or too uncertain (figure \ref{fig:GPET_prior}(C)) in its predictions around these observations. 
    
            \end{mysubsubsection}

        \end{mysubsection}

    \end{mysection}

    \begin{mysection}[]{GPET's Methodology} 

        This methodology works as an individual edge tracer. As such, the methodology is applied to the \acrshort{RPE}-Choroid and Choroid-Sclera boundaries in series. Given an \acrshort{OCT} B-scan with $M$ rows and $N$ columns, we consider each boundary as being modelled by an underlying true function $f$, whose input values $X = \{i : i=1, \dots, N\}$ are the column index, and function values $\{f(x_i) : x_i \in X, \hspace{2pt} i=1, \dots, N\}$ are row indexes, i.e. $f(x_i) \in [1, M]$.
    
        \begin{mysubsection}[]{Pre-processing}\label{subsec:GPET_method_prep}

            In order to perform edge-tracing for an image with GPET, we must first generate an edge map for each \acrshort{EOI}. An edge map gives an approximation to where the \acrshort{EOI} lies in the image, but not an indication on the specific coordinates --- this latter task is the edge tracing which \acrshort{GPET} performs. Edge maps are common to aid edge-based segmentation methods. Graph-based and level set-methods use approximate edge maps to define a cost function to optimise over \cite{dijkstra2022note, kass1988snakes}, and \acrshort{GPET} also leverages the edge map in a similar way. 
            
            The following steps are used to pre-process an \acrshort{OCT} B-scan to obtain approximate edge maps for the \acrshort{RPE}-Choroid and Choroid-Sclera boundaries:
            \begin{enumerate}\setlength\itemsep{0em}
                \item Denoising: a 5 $\times$ 5 median smoothing filter to reduce speckle noise inherent in an \acrshort{OCT} B-scan. This slides a 5 $\times$ 5 array (or window) across the image and replaces the pixel intensity which the window is centred on with the median value of the pixels in the window. Median smoothing is preferable because the median value is more robust to outliers in localised regions of the image relative to the mean value.
                
                \item Contrast enhancement: application of contrast limited adaptive histogram equalisation (\acrshort{CLAHE}). We break the image into 8 $\times$ 8 tiles, and apply histogram equalisation to each tile using a histogram clipping limit of 5. See section \ref{subsec:MMCQ_intro_CLAHE} for more details on this method.
                
                \item Gradient-based edge detection: a common way to obtain edge maps in an image is to apply a discrete derivative filter across the image, which have been well studied \cite{rosenfeld1969picture, rosenfeld1976digital} and include filters such as Sobel's \cite{danielsson1990generalized}, Robert's \cite{roberts1963machine}, Prewitt's \cite{prewitt1970object} and the Canny edge detector \cite{canny1986computational}. See below for more details.

                \item Edge map enhancement: once edge maps have been obtained using discrete derivative filters, these are post-processed to enhance each edge of interest by using minimum filters and morphological operations, as is discussed below in more detail.
            \end{enumerate}

            \begin{figure}[tb]
                \centering
                \includegraphics[width=\textwidth]{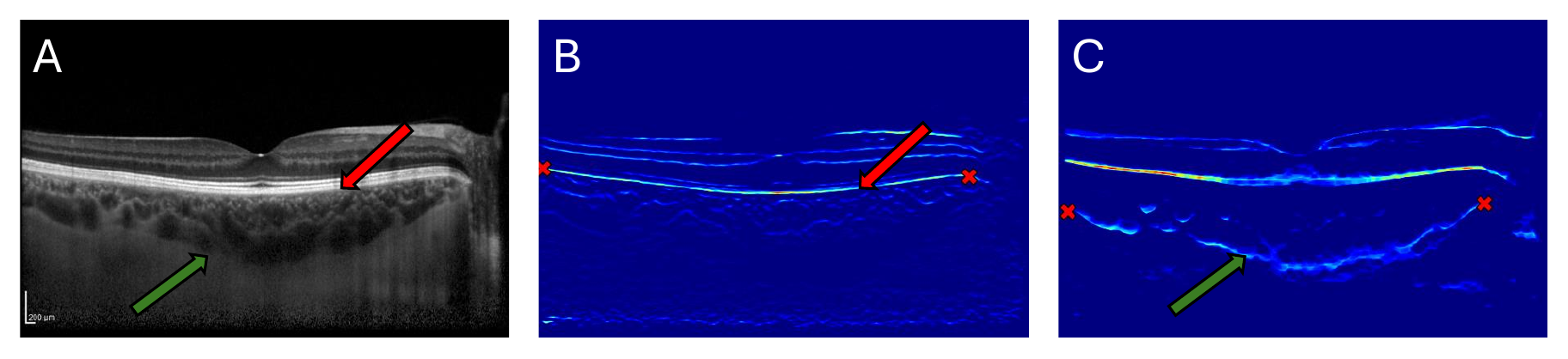}
                \caption[Pre-processing for \acrshort{GPET}'s image analysis pipeline.]{(A) OCT B-scan. (B -- C) Gradient-based edge maps for the \acrshort{RPE}-Choroid boundary (red arrows) and Choroid-Sclera boundary (green arrows). Red crosses overlaid represent the edge endpoints selected manually.}
                \label{fig:GPET_final_prep}
            \end{figure}

            Once the edge maps (also referred to as image gradients) have been generated, an additional pre-processing steps requires the end-user to manually detect the endpoints of the \acrshort{EOI} on the image gradient for both boundaries. This is done through a simple, high-level graphical user interface (GUI) developed using Python, where the image gradient appears on the end-user's home screen and they are prompted to select the endpoints using their track-pad or mouse. Note that this step could be automated to prevent manual intervention, and has indeed been done for a similar application related to a graph-based approach to retinal layer segmentation \cite{chiu2010automatic}. To enable endpoint detection for retinal layer segmentation, Chiu, et al. \cite{chiu2010automatic} leveraged the enhanced contrast between the innermost retinal layer (inner limiting membrane) relative to the vitreous body. While the \acrshort{RPE}-Choroid is typically well illuminated for this automated procedure, this is not the case for the Choroid-Sclera boundary. 
            
            Figure \ref{fig:GPET_final_prep} shows an \acrshort{OCT} B-scan in panel (A) with the approximate edge maps (image gradient) for the \acrshort{RPE}-Choroid and Choroid-Sclera boundaries of the choroid in panels (B) and (C), respectively. Red crosses show the edge endpoints manually selected for each \acrshort{EOI}. It is the image gradient and edge endpoints which initialise the algorithm.

            \begin{mysubsubsection}[]{Gradient-based edge detection}
    
                Many early approaches to image segmentation used edge detection methods based on zero-crossings, discrete derivative operators or histogram thresholding \cite{prewitt1970object, duda1970experiments, roberts1963machine, rosenfeld1969picture}. A variety of these early approaches to image segmentation can be found in Rosenfeld's work \cite{rosenfeld1976digital}. The problem of edge detection has been well studied and with methods such as Canny \cite{canny1986computational}, it can be straightforward to obtain some approximation of the edges in the image. 
                
                However, there is a distinct difference between identifying the edges in an image and linking them to form a meaningful contour, with direct knowledge of the pixel coordinates. Szeliski \cite{szeliski2010computer} notes that this latter task is far more useful than the former for strict boundary delineation and downstream feature measurement, but is more challenging because of its inter-dependency on the quality of the resulting edge map.
    
                \begin{figure}[tb]
                \centering
                \includegraphics[width=\textwidth]{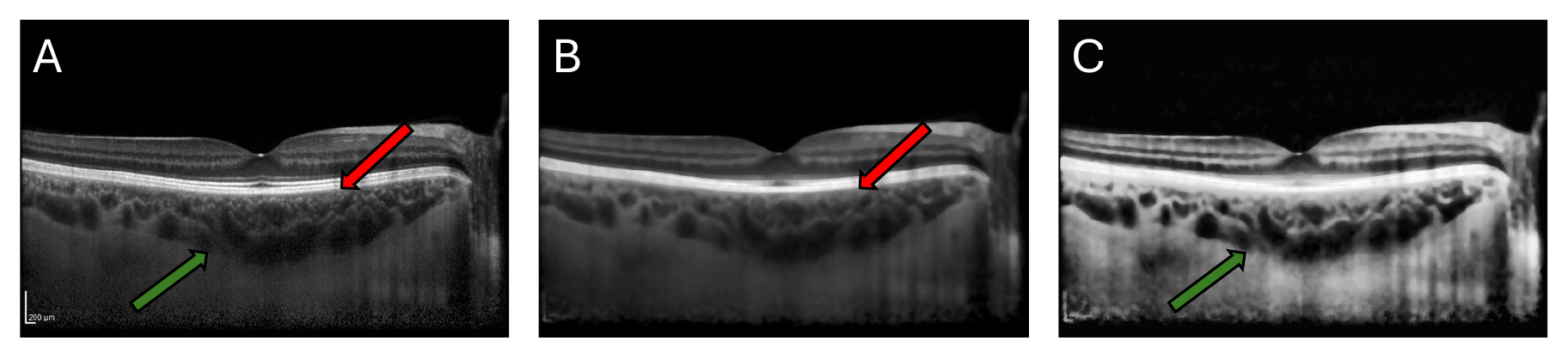}
                \caption[Denoising and contrast enhancement of \acrshort{RPE}-Choroid and Choroid-Sclera boundaries.]{(A) OCT B-scan. (B) Overly smoothed B-scan after median filtering. (C) Application of \acrshort{CLAHE} with high clipping limit for severe enhancement.}
                \label{fig:GPET_den_enh}
                \end{figure}
                
                Thus, it's important to extract a good quality edge map. In the first instance, we pre-process the \acrshort{OCT} B-scan to denoise and enhance the boundaries of interest. Figure \ref{fig:GPET_den_enh} shows the result of smoothing with a large window (5 $\times$ 5) in panel (B) which improves the contrast between the \acrshort{RPE} and Choroid junction (red arrow). In panel (C), this shows the application of \acrshort{CLAHE} with a reasonably large clip limit (5), which improves the contrast between the choroid's posterior vessels and sclera, subsequently improving the visible junction separating the posterior choroid from the sclera (green arrow). Thus, the \acrshort{RPE}-Choroid and Choroid-Sclera edge-maps are detected from the denoised and the enhanced B-scans, respectively. 
    
                Discrete derivative filters are small arrays which encode a set of values which, when \textit{convolved} with the input image, are able to quantify the strength of intensity transition over local regions of the image, such as the Sobel's \cite{danielsson1990generalized}, Robert's \cite{roberts1963machine} or Prewitt's \cite{prewitt1970object} filter. Given that the choroid boundaries are predominantly horizontal, our filters must be encoded to look for predominantly horizontal intensity transitions. Moreover, the \acrshort{RPE}-Choroid boundary transitions from the bright, hyperreflective \acrshort{RPE} to the darker choriocapillaris, while the Choroid-Sclera boundary transitions from the dark, posterior choroidal vessels in Haller's layer to the brighter, homogeneous sclera. Thus, the transitions are the opposite of one another. This is another important detail to encode into the filters. 
                
                A  5 $\times$ 5 extension to the discrete derivative filters proposed by Sobel \cite{danielsson1990generalized} for detecting horizontal edges with different transition patterns, i.e. for light-to-dark \acrshort{RPE}-Choroid and dark-to-bright Choroid-Sclera boundaries are shown below as arrays $f_\text{b2d}$ and $f_\text{d2b}$, respectively.
                
                \begin{equation}
                    f_{\text{b2d}} = \left(\begin{array}{ccccc}
                      -1 & -1 & -2 & -1 & -1 \\
                      -1 & -2 & -3 & -2 & -1 \\
                       0 & 0 & 0 & 0 & 0\\
                      1 & 1 & 2 & 1 & 1 \\
                      1 & 2 & 3 & 2 & 1 \\
                       \end{array}\right) 
                    \quad \text{and} \quad
                    f_{\text{d2b}} = - f_{\text{d2b}}.
                \end{equation}
                The larger filter compared to traditional 3 $\times$ 3 filters promotes smoothness and reduces noise, while the increasing integer values toward the filter's centre ensures that pixels local to the centre contribute more, preventing nearby artefacts from negatively influencing the gradient approximation.

                \begin{figure}[tb]
                \centering
                \includegraphics[width=\textwidth]{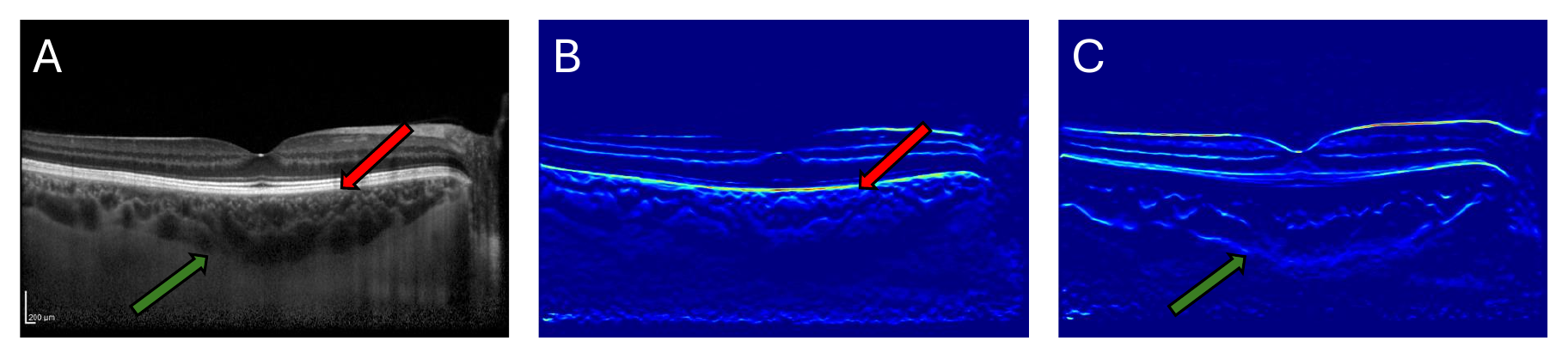}
                \caption[Gradient-based edge maps for \acrshort{RPE}-Choroid and Choroid-Sclera boundaries.]{(A) OCT B-scan. (B -- C) Initial, gradient-based edge maps for the \acrshort{RPE}-Choroid boundary (red arrows) and Choroid-Sclera boundary (green arrows).}
                \label{fig:GPET_init_grad}
                \end{figure}
                
                We obtain the image gradients by convolving these filters with the original image $I$, which are defined for the \acrshort{RPE}-Choroid and Choroid-Sclera edge maps as
                \begin{align}
                    \hat{G}_{\text{upper}} &= I * f_{\text{b2d}}\\
                    \hat{G}_{\text{lower}} &= I * f_{\text{d2b}}.
                \end{align}
                Figure \ref{fig:GPET_init_grad} show the approximate edge maps for the \acrshort{RPE}-Choroid and Choroid-Sclera boundaries in panels (B) and (C) from an \acrshort{OCT} B-scan.

                \begin{figure}[!b]
                    \centering
                    \includegraphics[width=\textwidth]{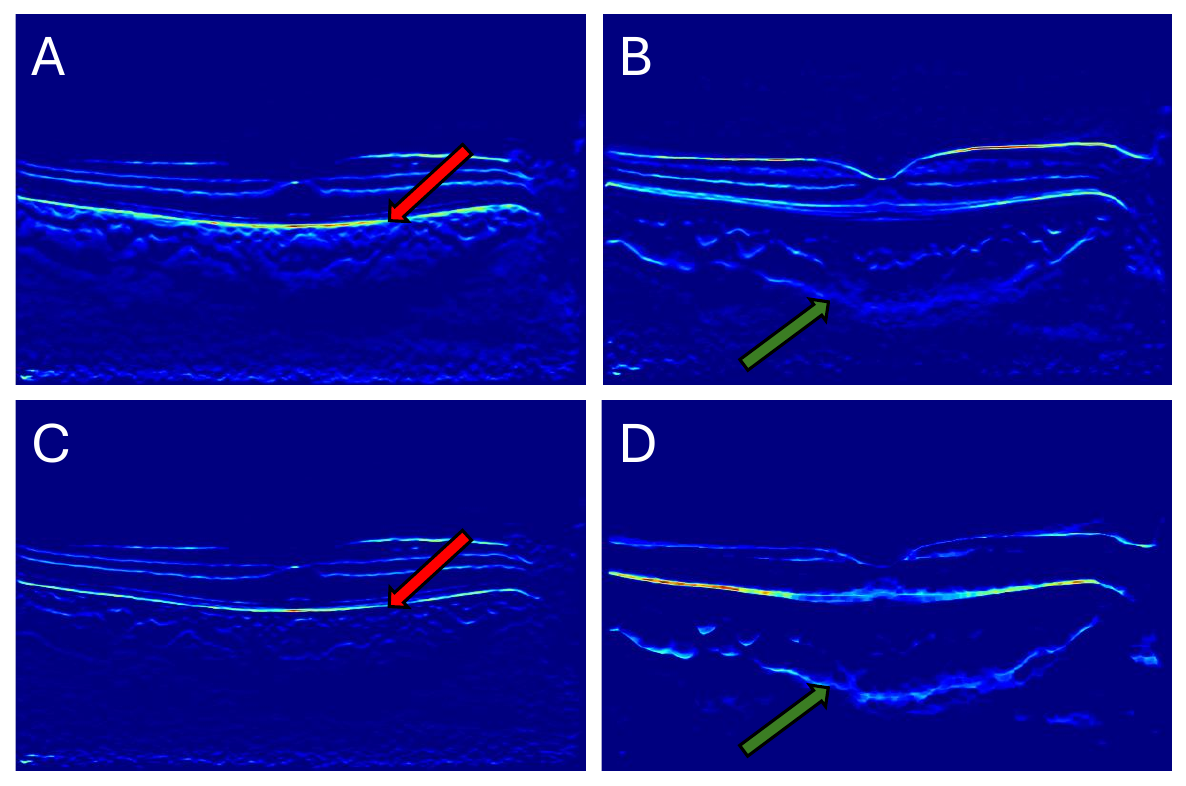}
                    \caption[Enhanced, gradient-based edge maps for \acrshort{RPE}-Choroid and Choroid-Sclera boundaries.]{(A--B) Naive edge maps $\hat{G}_{\text{upper}}$ and $\hat{G}_{\text{upper}}$. (C--D) Enhanced edge maps $G_{\text{upper}}$ and $G_{\text{upper}}$.}
                    \label{fig:GPET_final_grad}
                \end{figure}

                A final step is the enhancement of these edge maps. Note that in figures \ref{fig:GPET_init_grad}(B) and \ref{fig:GPET_init_grad}(C), while there is an obvious signal denoting the approximate location of the \acrshort{RPE}-Choroid and Choroid-Sclera boundaries, there is also signal coming from where individual choroidal vessels meet interstitial space. Moreover, the edge map in panel (C) shows the Choroid-Sclera boundary as quite a fuzzy object. 
                
                To improve the \acrshort{RPE}-Choroid edge map, we can actually apply the dark-to-bright filter because of the thickness of the approximated edge seen in figure \ref{fig:GPET_init_grad}(B). To improve the Choroid-Sclera edge map, we can apply \textit{morphological opening} using a 5 $\times$ 15 window, which removes small, bright and predominantly horizontal objects, and \textit{morphological erosion} using a 5 $\times$ 5 window which removes small objects. These steps are important for removing the signal from individual choroidal vessels and tightening up the Choroid-Sclera boundary signal in the edge map. Thus, the final gradient-based edge maps $G_{\text{upper}}$ and $G_{\text{lower}}$ for the \acrshort{RPE}-Choroid and Choroid-Sclera boundaries, respectively, are defined as
                \begin{align}
                    G_{\text{upper}} &= (I * f_{\text{b2d}}) * f_{\text{d2b}}\\
                    G_{\text{lower}} &= \text{erode}_{5 \times 5}\bigg(\text{open}_{5 \times 15}\big(I * f_{\text{d2b}}\big)\bigg).
                \end{align}
                Figure \ref{fig:GPET_final_grad} shows the original, naive edge maps in the top row (panels A--B), with the enhanced edge maps shown in the bottom row (panels C--D).
    
            \end{mysubsubsection}
            
        \end{mysubsection}

        \begin{mysubsection}[]{Initialisation} 

            For tracing each boundary of the choroid in series, GPET requires initialisation. To initialise the model for an \acrshort{EOI}, we assume that its edge endpoints and image gradient $G$ are known. Let the set of \acrshort{EOI} endpoints be defined as $\mathcal{D}^{(0)} = \big\{X^{(0)}, \bm{y}^{(0)}\big\}$. We assume for simplicity that the \acrshort{EOI} endpoints span the width of the image.
            
            \begin{mysubsubsection}[]{Prior distribution}
                \begin{figure}[tb]
                \centering
                \includegraphics[width=\textwidth]{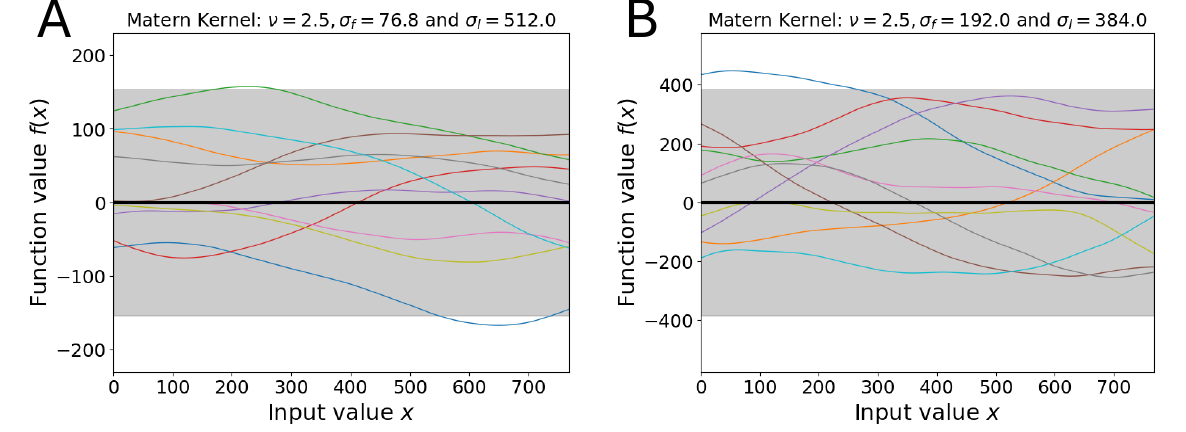}
                \caption[Gaussian process prior distributions for the \acrshort{RPE}-Choroid and Choroid-Sclera boundaries.]{10 random functions drawn from the initial prior distribution of Gaussian process regressors for \acrshort{RPE}-Choroid (A) and Choroid-Sclera (B) boundary detection. The mean function (black solid line) and 95\% credible intervals (grey shaded region) are also shown.}
                \label{fig:GPET_method_prior}
                \end{figure}

                At the beginning of the procedure, following a similar formulation to equation \eqref{eq:gp_defn}, our prior distribution comes from a zero mean Gaussian process with covariance function $k$ (with specified hyperparameters $\bm{\theta} = [\sigma_f, \sigma_l]$) such that 
                \begin{equation}
                    f^{(0)}(x) \sim \mathcal{GP}\big(0, k(x, x' ; \bm{\theta})\big).
                \end{equation}
                This induces a prior distribution where a random function $f^{(0)}$ can be sampled using
                \begin{equation}\label{eq:init_prior}
                    f^{(0)} \sim \mathcal{N}\big(0, K_{**}\big),
                \end{equation}
                where $K_{**} = K(X_*, X_*)$ and $X_* = \{i : i=1,\dots,N\}$. The choice of covariance function for the \acrshort{RPE}-Choroid and Choroid-Sclera boundary is typically dependent on the choroid. However, the \acrshort{RPE}-Choroid often appears as smooth and very slightly parabolic in shape, with little curvature. The Choroid-Sclera boundary is typically less smooth with more curvature and turning points. Thus, the default covariance functions for the upper and lower choroid boundaries --- for an OCT B-scan with pixel resolution 768 $\times$ 768 (height $\times$ width) --- are defined below as
                \begin{align}
                    k_{\text{upper}} &= \text{Matern}(\sigma_f=\nicefrac{M}{10}, \sigma_l=\nicefrac{2N}{3}; \nu=2.5)\\
                    k_{\text{lower}} &= \text{Matern}(\sigma_f=\nicefrac{M}{4}, \sigma_l=\nicefrac{N}{2}; \nu=2.5)
                \end{align} 

                Figure \ref{fig:GPET_method_prior} shows the initial prior distributions for the Gaussian process regressors modelling the \acrshort{RPE}-Choroid and Choroid-Sclera boundaries in panels (A) and (B), respectively.

            \end{mysubsubsection}
            
            \begin{mysubsubsection}[]{Posterior distribution} 

                \begin{figure}[tb]
                    \centering
                    \includegraphics[width=\textwidth]{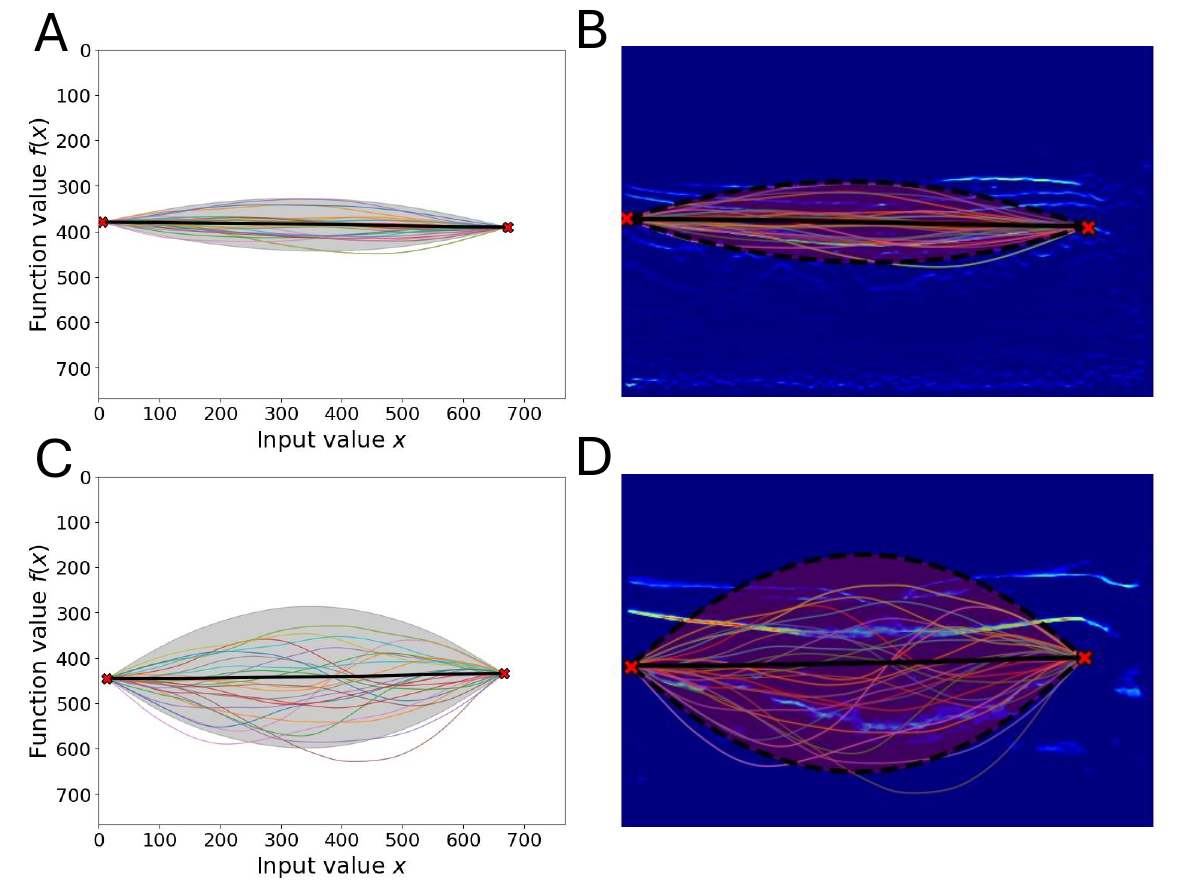}
                    \caption[Initial Gaussian process \acrshort{PPD}s for the \acrshort{RPE}-Choroid and Choroid-Sclera boundaries.]{25 random functions sampled from the initial \acrshort{PPD} of the Gaussian process regressors modelling the \acrshort{RPE}-Choroid (A--B) and Choroid-Sclera boundaries (C--D). Panels (B) and (D) plot the random functions (multicoloured), mean function (solid black) and 95\% credible interval (red shaded, with black dashed outlines) onto the respective edge maps.}
                    \label{fig:GPET_init_post_dist}
                \end{figure}
            
                To induce the initial posterior (predictive) distribution (\acrshort{PPD}), we condition the Gaussian process on the edge endpoints $\mathcal{D}^{(0)}$. These user-specified pixel coordinates act as domain-specific knowledge and are assumed to be true realisations of the \acrshort{EOI}. This step tells the algorithm which edge the end-user wishes to trace and fitting $\mathcal{D}^{(0)}$ induces the initial \acrshort{PPD}, $\bm{f}^{(1)} | \mathcal{D}^{(0)}$, from which we can draw samples evaluated at the image column indexes in $X_*$. 

                The objective is to now predict the missing pixels between the edge endpoints. The predictive variance (the 95\% credible intervals shown in figure \ref{fig:GPET_init_post_dist}) of the \acrshort{PPD} increases in regions far away from the edge endpoints. This allows the random functions we draw from the \acrshort{PPD} to evolve freely across the image --- according to the properties imposed by the selected covariance function --- as shown in figure \ref{fig:GPET_init_post_dist}(B, D). 
                
                In fact, the signal variance $\sigma_f^2$ dictates the size of this predictive variance. With enough distance from the edge endpoints, the initial \acrshort{PPD} will have a 95\% credible interval of $\overline{y}^{(0)} \pm $2$\sigma_f$, where $\overline{y}^{(0)}$ is the mean of the edge endpoint pixel row indexes. This predictive variance is something we take advantage of throughout the fitting procedure so that the random functions can explore around the \acrshort{EOI} and propose possible locations for it.

            \end{mysubsubsection}

            We will now refer to random functions as ``posterior curves'', and use the terminology ``predictive variance'' to mean the 95\% credible interval of the \acrshort{PPD}. The remainder of this section describes the methodology using an arbitrary iteration $n \geq 0$. 

        \end{mysubsection}
        
        \begin{mysubsection}[]{Curve scoring}

            $L$ posterior curves $f^{(n)}_l$, $l=1, \dots, L$ are drawn at random from the \acrshort{PPD} for the current iteration which are defined at the column indexes in $X_*$. These curves are a representative sample of the model's belief in where the \acrshort{EOI} lies, given the (current) observation set defined from the last iteration $\mathcal{D}^{(n-1)}$. 
            
            $L$ is set at 500 so that there are enough posterior curves to represent the \acrshort{PPD} to sufficiently explore the image gradient $G$. Lower values of $L$ will not explore the space well enough leading to an under-specified model. Additionally, $L$ should not be arbitrarily large due to the computational overhead of curve sampling.

            These curves are scored using the pre-computed image gradient $G$ and the highest scoring posterior curves are used to decide on the next set of pixels that will be added to the model's observation set. The score for each posterior curve computes the cumulative value of the gradient response, per unit length, along each posterior curve. For a given posterior curve $f^{(n)}_l$ the score $I(f^{(n)}_l)$ is
            \begin{equation}
                I(f^{(n)}_l) = \int_{f^{(n)}_l} G\big(\bm{r}(s)\big) \hspace{2pt} ds \bigg/ \int_{f^{(n)}_l} ds.\label{eq:LI_1}
            \end{equation}
            This score is analogous to computing the area under the posterior curve along the interpolated image gradient surface, divided by the posterior curve's arc length. 
            
            We use Simpson's and finite difference rules to estimate the line integral and arc length. The advantage of scaling a curve's score by their arc length is to penalise them for making longer excursions than necessary, as these are unlikely to happen for continuous edges such as the upper and lower choroid boundaries. Moreover, this can also be particularly helpful in the presence of noise in the image gradient. 
            
            We invert this gradient score to create a cost function such that
            \begin{equation}
                C(f^{(n)}_l) = \big(I(f^{(n)}_l)\big)^{-1}
            \end{equation}
            This is so that higher scoring curves are minimally costly and the minimisation of this cost becomes part of the optimisation scheme to model the edge. 
            
        \end{mysubsection}
        
        \begin{mysubsection}[]{Pixel scoring}
            the 50 most optimal posterior curves according to the cost function defined above are selected to determine the next observation set, with all other curves discarded. These 50 curves were defined as 10\% of $L$=500. Selecting fewer than 10\% risks under-specification, as the model is forced to rely on a small subset of posterior curves to choose pixels with. Moreover, too large a proportion may introduce unwanted noise into the fitting procedure from low-scoring posterior curves.
            
            These optimal curves provide a best guess at where the current \acrshort{PPD} believes the \acrshort{EOI} lies and their coordinates together map out which regions have been visited the most. Given these curves lie in the vicinity of the \acrshort{EOI},  we can obtain an empirical approximation for where the \acrshort{PPD} believes the \acrshort{EOI} is by generating a two dimensional density function which measures the location frequency of the most optimal posterior curves. Think of this as a two dimensional (continuous) histogram the size of the original image, keeping track of where and how many times a pair of optimal posterior curves intersected in the image.
            
            This frequency density function is defined by  
            \begin{equation}\label{eq:freq_kde}
                \phi^{(n)}(\bm{z}) = \frac{1}{\sum_{l, i}w(\bm{z}_{li})}\sum_{l, i}w(\bm{z}_{li})K_{\mathbb{I}}(\bm{z} - \bm{z}_{li}),
            \end{equation}
            where $\bm{z}, \bm{z}_{li}$ are two dimensional coordinates such that $z=[x,y]$ with $x \in [1, N]$ and $y \in [1, M]$. $\bm{z}_{li}$ corresponds to the $i^{\textrm{th}}$ coordinate of the $l^{\textrm{th}}$ optimal posterior curve, $f^{(n)}_l$. 
            
            The weighting for coordinate $\bm{z}_{li}$ scores the point according to how optimal the curve $f^{(n)}_l$ is relative to all other optimal curves, using $I(f^{(n)}_l)$. This weighting will result in a higher density for regions where most optimal curves intersected in the image. The weight function $w$ is defined by
            \begin{equation}\label{eq:kde_weight}
                w(\bm{z}_{li}) = \frac{I\big(f^{(n)}_l\big)}{\sum^{L/10}_{l=1}I\big(f^{(n)}_l\big)},
            \end{equation}
            for all coordinates $i$ for the $l^{\textrm{th}}$ optimal curve $f^{(n)}_l$.
            
            $K_{\mathbb{I}}(\bm{z}-\bm{z}_{li})$ is a symmetric, multivariate kernel (covariance function) of the density $\phi^{(n)}$. Here, we use the isotropic, two-dimensional Gaussian kernel with the identity matrix as its length-scale,
            \begin{equation}\label{eq:gauss_kernel}
                K_{\mathbb{I}}(\bm{z} - \bm{z}_{li}) = (2\pi)^{-1}\exp\left(-\frac{1}{2}\big(\bm{z} - \bm{z}_{li}\big)^{\text{T}}\big(\bm{z} - \bm{z}_{li}\big)\right).
            \end{equation}

            \begin{figure}[tb]
                \centering
                \includegraphics[width=\textwidth]{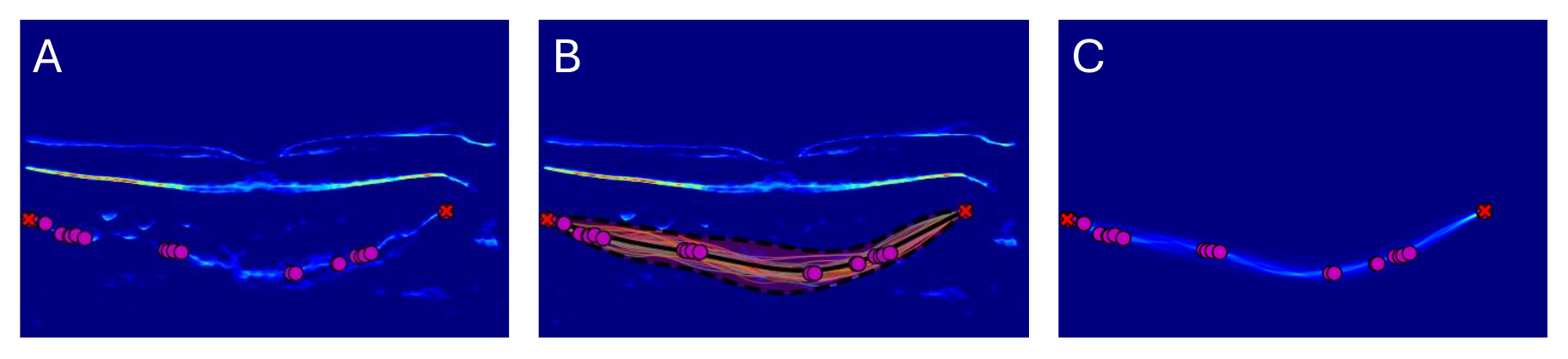}
                \caption[Frequency density from posterior curves of the \acrshort{PPD} during Choroid-Sclera boundary fitting.]{Frequency density and \acrshort{PPD} during an arbitrary iteration of the fitting procedure for the Choroid-Sclera boundary. (A) Image gradient. (B) \acrshort{PPD} with optimal posterior curves overlaid. (C) Weighted frequency density function. Current observation set $\mathcal{D}^{(n-1)}$ overlaid as purple dots, with edge endpoints as red crosses.}
                \label{fig:GPET_method_iter}
            \end{figure}
            
            The isotropic and unitary length-scale imposes no dominant orientation in the Gaussian kernel which is centred at each real-valued point $\bm{z}_{li}$. Moreover, the unitary property ensures each point's kernel contributes the majority of its density to only the 4 nearest pixel coordinates in the discretised space. 
            
            Figure \ref{fig:GPET_method_iter} illustrates the fitting procedure for the Choroid-Sclera boundary whose initial \acrshort{PPD} was shown in figure \ref{fig:GPET_init_grad}(D). In panel (A) we show the image gradient with the current observation set $\mathcal{D}^{(n-1)}$. This observation set has updated the \acrshort{PPD} around the \acrshort{EOI} as shown in panel (B), whose 95\% credible interval has shrunk and is centred around the \acrshort{EOI}. Note the central column of the image has the largest predictive variance (uncertainty) because there are no observations present in that region. The most optimal posterior curves, overlaid in panel (B) are able to collectively map out the approximate location of the \acrshort{EOI}, as shown by the weighted kernel density in panel (C). 
            
            Once $\phi^{(n)}$ is defined, individual pixel coordinates, $\bm{p} = [x, y]$, can be scored by combining their density, $\bm{\phi}^{(n)}(\bm{p})$ and gradient response, $G(\bm{p})$. This score, $s(\bm{p}) \in [0, 1]$, is defined by
            \begin{equation}\label{eq:score_eqn}
                s(\bm{p}) = \frac{1}{3}\left(\phi^{(n)}(\bm{p}) \cdot G(\bm{p}) + \phi^{(n)}(\bm{p}) + G(\bm{p})\right).
            \end{equation}
            This scoring function gives equal weighting to the local information supplied by the image gradient and the global perspective provided by the \acrshort{PPD}.
            
        \end{mysubsection}
        
        \begin{mysubsection}[]{Accept-discard scheme}
            The primary advantage of using a score function which includes the iteration-dependent frequency density $\phi^{(n)}$ is that densities are updated in each consecutive iteration using the new \acrshort{PPD}. This means that pixels fitted during previous iterations are re-scored using an updated density allowing a more accurate representation of how probable those pixels are of belonging to the \acrshort{EOI},  given the new set of optimal posterior curves. This allows the possibility of older, incorrectly fitted pixels to be discarded from the model if their score drops sufficiently in later iterations. 
            
            An adaptive, user-specified value $T \in$ [0, 1] is chosen to threshold these pixels such that
            \begin{equation}\label{eq:thresh_set}
               P^{(n)}_T = \Big\{\bm{p} \hspace{3pt} \Big| \hspace{3pt} s(\bm{p}) \geq T\Big\},
            \end{equation}
            where $T$ is initially set at 1 for every iteration (see below).

            The thresholded pixels are binned into $\ceil{\nicefrac{N}{\Delta x}}$ linearly spaced subsets $S_i$, $i=1,\dots,\ceil{\nicefrac{N}{\Delta x}}$, according to their horizontal column index, such that
            
            \begin{equation}\label{eq:binning}
               S_i = \Big\{\bm{p}=[x,y] \hspace{3pt} \Big| \hspace{3pt} \bm{p} \in P^{(n)}_T, \hspace{2pt} x \in \big[(i-1)\Delta x, \hspace{3pt} i\Delta x\big] \Big\}_{i=1}^{\ceil{\nicefrac{N}{\Delta x}}},
            \end{equation}
            where $\Delta x$ is the length of each sub-interval. That is, $S_i$ contains all thresholded pixel coordinates whose column index is in the interval $\left[(i-1)\Delta x, \hspace{3pt} i\Delta x\right]$. $\Delta x$ defines how many pixels are used to reconstruct the \acrshort{EOI} and the choice of $\Delta x$ depends on the \acrshort{EOI}'s complexity and number of columns it spans in the image. However, For practical purposes, $\Delta x = 10$ can reliably reconstruct the upper and lower choroid boundaries well. Lowering this value may be required when the \acrshort{EOI} is less smooth.
            
            As a continuous function is fitted, we enforce a property called injectivity. This is a property where two different row indexes uniquely corresponds to two different column indexes, such that for some function $f$, $f(x) \neq f(y)$ implies that $x \neq y$, i.e. the function $f$ does not have any regions which are entirely vertical, or loops back on itself. Injectivity is enforced by using non-max suppression, selecting the highest scoring pixel per sub-interval $S_i$, yielding
            \begin{equation}\label{eq:suppression} 
                P^{(n)}_A = \Big\{\displaystyle\argmax_{\bm{p}}\big\{s(\bm{p})\big\}\hspace{3pt} \Big| \hspace{3pt} \bm{p} \in S_i\Big\}_{i = 1}^{\ceil{\nicefrac{N}{\Delta x}}}.
            \end{equation}
            Coordinates are binned because it is unnecessary to fit the model with points which are very close in the input space, unless the underlying edge is very complex, in which case we reduce the value of $\Delta x$.
            
            The observation set for the next iteration $\mathcal{D}^{(n)}$ is formed by re-scoring the previous iteration's observation set according to the new set of optimal posterior curves, $\mathcal{D}^{(n-1)}$, and combining this with $P^{(n)}_A$ through binning (equation \eqref{eq:binning}) and non-max suppression (equation \eqref{eq:suppression}) yielding 
            \begin{equation}\label{eq:concat_prev_new_set}
                \mathcal{D}^{(n)} = \textrm{suppression}\left(\textrm{binning}\big(P^{(n)}_A \cup \mathcal{D}^{(n-1)}\big)\right).
            \end{equation}
            $P^{(n)}_A$ is combined with $\mathcal{D}^{(n-1)}$ because $\phi^{(n)}$ changes with each iteration, meaning that the highest scoring pixel coordinate in each set $S_i$ may change dependant on the score it obtains compared to other possible pixels in its neighbourhood. 

            The accept-discard procedure in this section is repeated if $|\mathcal{D}^{(n)}| \leq |\mathcal{D}^{(n-1)}|$ by reducing $T$ by 1\%. Thus, setting this adaptive threshold at 1 at every iteration ensures that only the best scoring pixels are selected since $s(\bm{p}) \leq 1$ for all pixel coordinates $\bm{p}$. This adaptive property ensures that $|\mathcal{D}^{(n)}| > |\mathcal{D}^{(n-1)}|$, and that the fitting procedure is guaranteed to converge in a finite number of iterations, which is exactly $\ceil{\nicefrac{N}{\Delta x}}$.
            
        \end{mysubsection}

        \begin{mysubsection}[]{Refitting}
        
            The observations in $\mathcal{D}^{(n)}$ represent the most probable pixel coordinates which form part of the \acrshort{EOI} for the next iteration. An improved \acrshort{PPD} can be formed using this new set of pixel coordinates, updating the model's belief in where the \acrshort{EOI} lies, such that 
            \begin{align}\label{eq:iter_posterior}
                \bm{f}^{(n+1)} \hspace{3pt}\big|\hspace{3pt}\mathcal{D}^{(n)} &\sim \mathcal{N}\big(\bm{\mu}^{(n+1)}, \bm{\sigma}^{(n+1)}\big);\\ \label{eq:iter_predictive_mean}
                \bm{\mu}^{(n+1)} &= K_{*n}\left[K_{nn}+ \sigma^2_y\mathbb{I} \right]^{-1}\bm{y}^{(n)};\\\label{eq:iter_predictive_cov} 
                \bm{\sigma}^{(n+1)} &= K_{**} - K_{*n}\left[K_{nn}+ \sigma^2_y\mathbb{I}\right]^{-1}K_{n*},
            \end{align}
            where $K_{nn} = K(X^{(n)}, X^{(n)})$ and similarly for $K_{n*}$ and $K_{*n}$. The posterior curves $f^{(n+1)}$ we can sample should now be more accurate in mapping regions of the \acrshort{EOI} than in previous iterations.

            \begin{figure}[tb]
                \centering
                \includegraphics[width=\textwidth]{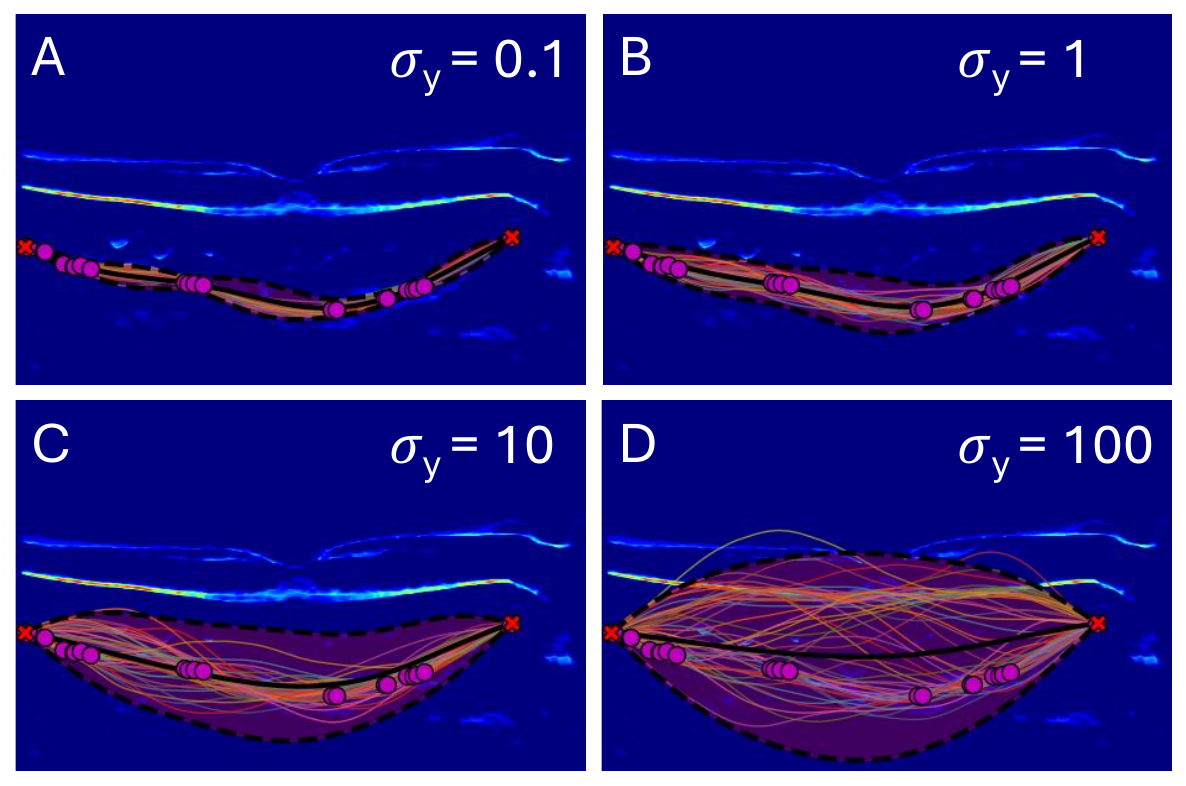}
                \caption[Influence of the observation noise during \acrshort{GPET}'s Choroid-Sclera boundary fitting.]{Influence of the observation noise during \acrshort{GPET}'s Choroid-Sclera boundary fitting. \acrshort{PPD}s (with the 50 optimal posterior curves overlaid with mean function and 95\% credible interval) for the same observation set as in figure \ref{fig:GPET_method_iter}, with different observation noise $\sigma_y = 0.1$ (A), $\sigma_y = 1$ (B),$\sigma_y = 10$ (C), $\sigma_y = 100$ (D).}
                \label{fig:GPET_method_noise}
            \end{figure}
            
            Observation noise $\sigma_y$ is user-specified and encodes the confidence in the models' observation set $\mathcal{D}^{(n)}$. Specifically, $\sigma_y$ is used to perturb the posterior curves from passing through the observation points exactly, allowing posterior curves to continue searching for pixels in the image that are in the vicinity of previously fitted pixels. This permits the possibility of accepting better edge pixels in each sub-interval in future iterations. The lower the value of $\sigma_y$ implies a lower predictive variance in regions near observations in the \acrshort{PPD}. Too large a value implies the model lacks confidence during pixel selection which can reduce accuracy. Figure \ref{fig:GPET_method_noise} shows the effect of $\sigma_y$ on the same observation set used in figure \ref{fig:GPET_method_iter}. Here, a value of 1 balances the degree of uncertainty away from $\mathcal{D}^{(n)}$ with the confidence near $\mathcal{D}^{(n)}$ most optimally.

            \begin{figure}[tb]
                \centering
                \includegraphics[width=\textwidth]{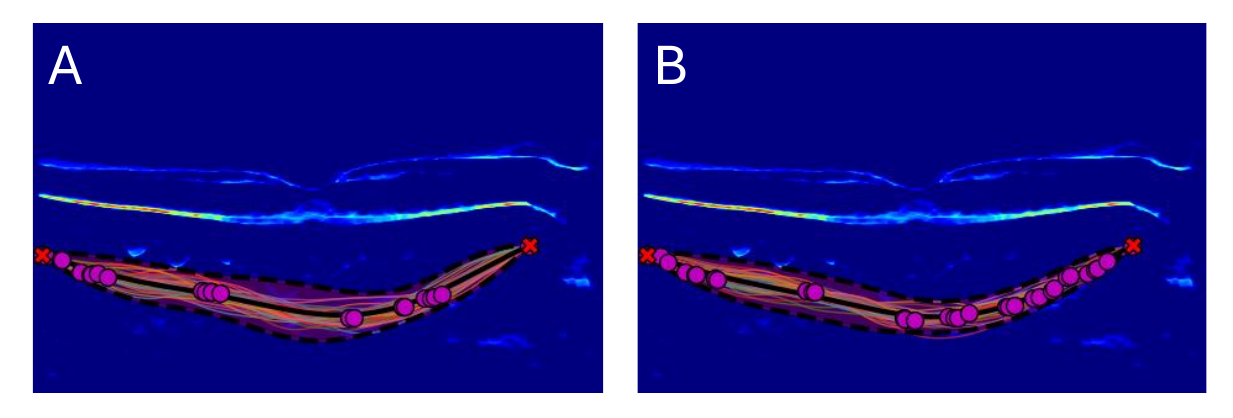}
                \caption[Influence of increasing the observation set on the \acrshort{PPD} during Choroid-Sclera boundary fitting.]{Influence of increasing the observation set on the \acrshort{PPD} during Choroid-Sclera boundary fitting. (A) Observation set $\mathcal{D}^{(n-1)}$ and corresponding \acrshort{PPD} shown from figure \ref{fig:GPET_method_iter}(B). (B) Observation set $\mathcal{D}^{(n)}$ and corresponding \acrshort{PPD} after re-fitting.}
                \label{fig:GPET_method_newset}
            \end{figure}

            Figure \ref{fig:GPET_method_newset} shows the \acrshort{PPD} with the observation set $\mathcal{D}^{(n-1)}$ used throughout this section in panel (A), with the new observation set $\mathcal{D}^{(n)}$ in panel (B). Clearly, there is smaller predictive variance as the number of observations grow, and the \acrshort{PPD} has converged closer to the Choroid-Sclera boundary.

        \end{mysubsection}

        \begin{mysubsection}[]{Convergence}

            \begin{figure}[tb]
                \centering
                \includegraphics[width=\textwidth]{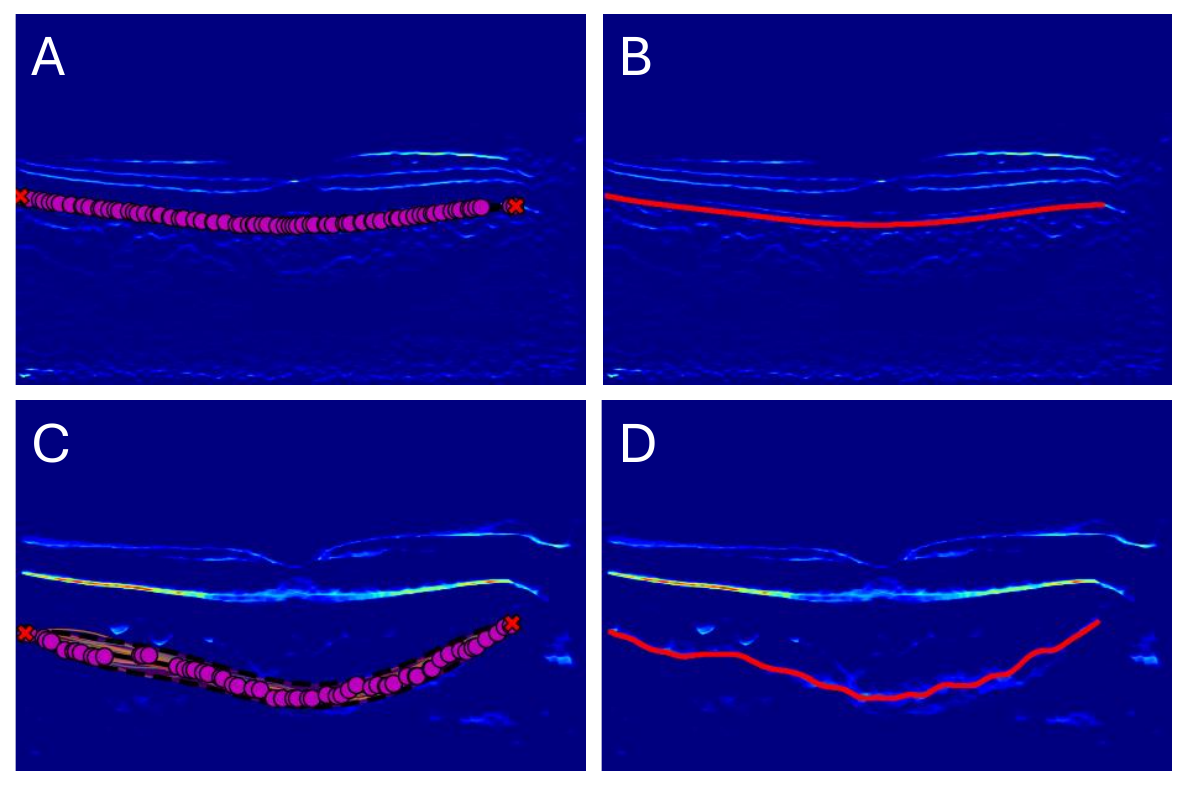}
                \caption[Convergence of \acrshort{GPET}'s image analysis pipeline.]{(A, C) Image gradients with final \acrshort{PPD} and observation set overlaid for \acrshort{RPE}-Choroid (A) and Choroid-Sclera boundary (C). Final edge trace overlaid onto image gradient for \acrshort{RPE}-Choroid (B) and Choroid-Sclera boundary (D).}
                \label{fig:GPET_method_converged}
            \end{figure}
            
            Convergence is when the final observation set $\mathcal{D}^{(n)}$ has enough pixels to reconstruct the \acrshort{EOI}. Specifically, the algorithm reaches termination once all sub-intervals have selected a high-scoring pixel, i.e. $|\mathcal{D}^{(n)}| = \ceil{\nicefrac{N}{\Delta x}}$. The model is refitted once more using $\mathcal{D}^{(n)}$, this time optimising $\bm{\theta}$ by maximising the log marginal likelihood according to equation \eqref{eq:optim_theta}, yielding $\hat{\bm{\theta}}$ and $\hat{\sigma}^2_y$. This maximisation scheme chooses kernel hyperparameters and noise variance which make $\mathcal{D}^{(n)}$ seem probable. 
            
            This optimisation drives $\sigma_y^2$ toward 0 because there are enough edge pixels in $\mathcal{D}^{(n)}$ to interpolate through the observations almost exactly. Note via equation \eqref{eq:gp_updated_mean} that the smaller $\sigma^2_y$ is to 0, the closer the posterior predictive mean is to all observations. It is this optimisation scheme which converges the final \acrshort{PPD} to the \acrshort{EOI},  using the optimised posterior predictive mean,
            \begin{equation}\label{eq:optimal_predict_mean}
                \hat{\bm{f}}^{(n+1)} \hspace{3pt}\big|\hspace{3pt}  \mathcal{D}^{(n)}, \hat{\bm{\theta}} = K_{*n}\left[K_{nn}+ \hat{\sigma}^2_y\mathbb{I}\right]^{-1}\bm{y}^{(n)},
            \end{equation}
            as the proposed edge. This, as well as a 95\% credible band, evaluated at $X_*$, is outputted to the end-user. Figure \ref{fig:GPET_method_converged} shows the final observation set and \acrshort{PPD} overlaid onto the \acrshort{RPE}-Choroid and Choroid-Sclera image gradients in panels (A) and (C), with the final edge trace (the optimised posterior predictive mean) shown in panels (B) and (D), respectively.

        \end{mysubsection}

        \begin{mysubsection}[]{Pipeline}
            
            \begin{figure}[!b]
            \begin{adjustwidth}{-0.5in}{-0.5in}  
            \centering
            \includegraphics[width=\linewidth]{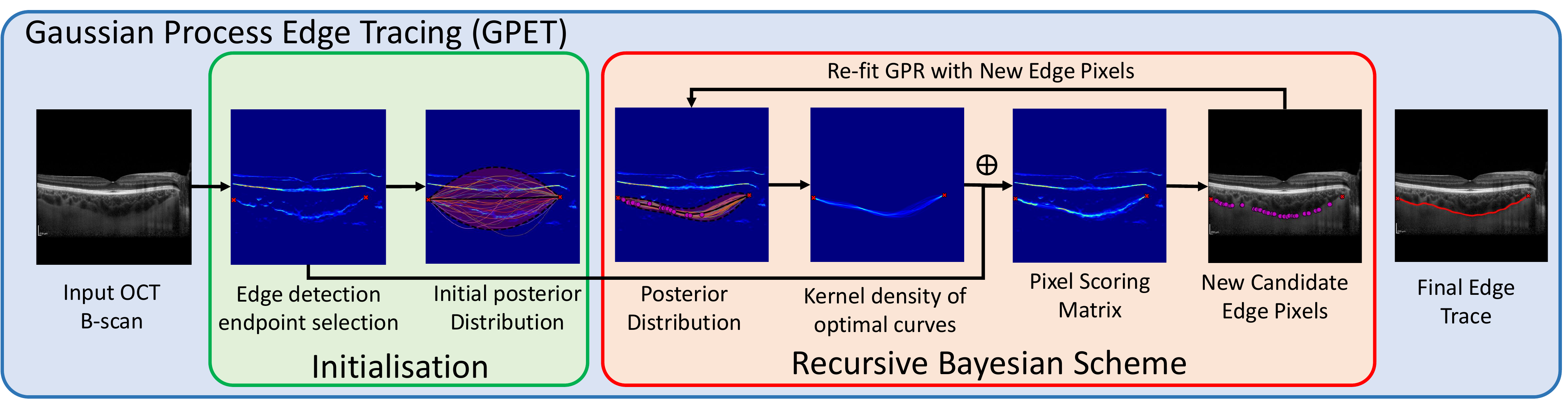}
            \end{adjustwidth}
            \caption[Schematic of \acrshort{GPET}'s image analysis pipeline.]{Schematic diagram of \acrshort{GPET} segmenting the Choroid-Sclera boundary of the choroid on an \acrshort{OCT} B-scan.}
            \label{fig:GPET_schematic_diag}
            \end{figure}
            
            \begin{algorithm}[!t]\footnotesize
            \caption{Edge Tracing using Gaussian Process Regression}
            \textbf{Inputs}: Image gradient $G$ and edge endpoints $\mathcal{D}^{(0)}$. Covariance function $k$ and its hyperparameters $\bm{\theta}$, including observation noise $\sigma_y$. \\
            \textbf{Output}: Optimal posterior predictive mean with a 95\% credible band, evaluated at $X_*$.
            \begin{algorithmic}[1]
                \State Induce prior predictive distribution, $f^{(0)}(x) \sim \mathcal{GP}\big(0, k(x, x' ; \bm{\theta})\big)$;
                \State Fit edge endpoints in $\mathcal{D}^{(0)}$ to the Gaussian process and induce the initial \acrshort{PPD}, $\bm{f}^{(1)} | \mathcal{D}^{(0)}$, corrupting edge endpoints with observation noise $\sigma^2_y$;
                \While{$|\mathcal{D}^{(n)}| \neq \ceil{\nicefrac{N}{\Delta x}}$}
                    \State \multiline{Draw $L$=500 posterior curves from the current \acrshort{PPD};}
                    \State \multiline{Compute cost of each posterior curve and select the most optimal 10\% (50) of curves;}
                    \State \multiline{Estimate frequency density function $\phi^{(n)}$ using optimal posterior curves, weighted by their curve score;}
                    \State \multiline{Score and threshold pixel coordinates with $T$ initialised as 1, yielding $P^{(n)}_T$;}
                    \State \multiline{Bin and use non-max suppression to obtain $P^{(n)}_A$;}
                    \State \multiline{Combine $\mathcal{D}^{(n-1)}$ and $P^{(n)}_A$ to obtain $\mathcal{D}^{(n)}$ through binning and non-max suppression;}
                    \State {While $|\mathcal{D}^{(n)}| \leq |\mathcal{D}^{(n-1)}|$, reduce $T$ by 1\% and repeat steps 7--9;}
                    \State \multiline{Retrain the model to induce an updated \acrshort{PPD}, $\bm{f}^{(n+1)} | \mathcal{D}^{(n)}$, corrupting $\mathcal{D}^{(n)}$ with observation noise $\sigma_y$.}
                \EndWhile
            \State \multiline{Optimise hyperparameters $\bm{\theta}$ and retrain the Gaussian process using final set of observations $\mathcal{D}^{(n)}$ with optimised hyperparameters $\hat{\bm{\theta}}$;}
            \State \textbf{return} optimised posterior predictive mean, $\hat{\bm{f}}^{(n+1)} | \mathcal{D}^{(n)}, \hat{\bm{\theta}}$, and 95\% credible band evaluated at $X_*$ .
            \end{algorithmic}
            \label{alg:gpet}
            \end{algorithm}

            Figure \ref{fig:GPET_schematic_diag} presents a schematic diagram of \acrshort{GPET}'s image analysis pipeline. Additionally, Algorithm \ref{alg:gpet} outlines the pseudocode of the proposed edge tracing algorithm.

            \vfill

        \end{mysubsection}
        
    \end{mysection}

    \begin{mysection}[]{GPET's evaluation} \label{sec:ch_gpet_eval}
        
        We evaluate \acrshort{GPET} by comparing it's performance against manual grading for measuring choroid thickness. This is accomplished using a longitudinal cohort of renal transplant and donor patients.

        \begin{mysubsection}[]{Data and methods}

            \begin{mysubsubsection}[]{Study population}
            
                We prospectively analysed a longitudinal cohort of healthy kidney donors and patients with end-stage kidney disease (\acrshort{CKD}) undergoing living donor kidney transplantation (NCT0213274) \cite{dhaun2014optical}. Data collection and subsequent analyses were conducted after ethical approval from the South East Scotland Research Ethics Committee, in accordance with the principles of the Declaration of Helsinki and all participants gave informed consent to recruitment. Eligibility criteria for recruitment were (1) donors must be living and healthy throughout the period of analysis, (2) recipients with end-stage \acrshort{CKD} must have had a functional kidney transplant and (3) must be aged 18 or over. Exclusion criteria were (1) any ocular pathology pre-transplant, (2) any previous eye surgery, (3) a refractive error exceeding ±6 dioptres or (4) a diagnosis of diabetes mellitus. We judged image quality by the OSCAR-IB criteria \cite{tewarie2012oscar} and excluded images with B-scan signal quality $\leq$ 15, indistinguishable Choroid-Sclera boundaries due to speckle noise, or partial image cropping of choroid. Figure \ref{fig:GPET_sample_derivation} presents a sample derivation flowchart on how the population was selected for the wider research study, and then used for \acrshort{GPET}'s evaluation. 
    
                \begin{figure*}[tb]
                    \centering
                    \includegraphics[width = \textwidth]{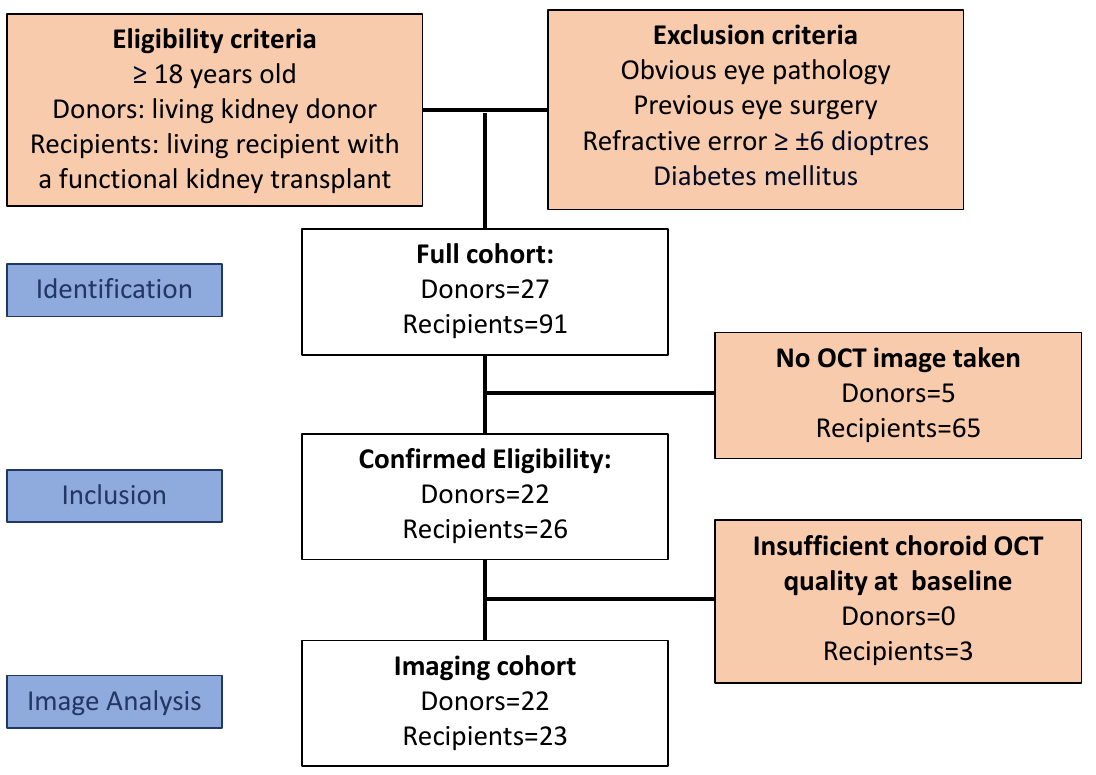}
                    \caption[Sample derivation flowchart of the renal cohort used to validate \acrshort{GPET}.]{Population derivation for \acrshort{CKD} sample used to assess \acrshort{GPET}'s performance against manual grading.}
                    \label{fig:GPET_sample_derivation}
                \end{figure*}
                
                At the time of data extraction, 22 donors and 23 recipients were eligible for analysis, after excluding a total of 65 recipients and 5 donors. The majority of exclusions were because there were no eye scans performed for these individuals. This was because, once recruited, they were scheduled for transplantation and then immediately returned to their referring institution after transplant away from Edinburgh, thus making it impractical to retain in the study and re-scan. Each participant was assessed on the day of transplant and then attempts were made at 7 other time periods over the following year (8 time points in total). We used the entire analysis cohort for performance evaluation of the automatic approach against manual choroid thickness, resulting in 480 choroid thickness measurements for comparison from 160 OCT B-scans in 45 eyes. Table \ref{tab:GPET_eval_pop} shows some population statistics for the analysis cohort at baseline, with table \ref{tab:GPET_eval_long_pop} showing sample size and time-point statistics across all other longitudinal time-points sampled.
                \begin{table}[tb]
                \centering
                \begin{tabular}{lll}
\toprule
\multirow{2}{*}{} & \multicolumn{2}{c}{Cohort} \\
\cmidrule(l){2-3}
 & Donors & Recipients \\
 \midrule
Sample, N & 22 & 23 \\
Age, years & 50 ± 11 & 47 ± 12 \\
Sex (male) & 9 (41) & 15 (65) \\
\bottomrule
\end{tabular}

                \caption[Demographics of renal cohort used to validate \acrshort{GPET}.]{Basic demographic information for the donors and recipients used to evaluate \acrshort{GPET}. Where appropriate, values are shown as mean ± SD.}
                \label{tab:GPET_eval_pop}
                \end{table}
    
                \begin{table}[tb]\footnotesize
                \centering
                \begin{tabular}{lp{1.5cm}p{2cm}p{1.5cm}p{1.5cm}p{2cm}p{1.5cm}}
\toprule
\multicolumn{1}{c}{\multirow{3}{*}{Timepoint}} & \multicolumn{6}{c}{Cohort} \\
\cmidrule(l){2-7}
\multicolumn{1}{c}{} & \multicolumn{3}{c}{Donors} & \multicolumn{3}{c}{Recipients} \\
\cmidrule(l){2-4}\cmidrule(l){5-7}
\multicolumn{1}{c}{} & Sample, n (\%) & Daytime (Hr:Min) & Time from baseline (weeks) & Sample, n (\%) & Daytime (Hr:Min) & Time from baseline (weeks) \\
\midrule
Baseline & 22 (100) & 13:50 ± 1:41 & 0 & 23 (100) & 13:59 ± 2:19 & 0 \\
1 Week & 11 (50) & 13:48 ± 1:20 & 1 ± 0 & 13 (61) & 12:44 ± 1:42 & 1 ± 0 \\
2 Weeks & 2 (9) & 12:04 ± 1:34 & 2 ± 0 & 5 (48) & 11:45 ± 1:50 & 2 ± 0 \\
4 Weeks & 8 (36) & 12:48 ± 3:47 & 4 ± 1 & 5 (48) & 12:23 ± 2:01 & 4 ± 1 \\
8 Weeks & 10 (45) & 13:43 ± 2:12 & 8 ± 2 & 10 (43) & 11:31 ± 1:13 & 8 ± 2 \\
16 Weeks & 8 (36) & 13:54 ± 2:00 & 16 ± 4 & 6 (26) & 12:29 ± 1:55 & 17 ± 4 \\
28 Weeks & 10 (45) & 13:30 ± 2:03 & 29 ± 3 & 8 (35) & 11:47 ± 1:39 & 29 ± 4 \\
52 Weeks & 11 (50) & 13:48 ± 2:07 & 53 ± 6 & 8 (35) & 12:25 ± 1:31 & 52 ± 4\\
\bottomrule
\end{tabular}

                \caption[Longitudinal sample information on renal cohort used to validate \acrshort{GPET}.]{Information on sample size, time of day and approximate time from baseline for donor and recipient samples. Values are shown as mean ± SD and those in parentheses are percentages of the total baseline sample size.}
                \label{tab:GPET_eval_long_pop}
                \end{table}

            \end{mysubsubsection}

            \begin{mysubsubsection}[]{Image acquisition} \label{subsubsec:GPET_eval_acq}
            
                We imaged each participant's right eye using the Heidelberg Engineering \acrshort{SD-OCT} Spectral \acrshort{OCT}1 Standard Module. Patients were examined between 9am and 5pm and were endeavoured to be followed up at around the same time of day at each time point so as to prevent any diurnal effects on downstream statistical analyses. The average (and standard deviation) daytime which scans were taken at each time point are listed in table \ref{tab:GPET_eval_long_pop}. Follow-up was attempted at 7 different time points, but circumstances related to the availability of the imaging device, technician and participant resulted in a number of follow-up imaging not taking place for many participants. 
                
                During each patient's examination, a single horizontal-line, \acrshort{EDI-OCT} B-scan centred at the foveal pit was taken. Each B-scan covered a 30$^{\circ}$ (approximately 9 mm laterally) region and was extracted as a 768 $\times$ 768 (pixel height $\times$ width) high-resolution image for downstream image processing. Each B-scan was captured using active eye tracking with an \acrshort{ART} of 100. Figure \ref{fig:example_oct_measure} shows an example \acrshort{EDI-OCT} B-scan with results from manual and automatic assessment. 
            \end{mysubsubsection}

            \begin{mysubsubsection}[]{Manual measurement of choroid thickness}

                \begin{figure}[tb]
                \centering
                \includegraphics[width = \textwidth]{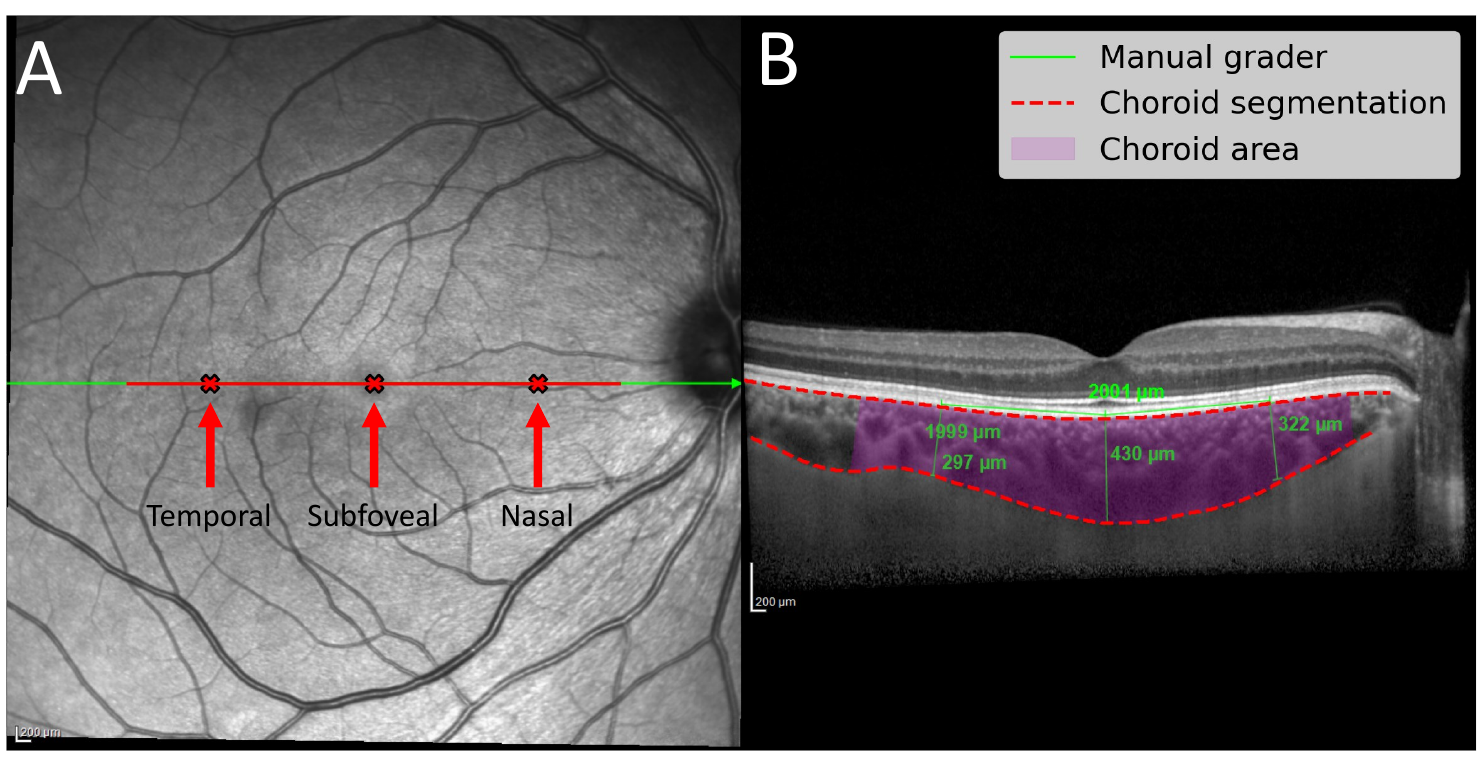}
                \caption[Demonstration of manual and automated methods to measure the choroid.]{(A) En face SLO image with the location of the B-scan in green and markers for measuring thickness (crosses) and area (line) in red. (B) B-scan showing chorioretinal structures with manual grading and automatic segmentation overlaid.}
                \label{fig:example_oct_measure}
                \end{figure}
            
                Manual grading for measuring choroid thickness was performed by a single, trained operator across all image sessions for each participant. Manual choroid thickness was measured using the calliper tool in the Heidelberg Eye Explorer (HEYEX) software (version 1.10.4.0, Heidelberg Engineering, Heidelberg, Germany). The calliper tool was used to mark the centre of the fovea at the level of the photoreceptor outer segment and mark 2000 microns temporal and nasal to this point, parallel to the \acrshort{RPE}-Choroid boundary. Choroid thickness was defined as the straight line distance between the \acrshort{RPE}-Choroid and Choroid-Sclera boundaries, measured locally \textit{perpendicular} to the \acrshort{RPE}-Choroid boundary. Choroid thickness was measured at 3 distinct locations across the macula, subfoveal and 2000 microns nasal and temporal to the fovea.

                Figure \ref{fig:example_oct_measure} shows three choroid thickness markings measured by the manual grader in green and the choroid boundaries delineated using \acrshort{GPET}. Note that the reference lines used by the manual grader are also visible, measuring approximately 2000 microns temporal and nasal to the foveal pit along the \acrshort{RPE}-Choroid boundary. 
                
            \end{mysubsubsection}

            \begin{mysubsubsection}[]{Automatic measurement of choroid thickness}

                \acrshort{GPET} was used to segment the \acrshort{RPE}-Choroid and Choroid-Sclera boundaries in every \acrshort{OCT} B-scan following the pipeline outlined in figure \ref{fig:GPET_schematic_diag}.

                Automated choroid thickness was measured at the same subfoveal, temporal and nasal locations as manual grading but using two separate approaches, yielding \textit{perpendicular} measurements and \textit{parallel} measurements. \textit{Perpendicular} choroid thickness was measured locally \textit{perpendicular} to the \acrshort{RPE}-Choroid boundary, while \textit{parallel} choroid thickness measured parallel to the line drawn by the manual grader. These lines were overlaid on each B-scan using the calliper tool (figure \ref{fig:example_oct_measure}, green), allowing an automated procedure to detect each lines' endpoints and measure thickness between the \acrshort{GPET}-detected boundaries, as well as measure the difference in angle between perpendicular and parallel lines drawn by \acrshort{GPET}. These parallel measurements mimic any potential angular error made by the manual grader, which we know can have an inordinate impact on choroid thickness measurements (section \ref{subsec:INTRO_MEASURE_ROI}).
  
            \end{mysubsubsection}

            \begin{mysubsubsection}[]{Statistical analysis}
    
                To determine any systematic or proportional differences between the two approaches, we measured agreement between approaches by comparing (parallel) automatic measurements with manual ones using correlation plots. We also investigated residuals, calculated by subtracting manual measurements from automatic ones (automatic - manual), using a Bland-Altman plot \cite{bland1986statistical}. Mean absolute error (\acrshort{MAE}) and Pearson correlation coefficient were also computed between both approaches, stratified by cohort and macular location. To compare the consistency of manual measurements across longitudinal data, we investigated the angular deviation made by the manual grader. This was done by computing angles between the manual calliper measurements made from manual grading against the perpendicular automatic choroid thickness measurements (whose perpendicularity was determined programmatically and thus acts as ground truth). Assessment of the choroid and quantitative analyses comparing manual and automatic measurements were done in Python (version 3.11.3).
    
                Rahman, et al. \cite{rahman2011repeatability} found an unsigned residual of 32 $\mu$m as a threshold suggested to exceed inter-observer variability between manual graders of \acrshort{OCT} B-scans. Therefore, we defined major discrepancies between automatic and manual choroid thickness to be any residual greater than 32 $\mu$m in absolute value. These were selected for external adjudication by a clinical ophthalmologist (Supervisor Dr. Ian J.C. MacCormick). For each discrepancy, the adjudicator (I.M.) was shown two choroid thickness measurements overlaid on the corresponding OCT B-scan using equipment described in appendix \ref{apdx:equipment_protocol}, and was blinded to which approach was used in each case. He was asked to rate each measurement in terms of its accuracy in measuring thickness from the upper to lower boundaries. He was also asked to rate the overall quality of the OCT B-scan regarding the visibility of the Choroid-Sclera boundary. Quality and ratings were graded using a 3-ordinal scale: ``bad'', ``okay'' and ``good''. The adjudicator followed a protocol set out in appendix \ref{apdx:quality_protocol} for measuring image quality, while the scales for rating the thickness measurements were based on the definitions set out in appendix \ref{apdx:definitions}. Finally, I.M. was asked to select the approach which was preferred --- options for both or neither approaches were also available.
    
            \end{mysubsubsection}

        \end{mysubsection}

        \begin{mysubsection}[]{Results}\label{subsec:GPET_eval_results}

            \begin{table}[tb]\footnotesize
            \centering
            \begin{tabular}{clllll}
\toprule
\multicolumn{2}{l}{} & GPET ($\mu$m) & Manual ($\mu$m) & Residual ($\mu$m) & MAE ($\mu$m) \\
\midrule
\multicolumn{2}{c}{All} & 278 ± 83 & 277 ± 79 & 1.8 ± 22.0 & 14.1 ± 16.9 \\
\multirow{3}{*}{Donors} & Temporal & 290 ± 65 & 285 ± 57 & 5.1 ± 18.4 & 13.8 ± 13.2 \\
 & Subfoveal & 287 ± 77 & 293 ± 71 & -5.9 ± 17.4 & 13.4 ± 12.5 \\
 & Nasal & 215 ± 71 & 213 ± 65 & 2.1 ± 18.8 & 12.9 ± 13.8 \\
\multirow{3}{*}{Recipients} & Temporal & 304 ± 61 & 302 ± 58 & 2.9 ± 24.3 & 13.5 ± 20.4 \\
 & Subfoveal & 338 ± 79 & 336 ± 77 & 3.0 ± 28.1 & 17.3 ± 22.3 \\
 & Nasal & 235 ± 78 & 233 ± 72 & 3.8 ± 21.0 & 13.8 ± 16.2 \\
 \bottomrule
\end{tabular}

            \caption[\acrshort{GPET}'s population-level agreement against manual choroid thickness.]{Performance evaluation between manual and automatic measurements, stratified by cohort and macular location. Where appropriate, values are shown as mean ± SD.}
            \label{tab:GPET_table_perf}
            \end{table}

            \begin{table}[tb]\footnotesize
            \centering
            \begin{tabular}{cllll}
\toprule
\multicolumn{2}{l}{Metric} & Slope {[}95\% CI{]} & Intercept {[}95\% CI{]} ($\mu$m) & Pearson correlation \\
\midrule
\multicolumn{2}{c}{All} & 1.02 {[}0.99, 1.04{]} & -3.34 {[}-10.77, 3.50{]} & 0.97 \\
\multirow{3}{*}{Location} & Temporal & 1.02 {[}0.96, 1.07{]} & -1.80 {[}-17.60, 16.60{]} & 0.94 \\
 & Subfoveal & 1.03 {[}0.98, 1.08{]} & -11.81 {[}-27.97, 3.36{]} & 0.96 \\
 & Nasal & 1.06 {[}1.01, 1.10{]} & -9.17 {[}-19.83, 0.83{]} & 0.97 \\
\multirow{2}{*}{Cohort} & Donors & 1.04 {[}1.01, 1.07{]} & -11.04 {[}-20.42, -3.00{]} & 0.97 \\
 & Recipients & 0.99 {[}0.96, 1.03{]} & 5.06 {[}-6.53, 16.56{]} & 0.96 \\
 \bottomrule
\end{tabular}

            \caption[\acrshort{GPET}'s cohort- and location-level agreement with manual choroid thickness.]{Correlation linear fits between manual and  automatic measurements. Where appropriate, values are shown as mean ± SD. All Pearson correlations were statistically significant with $p < 0.0001$.}
            \label{tab:GPET_corr_linfits}
            \end{table}

            \begin{figure}[!t]
                \centering
                \includegraphics[width=\textwidth]{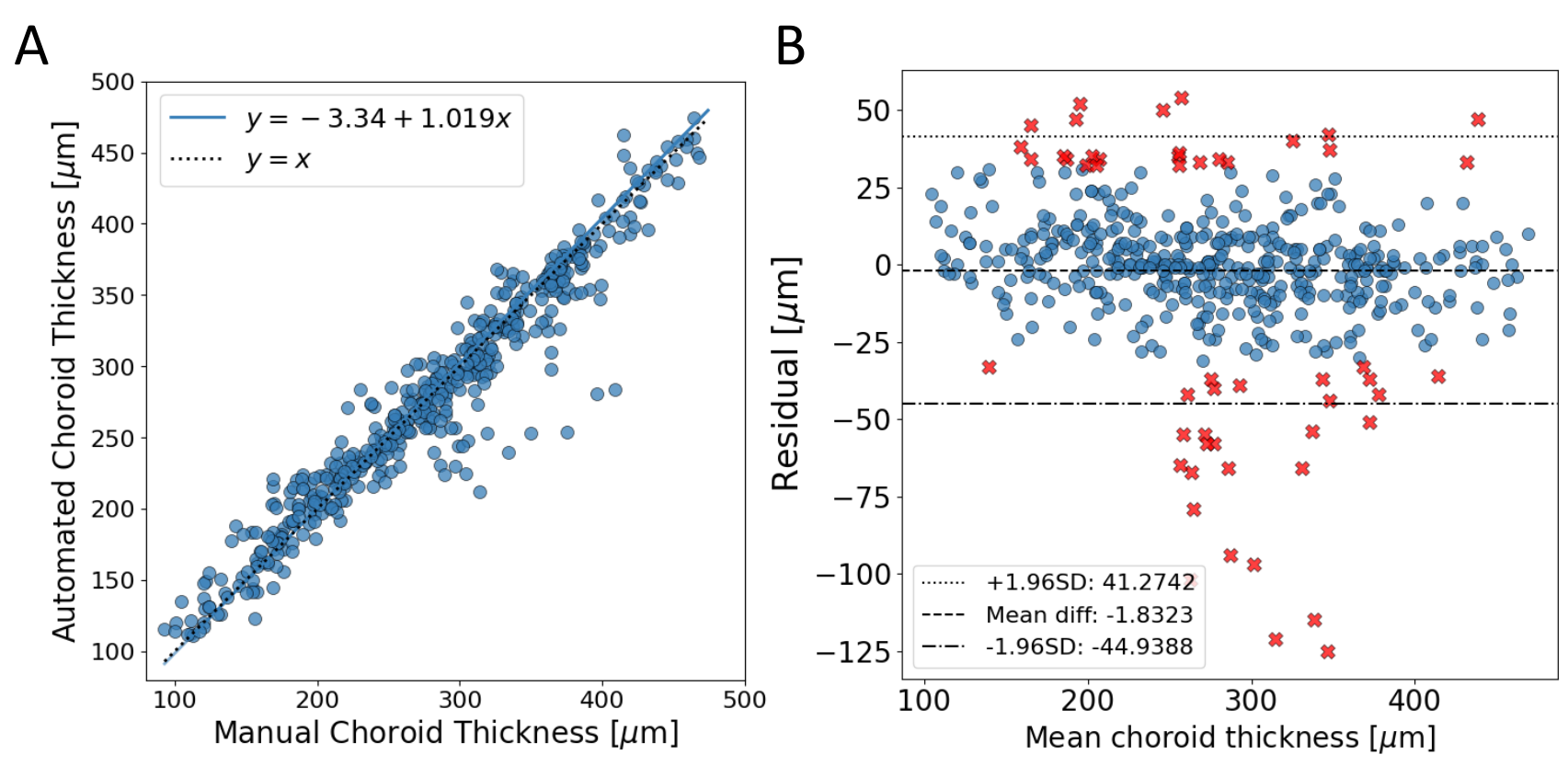}
                \caption[\acrshort{GPET}'s population-level agreement against manual choroid thickness.]{\acrshort{GPET}'s population-level agreement against manual choroid thickness. (A) Correlation plot of manual and (parallel) automatic measurements. (B) Bland-Altman plot showing residual values, with 52 major discrepancies highlighted as red crosses whose residual are greater than 32$\mu$m in absolute value.}
                \label{fig:GPET_BA_PB_plots}
            \end{figure}

            \begin{figure}[!t]
                \centering
                \includegraphics[width=0.9\textwidth]{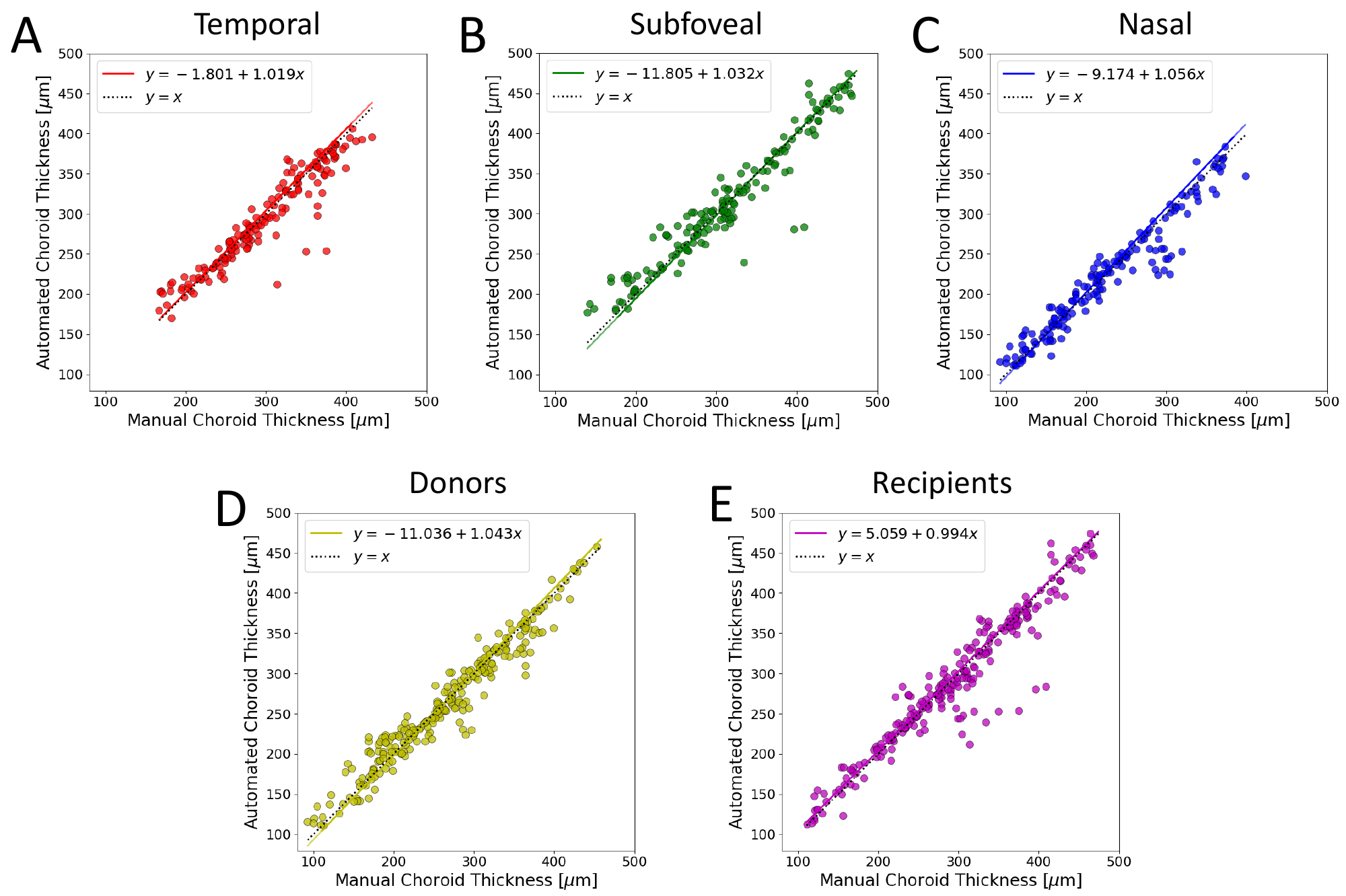}
                \caption[\acrshort{GPET}'s cohort- and location-level agreement with manual choroid thickness.]{Correlation plots comparing manual and automatic choroid thickness measurements, stratified by macular location and cohort. Macular locations shown in (A -- C), with (A) 2000 microns temporal to fovea, (B) subfoveal and (C) 2000 microns nasal to fovea. Cohorts shown in (D -- E), with (D) donors and (E) recipients.}
                \label{fig:GPET_corr_plot_loc}
            \end{figure}

            \begin{table}[tb]\footnotesize
                \centering
                \begin{tabular}{@{}lccc@{}}
\toprule
\multicolumn{1}{c}{\multirow{2}{*}{\begin{tabular}[c]{@{}c@{}}C-S Junction \\ Visibility\end{tabular}}} & \multicolumn{2}{c}{Score} \\ \cmidrule(l){2-3}
& Automated & Manual \\ 
\midrule
Good ($N=9$) & 8 & 2 \\
Okay ($N=15$) & 9 & 13 \\
Bad ($N=28$) & 24 & 15 \\
Total ($N=52)$ & 41 & 30 \\
\midrule
\midrule
Method & \multicolumn{2}{c}{Discrepancy Measurement} \\
\midrule
Automated & \multicolumn{2}{l}{Good: 32, Okay: 19, Bad: 1} \\
Manual   & \multicolumn{2}{l}{Good: 24, Okay: 18, Bad: 10} \\
\bottomrule
\end{tabular}%

                \caption[Blinded adjudication of major disagreements between manual and automatic choroid thickness.]{(Top) Numerical score of both approaches from masked adjudication of 52 major discrepancies, stratified by the visibility of the Choroid-Sclera junction. (Bottom) Qualitative preference scores for both approaches.}
                \label{tab:GPET_outlier_anal}
            \end{table}

             \begin{figure}[tb]
                \begin{adjustwidth}{-0.5in}{-0.5in}  
                \centering
                \includegraphics[width=\linewidth]{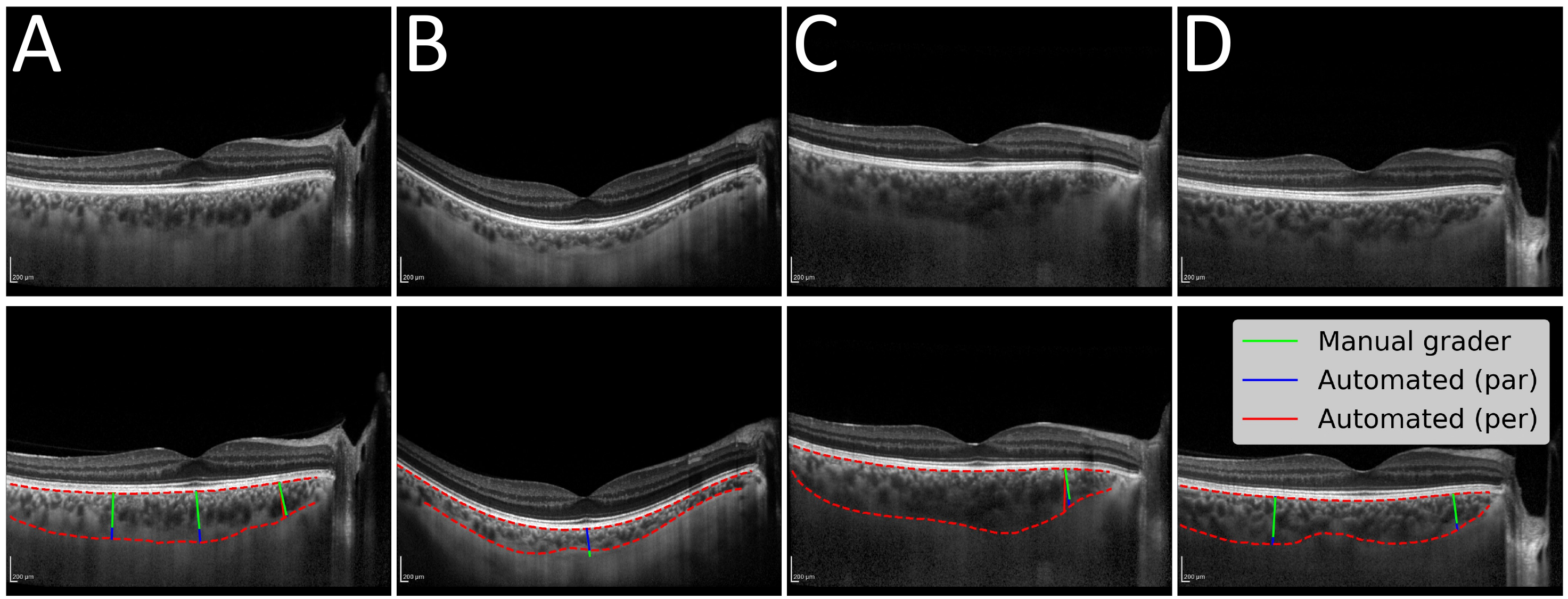}
                \end{adjustwidth}
                \caption[Examples of major disagreements between manual and automatic choroid thickness.]{A selection of major discrepancies between automatic and manual choroid thickness measurements.}
                \label{fig:GPET_major_disc_examples}
            \end{figure}

            \begin{figure}[tb]
                \centering
                \includegraphics[width=\textwidth]{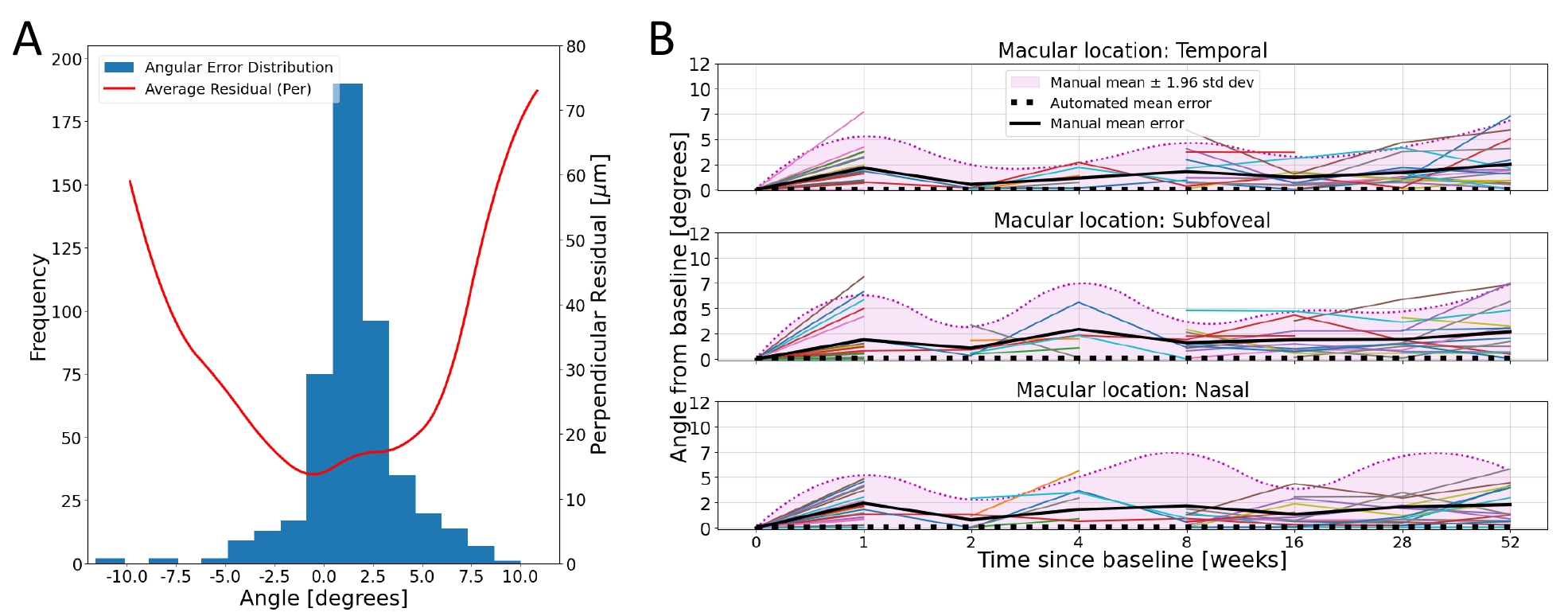}
                \caption[Measurement error in manual choroid thickness.]{(A) Distribution of angular errors from manual choroid thickness measurements (blue). Average unsigned difference between manual and automated perpendicular thickness measurements (red). (B) Unsigned, longitudinal within-patient angular deviations, relative to baseline measurements. Coloured lines represent distinct individuals using manual measurements.}
                \label{fig:GPET_angle_disc}
            \end{figure}

            Performance results between automatic and manual choroid thickness measurements are shown in table \ref{tab:GPET_table_perf}. There was good agreement between manual and automatic measures, with an average residual of 1.8 ± 22.0$\mu$m and \acrshort{MAE} of 14.1 ± 16.9$\mu$m across all 480 choroid thickness measurements. The largest residuals were at the subfoveal macular location. In table \ref{tab:GPET_corr_linfits} we show the estimated slope and intercept coefficients, as well as the Pearson correlation coefficient, which show excellent agreement after stratifying across location and cohorts. 

            Figure \ref{fig:GPET_BA_PB_plots} shows the correlation and Bland-Altman plots comparing \acrshort{GPET}'s measurements against manual ones. Across the whole sample of 480 measurements compared, the linear fit estimated a slope value of 1.02 (95\% confidence interval of 0.99 to 1.04) and an intercept of -3.34 $\mu$m (95\% confidence interval of -10.77 $\mu$m to 3.50 $\mu$m) (figure \ref{fig:GPET_BA_PB_plots}(A)). These 95\% confidence intervals include 1 and 0, respectively, suggesting that there is no real systematic bias or proportional bias between manual and automatic methods to measure choroid thickness (according to the automated, parallel measurements) \cite{passing1983new}. This was also the case when stratified across macular locations and cohorts, with exception for donors and nasal measurements, which showed slight proportional and systematic bias, as shown in table \ref{tab:GPET_corr_linfits} with supporting correlation plots shown in figure \ref{fig:GPET_corr_plot_loc}. 
            
            Bland-Altman analysis on figure \ref{fig:GPET_BA_PB_plots}(B) computed an average residual of -1.83 $\mu$m with limits of agreement of -44.94 $\mu$m to 41.27 $\mu$m. Approximately 11\% of all 480 choroid thickness measurements (52 / 480) exceeded an unsigned residual of 32 $\mu$m \cite{rahman2011repeatability}.

            Table \ref{tab:GPET_outlier_anal} presents the results from the external adjudication of all 52 major discrepancies. Overall, the automatic choroid thickness measurements scored higher in terms of preference, where 79\% (41/52) of the automatic thicknesses were preferred compared to 58\% (30/52) for manual thicknesses. Moreover, the automatic approach scored higher in terms of quality, with only 1 measurement judged as ``bad'', versus 10 measurements judged ``bad'' for the manual approach. In addition to manual measurements voted ``bad'' because they failed to identify the Choroid-Sclera junction, the adjudicator noted that the manual grader did not follow the protocol consistently and failed to measure locally perpendicular to the \acrshort{RPE}-Choroid boundary.  
            
            Figure \ref{fig:GPET_major_disc_examples} shows four major discrepancies from four different OCT B-scans. Red, blue and green lines represent perpendicular, parallel and manual measurements, respectively. The lines in red are shown in panels (A,C) to demonstrate the observed angular deviation between perpendicular automatic and manual measurements. Figure \ref{fig:GPET_major_disc_examples}(A) shows temporal and subfoveal major discrepancies from a donor, residuals 97 $\mu$m and 114 $\mu$m, respectively. These discrepancies were the largest, with adjudicator I.M. preferring \acrshort{GPET}'s thickness measurements. Additionally, figure \ref{fig:GPET_major_disc_examples}(A) shows manual angular deviation from perpendicular at the nasal macular location (5.4$^\circ$), resulting in perpendicular and parallel residuals of 22 $\mu$m and 2 $\mu$m, respectively. 
            
            In figure \ref{fig:GPET_major_disc_examples}(B), we observe a subfoveal major discrepancy in a recipient with residual -47 $\mu$m, of which adjudicator I.M. preferred the manual grader. In figure \ref{fig:GPET_major_disc_examples}(C), we observe angular deviation at the nasal macular location (8.1$^\circ$), resulting in perpendicular and parallel residuals of 106 $\mu$m and 79 $\mu$m, respectively. Finally, figure \ref{fig:GPET_major_disc_examples}(D) shows temporal and nasal major discrepancy from a donor, residuals 66 $\mu$m and 65 $\mu$m, respectively. The adjudicator preferred \acrshort{GPET}'s thickness measurements for the choroids in panels (C--D), but noted significant error in the manual grader failing to measure perpendicularly in panel (C).

            Interestingly, we found that deviation in the angle of manual measurement from perpendicular led to a disproportionate error between manual and perpendicular automatic choroid thickness measurements. This is illustrated in figure \ref{fig:GPET_angle_disc}(A) which shows the distribution (mean 0.5$^\circ$ and standard deviation 2.5$^\circ$) of angular errors made from manual grading. Absolute differences between manual and perpendicular automatic measurements grew quadratically as the angular error deviated from 0 (red curve). We also found large deviation in manual grading when tracking the longitudinal, within-patient angular difference from baseline, as shown in figure \ref{fig:GPET_angle_disc}(B). Distributions at each time point were estimated through empirical mean and standard deviation values. Note that the mean fluctuation in manual grading was always above 0 in comparison to the mean fluctuation from \acrshort{GPET} which was constant at 0 (by design). 
            
        \end{mysubsection}
        
    \end{mysection}

    \begin{mysection}[]{Discussion}

        Low signal quality, vessel shadowing, lack of signal penetration and the potential for extravascular fluid to obscure and occlude the choroid all contribute to the challenging task of tracing the Choroid-Sclera boundary in \acrshort{OCT} B-scans. However, \acrshort{GPET} has the ability to overcome many of these problems due its tunable hyperparameters and model uncertainty quantification, the latter of which many other semi-automatic methods based on graph theory and level sets cannot take into account (table \ref{tab:INTRO_region_methods}). 
        
        The algorithm can successfully trace discontinuous edges through its ability to interpolate between areas of high image gradient response. The hyperparameters of the selected covariance function governs the properties of the types of curves which are used to model the choroid boundaries, and can be chosen to reflect the underlying properties of the \acrshort{RPE}-Choroid and Choroid-Sclera boundaries. Combining this with being able to quantify uncertainty in regions of discontinuity, the model's curve sampling scheme can search for edge pixels which conform to the properties of these interpolates, particularly around areas of known edge pixels. Therefore, it is well adept to accurately quantifying and interpolating occluded regions caused by vessel shadowing or poor reflectivity. 
        
        A huge advantage to \acrshort{GPET} over larger models which require expertly annotated datasets, such as deep learning-based approaches, is that is requires no prior training. \acrshort{GPET} learns each boundary on-the-fly, enabling each fitting procedure to be tuned to the underlying properties of both choroidal boundaries. Moreover, the methodology provides an analytical and functional representation of the edge whose fitting procedure is both interpretable and explainable. These attributes are often lacking in many deep learning models employed in medical image segmentation. 
        
        The \acrshort{GPET} algorithm is also flexible in terms of input and usage. While the fitting procedure requires initial guidance from the edge endpoints inputted manually, \acrshort{GPET} can also be further initialised with intermediary pixels between the two endpoint edge pixels in cases where there is poor visualisation along the Choroid-Sclera boundary. Selecting the edge endpoints and an additional pixel to guide tracing around the central column of the edge would likely add a negligible amount of time for the end-user, but potentially lead to improved segmentation performance. Finally, while this was not the focus of this chapter, \acrshort{GPET} can also be used for processing an \acrshort{OCT} volume scan in a two-step procedure. Once the first B-scan in the stack is fully traced, a subset of the edge pixels in the output can be used as part of the initialisation stage for the next B-scan. Thus, by recursively propagating the output trace of one B-scan to be used as the initialisation for consecutive B-scans, an entire volume scan can be traced in a single sweep. However, the accuracy of such volumetric segmentations has not been fully investigated.

        \begin{mysubsection}[]{Evaluation}

            We found that \acrshort{GPET} could replace human measures of the choroid. We observed no systematic or proportional differences when comparing manual to automatic measurements, when measuring parallel to the grader. Advantageously, automatic choroid thickness eliminated angular error and therefore had greater precision when assessing within-patient longitudinal data. The elimination of inconsistent measurement angle was achieved by \acrshort{GPET}, since the automatic method is always able to measure choroid thickness locally perpendicular at the same location and across all time points for each patient. This is an important attribute of measurement protocols so that choroidal curvature or skew can be accounted for. In contrast, error in the angle of manual measurement had a disproportionate impact when comparing these to the perpendicular automated thickness measurements. 

            Measurement error is a major problem in research studies, and especially in longitudinal studies where repeated measurements are made from highly sensitive retinal imaging modalities. The \acrshort{RPE}-Choroid boundary is rarely parallel to the imaging axis across the width of the B-scan, making it a challenge for humans to measure perpendicular angles by eye across repeated measures, which is a major source of error. The extent of angular error from manual grading across individual patients (figure \ref{fig:GPET_angle_disc}(B)) necessarily adds noise into the signal being measured, and is likely to have a significant negative impact for predicting clinical outcomes.    

            What exacerbates this potential measurement error is the extent that choroid thickness can change as the angle deviates from perpendicular, as exemplified by the quadratic relationship between \acrshort{GPET} measurements made perpendicularly against the manual ones. The performance evaluation of the automatic method would have been unjustly skewed had only perpendicular measurements been used for the analyses seen in figure \ref{fig:GPET_BA_PB_plots}(A). This is because the default presentation for \acrshort{OCT} B-scans have different axial and lateral pixel length-scales, which can result in the slightest change in marking a lateral position during manual grading to misrepresent the true choroid thickness. While this can be accounted for by viewing the B-scan in `1:1 micron mode', this squashes the choroid in the axial direction, which can make manual measurements even more prone to error \cite{lawali2023measurement} (section \ref{subsec:INTRO_MEASURE_VIEW}).

            Rahman, et al. \cite{rahman2011repeatability} reported an average inter-rater agreement of 32 $\mu$m between manual raters for measuring choroid thickness, which sets a high threshold to exceed to ensure that the signal measured manually isn't driven by human error, but true biological change. These observations from manual grading ultimately highlight a potential issue of reporting manual measurements within choroid image analysis, and make the case for releasing methods which can provide consistent and clinically meaningful measurements of the choroid, particularly in cases where effect sizes are likely to be small, such as in tracking myopia progression, which has seen effect sizes approximately in the range of 20$\mu$m -- 30$\mu$m \cite{breher2019metrological, flores2013relationship}. 

            Much of the discrepancy between manual and automatic measurements arose from choroidal features that are inherently difficult to measure. For example, the Choroid-Sclera boundaries of larger choroids (figure \ref{fig:GPET_major_disc_examples}(A,C,D)) are prone to poor quality image acquisition due to higher incidence of poor signal penetration. However, the external adjudicator preferred the automatic measurements over the manual grader for the major discrepancies in panels (A, C, D). This was because of the propensity for the manual grader to fail to measure perpendicularly, as this inconsistency becomes more obvious (and larger in error) with larger choroids. However, the manual grader was preferred for the discrepancy shown in panel (B) because \acrshort{GPET} failed to identify the Choroid-Sclera boundary underneath the fovea because it was obscured as the posterior of Haller's layer was low contrast.
            
            The choroid in figure \ref{fig:GPET_major_disc_examples}(A) represents a source of major disagreement between the manual grader and the automatic approach. The manual grader defined the Choroid-Sclera boundary as the junction below the most visible posterior vessels (green), while the automatic approach has successfully identified the boundary as a smooth and continuous contour tracking posterior vessels with much lower visibility (red), of which the external adjudicator agreed with.
    
            The choroids in figure \ref{fig:GPET_major_disc_examples} represent just over half (52\%) of all 52 major discrepancies lying outside the threshold suggested to exceed inter-observer variability between graders of \acrshort{EDI-OCT} images \cite{rahman2011repeatability}. That is, most outliers came from the same four choroids at different time points. Moreover, of the 28 thickness measurements from choroids described as having bad visibility (table \ref{tab:GPET_outlier_anal}), 20 of these came from the larger choroids (figure \ref{fig:GPET_major_disc_examples}(A,C,D)). Thus, it is encouraging to see that most areas of poor disagreement come from large choroids with poor visibility due to degrading optical signal.

        \end{mysubsection}

        \begin{mysubsection}[]{Limitations and future work}

            \acrshort{GPET} provides a more consistent approach to measuring choroid thickness than manual grading, and permits measurement of more robust metrics such as choroid area and volume (across the macula). These are two and three dimensional measurements which are able to average out measurement error posed by one-dimensional point source metrics like thickness. Moreover, as discussed in section \ref{subsec:INTRO_MEASURE_ROI}, manual thickness is more sensitive to changes in scaling and the presentation of the \acrshort{OCT} B-scan \cite{lawali2023measurement}. Thus, we expect it to be a suitable replacement for manual methods for choroid image analysis in \acrshort{OCT} image sequences. \acrshort{GPET} has also been released as open-source on GitHub to enable the research community to promote the transition from manual annotation to automatic measurement. 

            However, while the fitting procedure of \acrshort{GPET} is fully automatic, the entire image analysis pipeline is not. The pipeline involves a series of pre-processing steps to obtain the desired choroid region segmentation. This includes denoising, contrast enhancement and edge map detection. Thus, the success rate of the fitting procedure is completely dependent on the quality of the resulting edge maps, which can have a significant effect on the final edge trace and thus downstream choroidal measurements. 
            
            Failure to trace the boundaries correctly in a one-shot fashion will require further manual intervention to tune the hyperparameters (of the covariance function), select intermediary pixels between the edge endpoints, or refine the pre-processing steps to enhance the image gradient. All these additional steps not only adds precious time to the end-user, but also requires domain knowledge of the algorithm's steps, such as image processing techniques for obtaining edge maps, and underlying mathematical theory of Gaussian process regression for hyperparameter tuning. The domain-specific knowledge necessary for applying \acrshort{GPET} ultimately prevents this method's accessibility to those who are likely to apply segmentation methods most to choroidal image analysis in \acrshort{OCT} image sequences, i.e. imaging technicians who may have little experience in mathematics, programming or image processing.
            
            Automated tools to assist prognosis and treatment will strengthen future studies, enabling the delivery of robust, reproducible and responsible research in areas which have previously been at risk of human error. However, the aforementioned limitations of which many semi-automatic approaches suffer contribute to the barrier between research software and their routine use in the broader research community (i.e. clinicians and technicians), regardless of their ability to produce more consistent measurement of the choroid compared to manual grading.
            
            While \acrshort{GPET} provides a progressive step forward in obtaining the goal of producing standardised choroid measurement, this method leaves a lot of space for developing better methods. These methods would be ones which are not multi-stage, inter-dependent pipelines or require manual intervention/initialisation, and whose accessibility does not require pre-requisite knowledge of theory in image analysis, mathematics or computer science. In an ideal world, these methods would be fully automatic and require very little effort to run regardless of the end-user's experience, and would be deterministic in nature such that their approach to choroidal image analysis in \acrshort{OCT} image sequences would be robust and reproducible.

        \end{mysubsection}

        \begin{mysubsection}[]{Outputs}

            In this chapter, there have been two publication outputs, both of which has been published and peer-reviewed (by December 2024). The author's name is in bold type.
    
            \begin{itemize}\setlength\itemsep{0em}
    
                \item \textbf{Burke, Jamie}, and Stuart King. ``\textit{Edge tracing using Gaussian process regression.}'' IEEE Transactions on Image Processing 31 (2021): 138-148.
    
                \item \textbf{Burke, Jamie}, Dan Pugh, Tariq Farrah, Charlene Hamid, Emily Godden, Thomas J. MacGillivray, Neeraj Dhaun, J. Kenneth Baillie, Stuart King, and Ian J.C. MacCormick. ``\textit{Evaluation of an automated choroid segmentation algorithm in a longitudinal kidney donor and recipient cohort.}'' Translational Vision Science \& Technology 12, no. 11 (2023): 19-19.
                
            \end{itemize}

            The software associated with this method, \acrshort{GPET}, has been published as open-source and is freely available on GitHub \href{https://github.com/jaburke166/gaussian_process_edge_trace}{here}.
        
        \end{mysubsection}

        \begin{mysubsection}[]{Executive summary}

            In this chapter, we introduced a semi-automatic, edge-based approach to choroid region segmentation using Gaussian process regression, Gaussian Process Edge Tracing (\acrshort{GPET}). Advantageously, \acrshort{GPET} is able to consider local and global optimisation through use of a pre-processed edge map and a Bayesian recursive scheme using Gaussian process regression. This is particularly relevant for detecting the Choroid-Sclera boundary, which suffers from optical signal degradation and choroid stroma obstructions. \acrshort{GPET} removes the need for intensive, manual segmentation of the choroid and reduces the degree of human subjectivity injected into the measurement procedure. We found that \acrshort{GPET} agreed strongly with manual measurement of choroid thickness and, importantly, was more consistent at measuring thickness in the presence of choroid skew and curvature. This is crucial for producing standardised measurements of the choroid in \acrshort{OCT} image sequences.
            
            However, \acrshort{GPET} is a multi-stage pipeline, requiring manual initialisation and crucial knowledge of the underlying theory of Gaussian processes to work effectively. These limitations significantly impact \acrshort{GPET}'s accessibility to the wider research community and limits its ability to produce reproducible measurements given the, albeit minimal, human subjectivity injected into the segmentation procedure. Therefore, future work should consider a more fully automatic approach which does not require specialist training, pre-processing or manual intervention.
            
        \end{mysubsection}        
        
    \end{mysection}
    
\end{mychapter}

\begin{mychapter}[]{MMCQ: Semi-automatic choroid vessel segmentation using multi-scale median cut quantisation} \label{chp:chapter-mmcq}

    \begin{mysection}[]{Introduction}  \label{subsec:ch3_mmcq_intro}
    
        \begin{figure}[tb]
            \centering
            \includegraphics[width=0.75\textwidth]{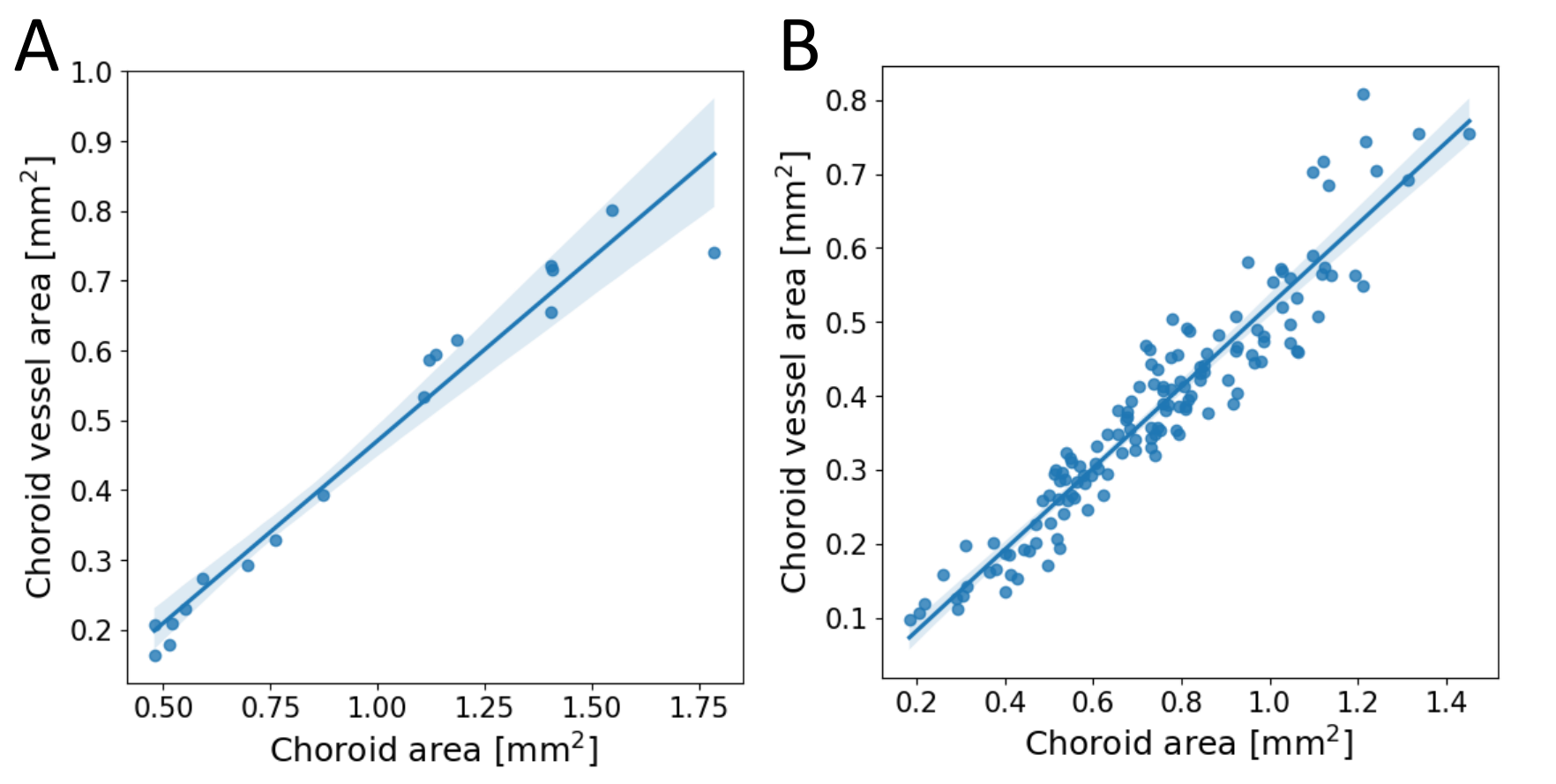}
            \caption[Relationship between total choroid area and vessel area.]{Relationship between choroidal total and vessel area for two cohorts described in sections \ref{sec:ch_app_sec_prevent} (A) and \ref{sec:ch_app_sec_shock} (B).}
            \label{fig:MMCQ_drisc_prevent_ca_cv}
        \end{figure}
    
        In the last chapter we developed a semi-automatic procedure to segment the choroidal space, which enables measurement of choroid-derived thickness and area in a single \acrshort{OCT} B-scan (and volume across the macula). However, the choroid is primarily a vascular layer, with the highest blood flow per unit tissue weight of any body organ \cite{nickla2010multifunctional}. Thus, enabling quantification of the size of choroidal vessels would be significantly more informative than the overall choroidal space. Understanding pathophysiology through vascular metrics, instead of regional metrics, may provide a more stable and representative way of studying pathophysiological changes in the presence and during treatment of systemic disease which affect the microvascular system \cite{agrawal2016choroidal}. Because of the historic challenges of measuring the choroidal vasculature from OCT reliably, choroidal thickness has been used instead as a surrogate.
        
        It may be that an increase in choroidal vessel density is directly proportional to changes in the choroidal space. In figure \ref{fig:MMCQ_drisc_prevent_ca_cv} we visualise the relationship between total choroid area against choroid vessel area for two populations discussed in more detail in sections \ref{sec:ch_app_sec_prevent} and \ref{sec:ch_app_sec_shock}. In each case, measurements were taken from a 6 mm, fovea-centred \acrshort{ROI} from horizontal-line \acrshort{EDI-OCT} B-scans and each scatter-point represents an individual eye. See section \ref{subsec:INTRO_MEASURE_ROI} for more details on how total area and vessel area were measured and sections \ref{sec:ch_app_sec_prevent} and \ref{sec:ch_app_sec_shock} for information on the relevant populations. In both cases, vessel area and total choroid area are significantly correlated with each other (panel A, Pearson $r$=0.99, P<0.0001; panel B, Pearson $r$=0.95, $p$<0.0001). Thus, information on the choroidal space may reflect that of the choroidal vasculature, but measuring the choroidal vasculature directly is likely to be more representative of the choroid if done accurately.
    
        A popular metric to quantify the choroidal vasculature in an eye is the choroid vascular index (\acrshort{CVI}), which is a measure of an individual choroid's vessel density in an \acrshort{OCT} B-scan. \acrshort{CVI} describes the proportion of the choroid in an \acrshort{OCT} B-scan made up of vessels, and is thus a dimensionless ratio, bound between 0 and 1. In an \acrshort{OCT} B-scan, \acrshort{CVI} is simply the number of pixels classified as choroid vessel divided by the number of pixels classified as choroid. The choroidal space is made up of interstitial fluid, or stroma, seen as brightly illuminated regions in the \acrshort{OCT} B-scan, with interspersed, irregular areas of darker intensity representing choroidal vasculature. This has been both empirically observed \cite{sohrab2012pilot, branchini2013analysis} and widely accepted among the research community \cite{agrawal2020exploring}. 

        \acrshort{CVI} has become particularly promising metric in choroidal image analysis and biomarker investigation because of it's potential invariance to certain physiological factors which often affect choroid thickness like systemic blood pressure \cite{agrawal2016choroidal}. This is a direct consequence of it's formula, as \acrshort{CVI} is inversely proportional to choroid regional metrics, and it's signal is primarily driven by choroidal vascular metrics. Thus, if the magnitude of change in choroidal thickness reflect a similar magnitude of change in choroidal vessel area, the change in \acrshort{CVI} will be minimal. Therefore, \acrshort{CVI} has been observed as a more robust index of choroidal variation compared with a one-dimensional, point-source measurement like subfoveal choroid thickness \cite{kim2021choroidal}. This relatively new choroidal measurement has been used among the research community to quantify the choroidal vasculature in ocular pathology --- see Agrawal, et al. \cite{agrawal2020exploring} for an extensive review on \acrshort{CVI} in chorioretinal disease. Figure \ref{fig:MMCQ_choroid_cvi} shows an OCT B-scan with region and vessel segmentations overlaid. 

        \begin{figure}[tb]
            \centering
            \includegraphics[width=\linewidth]{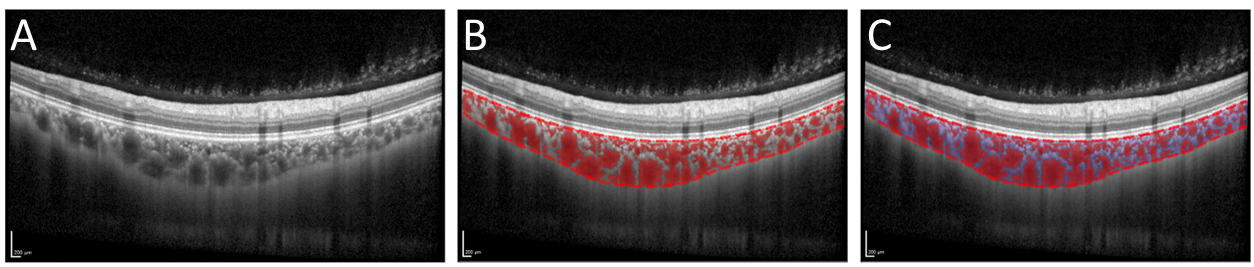}
            \caption[Vessel segmentation in an \acrshort{OCT} B-scan.]{Vessel segmentation in an \acrshort{OCT} B-scan. (A) \acrshort{OCT} B-scan. (B) Choroid region demarcated with dashed red lines and vessel segmented as red blobs. (C) \acrshort{CVI} is measured as the ratio of red pixels to the sum of red and shaded blue pixels.}
            \label{fig:MMCQ_choroid_cvi}
        \end{figure}
    
        Estimating \acrshort{CVI} requires segmentation of the choroidal vasculature, which is a significantly more challenging task than choroid region segmentation. Segmenting the choroid region only requires a single shape of heterogeneous vascular structure to be identified, which sits below the hyperreflective \acrshort{RPE} layer and above the sclera which appears relatively homogeneous in pixel intensity --- two well-defined landmarks. Conversely, segmenting the choroidal vessels is a much more ambiguous task. In terms of size and shape, the choroid contains a dense, heterogeneous population of vessels interspersed and suspended by connective tissue in its interstitial space. Cross-sectional image capture using \acrshort{OCT} of this microvascular bed often means that individual choroidal vessels are captured at oblique cross-sections and thus do not have clear vessel wall boundaries. This is exacerbated by poor illumination, contrast and inherent speckle noise. This can make manual segmentation prohibitively time consuming, and has the potential to be impacted by human subjectivity.
    
        \begin{figure}[tb]
            \centering
            \includegraphics[width=\linewidth]{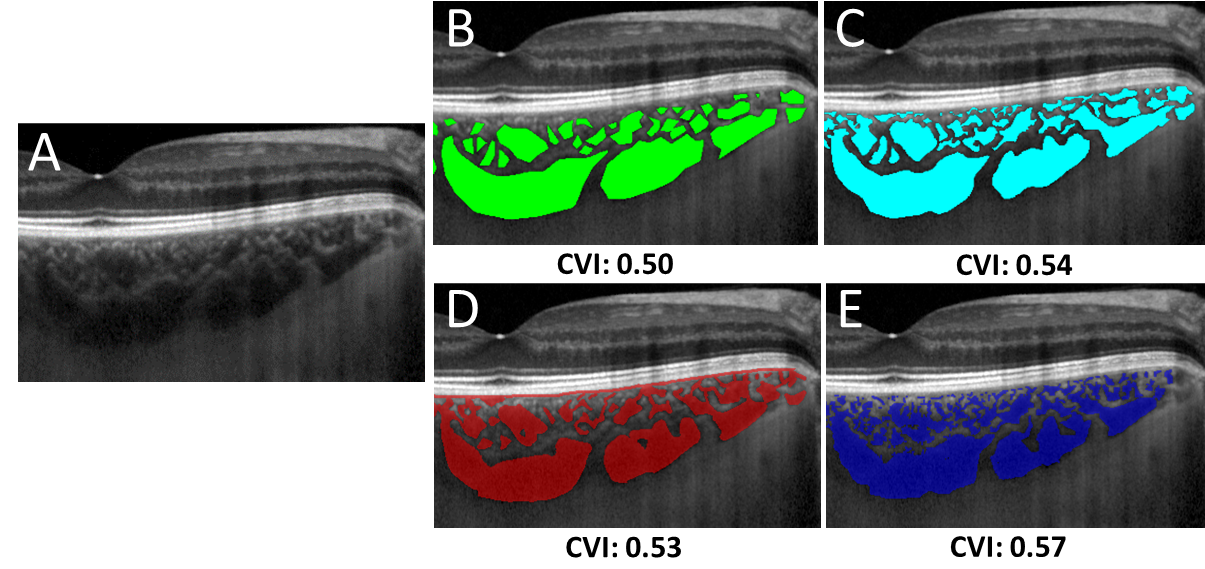}
            \caption[Poor agreement between manual vessel segmentation labels.]{Poor agreement between manual vessel segmentation labels. (A) \acrshort{OCT} B-scan. (B -- E) Manual choroid vessel segmentations from three raters, supervisor Ian J.C. MacCormick (B, E), image technician Charlene Hamid (C) and author Jamie Burke (D). The corresponding \acrshort{CVI} is annotated, using the same choroid region segmentation to facilitate interpretation.}
            \label{fig:MMCQ_manual_cvi}
        \end{figure}
    
        Figure \ref{fig:MMCQ_manual_cvi} highlights the potential variation in vascular metrics such as \acrshort{CVI} when manual segmentation is used for choroid vessel segmentation. Three raters were asked to manually segment the same choroid (supervisor Dr. Ian J.C. MacCormick, author Jamie Burke and image technician Charlene Hamid) using ITK-Snap, which permits pixel-level annotation \cite{py06nimg}. The segmentations and resulting \acrshort{CVI} are shown in panels (B -- E).
        
        While the segmentations from panels (B, C) were done according to each graders' predisposed notion of the choroidal vasculature and without measurement protocol, the segmentations in panels (D, E) followed the measurement protocol outlined in appendix \ref{apdx:seg_protocol} using definitions and equipment from appendices \ref{apdx:definitions} and \ref{apdx:equipment_protocol} (grader I.M. segmented the choroid twice, 2 years apart). 
        
        Note the significant variation in granularity from each segmentation, and resulting difference in \acrshort{CVI}. Importantly, even following the same protocol (panels (D, E)) led to a difference in CVI of 4\%, and the largest difference of 7\% came from the same individual (panels (B, E)) after a time difference of 2 years before and after a well defined measurement protocol. These error are likely to be clinically significant, given Agrawal, et al.'s review \cite{agrawal2020exploring} suggested effect sizes between 2\% -- 6\% between healthy and diseased eyes.
    
        While this only describes the potential error in manual vessel labelling for a single example, it alludes to a wider issue around manual labelling in and of itself. Thus, in lieu of manual annotation, local thresholding algorithms became commonplace for this segmentation task \cite{sonoda2014choroidal, branchini2013analysis, agrawal2016choroidal} (see section \ref{subsubsec:intro_litreview_vessel} for a brief literature review on the topic). Briefly, the most widely used approach is the Niblack algorithm \cite{agrawal2020exploring, betzler2022choroidal}. This adaptive thresholding algorithm is semi-automatic and requires parameter specification, and we have briefly described its potential problems (see section \ref{sec:INTRO_litreview_challenge_vessels}). Importantly, very few studies report the parameter configuration used for Niblack \cite{muller2022application, arian2023automatic}. Instead, many utilise the version supported by Fiji (ImageJ) \cite{schindelin2012fiji} and either manually tune the internal algorithm parameters for every B-scan, or use the default setting. Moreover, ad-hoc pre-processing steps such as manual demarcation of the \acrshort{ROI} \cite{agrawal2016choroidal} and manual measurement of the average pixel intensity of specific choroidal vessels \cite{sonoda2014choroidal} are typically used in research. These reasons alone can lead to under-specification of the algorithm, potentially poor quality segmentation and non-standardisation across studies and populations.

        Therefore, we propose that the inconsistent and widespread usage of Niblack's method has the potential to prevent the reporting of reproducible and standardised ocular measurements of the choroidal vasculature. We further propose that there is space for better methods which are able to target the core challenges of choroid vessel segmentation, which a fixed parameter setup such as Niblack, whose creation was designed around word document thresholding \cite{niblack1985introduction}, is not able to address. Finally, Niblack is likely the most widely used because of it's availability on Fiji (ImageJ) \cite{schindelin2012fiji}, or through membership-only access \cite{betzler2022choroidal}, while other state-of-the-art methods based on deep learning are not openly available \cite{zhu2022synergistically, wen2024transformer}. Thus, there is an unmet need for more standardised approaches to choroid vessel segmentation which are domain-specific and freely available to the research community.
    
        Because of the poor contrast and heterogeneity of the choroidal vasculature as seen on \acrshort{OCT}, we hypothesised that a multi-scale procedure based on enhancement would address these two issues. Here, we propose a multi-scale approach to choroidal vessel segmentation, after sufficiently enhancing the choroidal vasculature through an extension to contrast limited adaptive histogram equalisation. This method performs image quantisation, enhancement and pixel clustering at several scales designed for the accurate segmentation of both small- and large-scale vasculature in the choroid seen in \acrshort{EDI-OCT} B-scans. Importantly, we sought to develop a method which did not require a human element, and which was reproducible, high quality and aimed to release it as open-source.

        \begin{mysubsection}[]{Niblack autolocal thresholding}\label{subsec:MMCQ_intro_niblack}
    
            The Niblack algorithm \cite{niblack1985introduction} is a local thresholding technique which binarises individual pixels within an image based on the summary statistics of a patch centred on each pixel. For each pixel $(x, y)$, a threshold $T(x,y)$ is estimated using
            \begin{equation}
                T(x,y) = \mu(x,y) + k \cdot \sigma(x,y),
            \end{equation}
            where $\mu$ and $\sigma$ are the mean and standard deviation within a neighbourhood centred on pixel $(x,y)$. The size of this neighbourhood is defined by parameter $w$, which should be dependent on the size of the choroidal vessels being segmented. $k$ is the standard deviation offset, which is a constant that determines the weighting of the standard deviation against the overall mean intensity of the window. Thus, $k$ defines how sensitive the local threshold value is to local variance --- if $k$ is large and negative then the threshold will be very low and over-segmentation will occur, and vice versa for where $k$ is large and positive. Muller, et al. \cite{muller2022application} reported that this algorithm suffers when a neighbourhood around a pixel is homogeneous in pixel intensity, which can result in noise amplification. 
            
            There are two widely used approaches to semi-automatic choroid vessel segmentation which use Niblack's local thresholding method, Sonoda's \cite{sonoda2014choroidal} and Agrawal's \cite{agrawal2020exploring, agrawal2016choroidal, betzler2022choroidal}. Sonoda, et al. \cite{sonoda2014choroidal} is dependent on the end-user using Fiji (Imagej) \cite{schindelin2012fiji} and selecting the choroidal space, identifying the central location of three choroidal vessels (of size greater than 100 $\mu$m) and lowering the mean intensity of the B-scan to the minimum value of the three individual vessels' mean intensity. Only after this pre-processing is Niblack's method applied, with no details on parameter specification. This approach is described in detail in the supplementary material S1 of the original paper by Sonoda, et al. \cite{sonoda2014choroidal}. Agrawal's approach removed the need to select individual choroidal vessels, and used several morphological operations to enhance the vessel segmentation \cite{betzler2022choroidal}. In both cases, manual identification of the \acrshort{ROI} to generate measurement is required. To our knowledge, the most detailed description (which is publicly available) of their modified Niblack method ---  which is available under membership-only --- can be found in the work by Betzler, et al. \cite{betzler2022choroidal}. Importantly, in both cases \cite{sonoda2014choroidal, agrawal2020exploring, agrawal2016choroidal, betzler2022choroidal}, there is no mention of the window size $w$ or standard deviation offset $k$, which potentially means the default setting is used, or the choice is at the end-user's discretion.
    
            \begin{figure}[tb]
                \centering
                \includegraphics[width=\linewidth]{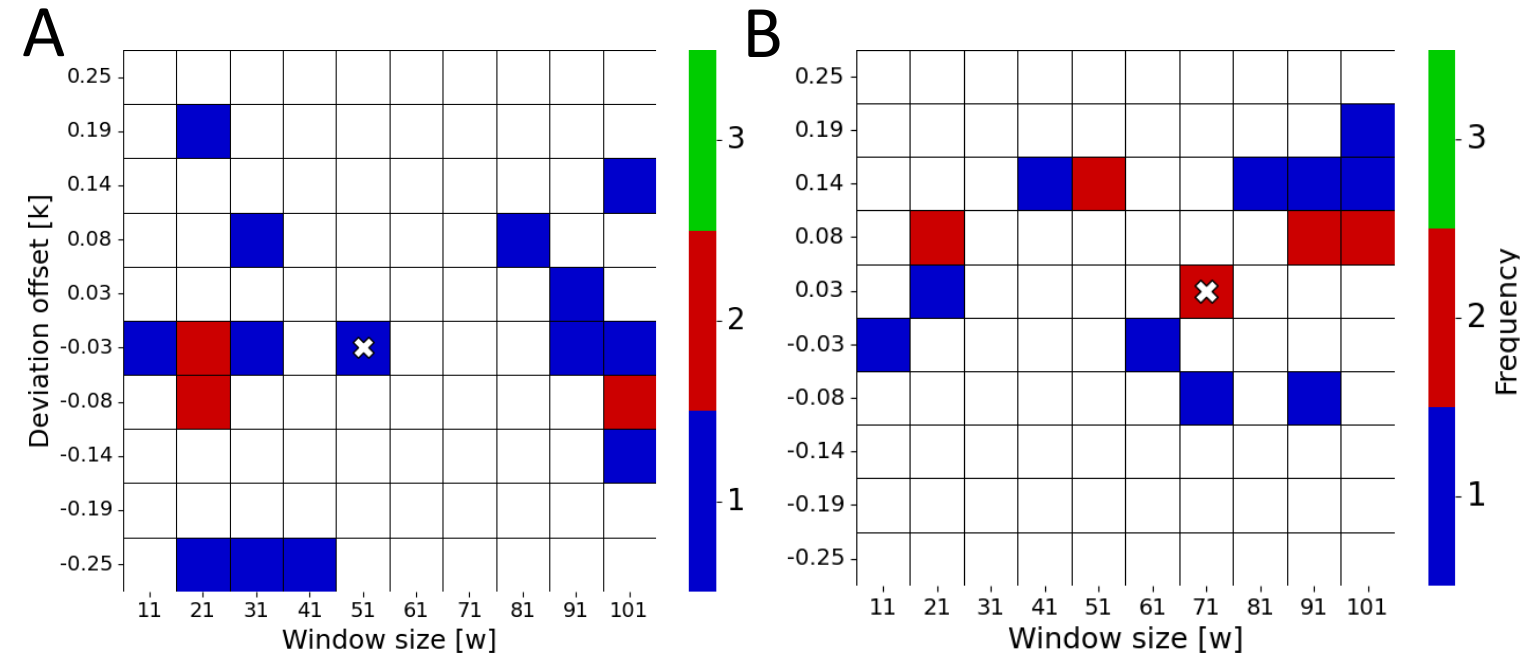}
                \caption[Grid-searching optimal thresholding parameter configurations for Niblack's method against manual labelling.]{Frequency map after grid-searching over 100 possible pairs of Niblack parameters $(w, k)$ and optimising for the best segmentation (using Dice) against raters M1 (A) and M2 (B). The white cross represents the optimal parameter configuration for the same B-scan which had the highest Dice similarity score between the manual segmentations performed by M1 and M2.}
                \label{fig:MMCQ_niblack_parameter_count}
            \end{figure}

            \begin{figure}[tb]
                \centering
                \includegraphics[width=\linewidth]{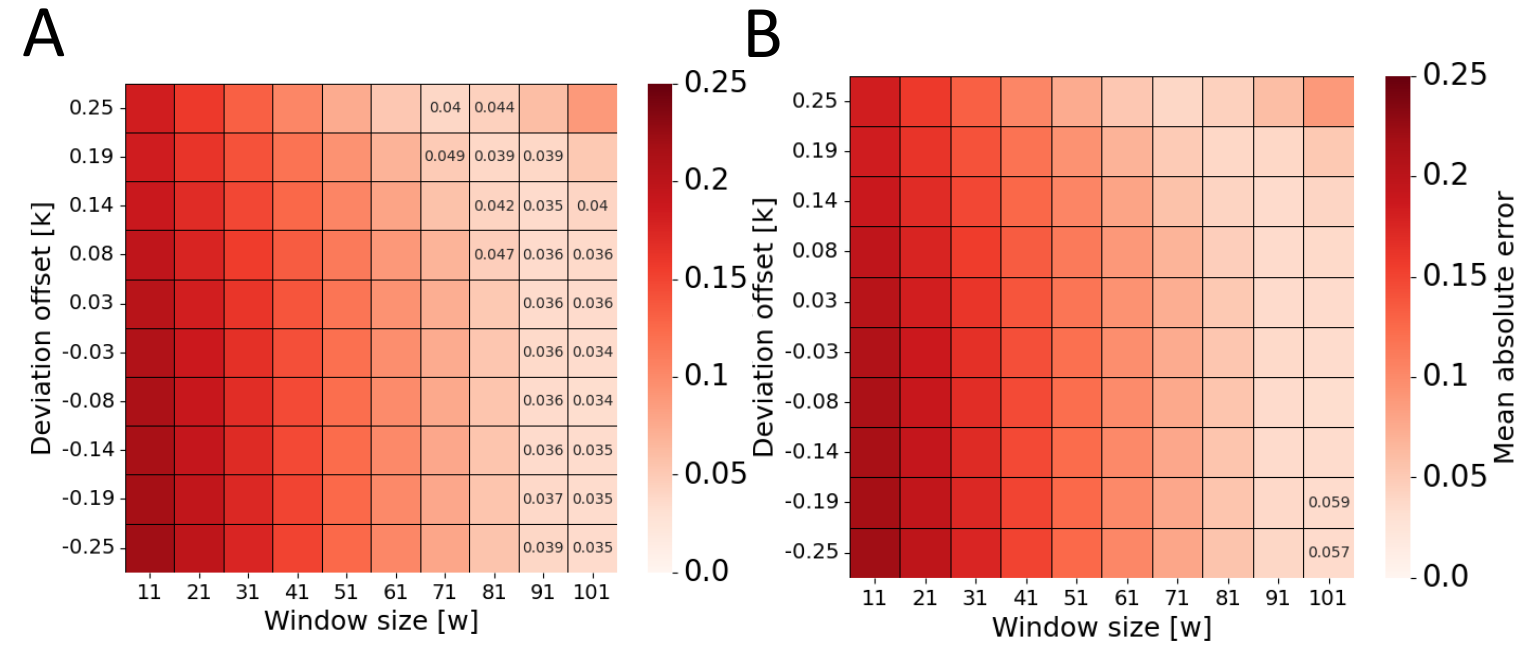}
                \caption[Error sensitivity of Niblack's thresholding parameters against manual labelling.]{Heat map of mean absolute errors (\acrshort{MAE}s) comparing \acrshort{CVI} computed with Niblack, using 100 possible Niblack parameters $(w, k)$, against manual graders M1 (A) and M2 (B). Cells annotated are those whose  \acrshort{MAE} was below 0.06 (6\%).}
                \label{fig:MMCQ_niblack_parameter_mae}
            \end{figure}

            \begin{figure}[tb]
                \centering
                \includegraphics[width=\linewidth]{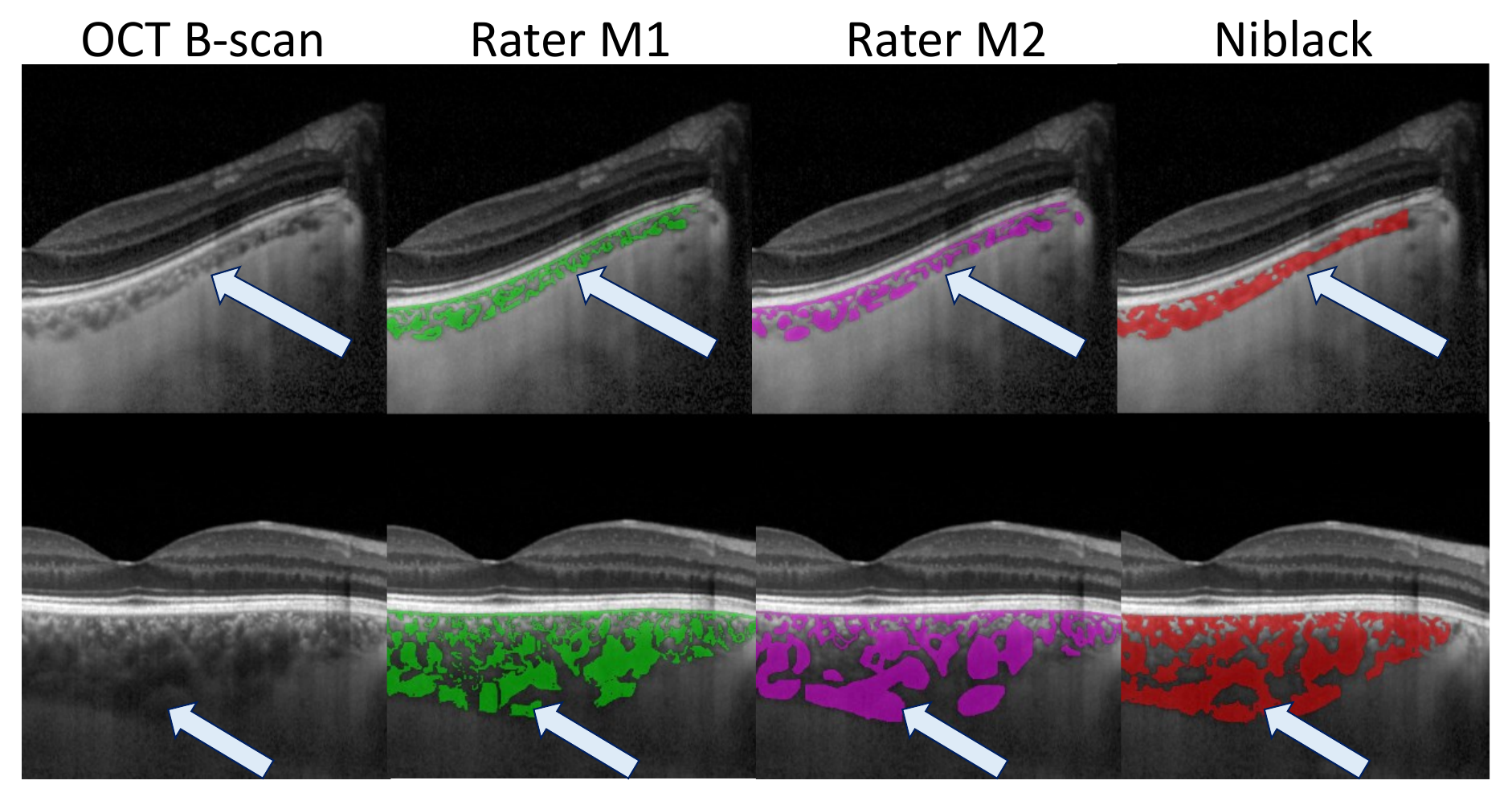}
                \caption[Qualitative comparison of the Niblack method against manual labelling.]{Choroid vessel segmentation for two choroids from raters M1, M2 and Niblack (with the same parameter configuration).}
                \label{fig:MMCQ_niblack_sameparams}
            \end{figure}   

            The rule of thumb for selecting $w$ is based on the size of the choroid and choroidal vessels of interest, while $k$ is more abstract and therefore more difficult to define as a sensitivity measure. Muller, et al. \cite{muller2022application} decided to fix $k$ and allow manual annotators to tune $w$, as this was a more intuitive tunable parameter. During the manual annotation procedure, Muller, et al. \cite{muller2022application} reported that varying $w$ between 20 and 75 with $k$ fixed as -0.05 resulted in a mean (standard deviation) difference in \acrshort{CVI} of 11.2\% $\pm$ 4.1\% (range between 2\% -- 21\%). To put this into context, Agrawal, et al.'s review \cite{agrawal2020exploring} discussed multiple studies which reported differences in \acrshort{CVI} values between healthy and diseased eyes in the range of 2\% -- 6\%. Thus, it would appear that variations of Niblack's internal parameter could have a clinically significant impact to measurements like \acrshort{CVI}. 
            
            To understand the importance of these two parameters in more detail, we investigated the possible variation of \acrshort{CVI} and segmentation performance for different paired parameter configurations $(w, k)$ in a small dataset of 20 horizontal-line, \acrshort{EDI-OCT} B-scans of 20 right eyes (10 healthy participants and 10 participants receiving renal transplantation --- data sourced from the larger cohort described in section \ref{sec:ch_gpet_eval}, table \ref{tab:GPET_eval_pop}).
    
            Choroidal vessels were manually segmented by two raters M1, supervisor Dr. Ian J.C. MacCormick; and M2, author Jamie Burke using ITK-Snap \cite{py06nimg}, which permits pixel-level annotation. Graders followed the measurement protocol outlined in appendix \ref{apdx:seg_protocol} and were blinded to each others segmentations. Graders used the definitions and equipment outlined in appendices \ref{apdx:definitions} and \ref{apdx:equipment_protocol} to aid in region and vessel detection. Niblack autolocal thresholding was applied to each B-scan using 100 different parameter configurations (following Agrawal's procedure \cite{betzler2022choroidal}), with window size $w$ varying between 11 $\times$ 11 and 101 $\times$ 101 over 10 equally spaced intervals, and similarly for $k$ from -0.25 to 0.25.
            
            An optimal parameter configuration was defined as the pair $(w, k)$ which had the highest Dice score comparison between the resulting segmentation and manual segmentation. As the parameter space of $w$ and $k$ are discretised, we can count the number of occurrences for each optimal parameter configuration. We also measured the \acrshort{CVI} in a standard 6 mm fovea-centred, \acrshort{ROI} using each paired parameter setting, and measured the mean absolute error (\acrshort{MAE}) between the \acrshort{CVI} outputted using Niblack and those outputted using the manual graders' segmentations. \acrshort{MAE}s were averaged across the 20 B-scans, creating an overall averaged, \acrshort{MAE} of \acrshort{CVI} per parameter configuration.
            
            Figure \ref{fig:MMCQ_niblack_parameter_count} shows the frequency count of optimal parameter settings for all 20 B-scans, when compared with manual segmentations from rater M1 (A) and rater M2 (B). While this is only a small dataset, what is clear is the extent of the variation of optimal parameter configurations across the B-scans, and across raters. In fact, for the choroid with the highest Dice similarity between raters (Dice score 0.83), the optimal parameter configuration for Niblack agreeing with M1 was (51, -0.03) and M2 was (71, 0.03) (white crosses). The extent of variation of optimal parameter configuration and the different pattern of occurrences both suggest that no one optimal configuration is applicable across all B-scans or applicable across different raters.
            
            Figure \ref{fig:MMCQ_niblack_parameter_mae} shows the heatmaps of averaged \acrshort{MAE} between \acrshort{CVI} measured using Niblack and raters M1 (A) and M2 (B). A similar pattern is observed across raters, which suggest that higher window sizes overall produce less error (when compared to manual segmentations), and that this is consistent across most standard deviation offsets. The minimum, averaged \acrshort{MAE} for M1 and M2 were 0.034 (3.4\%) and 0.057 (5.7\%), respectively. Thus, while expert manual grading is not a gold standard, this does raise some concern over the accuracy of Niblack's \acrshort{CVI}, particularly given the small effect sizes observed in the existing literature \cite{agrawal2020exploring}. Moreover, these minimum \acrshort{MAE} values corresponded to different parameter configurations (Minimum \acrshort{MAE} for M1, $w=$101, $k$ $\approx$ -0.055; for M2, $w=$101, $k=$-0.25.  

            Thus, no one parameter configuration is optimal across different choroids, and it should be tailored for the particular choroid. Figure \ref{fig:MMCQ_niblack_sameparams} shows the segmentations by raters M1 (green) and M2 (pink) as well as Niblack (red) using the same parameter setting ($w=51$, $k=-0.05$ \cite{muller2022application}). Compared with the choroid in the bottom row, Niblack over-segments the choroid in the top row (arrows). Quantitatively, Niblack over-segmented both choroids according to the manual graders, producing an average \acrshort{MAE} of 0.158 (15.8\%) and 0.104 (10.4\%) for the top and bottom choroids, respectively (averaged across raters).            
            
            Therefore, we propose that Niblack's autolocal threshold algorithm may be sufficient for individual choroid analysis, but likely requires significant tuning of parameters for optimal segmentation quality, thus rendering it unfeasible for large-scale choroidal image analysis. Moreover, we hypothesise that poor selection of parameters may result in over-segmentation of the interstitial space, particular in the anterior section of the choroid where smaller vessels and the choriocapillaris are situated. 
            
        \end{mysubsection}

    \begin{mysubsection}[]{Contrast limited adaptive histogram equalisation}\label{subsec:MMCQ_intro_CLAHE}

        Contrast limited adaptive histogram equalisation (\acrshort{CLAHE}) \cite{pizer1987adaptive} is an image enhancement method for increasing the contrast in an image. For the choroid, increased contrast can improve the definition of vessel walls, thus making the task of identifying vessel from interstitial space an easier one. Often the choroidal vessels have poor contrast with the interstitial space and significantly so for smaller vessels, and \acrshort{CLAHE} is an approach to rectify this.

        \acrshort{CLAHE} is a local enhancement method, which splits the image into patches and adjusts the contrast of each patch locally. For each patch, it adjusts the contrast by spreading out the brightness levels evenly using its histogram, known as histogram \textit{equalisation}. \acrshort{CLAHE} is particularly helpful for ensuring that different parts of the image aren't overly enhanced, as it is able to limit the redistribution of each patch's brightness levels. Each patch is then combined using interpolation to preserve continuity between the newly enhanced patches. \acrshort{CLAHE} has two parameters, one for deciding how many patches to generate (and consequently how local the enhancement will be) and another for deciding the strength of enhancement.
        
        In figure \ref{fig:MMMCQ_CLAHE} we show an \acrshort{OCT} B-scan with a particularly low contrast choroid with poor vessel wall definition. After cropping out only the choroid and applying \acrshort{CLAHE} by splitting the region into an 8 $\times$ 8 grid of patches and limiting each patch's histogram bins to 25 pixels, choroidal vessels suspended in connective tissue in the interstitial space have greater definition. However, note that the contrast of larger vessels is greater (red arrows) than for smaller smaller vessels (blue arrows).

        \begin{figure}[tbp]
            \centering
            \includegraphics[width=\linewidth]{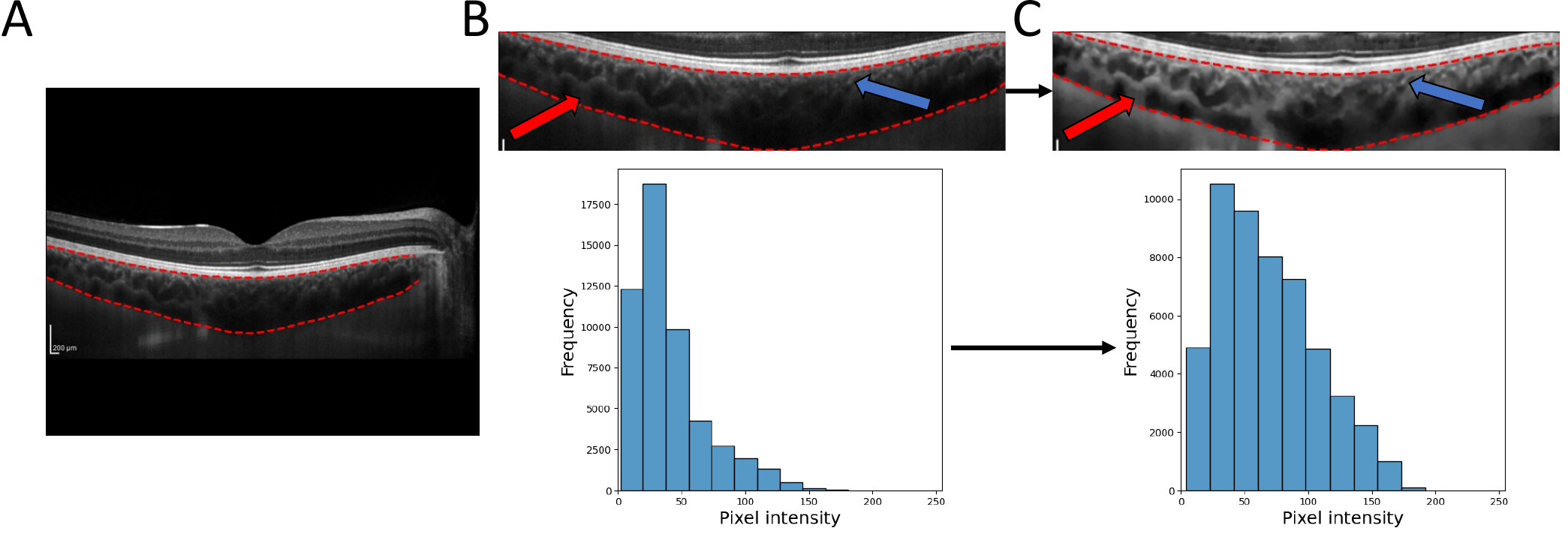}
            \caption[Demonstration of \acrshort{CLAHE} on an \acrshort{OCT} B-scan.]{Demonstration of \acrshort{CLAHE} on an \acrshort{OCT} B-scan. (A) \acrshort{OCT} B-scan with poor contrast choroidal vessels, with choroid demarcated in red. Cropped choroid and histogram of choroid pixels in raw B-scan (B) and after application of \acrshort{CLAHE} (C). Red arrows show good enhancement, blue arrows show poor enhancement.}
            \label{fig:MMMCQ_CLAHE}
        \end{figure}

        \begin{figure}[!t]
            \centering
            \includegraphics[width=0.75\linewidth]{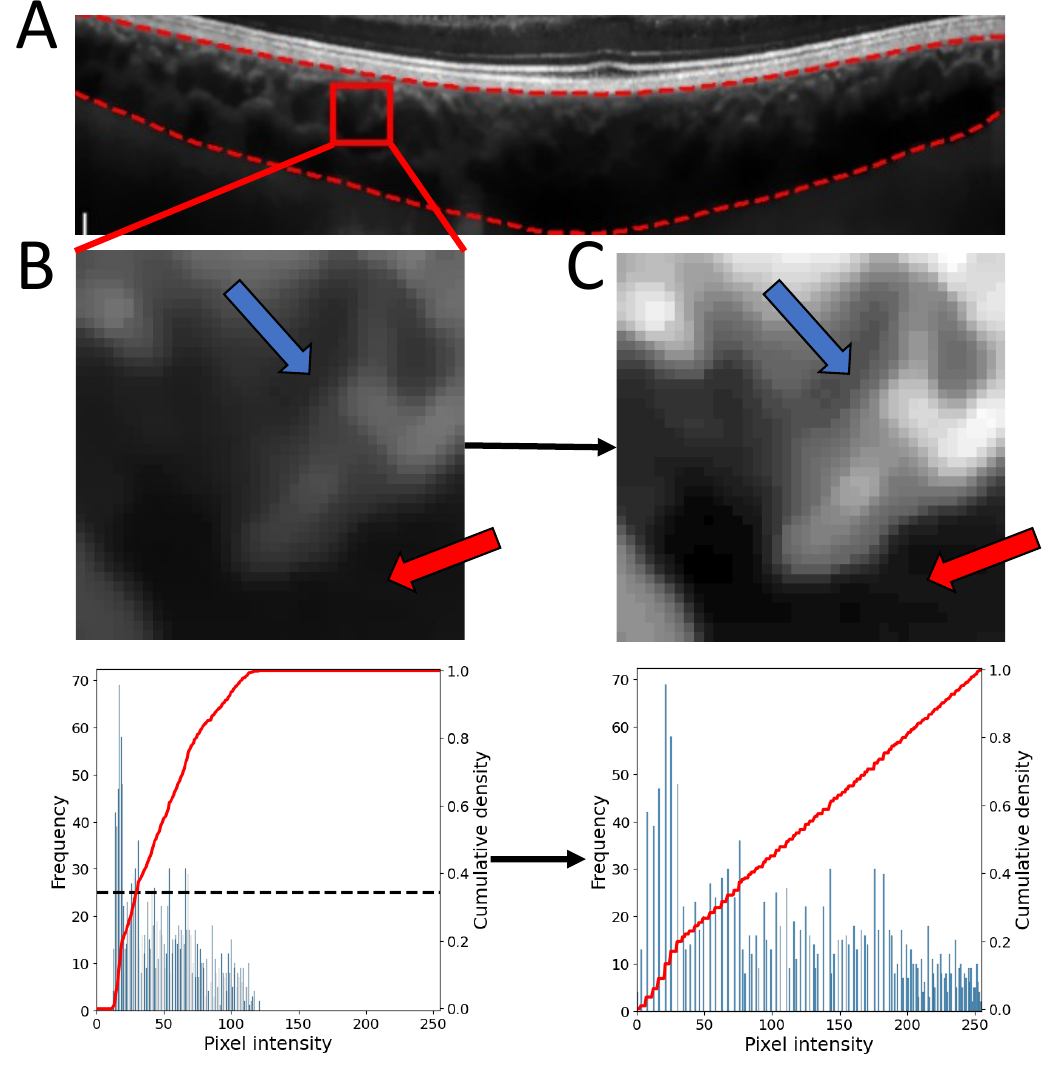}
            \caption[Demonstration of \acrshort{CLAHE} at the patch-level during pre-processing.]{Demonstration of \acrshort{CLAHE} at the patch-level during pre-processing. (A) Choroid cropped from \acrshort{OCT} B-scan in figure \ref{fig:MMMCQ_CLAHE} with boundaries annotated and a red square representing patch of interest in panels (B, C). Zoomed-in patch before (B) and after (C) contrast enhancement with \acrshort{CLAHE}. Below each patch is the patch histograms shown as blue bars with their normalised cumulative distribution function shown in red. In (B), the black dashed line represents the histogram clipping limit. Red arrows show good enhancement, blue arrows show poor enhancement.}
            \label{fig:MMCQ_CLAHE_patch}
        \end{figure}
        
        A closer inspection for a small, square patch in figure \ref{fig:MMCQ_CLAHE_patch} describes the \acrshort{CLAHE} process in more detail. The patch's histogram undergoes a transformation which results in clipping the distribution of histogram bins (brightness levels) and re-distributing excess pixels uniformly across the rest of the histogram bins. Excess pixels are defined by those in each histogram bin which fill it beyond the \textit{clip limit}, seen by the black dashed-line in panel (B). This process helps in preventing over-amplification of noise in relatively homogeneous areas (in pixel intensity) of the image.
        
        After uniform re-distribution of excess pixels, the resulting histogram is then stretched to cover the possible range of bins (typically 256 for an 8-bit image), which produces the adjusted patch you see in panel (C). From the perspective of cumulative distributions, \acrshort{CLAHE} promotes linearity in the cumulative density function for each patch's new histogram, as evidenced by the attempted flattening of the red curve between panels (B) and (C). The linearity of this cumulative density function represents the distributing of pixels evenly across the range of brightness levels. Similar to the global view from figure \ref{fig:MMMCQ_CLAHE}, \acrshort{CLAHE} is helpful in improving the contrast for larger vessels (red arrows) but not for smaller vessels (blue arrows), whose choroidal vessel walls still have relatively poor contrast relative to the choroidal interstitial space.

    \end{mysubsection}

    \begin{mysubsection}[]{Median cut image quantisation}\label{subsec:MMCQ_method_vesselshadow}

        \begin{figure}[tb]
        \centering
        \includegraphics[width=\linewidth]{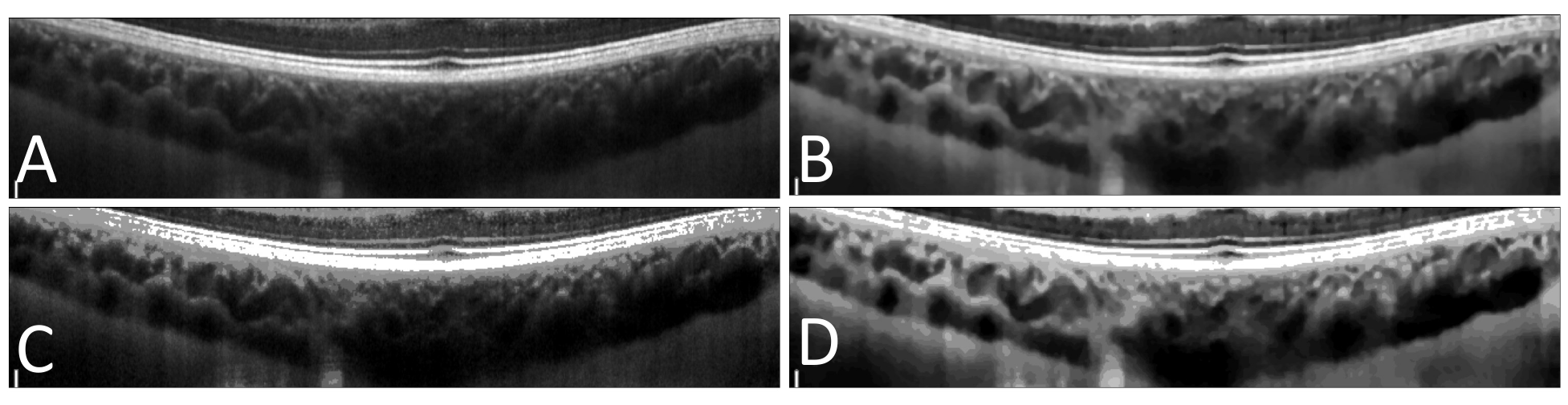}
        \caption[Demonstration of median cut quantisation on an \acrshort{OCT} B-scan.]{Demonstration of median cut quantisation on an \acrshort{OCT} B-scan. (A) Raw \acrshort{OCT} B-scan. (B) B-scan after median filter denoising and \acrshort{CLAHE} contrast enhancement. Median cut quantisation of raw (C) and enhanced (D) B-scans into 20 quantisation levels/clusters.}
        \label{fig:MMCQ_mediancut}
        \end{figure}

        A problem with application of \acrshort{CLAHE} is its inability to enhance smaller vessels. To overcome this limitation we propose that instead of enhancing the raw pixel intensities, we enhance a quantised equivalent. Quantisation is common in image compression and is simply a way of representing an image using fewer intensity values. Thus, instead of allowing all 256 brightness levels in an 8-bit image, we can reduce (or quantise) this to 16 levels (4-bit) instead. A popular approach to quantisation is the median cut algorithm \cite{heckbert1982color}, and for a grayscale image, this quantisation procedure is in a similar vein to pixel clustering but is far less computationally expensive.

        Median cut quantisation clusters are computed by repeatedly splitting the image histogram at it's median value into two groups, and splitting each resulting group by their own median value and so on recursively until a desired number of clusters are produced. The quantised image is the result of converting the original intensity of each pixel to the average value of all pixel intensities assigned the same cluster. This simplification of the original image's histogram is particularly useful for assigning pixels to categories of similar pixel value, and can be used at a local patch-based level, i.e. for small patches in the choroid of \acrshort{OCT} B-scans.

        The method developed in this chapter leverages median cut quantisation to enhance the choroid in \acrshort{OCT} B-scans at a local level, and across multiple scales. However, a slight variation made to this algorithm is that the median value for each cluster group is assigned rather than the average value. This is because, relative to the mean, the median value is more robust to outliers and is a better measure of centrality for asymmetrical distributions, such as those found here in the pixel intensity histograms of the choroid in \acrshort{OCT} B-scans. 
        
        Figure \ref{fig:MMCQ_mediancut} shows the raw \acrshort{OCT} B-scan and an enhanced B-scan in panels (A) and (B) with their quantised equivalents (using 20 quantisation levels/clusters) in panels (C) and (D). Note that the simpler representation is advantageous for vessel classification, and that the enhanced \acrshort{OCT} B-scan provides a suitable compression for differentiating large vessel from interstitial space, but not for smaller vessels.
        
    \end{mysubsection}
    
    \begin{mysubsection}[]{Vessel shadow compensation} \label{sec:MMCQ_intro_shadows}
            
        \begin{figure}[tb]
            \centering
            \includegraphics[width=\linewidth]{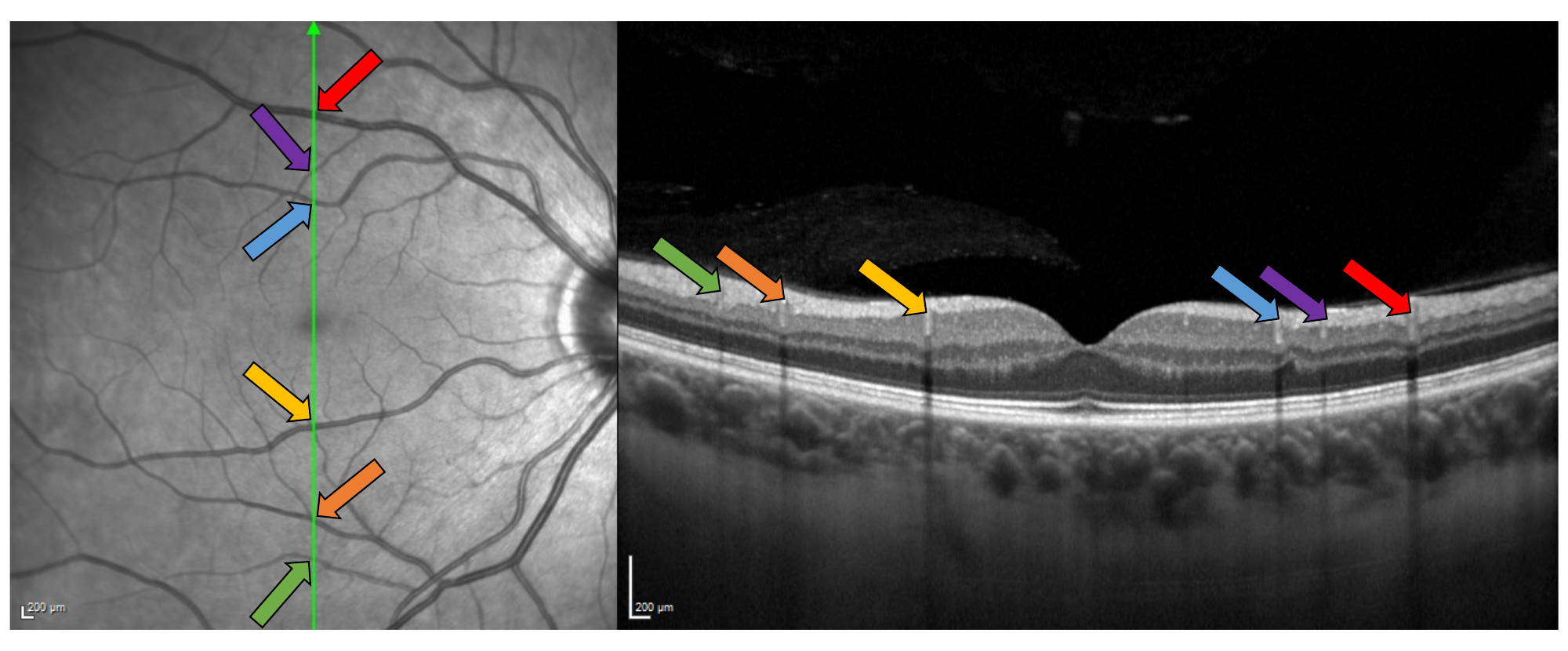}
            \caption[Correspondence between en face retinal vessels and A-scan shadowing on an \acrshort{OCT} B-scan.]{Demonstration of superficial retinal vessels on the localiser SLO causing shadows on the \acrshort{OCT} B-scan, corrupting the relevant A-scans of the choroid.}
            \label{fig:MMCQ_shadow_examples}
        \end{figure}

         \begin{figure}[tbhp]
            \centering
            \includegraphics[width=0.925\linewidth]{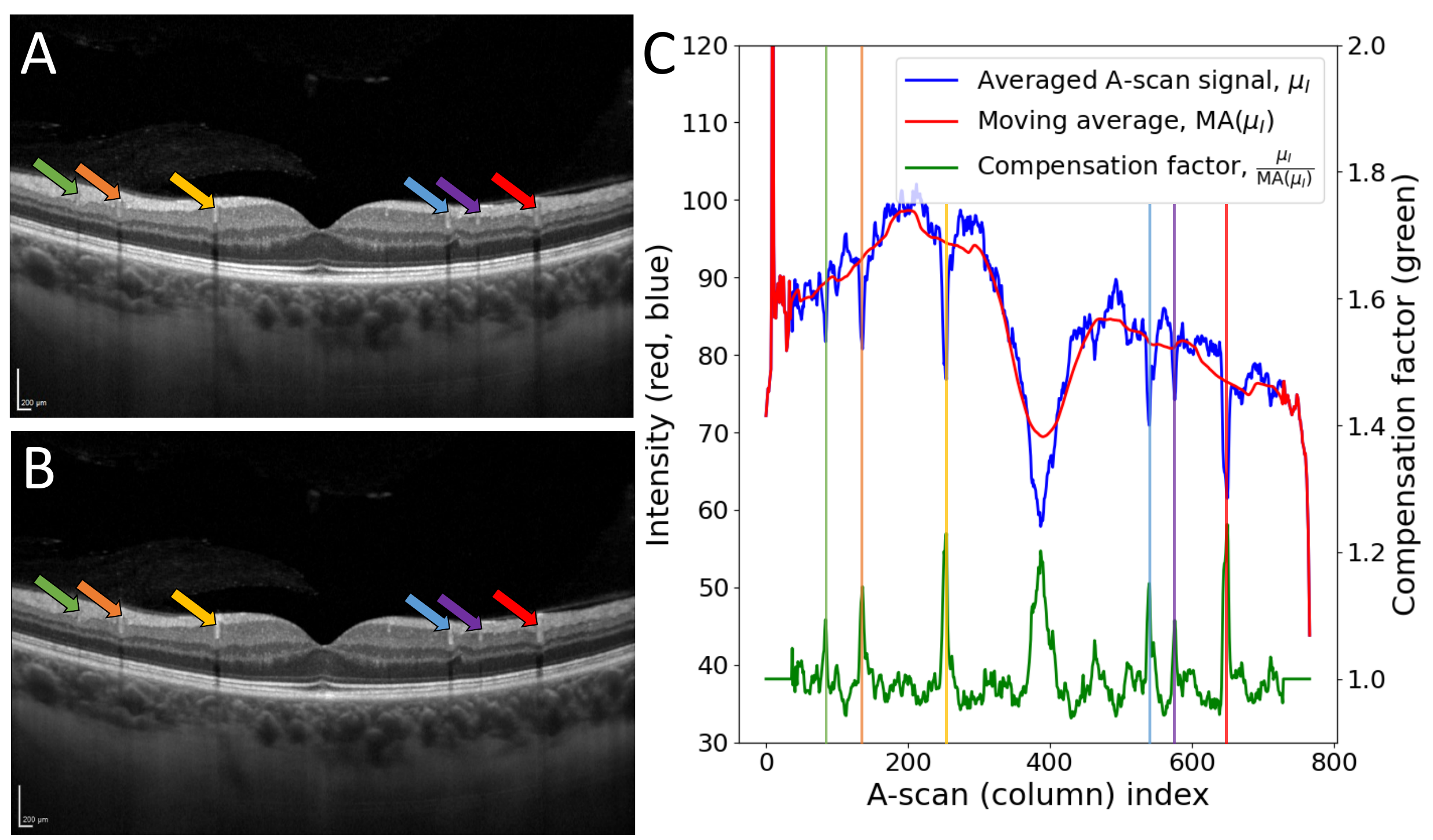}
            \caption[Superficial retinal vessel shadow compensation scheme.]{Demonstration of superficial retinal vessel shadow compensation. (A) Raw \acrshort{OCT} B-scan. (B) Shadow compensated B-scan. (C) Graphs of $\mu_I$, $\text{MA}(\mu_I))$ and the compensation factor. Coloured, shaded vertical lines correspond to shadow locations (coloured arrows) in panels (A) and (B).}
            \label{fig:MMCQ_shadow_compensate_graph}
        \end{figure}

        \begin{figure}[tbhp]
            \centering
            \includegraphics[width=0.925\linewidth]{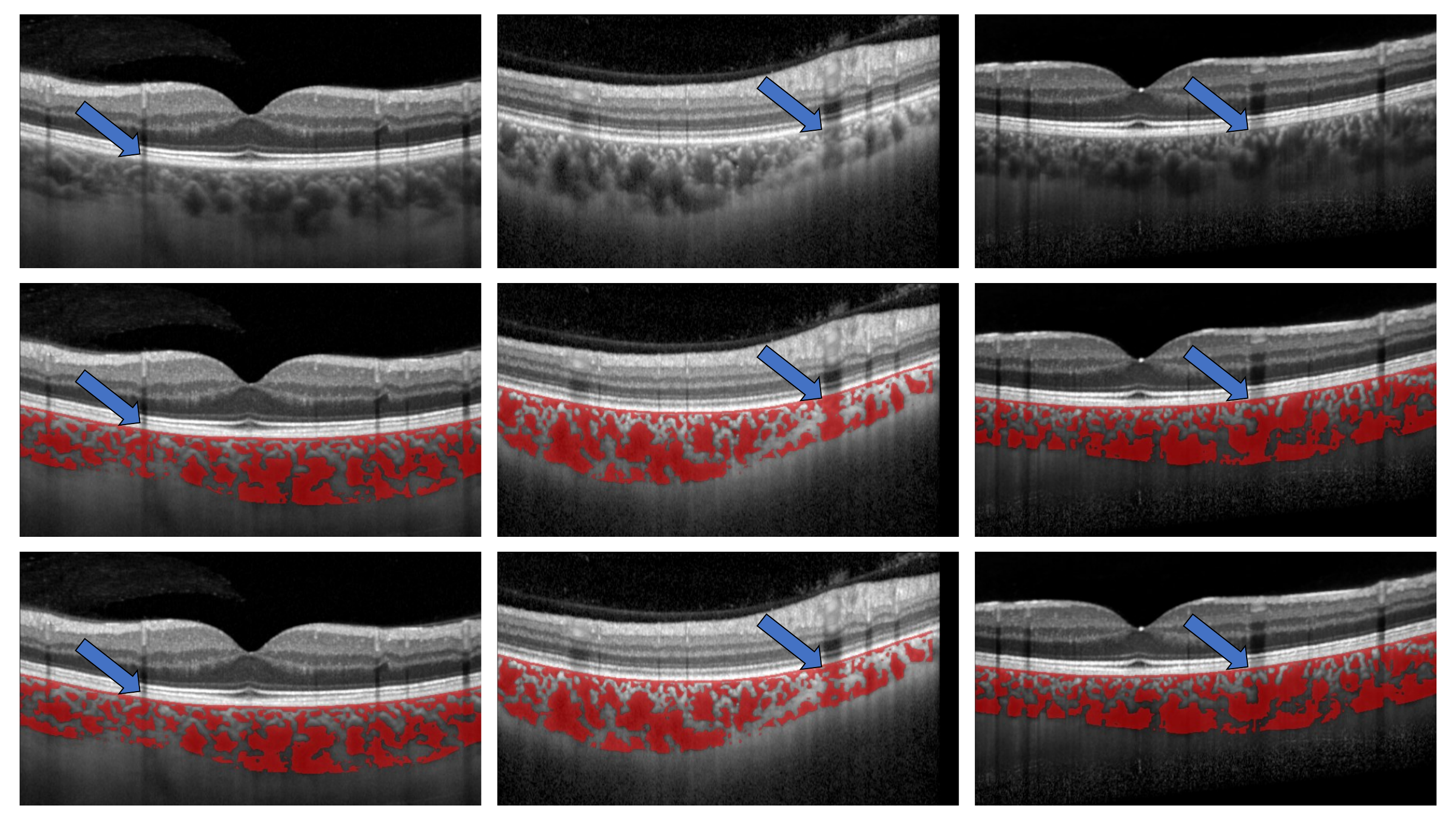}
            \caption[Influence of vessel shadow compensation on downstream vessel segmentation performance.]{Choroid vessel segmentation before and after retinal vessel shadow compensation. (Top) raw \acrshort{OCT} B-scans. (Middle) choroid vessel segmentation before shadow compensation. (Bottom) choroid vessel segmentation after shadow compensation. Arrows indicate locations of improved vessel segmentation.}
            \label{fig:MMCQ_shadow_compensate_segmentations}
        \end{figure}

        Choroidal vessel visualisation can be corrupted by superficial retinal vessel shadowing, where the \acrshort{OCT} beam penetrates through the retinal vasculature lying along the inner surface of the retina and sitting perpendicular to the incident laser light. This consequently darkens the corresponding A-scans of the deeper outer retinal and choroidal structures behind it \cite{schuman2024optical}. This can make choroidal stroma appear as vasculature due to the high contrast between A-scans adjacent to the shadowing. Figure \ref{fig:MMCQ_shadow_examples} shows an \acrshort{OCT} B-scan with it's corresponding en face localiser \acrshort{SLO} image with arrows on both scans showing the correspondence between superficial retinal vessels corrupting A-scans on the \acrshort{OCT} B-scan through light attenuation.
    
        These shadows can often confuse segmentation algorithms into incorrectly believing stromal area is part of the choroid's vascular compartment. To prevent this over-segmentation, we apply a multiplicative column-wise compensation factor to automatically brighten shadow-corrupted A-scans \cite{mao2019deep} as a pre-processing step before the vessel segmentation procedure. 
        
        Column-wise compensation factors (which are multiplied to each A-scan (column) of the raw \acrshort{OCT} B-scan) are computed as the element-wise ratio between each A-scans' averaged signal (axially), and the lateral moving average over these averaged A-scans. Note that the lateral moving average of A-scans smooths out any sudden dips in brightness from the shadow-corrupted A-scans. For an image array $I$ with $M$ rows indexed by $y$ and $N$ columns indexed as $x$, the pixel indexed by $(x,y)$ in the shadow compensated image $I_s$ is
        \begin{equation}
            I_s(x, y) = I(x, y) \cdot \frac{\mu_I(x)}{\text{MA}(\mu_I)(x)},
        \end{equation}
        where $\mu_I(x)$ is the average pixel intensity of the A-scan indexed at column $x$ such that
         \begin{equation}
            \mu_I(x) = \nicefrac{1}{M}\sum^{M}_j I(x, j).
        \end{equation}
        $\mu_I$ is thus a list of A-scan averages, indexed at all columns $x = 1, \dots, N$ and $\text{MA}(\mu_I)$ is the moving average of $\mu_I$, with $\text{MA}(\mu_I)(x)$ the value of this moving average at column $x$. 
        
        Figure \ref{fig:MMCQ_shadow_compensate_graph} shows the raw \acrshort{OCT} B-scan from figure \ref{fig:MMCQ_shadow_examples} in panel (A), with the shadow compensated result in panel (B) and the plots for the signals $\mu_I$, $\text{MA}(\mu_I)$ and the compensation factor signal $\nicefrac{\mu_I}{\text{MA}(\mu_I)}$ shown in blue, red and green, respectively. Note that the number of columns in this B-scan was 768. The arrows on the B-scans correspond to the shaded vertical lines overlaid onto panel (C). Figure \ref{fig:MMCQ_shadow_compensate_segmentations} shows the impact of vessel shadow compensation on the resulting choroid vessel segmentation using the methodology introduced in this chapter.
        
    \end{mysubsection}
    
    \end{mysection}

    \begin{mysection}[]{MMCQ's methodology}

        For choroid vessel segmentation, our task is to identify which pixels within the choroid belong to vasculature --- typically seen as dark blobs \cite{agrawal2020exploring, sohrab2012pilot, branchini2013analysis} --- and which do not. However, the problem is made difficult because of the poor visualisation of the vasculature within the choroid, particularly for small-scale vasculature, i.e. the choriocapillaris located at the anterior point of the choroid, and the smaller vessels which supply it. Larger vessels have greater contrast because there is more vessel lumen visible on the \acrshort{OCT} device --- contributing to a stronger and more consistent intensity transition between interstitial and intravascular compartments --- while smaller vessels do not. 

        In this section we describe a methodology for semi-automatic choroid vessel segmentation known as multi-scale median cut quantisation (\acrshort{MMCQ}). \acrshort{MMCQ} takes a multi-scale approach to choroid vessel segmentation by enhancing choroidal vessels at different scales, thus addressing their heterogeneity in both size and shape. We also combine the effects of quantisation with histogram equalisation at the local level to address the poor contrast often exhibited by the choroidal vasculature.
        
       \begin{mysubsection}[]{Pre-processing}

            Before applying \acrshort{MMCQ}, we apply some pre-processing to do some initial denoising and contrast enhancement. These steps are outlined below:
            \begin{enumerate}\setlength\itemsep{0em}
                \item Shadow compensation: we compensate for A-scans which have been corrupted by superficial retinal vessel shadowing using the compensation scheme described in section \ref{sec:MMCQ_intro_shadows} \cite{mao2019deep}.
                \item Cropping: we crop the \acrshort{OCT} B-scan using a previously calculated segmentation mask or edge-based segmentation of the choroidal space. This not only decreases execution time, but also means that superfluous regions outside the choroid do not influence the segmentation procedure.
                \item Denoising: we reduce the speckle noise by using a 3 $\times$ 3 median filter, which applies the same procedure as described in section \ref{subsec:GPET_method_prep}. The window size is smaller here so as to ensure smaller vessels are not overly smoothed.
                \item Contrast enhancement: we apply the \acrshort{CLAHE} algorithm by breaking the image into a set of 8 $\times$ 8 tiles with a clipping limit of 2. The clip limit is lower here than in section \ref{subsec:GPET_method_prep} so as to not over-enhance the smaller vessels, which would harm the enhancement procedure.
            \end{enumerate}
            
            Figure \ref{fig:MMCQ_preprocessing} shows a horizontal-line \acrshort{EDI-OCT} scan of the choroid with its region demarcated in panel (A), with the choroid region cropped out in the top row of panel (B) and the result of pre-processing in the bottom row of panel (B). In this example, the larger vessels have seen improved definition but less so for smaller vessels. 

            \begin{figure}[!b]
                \centering
                \includegraphics[width=\linewidth]{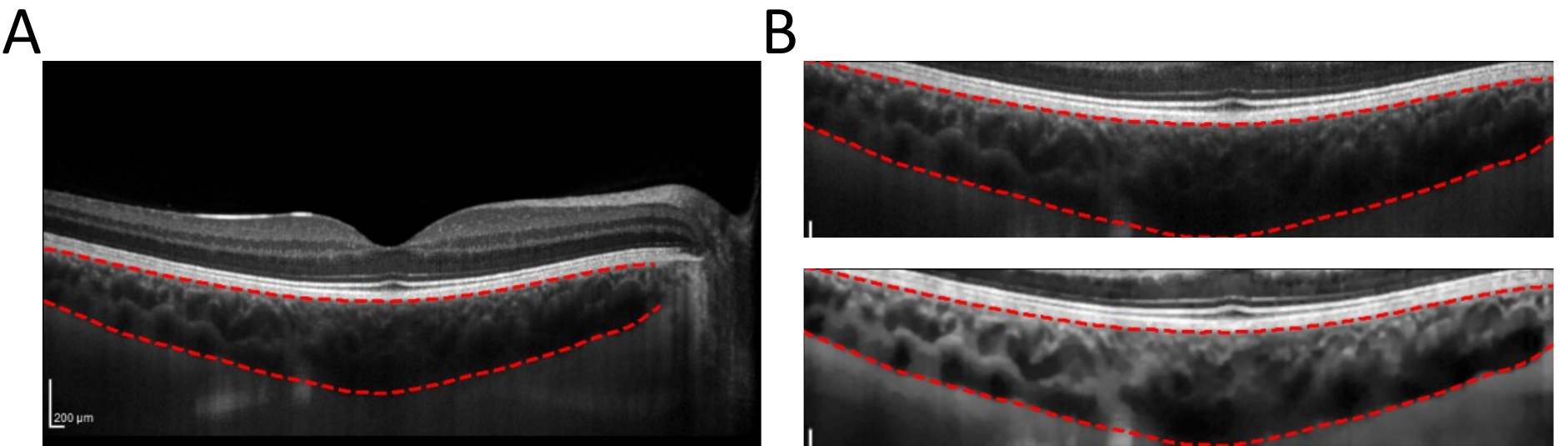}
                \caption[Initial pre-processing of an \acrshort{OCT} B-scan before \acrshort{MMCQ}'s pipeline.]{Initial pre-processing of an \acrshort{OCT} B-scan before \acrshort{MMCQ}'s pipeline. (A) \acrshort{OCT} B-scan with choroid region segmented using red-dashed lines. (B) the \acrshort{OCT} B-scan cropped to only show the choroid (top) and the result after pre-processing (bottom).}
                \label{fig:MMCQ_preprocessing}
            \end{figure}
            
        \end{mysubsection}
        
        \begin{mysubsection}[]{Patch quantisation and enhancement}
        
            \begin{mysubsubsection}[]{Patch size}\label{subsubsec:MMCQ_method_patch}
            
                The multi-scale procedure applies quantisation and histogram equalisation at a local, patch-based level. We select three distinct scales to extract patches from the choroid, based on the (image-aligned) pixel thickness of the choroid in the \acrshort{OCT} B-scan. The pixel thickness for an A-scan is defined here as the number of (row) pixels between the \acrshort{RPE}-Choroid and Choroid-Sclera boundary for that particular lateral position (column). The pixel thickness $T$ of the choroid is the average pixel thickness across all A-scans which have the choroid segmented.
                
                Given some pixel thickness $T$, we apply the quantisation and enhancement scheme by decomposing the cropped choroid into square patches of size $\{\nicefrac{T}{10}, \nicefrac{T}{5}, \nicefrac{T}{2}\}$. The average choroid thickness in a healthy population is approximately 300 microns \cite{kong2018measurable}. Thus, given that pixel length-scales for commercial \acrshort{OCT} devices (Heidelberg Engineering and Topcon) range axially between approximately 3 and 4 microns-per-pixel, this corresponds to patch size scales for the average choroid of approximately $\{10 \times 10, 20 \times 20, 50 \times 50\}$. These approximate scales assume coverage of the range of different sizes of choroidal vessels found in Sattler's and Haller's layer and, importantly, scales with the size of the choroid. For the remainder of this section, we will consider a patch size $s$, i.e. the cropped choroid will be divided into mutually exclusive, non-overlapping patches of size $s \times s$. 
                
            \end{mysubsubsection}

            \begin{mysubsubsection}[]{Quantisation} \label{subsubsec:MMCQ_method_quant}

                \begin{figure}[tb]
                    \centering
                    \includegraphics[width=0.65\linewidth]{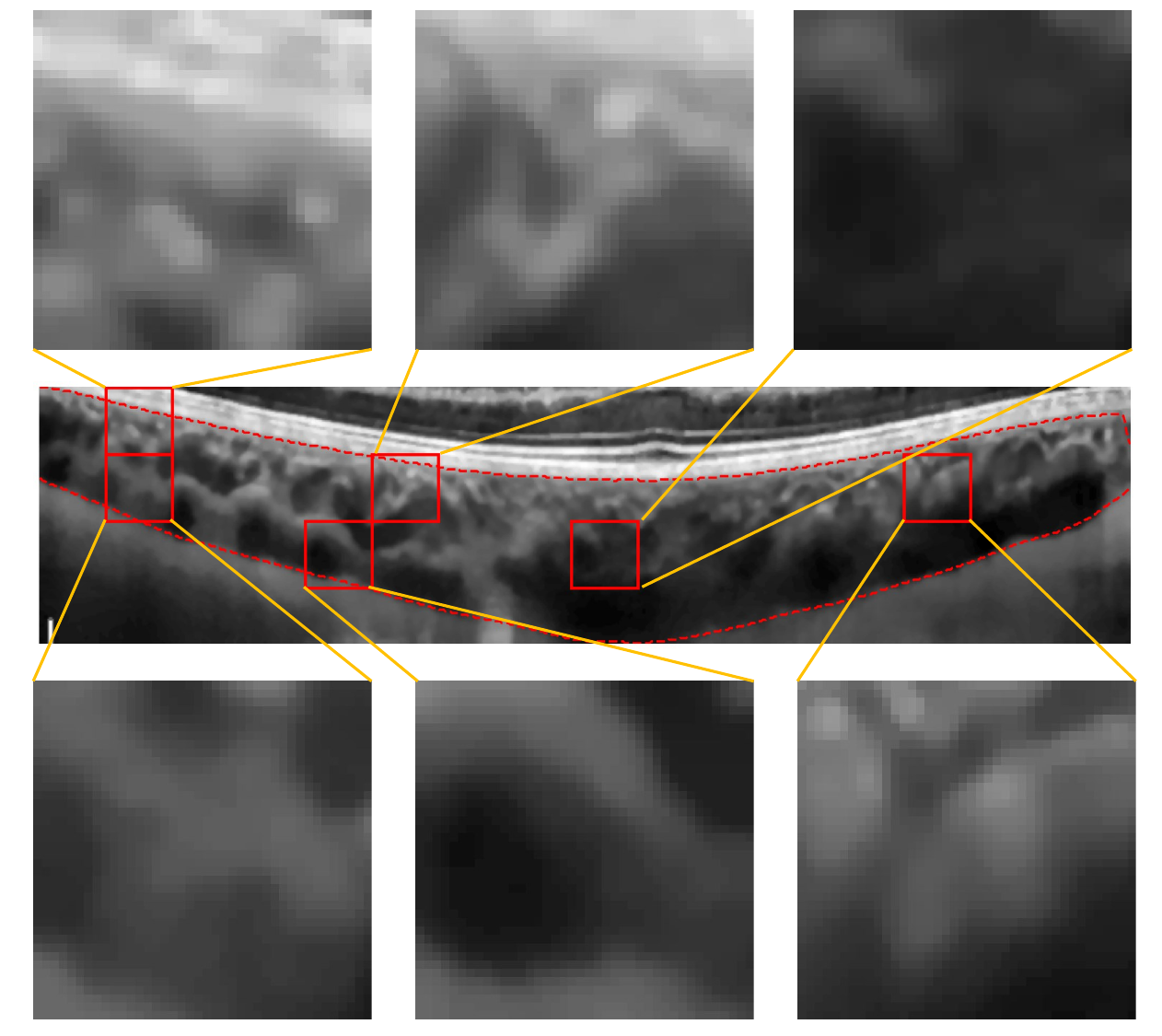}
                    \caption[Pre-processing at the patch-level before applying \acrshort{MMCQ}'s pipeline.]{Six patches extracted from the pre-processed choroid of size 40 $\times$ 40.}
                    \label{fig:MMCQ_patches}
                \end{figure}

                \begin{figure}[tb]
                    \centering
                    \includegraphics[width=\linewidth]{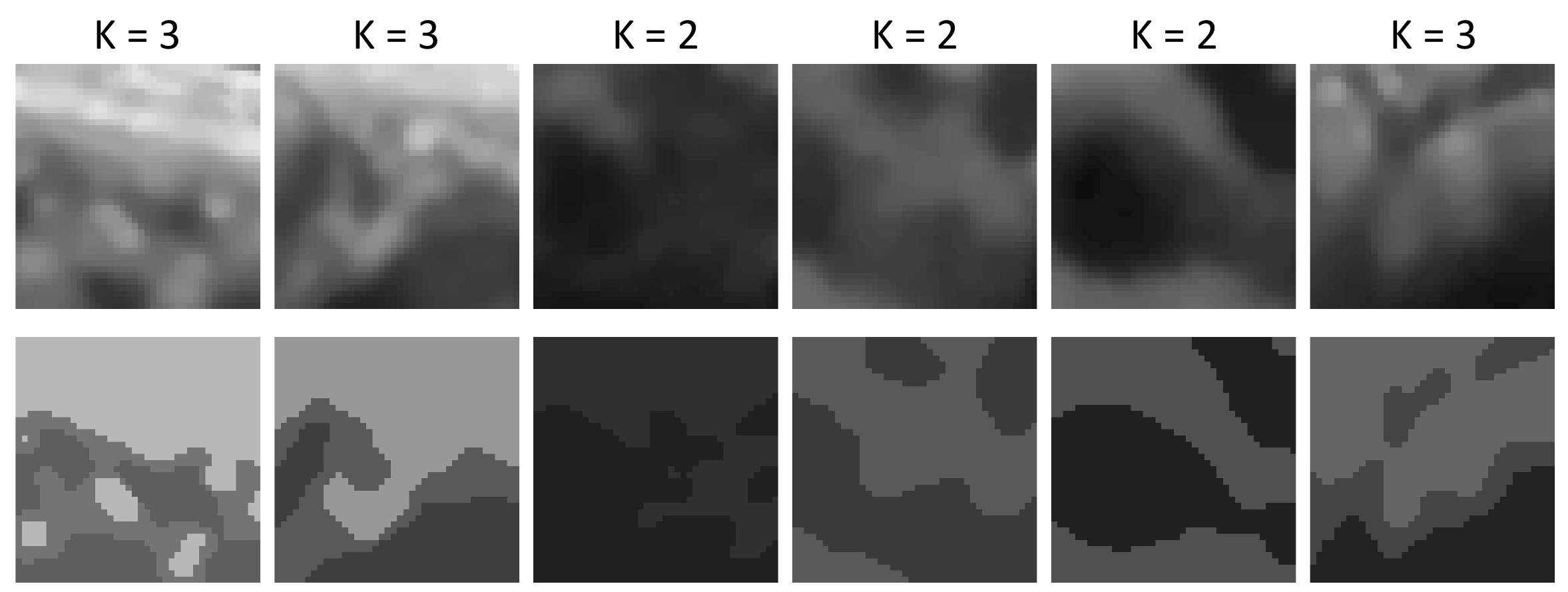}
                    \caption[Median cut quantisation at the patch-level.]{Six quantised patches using median cut clustering and replacing pixel intensities with median value for each cluster group.}
                    \label{fig:MMCQ_patches_quantised}
                \end{figure}
            
                Figure \ref{fig:MMCQ_patches} shows several patches extracted from the pre-processed choroid in figure \ref{fig:MMCQ_preprocessing}(B) (bottom) which we will use as running examples for the rest of this section.

                Each patch $P$ is quantised using median cut quantisation, $P_q$, replacing the intensity value of every pixel assigned to the same cluster group with their median value. The number of clusters, $K_P$ is patch-dependent and is based on the intensity dispersion of the patch histogram across the range of 256 brightness levels available for an 8-bit grayscale image. For a perfectly imaged choroid without any speckle noise, we would be able to set $K = 2$ for all patches, i.e. one cluster for vessel pixels and another for stromal pixels. In reality, we must allow for poor contrast, low resolution and speckle noise to create a heterogeneous visualisation of the choroid in terms of pixel intensity. Thus, we allow $K_P$ to vary between 2 and 5. 
                
                We first split the 256 brightness levels into a set of 5 equally spaced intervals, $S$ defined by
                \begin{align}
                    S &= \bigg\{I_i = \big[(i-1)\cdot\nicefrac{255}{5}, i\cdot\nicefrac{255}{5}\big) \hspace{3pt} \bigg| \hspace{3pt} i = 1,\dots,5\bigg\} \\
                    &= \{[0,51), [51,102), [102,153), [153,204), [204, 255]\},
                \end{align}
                where $[x_1, x_2)$ is the half-open interval of integers (brightness levels) from integers $x_1$ up to but excluding $x_2$ (with exception of the last interval which is closed, i.e. $[204, 255]$). $K_P$ is then the number of intervals in $S$ which the intensity range of patch $P$ overlaps with, i.e.
                \begin{equation}\label{eqref:MMCQ_cluster_K}
                    K_P = \big|\big\{I_i \hspace{2pt} \big| \hspace{2pt} I_i \in S, \hspace{2pt} \exists j \in I_i, \hspace{2pt} \hat{H}_P(j) > 0\big\}\big|,
                \end{equation}
                where $\hat{H}_P$ represents a modified histogram of patch $P$ such that $\hat{H}_P(j)$ is the number of pixels in $P$ which have intensity (brightness level) $j$, assuming that $j$ exists within the 99$^{\text{th}}$ percentile range of patch $P$. This is to prevent outliers in the patch from influencing the estimate for $K_P$. The modified histogram $\hat{H}_P$ is related to the raw histogram $H_P$ such that for brightness levels $j=0,\dots,255$
                 \begin{equation}\label{eqref:MMCQ_Qmap}
                    \hat{H}_P(j) = 
                    \begin{cases} 
                       0 & j < \text{perc}(P,0.005) \\
                       H_P(j) & \text{perc}(P,0.005) \leq j \leq \text{perc}(P,0.995) \\
                       0 & j > \text{perc}(P,0.995).
                   \end{cases}
                \end{equation} 
                Figure \ref{fig:MMCQ_patches_quantised} show the six patches extracted from figure \ref{fig:MMCQ_patches} in the top row, their quantised equivalents in the bottom row, with the number of quantisation levels estimated above as $K$.

            \end{mysubsubsection}

            \begin{mysubsubsection}[]{Enhancement}

                \begin{figure}[tb]
                    \centering
                    \includegraphics[width=\linewidth]{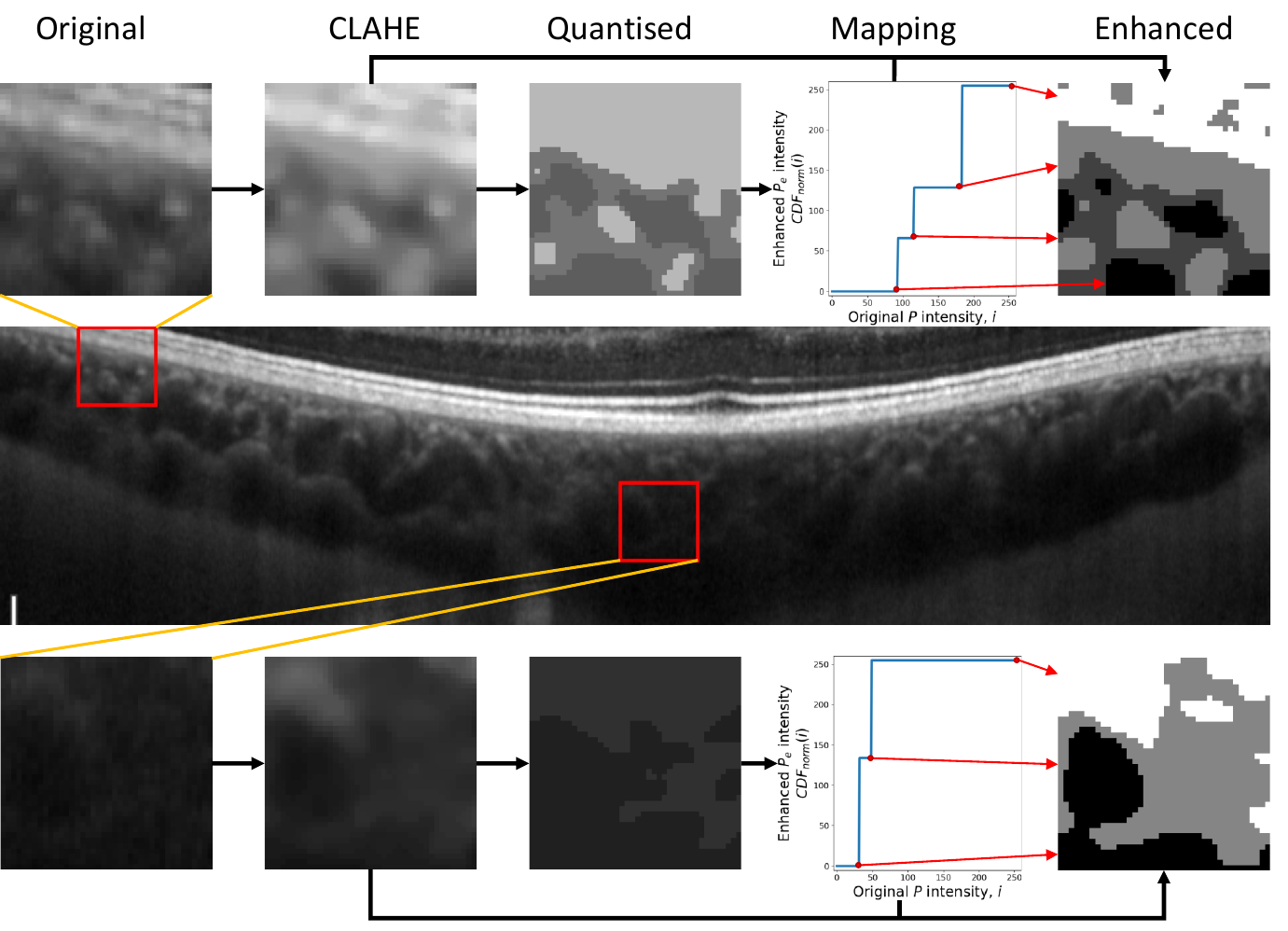}
                    \caption[Patch-level enhancement scheme.]{The quantisation and enhancement procedure for two 40 $\times$ 40 patches from the choroid in the raw \acrshort{OCT} B-scan. Red arrows represent the pixel intensity on the new patch which the $CDF_{\text{norm}}$ maps to from the original pixel intensities of patch $P$.}
                    \label{fig:MMCQ_quant_enh_patch_example}
                \end{figure}
            
                The enhancement procedure performs histogram equalisation on each quantised patch. We first construct the cumulative distribution function, $CDF$ of the quantised patch $P_q$. This function is defined at brightness level $i$ as 
                \begin{equation}\label{eqref:MMCQ_cdf_def}
                    CDF(i) = \displaystyle\sum_{j=0}^i H_q(j),
                \end{equation}
                where $H_q(j)$ is the histogram of quantised patch $P_q$ at brightness level $j$. This is normalised such that
                \begin{equation}\label{eqref:MMCQ_cdf_norm_def}
                    CDF_{\text{norm}}(i) = \frac{CDF(i) - CDF(0)}{s^2 - CDF(0)} \cdot (L - 1).
                \end{equation}
                Here, the $CDF$ is normalised between [0, 1] and then scaled to the bit depth $L$, i.e. 8-bit depth ($L$ = 256). This provides a brightness level mapping for the original patch's pixels, spreading the distribution of brightness levels across the full range of available levels, i.e. between 0 and 255. Thus, the enhanced patch $P_e$ is computed by mapping the pixel intensities from the original patch $P$ using $CDF_{\text{norm}}$ such that
                \begin{equation}
                    P_e(x,y) = CDF_{\text{norm}}(P(x,y)),
                \end{equation}
                where $x$, $y$ are the column and row index locations of the patch. Since the quantisation is preserved by $CDF_{\text{norm}}$ having been generated using the quantised patch $P_q$, $P_e$ is thus both quantised and enhanced. Figure \ref{fig:MMCQ_quant_enh_patch_example} shows two patches from the original choroid and the evolution of their pre-processing, quantisation and enhancement.
                
                This enhancement is suitable for differentiating vessel from stroma for heterogeneous patches like the one in the top row of figure \ref{fig:MMCQ_quant_enh_patch_example}. However, the same enhancement for relatively homogeneous patches, i.e. those which are primarily vessel or stroma, has the effect of over-enhancing small pixel intensity variations which introduces unwanted noise into the enhancement procedure. This is evident in the patch shown in the bottom row of figure \ref{fig:MMCQ_quant_enh_patch_example}. Thus, homogeneous patches should not be enhanced as much as heterogeneous patches. In \acrshort{CLAHE}, preventing homogeneous areas from being over-enhanced is addressed through the \textit{clip limit}. However, clipping the quantised patch's histogram will redistribute any excess pixels uniformly across the enhanced histogram, thus rendering the quantisation step meaningless. Moreover, selecting the optimal clipping limit is challenging for multiple patches at different sizes across different choroids.

                \begin{figure}[!b]
                    \centering
                    \includegraphics[width=0.5\linewidth]{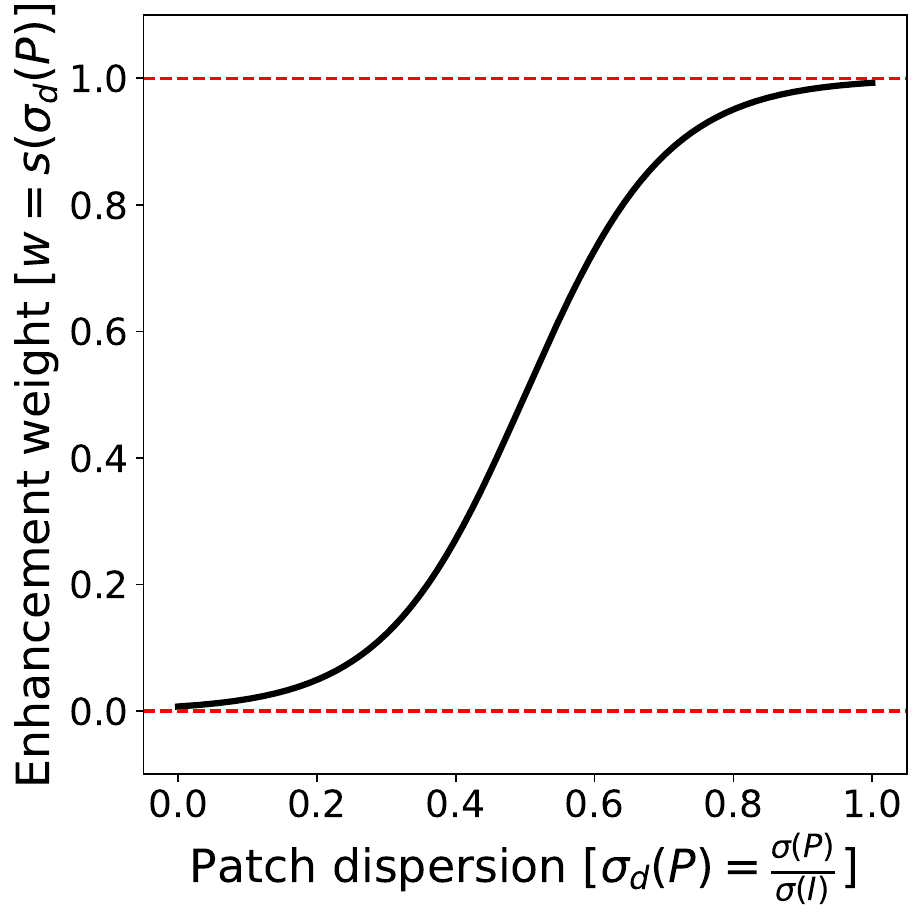}
                    \caption[Custom Sigmoid function to weight enhancement strength.]{Sigmoid function evolving over input range $[0, 1]$ following equation \eqref{eqref:MMCQ_sigmoid}.}
                    \label{fig:MMCQ_sigmoid_curve}
                \end{figure}
                Therefore, we apply a simple weighting scheme to control the strength of enhancement. We do this by measuring the proportion of the patch $P$'s intensity dispersion relative to the intensity dispersion of the cropped image (only showing the choroid), $I$. We characterise dispersion as the standard deviation over the region, so that patch $P$'s \textit{relative} dispersion, $\sigma_d(P)$ is
                \begin{equation}\label{eqref:MMCQ_dispersion}
                    \sigma_d(P) = \frac{\sigma(P)}{\sigma(I)},
                \end{equation}                
                where $\sigma(\cdot)$ is just the standard deviation of the flattened array $P$ or $I$. $\sigma_d(P)$ is bound between 0 and 1, and to promote minimal enhancement for low values of $\sigma_d(P)$ and maximum enhancement for high value of $\sigma_d(P)$ we use the Sigmoid function to dictate the weight of enhancement strength. The Sigmoid function is ideal for computing weight $w$ due to the sharp plateaus at 0 and 1 at either end of the input range. The enhancement weight for patch $P$, $w(P)$ is defined using the Sigmoid formula
                \begin{equation}\label{eqref:MMCQ_sigmoid}
                    w(P) = s(\sigma_d(P)) = \large(1+e^{-2\pi(\sigma_d(P) - \nicefrac{1}{2})}\large)^{-1},
                \end{equation}
                which yields an output range $w(P)$ in the interval $[0, 1]$, given an input interval $[0, 1]$ of which $\sigma_d(P)$ of bound by. The Sigmoid function is plotted in figure \ref{fig:MMCQ_sigmoid_curve}. The enhancement of patch $P$ is then scaled by $w(P)$, with the remaining $(1-w(P))$ factor made up by an intensity mapping equivalent to the quantised patch $P_q$, called $Q$. $Q$ converts the pixel intensities in patch $P$ to the brightness levels dictated by the quantised patch $P_q$. The piece-wise formula for mapping $Q$ for a patch quantised into $n$ levels $\{q_1, \dots, q_n\}$ is
                \begin{equation}\label{eqref:MMCQ_Qmap}
                    Q(i) = 
                    \begin{cases} 
                      0 & i < q_1 \\
                       q_1 & q_1 \leq i < q_2 \\
                       &\vdots \\
                      q_n & i \geq q_n,
                   \end{cases}
                \end{equation}    
                where $i$ and $q_j$ are both in the interval $[0, 255]$ for all $j = 1, \dots, n$. This mapping is equivalent to zero enhancement and only quantisation. The enhanced patch $P_e$ is thus a linear combination between $Q$ and $CDF_{\text{norm}}$ with $w(P)$ as the multiplicative factor balancing the strength of enhancement, such that 
                \begin{equation}\label{eqref:MMCQ_enh_patch}
                    P_e(x, y) = w(P)\cdot CDF_{\text{norm}}(P(x, y)) + (1 - w(P))\cdot Q(P(x, y)).
                \end{equation}

                \begin{figure}[tbp]
                    \centering
                    \includegraphics[width=0.925\linewidth]{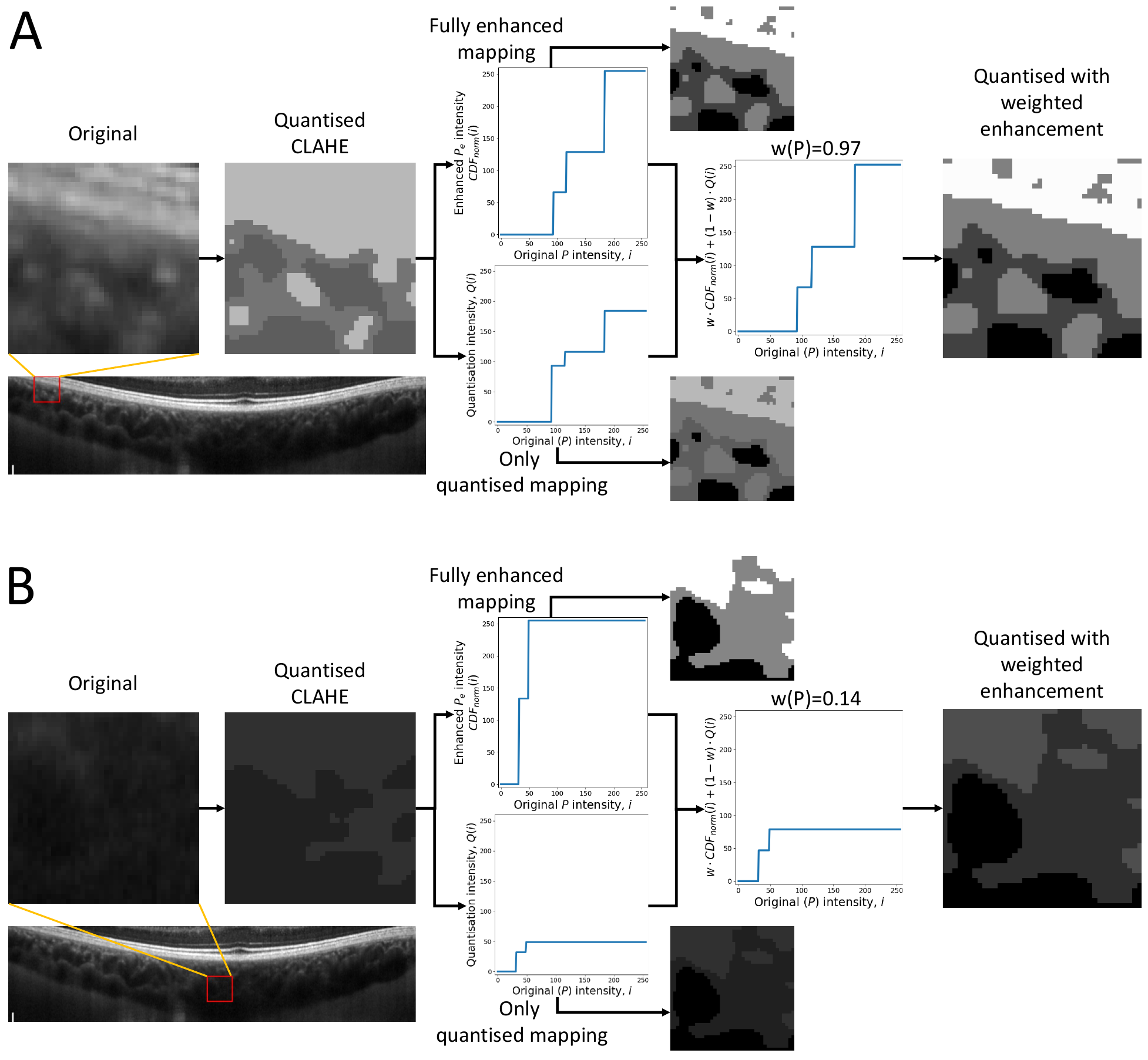}
                    \caption[Schematic of quantisation and enhancement at the patch-level.]{The weighted, quantisation and enhancement procedure for two 40 $\times$ 40 patches from the choroid in the raw \acrshort{OCT} B-scan.}
                    \label{fig:MMCQ_quant_enh_weight_patch_example}
                \end{figure}

                \begin{figure}[tbp]
                    \centering
                    \includegraphics[width=0.95\linewidth]{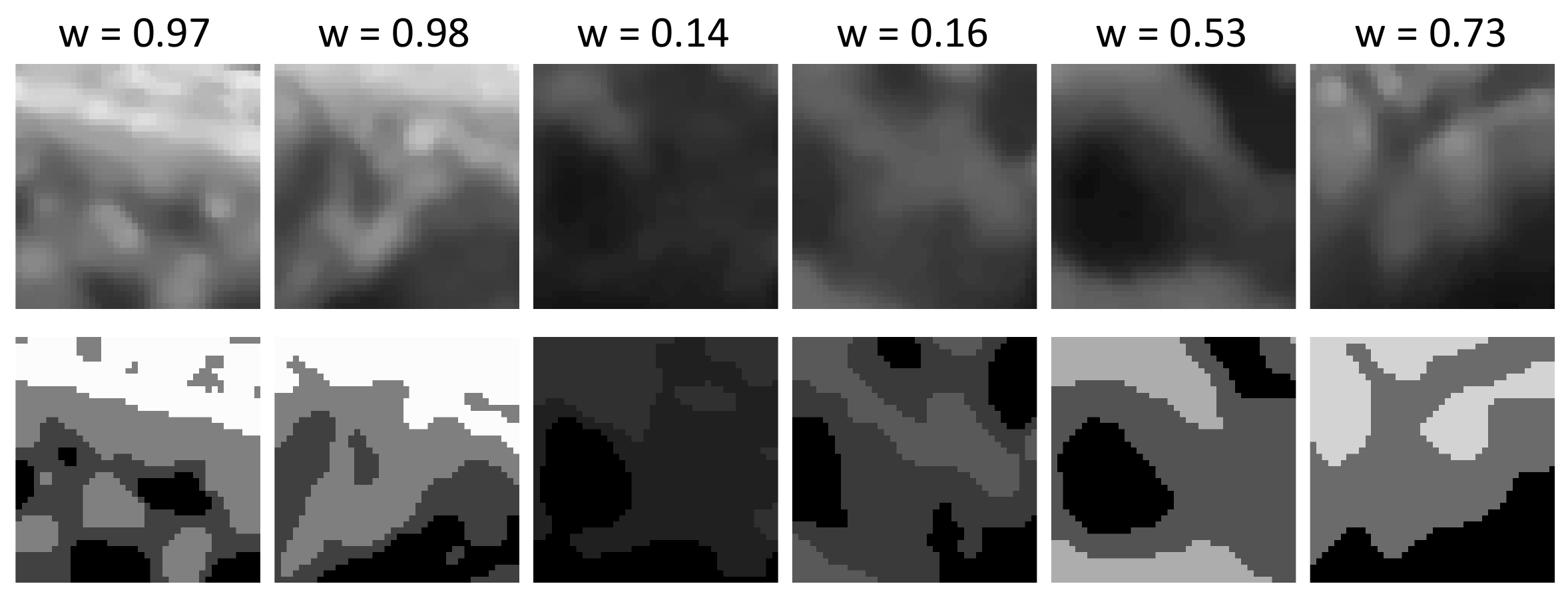}
                    \caption[Patch-level result after \acrshort{MMCQ}'s enhancement.]{Enhanced patches extracted from the raw \acrshort{OCT} B-scan for a single scale (40 $\times$ 40).}
                    \label{fig:MMCQ_all_enh_patches}
                \end{figure}
                
                The computation of $P_e$ is shown for the two demonstrative patches from figure \ref{fig:MMCQ_quant_enh_patch_example} in figure \ref{fig:MMCQ_quant_enh_weight_patch_example}. In panel (A) we can see the heterogeneous patch is mostly fully enhanced as the weight is 0.97, while the homogeneous patch in panel (B) is mostly quantised as the weighting is only 0.14. For completeness, figure \ref{fig:MMCQ_all_enh_patches} shows all six patches extracted from the raw \acrshort{OCT} B-scan in the top row and their weighted, quantised and enhanced patches in the bottom row for a single scale (40 $\times$ 40). 
                
            \end{mysubsubsection}
        
        \end{mysubsection}
        
        \begin{mysubsection}[]{Merging}

            \begin{figure}[!b]
                \centering
                \includegraphics[width=0.9\linewidth]{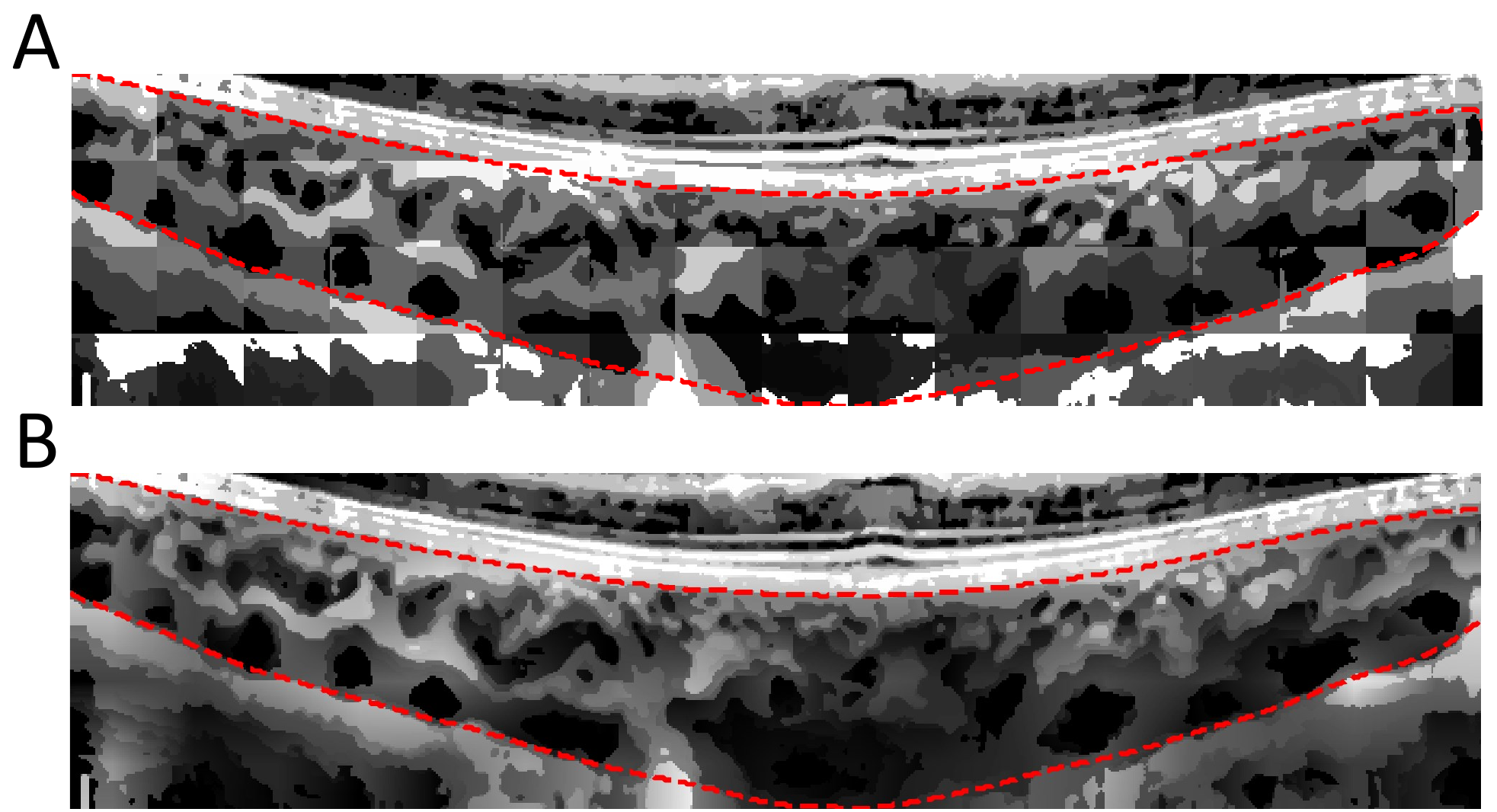}
                \caption[\acrshort{MMCQ}'s enhancement at 40 $\times$ 40 scale.]{(A) Enhanced patches stacked in order to recreate a discontinuously enhanced choroid. (B) Final enhanced choroid at scale 40 $\times$ 40 after bi-linear interpolation.}
                \label{fig:MMCQ_patch_merge_final}
            \end{figure}

            \begin{figure}[tbp]
                \centering
                \includegraphics[width=0.825\linewidth]{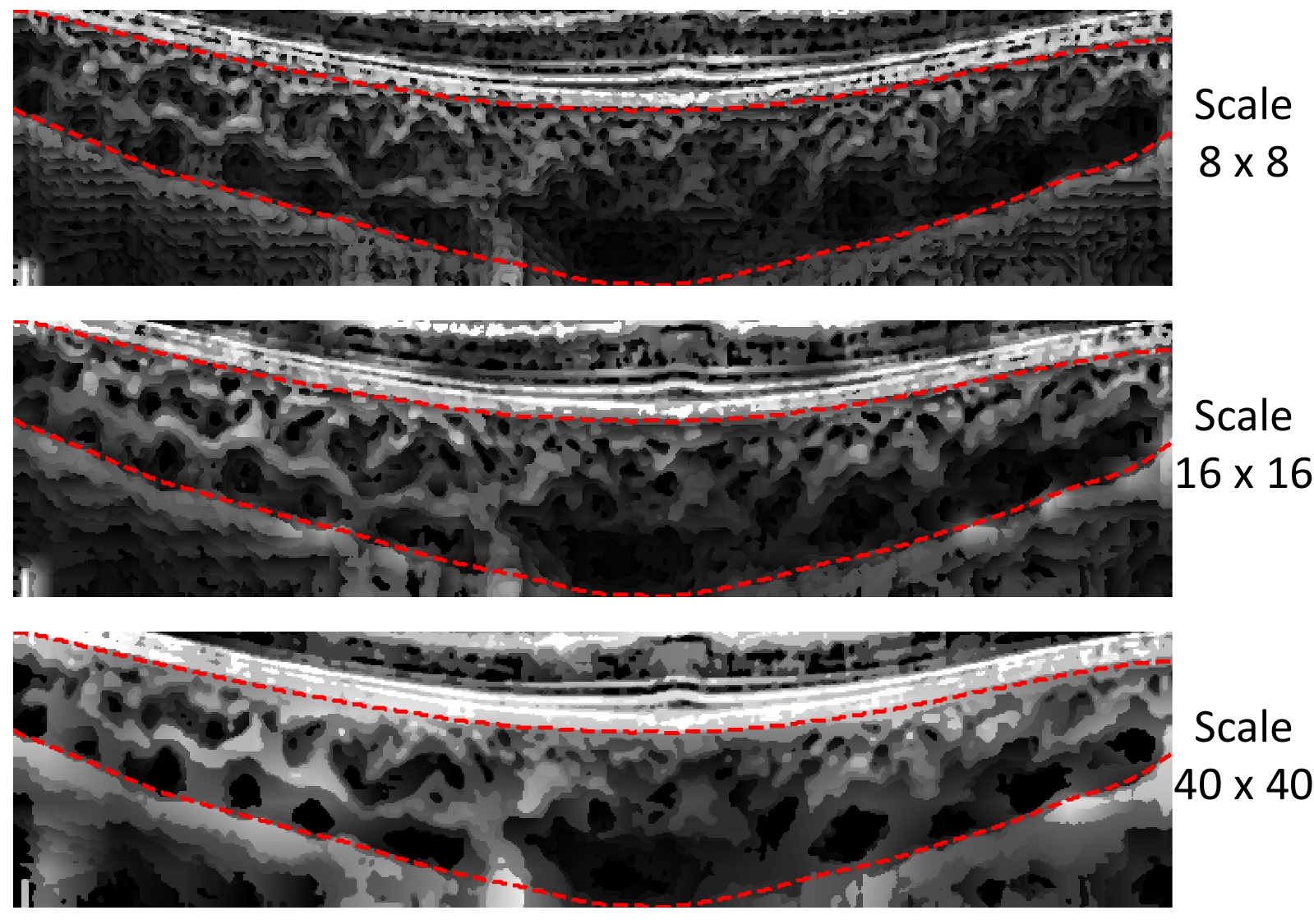}
                \caption[\acrshort{MMCQ}'s enhancement at choroid-dependent scales.]{Enhanced choroid at scales 8 $\times$ 8, 16 $\times$ 16 and 40 $\times$ 40 for a choroid whose pixel thickness was 80 pixels.}
                \label{fig:MMCQ_enh_chors_scales}
            \end{figure}

            \begin{figure}[tbp]
                \centering
                \includegraphics[width=0.825\linewidth]{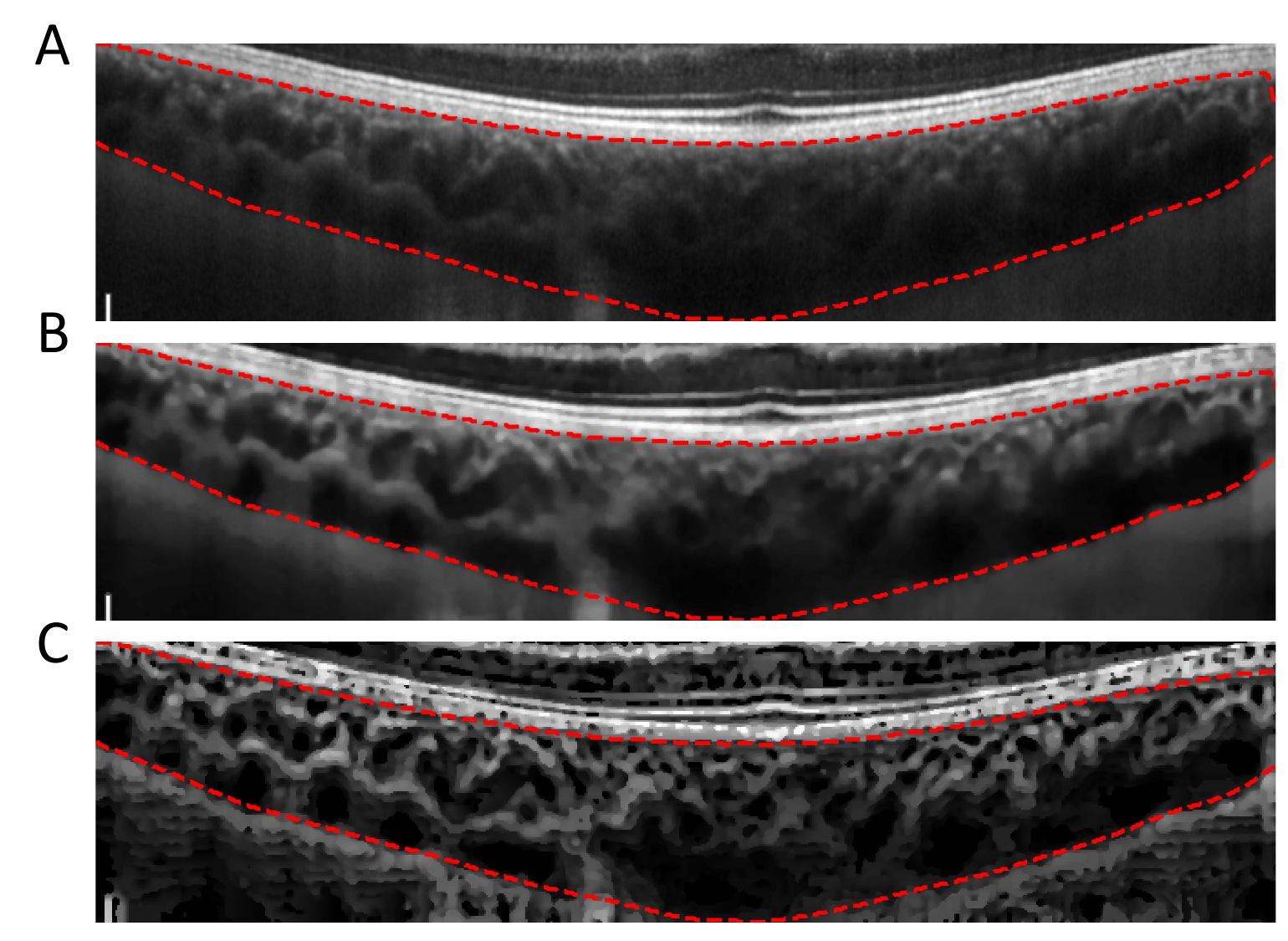}
                \caption[\acrshort{MMCQ}'s enhancement compared to simple pre-processing.]{(A) Original choroid from raw \acrshort{OCT} B-scan. (B) Pre-processed choroid after pixel smoothing and simple \acrshort{CLAHE} application. (C) Enhanced choroid using \acrshort{MMCQ}.}
                \label{fig:MMCQ_chor_enh}
            \end{figure}

            \begin{mysubsubsection}[]{Patch merging}

                After all patches have been enhanced according to the procedure in the previous section, adjacent patches are blended together in both axial and lateral directions using bi-linear interpolation. Figure \ref{fig:MMCQ_patch_merge_final} shows all of the non-overlapping patches at scale 40 $\times$ 40 stacked together in panel (A) with the final result after bi-linear interpolation in panel (B). Note that in panel (A) there are discontinuities at the patch edges. 
        
            \end{mysubsubsection}
            
            \begin{mysubsubsection}[]{Scale merging}

                After enhancement at all scales has been performed we are left with several quantised and enhanced choroids. At each scale, a particular scale of vasculature is prioritised in the enhancement and quantisation procedure. Figure \ref{fig:MMCQ_enh_chors_scales} shows the enhanced choroids for each of the three scales, 8 $\times$ 8, 16 $\times$ 16 and 40 $\times$ 40. These scales were dictated by the pixel thickness of this particular choroid which was $T=$ 80 pixels, i.e. patch sizes were chosen as $\{\nicefrac{T}{10}, \nicefrac{T}{5}, \nicefrac{T}{2}\} = \{\nicefrac{80}{10}, \nicefrac{80}{5}, \nicefrac{80}{2}\} = \{8, 16, 40\}$.
               
                We synthesise the enhanced choroids by taking the element-wise minimum across the scale dimension. This ensures that we prioritise the darkest pixels systematically across all enhancements. Figure \ref{fig:MMCQ_chor_enh} shows the raw choroid, the pre-processed choroid and the \acrshort{MMCQ}-enhanced choroid in series.
                
            \end{mysubsubsection}
            
        \end{mysubsection}
        
        \begin{mysubsection}[]{Vessel segmentation}\label{subsec:MMCQ_method_vesselseg}

            \begin{figure}[tb]
                \centering
                \includegraphics[width=\linewidth]{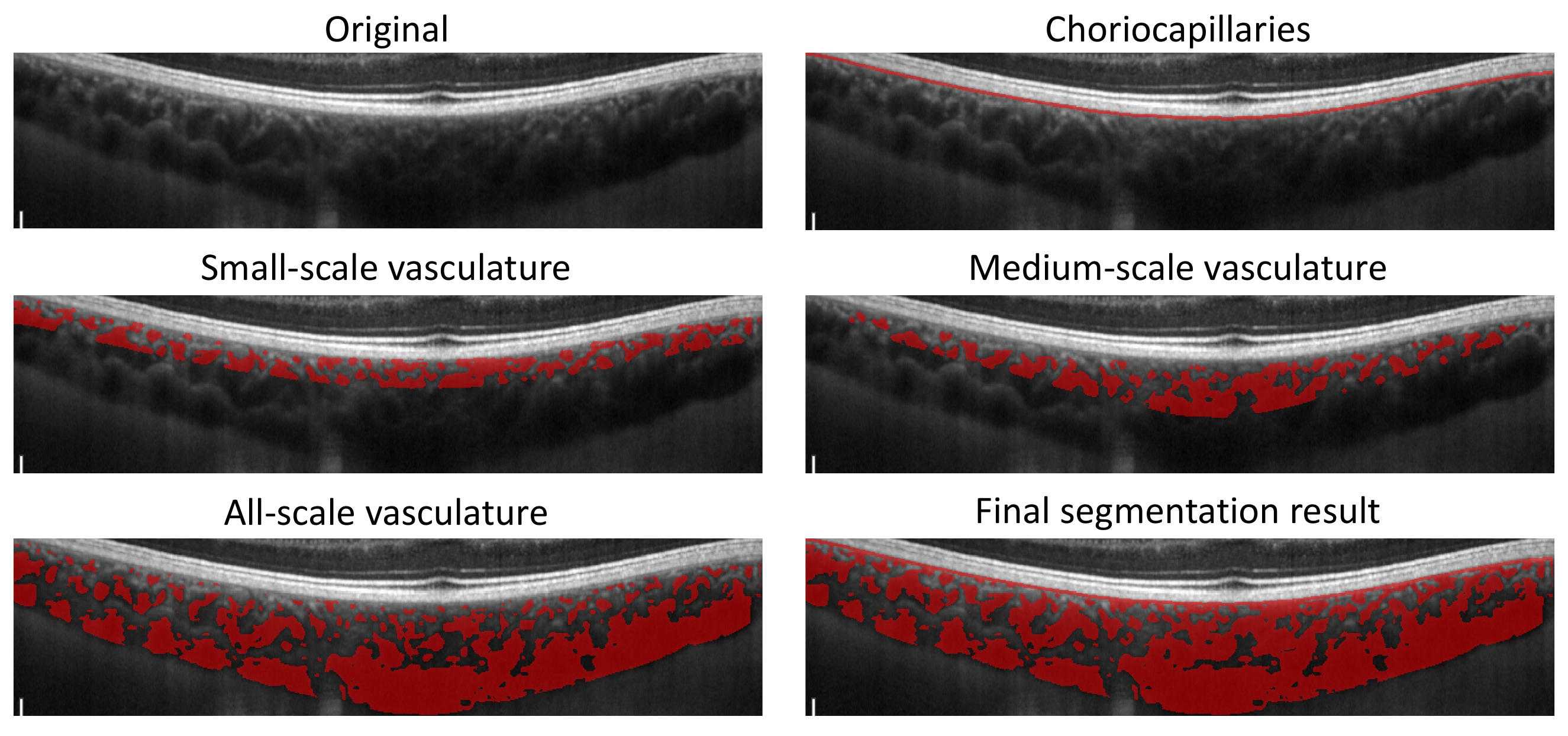}
                \caption[\acrshort{MMCQ}'s multi-depth segmentation scheme.]{The choroid segmented at different depths of the choroid. The segmentations are combined by taking their union (bottom right).}
                \label{fig:MMCQ_seg_vessel}
            \end{figure}
            
            We use a multi-depth approach to vessel segmentation on the MMCQ-enhanced choroid using the median cut algorithm. To ensure that small- and large-scale vessels are clustered together, we repeatedly apply the median cut algorithm at different depths of the choroid, starting from the choriocapillaris, then the small-scale choroidal vessels sitting posterior to these, then the smaller and medium-sized vessels and finally the whole choroid. 

            The choriocapillaris layer sits anterior-ward in the choroid, and is approximately 10 microns in thickness at the posterior pole \cite{nickla2010multifunctional}. However, the individual capillaries in this layer are too small to resolve given current \acrshort{OCT} resolution limitations. Given how densely packed the choriocapillaries are \cite{mrejen2013optical}, we assume in our vessel segmentation procedure that the first 10 microns of depth is vascular, and since the axial and lateral pixel length scales are known, this can be approximated well enough. For example, for Heidelberg Engineering \acrshort{OCT}1 devices, the axial length-scale is 3.87 microns, and so we assign the first 3 rows posterior and parallel to the \acrshort{RPE}-Choroid boundary as vessel.

            For the remaining depths of the choroid, we apply median cut clustering to the grayscale values of only the choroid pixels (those between the segmented boundaries). Fortunately, due to the nature of the median cut and \acrshort{MMCQ} algorithms, clusters can be ordered via their average intensity to work out which are vascular and which are not. The only set of parameters to define are the number of clusters $K$ to generate and how many to assign as choroidal vasculature, $k$. While these parameters can be left free to the end-user, we have empirically found $(K, k) = (20, 11)$ to be sensible given earlier parameter choices.
            
            We use the average pixel thickness $T$ to determine the different depth-levels, segmenting small- and medium-scale vessels from the choriocapillaris to $\nicefrac{T}{3}$ and $\nicefrac{3T}{5}$ posterior to the \acrshort{RPE}-Choroid boundary, respectively. Finally, the whole choroid is segmented for large-scale vessels. Figure \ref{fig:MMCQ_seg_vessel} shows the choroid used throughout this section, with segmentations for each of the depth-levels. In this case, the setting $(K, k) = (20, 11)$ was used. The final segmentation result (bottom-right panel) is the element-wise union of all segmentations.
        
        \end{mysubsection}

        \begin{mysubsection}[]{Pipeline}

            Figure \ref{fig:MMCQ_schematic_diag} presents a schematic of \acrshort{MMCQ}'s pipeline with figure \ref{fig:MMCQ_examples} showing the output segmentations on a selection of choroids using $(K, k) = (20, 11)$. Additionally, algorithm \ref{alg:mmcq} outlines the pseudocode for the \acrshort{MMCQ} algorithm.

            \begin{figure}[tbph]
            \centering
            \includegraphics[width=\textwidth]{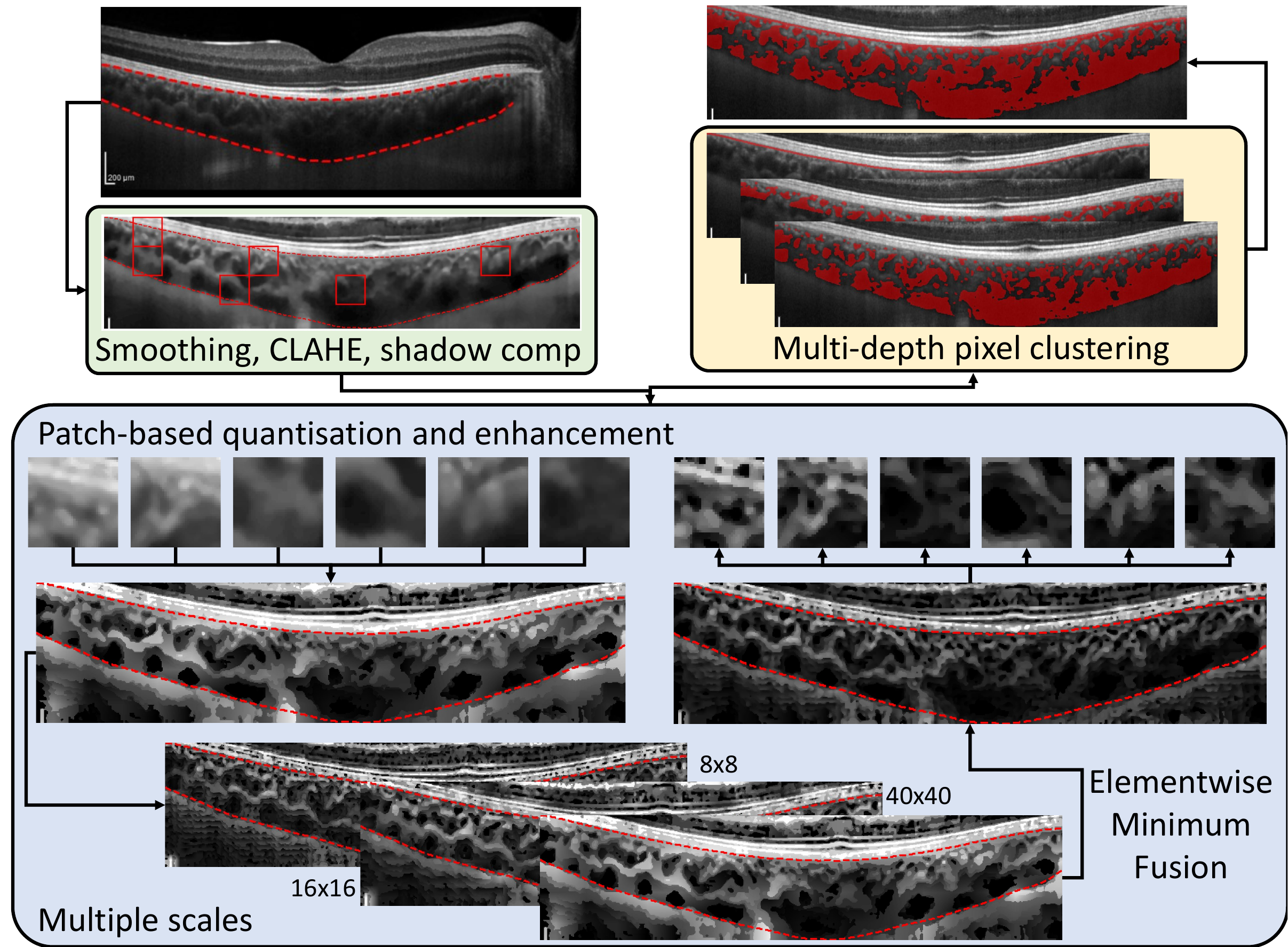}
            \caption[Schematic diagram of \acrshort{MMCQ}'s image analysis pipeline.]{Schematic diagram of \acrshort{MMCQ} segmenting the choroidal vessels in the \acrshort{EDI-OCT} B-scan used throughout this section.}
            \label{fig:MMCQ_schematic_diag}
            \end{figure}

            \begin{figure}[tbph]
                \centering
                \includegraphics[width=\linewidth]{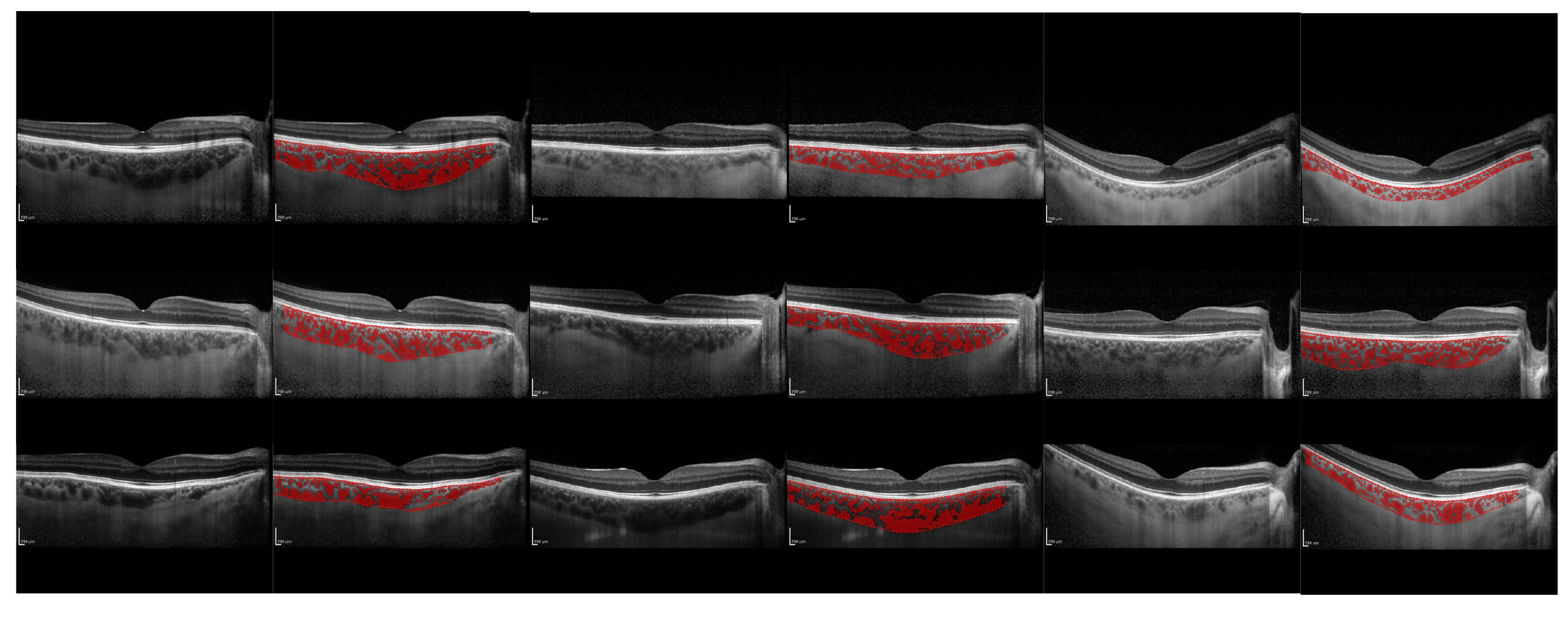}
                \caption[\acrshort{MMCQ}'s image analysis pipeline applied to several exemplar \acrshort{OCT} B-scans.]{Vessel segmentations using \acrshort{MMCQ} for a variety of choroids large and small with varying contrast, vessel distribution and speckle noise.}
                \label{fig:MMCQ_examples}
            \end{figure} 

            \begin{algorithm}[tbp]\footnotesize
            \caption{Multi-scale Median Cut Quantisation}
            \textbf{Inputs}: 8-bit \acrshort{OCT} B-scan $I$, choroid region segmentation boundaries, pixel length-scales and cluster configuration $(K, k$). \\
            \textbf{Output}: Segmented image $S$ of choroid vessels.
            \begin{algorithmic}[1]
                \State \multiline{Crop image $I$ to only choroid $I_c$, and pre-process $I_c$ using vessel shadow compensation, a simple 3 $\times$ 3 median filter and \acrshort{CLAHE} application with a tile size of 8 and clip limit of 2;}
                \State \multiline{Measure average choroid thickness $T$ (in pixels) between upper and lower choroid boundaries;}
                \State Three square patch sizes defined according to $\{\nicefrac{T}{10}, \nicefrac{T}{5}, \nicefrac{T}{2}\}$;
                \For{square patch size $p_i \in \{\nicefrac{T}{10}, \nicefrac{T}{5}, \nicefrac{T}{2}\}$}
                    \State \multiline{Split choroid image $I_c$ into non-overlapping patches of size $p_i \times p_i$;}
                    \For{patch $P$}
                        \State \multiline{Number of quantisation levels $K_P$ (clusters) via equation \eqref{eqref:MMCQ_cluster_K};}
                        \State \multiline{Quantise patch $P$ into $K_P$ levels using median cut quantisation, called $P_q$, replacing cluster group pixel intensities with their collective median value;}
                        \State \multiline{Compute the normalised cumulative distribution $CDF_{\text{norm}}$ via equation \eqref{eqref:MMCQ_cdf_norm_def};}
                        \State \multiline{Enhancement weight $w=w(P)=s(\sigma_d(P_q))$ given by equation \eqref{eqref:MMCQ_sigmoid} and $\sigma_d(P_q)$ given by equation \eqref{eqref:MMCQ_dispersion};}
                        \State \multiline{Generate intensity mapping $Q$ of the quantised patch $P_q$ representing only quantisation and zero enhancement via equation \eqref{eqref:MMCQ_Qmap};}
                        \State \multiline{Compute enhanced patch $P_e$ via equation \eqref{eqref:MMCQ_enh_patch};}
                    \EndFor
                    \State \multiline{Enhanced patches are combined and smoothed using bi-linear interpolation to create the enhanced choroid at patch size $p_i$, $I_{p_i}$;}
                \EndFor
                \State \multiline{Final, enhanced choroid $E(x,y) = \min\{{I_{p_i}(x,y) : i=0,1,2\}}$, i.e. fuse enhancements by taking element-wise minimum value across all $I_{p_i}$ across all scale dimensions;}
                \State \multiline{Thin, 10$\mu$m layer sub-\acrshort{RPE}-Choroid boundary segmented as choriocapillaris $S_c$ using axial pixel length-scale;}
                \For{depth-level $d \in \{\nicefrac{T}{3}, \nicefrac{3T}{5}\}$}
                    \State \multiline{Duplicate \acrshort{RPE}-Choroid boundary and shift $d$-pixels vertically downward, cropping out all choroid pixels outside of region in between these two boundaries;}
                    \State \multiline{Quantise choroid pixels within region of enhanced choroid $E$, using median cut quantisation into $K$ levels, keeping $k$ as vessel pixels, $S_d$;}
                \EndFor
                \State \multiline{Quantise whole enhanced choroid $E$ using median cut quantisation into $K$ levels, keeping $k$ as vessel pixels $S_w$;}
                \State \multiline{Segmented choroid $S = S_c + \displaystyle\sum_d S_d$ + $S_w$;}
                \State \multiline{\textbf{return} $S$ after clipping values such that $S$ only takes on values 0 and 1 to represent interstitial and vessel pixels.}
            \end{algorithmic}
            \label{alg:mmcq}
            \end{algorithm}
            
        \end{mysubsection}
                
    \end{mysection}    

    \begin{mysection}[]{MMCQ's evaluation}\label{sec:MMCQ_eval}
    
        We evaluate \acrshort{MMCQ} by comparing its segmentation performance against two manual graders, and Agrawal's modified Niblack method \cite{agrawal2020exploring}, as this is the most widely adopted approach for choroid vessel segmentation in the research community.

         \begin{table}[tb]\footnotesize
            \centering
            \begin{tabular}{llll}
\toprule
\multirow{2}{*}{} & \multicolumn{2}{c}{Cohort} & \multirow{2}{*}{Total} \\
\cmidrule(l){2-3}
 & Donors & Recipients &  \\
 \midrule
Sample, N & 10 & 10 & 20 \\
Right eye, N  & 10 & 10 & 20 \\
Age, years & 49 $\pm$ 14 & 44 $\pm$ 12 & 47 $\pm$ 13 \\
Sex (male \%) & 5 (50) & 7 (70) & 12 (60)\\
Pixel resolution & & & 768 $\times$ 768 \\
\acrshort{EDI} mode & & & Yes \\
\acrshort{ART} & & & 100 \\
\bottomrule
\end{tabular}

            \caption[Demographic information on cohort used to evaluate \acrshort{MMCQ}.]{Population demographics and image characteristics of dataset used to evaluate \acrshort{MMCQ} against manual grading and Agrawal's Niblack method \cite{agrawal2016choroidal}.}
            \label{tab:MMCQ_eval_pops}
        \end{table}

        \begin{mysubsection}[]{Study population}

            We used a subsample of the dataset used to evaluate \acrshort{GPET}'s choroid region segmentation described in section \ref{sec:ch_gpet_eval}. Briefly, the cohort \cite{dhaun2014optical} was collected to investigate chorioretinal changes in response to kidney transplantation in healthy donors and recipients with end-stage \acrshort{CKD}.
            
            For this evaluation, as we required manual segmentation of the choroidal vessels, we selected a random subsample of 20 B-scans at the patient-level from 10 renal transplant recipients and 10 donors. As the temporal nature of this cohort was longitudinal, we only selected images at baseline for this evaluation to remain objective in our analysis. Thus, there is equal proportion of cases and controls in our evaluation (10 cases of end-stage \acrshort{CKD} about to undergo kidney transplant and 10 controls about to undergo kidney donation).
            
            The basic demographics and image characteristics for this subsample are shown in table \ref{tab:MMCQ_eval_pops}. Please see section \ref{subsubsec:GPET_eval_acq} for a description of the image acquisition for this subsample.
        
        \end{mysubsection}

        \begin{mysubsection}[]{Statistical analysis}\label{subsec:MMCQ_eval_statanal}

            Two manual graders segmented the choroidal vessels in each \acrshort{OCT} B-scan, M1 and M2. M1 was a clinical ophthalmologist (Supervisor Dr. Ian J.C. MacCormick) and M2 was the author of this thesis (Jamie Burke). Graders segmented the choroid region and vessels using ITK-Snap \cite{py06nimg} and were blinded to each others segmentations, following the pre-specified measurement protocol outlined in appendix \ref{apdx:seg_protocol} using the definitions and equipment outlined in appendices \ref{apdx:definitions} and \ref{apdx:equipment_protocol}. We also applied \acrshort{MMCQ} and Agrawal's modified Niblack method \cite{agrawal2020exploring} to these 20 \acrshort{OCT} B-scans for quantitative comparison with the manual annotations.

            To keep the use of \acrshort{MMCQ} and Niblack consistent, we fixed the free parameters of both algorithms. For \acrshort{MMCQ}, we used the segmentation cluster configuration $(K, k) = (20, 11)$ (see section \ref{subsec:MMCQ_method_vesselseg}) and for Niblack, we fixed the window size $w$ to 51 and the standard deviation offset $k$ to -0.05 --- values proposed by Muller, et al. \cite{muller2022application} as the mean values of $w$ and $k$ when the algorithm was used manually to create a ground truth label dataset for a separate deep learning model.
                
            We used the Dice coefficient to measure overall segmentation agreement between the automatic methods (\acrshort{MMCQ}/Niblack) and the manual graders (M1/M2), and also used Precision and Recall to quantify the positive predictive value and sensitivity of the two approaches for choroid vessel segmentation. These metrics will quantify miss-classification of interstitial and vessel pixels, respectively. Segmentation metrics were measured only for the region in the image which constituted the choroidal space, dictated by Choroidalyzer, a method described later in chapter \ref{chp:chapter-choroidalyzer} for segmenting the choroidal space, vessel and detecting the fovea on \acrshort{OCT} B-scans. The segmentation metrics used here have been previously defined in section \ref{subsec:INTRO_metrics}.
            
            We also compare \acrshort{MMCQ} and Niblack to the manual grader on downstream choroidal measurements including vessel area and \acrshort{CVI} measured using a 6 mm, fovea-centred \acrshort{ROI}. The fovea was detected also through Choroidalyzer to remain objective in our analyses. These metrics have previously been defined in section \ref{subsec:ch1_INTRO_measure_choroid}. To compare derived metrics, we use mean absolute error (\acrshort{MAE}) as well as Pearson and Spearman correlation coefficients previously defined in section \ref{subsec:INTRO_metrics}. 
            
            Finally, we also compared the execution time of each approach for region and vessel segmentation, and report mean and standard deviation in seconds. \acrshort{MMCQ} and Niblack were run on a laptop with a 4 year old Intel Core i5 (8$^{\text{th}}$ generation) CPU and 16 Gb of RAM. Manual graders timed themselves during manual vessel segmentation using ITK-Snap.

        \end{mysubsection}

        \begin{mysubsection}[]{Results} \label{sec:MMCQ_eval_results}
        
            \begin{table}[tb]\footnotesize
                \begin{adjustwidth}{-1in}{-1in}  
                \centering
                \scalebox{0.9}{\begin{tabular}{@{}p{3cm}llllllllll@{}}
\toprule
  \multicolumn{1}{c}{\multirow{2}{*}{Comparison}} &
  \multicolumn{3}{c}{Segmentation} &
  \multicolumn{3}{c}{Vessel area} &
  \multicolumn{3}{c}{Vascular Index}\\ 
  \cmidrule(l){2-4}\cmidrule(l){5-7}\cmidrule(l){8-10}
  \multicolumn{1}{c}{} &
  \multicolumn{1}{c}{Dice} &
  \multicolumn{1}{c}{Precision} &
  \multicolumn{1}{c}{Recall} &
  \multicolumn{1}{c}{Pearson} &
  \multicolumn{1}{c}{Spearman} &
  \multicolumn{1}{c}{MAE (mm$^2$)} &
  \multicolumn{1}{c}{Pearson} &
  \multicolumn{1}{c}{Spearman} &
  \multicolumn{1}{c}{MAE}\\ \midrule
MMCQ vs. M1 & \textbf{0.7148} & \textbf{0.6921} & 0.7486 & \textbf{0.9622}** & \textbf{0.9564}** & \textbf{0.1080} & 0.5503* & \textbf{0.5242}* & \textbf{0.0759} \\
Niblack vs. M1 & 0.6838 & 0.6238 & \textbf{0.7644} & 0.9187** & 0.9113** & 0.1153 & \textbf{0.5958*} & 0.5092* & 0.1901 \\
\midrule
MMCQ vs. M2 & \textbf{0.7350} & \textbf{0.6890} & 0.7911 & \textbf{0.9873}** & \textbf{0.9700}** & \textbf{0.1182} & 0.5590* & 0.5323* & \textbf{0.0816} \\
Niblack vs. M2 & 0.7065 & 0.6262 & \textbf{0.8136} & 0.9643** & 0.9474**  & 0.1435 & \textbf{0.6068*} & \textbf{0.5353}* & 0.1464 \\
\midrule
M1 vs. M2 & 0.7699 & 0.7937 & 0.7479 & 0.9718** & 0.9504** & 0.0637 & 0.8074** & 0.6857** & 0.0618 \\
MMCQ vs. Niblack & 0.7366 & 0.7866 & 0.6959 & 0.9851** & 0.9744** & 0.0568 & 0.1637\textsuperscript{\textdagger} & 0.0903\textsuperscript{\textdagger} & 0.0461 \\
\bottomrule
\end{tabular}}

                \end{adjustwidth}
                \caption[Segmentation performance of \acrshort{MMCQ} and Niblack against manual labels.]{Comparison of all four approaches in a pairwise fashion. Bold values represents the best method (of MMCQ/Niblack) for each comparison with the manual grader. **: P < 0.0001, *: P < 0.05, \textsuperscript{\textdagger}: P > 0.05. }
                \label{tab:MMCQ_eval_results}
            \end{table}
            
            \begin{table}[tb]
                \centering
                \begin{tabular}{lc}
\hline
Method & Execution time (s) \\ \hline
M1 & 1506.000 $\pm$ 744.073 \\
M2 & 1176.700 $\pm$ 263.968 \\
Niblack & 0.370 $\pm$ 0.105 \\
MMCQ & 2.148 $\pm$ 0.738 \\ \hline
\end{tabular}

                \caption[Execution times of \acrshort{MMCQ}, Niblack and the manual graders.]{Execution time (in seconds) for the manual graders and semi-automatic approaches for choroid vessel segmentation.}
                \label{tab:MMCQ_eval_times}
            \end{table}

            Table \ref{tab:MMCQ_eval_results} shows the main results of comparing all four methods with each other in a pairwise fashion. Overall, \acrshort{MMCQ} had better agreement with both manual graders, both from the segmentation and choroid-derived evaluation metrics. For segmentation metrics, \acrshort{MMCQ} had a Dice score with graders M1 and M2 of 0.7148 and 0.7350, respectively, while Niblack had 0.6838 and 0.7065, respectively. Interestingly, \acrshort{MMCQ} achieved higher precision than Niblack (0.6921 vs. 0.6238 for comparisons with rater M1) with similarly comparative values for rater M2. Conversely, Niblack had higher recall (0.7486 vs. 0.7644 for comparisons with rater M1) with similarly comparative values with rater M2. This would suggest that \acrshort{MMCQ}-segmented choroids have higher positive predictive value, with fewer false positives than Niblack, at the cost of slight under-segmentation (more false negatives). Thus, according to the manual segmentations, Niblack over-segments the choroidal space, classifying more interstitial pixels as vessel compared with \acrshort{MMCQ}. However, Niblack's cost of over-segmenting the choroidal space is somewhat balanced by it's ability to have fewer false negatives, i.e. is able to correctly identify vessel pixels better than \acrshort{MMCQ}.

            \begin{figure}[tbhp]
                \centering
                \includegraphics[width=0.95\linewidth]{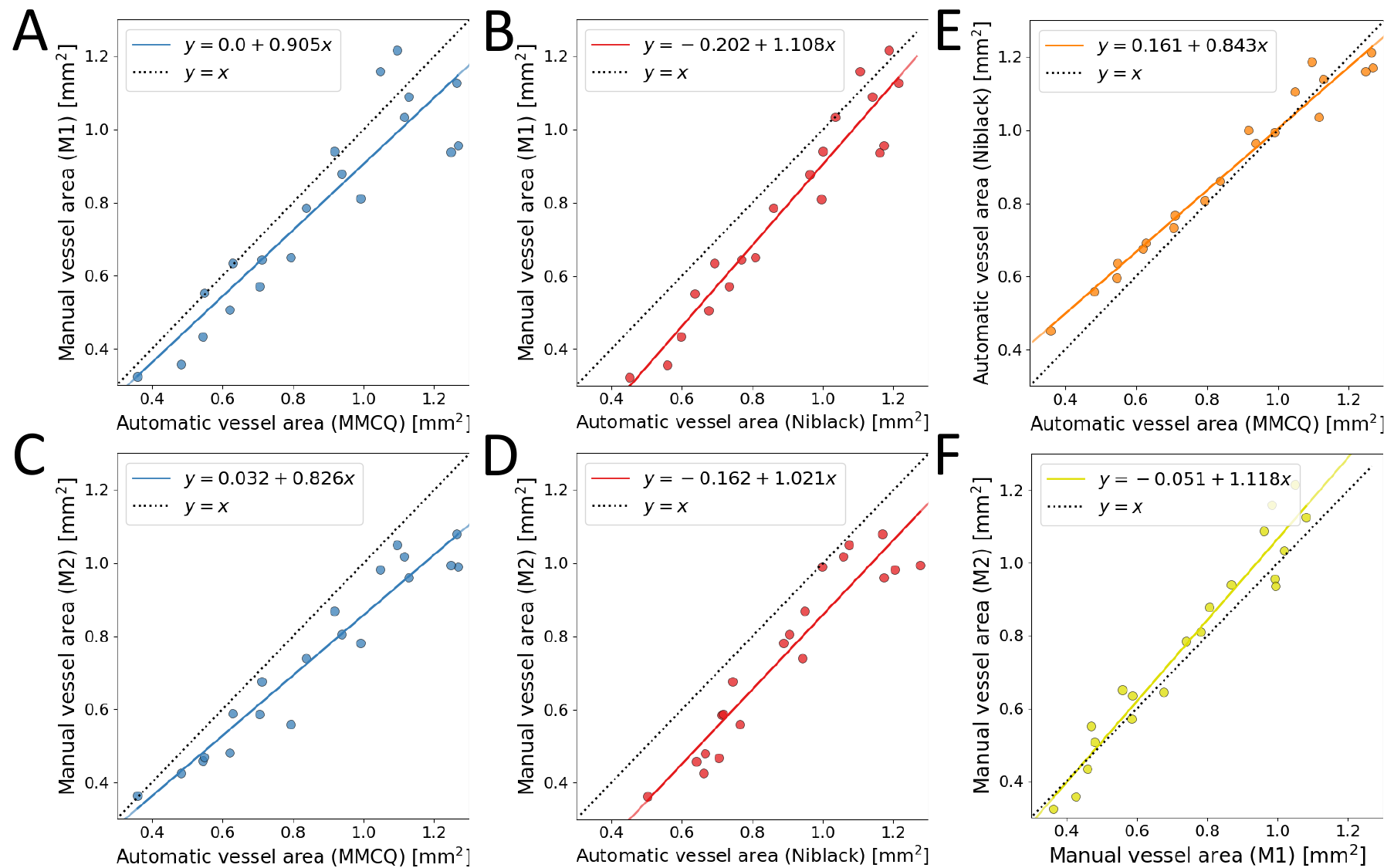}
                \caption[Correlation in choroid vessel area between \acrshort{MMCQ}, Niblack and the manual graders.]{Scatter plots demonstrating correlation in choroid vessel area between different methods. (A) \acrshort{MMCQ} vs. M1, (B) Niblack vs. M1, (C) \acrshort{MMCQ} vs. M2, (D) Niblack vs. M2, (E) \acrshort{MMCQ} vs. Niblack, (F) M1 vs. M2.}
                \label{fig:MMCQ_va_corrplots}
            \end{figure}

            \begin{figure}[tbhp]
                \centering
                \includegraphics[width=0.95\linewidth]{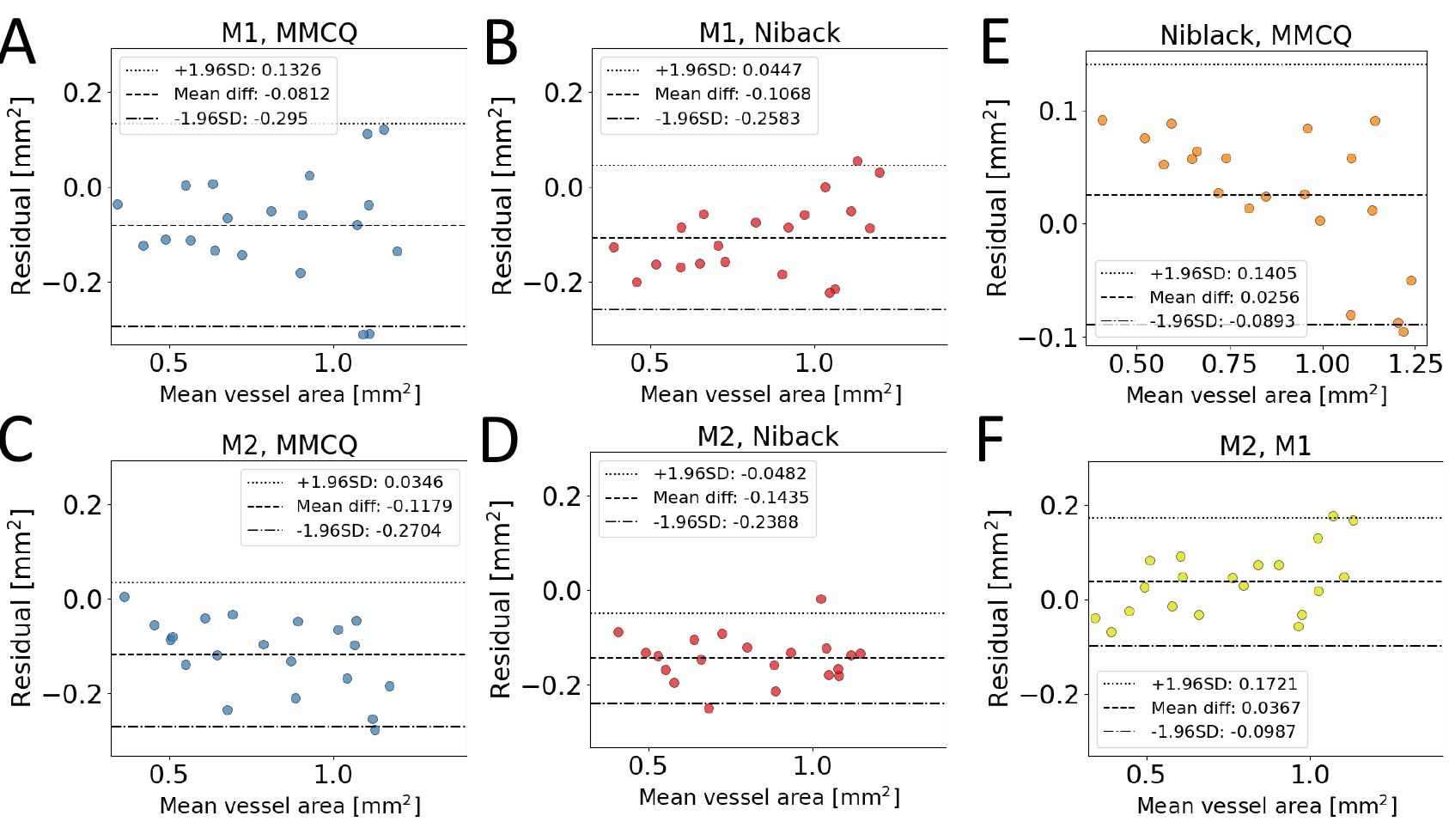}
                \caption[Agreement in choroid vessel area between \acrshort{MMCQ}, Niblack and the manual graders.]{Bland-Altman plots of agreement in choroid vessel area between different methods. (A) \acrshort{MMCQ} vs. M1, (B) Niblack vs. M1, (C) \acrshort{MMCQ} vs. M2, (D) Niblack vs. M2, (E) \acrshort{MMCQ} vs. Niblack, (F) M1 vs. M2.}
                \label{fig:MMCQ_va_BAplots}
            \end{figure}

            \begin{figure}[tb]
                \centering
                \includegraphics[width=\linewidth]{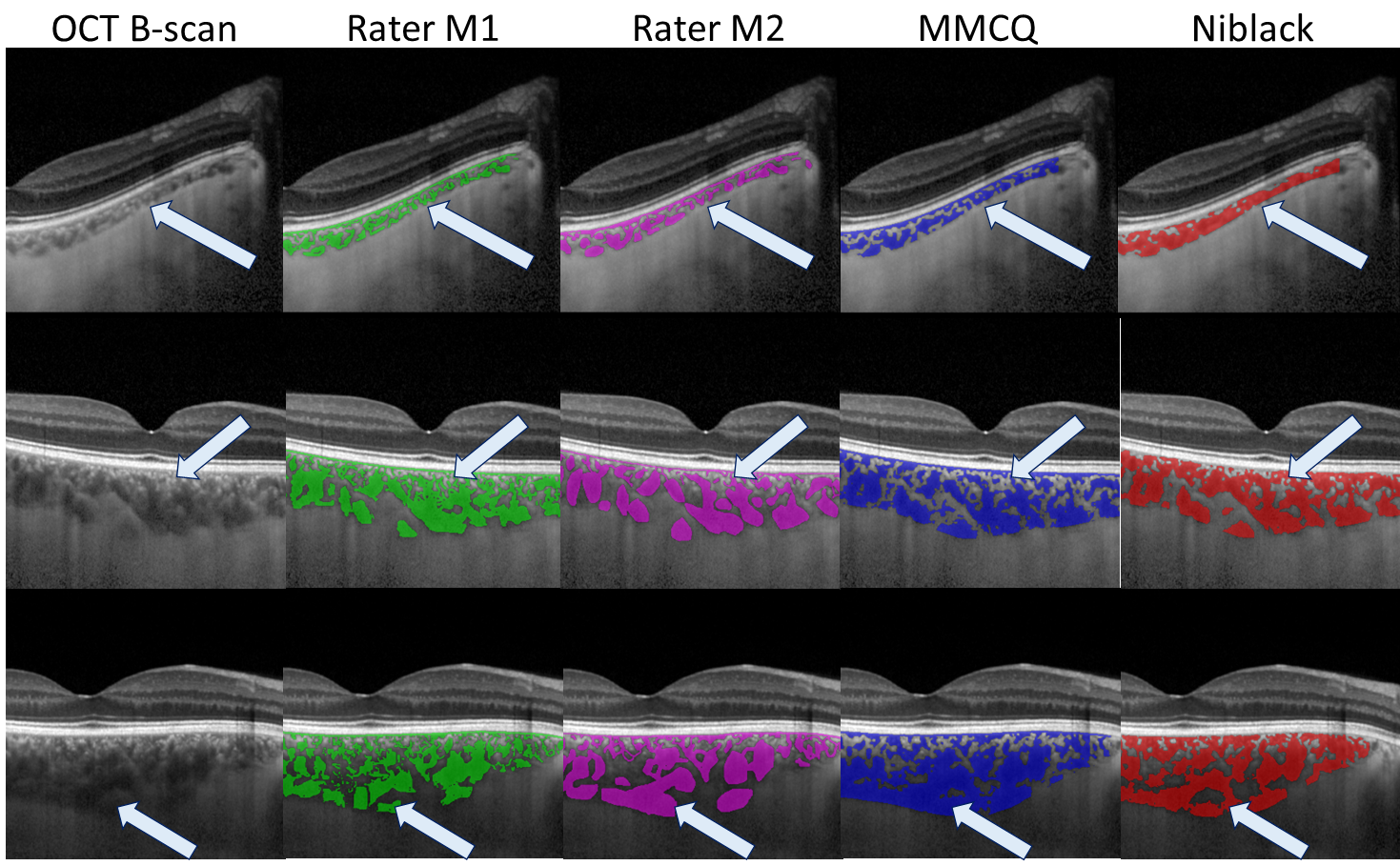}
                \caption[Qualitative comparison between \acrshort{MMCQ}, Niblack and manual graders.]{Qualitative comparison of the vessel segmentations for semi-automatic and manual approaches for the smallest and largest choroid in the dataset. Light blue arrows indicate sources of disagreement.}
                \label{fig:MMCQ_eval_choroids}
            \end{figure}

            There were strong and statistically significant correlations between both semi-automatic methods with the manual graders for choroid vessel area. \acrshort{MMCQ}-derived vessel area had higher correlations and lower \acrshort{MAE} than Niblack (Pearson correlation between \acrshort{MMCQ} and raters M1 and M2 was 0.9622 and 0.9873, respectively, while for Niblack this was 0.9113 and 0.9474, respectively). These relationships can are visualised through scatter plots and Bland-Altman plots of choroid vessel area for each of the comparisons made, as shown in figures \ref{fig:MMCQ_va_corrplots} and \ref{fig:MMCQ_va_BAplots}.
            
            Both figures help describe why the there are such large \acrshort{MAE}s when comparing the semi-automatic methods with the manual graders (0.1080 and 0.1182 for \acrshort{MMCQ} vs. M1 and M2, respectively). In all cases where we compare semi-automatic to manual, the semi-automatic method segments significantly more of the choroid than the manual grader. Interestingly, \acrshort{MMCQ} and Niblack have opposite effects when compared to manual segmentations, i.e. \acrshort{MMCQ} is closest to the manual graders for smaller choroids, and vice versa for Niblack. This may suggest, on the assumption of manual grading reflecting accurate segmentation, that \acrshort{MMCQ} could be better suited for smaller choroids than Niblack, and vice versa for larger choroids.
            
            Figure \ref{fig:MMCQ_eval_choroids} shows three \acrshort{OCT} B-scans of which have been magnified to show only half of the choroid, with segmentations from all four methods (rater M1, green; rater M2, pink; \acrshort{MMCQ}, blue; Niblack, red) overlaid in series. The choroids were selected based on their total choroid area, representing the smallest (top), median (middle) and largest (bottom) choroids in the dataset. Blue arrows indicate sources of disagreement between the methods.
            
            In the top row, we can see that Niblack over-segments the choroidal space significantly more than the other three methods, which are all able to identify the smaller vessels with precision. This translated into an overestimation of \acrshort{CVI} of 0.66, compared with 0.47, 0.54 and 0.53 for raters M1, M2 and MMCQ, respectively. In the middle row, rater M1 segmented the choroidal vessels in more detail (\acrshort{CVI}, 0.57) compared with rater M2 (\acrshort{CVI}, 0.52) translating to a 0.05 discrepancy in \acrshort{CVI}, while Niblack (\acrshort{CVI}, 0.62) appeared to over-segment the smaller vessels nearer the \acrshort{RPE}-Choroid boundary, unlike \acrshort{MMCQ} (\acrshort{CVI}, 0.59). In the bottom row, \acrshort{MMCQ} was subject to over-segmentation in this case because of the poor vessel boundary definition in the posterior segment of the choroidal space. This translated to \acrshort{MMCQ} and Niblack both over-estimating the \acrshort{CVI} as 0.63 and 0.61, respectively, with raters M1 and M2 estimating \acrshort{CVI} as 0.50 and 0.51, respectively. Thus, we propose that \acrshort{MMCQ} is more faithful to the manual raters for smaller choroids, and produces higher quality segmentations over the Niblack algorithm, on the assumption these manual graders represent accurate ground truths. 

            While the Bland-Altman plot in figure \ref{fig:MMCQ_va_BAplots} demonstrate significant systematic bias between semi-automatic and manual methods, it is encouraging to observe from figure \ref{fig:MMCQ_va_corrplots} strong linear and monotonic relationships between semi-automatic and manual methods in this small dataset. Based on this evaluation, \acrshort{MMCQ} provides more accurate segmentations relative to Niblack, promoting vessel fidelity and positive predictive value over vessel pixel completeness at the cost of over-segmentation. However, the differences in these population-level performance metrics between \acrshort{MMCQ} and Niblack are not large in absolute terms, and actually tend to average over specific, eye-level differences between the methods, such as Niblack's tendency to over-segment small-scale choroidal vasculature (figure \ref{fig:MMCQ_eval_choroids}).

            \begin{figure}[tbhp]
                \centering
                \includegraphics[width=0.95\linewidth]{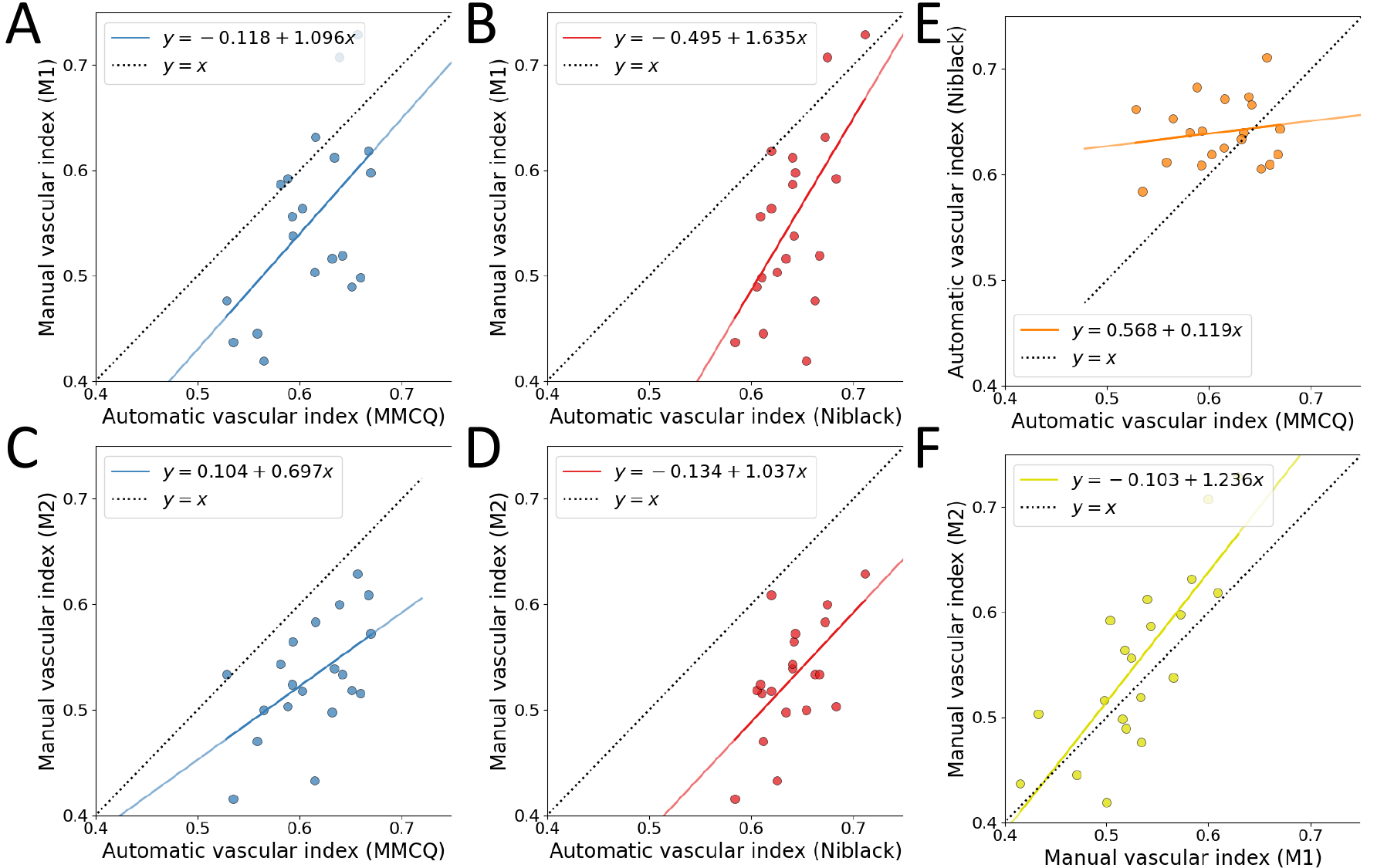}
                \caption[Correlation in \acrshort{CVI} between \acrshort{MMCQ}, Niblack and the manual graders.]{Scatter plots demonstrating correlation in \acrshort{CVI} between different methods. (A) \acrshort{MMCQ} vs. M1, (B) Niblack vs. M1, (C) \acrshort{MMCQ} vs. M2, (D) Niblack vs. M2, (E) \acrshort{MMCQ} vs. Niblack, (F) M1 vs. M2.}
                \label{fig:MMCQ_cvi_corrplots}
            \end{figure}

            \begin{figure}[tbhp]
                \centering
                \includegraphics[width=0.95\linewidth]{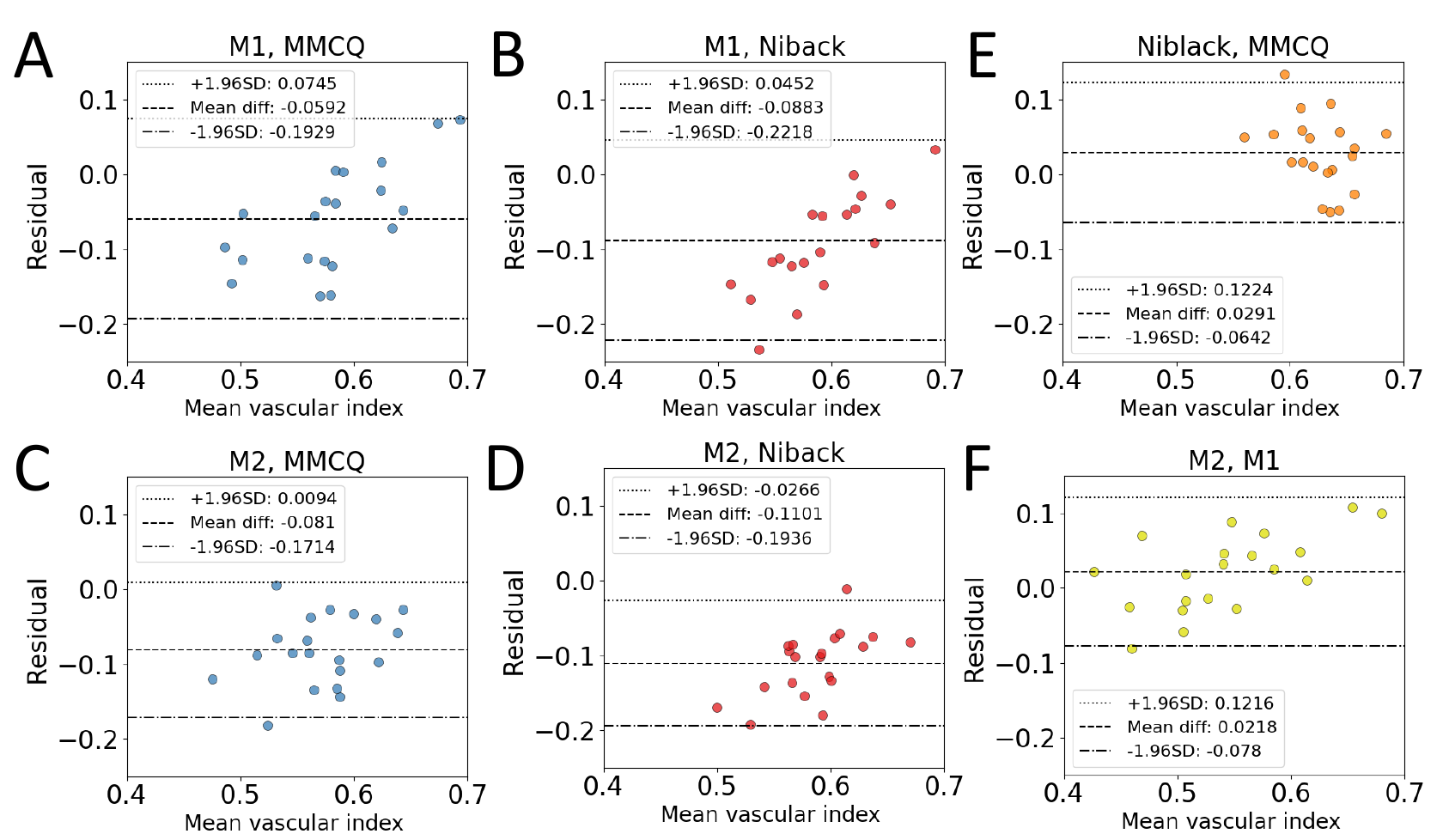}
                \caption[Agreement in \acrshort{CVI} between \acrshort{MMCQ}, Niblack and the manual graders.]{Bland-Altman plots of agreement in \acrshort{CVI} between different methods. (A) \acrshort{MMCQ} vs. M1, (B) Niblack vs. M1, (C) \acrshort{MMCQ} vs. M2, (D) Niblack vs. M2, (E) \acrshort{MMCQ} vs. Niblack, (F) M1 vs. M2.}
                \label{fig:MMCQ_cvi_BAplots}
            \end{figure} 

            For \acrshort{CVI}, semi-automatic methods had poor agreement with the manual methods, both observed by Pearson/Spearman correlation (\acrshort{MMCQ} vs M1, 0.5503/ 0.5242 and Niblack vs. M1, 0.5958/0.5092) which were weakly significant (P < 0.05) and also exhibited large \acrshort{MAE}s (\acrshort{MMCQ} vs M1, 0.0759 and Niblack vs. M1, 0.1901). While \acrshort{MMCQ} performed better than Niblack according to these manual labels, agreement was poor in absolute terms. The discrepancy between the correlations observed with choroid vessel area and \acrshort{CVI} is an interesting contradiction. Figures \ref{fig:MMCQ_cvi_corrplots} and \ref{fig:MMCQ_cvi_BAplots} show scatter and Bland-Altman plots in a similar style to figure \ref{fig:MMCQ_va_corrplots} but for \acrshort{CVI}. Overall, there was poor agreement, with exception for the two manual graders, which reported reasonable agreement (Pearson/Spearman correlation, 0.7937/0.6857). However, the manual graders do not represent a gold standard approach, especially for larger choroids as shown in the figures \ref{fig:MMCQ_va_corrplots}(F) and \ref{fig:MMCQ_cvi_corrplots}(F).
            
            \acrshort{MMCQ} and Niblack had very poor agreement for \acrshort{CVI} (Pearson/Spearman correlation, 0.1637/0.0903, figure \ref{fig:MMCQ_cvi_corrplots}(E)). While at the population-level, the extent of disagreement in \acrshort{CVI} appears random, the disagreement is actually ordered by the choroid vessel area (and thus the size of the choroid). The largest disagreement between semi-automatic methods in \acrshort{CVI} (absolute difference of 0.13, 13\%) corresponded to the choroid with the smallest vessel area, of which Niblack segmented a significant proportion more than MMCQ (figure \ref{fig:MMCQ_va_BAplots}(E)). This pattern was consistent such that smaller choroids corresponded to proportionately greater disagreement (relative to the size of choroid) in both CVI and vessel area between the two semi-automatic methods. However, unlike the linear relationship observed for choroid vessel area across the population (figure \ref{fig:MMCQ_va_corrplots}(E)), once an eye's choroid vessel area is normalised by its total choroid area (calculating \acrshort{CVI}), this relationship disappears quite dramatically (figure \ref{fig:MMCQ_cvi_corrplots}(E)). This is likely because any error in vessel area in the numerator of \acrshort{CVI} is further exacerbated by any error in the denominator, given the size of the choroid. Larger error in vessel area in smaller choroids introduces proportionately larger error in \acrshort{CVI}. Thus, measurement error in vessel area is made significantly worse when translated into \acrshort{CVI} for smaller choroids.

            In terms of execution time, manual graders M1 and M2 took approximately 25 and 20 minutes to segment the choroidal vessels in an \acrshort{OCT} B-scan, respectively, while Niblack and \acrshort{MMCQ} took 0.37 and 2.14 seconds, respectively, with no human input. While Niblack has a faster execution time, by virtue of only working at a single scale and measuring summary statistics, both execution times are entirely reasonable for large-scale \acrshort{OCT} image analysis. For a typical \acrshort{OCT} volume scan with 61 successive B-scans, the manual graders would take approximately 23 hours on average to complete, compared with Niblack and \acrshort{MMCQ} which would take 23 seconds and 131 seconds, respectively. This translates to significant time saving for the researcher, allowing them to focus on the science rather than the image analysis. Note that Agrawal's original method \cite{agrawal2016choroidal} took approximately 1 minute per B-scan due to additional manual identification of the \acrshort{ROI} to measure the choroid. This execution time has been significantly reduced by virtue of automatically detecting the fovea to produce standardised, choroid-derived measurements. 

            \vfill
            
        \end{mysubsection}

    \end{mysection}

    \begin{mysection}[]{MMCQ's Reproducibility} 

        We investigate the reproducibility of \acrshort{MMCQ} and Agrawal's Niblack method \cite{agrawal2020exploring} using two datasets which contain repeated \acrshort{OCT} samples.

        \begin{table}[tb]\footnotesize
            \begin{adjustwidth}{-1in}{-1in}  
            \centering
            \scalebox{0.8}{\begin{tabular}{p{4cm}p{5cm}p{5cm}}
\hline
\multirow{2}{*}{} & \multicolumn{2}{c}{Study} \\ \cline{2-3} 
 & i-Test \cite{dhaun2014optical} & \acrshort{GCU} Topcon \cite{moukaddem2022comparison} \\ \hline
Cohort demographics &  &  \\ \cline{1-1}
Participants (Eyes) & 60 (120) & 21 (33) \\
Right eyes (\%) & 60 (50) & 15 (45.5) \\
Age (SD) & 34.7 (5.2) & 23.9 (4.2) \\
Sex, F (\%) & 60 (100) & 9 (43) \\
Ethnicity & 52 White, 6 Asian, 2 Mixed & 12 White, 5 Asian, 2 Black, 2 Middle Eastern \\
Gestation & 35.6 (3.4) & NA \\
Refractive status\textsuperscript{\textdagger} & 3 hyperopes, 31 emmetropes, 26 myopes & 9 hyperopes, 8 emmetropes, 3 myopes, 1 missing \\
Study purpose & Pre-eclamptic / normative pregnancy at late gestation & Diurnal variation \\
Control/Case & 45/15 & 21/0 \\
 &  &  \\
Image characteristics &  &  \\ \cline{1-1}
Device & Spectralis (Heidelberg) & DRI Triton Plus (Topcon) \\
\acrshort{OCT} Type & Spectral-domain & Swept-source \\
Scan Pattern & Macular volume & Macular radial \\
Mode & \acrshort{HRA}+\acrshort{OCT} & \acrshort{CFP}+\acrshort{OCT} \\
Time of day (Interval) & All in afternoon (1 minute) & 13 morning, 12 afternoon, 8 evening (5 minutes) \\
B-scans per eye & 31 (\acrshort{EDI}) / 61 (non-\acrshort{EDI}) & 12 \\
\acrshort{ART} & 50 (\acrshort{EDI}) / 12 (non-\acrshort{EDI}) & NA \\
\acrshort{SNR} & 35.61 & >88 \\
Pixel resolution & 496 $\times$ 768 & 992 $\times$ 1024 \\ \hline
\end{tabular}
}

            \end{adjustwidth}
            \caption[Demographics of samples used to assess the reproducibility of \acrshort{MMCQ} and Niblack.]{Population demographics and image characteristics of the datasets used to assess the reproducibility of \acrshort{MMCQ} and Niblack. \textsuperscript{\textdagger}: myopic/hyperopic status defined as $<-1 / >1$ dioptres. Dioptre measurements for i-Test sample was scan focus collected from the \acrshort{OCT} imaging device metadata, and this was the spherical equivalent for the \acrshort{GCU} Topcon sample.}
            \label{tab:MMCQ_repr_pops}
        \end{table}
    
        \begin{mysubsection}[]{Study population} \label{subsec:MMCQ_repr_pops}
            Two samples were used in this section, the i-Test and Glasgow Caledonian University (\acrshort{GCU}) Topcon datasets. The i-Test cohort \cite{dhaun2014optical} is being collected to investigate the maternal vascular response to pregnancy that may manifest within the eye. Specifically, the i-Test study is interested in the utility of ocular biomarkers being predictive of late-gestation pre-eclampsia. The \acrshort{GCU} Topcon cohort \cite{moukaddem2022comparison} was collected to investigate diurnal variation of the choroid in eyes of varying refractive status, and had a preference for recruiting hyperopic eyes. The study population and imaging characteristics of the two samples used to assess the reproducibility of \acrshort{MMCQ} (and Niblack) are shown in table \ref{tab:MMCQ_repr_pops}. Briefly, there were 120 eyes from 60 participants in the i-Test sample and 33 eyes from 21 participants in the \acrshort{GCU} Topcon sample with repeated data available at the point of analysis.

            Acquisition for i-Test \cite{dhaun2014optical} used the \acrshort{SD-OCT} Heidelberg \acrshort{OCT}1 Spectralis Standard and FLEX modules (Heidelberg Engineering, Heidelberg, Germany). During acquisition two unregistered, macula-centred volume scans of both eyes were taken covering a 30 $\times$ 20 degree (approximately 9 $\times$ 6.6 mm, horizontal $\times$ vertical) field of view (\acrshort{FOV}). Volume scans were taken immediately (within 1 minute) after one another, with \acrshort{EDI} mode toggled on and off. For each \acrshort{EDI-OCT} volume, 31 parallel line-scans were collected, linearly spaced approximately 240 microns apart with an \acrshort{ART} of 50. For each non-\acrshort{EDI} \acrshort{OCT} volume, 61 parallel line-scans were collected, linearly spaced approximately 120 microns apart with an \acrshort{ART} of 12. \acrshort{OCT} B-scans were of pixel resolution 496 $\times$ 768 (pixel height $\times$ width). 

            Acquisition for the \acrshort{GCU} Topcon sample used the \acrshort{SS-OCT} Topcon DRI Triton Plus device (Topcon, Tokyo, Japan). During acquisition, a twelve-line, fovea-centred radial scan was captured, starting from a horizontal-line B-scan and rotating clockwise in 30 degree intervals. Each B-scan covered approximately a 9 mm \acrshort{FOV} laterally and had a pixel resolution of 992 $\times$ 1024 pixels. Any set of 12 B-scans with an average image quality score less than 88 --- reported by the \acrshort{SNR} index built-in to the Topcon scanning device --- was excluded. Repeated acquisitions were taken at the same location during each visit within 5 minutes. Given repeated acquisition was not the primary aim of the \acrshort{GCU} Topcon study, only 21 participants had repeated \acrshort{OCT} data, with nine only in one eye, and twelve in both eyes.
        
        \end{mysubsection}
        
        \begin{mysubsection}[]{Statistical analysis}\label{subsec:MMCQ_repr_statanal}
        
            \begin{figure}[tb]
                \centering
                \includegraphics[width=\linewidth]{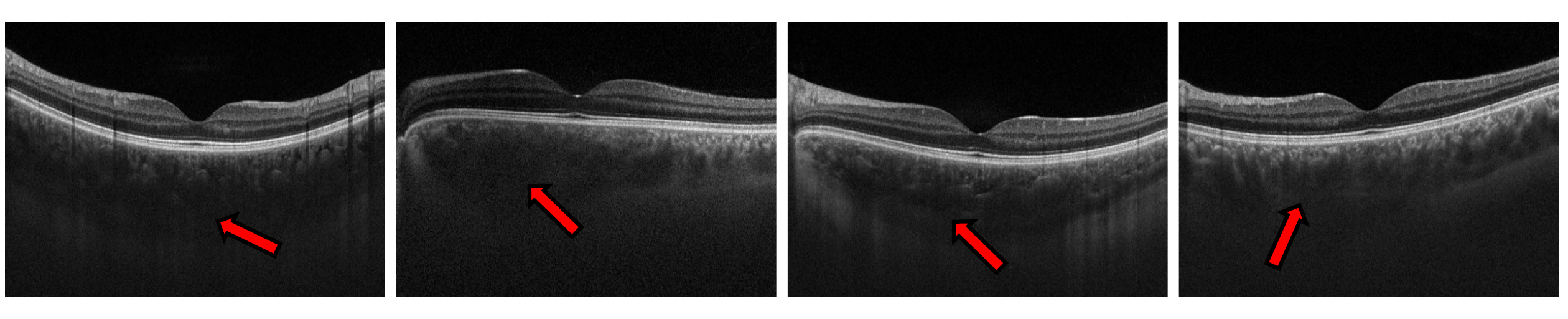}
                \caption[\acrshort{OCT} B-scans from the excluded eyes with poor vessel visualisation.]{OCT B-scan slices from each eye excluded from the vessel segmentation reproducibility analysis due to poor vessel definition (arrows).}
                \label{fig:MMCQ_repr_exc}
            \end{figure}
        
            For all repeated pairs, we segmented the choroidal vessels using both algorithms, after segmenting the choroidal space and detecting the fovea using Choroidalyzer (see chapter \ref{chp:chapter-choroidalyzer}, as this method is out of the scope of this chapter). Similar to section \ref{sec:MMCQ_eval}, we fixed the parameter configuration for \acrshort{MMCQ} and Niblack such that MMCQ's vessel segmentation cluster configuration was $(K, k)$ = (20, 11) and for Niblack we fixed the window size $w$ to 51 and the standard deviation offset $k$ to -0.05 \cite{muller2022application}.
            
            For the i-Test sample, we measured the choroid vessel volume (in mm$^3$) and \acrshort{CVI} for the nine sub-fields in the \acrshort{ETDRS} grid \cite{early1991grading} for each \acrshort{EDI-OCT} and non-\acrshort{EDI} \acrshort{OCT} volume pair. Thus, there were 1080 comparisons made per choroid metric for this sample (9 values $\times$ 120 eyes). Details on measuring the vessel and \acrshort{CVI} maps can be found in section \ref{subsubsec:ch1_INTRO_oct_volume}.
            
            For each participant in the \acrshort{GCU} Topcon sample, \acrshort{OCT} B-scan capture was collected throughout the day, with some examinations containing repeated acquisitions. There were 79 instances of repeated \acrshort{OCT} B-scan data overall, and approximately 25\% of these included more than two acquisitions per examination. Additionally, each acquisition included 12 \acrshort{OCT} B-scans per eye. To prevent any sampling bias and remain objective in our reproducibility analysis, we selected one repeated pair per eye to maximise equal sample size for morning, afternoon and evening acquisition where possible. This led to selecting 13 repeated pairs from the morning, 12 from the afternoon and 8 from the evening. Upon further inspection of each participant's choroid, four eyes from two participants were excluded because of poor vessel visualisation. An \acrshort{OCT} B-scan slice from each eye can be seen in figure \ref{fig:MMCQ_repr_exc}, with arrows indicating very poor vessel definition in the choroidal space. Thus, there were 348 comparisons made (29 eyes $\times$ 12 B-scans), using choroid vessel area (in mm$^2$) and \acrshort{CVI} measured in a 6 mm, fovea-centred \acrshort{ROI}. Details on making these choroid measurements can be found in section \ref{subsec:ch1_INTRO_measure_bsan}.

            We measure the reproducibility of \acrshort{MMCQ} and Niblack at the population-level, reporting population mean and standard deviation (\acrshort{SD}), as well as mean absolute error (\acrshort{MAE}) and Pearson and Spearman correlation coefficients. Additionally, we show correlation plots and use Bland-Altman \cite{bland1986statistical} plots to assess the relationship between repeated measurements and the distribution of residuals. We also report the reproducibility of \acrshort{MMCQ} and Niblack at the eye-level using the measurement noise parameter $\lambda$ \cite{engelmann2024applicability}. All aforementioned metrics have been defined and described earlier in section \ref{subsec:INTRO_metrics}.

        \end{mysubsection}

        \begin{mysubsection}[]{Results}

        \begin{mysubsubsection}[]{Population-level}
                    
            \begin{table}[tb]\footnotesize
                \centering
                
\begin{tabular}{rp{2cm}lllll}
\toprule
\multicolumn{1}{l}{Dataset} & Metric {[}unit{]} & Method & Mean (\acrshort{SD}) & \acrshort{MAE} & Pearson & Spearman  \\ 
\midrule
\multirow{4}{1.75cm}{i-Test (\acrshort{SD-OCT} \acrshort{ETDRS})} & \multirow{2}{2cm}{\acrshort{CVI}} & \acrshort{MMCQ} & 0.6121 (0.0351) & \textbf{0.0112} & \textbf{0.9633} & \textbf{0.9600} \\
 &  & Niblack & 0.6335 (0.0333) & 0.0134 & 0.8993 & 0.9022\\ \cmidrule(l){2-7} 
 & \multirow{2}{2.75cm}{vessel volume {[}mm$^3${]}} & \acrshort{MMCQ} & 0.9424 (0.3233) & \textbf{0.0251} & 0.9994 & 0.9987 \\
 &  & Niblack & 0.9662 (0.2741) & 0.0257 & \textbf{0.9995} & \textbf{0.9990} \\ 
 \midrule
\multirow{4}{1.75cm}{\acrshort{GCU} Topcon (\acrshort{SS-OCT} Foveal)} & \multirow{2}{2cm}{\acrshort{CVI}} & \acrshort{MMCQ} & 0.6332 (0.0332) & 0.0113 & \textbf{0.9266} & \textbf{0.9226} \\
 &  & Niblack & 0.5899 (0.0209) & \textbf{0.0083} & 0.9084 & 0.8949\\ \cmidrule(l){2-7} 
 & \multirow{2}{2.25cm}{vessel area {[}mm$^2${]}} & \acrshort{MMCQ} & 1.1994 (0.3522) & 0.0400 & \textbf{0.9871} & \textbf{0.9881} \\
 &  & Niblack & 1.0994 (0.2522) & \textbf{0.0363} & 0.9796 & 0.9806 \\ 
 \bottomrule
\end{tabular}

                \caption[Reproducibility performance at the population-level for \acrshort{MMCQ} and Niblack.]{Reproducibility performance for \acrshort{MMCQ} and Niblack. All Pearson and Spearman correlations were statistically significant with P-values P < 0.0001. Bold values represent where \acrshort{MMCQ} or Niblack had the better score, per metric.}
                \label{tab:MMCQ_repr_results}
            \end{table}

            \begin{figure}[tb]
                \centering
                \includegraphics[width=\linewidth]{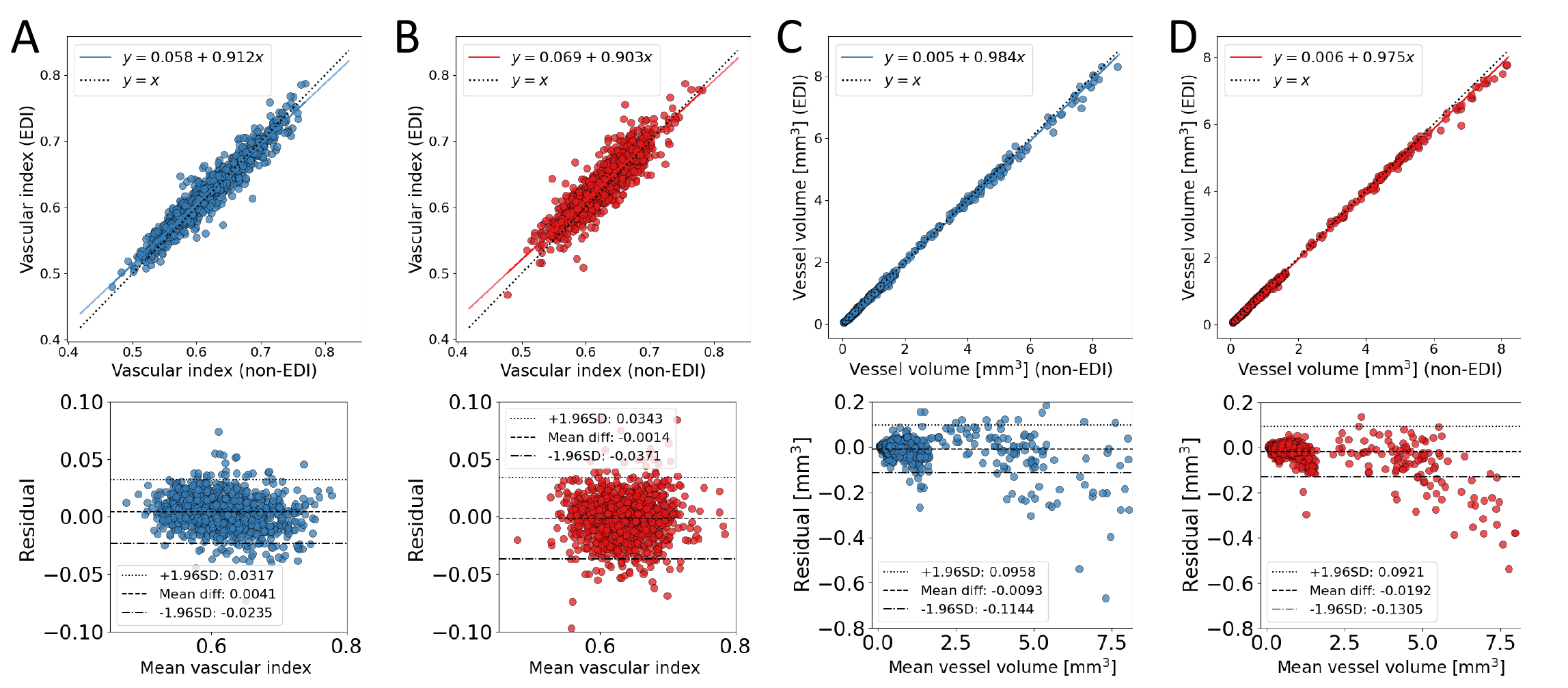}
                \caption[Population-level agreement in repeated \acrshort{SD-OCT} volume data for \acrshort{MMCQ} and Niblack.]{Correlation and Bland-Altman plots for assessing the reproducibility of \acrshort{MMCQ} and Niblack using repeated, \acrshort{SD-OCT} volume data (i-Test). (A) \acrshort{MMCQ} \acrshort{CVI}, (B) Niblack \acrshort{CVI}, (C) \acrshort{MMCQ} vessel volume, (D) Niblack vessel volume.}
                \label{fig:MMCQ_repr_sdoct_corrplots}
            \end{figure}

            \begin{figure}[tb]
                \centering
                \includegraphics[width=\linewidth]{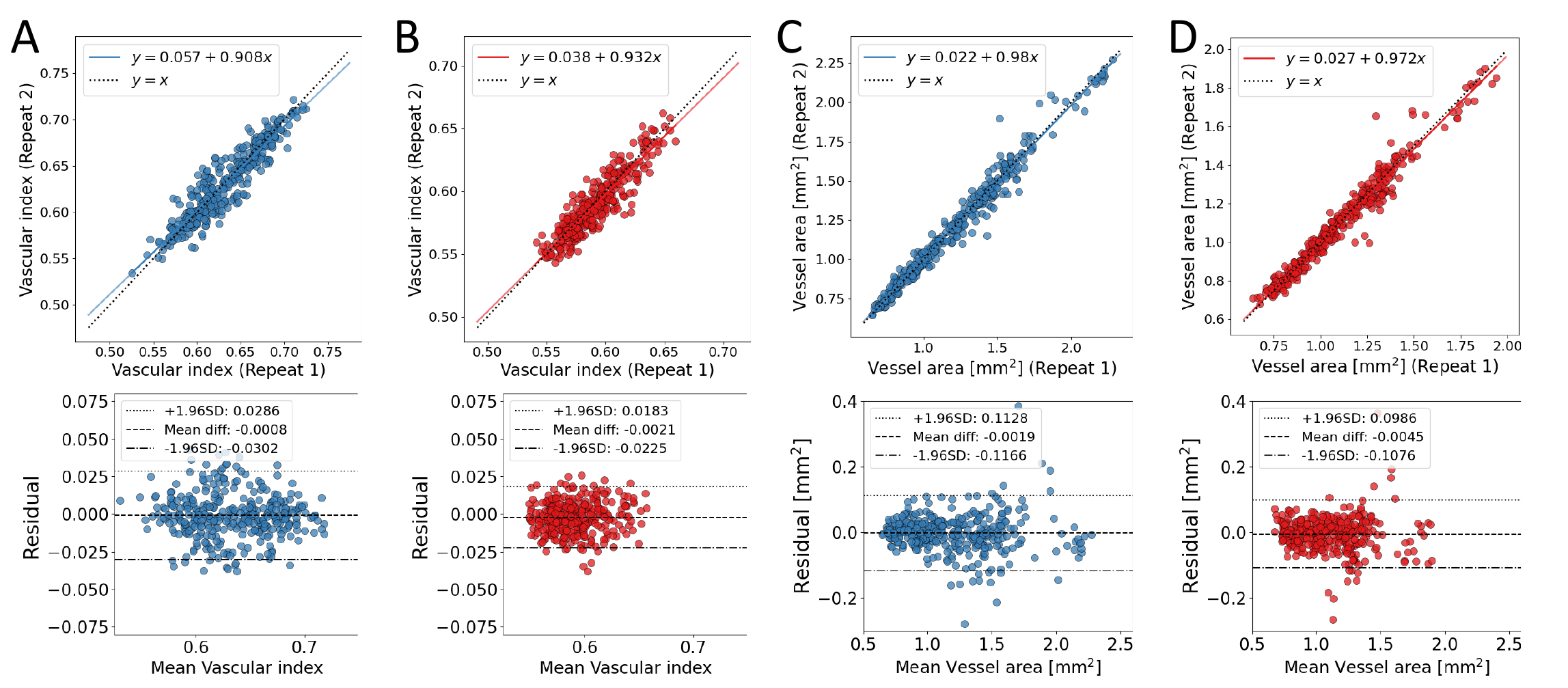}
                \caption[Population-level agreement in repeated \acrshort{SS-OCT} B-scan data for \acrshort{MMCQ} and Niblack.]{Correlation and Bland-Altman plots for assessing the reproducibility of \acrshort{MMCQ} and Niblack using repeated, \acrshort{SS-OCT} B-scan data (\acrshort{GCU} Topcon). (A) \acrshort{MMCQ} \acrshort{CVI}, (B) Niblack \acrshort{CVI}, (C) \acrshort{MMCQ} vessel area, (D) Niblack vessel area.}
                \label{fig:MMCQ_repr_ssoct_corrplots}
            \end{figure}

            Table \ref{tab:MMCQ_repr_results} presents the population-level reproducibility results for \acrshort{MMCQ} and Niblack. Both \acrshort{MMCQ} and Niblack had strong levels of reproducibility for vessel metrics across the macula and at the fovea. For \acrshort{MAE} in the i-Test \acrshort{SD-OCT} sample, \acrshort{MMCQ} had greater reproducibility than Niblack (\acrshort{ETDRS} \acrshort{CVI} for \acrshort{MMCQ} vs. Niblack was 0.0112 vs. 0.0134; \acrshort{ETDRS} vessel volume for \acrshort{MMCQ} vs. Niblack was 0.0251 vs 0.0257 mm$^3$), while the opposite was observed for the \acrshort{GCU} Topcon \acrshort{SS-OCT} dataset (fovea \acrshort{CVI} for \acrshort{MMCQ} vs. Niblack was 0.0113 vs. 0.0083; foveal vessel area for \acrshort{MMCQ} vs. Niblack was 0.0400 vs 0.0363 mm$^2$). This could be because, at the time of \acrshort{MMCQ}'s methodology development, there was only access to \acrshort{SD-OCT} data, which is also why the evaluation in section \ref{sec:MMCQ_eval} only featured \acrshort{SD-OCT} data. 
            
            Encouragingly, both methods had strong linear and monotonic correlations between repeated samples (All Pearson and Spearman correlations were greater than 0.96 and 0.89 for \acrshort{MMCQ} and Niblack, respectively). The \acrshort{MAE} for \acrshort{ETDRS} \acrshort{CVI} of approximately 0.01 corresponds to a 1\% error on average across individual sub-fields of the \acrshort{ETDRS} grid. Additionally, a similarly low \acrshort{CVI} \acrshort{MAE} was found for fovea-centred \acrshort{SS-OCT} B-scans in \acrshort{GCU} Topcon sample, in which we report a 1\% and 0.8\% \acrshort{MAE} for \acrshort{MMCQ} and Niblack, respectively.
            
            To put these errors into context, Breher, et al. \cite{breher2020choroidal} tested the reproducibility of Sonoda's Niblack method \cite{sonoda2014choroidal} across different sub-fields of the \acrshort{ETDRS} macular region and reported a mean difference ranging from 3.9\% to 5.1\%. Additionally, Agrawal's major review \cite{agrawal2020exploring} on \acrshort{CVI} as a biomarker in retinal pathology reported changes between healthy and diseased eyes between 2\% and 6\% \cite{agrawal2020exploring}. Therefore, it is unlikely that our \acrshort{MAE}s of approximately 1\% will be clinically significant. 
            
            Figures \ref{fig:MMCQ_repr_sdoct_corrplots} and \ref{fig:MMCQ_repr_ssoct_corrplots} show the population-level reproducibility for \acrshort{MMCQ} (blue) and Niblack (red) as correlation and Bland-Altman plots for the \acrshort{SD-OCT} i-Test and \acrshort{SS-OCT} \acrshort{GCU} Topcon repeated samples, respectively. Across the board, \acrshort{SD-OCT} vessel volume and \acrshort{SS-OCT} vessel area had strong linear fits using both methods, with residuals centred around 0. The absolute value of residuals for vessel area increase with the magnitude of vessel pixels in the choroid, and this observation is exacerbated for vessel volume which have an obvious negative skew, especially for the Niblack method. This is perhaps unsurprising because one member of the i-Test volume pair was collected with \acrshort{EDI} mode toggled off, which has poorer image quality than the corresponding \acrshort{EDI-OCT} member of the pair. In these figures, we can also observe \acrshort{MMCQ}'s stronger performance on the \acrshort{SD-OCT} dataset (figure \ref{fig:MMCQ_repr_sdoct_corrplots}) and, conversely, Niblack's superior performance on the \acrshort{SS-OCT} dataset (figure \ref{fig:MMCQ_repr_ssoct_corrplots}).
            
            Finally, \acrshort{CVI} had a reasonably strong linear fit for both methods across both datasets, but was weaker than vessel area and volume by virtue of errors in both numerator and denominator having an exacerbated effect on overall \acrshort{CVI} error (as discussed in section \ref{sec:MMCQ_eval_results}). However, their residual distributions appeared to show no apparent trend across both datasets, and were centred around 0, albeit Niblack's residual distributions were more spherical and tighter than MMCQ. Note also that \acrshort{CVI} was significantly less variable than vessel volume (\acrshort{MMCQ} standard deviation for \acrshort{SD-OCT} (\acrshort{ETDRS}) \acrshort{CVI} and vessel volume was 0.0351 and 0.2741 mm$^3$, respectively, and for \acrshort{SS-OCT} (foveal) \acrshort{CVI} and vessel area was 0.0332 and 0.3522 mm$^2$, respectively). 
        
            \end{mysubsubsection}

            \begin{mysubsubsection}[]{Eye-level}
            
                 \begin{figure}[tb]
                    \centering
                    \includegraphics[width=\linewidth]{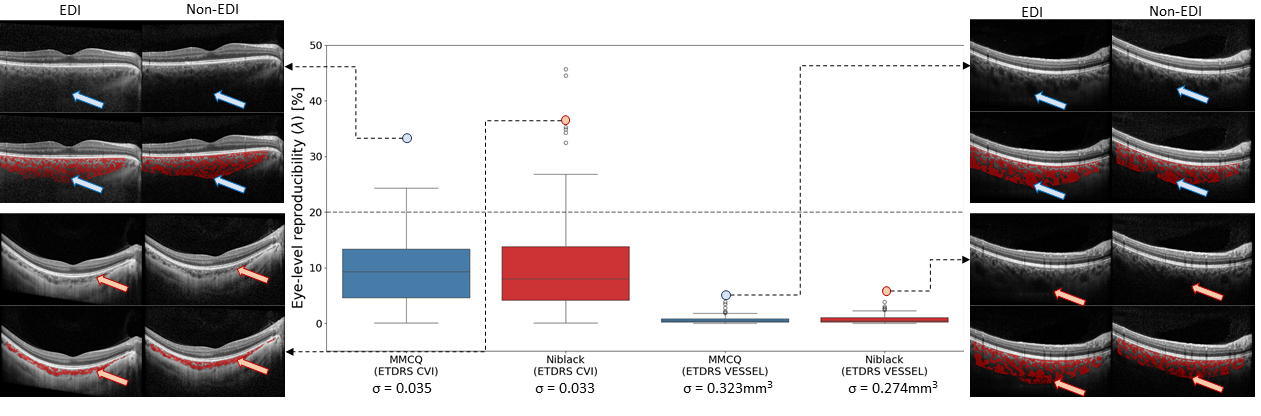}
                    \caption[Eye-level agreement in repeated \acrshort{SD-OCT} volume data for \acrshort{MMCQ} and Niblack.]{Reproducibility for \acrshort{MMCQ} (blue) and Niblack (red) at the eye-level for the \acrshort{SD-OCT} volume data (i-Test). (A) \acrshort{MMCQ} \acrshort{CVI}, (B) Niblack \acrshort{CVI}, (C) \acrshort{MMCQ} vessel volume, (D) Niblack vessel volume.}
                    \label{fig:MMCQ_ind_repr_sdoct}
                \end{figure}
    
                \begin{figure}[tb]
                    \centering
                    \includegraphics[width=\linewidth]{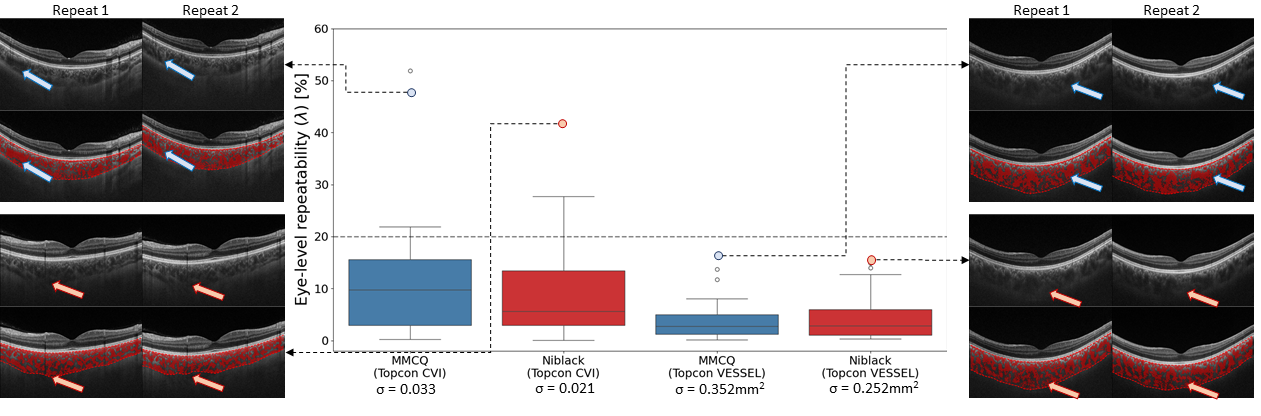}
                    \caption[Eye-level agreement in repeated \acrshort{SS-OCT} B-scan data for \acrshort{MMCQ} and Niblack.]{Eye-level for assessing the reproducibility of \acrshort{MMCQ} and Niblack using repeated, \acrshort{SS-OCT} B-scan data (\acrshort{GCU} Topcon). (A) \acrshort{MMCQ} \acrshort{CVI}, (B) Niblack \acrshort{CVI}, (C) \acrshort{MMCQ} vessel area, (D) Niblack vessel area.}
                    \label{fig:MMCQ_ind_repr_ssoct}
                \end{figure}

                Eye-level reproducibility helps quantify the effect size of your measurement error relative to the overall variability of the population. Effect size is critical in understanding the degree to which measurements correspond to real biological change compared to measurement error. Figures \ref{fig:MMCQ_ind_repr_sdoct} and \ref{fig:MMCQ_ind_repr_ssoct} show the eye-level reproducibility of \acrshort{MMCQ} (blue) and Niblack (red) for the \acrshort{SD-OCT} i-Test and \acrshort{SS-OCT} \acrshort{GCU} Topcon samples, respectively. In each case, we show distributions of eye-level measurement noise $\lambda$ \cite{engelmann2024applicability}. A value of $\lambda = 20\%$ has been overlaid as a black, horizontal dashed line to aid interpretation and readability, with representative B-scans of major outliers shown with their corresponding vessel segmentations overlaid in red, with arrows indicating the source of the error. The between-eye standard deviations are shown below each box-plot as $\sigma$.

                At the eye-level, vessel area and volume showed significantly lower measurement variability across the population compared with \acrshort{CVI} for both \acrshort{MMCQ} and Niblack methods. Specifically, 100\% and 75\% of the eyes in the i-Test and \acrshort{GCU} Topcon samples were below about 7\% of the overall population variability for vessel volume, respectively. Conversely, for \acrshort{CVI}, 75\% of eyes for both datasets were below 15\% of the overall population variability. It's perhaps unsurprising that \acrshort{CVI} is less reproducible than vessel area or volume because it is more prone to larger error from both the numerator (vessel area) and denominator (choroid area). \acrshort{CVI} is dependent on both vessel and region segmentation. Thus, error in \acrshort{CVI} incorporates measurement error from both vessel and total choroid area in the numerator and denominator of its formula, and the size of the choroid can also disproportionately inflate error. 
                
                Moreover, \acrshort{CVI} had significantly smaller population variability than vessel area/volume (also observed by Agrawal, et al. \cite{agrawal2016choroidal}). This smaller variability exacerbates this potential for larger error even further. Additionally, smaller variability implies that effect sizes are likely to be small, which consequently means that reported differences between groups need to be interpreted with care. All of these points indicate that vessel area and volume are more reliable a measurement of the choroidal vasculature than \acrshort{CVI}, at least according to \acrshort{MMCQ} and Niblack.

                Figures \ref{fig:MMCQ_ind_repr_sdoct} and \ref{fig:MMCQ_ind_repr_ssoct} also show the major outliers according to the eye-level reproducibility analysis. For the i-Test sample, unsurprisingly, the greatest outliers from both methods for \acrshort{ETDRS} \acrshort{CVI} and vessel volume came from the same two eyes, which both had large choroids. For \acrshort{ETDRS} \acrshort{CVI}, even the \acrshort{EDI-OCT} scan did not visualise the choroid particularly well (top-left \acrshort{OCT} B-scans in figure \ref{fig:MMCQ_ind_repr_sdoct}), leading to unwanted noise in the output vessel segmentations which translated to high error. For \acrshort{ETDRS} vessel volume (right-hand side \acrshort{OCT} B-scans in figure \ref{fig:MMCQ_ind_repr_sdoct}), under-segmentation from both approaches for the non-\acrshort{EDI} volume scan lead to large error in vessel volume. This highlights a significant limitation behind choroid vessel segmentation which is the pre-requisite segmentation of the choroidal space. Thus, the reproducibility of choroid-derived vessel metrics are dependent on the reproducibility of choroid-derived regional metrics \cite{breher2020choroidal}.

                Interestingly, one of the major outliers for Niblack's \acrshort{ETDRS} \acrshort{CVI} was an eye whose choroid was reasonably small and visualised well, as shown in the bottom-left set of \acrshort{OCT} B-scans in figure \ref{fig:MMCQ_ind_repr_sdoct}. Here we observe major sources of over-segmentation in both sets of scans. This highlights Niblack's inability to segment smaller choroids well, which is likely a consequence of poorly selected window size $w$, which translated to a $\lambda$ value of approximately 36$\%$ (of the population's variability). Conversely, \acrshort{MMCQ} reported only a 3$\%$ $\lambda$ value for this eye, suggesting \acrshort{MMCQ}'s precision for smaller choroids.

                The major outliers for the \acrshort{SS-OCT} \acrshort{GCU} Topcon sample in figure \ref{fig:MMCQ_ind_repr_ssoct} follow suit to the \acrshort{SD-OCT} i-Test sample for foveal \acrshort{CVI}, i.e. the same eye corresponded to the largest eye-level reproducibility error for both methods (bottom-left set of B-scans in figure \ref{fig:MMCQ_ind_repr_ssoct}). For this particular choroid, the region was not segmented correctly for the repeated scan, thus overestimating the \acrshort{CVI} in both methods leading to large measurement error, relative to the population variability (\acrshort{MMCQ} and Niblack $\lambda$ were 52\% and 42\%, respectively). This further highlights the dependence that choroid-derived vessel metrics have on choroid-derived regional metrics \cite{breher2020choroidal}.
                
                Interestingly, the second largest outlier for foveal \acrshort{CVI} for both methods (top-left set of B-cans in figure \ref{fig:MMCQ_ind_repr_ssoct}) came from a choroid with good visualisation of the Choroid-Sclera boundary and vessels. However, upon comparing the appearance of the vessels in the pair it was found that they were not registered properly. Topcon imaging devices do not have the same advanced eye tracking system which Heidelberg Engineering imaging devices do. Thus, there is potential for repeated acquisitions to not be perfectly registered, which can lead to a very slightly different cross-sectional visualisation of the choroidal vasculature. While emerging research suggests that macular \acrshort{CVI} is stable \cite{agrawal2017influence, goud2019new, kim2022influence}, in this analysis this actually produced large measurement error. Although, the measurement error was worse in \acrshort{MMCQ} than it was for Niblack ($\lambda$ value of 47\% and 28$\%$, respectively) suggesting that Niblack may be marginally more generalisable to macular scanning locations than \acrshort{MMCQ} \cite{agrawal2017influence,goud2019new, kim2022influence}.  
                
                For choroid vessel area, the same eye produced the largest measurement error $\lambda$ for both methods (top- and bottom-right side \acrshort{OCT} B-scans in figure \ref{fig:MMCQ_ind_repr_ssoct}), which was 16\% and 15\% for \acrshort{MMCQ} and Niblack, respectively. While the \acrshort{SS-OCT} imaging properties enables sufficient Choroid-Sclera boundary visualisation, this comes at decreased \acrshort{SNR} for deeper ocular structures (larger choroids) \cite{teussink2019state}, which can result in poorly defined choroidal vessels with a lack of contrast and vessel boundary definition. While \acrshort{MMCQ} was able to make a reasonable effort of segmenting whole vessels, Niblack struggled to segment the posterior segment of the choroid, leading to a noisy segmentation (red arrows). Moreover, the representative pair of \acrshort{OCT} B-scans shown in the right-hand side of figure \ref{fig:MMCQ_ind_repr_ssoct} do not appear to be perfectly registered (blue arrows) which also likely played a role in the reported measurement error.
    
            \end{mysubsubsection}
        
        \end{mysubsection}
    
    \end{mysection}

    \begin{mysection}[]{Discussion}

        \begin{mysubsection}[]{Evaluation}

            We showed that the proposed method, \acrshort{MMCQ}, agreed better with the segmentations from two manual graders compared to Agrawal's Niblack method \cite{agrawal2020exploring}. Our method achieved lower \acrshort{MAE}s and stronger correlations for the majority of comparisons with manual labelling on choroid-derived measurements of vessel area and \acrshort{CVI}. \acrshort{MMCQ} also had greater segmentation agreement, albeit with lower sensitivity than Niblack. However, while our comparison suggests \acrshort{MMCQ}'s superiority, these are according to manual vessels labels and the observed differences in these population-level performance metrics are small in absolute terms.  
            
            There does not yet exist a gold standard for choroid vessel segmentation, which makes quantitative evaluation of automatic methods for vessel segmentation a challenge (hence the lack thereof in table \ref{tab:INTRO_vessel_methods}). To attempt evaluation rigour and facilitate segmentation consistency, our evaluation procedure used two expert analysts to conduct segmentation, both following the same pre-specified protocol which ensured a similar setup of screen size, image resolution and brightness, and application-dependent instructions on ITK-Snap \cite{py06nimg}. This protocol ultimately helped facilitate reasonable agreement in segmentation and measurements between graders M1 and M2. 
            
            Relative to the manual segmentations in our evaluation, we showed that Agrawal's Niblack method \cite{agrawal2016choroidal} has an inherent problem of over-segmentation, particularly for small choroids and small-scale vessels. This is an observation which, to our knowledge, is only mentioned very briefly in the original study proposing Agrawal's method for choroid vessel segmentation using Niblack \cite{agrawal2016choroidal}. While a reasonable circumvention would likely be to tune the window size $w$ and standard deviation offset $k$, this becomes unfeasible as datasets grow in size. 

            While \acrshort{MMCQ} was superior to Niblack in this evaluation, the magnitude of agreement in \acrshort{CVI} overall between the semi-automatic methods and the manual segmentations was generally poor and exhibited significant cases of systematic bias. Encouragingly, the semi-automatic methods did exhibit a linear and monotonic relationship with the manual methods for choroid vessel area, but this was not observed for \acrshort{CVI}.

            The poor agreement observed between the \acrshort{MMCQ} and Niblack methods, and their tendency to agree for choroids with more vessel area (or thicker choroids) is an essential observational difference between the two approaches (figure \ref{fig:MMCQ_va_corrplots}(E)). \acrshort{MMCQ} and Niblack represent two common ideologies which are akin to the precision-recall trade-off that exists within machine learning. Niblack prioritises recall over precision, which ensures all vessels are classified, at the cost of miss-classifying some of the interstitial space as vessel. On the other hand, \acrshort{MMCQ} prioritises precision (or positive predictive value), which promotes vessel pixel fidelity, i.e. at the cost of under-segmentation of the choroidal vessels, \acrshort{MMCQ} minimises miss-classifying interstitial pixels as choroid vessel. This was exemplified by \acrshort{MMCQ} scoring higher precision and Niblack scoring higher recall against the manual graders (table \ref{tab:MMCQ_eval_results}). 

            An important consequence of evaluating against manual vessel labels in this chapter is the assumption that the manual vessel labels are accurate and of a gold standard. While there was reasonable agreement between graders, it was by no means clinically insignificant \cite{agrawal2020exploring}. This highlights the potential variation that exists within manual choroid vessel segmentation in \acrshort{OCT} B-scans, given the potential for different definitions, experience and effort that two different human observers may apply to the same \acrshort{OCT} B-scan. While our evaluation shows \acrshort{MMCQ} as a more faithful methodology according to these manual graders, the lack of gold standard for this segmentation task, accompanied by human subjectivity and poor vessel visualisation, contribute to the overall problem of identifying a singularly superior approach to choroid vessel segmentation. 
            
            Agrawal's modified Niblack method \cite{agrawal2016choroidal} (and its older variant \cite{sonoda2014choroidal}) is clearly the most widespread approach for choroid vessel segmentation in \acrshort{OCT} image analysis \cite{agrawal2020exploring}. However, it remains subjective in nature due to its poor agreement \cite{wei2018comparison} and its ambiguity around parameter selection (section \ref{subsec:MMCQ_intro_niblack}). There should be more concern among the research community around this lack of parameter reporting and its subsequent improper use by the research community. To our best knowledge, only two studies report the values for $w$ and $k$ \cite{arian2023automatic, muller2022application} for ground truth label generation of deep learning-based algorithms, with other studies \textit{presumably} resorting to default application from Fiji (ImageJ) or from Agrawal's implementation \cite{agrawal2020exploring}. However, default application fixes $w$ and $k$ which is not optimal across choroids of different sizes as previously observed (figures \ref{fig:MMCQ_niblack_parameter_count}, \ref{fig:MMCQ_niblack_parameter_mae} and \ref{fig:MMCQ_niblack_sameparams}). This has the potential to impact standardisation and reproducibility of choroid-derived vascular metrics in ocular and systemic health.

            Various image quality issues contribute to poor vessel visualisation, including speckle noise, poor eye tracking, difficult patient fixation and poor acquisition from image technicians. There is an inherent assumption made by manual graders, and indeed the semi-automatic methods discussed in this chapter, which is that vessel segmentation in \acrshort{OCT} images is seen as a binary classification problem. However, poor vessel visualisation tends to create an ambiguous choroidal space where smaller vessels, and large vessel boundaries may be misconstrued as interstitial fluid, or vice versa for interstitial space around vessels which are visualised obliquely at image capture. This all carries major challenges to produce an accurate binary classification of the choroidal vessels. Thus, we suggest that future work would make significant progress by providing a probabilistic solution to account for the uncertainty of low contrast vessel boundary regions.
            
        \end{mysubsection}

        \begin{mysubsection}[]{Reproducibility}

            Both \acrshort{MMCQ} and Niblack showed strong levels of reproducibility at the population-level, which was lower than previously reported effect sizes between healthy and diseased eyes (between 2\% and 6\%) \cite{agrawal2020exploring}, and previously reported studies of reproducibility \cite{breher2020choroidal} (3.9\% to 5.1\%). Thus, regardless of their ability to accurately segment the choroidal vasculature, their measurement error is lower than what current effect sizes in disease are. This suggests that these approaches should not significantly diminish the biological signal being measured, assuming they can do so accurately in the first place, as the measured effect should be larger than any measurement variability inherent in the method. 
            
            Moreover, at the eye-level, it's encouraging to see that this reproducibility analysis showed major outliers representing challenging cases of thick choroids with poor Choroid-Sclera boundary and/or vessel visualisation, or where \acrshort{OCT} B-scans did not appear to be sufficiently registered in the case of the \acrshort{GCU} Topcon sample. Heidelberg Engineering are able to leverage the confocal \acrshort{SLO} image to enable real time eye tracking \cite{teussink2019state}, unlike Topcon imaging devices like the DRI Triton Plus, which use \acrshort{CFP} as the corresponding en face localiser during image acquisition. 

            Finally, we concluded at the eye-level that choroid-derived vessel area and vessel volume appeared to be significantly more reliable than \acrshort{CVI}. This was because of the dependence \acrshort{CVI} had on both the reproducibility of region and vessel segmentation which has the propensity to exacerbate measurement error unlike metrics like vessel area and volume. Additionally, the significantly lower population variability in \acrshort{CVI}, whose standard deviation across both datasets was approximately 10\% of the standard deviation reported for vessel area and vessel volume, also played a contributory role in exacerbating measurement variability.

            Thus, it is imperative that results from \acrshort{CVI} measurements are interpreted according to the effect size associated to real biological change, which is dependent on the method used to estimate it \cite{breher2020choroidal}. Fortunately, the reproducibility analysis conducted for \acrshort{MMCQ} and Niblack help provide a interpretable metric for end-users to  apply to their own studies to help differentiate measurement error from true biological change. Ultimately, the potential for higher measurement error in \acrshort{CVI} suggests that predominantly vascular metrics like area and volume are more reliable and should be considered just as important to report in future studies alongside \acrshort{CVI}.
            
        \end{mysubsection}

        \begin{mysubsection}[]{Limitations and future work}

            While we propose \acrshort{MMCQ} as a promising approach for choroidal vessel segmentation, it still faces several limitations that affect its current applicability. The very nature of the choroidal vasculature presents a challenge to segment accurately, given its complex and dense mesh network. This means that \acrshort{OCT} B-scans only provide oblique cross-sectional visualisations of a microvascular bed which is highly heterogeneous in terms of size and shape. This makes it difficult to detect individual vessels and their morphology. As a result, while \acrshort{MMCQ} can measure overall vessel area and density within the choroid, it falls short in offering more localised metrics, such as vessel calibre distributions or 3D rendering of the entire vascular network across the macula.

            Another potential limitation of \acrshort{MMCQ} is its inability to accurately quantify individual vascular layers within the choroid, particularly Sattler's and Haller's layers. However, while previous research has attempted this \cite{esmaeelpour2014choroidal, li2021automated}, there is currently no agreed-upon and standardised protocol in the literature. Sim, et al. \cite{sim2013repeatability} did draft a protocol for multi-layer choroid segmentation, but only in a relatively small sample looking at diabetic eyes. There is also currently no reliable ground truth/gold standard data which exists to differentiate the choroidal vasculature into these individual vascular layers, or distinguish large vessels from medium-sized ones \cite{mrejen2013optical}. It's possible that histological samples of the choroid may help with developing an automatic procedure in the future, but validating any such approach would prove a significant challenge compared with only identifying vasculature across the choroidal space alone.
            
            \acrshort{MMCQ} is a pipeline with multiple stages, which introduces additional complications. For example, the pipeline's performance is highly interdependent, with each stage influencing the subsequent one. Sub-optimal performance during pre-processing or enhancement can negatively impact the subsequent vessel segmentation for example. Moreover, while domain-specific decisions such as scaling configurations (section \ref{subsubsec:MMCQ_method_patch}) and patch-based quantisation level estimation (section \ref{subsubsec:MMCQ_method_quant}) were made with experienced judgement in choroid image analysis, these were not rigorously validated across a diverse source of \acrshort{OCT} choroid data (such as retinal pathology). Thus, it would be pertinent to conduct rigorous testing on each stage of the pipeline to assess its feasibility across different choroids in both health and retinal disease. 
            
            Additionally, the algorithm was primarily designed and tested on \acrshort{SD-OCT} data from Heidelberg Engineering OCT devices due to a lack of access to \acrshort{SS-OCT} data at the time of algorithm development. This likely contributed to the lower reproducibility observed in the \acrshort{GCU} Topcon \acrshort{SS-OCT} sample, relative to the i-Test \acrshort{SD-OCT} sample. This further highlights the need for additional testing and validation of \acrshort{MMCQ} across different \acrshort{OCT} systems to ensure generalisability.
            
            A key assumption in \acrshort{MMCQ} is its focus on improving precision (positive predictive value) in choroidal vessel segmentation by avoiding over-segmentation of connective tissue within the choroid. While this approach is sensible given the ambiguous nature of the choroidal vessel walls appearing on \acrshort{OCT} B-scans, it may result in under-segmentation (lower recall, or sensitivity). This would render vessel wall tissue potentially miss-classified, particularly for larger choroidal vessels.
            
            Another limitation is MMCQ's reliance on prior segmentation of the choroidal space --- \acrshort{MMCQ} requires another approach to quantify the choroidal space before \acrshort{MMCQ} can be applied. Ideally, an end-to-end procedure able to simultaneously segment the choroidal space and vessels would be advantageous for choroidal image analysis in \acrshort{OCT} image sequences. Moreover, \acrshort{MMCQ} also requires an element of human subjectivity by requiring detection of the fovea on \acrshort{OCT} B-scans. This is so that \acrshort{CVI} and vessel area can be reliably measured in a standardised, fovea-centred \acrshort{ROI}. These limitations add technical and clinical complexity, decreasing its end-user accessibility for automatic choroidal vasculature measurement.
            
            Both \acrshort{MMCQ} and Niblack operate under the assumption that the choroidal vasculature can be classified into binary ``vessel''/``not-vessel'' categories in \acrshort{OCT} B-scans. However, this approach is misguided due to the inherent uncertainty and non-binary nature of the choroidal vessels on \acrshort{OCT} B-scans, compounded by their limited resolution and oblique appearance. \acrshort{MMCQ} begins to address this issue by allowing researchers to adjust the number of quantised levels $k$ to assign as choroidal vasculature (out of a total of $K$ clusters). However, a more robust solution would be to characterise the uncertainty of choroid pixels, and particularly vessel wall pixels.
            
            Ultimately, the future direction for \acrshort{MMCQ} lies in developing an approach that not only handles this uncertainty but also provides a fully automatic, end-to-end solution. Such a method would enable reproducible and accurate choroid-derived measurements directly from an \acrshort{OCT} B-scan, through the automatic segmentation of both the choroidal space and its vessels.

        \end{mysubsection}

        \begin{mysubsection}[]{Outputs}

            In this chapter, there were no publication outputs, but this method was used to generate ground truth labels for the method introduced in chapter \ref{chp:chapter-choroidalyzer}, and is briefly described in the publication therein \cite{engelmann2024choroidalyzer}. 
            
            The software associated with this method, \acrshort{MMCQ}, has been published as open-source and is freely available on GitHub \href{https://github.com/jaburke166/mmcq}{here}.
        
        \end{mysubsection}

        \begin{mysubsection}[]{Executive summary}

            In this chapter, we discussed how quantifying choroidal vasculature is likely more informative than the choroidal space alone, allowing measurement of vessel area and \acrshort{CVI}. We found that the current reference standard approach, the Niblack method, has internal parameters which greatly influence the output segmentation and are not reported in the literature, which can lead to poor standardisation and reproducibility. We introduced a domain-specific, semi-automatic approach to choroid vessel segmentation, Multi-scale Median-cut Quantisation (\acrshort{MMCQ}), to replace this current reference standard, and address the heterogeneity and contrast issues of the choroidal vasculature in \acrshort{OCT} B-scans.

            There is currently no gold standard to choroid vessel segmentation, thus making it a challenge to evaluate automatic methods accurately. According to manual ground truth labels from two experienced raters, \acrshort{MMCQ} was more precise but Niblack was more sensitive. However, these differences were small, and inherently biased by comparing to segmentations flawed by human bias. \acrshort{MMCQ} and Niblack together agreed reasonably well in vessel area and volume, but not for \acrshort{CVI}. We found that both methods were reproducible at the population- and eye-level, with errors below reported effect sizes from the literature in retinal disease. However, measurement error in vessel segmentation in both methods was exacerbated more in \acrshort{CVI} than in purely vascular metrics like vessel area and volume. This was due to error exacerbation in the formula for \acrshort{CVI}. Thus, while it remains a popular metric in ocular and systemic health, \acrshort{CVI} should be interpreted with care and reported alongside more reliable vascular metrics like vessel area and volume.

        \end{mysubsection}

    \end{mysection}
	
\end{mychapter}

\begin{mychapter}[]{DeepGPET: Efficient and fully-automatic choroid region segmentation} \label{chp:chapter-deepgpet}

    \begin{mysection}[]{Introduction} 

        Earlier in this thesis we introduced \acrshort{GPET}, a semi-automatic approach to pixel-based edge tracing (edge-based image segmentation). We showed that \acrshort{GPET} was suitable for choroid region segmentation through detection of the \acrshort{RPE}-Choroid and Choroid-Sclera boundaries, and that this method was a significant improvement over manual measurement for downstream measurement of the choroid.

        \begin{figure}[tb]
            \centering
            \includegraphics[width=\linewidth]{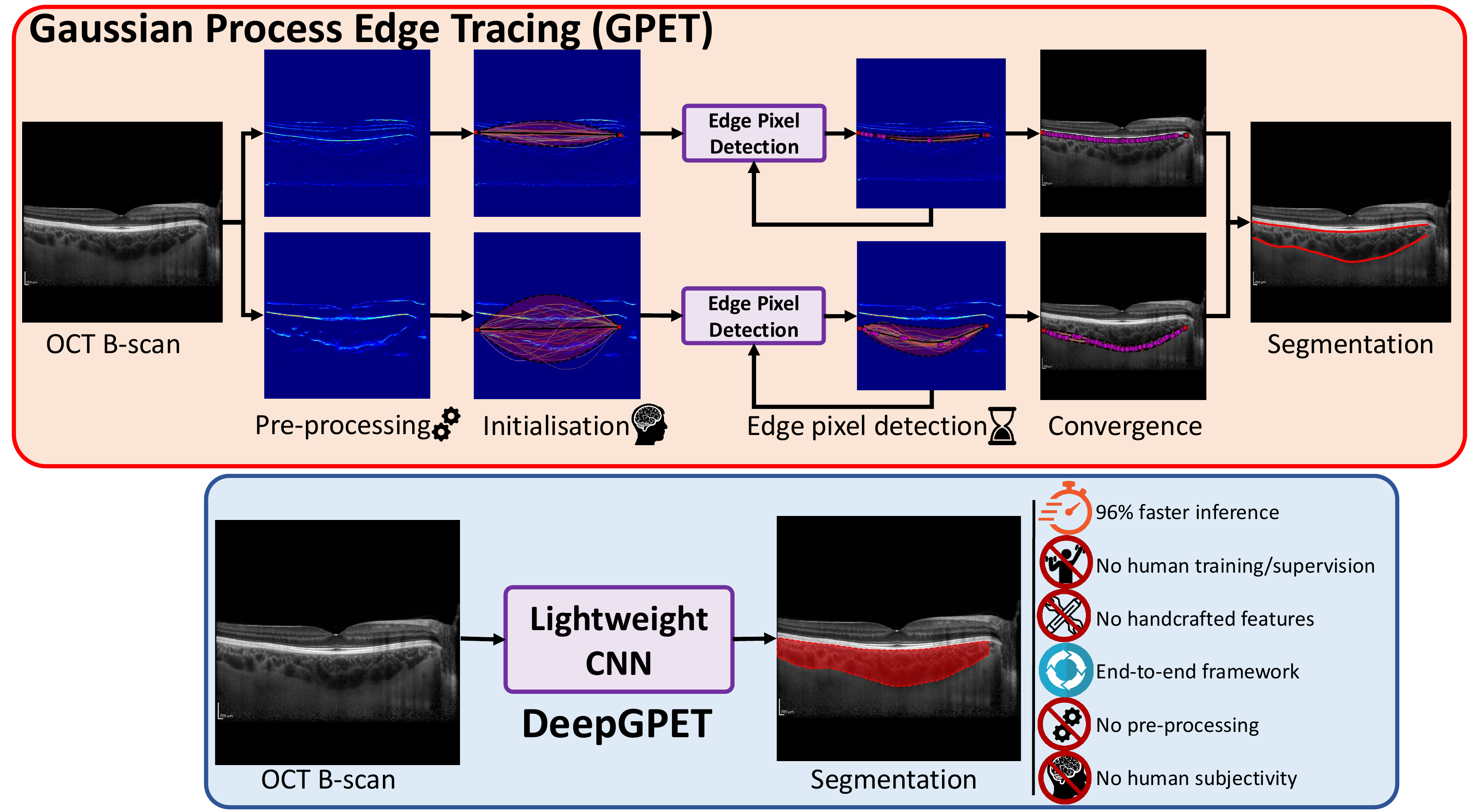}
            \caption[Comparison between \acrshort{GPET}'s complexity and DeepGPET's simplicity.]{Detailed schematic comparison between \acrshort{GPET} and the proposed deep learning alternative, DeepGPET.}
            \label{fig:DEEPGPET_detailed_schematic}
        \end{figure}
        
        However, the semi-automatic nature of \acrshort{GPET} requires time-consuming manual interventions by specifically trained personnel which introduces subjectivity. This limits \acrshort{GPET}'s potential for processing larger datasets or being used in research studies by those without experience in image processing or mathematics. Unfortunately, the training required for this approach extends not only from computer vision and image processing but also a sufficient theoretical background in Gaussian process regression. Additionally, execution time depends on image quality, with successful segmentation requiring manual identification of boundary endpoints for each boundary for a single B-scan, and potentially requires a range of pre-processing techniques to overcome noise to provide a target edge for \acrshort{GPET} to trace in the first instance.
    
        Thus, providing a fully automatic solution to choroid region segmentation in substitute of \acrshort{GPET} which not only has quicker execution time but is end-to-end, i.e. without manual input or pre-processing, would be optimal. A solution based on deep learning methods has several advantages over semi-automatic solutions. Figure \ref{fig:DEEPGPET_detailed_schematic} presents a direct comparison between \acrshort{GPET}'s semi-automatic solution (A) and the proposed deep learning alternative (B). The deep learning alternative removes the multi-stage procedure of semi-automatic \acrshort{GPET}, requiring no pre-processing to prepare images for downstream segmentation, and removes any need for human intervention or initialisation at the point of image analysis. 
        
        A common way to replace an existing method with one based on deep learning is through \textit{distillation}. Deep learning has been used to distil existing semi-automatic traditional image processing pipelines into fully-automatic methods. That is, the semi-automatic method has been used to generate ground truth labels which are used to train and evaluate the corresponding deep learning model. Recently, Engelmann, et al. \cite{engelmann2022robust} used deep learning to emulate a previous semi-automatic approach \cite{perez2011vampire} for retinal fractal dimension approximation, the aim to overcome image quality in \acrshort{CFP}. 
        
        Following a similar suit, we expect the proposed deep learning-based solution to be more robust to poorer image quality than \acrshort{GPET} because of the lack of necessary human input and subjectivity, and the substitution of handcrafted features for automatic, hierarchical feature learning. Additionally, a deep learning alternative will be able to take into account the entire B-scan for training and inference, segmenting the entire choroidal space in a single sweep, rather than segmenting each individual upper and lower boundary. The former only requires identification of a reasonably large area of the B-scan, rather than the latter which requires precise contour tracing and linkage of individual and adjacent pixels in series, thus treating the choroids' anterior and posterior surfaces effectively as independent entities.
        
        Thus, we aimed to develop a fully-automatic method for end-to-end choroid region segmentation that can be easily used without specialised training. We do this through distilling \acrshort{GPET} into a deep learning algorithm, DeepGPET, which can process images without human supervision in a fraction of the time. These attributes permit analysis of large-scale datasets and potential deployment into research practice without prerequisite knowledge in image processing or computer science. Moreover, DeepGPET will be released to address the distinct lack of open-source methods for automatic choroid region segmentation, and in order to replace pre-existing manual measurement of the choroid as the current reference standard. 

        \begin{mysubsection}[]{Deep learning for image segmentation}\label{subsec:DEEPGPET_intro_DL}  
        
            Deep learning has revolutionised the field of image segmentation, enabling more accurate and efficient delineation of objects within images than traditional methods \cite{o2020deep}. Traditional image segmentation methods often rely on hand-crafted features, which require domain expertise and are not always effective across different datasets or image modalities. In contrast, deep learning models can automatically extract relevant features from raw image data without the need for manual intervention. This automation not only reduces the time and effort required for feature engineering but also enables the models to adapt to a wide range of segmentation tasks and domains. Deep learning's hierarchical model architectures (such as \acrshort{CNN}s) can automatically learn hierarchical features (from low-level to high-level) from the input data. This capability allows them to capture complex patterns and details within images, leading to more robust performance compared to traditional imaging methods \cite{o2020deep}.

            \begin{figure}[tb]
                \centering
                \includegraphics[width=\linewidth]{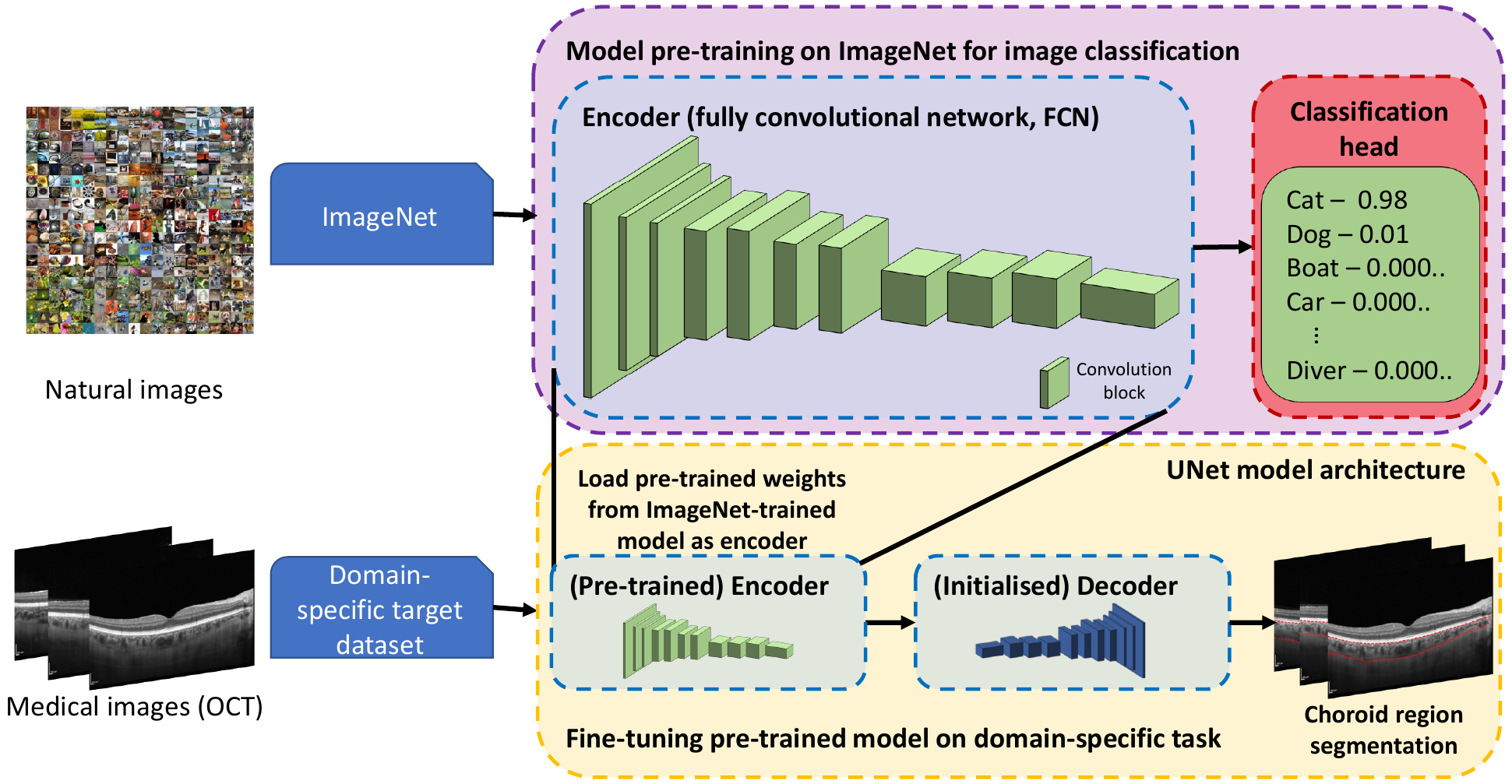}
                \caption[The concept of fine-tuning in deep learning for semantic image segmentation.]{Diagram describing the process of fine-tuning a model for image segmentation using a pre-trained model from a large natural image dataset, such as ImageNet \cite{deng2009imagenet}. ImageNet demonstrative image courtesy of \cite{ss2019imagenet_visualisation}.}
                \label{fig:DEEPGPET_finetuning}
            \end{figure}
            
            Among the numerous models developed for this purpose, the UNet \cite{ronneberger2015u} model stands out as one of the most influential and widely used architectures \cite{williams2023unified}, originally introduced for biomedical image segmentation. The model’s architecture is characterised by its U-shaped structure, comprising a contracting path (known as the encoder arm) to capture context and a symmetric expanding path (known as the decoder arm) that enables precise localisation \cite{williams2023unified}. This design allows UNet to excel in segmenting images with complex structures, even when training data is limited \cite{ronneberger2015u}. A more detailed description of the UNet deep learning architecture can be found in section \ref{subsec:CHOROID_Intro_UNet}.

            In recent years, the practice of fine-tuning pre-trained models has been a standard approach to model training in the field of image segmentation \cite{cheplygina2019not}. Fine-tuning involves taking a model that has been pre-trained on a large dataset and adapting it to a specific task by training it on a smaller, domain-specific dataset. For fine-tuning semantic segmentation models, the encoder arm has typically been pre-trained on some large dataset like ImageNet \cite{deng2009imagenet}, such that the weights already represent low- and high-level features of some natural image dataset, while the weights in the decoder arm are semi-randomly initialised --- care is given to ensure random initialisation does not harm training \cite{narkhede2022review}. Fine-tuning then constitutes training the decoder weights from scratch, while the encoder weights can either be trained simultaneously, or frozen altogether. In the latter, this is so that the learned representations from pre-training are not hindered by fine-tuning on an image domain entirely different to the larger dataset it was previously trained on. Although, the former allows the encoder to be tuned to the domain of interest.
            
            Figure \ref{fig:DEEPGPET_finetuning} shows the typical setup for fine-tuning a deep learning model pre-trained on a large and diverse dataset such as ImageNet \cite{deng2009imagenet}. Note that the convolutional block shown in green is a combination of linear and non-linear mappings which enable hierarchical feature representation learning. This approach is particularly beneficial in scenarios where annotated data is scarce, as it allows leveraging the learned features from the pre-trained model, reducing the need for large amounts of labelled data. Fine-tuning can be used in combination with models like UNet \cite{ronneberger2015u}, as the model's architecture can be pre-trained on generic image datasets and later fine-tuned on domain-specific data, such as medical images for specific segmentation tasks, enhancing both the speed and accuracy of the training process. The combination of UNet's robust architecture and the fine-tuning of pre-trained models offers a powerful strategy for image segmentation, particularly in medical imaging and other domains where data is limited. As deep learning continues to evolve, these methods are likely to remain at the forefront of image segmentation research, driving further advancements in accuracy, efficiency, and applicability across various fields. 
            
        \end{mysubsection}
        
    \end{mysection}

    \begin{mysection}[]{Study populations}

        \begin{mysubsection}[]{Model training \& evaluation} \label{subsec:chp4_DEEPGPET_pop}

            We used 715 \acrshort{OCT} B-scans belonging to 82 subjects from three studies of healthy and diseased individuals (unrelated to ocular pathology):
            \begin{itemize}\setlength\itemsep{0.1em}
                \item \textbf{\acrshort{OCTANE}} \cite{dhaun2014optical}, a longitudinal cohort ($N$=47) study investigating choroidal changes in renal transplant recipients and donors. Note that this is 2 participants and 6 B-scans more than the dataset in \acrshort{GPET}'s evaluation. This was because of additional data found available on the \acrshort{OCT} imaging device during data extraction;
                \item \textbf{i-Test} \cite{dhaun2014optical}, a cohort ($N$=5) of pregnant women evaluating retinochoroidal changes in normative, pre-eclamptic or fetal growth restricted pregnancies;
                \item \textbf{Normative}, a cohort ($N$=30) of healthy volunteers collected as a control group for a study investigating retinal biomarkers in multiple sclerosis \cite{pearson2022multi}.
            \end{itemize}
            All studies conformed with the Declaration of Helsinki and received relevant ethical approval and informed consent from all subjects.
    
            Two Heidelberg Engineering \acrshort{SD-OCT} Spectralis \acrshort{OCT}1 modules were used for image acquisition: the Spectralis Standard and FLEX modules. Horizontal- and vertical-line scans with a 30$^\circ$ (approximately 9 mm laterally) fovea-centred \acrshort{FOV} were captured using active eye tracking with an \acrshort{ART} of 100. Posterior pole macular line scans covered a 30 $\times$ 20 degree rectangular \acrshort{ROI} using 31 consecutive line scans using an \acrshort{ART} of 50 for the i-Test cohort, and 61 consecutive line scans using an \acrshort{ART} of 9 for the Normative and \acrshort{OCTANE} cohorts. 
            
            \acrshort{OCT} B-scans were captured at a pixel resolution of 768 $\times$ 768 pixels using the Standard Spectralis module and 496 $\times$ 768 using the FLEX module which was then padded vertically to a common resolution of 768 $\times$ 768. To reduce processing, images were then cropped to remove unnecessary blank space above and below where the choroid and retina appear on the \acrshort{OCT} B-scan such that input images for training were size 544 $\times$ 768. 

            \begin{mysubsubsection}[]{Sample derivation}
            
                \begin{figure}[!b]
                    \centering
                    \includegraphics[width=\linewidth]{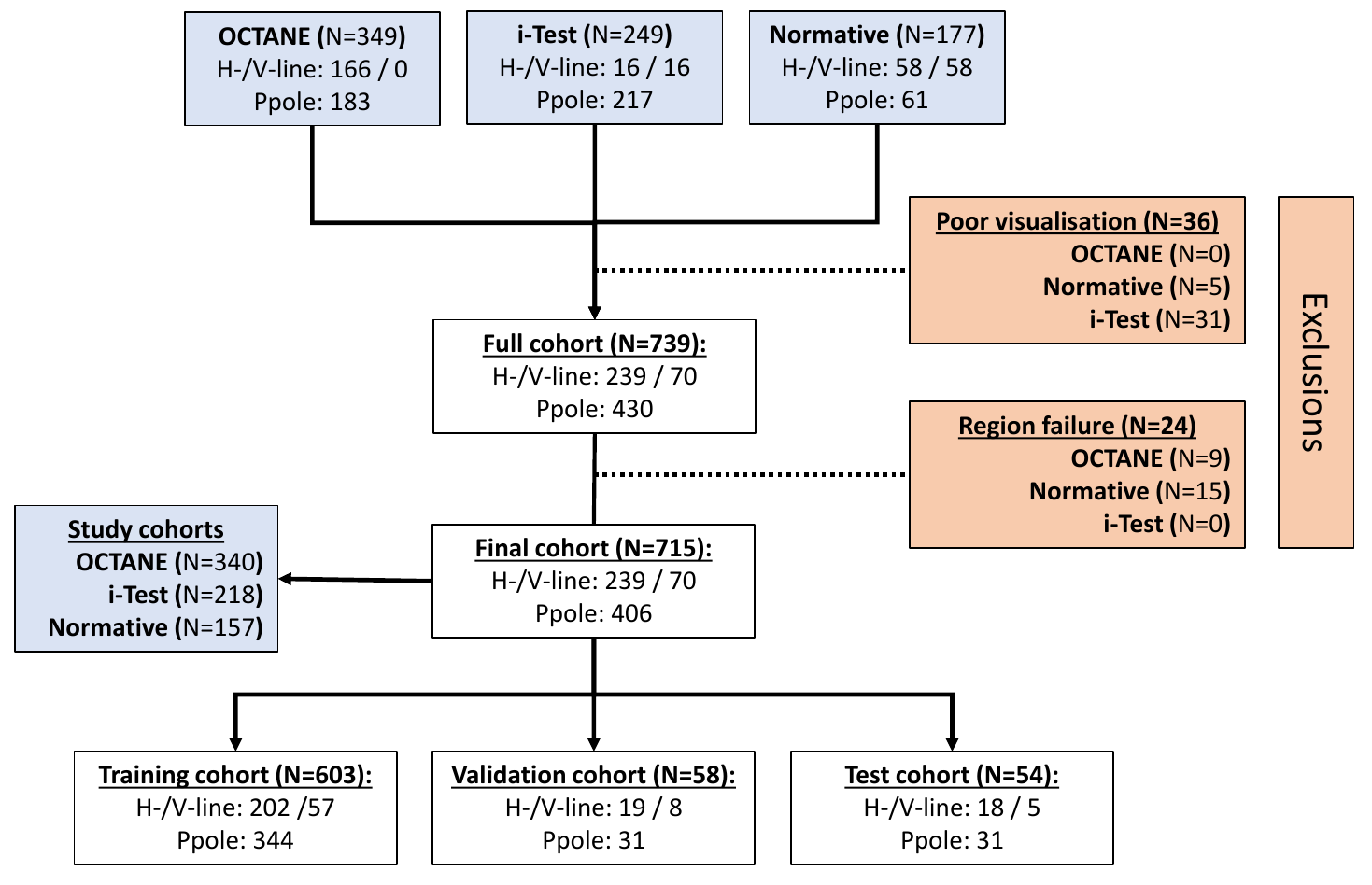}
                    \caption[Sample derivation for DeepGPET's modelling dataset.]{Sample derivation flowchart of how \acrshort{OCT} data from three cohorts were combined to create DeepGPET's modelling dataset.}
                    \label{fig:deepgpet_samplederiv}
                \end{figure}

                \begin{table}[tb]\footnotesize
                    \centering
                    \begin{tabular}{@{}lccc|c@{}}
\toprule
\hspace{3em}                    & \hspace{1em}OCTANE\hspace{1.3em}       & \hspace{1em}i-Test\hspace{1.3em}     &\hspace{1em}Normative\hspace{1.3em}  & Total        \\ \midrule
Subjects            & 47           & 5          & 30         & 82           \\
Male/Female         & 24 / 23     & 0 / 5      & 20 / 10    & 44 / 38     \\
Right/Left eyes     & 47 / 0       & 5 / 5      & 29 / 29    & 81 / 34      \\
Age (mean (SD))     & 48.8 (12.9) & 34.4 (3.4) & 49.1 (7.0) & 48.0 (11.2) \\
Machine             & Spectralis            & FLEX          & Spectralis          &    Both          \\
Horizontal/Vertical scans & 166 / 0      & 16 / 16    & 57 / 54    & 239 / 70     \\
Volume scans           & 174          & 186        & 46         & 406          \\
EDI / non-EDI         & 340 / 0          & 218 / 0        & 57 / 100  & 615 / 100         \\
Total B-scans         & 340          & 218        & 157    & 715          \\ \bottomrule
\end{tabular}%

                    \caption[Cohort demographics of DeepGPET's modelling dataset.]{Overview of population characteristics for DeepGPET's training and evaluation dataset.}
                    \label{tab:DEEPGPET_pop}
                \end{table}
                                                    
                \begin{table}[tb]\footnotesize
                    \centering
                    \begin{tabular}{@{}lccc|c@{}}
\toprule
\hspace{4em}                    & \hspace{1em}Training\hspace{1.3em}       & \hspace{1em}Validation\hspace{1.3em}     &\hspace{1em}Testing\hspace{1.3em}  & Total        \\ \midrule
\multicolumn{1}{l}{Subjects, N} & 66 & 9 & 7 & 82\\
\multicolumn{1}{r}{OCTANE cohort} & 38 & 5 & 4 & 47 \\
\multicolumn{1}{r}{i-Test cohort} & 3 & 1 & 1 & 5 \\
\multicolumn{1}{r}{Normative cohort} & 25 & 3 & 2 & 30 \\
\multicolumn{1}{r}{Male/Female} & 32 / 34 & 7 / 2 & 5 / 2& 44 / 38     \\
\multicolumn{1}{r}{Age (mean (SD))}   & 49.2 (11.4) & 40.2 (8.2) & 44.9 (8.2) & 48.0 (11.2) \\
\multicolumn{1}{r}{Right/Left eyes}     & 66 / 27       & 9 / 4      & 6 / 3    & 81 / 34      \\
\multicolumn{1}{r}{Spectralis/FLEX Device} & 63 / 3 & 8 / 1 & 6 / 1 & 77 / 5 \\
\multicolumn{1}{r}{EDI/non-EDI} & 51 / 15 & 6 / 3 & 5 / 2 & 62 / 20 \\
B-scans & & & & \\
\multicolumn{1}{r}{Horizontal/Vertical} & 202 / 57      & 19 / 8    & 18 / 5    & 239 / 70     \\
\multicolumn{1}{r}{Volume}           & 344          & 31        & 31         & 406          \\
\multicolumn{1}{r}{Total}         & 603          & 58        & 54        & 715          \\ 
\bottomrule
\end{tabular}%

                    \caption[Demographics of the train, validation and test split for DeepGPET's modelling dataset.]{Overview of population and image characteristics of the training, validation and test sets for DeepGPET's model training and evaluation.}
                    \label{tab:DEEPGPET_tvt_demo_tab}
                \end{table}

                Figure \ref{fig:deepgpet_samplederiv} shows a sample derivation flowchart of how DeepGPET's modelling dataset was constructed. At the time of DeepGPET's development, there were a total of $N$=775 \acrshort{OCT} B-scans available from 82 participants. Upon visual inspection, $N$=36 B-scans were removed from the dataset due to poor quality image acquisition, which resulted in an invisible Choroid-Sclera boundary. Of the remaining $N$=739 \acrshort{OCT} B-scans, ground-truth labelling for the choroidal space was provided by \acrshort{GPET} (chapter \ref{chp:chapter-GPET}), after manual selection of each boundary's edge endpoints using a custom-built graphical user interface in Python (version 3.11). Upon manual adjudication of these ground-truth labels, $N$=24 B-scans constituting posterior pole volume B-scans, were excluded. This was because \acrshort{GPET} failed to segment the Choroid-Sclera boundary in regions of low \acrshort{SNR}, as a result of low-\acrshort{ART} volume B-scans. 

                Table \ref{tab:DEEPGPET_pop} provides an overview of basic population characteristics and number of subjects/images of these studies. Notably, 14\% of the \acrshort{OCT} B-scans were non-\acrshort{EDI} and thus presented more challenging cases with lower \acrshort{SNR}. 
                
                We split the data into approximately an 85:8:7 ratio split between training (603 B-scans, 66 subjects), validation (58 B-scans, 9 subjects) and test sets (54 B-scans, 7 subjects). To avoid within-patient data leakage between sets, when splitting the data we did so at the patient-level, i.e. each subject's \acrshort{OCT} images were present in only one set. Moreover, the sets were selected so that each set had proportionally equal amounts of scan types (\acrshort{EDI}/non-\acrshort{EDI}) to best represent image quality. Table \ref{tab:DEEPGPET_tvt_demo_tab} provides for an overview of basic population and imaging characteristics for each set.
                
            \end{mysubsubsection}

        \end{mysubsection}

        \begin{mysubsection}[]{Reproducibility}\label{subsec:ch4_DEEPGPET_repr_pop}
        
            We also assess the reproducibility of DeepGPET on downstream choroidal measurements on two types of \acrshort{OCT} data: macular volume scans and peripapillary B-scans. Two cohorts were used in this analysis, i-Test\footnote{Note that this cohort was collected after developing DeepGPET, hence the increase in sample size from table \ref{tab:DEEPGPET_pop}.}\cite{dhaun2014optical} and Diurnal Variation for Chronic Kidney Disease (\acrshort{DVCKD}) \cite{dhaun2014optical}. We used all available eyes which had repeated data available (at the point of analysis) from each cohort (120 eyes from 60 participants from the i-Test cohort and 22 eyes from 22 participants in the \acrshort{DVCKD} cohort). Table \ref{tab:DEEPGPET_repr_pop} describes their population and imaging characteristics. Note that the i-Test sample of repeated data is the same dataset used to assess the reproducibility of \acrshort{MMCQ} and Niblack in chapter \ref{chp:chapter-mmcq}, which was described in section \ref{subsec:MMCQ_repr_pops}.

            \begin{table}[tb]\footnotesize
                \begin{adjustwidth}{-1in}{-1in}  
                \centering
                \begin{tabular}{p{4cm}p{5cm}p{5cm}}
\toprule
\multirow{2}{*}{} & \multicolumn{2}{c}{Study} \\ \cmidrule(l){2-3}
 & i-Test \cite{dhaun2014optical} & \acrshort{DVCKD} \cite{dhaun2014optical, farrah2023choroidal} \\ 
 \midrule
Cohort demographics &  &  \\ \cmidrule(l){1-1}
Participants (Eyes) & 60 (120) & 22 (22) \\
Right eyes (\%) & 60 (50) & 22 (100) \\
Age (\acrshort{SD}) & 34.7 (5.2) & 21.3 (2.2) \\
Sex, F (\%) & 60 (100) & 10 (45.5) \\
Ethnicity & 52 White, 6 Asian, 2 Mixed & Unknown \\
Gestation & 35.6 (3.4) & NA \\
Study purpose & Pre-eclamptic / normative pregnancy at late gestation & Diurnal variation \\
Control/Case & 45/15 & 22/0 \\
 &  &  \\
Image characteristics &  &  \\ \cmidrule(l){1-1}
Device & Spectralis (Heidelberg) & Spectralis (Heidelberg) \\
\acrshort{OCT} Type & Spectral-domain & Spectral-domain \\
Scan Pattern & Macular volume & Peripapillary \\
Mode & \acrshort{HRA}+\acrshort{OCT} & \acrshort{HRA}+\acrshort{OCT} \\
Time of day (Interval) & All in afternoon (1 minute) & Each at 9am, 12:30pm, 4pm ($\pm$ 40 minutes) \\
B-scans per eye & 31 (\acrshort{EDI}) / 61 (non-\acrshort{EDI}) & 1 \\
\acrshort{ART} & 50 (\acrshort{EDI}) / 12 (non-\acrshort{EDI}) & 100 \\
Pixel resolution & 496 $\times$ 768 & 496 $\times$ 1536 \\ 
\bottomrule
\end{tabular}

                \end{adjustwidth}
                \caption[Cohort demographics for DeepGPET's reproducibility analysis.]{Population demographics and image characteristics of the reproducibility cohorts to evaluate DeepGPET's reproducibility.}
                \label{tab:DEEPGPET_repr_pop}
            \end{table}

            In brief, the i-Test sample collected paired \acrshort{EDI} and non-\acrshort{EDI} macular \acrshort{SD-OCT} volume scans (acquired within 1 minute of one another), both covering a 30 $\times$ 20 degree (approximately 9 $\times$ 6.6 mm) \acrshort{FOV}. \acrshort{EDI} volume scans contained 31 line-scans with an \acrshort{ART} of 50 while non-\acrshort{EDI} scans contained 61 line-scans with an \acrshort{ART} of 9. Line-scans had a pixel resolution of 496 $\times$ 768.
    
            \acrshort{DVCKD} participants were recruited to assess diurnal variation of the choroid in healthy volunteers to support research in \acrshort{CKD} \cite{farrah2023choroidal}. During acquisition, the \acrshort{SD-OCT} Heidelberg Spectralis Standard Module was used to collect horizontal-line \acrshort{OCT} B-scans, \acrshort{OCT} volume scans and \acrshort{OCT} peripapillary scans of the right eye only. To assess DeepGPET's generalisability to other locations in the posterior pole, and to minimise the affect of diurnal variation on this reproducibility analysis, the macular \acrshort{OCT} scans were not selected for analysis, and only the peripapillary B-scans were. Peripapillary B-scans collected for the \acrshort{DVCKD} study were circular cross-sections of the retina and choroid and centred on the optic disc. B-scans were collected with \acrshort{EDI} mode activated using an \acrshort{ART} of 100 and had a pixel resolution of 496 $\times$ 1536. The study collected scans at three time points during daylight, in the morning (09:12 $\pm$ 12 minutes), early afternoon (12:36 $\pm$ 7 minutes) and early evening (16:08 $\pm$ 8 minutes).

        \end{mysubsection}

        \begin{mysection}[]{DeepGPET's deep learning model}

            DeepGPET used a UNet deep learning architecture \cite{ronneberger2015u} and we employed fine-tuning. DeepGPET's encoder arm is a \acrshort{CNN} architecture called MobileNetV3 \cite{howard2019searching} whose model weights were pre-trained using the large bank of natural images in ImageNet \cite{deng2009imagenet}. DeepGPET's decoder arm was symmetrical in terms of blocks and spatial resolution to the encoder arm, with default random initialisation of model weights \cite{Iakubovskii2019}. The MobileNet \acrshort{CNN} architectures were designed to be efficient, scalable and lightweight in terms of parameter (model weights) size and inference time \cite{howard2019searching}, with the overall aim of these architectures being used for mobile applications. 
            
            DeepGPET was fine-tuned by retraining the encoder weights, as well as the decoder weights from scratch, for 60 epochs --- an epoch representing one entire pass through the training dataset --- using a batch size of 16. Model training leveraged the AdamW optimiser (using standard parameters $\beta_1$=0.9, $\beta_2$=0.999 and weight decay 10$^{-2}$) \cite{loshchilov2017decoupled}, and the learning rate was 10$^{-3}$. The Adam optimiser \cite{kingma2014adam} combines older concepts around momentum \cite{qian1999momentum} and parameter-specific adapting learning rates \cite{duchi2011adaptive} to improve convergence rates \cite{ruder2016overview}, and AdamW extends this further by using additional weight decay as implicit regularisation to help prevent overfitting \cite{loshchilov2017decoupled}.
            
            Exponential moving average (EMA) is a simple yet effective approach to overcome occasionally poor quality ground truth labels, prevent noisy predictions during training and enhance transfer learning \cite{morales2024exponential}. EMA of the model weights simply computes a moving average for each model parameter across epochs, placing higher emphasis to more recent epochs --- this is in contrast to the linear moving average applied to retinal vessel shadow compensation in section \ref{subsec:MMCQ_method_vesselshadow}. Thus, to enhance model performance and help overcome any poorly labelled ground-truths, we utilised EMA across model parameters which started after epoch 30. 
            
            We employ common data augmentation techniques such as brightness and contrast alterations ($p=0.5$), Gaussian blur ($p=0.25$) and scale, rotation and translation transformations ($p=0.25$). More realistic augmentations include horizontal flipping ($p=0.5$) to synthetically produce a right eye scan from a left eye scan and vice versa, and applying Gaussian noise followed by multiplicative noise ($p=0.5$) to corrupt images with custom speckle noise to simulate the natural noise degradation seen in \acrshort{OCT} image data. Data augmentation is very useful for artificially obtaining a larger, and more diverse dataset. Data augmentation can be done off-line in advance, but is better applied during training where randomly augmented samples are generated anew within an epoch, and therefore are different per epoch. While this approach to generating `new' data is only artificial augmentations from pre-existing data, empirically this does improve the model's robustness and generalisation to augmented instances not readily available in the original dataset \cite{shorten2019survey}. 
            
            DeepGPET was developed, trained and evaluated using Python (3.11), PyTorch (2.0) and Segmentation Models PyTorch \cite{Iakubovskii2019}.

        \end{mysection}
        
    \end{mysection}

    \begin{mysection}[]{DeepGPET's evaluation} \label{subsec:ch4_DEEPGPET_eval}

        \begin{mysubsection}[]{Statistical analysis}

            For evaluation, a threshold of $p$=0.5 was used to convert the model's probabilistic prediction map to a binary segmentation mask. 

            We used the Dice similarity score and Area Under the Receiver Operating Characteristic Curve (\acrshort{AUC}) for evaluating segmentation agreement, as well as the Pearson correlation $r$ and mean absolute error (\acrshort{MAE}) for segmentation-derived choroid thickness and area. These metrics have been defined previously in this thesis in section \ref{subsec:INTRO_metrics}. We measure choroid thickness and area following the definitions outlined in section \ref{subsec:ch1_INTRO_measure_bsan}, take a representative thickness value as the average of three locations; subfoveal and 2000 microns temporal and nasal to the fovea, while area is measured in a 6 mm fovea-centred \acrshort{ROI}.

            To aid comparison of DeepGPET and GPET, we compare DeepGPET's agreement with \acrshort{GPET}'s ground-truth segmentations, and this is further compared against the \textit{repeatability} of \acrshort{GPET}. \acrshort{GPET} was used to process all 54 test images 2 months after it was used to generate the initial set of segmentations used for DeepGPET's model evaluation. Both times, the same experienced analyst who developed the \acrshort{GPET} method (author J.B.) segmented the choroidal space, which acts as an upper bound of performance of \acrshort{GPET} which we compare DeepGPET with. 
            
            We define \textit{repeatability} here as the ability to reproduce segmentations (and choroid-derived measurements) on the same input, i.e. when the same OCT B-scan is processed by GPET twice. Conversely, \textit{reproducibility} --- used elsewhere in this thesis --- is defined as the ability to reproduce segmentations (and choroid-derived measurements) from different inputs of the same object, i.e. when two OCT B-scans are taken 5 minutes apart for the same location of the same eye. Thus, the expected output should be the same between repeatability and reproducibility, but the inputs are different.
            
            We also measured the execution time of both approaches on the test set and reported mean and standard deviation values in seconds. The execution time for \acrshort{GPET} was measured on the re-segmentations, with the author timing himself from when each image was first loaded up on the screen, to when a successful choroid segmentation was obtained after application of the \acrshort{GPET} image analysis pipeline (figure \ref{fig:GPET_schematic_diag}). Experiments were run on a standard, Windows laptop with a 3.5 year old Intel Core i5 (8$^{\text{th}}$ generation) CPU with 16Gb of RAM, but no graphics processing unit (\acrshort{GPU}). 

            In addition to quantitative evaluations, we also compared segmentations by \acrshort{GPET} and DeepGPET for 20 test set \acrshort{OCT} B-scans qualitatively by having them rated by an experienced clinical ophthalmologist (supervisor Dr. Ian J.C. MacCormick, I.M.). We selected 7 examples with the highest disagreement in thickness and area, 7 examples with disagreement closest to the median, and 6 examples with the lowest disagreement. Thus, these 20 examples cover cases where both methods are very different, cases of typical disagreement, and cases where both methods are very similar. In each instance, I.M. was shown the segmentations of both methods overlaid on the \acrshort{OCT} --- blinded to which method produced which segmentation --- and also provided with the raw, full-resolution B-scan. I.M. was then asked to rate each one along three dimensions: Quality of the upper boundary, quality of the lower boundary and overall smoothness using an ordinal scale: ``Very bad'', ``Bad'', ``Okay'', ``Good'', ``Very good''. Quality was defined by how well the detected boundary traced along each true boundary, given the definitions set out in appendix \ref{apdx:definitions} using equipment specified in appendix \ref{apdx:equipment_protocol}. Smoothness was defined by a lack of bumps or jagged edges in the predicted boundary segmentations for the \acrshort{RPE}-Choroid and Choroid-Sclera boundaries.
    
        \end{mysubsection}
    
        \begin{mysubsection}[]{Results}

            \begin{mysubsubsection}[]{Quantitative evaluation}
    
               Table \ref{tab:DEEPGPET_results} shows the results for DeepGPET and a repeat \acrshort{GPET}, compared to the initial \acrshort{GPET} segmentation as ground truth.
    
               Both methods have excellent agreement with the original segmentations. DeepGPET's agreement is comparable to the repeatability of \acrshort{GPET} itself, with DeepGPET's \acrshort{AUC} being slightly higher (0.9994 vs 0.9812) and Dice coefficient slightly lower (0.9664 vs 0.9672). DeepGPET performing better in terms of \acrshort{AUC} but worse in terms of Dice suggests that there is higher uncertainty (lower confidence) for pixels where it disagrees with \acrshort{GPET} after thresholding, compared with pixels where it agrees with \acrshort{GPET}. This in turn suggests that the raw (probabilistic) predictions made by DeepGPET are well-calibrated to the original \acrshort{GPET} segmentations for each pixel. Thus, the raw DeepGPET probabilities provide a reasonable measure of uncertainty of the choroidal space.
    
               \begin{table}[tb]\footnotesize
               \begin{adjustwidth}{-1in}{-1in}  
                \centering
                \begin{tabular}{@{}llllllll@{}}
\toprule
\multicolumn{1}{c}{\multirow{2}{*}{Method}} &
  \multicolumn{1}{c}{\multirow{2}{*}{\acrshort{AUC}}} &
  \multicolumn{1}{c}{\multirow{2}{*}{Dice}} &
  \multicolumn{1}{c}{\multirow{2}{*}{\begin{tabular}[c]{@{}c@{}}Time \\ (s/img)\end{tabular}}} &
  \multicolumn{2}{c}{Thickness} &
  \multicolumn{2}{c}{Area} \\ \cmidrule(l){5-6}\cmidrule(l){7-8} 
\multicolumn{1}{c}{} &
  \multicolumn{1}{c}{} &
  \multicolumn{1}{c}{} &
  \multicolumn{1}{c}{} &
  \multicolumn{1}{c}{Pearson $r$} &
  \multicolumn{1}{c}{\acrshort{MAE} [$\mu$m]} &
  \multicolumn{1}{c}{Pearson $r$} &
  \multicolumn{1}{c}{\acrshort{MAE} [mm$^2$]} \\ \midrule
DeepGPET &\hspace{0.5em} 0.9994\hspace{0.5em} & \hspace{0.5em}0.9664\hspace{0.5em} & \hspace{0.85em}1.25 $\pm$ 0.10\hspace{0.5em}   & 0.8908 & 13.3086 & 0.9082 & 0.0699 \\ \midrule
Repeat \acrshort{GPET}    & \hspace{0.8em}0.9812\hspace{0.5em} & \hspace{0.5em}0.9672\hspace{0.5em} & \hspace{0.5em}34.49 $\pm$ 15.09\hspace{0.5em} & 0.9527 & 10.4074 & 0.9726 & 0.0486 \\ \bottomrule
\end{tabular}%

                \end{adjustwidth}
                \caption[DeepGPET's modelling performance on \acrshort{GPET}'s ground truth labels.]{Metrics for DeepGPET and repeated \acrshort{GPET} using the initial \acrshort{GPET} annotation as ground truth. Time given as mean $\pm$ standard deviation.}
                \label{tab:DEEPGPET_results}
                \end{table}
    
                DeepGPET processed the images in only 3.6\% of the time that \acrshort{GPET} needed. DeepGPET ran fully-automatic and successfully segmented all images in 1.25 seconds on average, whereas \acrshort{GPET} took 34.50 seconds and required 1.27 manual interventions on average. Manual interventions were defined where the initial segmentation failed because of poor pre-processing or initialisation of \acrshort{GPET} parameters (covariance function hyperparameters). Moreover, each additional manual intervention per B-scan corresponded to re-selecting the initial edge endpoints as well as intermediary pixels between the edge endpoint pixels to help guide tracing for the Choroid-Sclera boundary. In contrast, DeepGPET ran end-to-end without the need for pre-processing or manual input.
                
                \begin{figure}[tb]
                    \centering
                    \includegraphics[width=\linewidth]{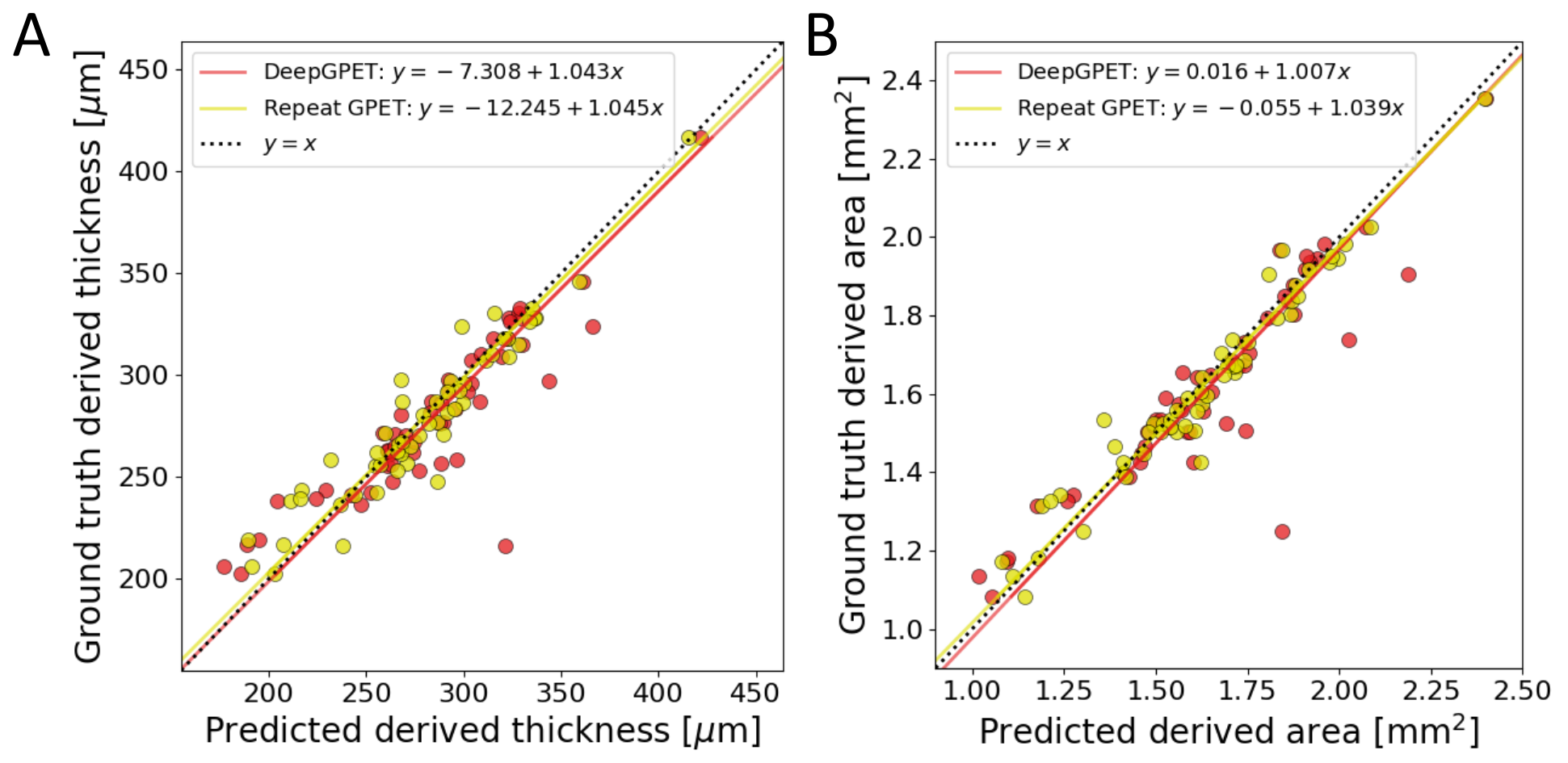}
                    \caption[DeepGPET's agreement with \acrshort{GPET}'s ground truth labels.]{Correlation plots comparing derived measures of mean choroid thickness (A) and choroid area (B) using DeepGPET and the re-segmentations using \acrshort{GPET}.}
                    \label{fig:DEEPGPET_corr_plot}
                \end{figure}
    
                \begin{figure}[!t]
                    \centering
                    \includegraphics[width=0.8\linewidth]{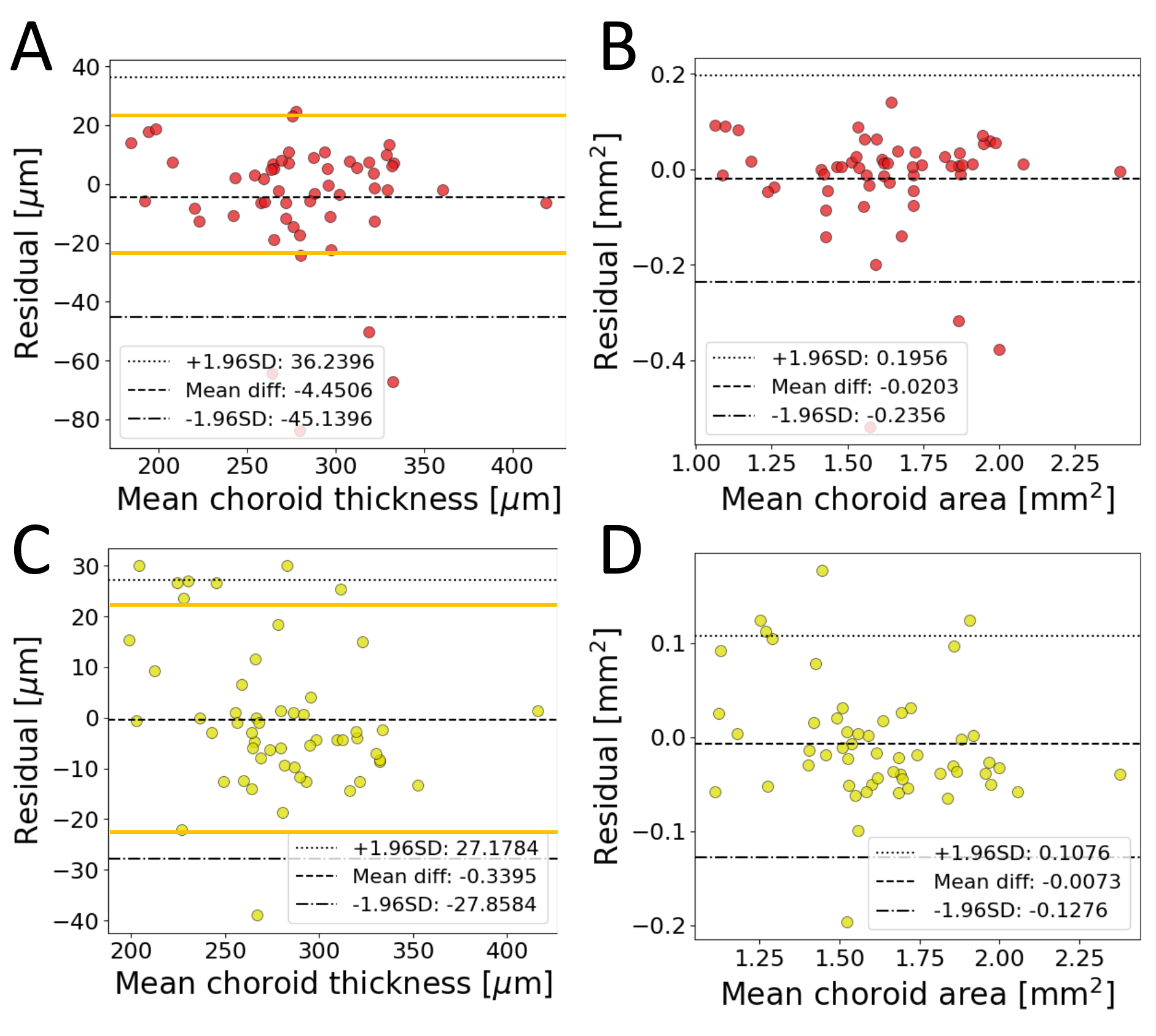}
                    \caption[DeepGPET's error distribution in thickness and area  against \acrshort{GPET}'s ground truth labels.]{Bland-altman plots comparing agreement in thickness and area for DeepGPET (A -- B) and \acrshort{GPET}'s repeatability (C -- D). Orange lines in panels (A, C) at $\pm$ 23 $\mu$m represent intra-rater repeatability threshold \cite{rahman2011repeatability}.}
                    \label{fig:DEEPGPET_ba_plot}
                \end{figure}
    
                \acrshort{GPET} showed very high repeatability for thickness (Pearson $r$=0.9527, \acrshort{MAE} = 10.4074 $\mu$m) and area (Pearson $r$=0.9726, \acrshort{MAE} = 0.0486 mm$^2$). DeepGPET achieved slightly lower, yet also very high agreement for both thickness (Pearson $r$=0.8908, \acrshort{MAE} = 13.3086 $\mu$m) and area (Pearson $r$=0.9082, \acrshort{MAE} = 0.0699 mm$^2$). Figure \ref{fig:DEEPGPET_corr_plot} shows correlation plots for thickness and area. DeepGPET's agreement with \acrshort{GPET} does not quite reach the repeatability of \acrshort{GPET} itself, when used by the same experienced analyst, but it is quite comparable and high in absolute terms. 
    
                Furthermore, the regression fits in both derived measures for DeepGPET are closer to the identity line than for the repeated \acrshort{GPET} measurements. For choroid thickness, the linear fit estimated a slope value of 1.043 (95\% confidence interval of 0.895 to 1.192) and intercept of -7.308 $\mu$m (95\% confidence interval of -48.967 $\mu$m to 34.350 $\mu$m). For choroid area, the linear fit estimated a slope value of 1.01 (95\% confidence interval of 0.878 to 1.137) and an intercept of 0.016 mm$^2$ (95\% confidence interval of -0.195 mm$^2$ to 0.226 mm$^2$). All confidence intervals contained 1 and 0 for the slope and intercepts, respectively, suggesting no systematic bias or proportional difference between \acrshort{GPET} and DeepGPET \cite{passing1983new, ranganathan2017common}.
                
                Figure \ref{fig:DEEPGPET_ba_plot} shows the residuals between DeepGPET and the re-segmentations of \acrshort{GPET} against the original \acrshort{GPET} ground truth labels from the held-out test set using Bland-Altman plots \cite{bland1986statistical}. Rahman, et al. \cite{rahman2011repeatability} found that intra-rater agreement and inter-rater agreement of manual subfoveal choroidal thickness measurements were 23 $\mu$m and 32 $\mu$m, respectively. For DeepGPET's residual error in choroid thickness, only 9.3\% (5 / 54) were greater than 23$\mu$m in absolute value, while 15\% (8 / 54) of the \acrshort{GPET} re-segmentations exceeded this threshold (figure \ref{fig:DEEPGPET_ba_plot}). For choroid area, the majority of residuals were centred around 0 (mean residual for DeepGPET: -0.02mm$^2$, \acrshort{GPET} re-segmentations: -0.0073 mm$^2$), however, \acrshort{GPET} had tighter distribution of residuals, and were better centred around 0 for both thickness and area (95\% confidence intervals for thickness: DeepGPET, (-45.14, 36.24) $\mu$m; \acrshort{GPET}, (-27.17, 27.86) $\mu$m; and area: DeepGPET, (-0.24, 0.20) mm$^2$; \acrshort{GPET}, (-0.13, 0.11) mm$^2$).

            \end{mysubsubsection}

            \begin{mysubsubsection}[]{Qualitative evaluation}

                Table \ref{tab:DEEPGPET_qualitative} shows the results of the manual adjudication between different cases comparing the segmentations outputted from \acrshort{GPET} and DeepGPET. The upper boundary was rated as ``Very good'' for both methods in all 20 cases. However, for the lower boundary, DeepGPET was rated as ``Bad'' in 2 cases for the lower boundary and 1 case for smoothness. Otherwise, both methods performed very similarly. 
    
                \begin{table}[tb]\footnotesize
                \begin{adjustwidth}{-1.5in}{-1.5in}  
                \centering
                \begin{tabular}{@{}lccc@{}}
\toprule
Method   & Upper boundary & Lower boundary & Smoothness\\ 
\midrule
DeepGPET & \hspace{0.5em}Very good: 20  \hspace{0.5em}& \hspace{0.5em}Very good: 4, Good: 10, Okay: 4, Bad: 2\hspace{0.5em} & \hspace{0.5em}Very good: 5, Good: 12, Okay: 2, Bad: 1 \\
\acrshort{GPET}     & \hspace{0.5em}Very good: 20  \hspace{0.5em}& \hspace{0.5em}Very good: 6, Good: 6, Okay: 8, Bad: 0\hspace{0.5em}  & \hspace{0.5em}Very good: 6, Good: 13, Okay: 1, Bad: 0 \\ 
\bottomrule
\end{tabular}%

                \end{adjustwidth}
                \caption[Blinded adjudication between DeepGPET and \acrshort{GPET} for different cases of error.]{Qualitative ratings of 20 test set segmentations along 3 key dimensions. The rater was blinded to the identity of the methods and their order was randomised for every example.}
                \label{tab:DEEPGPET_qualitative}
                \end{table}

                \begin{figure}[!t]
                    \centering
                    \includegraphics[width=\linewidth]{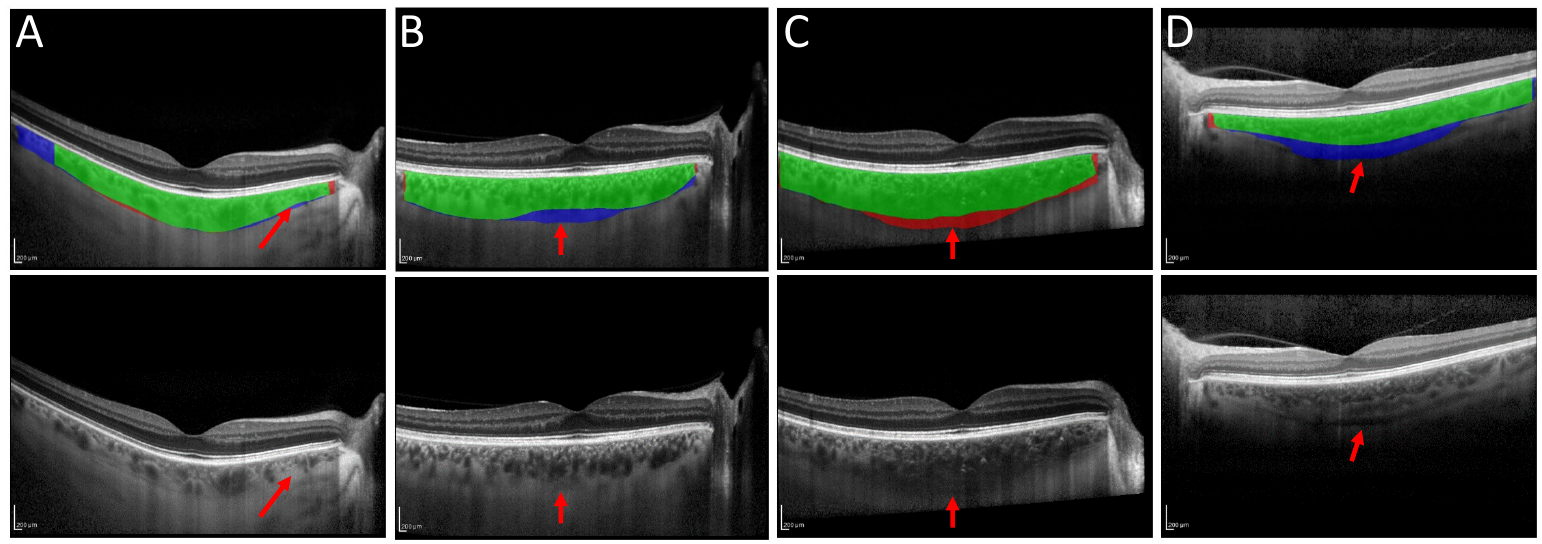}
                    \caption[Adjudication examples between DeepGPET and \acrshort{GPET}.]{Four examples from the adjudication. The rater preferred DeepGPET for (A -- B) and \acrshort{GPET} for (C -- D). Top row: green, segmented by both \acrshort{GPET} and DeepGPET; red, \acrshort{GPET} only; and blue, DeepGPET only. Bottom row: arrows indicate important choroidal features which can make segmentation challenging.}
                    \label{fig:DEEPGPET_adjudication_examples}
                \end{figure}

                Figure \ref{fig:DEEPGPET_adjudication_examples} shows some examples. In panel (A), DeepGPET segmented more of the temporal region than \acrshort{GPET} did, providing a full width segmentation which was preferred by the adjudicator I.M. Additionally, both approaches were able to segment a smooth boundary, even in regions with stroma fluid obscuring the lower boundary (red arrow). In panel (B), the lower boundary for this choroid is very faint and is actually below the majority of the vessels sitting most posterior (red arrow). DeepGPET produced a smooth and concave boundary preferred by the adjudicator, while \acrshort{GPET} fell victim to hugging the posterior most vessels in the subfoveal region. In panel (C), DeepGPET rejected the true boundary in the low contrast region (red arrow) and opted for a more well-defined one, while \acrshort{GPET} segmented the more uncertain path. Since \acrshort{GPET} permits human intervention, there is more opportunity to fine tune it's parameters to fit what the analyst believes is the true boundary. Here, the adjudicator preferred \acrshort{GPET}, while DeepGPET's under-confidence led to under-segmentation and to a bad rating. 
                
                In panel (D), the Choroid-Sclera boundary is difficult to delineate due to a thick suprachoroidal space (red arrow) and thus a lack of lower boundary definition. In this case, DeepGPET segmented the choroidal space, as well as the suprachoroid which became more obvious near the centre of the B-scan. Thus, DeepGPET over-segmented the scan overall, as we exclude the suprachoroidal space from our definition of the choroid (section \ref{subsec:ch1_INTRO_measure_bsan}). While \acrshort{GPET} excluded the suprachoroidal space, it actually segmented through posterior choroid vessels, thus under-segmenting. Nevertheless, the adjudicator preferred \acrshort{GPET} in this case because we considered the suprachoroid and choroid itself as separate spaces (according to the definitions set out in appendix \ref{apdx:definitions}). Note that the choroids in figure \ref{fig:DEEPGPET_adjudication_examples}(B -- D) are the choroids with the largest choroid thickness and area disagreement between DeepGPET and \acrshort{GPET} as observed in figure \ref{fig:DEEPGPET_ba_plot}.

            \end{mysubsubsection}

            \begin{mysubsubsection}[]{Potential generalisability}\label{subsec:deepgpet_eval_general}

                \begin{figure}[tb]
                    \centering
                    \includegraphics[width=\linewidth]{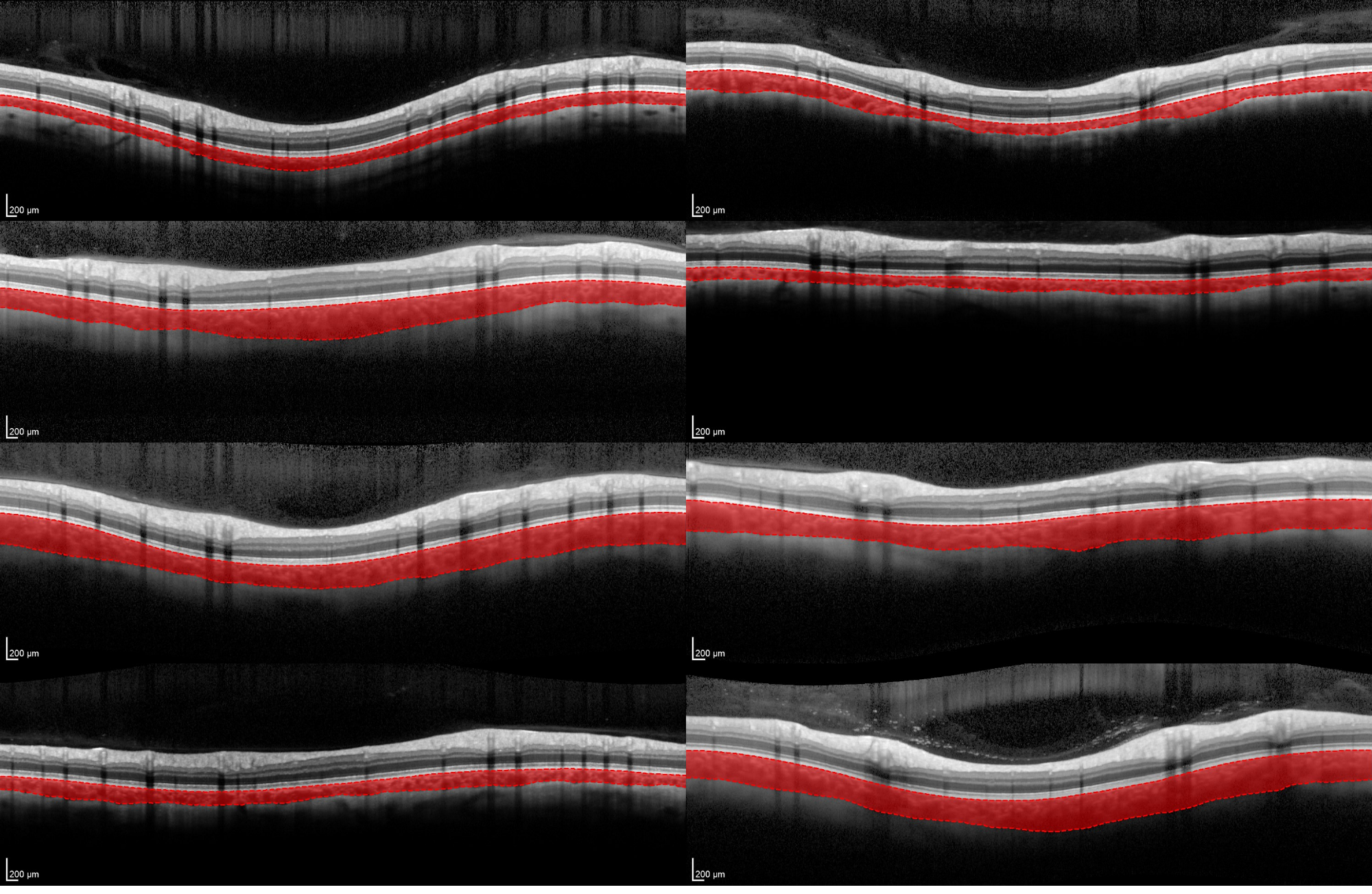}
                    \caption[DeepGPET's potential generalisability to peripapillary \acrshort{OCT} B-scans.]{A selection of peripapillary choroids from 8 different eyes in the \acrshort{DVCKD} cohort, with successful choroid region segmentation from DeepGPET.}
                    \label{fig:DEEPGPET_peripapillary}
                \end{figure}
        
                \begin{figure}[tb]
                    \centering
                    \includegraphics[width=\linewidth]{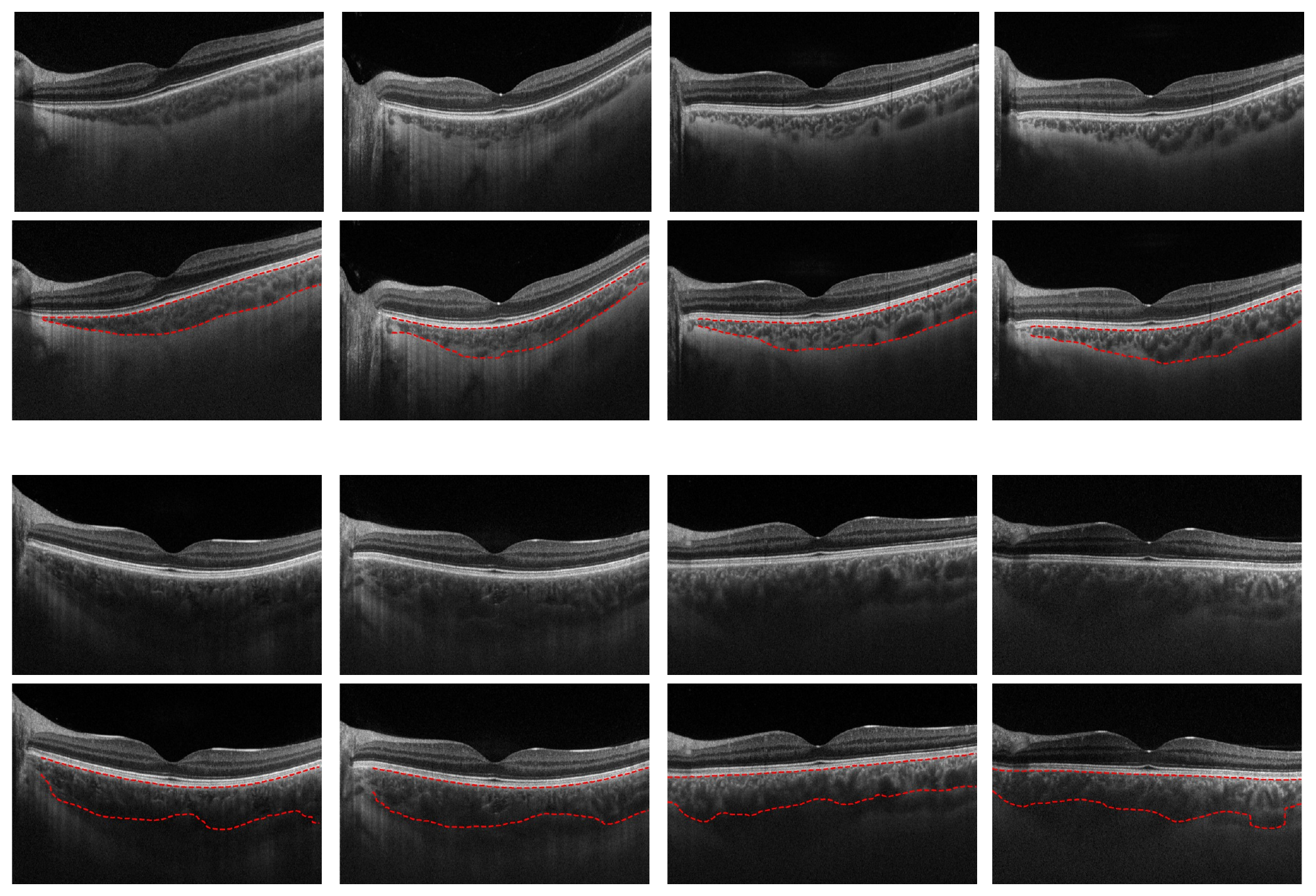}
                    \caption[DeepGPET's qualitative performance on a small set of \acrshort{SS-OCT} B-scans.]{A selection of small and large choroids from \acrshort{OCT} B-scans collected using the Topcon \acrshort{SS-OCT} DRI Triton plus (Topcon, Tokyo, Japan), with segmentations of the upper and lower boundaries as outputted by DeepGPET.}
                    \label{fig:DEEPGPET_topcon_large_small}
                \end{figure}

                DeepGPET was only trained on macular \acrshort{SD-OCT} B-scans using data from Heidelberg Engineering imaging devices. Thus, we do not know it's performance on \acrshort{SS-OCT} data, or at locations around the periphery of the posterior pole. While not explicitly part of the evaluation of DeepGPET, we conducted some initial experimentation on DeepGPET's potential to generalise to peripapillary \acrshort{SD-OCT} and macular \acrshort{SS-OCT} data. Note that while we highlight the following potential functionality of DeepGPET, these are not officially validated.

                Firstly, we considered DeepGPET's potential generalisability to peripapillary \acrshort{OCT} B-scans. These are circular-shaped B-scans centred on the optic nerve head and particularly useful for assessing retinal nerve fiber layer thickness. Figure \ref{fig:DEEPGPET_peripapillary} shows eight peripapillary choroids using the Heidelberg Engineering \acrshort{SD-OCT} Spectralis \acrshort{OCT}1 device. These B-scans are from participants in the \acrshort{DVCKD} cohort, and are used to assess DeepGPET's reproducibility. In this exploration, the peripapillary choroids were selected based on image quality, choroid size and extent of sinuosity of the circular peripapillary B-scan. 
                
                While DeepGPET was only trained on macular \acrshort{OCT} B-scans, it appeared to generalise well to segmenting peripapillary choroids, but tended to not fully segment the whole width. Peripapillary B-scans from Heidelberg Engineering \acrshort{OCT} imaging devices had a lateral pixel resolution of 1536 pixels, while the training data only had a lateral pixel resolution of 768 pixels. Ensuring the peripapillary choroid is segmented from edge-to-edge is particularly important since the peripapillary B-scan is circular and is thus continuous at either edge of the image. Therefore, a full segmentation is required to ensure consistent measuring of the different peripapillary sub-fields (section \ref{subsec:ch1_INTRO_measure_peri}). However, a suitable remedy is to pad the B-scan horizontally by padding each edge with it's opposite edge by a few hundred pixels, thus artificially elongating the B-scan. This is a valid approach because of the circular and continuous nature of the B-scan and is sufficient for DeepGPET to segment the entire peripapillary choroid, allowing average thickness in the peripapillary sub-fields to be measured. 

                Secondly, we wanted to experiment with how robust DeepGPET was to processing macular \acrshort{SS-OCT} B-scans with choroids of various size. We selected 8 \acrshort{OCT} B-scans from eight eyes from the \acrshort{GCU} Topcon cohort used to assess \acrshort{MMCQ}'s reproducibility from chapter \ref{chp:chapter-mmcq}. Figure \ref{fig:DEEPGPET_topcon_large_small} shows a row of four small and large choroids, with DeepGPET's region segmentations overlaid in the subsequent rows. While DeepGPET appeared to be robust to smaller choroids from these \acrshort{SS-OCT} B-scans (top rows), for larger choroids whose Choroid-Sclera boundaries had poor visibility, DeepGPET's robustness failed (bottom rows). For these cases, DeepGPET is simply under-segmenting the choroid, and failing to detect the Choroid-Sclera boundary with any certainty. 
            
            \end{mysubsubsection}
  
        \end{mysubsection}

    \end{mysection}
    
    \begin{mysection}[]{DeepGPET's reproducibility} \label{sec:DeepGPET_REPR}

        \begin{mysubsection}[]{Statistical analysis}

            For the \acrshort{OCT} volume data in the i-Test sample, we measured average choroid thickness in the 9 sub-fields in the \acrshort{ETDRS} grid \cite{early1991grading}. For each peripapillary \acrshort{OCT} B-scan in the \acrshort{DVCKD} sample, we measured average choroid thickness in the 6 peripapillary sub-fields, and the papillomacular bundle. The methods used to generate these sub-field measurements for volume and peripapillary data have been described previously in sections \ref{subsubsec:ch1_INTRO_oct_volume} and \ref{subsec:ch1_INTRO_measure_peri}, respectively. For the \acrshort{DVCKD} sample, there are three images per eye so we compare pairwise, chronological measurements (morning -- afternoon, afternoon -- evening). Thus, there were 1080 (120 eyes $\times$ 9 locations) and 308 (22 eyes $\times$ 7 locations $\times$ 2 pairwise comparisons) repeated samples to compare for the i-Test and \acrshort{DVCKD} samples, respectively. 
            
            As we are comparing measurements across the day in the \acrshort{DVCKD} sample, diurnal variation of the choroid is a potential confounding factor for this sample. Thus, we address the longitudinal evolution of the peripapillary choroid too. 

            We measure the reproducibility of DeepGPET at the population-level, reporting population mean and standard deviation (\acrshort{SD}), as well as mean absolute error (\acrshort{MAE}) and Pearson and Spearman coefficients. These metrics have been defined previously in section \ref{subsec:INTRO_metrics}. Additionally, we show correlation plots and use Bland-Altman \cite{bland1986statistical} plots to assess the relationship between repeated measurements and the distribution of residuals. We also report the reproducibility of DeepGPET at the eye-level using the measurement noise parameter $\lambda$ \cite{engelmann2024applicability} described in section \ref{subsec:INTRO_metrics}. 
                
        \end{mysubsection}

        \begin{mysubsection}[]{Results}

            \begin{mysubsubsection}[]{Population-level}

                \begin{table}[tb]
                \begin{adjustwidth}{-1in}{-1in}  
                \centering
                
\begin{tabular}{rllll}
\toprule
\multicolumn{1}{l}{Study} & Mean (\acrshort{SD}) [$\mu$m] & \acrshort{MAE} [$\mu$m] & Pearson & Spearman \\
\midrule
\multirow{1}{*}{i-Test (\acrshort{ETDRS})} & 275.68 (85.69) & 5.5510 & 0.9984 & 0.9979\\
\multirow{1}{*}{\acrshort{DVCKD} (Peripapillary)} & 169.60 (54.07) & 5.0320 & 0.9942 & 0.9940 \\
 \bottomrule
\end{tabular}

                \end{adjustwidth}
                \caption[DeepGPET's population-level reproducibility performance.]{Reproducibility performance for DeepGPET using macular and peripapillary choroid thickness. All Pearson and Spearman correlations were statistically significant with P-values P < 0.0001.}
                \label{tab:DEEPGPET_repr_tab}
                \end{table}
    
                \begin{figure}[tb]
                    \centering
                    \includegraphics[width=\linewidth]{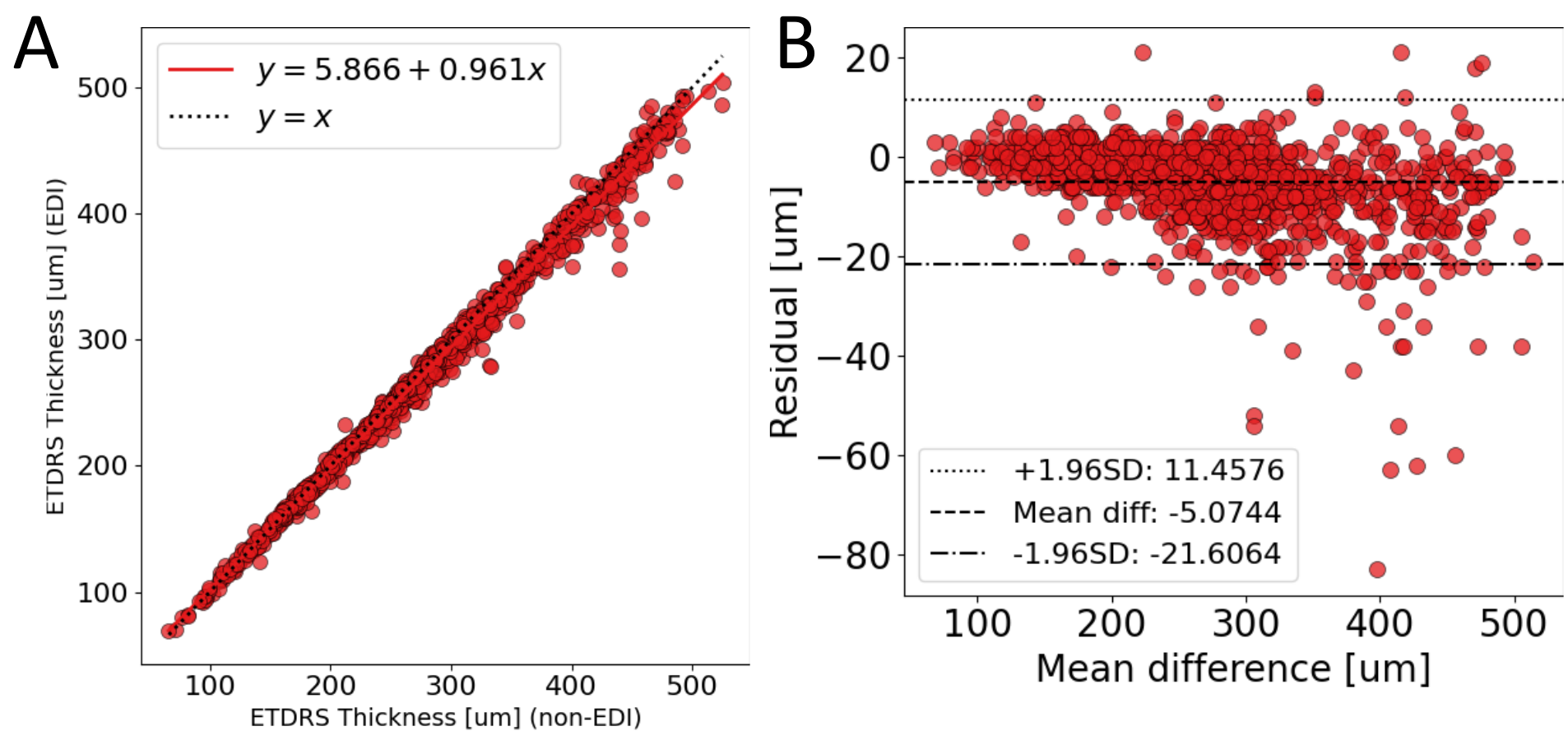}
                    \caption[DeepGPET's population-level reproducibility performance on \acrshort{SD-OCT} volume scans.]{Correlation (A) and Bland-Altman (B) plots for assessing the reproducibility of DeepGPET on macular \acrshort{OCT} volume data from the i-Test sample.}
                    \label{fig:DEEPGPET_pop_repr_macular}
                \end{figure}
    
                \begin{figure}[tb]
                    \begin{adjustwidth}{-0.5in}{-0.5in}  
                    \centering
                    \includegraphics[width=\linewidth]{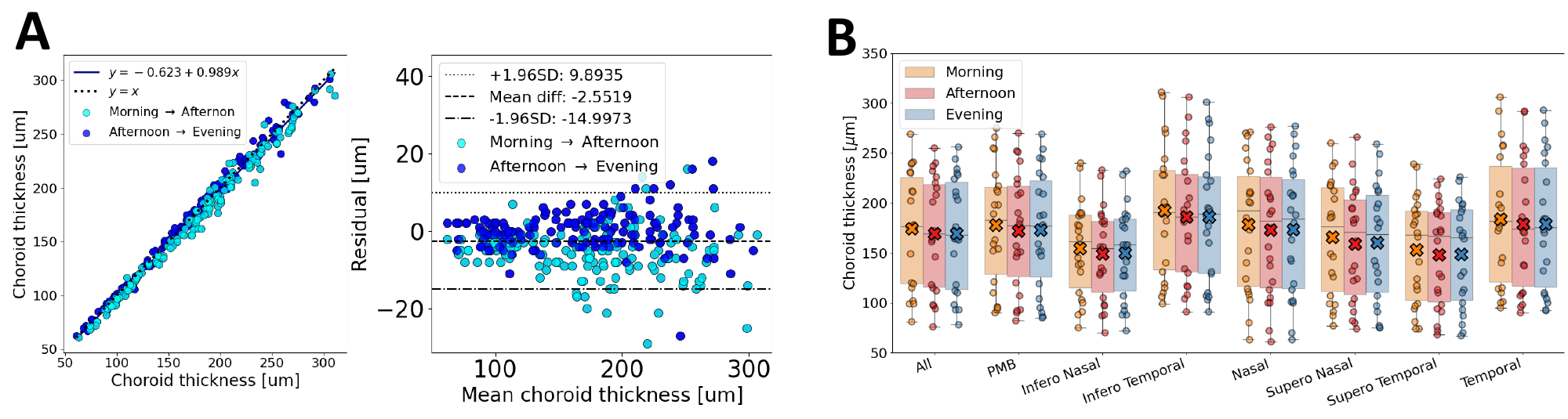}
                    \end{adjustwidth}
                    \caption[DeepGPET's population-level reproducibility performance on \acrshort{SD-OCT} peripapillary B-scans.]{(A) Correlation and Bland-Altman plot for all chronologically paired (morning -- afternoon, afternoon -- evening) choroid thickness measurements in all peripapillary sub-fields. (B) Longitudinal evolution of average choroidal thickness in each peripapillary sub-field shown as box-plots, with mean values overlaid as crosses.}
                    \label{fig:DEEPGPET_pop_peri_repr}
                \end{figure}
    
                Table \ref{tab:DEEPGPET_repr_tab} presents the reproducibility performance of DeepGPET at the population-level. Peripapillary choroids were thinnest in the \acrshort{DVCKD} sample (mean $\pm$ \acrshort{SD} of 169.9 $\pm$ 54.1 $\mu$m) compared to the i-Test sample (mean $\pm$ \acrshort{SD} of 275.68 $\pm$ 91.8 $\mu$m). This makes sense given peripapillary choroids are thinner near the optic disc than in the central macula \cite{yang2019factors}.
                
                In figure \ref{fig:DEEPGPET_pop_repr_macular} we show the correlation and Bland-Altman plots for the i-Test sample. DeepGPET showed very strong and statistically significant correlations between non-\acrshort{EDI} and \acrshort{EDI} paired \acrshort{ETDRS} measurements (Pearson/Spearman for choroid thickness in i-Test: 0.9965/0.9968) and achieved an \acrshort{MAE} of 5.5 $\mu$m. Compared to manual thickness measurement, this \acrshort{MAE} is well within the threshold expected to exceed manual inter- and intra-rater repeatability \cite{rahman2011repeatability} (32 and 23 $\mu$m, respectively). Additionally, from a semi-automatic perspective, an \acrshort{MAE} of 5.5 $\mu$m was almost half of \acrshort{GPET}'s repeatability (table \ref{tab:DEEPGPET_results}). Moreover, effect size is dependent on the context, and in systemic health this \acrshort{MAE} is still below reported differences in choroid thickness which can be as low as 20 -- 30 $\mu$m such as in myopia progression \cite{breher2019metrological, flores2013relationship}.
    
                Generally, we see that for larger choroids there is larger error in choroid thickness. In the case for i-Test, this is entirely expected because the optical signal for non-\acrshort{EDI} \acrshort{OCT} acquisition degrades the deeper through the eye it penetrates, thus the visibility of the deeper choroidal vasculature decays rapidly also. This larger error translated into very slight under-estimation of choroid thickness for the non-\acrshort{EDI} \acrshort{OCT} volume scans of each pair, as seen in the Bland-Altman plot in figure \ref{fig:DEEPGPET_pop_repr_macular}(B). However, this under-prediction with mean residual of -5.07 $\mu$m (95\% confidence interval of -21.61 $\mu$m to 11.46 $\mu$m) is by no means large or clinically significant, highlighting the potential robustness to non-\acrshort{EDI} scans of the choroid in macular \acrshort{OCT} data.
    
                Figure \ref{fig:DEEPGPET_pop_peri_repr}(A) show the correlation and Bland-Altman plots for comparing pairwise peripapillary sub-field thickness measurements for morning vs. afternoon and afternoon vs. evening (with scatter points coloured accordingly). The first striking observation is the degree of separation which measurements have between pairwise time-point comparisons. The residuals of measurements comparing afternoon and evening are centred around 0, while residuals comparing morning and afternoon appear to be systematically below the origin (and the identity line in the correlation plot). 
    
                Figure \ref{fig:DEEPGPET_pop_peri_repr}(B) shows the longitudinal evolution of peripapillary choroid thickness for each peripapillary sub-field between morning, afternoon and evening, with the mean value per time point and sub-field shown as coloured crosses. Across the board, morning measurements appear larger than afternoon and evening. This corresponds to the trend observed in measuring the same choroids manually in the macula \cite{farrah2023choroidal} (shown in their Supplementary Figure 8C), as well as the natural diurnal variation of the choroid observed in other studies \cite{chakraborty_diurnal_2011, tan2012diurnal, usui_circadian_2012, kinoshita_diurnal_2017, singh2019diurnal, ostrin_imi-dynamic_2023}. These previous studies have reported choroidal fluctuations over the course of day (in the macula and peripapillary regions) with an amplitude of approximately 30 $\mu$m, and the majority reporting larger choroidal variation in the morning, compared to the afternoon and evening. Thus, given that the largest residuals were those observed between morning and afternoon, and that we know the choroid thins between this time period, we propose that the diurnal variation of the choroid is the primary driver of the largest residuals in measuring peripapillary choroid thickness with DeepGPET, and not segmentation error.

            \end{mysubsubsection}

            \begin{mysubsubsection}[]{Eye-level}

                \begin{figure}[tb]
                    \centering
                    \includegraphics[width=\linewidth]{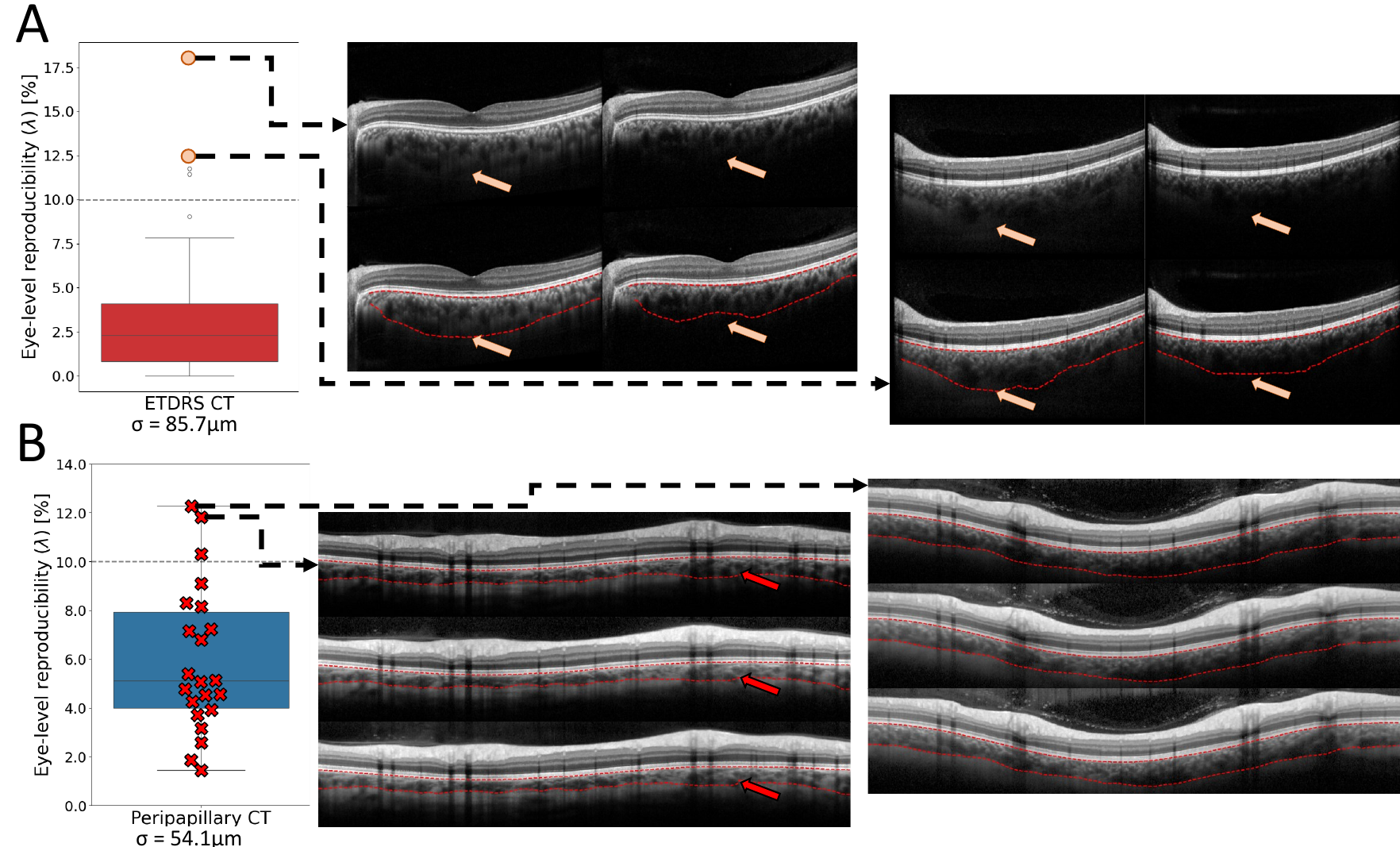}
                    \caption[DeepGPET's eye-level reproducibility performance.]{Eye-level reproducibility of DeepGPET for \acrshort{ETDRS} (A) and peripapillary (B) choroid thickness. We plot distributions of measurement noise $\lambda$ \cite{engelmann2024applicability}, with a horizontal dashed-line marking 10\% to aid readability and interpretation. Representative B-scans of major outliers are shown with segmentations overlaid in red and arrows indicating the source of the error. The between-eye standard deviations are shown below each box-plot.}
                    \label{fig:DEEPGPET_ind_repr_macular}
                \end{figure}
    
                In figure \ref{fig:DEEPGPET_ind_repr_macular}, we present the eye-level reproducibility of DeepGPET for macular \acrshort{OCT} volume data (A) and peripapillary \acrshort{OCT} B-scan data (B). DeepGPET's eye-level measurement noise was very low compared to the population's variability, with the upper quartile of the box-plot distributions for \acrshort{ETDRS} thickness sitting below 5\% of the population's (i-Test) variability, and for peripapillary thickness this was below 10\% of the population's (\acrshort{DVCKD}) variability. Measurement error was higher for peripapillary thickness likely due to the lower population variability in the cohort relative to i-Test ($\sigma$=54.1 $\mu$m for \acrshort{DVCKD} sample which was 63\% of the i-Test macular thickness population variability, $\sigma$=85.7 $\mu$m). Nevertheless, for choroid region metrics segmentation error contributed to only 5--10\% of the overall population variability in both cohorts. 
                
                In figure \ref{fig:DEEPGPET_ind_repr_macular}(A), we show two pairs of representative B-scans from the \acrshort{OCT} volume scan data which were major outliers. The major outliers were primarily driven by large choroids whose non-\acrshort{EDI} counterpart had poor Choroid-Sclera boundary visibility. This optical signal degradation had a significant impact on DeepGPET's ability to detect this lower boundary, especially underneath the fovea. The arrows indicate the location of the major sources of error, which highlight the under-prediction of \acrshort{ETDRS} thickness observed from the population-level reproducibility analysis. Nevertheless, DeepGPET is able to approximate the shape of the choroid well in regions of low visibility, albeit under-segmenting the choroid in regions of invisible Choroid-Sclera boundary.
    
                In figure \ref{fig:DEEPGPET_ind_repr_macular}(B), we show the B-scans of the major outliers at each time point with the segmentations overlaid using DeepGPET. Any differences (red arrows) are minimal, and the segmentations do not look qualitatively different. These two cases were the major outliers due to the difference in choroid thickness observed from morning -- afternoon (a negative differential of 15 and 14 $\mu$m in average thickness, respectively). Comparing this to the difference observed between afternoon -- evening, which was a positive differential of 1 and 2 $\mu$m, we propose the major outliers are primarily driven by marginal diurnal variation of the choroid, and not segmentation error.

            \end{mysubsubsection}
            
        \end{mysubsection}
        
    \end{mysection}

    \begin{mysection}[]{Discussion}\label{sec:ch4_DEEPGPET_discuss}

        \begin{mysubsection}[]{Evaluation}

            DeepGPET is a fully-automatic and efficient method for choroid region segmentation. DeepGPET achieved excellent agreement with \acrshort{GPET} on held-out data in terms of segmentation and choroid-derived measurements, approaching the repeatability of \acrshort{GPET} itself, i.e. best case repeatability where the author (who developed \acrshort{GPET} and arguably has the most domain knowledge around it's usability) did both sets of annotation. Since the author generated both the original and repeated \acrshort{GPET} segmentations for evaluating DeepGPET's performance, there was no inter-rater subjectivity at play. Thus, the intra-rater agreement measured in this work is a best case scenario and forms an upper-bound for agreement with the original segmentations.
            
            However, note that table \ref{tab:DEEPGPET_results} and figures \ref{fig:DEEPGPET_corr_plot} and \ref{fig:DEEPGPET_ba_plot} are comparing \acrshort{GPET}'s \textit{repeatability} with DeepGPET's \textit{generalisability}. Here, \acrshort{GPET} is being measured against it's own performance on the same 54 images it processed 2 months prior with manual input. Thus, \acrshort{GPET}'s \acrshort{MAE} of 10.41 $\mu$m provides an upper bound for it's segmentation performance when applied to the exact same input. Conversely, DeepGPET's model evaluation shown in table \ref{tab:DEEPGPET_results} represents its generalisability to entirely new \acrshort{OCT} data, which was slightly higher at 13.31 $\mu$m. However, DeepGPET's repeatability, i.e. application to the same image twice is 100\% accurate and will output the same prediction when given the same input. This is because hyperparameter-free deep learning models, post-training, like DeepGPET are \textit{deterministic}. 
            
            In contrast, \acrshort{GPET} has many tunable parameters and relies heavily on handcrafted edge maps, manual input and image pre-processing. Holding all of \acrshort{GPET}'s free parameters and pre-processing steps fixed does make its output deterministic. However, the end-user is free to tune its parameters to fit what they believe is the true boundary --- by selecting a different covariance function of the Gaussian process for example --- necessarily injecting human subjectivity into the segmentation procedure. This subjectivity was ultimately quantified as the \acrshort{MAE} of 10.41 $\mu$m from \acrshort{GPET}'s repeatability analysis.
            
            The improvement on segmentation inference time with DeepGPET over \acrshort{GPET} results in massive time savings. A standard \acrshort{OCT} volume scan collected on a Heidelberg Engineering imaging device usually consists of 61 B-scans. With \acrshort{GPET}, processing such a volume for a single eye takes about 35 minutes during which the operator has to select initial pixels to guide tracing and adjust parameters if \acrshort{GPET} initially failed (which may happen for about 25\% of images given 1.27 manual interventions were required across 54 test images in section \ref{subsec:ch4_DEEPGPET_eval}). In contrast, DeepGPET could do the same processing in about 76 seconds on the same hardware, during which no manual input is needed. DeepGPET could even be \acrshort{GPU}-accelerated to cut the processing time by another order of magnitude. 
            
            The lack of manual interventions required by DeepGPET means that no subjectivity is introduced unlike \acrshort{GPET}, particularly when used by different end-users. Importantly, DeepGPET does not require specialist training in image processing and can segment images fully-automatically in a fraction of the time, enabling analysis of large-scale datasets and greater potential for its applicability in future research studies. Thus, \acrshort{GPET} can be replaced by a much quicker and more effective end-to-end segmentation pipeline which is significantly more accessible for those without a background in image processing (figure \ref{fig:DEEPGPET_info_fig}). We anticipate DeepGPET to facilitate efficient knowledge transfer of expertise in choroidal image analysis to the wider research community, ultimately helping improve the standardisation of segmentation-derived choroidal measurements.
    
            \begin{figure}[tb]
                \centering
                \includegraphics[width=0.9\linewidth]{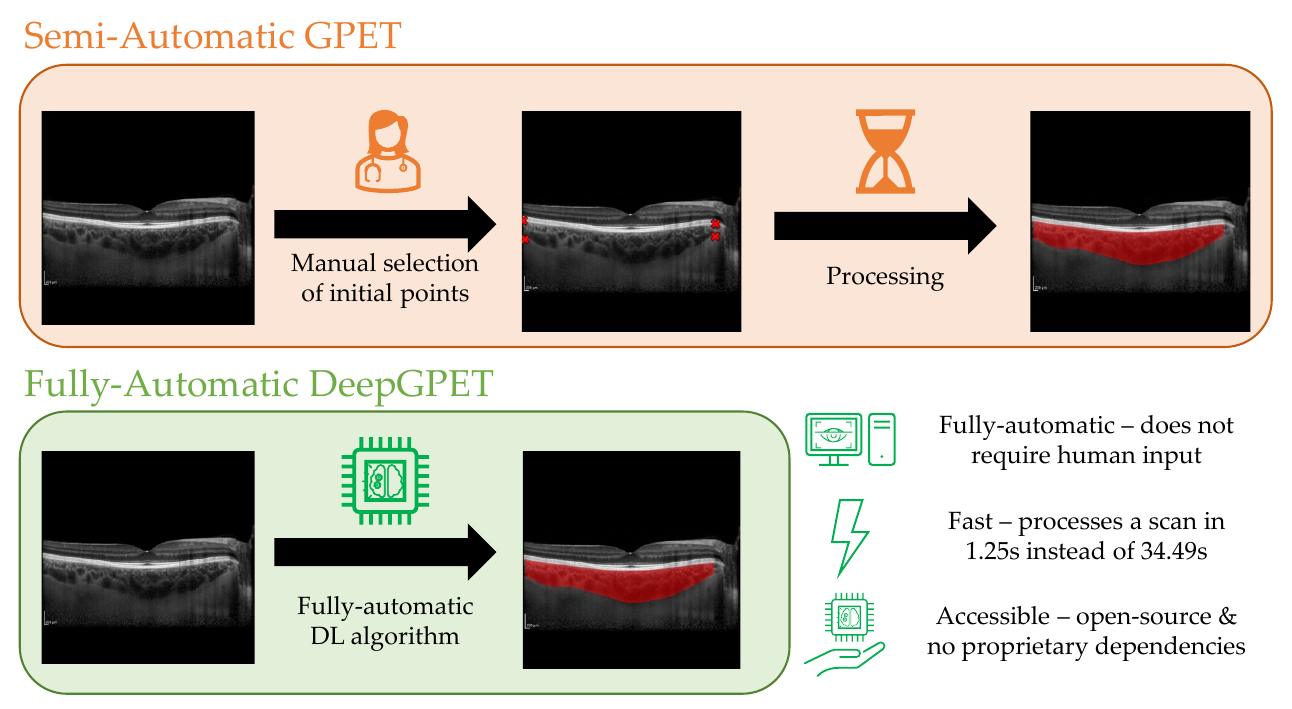}
                \caption[High-level, qualitative comparison between DeepGPET and \acrshort{GPET}.]{Comparison between the semi-automatic \acrshort{GPET} (top) and fully-automatic DeepGPET (bottom).}
                \label{fig:DEEPGPET_info_fig}
            \end{figure}
    
            While the observed agreement was very high between DeepGPET and \acrshort{GPET}, it was not perfect. However, even higher agreement with \acrshort{GPET} would not necessarily produce a more generalisable method as \acrshort{GPET} itself is not perfect, i.e. a perfect correspondence between DeepGPET and \acrshort{GPET} would mean that DeepGPET learns the biases of \acrshort{GPET}. For example, in figure \ref{fig:DEEPGPET_adjudication_examples}(A) GPET fails to segment the entire width of the choroid, and in figure \ref{fig:DEEPGPET_adjudication_examples}(B) under-segments the large choroid. 
            
            As described earlier (section \ref{subsec:ch1_INTRO_measure_choroid}), the definition of the Choroid-Sclera boundary can vary depending on the end-user and application \cite{yiu2014characterization}. This poses a huge challenge for manual segmentation of the choroid and standardisation of choroid-derived measurements \cite{boonarpha2015standardization, xie2021evaluation}. The variable definition, coupled with potentially low \acrshort{SNR},  superficial retinal vessel shadowing, vessels passing from the sclera into the choroid and stroma fluid obscurations all contribute to the ambiguous nature of defining and detecting the Choroid-Sclera boundary \cite{chandrasekera2018posterior}. 

            However, for quantitative analysis of choroidal measures, the specific definition of the Choroid-Sclera boundary is secondary to applying the same, consistent definition across and within patients. Here, fully-automatic methods like DeepGPET provide a significant benefit by removing the subjectivity present in semi-automatic methods. Where semi-automatic methods require manual input, two analysts with different understandings of the Choroid-Sclera boundary could produce vastly different segmentations. With DeepGPET as a deterministic and automatic approach, the same image is always segmented in the same way, removing subjectivity. 

            Finally, while DeepGPET was only trained on macular \acrshort{SD-OCT}, we found the model to be robust to peripapillary \acrshort{SD-OCT} data after suitable opposite-edge padding to ensure the whole width was segmented. This suggests DeepGPET's potential to generalise to different locations in the posterior pole, and the excellent reproducibility on the wider \acrshort{DVCKD} cohort further provides evidence for this. Additionally, we observed that DeepGPET was able to successfully detect the Choroid-Sclera boundary in certain choroids of macular \acrshort{SS-OCT} data which weren't too large. For large-choroid \acrshort{SS-OCT} B-scans, DeepGPET showed poor performance (figure \ref{fig:DEEPGPET_topcon_large_small}). This is perhaps unsurprising given that DeepGPET was only trained on \acrshort{SD-OCT} images. Thus, we recommend those using DeepGPET on \acrshort{SS-OCT} images to review the segmentations afterwards as a sanity check.

        \end{mysubsection}

        \begin{mysubsection}[]{Reproducibility}\label{subsec:ch4_deepgpet_rpr_lims}

            We found DeepGPET to be highly reproducible for measuring downstream choroidal measurements using macular \acrshort{SD-OCT} volume and peripapillary B-scan data. DeepGPET reported \acrshort{MAE}s well within thresholds expected to exceed manual, inter-rater agreement \cite{rahman2011repeatability}, \acrshort{GPET}'s repeatability and small effect sizes observed from the literature \cite{breher2019metrological, flores2013relationship}. Moreover, DeepGPET achieved an eye-level measurement error which contributed to a very small proportion (5 -- 10\%) of the overall population variability in both cohorts. Advantageously, this eye-level reproducibility helps provides an upper bound on DeepGPET's measurement variability which can be used to differentiate between measurement error and true biological change, a key step in the interpretation of choroidal measurements in research studies \cite{breher2020choroidal}.
    
            The reproducibility of DeepGPET for \acrshort{SD-OCT} peripapillary B-scans were confounded by natural diurnal variation of the choroid. Analysis of residuals suggested that the major outliers were sourced primarily between morning -- afternoon comparisons, rather than afternoon -- evening comparisons, and major outliers at the eye-level showed no qualitative difference in segmentation error. Manual measurements of fovea-centred choroidal thickness showed a similar longitudinal trend \cite{farrah2023choroidal} (supplementary figure 8C), as well as a wealth of literature suggesting choroidal thinning throughout daytime, and especially between the morning and afternoon \cite{chakraborty_diurnal_2011, tan2012diurnal, usui_circadian_2012, kinoshita_diurnal_2017, singh2019diurnal, ostrin_imi-dynamic_2023}. Thus, this suggests the error observed was primarily driven by diurnal variation and not segmentation error.
    
            While the reproducibility was high, the values compared were averages across regions around the posterior pole, and so reproducibility for point-source measurements like subfoveal choroidal thickness has not been measured. Thus, there is potential for the values reported in each macular/peripapillary region to be averaging over local errors. However, the \acrshort{ETDRS} and peripapillary sub-fields are commonly used by the research community and provide better ways to characterise the retina and choroid rather than extremely localised, point-source measurements. Therefore, assessing DeepGPET's reproducibility according to these spatial regions is an entirely valid approach to investigate the method's reproducibility.
    
            However, we must make the reader aware of any potential data leakage in the reproducibility analysis. For the i-Test sample, approximately 7\% (4 / 60) of women were part of the model training and validation for DeepGPET. However, only the \acrshort{EDI} counterpart of their \acrshort{SD-OCT} volume pair were used, and not their non-\acrshort{EDI} volume scan. Thus, while the eye itself had been observed before by DeepGPET, the image artefacts introduced by deactivating \acrshort{EDI} mode still posed a significant challenge to DeepGPET's reproducibility as a method.

        \end{mysubsection}

        \begin{mysubsection}[]{Limitations and Future work}\label{subsec:chp4_limitations}

            There are some limitations associated with DeepGPET as a method. Firstly, DeepGPET was trained on \acrshort{OCT} B-scans from examples related to systemic health, and thus its generalisability to ocular health is unknown. It would be important to quantify the reliability of DeepGPET in extreme scenarios of pathology which affect the choroid, such as choroidal thinning from pathological myopia \cite{ohno2021imi}, thickening from central serous chorioretinopathy \cite{fung2023central} or optical disturbance from photoreceptor degeneration in geographic atrophy \cite{vallino2024structural}. Nonetheless, DeepGPET remains a highly reproducible approach for generating choroidal region metrics from \acrshort{OCT} B-scans for the nascent field of oculomics \cite{wagner2020insights}, and we anticipate it will be a useful tool therein.

            Additionally, by distilling \acrshort{GPET} into DeepGPET, there is potential that it may have learned certain biases from \acrshort{GPET}'s semi-automatic procedure. For example, \acrshort{GPET} does not always segment the whole width of the choroid, as it's dependent on the user's manual edge endpoint selection. Currently, area and thickness are only calculated for a specific fovea-centred ROI, so this is not a problem. However, segmenting the whole visible choroid would be desirable when generating maps across the macula from an \acrshort{OCT} volume scan. Interestingly, DeepGPET is already capable of segmenting the whole choroid in most cases, i.e. figure \ref{fig:DEEPGPET_adjudication_examples}(A), but as some of the training data did not cover the full width, DeepGPET sometimes emulates this, too.
    
            Finally, and perhaps most importantly, DeepGPET is only focused on segmentation of the choroidal layer. While this is the most important and time-consuming step for measuring choroidal thickness, area and volume, the location of the fovea on \acrshort{OCT} B-scans is required to define the fovea-centred \acrshort{ROI} for choroid-derived measurements. This is critical for producing standardised measurements across and within patients. Given that identifying the fovea is less time-consuming or ambiguous than choroid segmentation, extending DeepGPET to output the fovea location is a natural next step. 
            
            Fortunately, DeepGPET's robustness to peripapillary B-scans permit fully automatic, segmentation-derived measurements of peripapillary choroids, as these B-scans do not require fovea detection for derived measurements. However, while this is the case for end-users who are reasonably familiar with the Python programming language, this may not be the case to all members of the research community. Therefore, it would be wise to wrap DeepGPET into a truly automatic `raw data to measurement' toolkit which end-users can run simply from the terminal on \acrshort{OCT} data extracted from an imaging device. In future work, we plan to release such a toolkit, OCTolyzer, which would leverage DeepGPET's ability to provide reproducible and clinically meaningful measurements of choroids in peripapillary B-scans. 
    
            Another limitation of DeepGPET is that it also does not quantify the choroidal vasculature, which would arguably be a more representative measure of the choroid (figure \ref{fig:MMCQ_drisc_prevent_ca_cv}). Segmenting the choroidal vessels is a very challenging task even for humans and is prohibitively time-consuming to do manually (table \ref{tab:MMCQ_eval_times}). Permitting additional functionality for DeepGPET to automatically segment the vasculature within the choroid as well would make DeepGPET a fast and efficient end-to-end framework capable of fully segmenting the choroidal space and vessels from a raw \acrshort{OCT} B-scan, converting the B-scan into a set of standardised, clinically meaningful and reproducible measurements. Fortunately, the semi-automatic approach defined in chapter \ref{chp:chapter-mmcq}, \acrshort{MMCQ}, could be used for ground truth label generation.
            
            DeepGPET was the first foray into deep learning during this doctorate degree, and while deep learning in general remains as the current state-of-the-art for medical image segmentation \cite{o2020deep, williams2023unified}, there are a number of drawbacks. For example, traditional image processing methods like \acrshort{GPET} and \acrshort{MMCQ} did not require any prior training data, while model training in deep learning requires expertly annotated datasets which can be scarce in the medical imaging domain. Although techniques like transfer learning and data augmentation can mitigate this issue \cite{cheplygina2019not}, the reliance on large datasets remains a fundamental limitation. 
            
            Furthermore, model training usually requires \acrshort{GPU} resources which may not necessarily be at everyone's disposal, and while DeepGPET is able to be run on a standard laptop CPU reasonably fast (at 1.25 seconds/image), much larger models such as vision transformers \cite{dosovitskiy2020image} or parameter-intense architectures like DenseNet \cite{huang2017densely} require \acrshort{GPU}-accelerated model training and inference to be run effectively, utilising large amounts of temporary memory (RAM). This complexity can make it difficult to run on edge compute devices, with limited access to high-performance compute resources to deploy deep learning-based segmentation models. Nevertheless, we anticipate that DeepGPET will be able to cover a broad range of end-users with different levels of hardware.

            Additionally, DeepGPET, unlike traditional image processing methods like \acrshort{GPET} and \acrshort{MMCQ}, is seen as a ``black box'' where the hierarchical nature of DeepGPET's \acrshort{CNN} makes it difficult for humans to interpret the decision making during inference. This is because each layer deeper into the model abstracts away the tangible relationship we end-users can attribute to the groups of pixels in the original input image. In critical applications like medical imaging, where understanding the reasoning behind a model's decision is essential, this lack of transparency can be a significant drawback, especially when compared to traditional image processing methods which have a clear mechanistic procedure which is both interpretable and explainable, like \acrshort{GPET} and \acrshort{MMCQ}. However, segmentation models like DeepGPET are still more interpretable than deep learning-based disease detection models, since DeepGPET outputs a visual result which can be critiqued, unlike the output of, say, a continuous risk score for disease.
            
            Finally, while fine-tuning from large natural image datasets such as ImageNet remains at the forefront of state-of-the-art deep learning model training \cite{cheplygina2019not}, foundational models are making traction in the field of medical image segmentation \cite{ma2024segment, zhou2023foundation}. Pre-trained models from datasets like ImageNet use supervised learning to train models on millions of natural images for classification. This ultimately means the features learned by the model are attributed to natural images. Conversely, foundational models take a self-supervised approach to model training, learning feature representations without distinct labels and, compared to ImageNet-based model pre-training, are more task- and domain-specific to medical images \cite{ma2024segment} and indeed retinal images \cite{zhou2023foundation}. These models can utilise large label-free datasets and have previously been shown to outperform similar models following supervised learning \cite{he2022masked}. 
            
            Zhou, et al. \cite{zhou2023foundation} found that their foundational model, RETFound, already showed superior performance over traditional ImageNet-based supervised (and self-supervised) learning for the detection of ocular and systemic diseases using \acrshort{OCT} images \cite{zhou2023foundation}. With freely available models pre-trained on retinal images \cite{zhou2023foundation}, we suspect that foundational models will undoubtedly improve the performance of future OCT choroid segmentation models. Additionally, while these models are traditionally thought to require significant compute resources to train, recent evidence suggests that this cost can be significantly reduced without harming performance and generalisation \cite{engelmann2024training}. Thus, this approach may be worthwhile for \acrshort{OCT} image segmentation problems.

        \end{mysubsection}

        \begin{mysubsection}[]{Outputs}

            In this chapter, there have been two publication outputs, one of which has been published and peer-reviewed (by December 2024) and another which is currently in peer-review but has been uploaded as a pre-print on Arxiv. The author's name is in bold type, with the names of any co-lead authors underlined.
    
            \begin{itemize}\setlength\itemsep{0em}
    
                \item \underline{\textbf{Burke, Jamie}}, \underline{Justin Engelmann}, Charlene Hamid, Megan Reid-Schachter, Tom Pearson, Dan Pugh, Neeraj Dhaun et al. ``\textit{An Open-Source Deep Learning Algorithm for Efficient and Fully Automatic Analysis of the Choroid in Optical Coherence Tomography.}'' Translational Vision Science \& Technology 12, no. 11 (2023): 27-27.
            
                \item \underline{\textbf{Burke, Jamie}}, Justin Engelmann, Charlene Hamid, Diana Moukaddem, Dan Pugh, Neerah Dhaun, Niall Strang et al.  ``\textit{OCTolyzer: Fully automatic analysis toolkit for segmentation and feature extracting in optical coherence tomography and scanning laser ophthalmoscopy data.}'' arXiv preprint arXiv:2407.14128 (2024). Submitted and in peer-review at Elsevier's Medical Image Analysis
                
            \end{itemize}
    
             The software associated with this method, DeepGPET, has been published as open-source and is freely available on GitHub \href{https://github.com/jaburke166/deepgpet}{here}. Similarly, OCTolyzer is freely available on GitHub \href{https://github.com/jaburke166/OCTolyzer}{here}.
        
        \end{mysubsection}

        \begin{mysubsection}[]{Executive summary}

            In this chapter, we distilled the previous semi-automatic approach to choroid region segmentation, \acrshort{GPET}, into a fully automatic deep learning-based method, DeepGPET. DeepGPET removes the need for manual segmentation of the choroidal space, and prevents any human subjectivity from being injected into the segmentation procedure. DeepGPET performed well on unseen test data related to systemic health, and was preferred by a clinical ophthalmologist when compared with it's predecessor, \acrshort{GPET}, on various examples of disagreement. We also found that DeepGPET was reproducible across \acrshort{SD-OCT} macular data and, interestingly, peripapillary data. This was observed both at the population- and eye-level, with clinically insignificant errors. Importantly, the reproducibility analysis conducted provides an interpretable measure of DeepGPET's measurement variability, enabling the research community to differentiate between measurement error and true biological change in future studies.

            While DeepGPET was a marked improvement over \acrshort{GPET}, it was only trained on examples related to systemic disease and thus has not been evaluated against abnormal retinae. Moreover, while initial experimentation showed DeepGPET's robustness to macular, \acrshort{SS-OCT} with good choroid visualisation, this was not observed for more challenging cases. Importantly, DeepGPET was unable to automate computation of segmentation-derived measurements as the fovea required detection --- typically manually --- thus injecting human subjectivity into the measurement procedure. Future work would look to validate DeepGPET on examples related to retinal pathology, extend the model to be generalisable to \acrshort{SS-OCT} data and automate fovea detection to permit end-to-end measurement of the choroidal space from raw \acrshort{OCT} B-scans.

        \end{mysubsection}
    
    \end{mysection}
	
\end{mychapter}

\begin{mychapter}[]{Choroidalyzer: End-to-end choroidal analysis in optical coherence tomography} \label{chp:chapter-choroidalyzer}

    \begin{mysection}[]{Introduction}

        In this work we have so far considered solutions for choroid region and vessel segmentation separately (choroid region: \acrshort{GPET}, DeepGPET and choroid vessel: \acrshort{MMCQ}). While these approaches are sufficient for their own use-case, they do not leverage information shared between both segmentation tasks, given that the choroidal vasculature is interspersed within the choroidal space. In an optimal setting, a unified approach to choroid analysis through solving both tasks simultaneously in an end-to-end framework would be desirable (and significantly more practical), thus utilising information from both the choroidal space and vasculature during model inference. 
        
        Furthermore, in addition to accurate and reliable segmentation of the choroidal region and vessels, there are two more necessary steps that are often overlooked, namely fovea detection and computation of choroidal metrics. Choroid-derived measurements require careful consideration of the \acrshort{ROI} used for analysis, typically using the location of the fovea as a reference guide. This is to ensure measurements are standardised for proper comparison across different studies, populations and imaging devices. Moreover, \acrshort{OCT} B-scans are not necessarily perfectly centred, with the chorioretinal structure presented at a skew or appearing curved with respect to the image axis, which is often ignored while deriving measurements \cite{cheong2018novel, yiu2014characterization, aksoy2023choroidal, shoshtari2021impact, rahman2011repeatability, cho2014influence}. We know that this can have an inordinate impact on downstream measurements given the 1:3 ratio between axial and lateral pixel length-scales in commercial \acrshort{OCT} imaging devices (section \ref{subsec:ch1_INTRO_measure_choroid}). Thus, once the region, vessels, and fovea are extracted, choroidal metrics should be computed in a fovea-centred \acrshort{ROI} which must account for key details like the skew or curvature of the chorioretinal structure and the pixel length-scales of each scan.
    
        While DeepGPET and \acrshort{MMCQ} are able to provide fast, accessible, reproducible and clinically meaningful solutions to choroid region and vessel segmentation, respectively, they ultimately only solve a fraction of the problem for choroidal analysis in \acrshort{OCT} image sequences. 
        
        As discussed in section \ref{subsec:intro_litreview}, the application of deep learning based algorithms to the problem of choroidal segmentation in \acrshort{OCT} image sequences is not new (reviews: Eghtedar, et al. \cite{eghtedar2022update} and Selvam, et al. \cite{selvam2023artificial}). However, in the wider research community the steps required for measuring the choroid and locating the fovea are commonly performed using separate tools for region (table \ref{tab:INTRO_region_methods}) and vessel segmentation (table \ref{tab:INTRO_vessel_methods}), often using non-standardised manual approaches for fovea detection or \acrshort{ROI} definition.

        \begin{figure}[tb]
        \begin{minipage}{\textwidth}
            \centering
            \includegraphics[width=\textwidth]{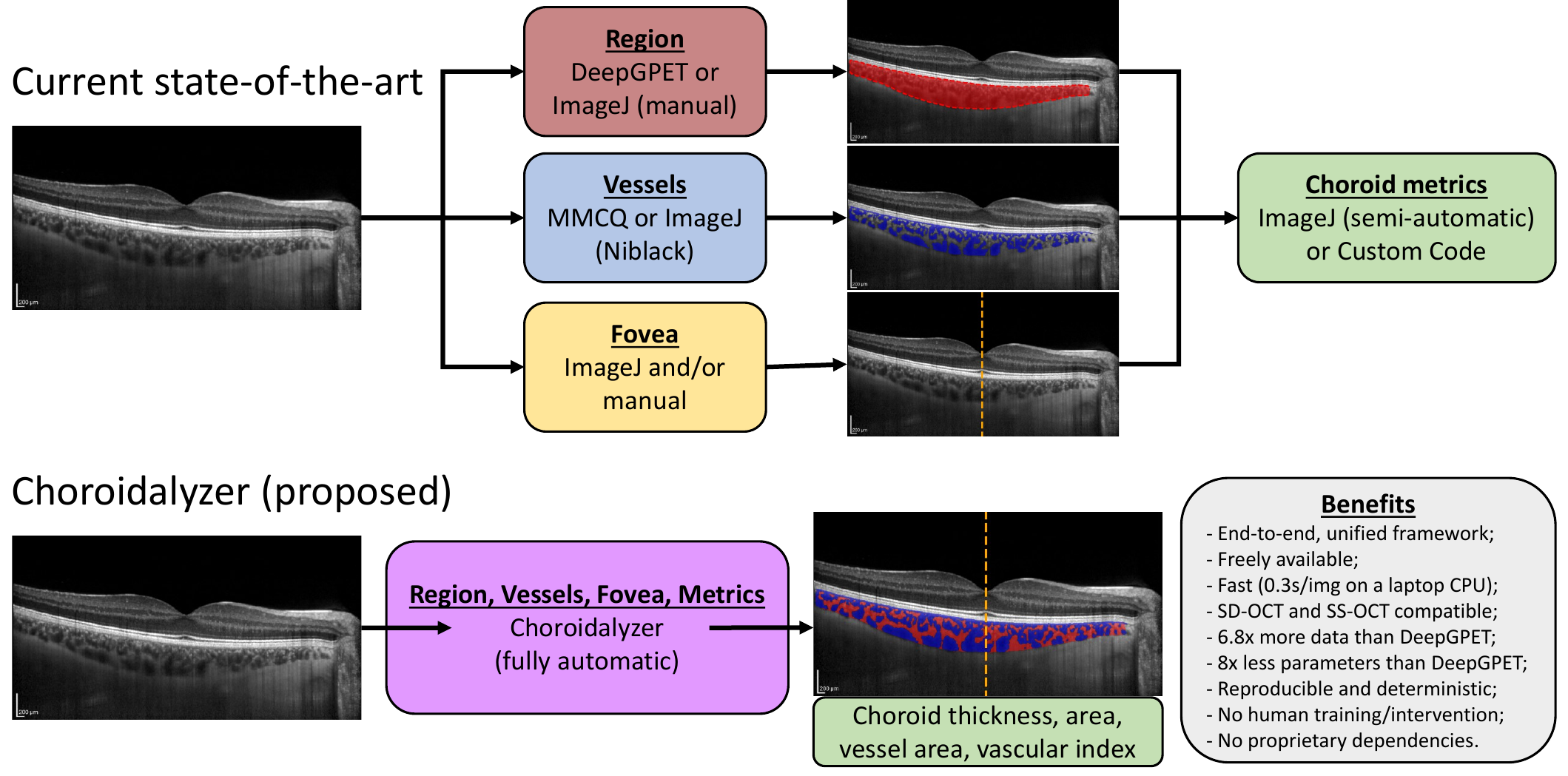}
            \caption[Choroidalyzer's proposed benefits over current state-of-the-art.]{A comparison between the current state-of-the-art (top) --- defined here as what's most widespread and available to the research community --- in choroidal image analysis in \acrshort{OCT} image sequences (top), and the proposed method, Choroidalyzer (bottom). To obtain choroidal metrics in a fovea-centred \acrshort{ROI}, researchers currently need to combine many different tools. Choroidalyzer unifies everything into an end-to-end pipeline that is very fast and convenient to use.}
            \label{fig:CHOROID_schematic}
        \end{minipage}
        \end{figure}
        
        While recently there has been a push for synthesising the two segmentation tasks into a single model (table \ref{tab:INTRO_region_vessel_methods}), there are still inherent issues in these methods. For example, most are either closed-source or are not packaged well for end-user accessibility to the research community \cite{xuan2023deep, arian2023automatic}. Only a small handful of these approaches report their reproducibility \cite{xuan2023deep, zheng2021deep}, and some still consider each task as separate ones but package them together \cite{xuan2023deep, zheng2021deep, arian2023automatic}, depending on semi-automatic Niblack for vessel segmentation. In the case for using Niblack, only two studies reports the parameters they use \cite{muller2022application, arian2023automatic}, making it impossible to reproduce the results of others in confidence. For those which do provide end-to-end region and vessel analysis using deep learning, their ground truths were generated manually \cite{khaing2021choroidnet, zhu2022synergistically, wang2023choroidal, wen2024transformer}, which can be subject to human bias \cite{maloca2023human}. Importantly, very few consider the importance of automatically measuring fovea-centred, choroid-derived measurements through detecting the fovea on the B-scan \cite{xuan2023deep}. All of these aforementioned reasons contribute to the current lack of standardisation of choroid-derived measurements in \acrshort{OCT} image sequences.

        Thus, we address these current issues by proposing Choroidalyzer, an end-to-end pipeline for choroidal analysis. Choroidalyzer consists of a single deep learning model that simultaneously segments the choroidal region and vessels and detects the fovea location, combined with all the code needed to extract choroidal thickness, area, and \acrshort{CVI} in a fovea-centred \acrshort{ROI}. Figure \ref{fig:CHOROID_schematic} shows how Choroidalyzer improves on the current state-of-the-art by providing a comprehensive solution for all elements of choroidal analysis. To our knowledge, Choroidalyzer is the first open-source tool for fully automatic, end-to-end choroidal analysis which reports interpretable reproducibility metrics, and is developed using a large dataset from 6 datasets related to systemic health, utilising both \acrshort{SD-OCT} and \acrshort{SS-OCT} data from two market-leading manufacturers for \acrshort{OCT} imaging systems (Heidelberg Engineering, Topcon). 

        \begin{mysubsection}[]{The UNet Deep Learning Model}\label{subsec:CHOROID_Intro_UNet}
            \begin{figure}[tb]
                \centering
                \includegraphics[width=\linewidth]{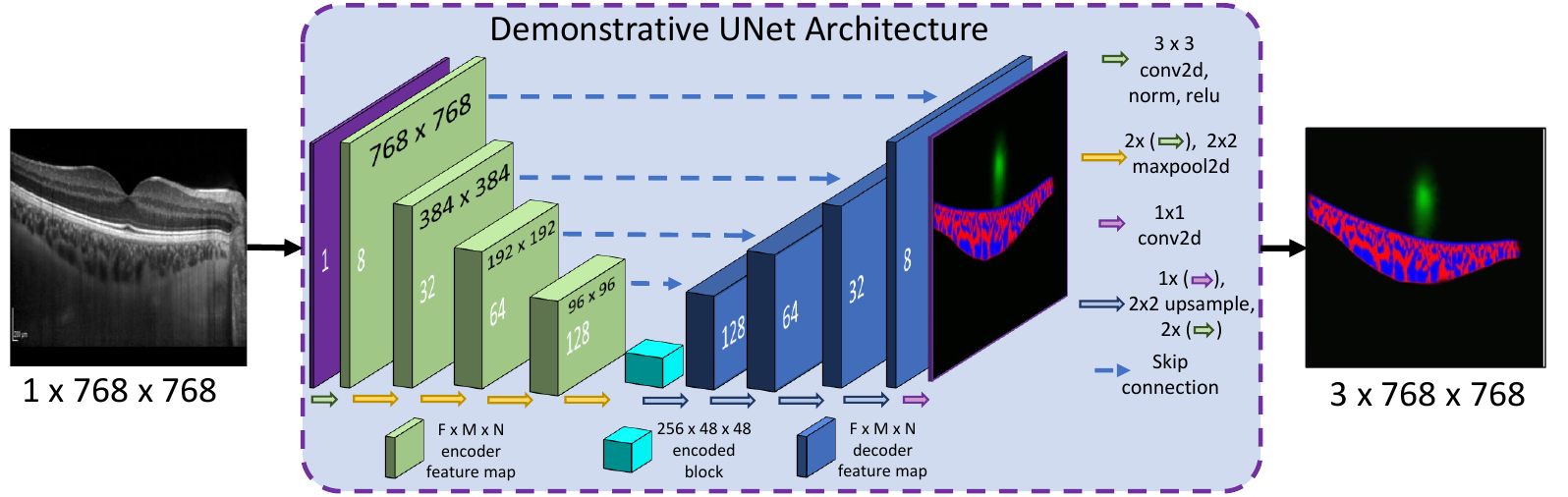}
                \caption[Demonstrative UNet architecture for choroid segmentation.]{Demonstrative UNet deep learning architecture which is used in this chapter, taking an input \acrshort{OCT} B-scan of resolution 1 $\times$ 768 $\times$ 768 and outputting a probability segmentation map for three segmentation tasks, choroid region, vessel and fovea detection with resolution 3 $\times$ 768 $\times$ 768.}
                \label{fig:CHOROID_UNet_demo}
            \end{figure}
        
            We briefly discussed deep learning for image segmentation and the UNet architecture in section \ref{subsec:DEEPGPET_intro_DL}, focusing on the general advantages and disadvantages which deep learning methods provide for medical image segmentation over traditional image processing methods. DeepGPET was a pre-trained UNet architecture which was trained from scratch on ImageNet \cite{deng2009imagenet}, while in this chapter, we develop a custom UNet deep learning architecture which Choroidalyzer leverages to segment the choroidal space, vessels and simultaneously detect the fovea on \acrshort{OCT} B-scans. 
            
            The UNet architecture has become a cornerstone in deep learning for medical image segmentation due to several key factors that make it particularly well-suited for this domain. Consequently, the UNet architecture has been successfully applied across multiple medical imaging modalities, such as in MRI \cite{walsh2022using}, CT \cite{manjunath2022modified}, X-ray \cite{liu2022automatic}, ultrasound \cite{ashkani2022fast} and indeed \acrshort{OCT} \cite{viedma2022deep}.

            Medical image segmentation requires pixel-level precision to accurately delineate anatomical structures. The UNet deep learning architecture is a type of \acrshort{CNN} for pixel-level semantic image segmentation, and excels in this regard due to its design, which captures both the global context and fine-grained details within images. The \acrshort{CNN} architecture is constructed in a U-shape with two distinctive arms, the \textit{encoder} and \textit{decoder} arms. The encoder (contracting) arm takes the input image and sequentially applies blocks of convolutional filters, learning low- to high-level descriptor features at every progressive block, and simultaneously pooling these features into smaller spatial dimensions to collect higher-level, global contextual features. After encoding, what is left is a number of coarse grids (feature maps) representing abstract, space-invariant and global features of interest from the input --- note that these are large in the number of feature map, but small in spatial dimension \cite{ronneberger2015u}. 
             
            The decoder (expanding) arm uses upsampling (via transposed convolutions or bi-linear interpolation) to increase the spatial dimensions while reducing the number of feature maps in a symmetrical and sequential manner to the encoder arm. This has the effect of decoding the global representations learned by the encoder into local ones, eventually outputting a pixel-level probability map of the model's belief at the original input pixel resolution. The symmetry is leveraged by combining (simply through addition) the upsampled features with corresponding high-resolution features from the encoder arm (through skip connections), enabling precise localisation and facilitating the output of high-fidelity segmentation maps. The additional U-shaped architecture is a significant improvement over traditional, fully convolutional networks which only have a contracting (encoder) path \cite{ronneberger2015u}. 

            The inclusion of skip connections is one of UNet's most distinctive features. These connections directly link the encoder layers to the corresponding decoder layers, allowing the model to bypass certain levels of abstraction and preserve spatial information from earlier layers. In medical image segmentation fine details, such as the boundaries between tissues like the choroid extravascular and vascular compartments, need to be accurately segmented and these skip connections are vital in facilitating this. They help the model retain high-resolution features that might otherwise be lost, leading to more accurate and detailed segmentations. Thus, this architecture ensures that important spatial details are preserved, which is crucial for the high accuracy required in medical applications \cite{ronneberger2015u}. 
             
            Figure \ref{fig:CHOROID_UNet_demo} shows a detailed and demonstrative diagram of the UNet deep learning architecture for the method proposed in this chapter. This UNet architecture presented here is a deep learning model with a depth of 4, i.e. has four distinct encoder (green) and decoder (blue) blocks. Note the symmetrical nature of the architecture, where each encoder block has a corresponding decoder block at the same spatial resolution, which allows skip connections (blue dashed lines). The black text denotes the spatial resolution of the encoder and decoder feature maps, which are scaled by $\nicefrac{1}{2}$ and 2 as you move between encoder blocks and decoder blocks, respectively. As the spatial resolution changes in one direction, the number of feature maps (white text) changes in the opposite direction, i.e. spatial maps become coarser and increase in quantity with depth into the encoder arm, as seen by the final encoded (cyan) block with 256 feature maps at a spatial resolution of 48 $\times$ 48. 
             
            In practice, scaling the feature maps by a higher magnitude would be more optimal in order to reduce memory and improve training/inference time. Additionally, having a higher depth (more convolutional blocks in the encoder and decoder arms) would likely lead to improved performance by having a very coarse, final encoded block to maximise spatial invariance and global representation learning \cite{loos2024demystifying}. This is entirely to do with the receptive field of each neuron in the network \cite{loos2024demystifying}, which is not within the scope of this thesis.

        \end{mysubsection}
        
    \end{mysection}

    \begin{mysection}[]{Study populations}

        \begin{mysubsection}[]{Model training and evaluation} \label{subsec:chp5_CHOROID_tvt_pop}
        
            The dataset used for Choroidalyzer's training and evaluation contains 5,600 \acrshort{OCT} B-scans of 233 participants from 6 cohorts of healthy and diseased individuals (unrelated to ocular pathology):
            \begin{itemize}\setlength\itemsep{0.1em}
                \item \textbf{\acrshort{OCTANE}} \cite{dhaun2014optical}, a longitudinal cohort ($N$=46) study investigating choroidal changes in renal transplant recipients and donors. Note that this cohort is one participant fewer and 3 H-line scans greater than DeepGPET's \acrshort{OCTANE} cohort from table \ref{tab:DEEPGPET_pop}. This is because of an administrative error during data curation and does not impact the rest of the analysis;
                \item \textbf{Diurnal Variation for Chronic Kidney Disease (\acrshort{DVCKD})} \cite{dhaun2014optical}, a cohort ($N$=20) of young participants collected as part of the \acrshort{OCTANE} study (but mutually exclusive to the longitudinal data above), for assessing diurnal variation in the choroid of healthy volunteers to support research related to chronic kidney disease \cite{farrah2023choroidal}; 
                \item \textbf{Normative}, a detailed \acrshort{OCT} examination of the author of this thesis with informed consent ($N$=1);
                \item \textbf{i-Test} \cite{dhaun2014optical}, a cohort ($N$=21) of pregnant women evaluating retinochoroidal changes in normative, pre-eclamptic or fetal growth restricted pregnancies;
                \item \textbf{Prevent Dementia} \cite{ritchie_prevent_2012}, a mid-life cohort ($N$=121), approximately half of whom have risk of developing later life Alzheimer's disease; 
                \item \textbf{Glasgow Caledonian University Topcon (\acrshort{GCU} Topcon)} \cite{moukaddem2022comparison}, a cohort ($N$=24) of healthy participants investigating diurnal variation of the choroid in individuals of varying refractive status.
            \end{itemize}
            All studies adhered to the Declaration of Helsinki and received relevant ethical approval, and informed consent from all subjects was obtained in all cases from the host institution.
    
            Three \acrshort{OCT} device types were used from two device manufacturers: the \acrshort{SD-OCT} Spectralis Standard and FLEX Modules (both \acrshort{OCT}1 systems from Heidelberg engineering, Heidelberg, Germany), and the \acrshort{SS-OCT} DRI Triton Plus (Topcon, Tokyo, Japan). 
            
            The majority of the Heidelberg data acquisition has been described previously in section \ref{subsec:chp4_DEEPGPET_pop}. Briefly, fovea-centred horizontal- and vertical-line scans were collected with active eye tracking using an \acrshort{ART} of 100, covering a 30-degree \acrshort{FOV}. Posterior pole volume scans covered a 30 $\times$ 20 degree rectangular \acrshort{ROI} using 31 consecutive scans with an \acrshort{ART} of 50 for the i-Test cohort and 61 line scans with an \acrshort{ART} of 9 for the \acrshort{OCTANE} and Normative cohorts. All Heidelberg data had signal quality $\geq$ 15 and was collected at a pixel resolution of 768 $\times$ 768 pixels, except for i-Test which had a native resolution of 496 $\times$ 768, which was then padded vertically to 768 $\times$ 768. Additionally, we collected peripapillary \acrshort{OCT} B-scans in the \acrshort{DVCKD} cohort, whose acquisition has been previously described in section \ref{subsec:ch4_DEEPGPET_repr_pop}. Briefly, \acrshort{EDI-OCT} circular B-scans centred at the optic nerve head were collected for each participant's right eye at three times on the same day, using active eye tracking with an \acrshort{ART} of 100.
            
            For the Topcon device, during acquisition, a twelve-line, fovea-centred \acrshort{OCT} radial scan was captured, starting from a horizontal-line B-scan and rotating clockwise in 30 degree intervals. Each B-scan covered approximately a 9 mm \acrshort{FOV} laterally and had a resolution of 992 $\times$ 1024 pixels which was cropped horizontally by 32 pixels and resized to the resolution of the Heidelberg scans of 768 $\times$ 768. Any set of 12 B-scans with an average image quality score less than 88 --- reported by the \acrshort{SNR} index built-in to the Topcon scanning device (Topcon, Tokyo, Japan) --- was excluded. 

            \begin{mysubsubsection}[]{Ground truth labels} \label{subsubsec:CHOROID_gtlabels}

                \begin{figure}[tb]
                    \centering
                    \includegraphics[width=\linewidth]{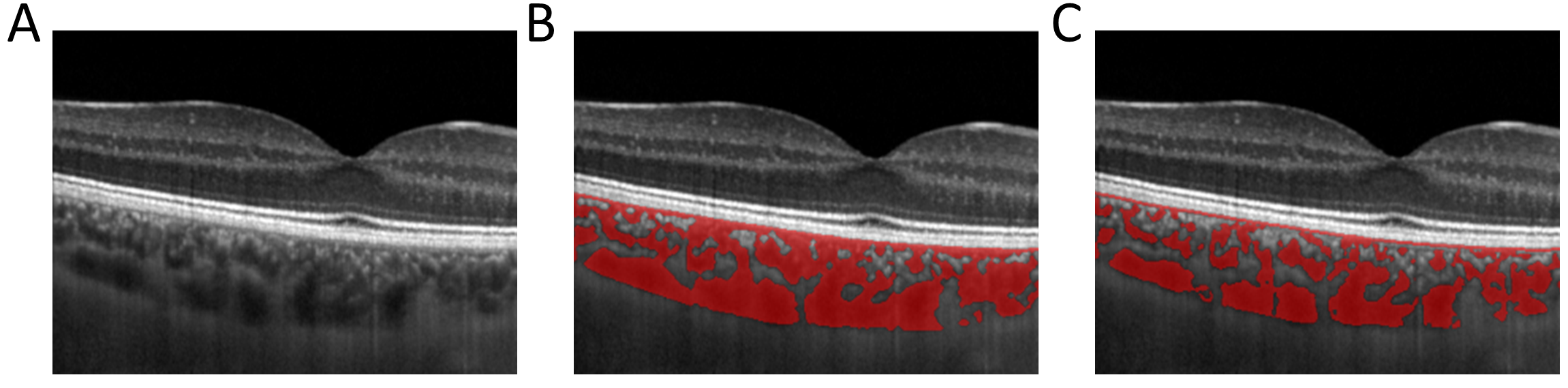}
                    \caption[Improved \acrshort{MMCQ} vessel segmentation by varying brightness/contrast and taking majority vote.]{A comparison between applying \acrshort{MMCQ} with and without the majority voting scheme. (A) \acrshort{OCT} B-scan snippet, (B) default \acrshort{MMCQ} application, (C) majority voting scheme \acrshort{MMCQ} application.}
                    \label{fig:CHOROID_better_vessels}
                \end{figure}
                 
                The choroidal region in an \acrshort{OCT} B-scan has been defined previously in section \ref{subsec:ch1_INTRO_measure_choroid}, as well as the suprachoroidal space. Because of the inconsistent appearance of the suprachoroidal space \cite{yiu2014characterization, chandrasekera2018posterior}, we consider this space not to be part of the choroid itself, which ensures that Choroidalyzer learns a consistent definition for the choroidal space. Ground truth labels for the choroidal space were generated using DeepGPET after thresholding the resulting binary segmentation maps with a value of 0.5. The choroidal vasculature in an \acrshort{OCT} B-scan has already been defined in section \ref{subsec:ch1_INTRO_measure_choroid}, and the ground truth labels were generated using \acrshort{MMCQ}. 
                
                Figure \ref{fig:CHOROID_better_vessels} shows a snippet from an \acrshort{OCT} B-scan (A), with the original vessel segmentation from \acrshort{MMCQ} (B) and after applying the majority voting scheme based on varying contrast and brightness adjustments (C). To improve the fidelity and robustness of our vessel segmentation labels, we randomly varied the brightness and contrast of each \acrshort{OCT} B-scan before application of \acrshort{MMCQ}. We used 5 linearly spaced gamma levels to fix the mean brightness of each image between 0.2 and 0.5 and simultaneously altered the contrast using 5 linearly spaced factors between 0.5 and 3. A 3:2 majority vote for vessel label classification was used across all 25 variants. Robustness is achieved by averaging out any erroneous over- or under-segmented pixels produced from each brightness and contrast combination (which produce a different set of image statistics and noise distribution). 

                 \begin{figure}[!t]
                    \centering
                    \includegraphics[width=\linewidth]{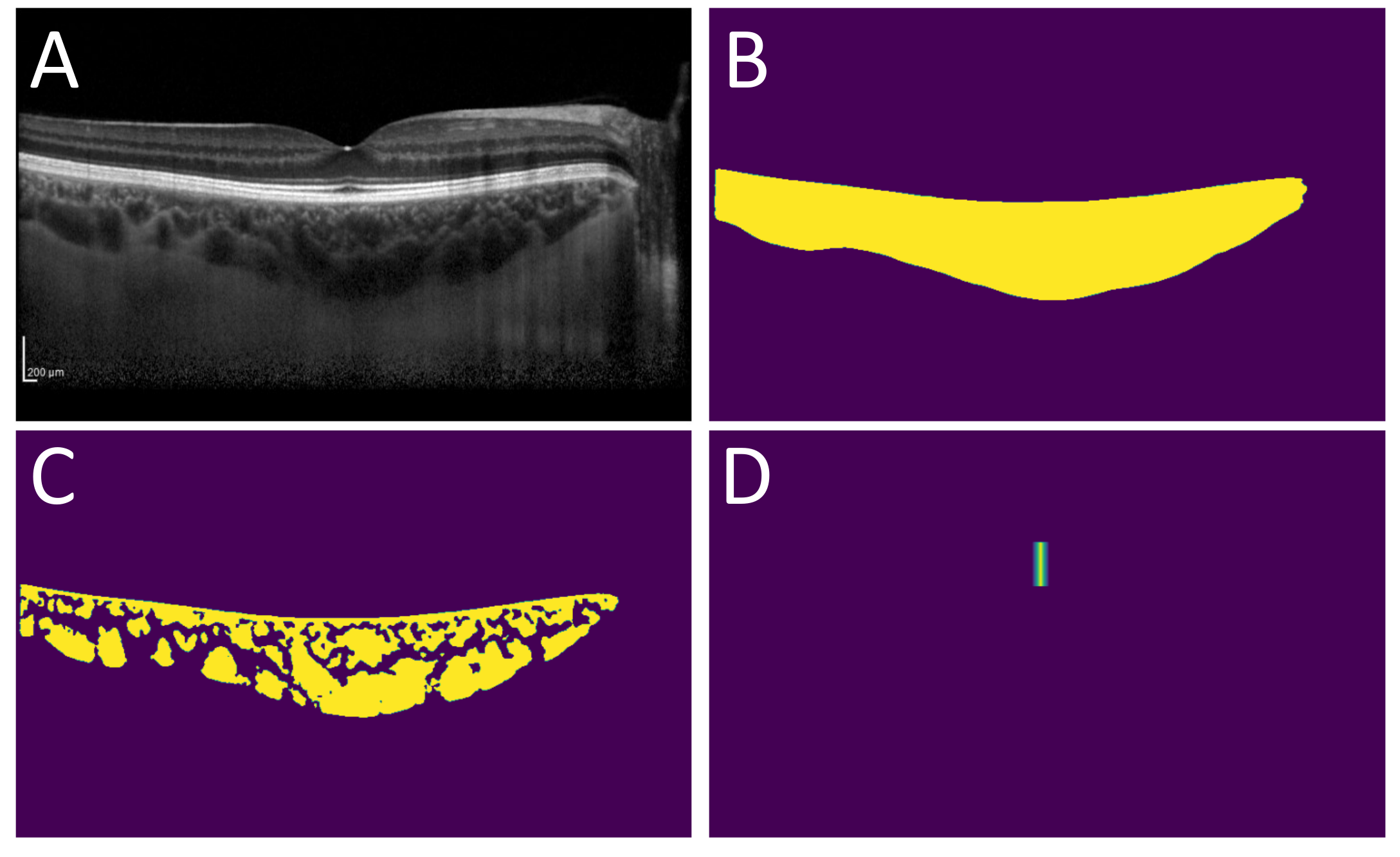}
                    \caption[Choroidalyzer's ground truth labels.]{Ground truth segmentation labels for an \acrshort{OCT} B-scan (A), showing region label (B), vessel label (C) and fovea label (D).}
                    \label{fig:CHOROID_gt_labels}
                \end{figure}
                
                The foveal pit has been defined previously in section \ref{subsec:ch1_INTRO_measure_choroid}. For a fovea-centred \acrshort{OCT} B-scan, the lateral position of the foveal pit was defined as the horizontal (column) pixel index which aligned with the deepest point of the foveal pit depression \cite{xuan2023deep}. For \acrshort{OCT} B-scans centred at the fovea (i.e. horizontal, vertical and radial scans), the foveal column location was detected manually using a custom-built graphical user interface in Python (version 3.11). Those not centred at the fovea had their fovea label set as the origin $(0,0)$ ($N$=2178). The fovea is only a single point which would be difficult for a segmentation model to learn, as predicting close to 0 for all pixels would yield virtually the same loss as a perfect prediction. Thus, we create a target 51 pixels high and 19 pixels wide centred at the fovea location. The exact fovea location is set to 1, the column of the fovea location to 0.95 for 25 pixels in both vertical directions, and adjacent columns to 0.95-0.1$d$, where d is the column distance from the fovea, for $d=1,\dots,9$.

                Figure \ref{fig:CHOROID_gt_labels} shows the region, vessel and fovea ground truth segmentation labels for an \acrshort{OCT} B-scan. 
        
            \end{mysubsubsection}

            \begin{mysubsubsection}[]{Sample derivation} \label{subsec:CHOROID_sample_deriv}
            
                \begin{figure}[tb]
                \centering
                \includegraphics[width=\linewidth]{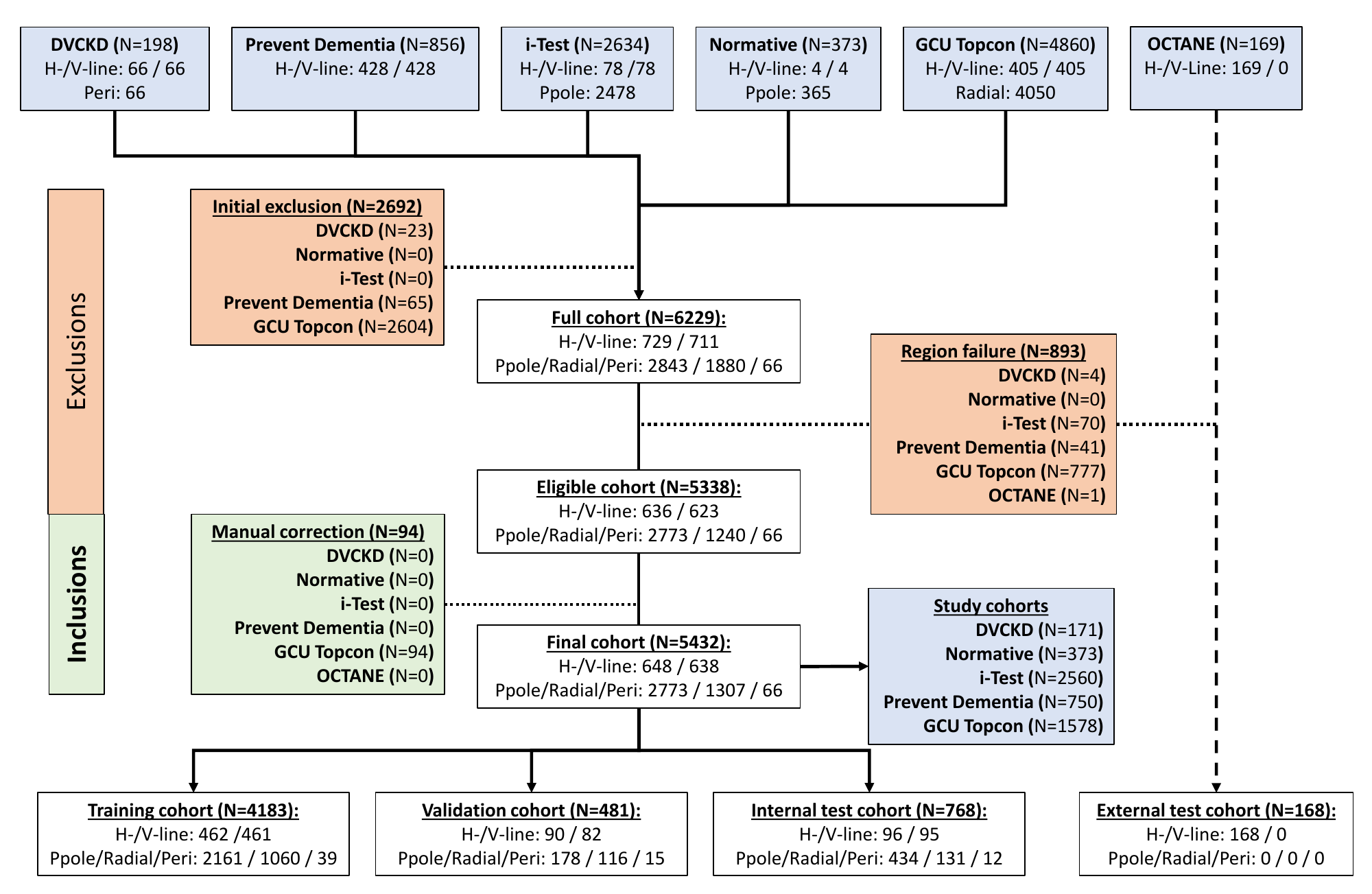}
                \caption[Sample derivation flowchart for Choroidalyzer's modelling dataset.]{Sample derivation flowchart of how \acrshort{OCT} data from six cohorts were combined to create Choroidalyzer's modelling dataset.}
                \label{fig:CHOROID_sample_deriv}
                \end{figure}

                Figure \ref{fig:CHOROID_sample_deriv} illustrates the sample derivation for the dataset used to train and evaluate Choroidalyzer. A total of 9090 \acrshort{OCT} B-scans were available from all six cohorts. Before ground truth labels were generated, an initial exclusion criteria was set as 1) Choroid-Sclera boundary was invisible, 2) the choroid was clipped or cropped and 3) initial experimentation with DeepGPET in \acrshort{SS-OCT} resulted in segmentation failure. During data extraction of the six cohorts, 23 and 65 B-scans from the \acrshort{DVCKD} and Prevent Dementia cohorts, respectively, were excluded due to invisible Choroid-Sclera boundary and choroid clipping/cropping. 
                
                DeepGPET's generalisability on \acrshort{SS-OCT} data was tested on a small sample of \acrshort{SS-OCT} data from the \acrshort{GCU} Topcon cohort (see section \ref{subsec:deepgpet_eval_general}). Through randomly sampling 6 \acrshort{OCT} B-scans per eye, per participant, manual adjudication (from author J.B.) found that DeepGPET failed to segment the choroids from 13 out of 24 participants. Failure was defined as significant under-/over-segmentation of the choroidal space (figure \ref{fig:DEEPGPET_topcon_large_small}, bottom). All data from these participants were excluded from Choroidalyzer's datasets ($N$=2604). Thus, initial exclusion saw a total of $N$=2692 B-scans removed, leaving $N$=6229 B-scans to be processed by DeepGPET and \acrshort{MMCQ} for choroidal space and vessel segmentation. 

                Ground truth labels for the choroidal space were reviewed by author J.B. and $N$=893 B-scans were excluded because of segmentation failure from DeepGPET. The majority of these failures were from large choroids in the \acrshort{GCU} Topcon cohort ($N$=777). The remaining exclusions ($N$=116) from the five other cohorts were also from large choroids with poor Choroid-Sclera boundary visibility. This left $N$=5506 B-scans to train and evaluate Choroidalyzer with.
                
                DeepGPET's lack of generalisability to \acrshort{SS-OCT} choroids of larger size resulted in an inordinately large exclusion rate of GCU Topcon data. Since the GCU Topcon cohort primarily targetted hyperopes during recruitment, many of the choroids were thus large and grainy in appearance at the point of the Choroid-Sclera boundary due to natural optical signal degradation and higher speckle noise at depth compared to \acrshort{EDI-OCT} data (figure \ref{fig:INTRO_SD_SS_SNR}), of which DeepGPET was primarily trained. Some examples of these large choroids can be found in the bottom row of figure \ref{fig:DEEPGPET_topcon_large_small} discussed in chapter \ref{chp:chapter-deepgpet} and also in figure \ref{fig:MMCQ_repr_exc} discussed in chapter \ref{chp:chapter-mmcq}. While disappointing to see such a large number of \acrshort{SS-OCT} data excluded from Choroidalyzer's model development, it also represents an obvious motivation for developing Choroidalyzer.

                Therefore, in order for Choroidalyzer to generalise well to large \acrshort{SS-OCT} choroids, author J.B. manually segmented 94 random \acrshort{OCT} B-scans from the 13 participants initially excluded from the \acrshort{GCU} Topcon cohort, and this was added to the overall dataset. Note that the initial expectation was to manually segment 100 B-scans, but a simple administrative error on behalf of the author lead to only 94 B-scans being segmented. Manual segmentation was performed semi-automatically using a custom-built graphical user interface in Python (version 3.11). For each of the \acrshort{RPE}-Choroid and Choroid-Sclera boundaries, the end-user selected points along the boundary and these were interpolated using a univariate spline whose polynomial degree was 3. 

                \begin{table}[tb]
                    \begin{adjustwidth}{-1in}{-1in}  
                    \centering
                    \scalebox{0.6}{\begin{tabular}{@{}lcccccc|c@{}}
\toprule
 & \acrshort{DVCKD} \cite{dhaun2014optical} & Normative & i-Test \cite{dhaun2014optical} & Prevent Dementia \cite{ritchie_prevent_2012} & \acrshort{GCU} Topcon \cite{moukaddem2022comparison} & OCTANE \cite{dhaun2014optical} & Total \\
  \midrule
\multicolumn{1}{l}{Device manufacturer}& Heidelberg & Heidelberg & Heidelberg & Heidelberg & Topcon & Heidelberg & All \\
\multicolumn{1}{l}{Device type}  & Spectralis & FLEX & FLEX & Spectralis & DRI Triton Plus & Spectralis & All \\
\multicolumn{1}{l}{Subjects, N} & 20 & 1 & 21 & 121 & 24 & 46 & 233 \\
\multicolumn{1}{r}{Control/Case} & 20 / 0 & 1 / 0 & 11 / 10 & 56 / 65 & 24 / 0 & 0 / 46 & 112 / 121\\
\multicolumn{1}{r}{Male/Female} & 11 / 9 & 1 / 0 & 0 / 21 & 66 / 55 & 14 / 9* & 24 / 22 & 116 / 116* \\
\multicolumn{1}{r}{Right/Left eyes} & 20 / 0 & 1 / 1 & 21 / 21 & 117 / 115 & 22 / 21 & 46 / 0 & 227 / 158 \\
\multicolumn{1}{r}{Age (mean (SD))} & 21.4 (2.3) & 23.0 (0.0) & 32.8 (5.4) & 50.8 (5.6) & 21.8 (7.9) & 42.9 (13.7) & 47.5 (12.3) \\
\multicolumn{1}{l}{B-scans} &  &  &  &  &  &  & \\
\multicolumn{1}{r}{\acrshort{EDI} / non-\acrshort{EDI} / \acrshort{SS-OCT}}  & 171 / 0 / 0 & 247 / 126 / 0 & 2438 / 122 / 0 & 394 / 356 / 0 &  0 / 0 / 1,578 & 168 / 0 / 0 &  3,418 / 604 / 1,578 \\
 & & & & & & & \\
\multicolumn{1}{r}{Horizontal/Vertical} & 55 / 50 & 4 / 4 & 76 / 76 & 381 / 369 & 132 / 139 & 168 / 0 & 816 / 638 \\
\multicolumn{1}{r}{Volume/Radial/Peripapillary} & 0 / 0 / 66 & 365 / 0 / 0 & 2,408 / 0 / 0 & 0 / 0 / 0 & 0 / 1,307 / 0 & 0 / 0 / 0 & 2,773 / 1,307 / 66 \\
\multicolumn{1}{r}{Total} & 171 & 373 & 2,560 & 750 & 1,578 & 168 & 5,600\\
\bottomrule
\end{tabular}}

                    \end{adjustwidth}
                    \caption[Choroidalyzer's modelling dataset used for training and evaluation.]{Overview of population characteristics. *: One participant's sex from the \acrshort{GCU} Topcon cohort was not recorded.}
                    \label{tab:CHOROID_pop}
                \end{table}

                \begin{table}[tb]
                    \begin{adjustwidth}{-1in}{-1in}  
                    \centering
                    \scalebox{0.6}{\begin{tabular}{@{}lcccc|c@{}}
\toprule
 & Training & Validation & Testing & External test & Total \\
 \midrule
\multicolumn{1}{l}{Subjects} & 122 & 28 & 37 & 46 & 233 \\
\multicolumn{1}{r}{Male/Female} & 64 / 57* & 12 / 16 & 16 / 21 & 24 / 22 & 116 / 116* \\
\multicolumn{1}{r}{Control/Case} & 76 / 46 & 16 / 12 & 20 / 17 & 0 / 46 & 112 / 121 \\
\multicolumn{1}{r}{Right/Left eyes} & 117 / 107 & 27 / 23 & 37 / 28 & 46 / 0 & 227 / 158 \\
\multicolumn{1}{r}{Standard/FLEX/DRI Triton Plus} & 88 / 14 / 20 & 24 / 2 / 2 & 29 / 6 / 2 & 46 / 0 / 0 & 187 / 22 / 24 \\
\multicolumn{1}{r}{Heidelberg/Topcon} & 102 / 20 & 26 / 2 & 35 / 2 & 46 / 0 & 209 / 24 \\
\multicolumn{1}{r}{Age (mean (\acrshort{SD}))} & 40.7 (14.2) & 42.5 (11.9) & 44.5 (13.4) & 47.5 (12.3) & 42.9 (13.4) \\
\multicolumn{1}{l}{Cohort} &  &  &  &  &  \\
\multicolumn{1}{r}{OCTANE \cite{dhaun2014optical}} & 0 & 0 & 0 & 46 & 46 \\
\multicolumn{1}{r}{\acrshort{DVCKD} \cite{dhaun2014optical}} & 12 & 4 & 4 & 0 & 20 \\
\multicolumn{1}{r}{Normative} & 1 & 0 & 0 & 0 & 1 \\
\multicolumn{1}{r}{i-Test \cite{dhaun2014optical}} & 13 & 2 & 6 & 0 & 21 \\
\multicolumn{1}{r}{Prevent Dementia \cite{ritchie_prevent_2012}} & 76 & 20 & 25 & 0 & 121 \\
\multicolumn{1}{r}{\acrshort{GCU} Topcon \cite{moukaddem2022comparison}} & 20 & 2 & 2 & 0 & 24 \\
\multicolumn{1}{l}{B-scans} &  &  &  &  &  \\
\multicolumn{1}{r}{Spectralis/Flex/DRI Triton Plus} & 582 / 2,281 / 1,281 & 136 / 190 / 140 & 137 / 462 / 157 & 168 / 0 / 0 & 1,023 / 2,933 / 1,578 \\
\multicolumn{1}{r}{Heidelberg/Topcon} & 2,863 / 1,281 & 326 / 140 & 599 / 157 & 168 / 0 & 3,956 / 1,578 \\
&  &  &  &  & \\
\multicolumn{1}{r}{Horizontal/Vertical scans} & 462 / 461 & 90 / 82 & 96 / 95 & 168 / 0 & 816 / 638 \\
\multicolumn{1}{r}{Volume/Radial/Peripapilary scans} & 2,161 / 1,060 / 39  & 178 / 116 / 15 & 434 / 131 / 12 & 0 / 0 / 0 & 2,773 1,307 / 66\\
\multicolumn{1}{r}{Total B-scans} & 4,183 & 481 & 768 & 168 & 5,600 \\
\bottomrule
\end{tabular}}

                    \end{adjustwidth}
                    \caption[Choroidalyzer's training, validation and testing dataset split.]{Overview of population and image characteristics of the internal training, validation and test sets, and also the external test set. *: One participant's sex from the \acrshort{GCU} Topcon cohort was not recorded.}
                    \label{tab:CHOROID_pop_tvt_pop}
                \end{table}
                
                Thus, Choroidalyzer's dataset for training and evaluation consists of $N$=5600 B-scans. Table \ref{tab:CHOROID_pop} gives a detailed overview of the population and image characteristics of the final dataset used to develop Choroidalyzer, comprising 648 horizontal-line (H-line), 638 vertical-line (V-line), 2,773 posterior pole volume (Ppole), 1,307 radial and 66 peripapillary (Peri) scans. Notably, 15\% of the Heidelberg \acrshort{SD-OCT} data ($N$=4022) was captured without \acrshort{EDI} mode ($N$=604). 
                
                Approximately 10\% of the entire dataset was randomly sampled (495 B-scans) and stratified to represent different eyes and locations on the macula. This subsample was used to assess Choroidalyzer's fovea detection accuracy.

                Five of the six cohorts were split into training (4,183 B-scans, 122 subjects), validation (481 B-scans, 28 subjects) and \textit{internal} test sets (768 B-scans, 37 subjects) containing approximately 75, 10, and 15\% of the B-scans. We split the data at the subject-level, such that no individual ended up in more than one set to prevent data leakage during training and evaluation. The remaining cohort, \acrshort{OCTANE}, was entirely held-out as an \textit{external} test set (168 B-scans, 46 subjects). Table \ref{tab:CHOROID_pop_tvt_pop} presents the population and image characteristics for each of the four sets and describes their population and image acquisition statistics.

                External validation is helpful for testing the generalisability of a model using out-of-domain (out-of-distribution) data, i.e. \acrshort{OCT} images which are from a source external to the internal set \cite{cabitza2021importance}. In this particular case, the external test set here indicates how well Choroidalyzer might generalise to an entirely different cohort and disease context (\acrshort{CKD}). 
                
            \end{mysubsubsection}

        \end{mysubsection}
        
       \begin{mysubsection}[]{Reproducibility}

            \begin{table}[h]
            \begin{adjustwidth}{-1in}{-1in}  
            \centering
            \scalebox{0.8}{\begin{tabular}{p{4cm}p{5cm}p{5cm}}
\hline
\multirow{2}{*}{} & \multicolumn{2}{c}{Study} \\ \cline{2-3} 
 & i-Test \cite{dhaun2014optical} & \acrshort{GCU} Topcon \cite{moukaddem2022comparison} \\ \hline
Cohort demographics &  &  \\ \cline{1-1}
Participants (Eyes) & 60 (120) & 21 (33) \\
Right eyes (\%) & 60 (50) & 15 (45.5) \\
Age (\acrshort{SD}) & 34.7 (5.2) & 23.9 (4.2) \\
Sex, F (\%) & 60 (100) & 9 (43) \\
Ethnicity & 52 White, 6 Asian, 2 Mixed & 12 White, 5 Asian, 2 Black, 2 Middle Eastern \\
Gestation & 35.6 (3.4) & NA \\
Refractive status\textsuperscript{\textdagger} & 3 hyperopes, 31 emmetropes, 26 myopes & 9 hyperopes, 8 emmetropes, 3 myopes, 1 missing \\
Study purpose & Pre-eclamptic / normative pregnancy at late gestation & Diurnal variation \\
Control/Case & 45/15 & 21/0 \\
 &  &  \\
Image characteristics &  &  \\ \cline{1-1}
Device & Spectralis (Heidelberg) & DRI Triton Plus (Topcon) \\
\acrshort{OCT} Type & Spectral-domain & Swept-source \\
Scan Pattern & Macular volume & Macular radial \\
Mode & \acrshort{HRA}+\acrshort{OCT} & \acrshort{CFP}+\acrshort{OCT} \\
Time of day (Interval) & All in afternoon (1 minute) & 13 morning, 12 afternoon, 8 evening (5 minutes) \\
B-scans per eye & 31 (\acrshort{EDI}) / 61 (non-\acrshort{EDI}) & 12 \\
\acrshort{ART} & 50 (\acrshort{EDI}) / 12 (non-\acrshort{EDI}) & NA \\
\acrshort{SNR} & 35.61 & >88 \\
Pixel resolution & 496 $\times$ 768 & 992 $\times$ 1024 \\ \hline
\end{tabular}
}

            \end{adjustwidth}
            \caption[Choroidalyzer's reproducibility dataset.]{Population and image characteristics of the three samples used for assessing Choroidalyzer's reproducibility. \textsuperscript{\textdagger}: myopic/hyperopic status defined as $<-1 / >1$ dioptres. Dioptre measurements for i-Test sample was scan focus collected from the \acrshort{OCT} imaging device metadata,, and this was the spherical equivalent for the \acrshort{GCU} Topcon sample.}
            \label{tab:CHOROID_repr_pop}
            \end{table}
       
            Two cohorts were used to assess Choroidalyzer's reproducibility, the i-Test \cite{dhaun2014optical} and \acrshort{GCU} Topcon \cite{moukaddem2022comparison} cohorts. Note that these were two cohorts used for training Choroidalyzer, and that the reproducibility analysis was performed after Choroidalyzer was constructed and during continued recruitment for the i-Test study. Thus, the reproducibility analysis contains more i-Test participants than the dataset used for training. We used all available eyes which had repeated data available (at the point of analysis) from each study (120 eyes from 60 participants from the i-Test study and 33 eyes from 21 participants in the \acrshort{GCU} Topcon study). Table \ref{tab:CHOROID_repr_pop} describes the population and image characteristics for these two samples.
    
            The image acquisition for the i-Test reproducibility sample was previously described in section \ref{subsec:ch4_DEEPGPET_repr_pop}, which used an \acrshort{SD-OCT} Spectralis \acrshort{OCT}1 imaging device (Heidelberg Engineering, Heidelberg, Germany). Briefly, acquisition for an i-Test participant included a repeated \acrshort{OCT} volume scan (collected within 1 minute) per eye, with \acrshort{EDI} mode toggled on and off using active eye tracking with an \acrshort{ART} of 50 and 9, respectively, collecting 31 and 61 parallel line-scans, respectively. 
            
            Acquisition for the \acrshort{GCU} Topcon sample was described in section \ref{subsec:chp5_CHOROID_tvt_pop}, which used an \acrshort{SS-OCT} imaging device (Topcon, Tokyo, Japan). Briefly, acquisition for a \acrshort{GCU} Topcon participant collected a twelve-line, fovea-centred \acrshort{OCT} radial scan --- with the native resolution kept fixed at 992 $\times$ 1024, unlike in Choroidalyzer's modelling dataset. Repeated acquisitions were taken at approximately the same location during each visit and within 5 minutes of each other. Given repeated acquisition was not the primary aim of the \acrshort{GCU} Topcon study (of which 24 participants were originally included in Choroidalyzer's modelling dataset), only 21 participants had repeated \acrshort{OCT} data, with nine only in one eye and twelve in both eyes.
       \end{mysubsection}
        
    \end{mysection}

    \begin{mysection}[]{Choroidalyzer's deep learning model} \label{subsubsec:CHOROID_model}

        \begin{figure}[tb]
            \centering
            \includegraphics[width=\linewidth]{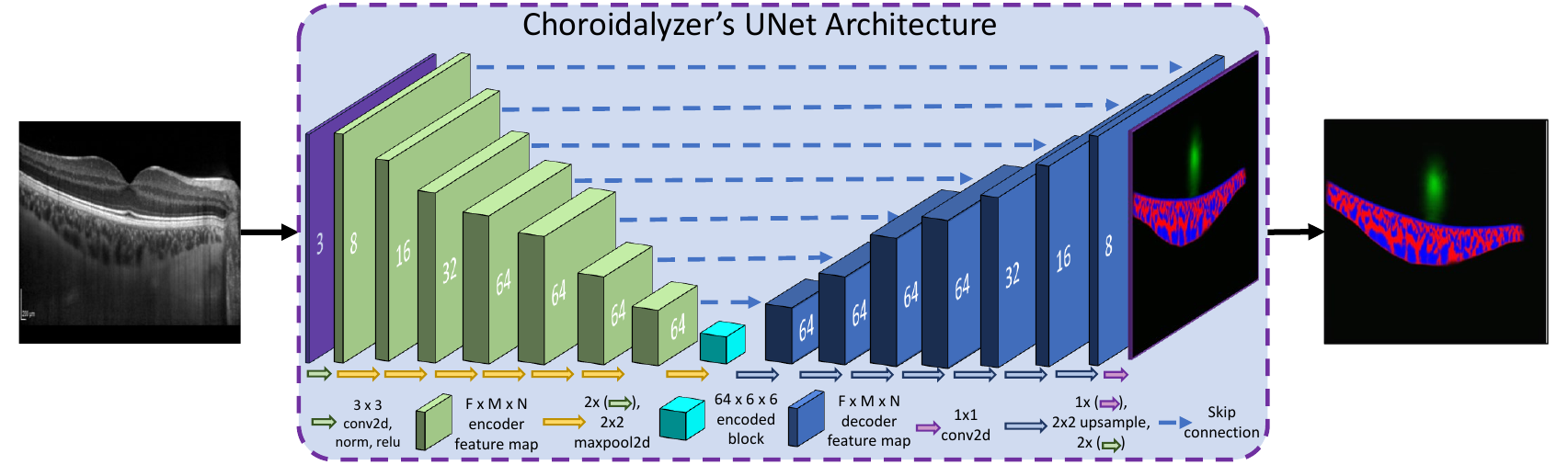}
            \caption[Choroidalyzer's UNet deep learning architecture.]{Diagram of Choroidalyzer's UNet deep learning architecture. An input \acrshort{OCT} B-scan is fed into Choroidalyzer's UNet architecture and outputs three probability segmentation maps for choroid region, vessel and fovea detection. The UNet architecture has 7 encoder (green) and 7 decoder (blue) blocks, with a 6 $\times$ 6 encoded block (cyan) of dense features for scale invariance and global feature representation learning. Block transitions in the encoder and decoder are annotated below the diagram by green, yellow, blue and purple arrows. The white text overlaid on each block describes the number of feature maps per block, with the spatial dimension halving and doubling per block transition in the encoder and decoder blocks, respectively. conv2d, two dimensional convolution filter; norm, batch normalisation \cite{ioffe2015batch}; relu, non-linear rectified linear unit activation; maxpool2d, two-dimensional maximum pooling filter; upsample, bi-linear interpolation.}
            \label{fig:CHOROID_UNet}
        \end{figure}

        Choroidalyzer segments the choroid region and vessels, and detects the fovea using a UNet deep learning model \cite{ronneberger2015u} with a depth of 7, utilising skip connections between symmetrical encoder (down) and decoder (up) blocks to aid information flow and enhance representation power. This depth allows Choroidalyzer to better consider the global context. The first 3 decoder blocks increase the internal channel dimension (number of feature maps) from 8 to 64, after which it is kept constant to reduce memory consumption and parameter count. The spatial dimensions of the feature maps are scaled by half at each layer which reduces the feature map of the deepest block to 6 $\times$ 6 (for an input resolution of 768 $\times$ 768) with 64 feature maps which enables spatial invariant feature extraction and global contextual learning.
        
        Convolutions blocks (in both encoder and decoder arms) consist of two convolutional layers, each followed by batch normalisation \cite{ioffe2015batch} and non-linear rectified linear unit (ReLU) activation. Choroidalyzer's down-blocks (in the encoder arm) use maximum pooling at the end of each convolutional block to scale the spatial dimension of each feature map by $\nicefrac{1}{2}$. Before every convolutional block, Choroidalyzer's up-blocks (in the decoder arm) use a 1 $\times$ 1 convolution to reduce the channel dimension followed by bi-linear interpolation to increase the spatial dimension, which is more compute and memory efficient than the standard transposed convolutions, as there are no model parameters associated with this interpolation. 
        
        Figure \ref{fig:CHOROID_UNet} shows the UNet deep learning architecture designed for Choroidalyzer. Given an input \acrshort{OCT} B-scan of 1 $\times$ 768 $\times$ 768, we expect Choroidalyzer's inference procedure to output a probability segmentation map of resolution 3 $\times$ 768 $\times$ 768 --- encoding each segmentation task's probability map as a separate colour channel: choroid region in red, vessels in blue and fovea detection in green.
        
        We trained Choroidalyzer for 40 epochs after establishing convergence through negligible improvement on the validation set. Training was conducted using the AdamW optimizer \cite{loshchilov2017decoupled} with a learning rate of 5 $\times$ 10$^{-4}$ and weight decay of 10$^{-8}$ to minimise binary crossentropy loss, clamping the maximum gradient norm to 3 before each step to maintain numerical stability at back-propagation during training. We use automatic mixed precision to speed up training dramatically while reducing memory consumption by almost half. Forward pass and loss computation are done in bfloat16, a half-precision datatype optimised for machine learning.
    
        During training, we apply the following data augmentations in random order per sample: Horizontal flip ($p=0.5$), changing the brightness and contrast independently (factors $\sim U(0.5, 1.5)$, $p=0.95$), random rotation and shear (degrees $\sim U(-25, 25)$ and $\sim U(-15, 15)$ respectively, $p=\nicefrac{1}{3}$), scaling the image (factor $ \sim U(0.8, 1.2)$, $p=\nicefrac{1}{3}$), where $U(a,b)$ denotes a uniform distribution between $a$ and $b$, and $p$ the probability of the transform being applied. For peripapillary B-scans which have a resolution of 768 $\times$ 1536, we place a window of size 768 $\times$ 768 to crop the scan to, shifted horizontally at random multiples of 192 per example and epoch. For effective back-propagation of gradients during training, Choroidalyzer's fovea detection module, we employ one-sided label smoothing and set all other pixels outside the rectangular segmentation mask (figure \ref{fig:CHOROID_gt_labels}(D)) to 0.01 instead of 0, stabilising training.
                
        At inference, Choroidalyzer outputs three probability maps, one each for the choroidal region, vessels and fovea. The choroidal region is thresholded at 0.5 to create a binary region mask. Note that this threshold can be fine-tuned by the end-user depending on the definition of the Choroid-Sclera boundary they prescribe to (see section \ref{subsec:ch1_INTRO_measure_choroid} and figure \ref{fig:INTRO_CS_definition}). As detection of choroidal vessels is not a binary classification problem (as previously discussed in chapter \ref{chp:chapter-mmcq}), we leverage the output probabilities as a measure of uncertainty and leave the vessel mask as the raw probability map, after element-wise multiplication with the binarised region mask to remove any potential erroneously segmented areas outside the detected choroidal space. Finally, Choroidalyzer extracts the lateral position of the fovea by applying a 21-wide triangular filter to the column-wise sums of our model's raw probability map, taking the column with the highest value. While we only evaluate the predicted lateral location of the fovea, Choroidalyzer can output the axial location too by passing 51-long triangular filters over the raw probability map and similarly taking the row with the highest value.
        
    \end{mysection}

    \begin{mysection}[]{Choroidalyzer's evaluation}

        \begin{mysubsection}[]{Statistical analysis}

            \begin{mysubsubsection}[]{Quantitative evaluation}   
            
                In this evaluation, we compare Choroidalyzer's performance to the ground-truth labels generated using DeepGPET, \acrshort{MMCQ} and manual labels for the choroidal space, vessels and fovea (section \ref{subsubsec:CHOROID_gtlabels}), respectively. To evaluate agreement in segmentations, we use the Dice coefficient and \acrshort{AUC}, with the former metric used after applying a fixed threshold of 0.5 to binarise our model's predictions. These metrics were defined previously in section \ref{subsec:INTRO_metrics}.

                Alongside standard segmentation metrics, we evaluate Choroidalyzer using choroid-derived metrics of thickness, area and \acrshort{CVI} in a 6 mm, fovea-centred \acrshort{ROI}, corresponding to the entire \acrshort{ETDRS} macular grid \cite{early1991grading}. The \acrshort{ROI} for volume B-scans without the fovea present were centred at the middle column index of the image. The definition of these metrics and how they are computed can be found earlier in this thesis in section \ref{subsec:ch1_INTRO_measure_bsan}. As Choroidalyzer outputs raw probabilities for vessel segmentation predictions, rather than a binary mask, we propose a `soft' \acrshort{CVI} which takes the ratio between the predicted probabilities of the vessel map against the binarised region map. We evaluate the agreement of Choroidalyzer's predicted metrics against ground truth-derived metrics using Pearson and Spearman correlations and mean absolute error (\acrshort{MAE}), previously defined in section \ref{subsec:INTRO_metrics}. 
                
                To measure agreement of fovea column predictions, we use \acrshort{MAE} and median absolute error (Median AE). Additionally, we tested the effect of perturbing the fovea column laterally on segmentation-derived metrics. This was done by comparing fovea-centred metrics with metrics derived after the fovea column was randomly perturbed using a discretised uniform distribution $ \sim U(-6, 6)$ (excluding 0). 50 simulations were run on a random sample of 495 \acrshort{OCT} B-scans, previously described in section \ref{subsec:CHOROID_sample_deriv}.
    
            \end{mysubsubsection}

            \begin{mysubsubsection}[]{Qualitative evaluation}

                In addition to quantitative evaluation against previously derived values from DeepGPET and \acrshort{MMCQ}, we also used human graders to examine and characterise the behaviour of Choroidalyzer in cases of high error. Concretely, for each of the three tasks (region and vessel segmentation, fovea detection), we selected the 30 examples from both test sets where Choroidalyzer produced the highest errors. For redundant cases --- such as adjacent, highly-similar slices from a volume scan --- only one was retained. This left 28 cases for region, 29 for vessel, and 25 for fovea. 

                An adjudicating clinical ophthalmologist (Supervisor Dr. Ian J.C. MacCormick) was provided with the original image, Choroidalyzer's prediction and the ground truth label while being masked to the identity of the methods. Images and labels were provided individually and as composites and the images were viewed according to the equipment listed in appendix \ref{apdx:equipment_protocol}. For each example, the adjudicator was asked which label they preferred. They also rated each label qualitatively on a 5-level ordinal scale (``Very bad'', ``Bad'', ``Okay'', ``Good'' and ``Very good'') for region segmentation quality, as well as intravascular and interstitial vessel segmentation quality according to the definitions in appendix \ref{apdx:definitions}. The latter two were used to quantify any potential under-segmentation of vessels (false negative rate) and over-segmentation of the interstitial space (false positive rate), and guidance was provided using a pre-specified protocol in appendix \ref{apdx:vessel_protocol}.
                
                Finally, we selected a random subsample of 20 B-scans at the subject-level from the external test set to be manually segmented by two graders, M1 and M2. M1 was a clinical ophthalmologist (Supervisor Dr. Ian J.C. MacCormick) and M2 was the author of this thesis. Manual graders segmented the region and choroidal vessels using ITK-Snap \cite{py06nimg}, which permits pixel-level annotation, and were blinded to each others segmentations following a pre-specified protocol outlined in appendix \ref{apdx:seg_protocol}. The manual segmentations were compared to Choroidalyzer and to the current state-of-the-art, namely DeepGPET for region segmentation and Niblack for vessel segmentation. To keep Niblack's use consistent, we fixed the window size to 51 and the standard deviation offset to -0.05, as proposed by Muller, et al. \cite{muller2022application}. We used Dice and \acrshort{AUC} to measure segmentation agreement, as well \acrshort{MAE}, and Pearson and Spearman correlation coefficients to measure agreement of choroidal metrics between the four methods for both region and vessel segmentation tasks. Evaluation metrics have been previously defined in section \ref{subsec:INTRO_metrics}. We also compared the execution time of each approach for region and vessel segmentation, and report mean and standard deviation (\acrshort{SD}) in seconds. Automatic methods were run on a laptop with a 4 year old Intel Core i5 (8$^{\text{th}}$ generation) CPU and 16 Gb of RAM, but no \acrshort{GPU}.
                
                The small sample size for these evaluations were chosen to strike a balance between the involvement of human experts to qualitatively assess Choroidalyzer's performance, and the time intensive nature of manual grading and segmentation of the choroid in \acrshort{OCT} B-scans. Thus, this qualitative evaluation is able to make optimal use of a costly resource (human time). Moreover, the use of measurement and adjudication protocols promoted consistency between manual labelling, and ensured alignment between the definitions imposed by Choroidalyzer and applied by the adjudicator.

            \end{mysubsubsection}
        
        \end{mysubsection}
    
        \begin{mysubsection}[]{Results}

            \begin{mysubsubsection}[]{Performance on internal and external test sets} \label{sec:CHOROID_RESULTS_QUANT}

                \begin{table}[tb]
                \centering
                \begin{tabular}{@{}lllllll@{}}
\toprule
  \multicolumn{1}{c}{\multirow{2}{*}{Set}} &
  \multicolumn{2}{c}{Region} &
  \multicolumn{2}{c}{Vessel} &
  \multicolumn{2}{c}{Fovea} \\
  \cmidrule(l){2-3}\cmidrule(l){4-5}\cmidrule(l){6-7}
  \multicolumn{1}{c}{} &
  \multicolumn{1}{c}{\acrshort{AUC}} &
  \multicolumn{1}{c}{Dice} &
  \multicolumn{1}{c}{\acrshort{AUC}} &
  \multicolumn{1}{c}{Dice} &
  \multicolumn{1}{c}{\acrshort{MAE} [px]} &
  \multicolumn{1}{c}{Median AE [px]} \\
  \midrule
Internal test & 0.9998 & 0.9789 & 0.9982 & 0.8817 & 3.9 & 3.0 \\
External test & 0.9998 & 0.9749 & 0.9980 & 0.8703 & 3.4 & 3.0 \\ \bottomrule
\end{tabular}

                \caption[Choroidalyzer's segmentation performance across internal and external test sets.]{Segmentation metrics for Choroidalyzer against ground-truth annotations from DeepGPET (region), \acrshort{MMCQ} (vessel) and manual fovea detection for the internal and external test sets. Fovea error is in pixels (px).}
                \label{tab:CHOROID_main_segresults}
                \end{table}

                \begin{table}[tb]
                \begin{adjustwidth}{-1in}{-1in}  
                \centering
                \scalebox{0.65}{\begin{tabular}{@{}llllllllll@{}}
\toprule
  \multicolumn{1}{c}{\multirow{2}{*}{Set}} &
  \multicolumn{3}{c}{Thickness} &
  \multicolumn{3}{c}{Area} &
  \multicolumn{3}{c}{\acrshort{CVI}}\\ 
  \cmidrule(l){2-4}\cmidrule(l){5-7}\cmidrule(l){8-10}
  \multicolumn{1}{c}{} &
  \multicolumn{1}{c}{Pearson} &
  \multicolumn{1}{c}{Spearman} &
  \multicolumn{1}{c}{\acrshort{MAE} [$\mu$m]} &
  \multicolumn{1}{c}{Pearson} &
  \multicolumn{1}{c}{Spearman} &
  \multicolumn{1}{c}{\acrshort{MAE} [mm$^2$]} &
  \multicolumn{1}{c}{Pearson} &
  \multicolumn{1}{c}{Spearman} &
  \multicolumn{1}{c}{\acrshort{MAE}}\\ \midrule
Internal test & 0.9754 & 0.9692 & 8.2252 & 0.9815 & 0.9786 & 0.0385 & 0.8285 & 0.8097 & 0.0206 \\
External test & 0.9831 & 0.9868 & 8.0888 & 0.9779 & 0.9848 & 0.0487 & 0.7948 & 0.7991 & 0.0306\\ \bottomrule
\end{tabular}}

                \end{adjustwidth}
                \caption[Choroidalyzer's derived-measurement performance across internal and external test sets.]{Agreement in choroid-derived metrics for Choroidalyzer against ground-truth annotations from DeepGPET (thickness, area) and \acrshort{MMCQ} (\acrshort{CVI}) from the internal and external test sets. All correlation coefficients were statistically significant such that P < 0.0001.}
                \label{tab:CHOROID_main_results}
                \end{table}

                \begin{table}[tb]
                \centering
                \begin{tabular}{@{}llllllllllllllll@{}}
\toprule
   &
 \multicolumn{2}{c}{Region} &
  \multicolumn{2}{c}{Vessel}\\
  \cmidrule(l){2-3}\cmidrule(l){4-5}
  \multicolumn{1}{c}{} &
  \multicolumn{1}{c}{\acrshort{AUC}} &
  \multicolumn{1}{c}{Dice} &
  \multicolumn{1}{c}{\acrshort{AUC}} &
  \multicolumn{1}{c}{Dice}\\ \midrule
 Internal validation & 0.9983 & 0.9910 & 0.9514 & 0.7315 \\
 Internal test & 0.9996 & 0.9636 & 0.9925 & 0.7155 \\ \bottomrule
\end{tabular}

                \caption[Choroidalyzer's segmentation performance metrics for peripapillary B-scans.]{Segmentation metrics for peripapillary \acrshort{OCT} B-scans for Choroidalyzer against ground-truth annotations from DeepGPET (region) and \acrshort{MMCQ} (vessel) from the internal validation and test sets.}
                \label{tab:CHOROID_per_results}
                \end{table}
    
                Tables \ref{tab:CHOROID_main_segresults} and \ref{tab:CHOROID_main_results} shows the performance of Choroidalyzer on the internal and external test sets for the pixel-wise segmentations and choroid-derived measurements, respectively. Choroidalyzer achieved very good performance in terms of \acrshort{AUC} and Dice for region and vessels on both sets. Metrics for region are higher than for vessels, which is expected as vessel segmentation is much more difficult and ambiguous than region segmentation, and thus the ground truths (generated using \acrshort{MMCQ}) are themselves not representative of a gold-standard. Performance was slightly higher for the internal test set than the external test set, which is expected, but only marginally so, indicating that our model potentially generalises well to this new cohort and potentially others related to systemic disease. 
                
                Table \ref{tab:CHOROID_per_results} shows the segmentation results when isolating the internal test set to only peripapillary B-scans. Our model achieved an \acrshort{AUC} of 0.9996 (region) / 0.9925 (vessel) and Dice of 0.9636 (region) / 0.7155 (vessel). This is reasonable performance but lower than for other scans. Upon further investigation, we found that Choroidalyzer's performance on peripapillary B-scans was worse than DeepGPET's performance, even though DeepGPET was not trained on peripapillary choroids. Figure \ref{fig:CHOROID_peri_deepgpet} shows a qualitative comparison of peripapillary choroids segmented by Choroidalyzer (left) and DeepGPET (right). Yellow arrows indicate area of disagreement, where Choroidalyzer either over- or under-segments, unlike DeepGPET.

                \begin{figure}[tbp]
                \centering
                \includegraphics[width=\textwidth]{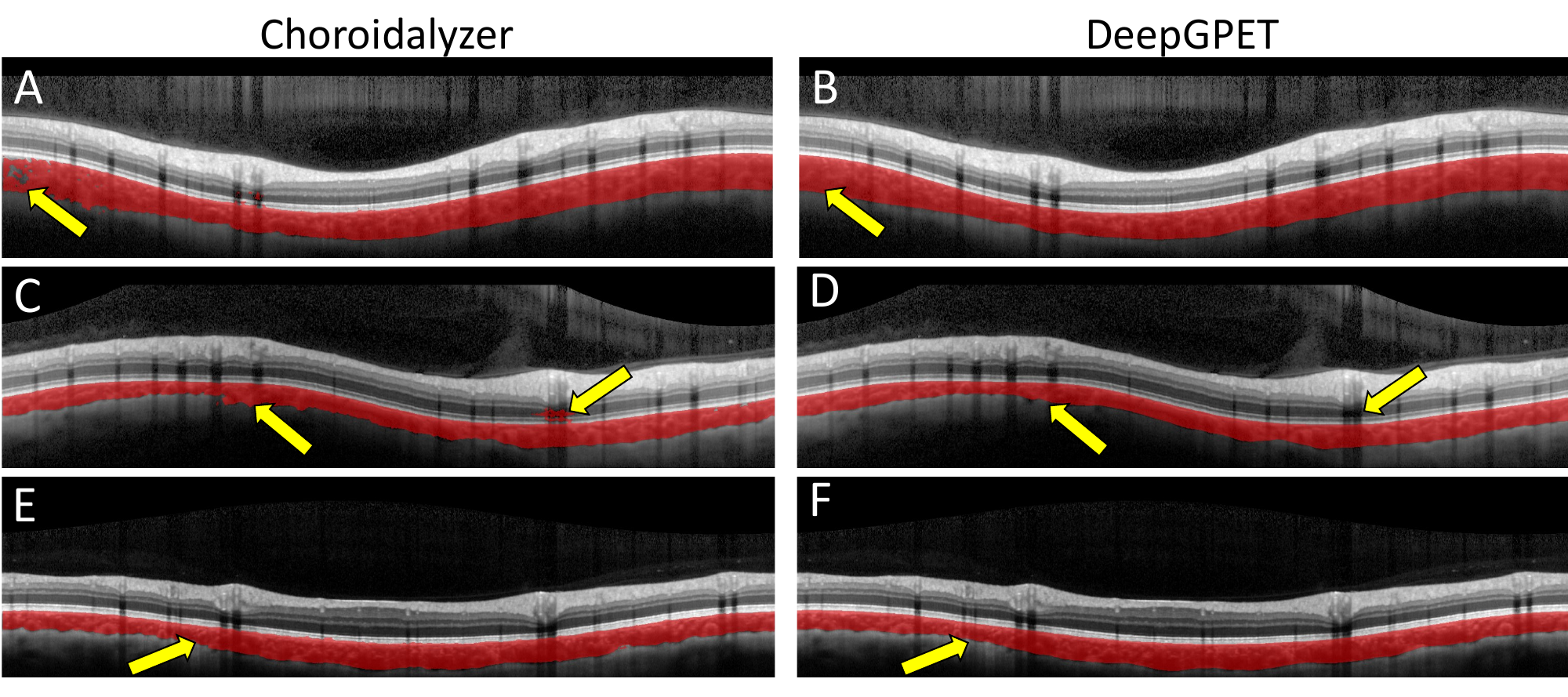}
                \caption[Choroidalyzer's segmentation performance compared to DeepGPET.]{Peripapillary B-scans where DeepGPET (right) performed better than Choroidalyzer (left). (A, C, E) Choroid region segmentations after application of Choroidalyzer, and similarly for DeepGPET (B, D, F), Yellow arrows in each row represent sources of poor quality segmentation for Choroidalyzer compared to DeepGPET.}
                \label{fig:CHOROID_peri_deepgpet}
                \end{figure}
                
                \begin{figure}[tbp]
                \centering
                \includegraphics[width=0.9\textwidth]{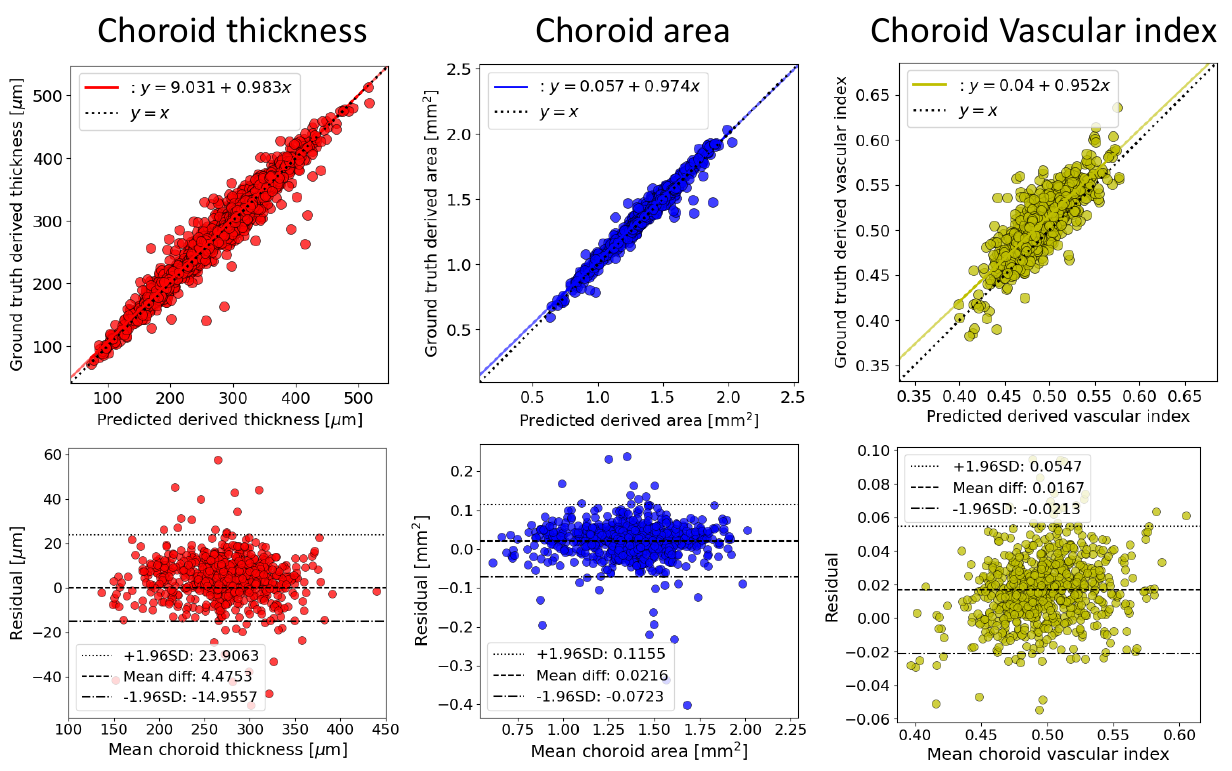}
                \caption[Choroidalyzer's agreement in choroid-derived metrics in the internal test set.]{Agreement in choroid-derived metrics against ground truth labels from DeepGPET (thickness, area) and \acrshort{MMCQ} (\acrshort{CVI}) for the internal test set. Top row are scatter plots with best regression fit and identity lines, bottom row are Bland-Altman plots.}
                \label{fig:CHOROID_int_corr_plots}
                \end{figure}
    
                \begin{figure}[tbp]
                \centering
                \includegraphics[width=0.9\textwidth]{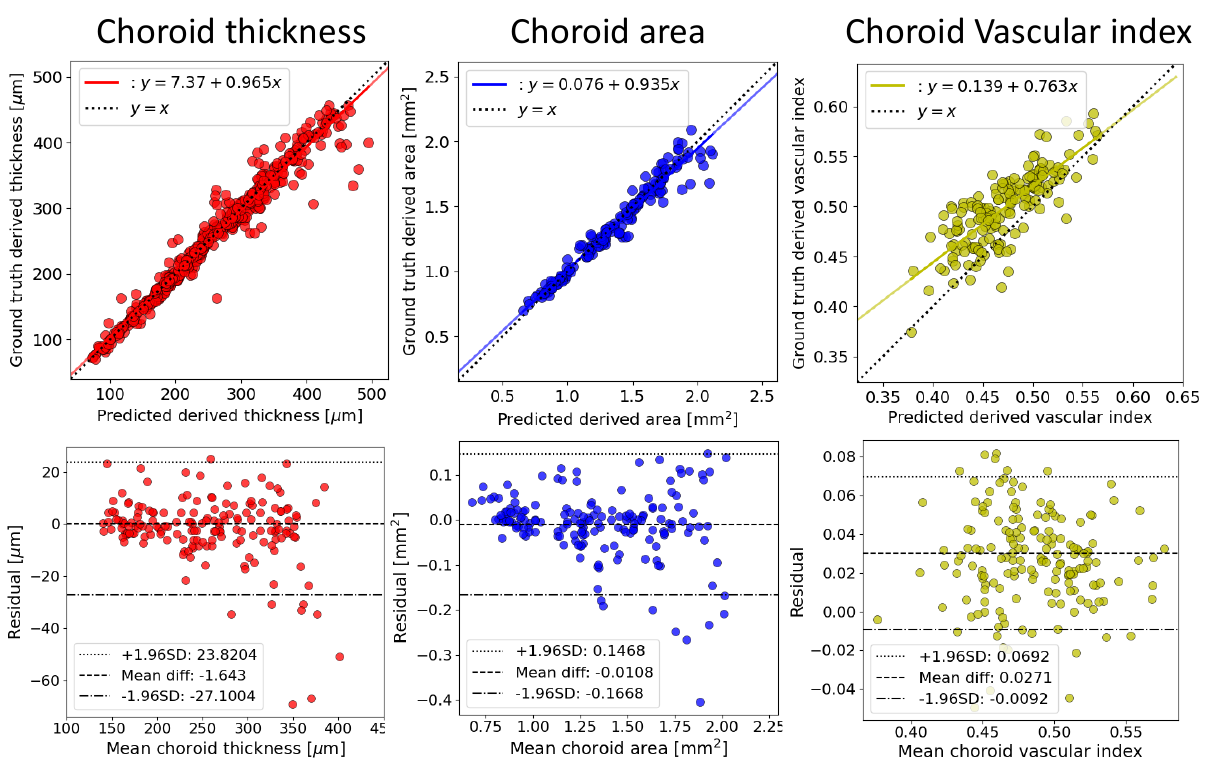}
                \caption[Choroidalyzer's agreement in choroid-derived metrics in the external test set.]{Agreement in choroid-derived metrics against ground truth labels from DeepGPET (thickness, area) and \acrshort{MMCQ} (\acrshort{CVI}) for the external test set. Top row are scatter plots with best regression fit and identity lines, bottom row are Bland-Altman plots.}
                \label{fig:CHOROID_ext_corr_plots}
                \end{figure}
    
                \begin{figure}[tbp]
                \centering
                \includegraphics[width=\textwidth]{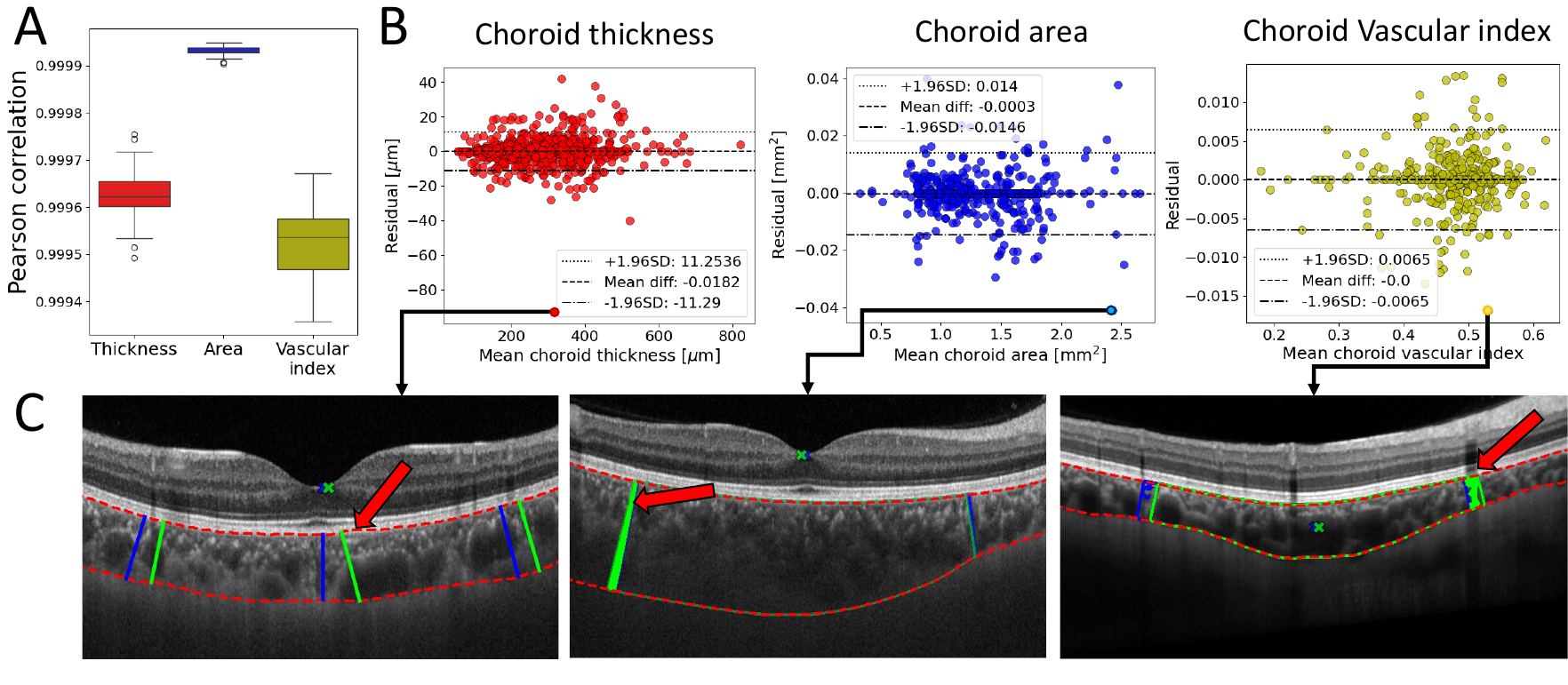}
                \caption[Choroidalyzer's fovea detection error analysis results.]{Results from comparing choroid-derived measurements after simulating 50 perturbations in the lateral position of the fovea using a uniform distribution $\mathbb{U}(-6,6) / \{0\}$. (A) Distribution of Pearson correlation coefficients. (B) Bland-Altman plots for the worst performing simulation. (C) \acrshort{OCT} B-scans with the highest error from the worst performing simulation. Original and perturbed fovea and measurements in blue and green, respectively. Red arrows to show the major source of the error.}
                \label{fig:CHOROID_fovea_perturb}
                \end{figure}

                For the derived choroid metrics, Choroidalyzer shows excellent agreement with the ground truth labels derived from DeepGPET for thickness and area, with Pearson and Spearman correlations of 0.9692 or greater for both internal and external test sets. For \acrshort{CVI}, performance is a bit lower, with correlations between 0.7948 and 0.8285. Although \acrshort{CVI} depends on both region and vessel segmentation, the other metrics indicate that the error in \acrshort{CVI} are driven primarily by differences in the vessel segmentation. Still, the observed correlations are high in absolute terms. Figures \ref{fig:CHOROID_int_corr_plots} and \ref{fig:CHOROID_ext_corr_plots} shows correlation and Bland-Altman plots for the three derived metrics on internal and external test sets, respectively, which likewise indicate generally very good agreement. 

                For detection of the lateral position of the fovea, the model had an \acrshort{MAE} of 3.9 pixels for the internal test set and 3.4 pixels for the external test set, with the median absolute error of 3 pixels for both. This is excellent performance, as an error of 3 pixels on a 768 pixel-wide image will not meaningfully change our \acrshort{ROI} or choroid-derived metrics. Testing this formally on approximately 10\% of the data by randomly perturbing the fovea column $\pm$ 6 pixels for 50 simulations yielded excellent Pearson correlation ($r$ > 0.99, P<0.001) across all derived choroid-derived metrics. 
                
                Figure \ref{fig:CHOROID_fovea_perturb} shows the distribution of correlations across all simulations for each metric in panel (A), with residual distributions for the poorest performing simulation per metric shown in panel (B). Even the poorest performing simulation generally showed strong agreement, with zero-centred residual distributions, and limits of agreement in the Bland-Altman plots well within acceptable bounds \cite{rahman2011repeatability, agrawal2020exploring, breher2020choroidal}. 
                
                Figure \ref{fig:CHOROID_fovea_perturb}(C) shows the major outliers for each metric in the worst performing simulation. For choroid thickness, a kink in the upper boundary segmentation (red arrow) skewed the subfoveal thickness measurement, since thickness is measured \textit{locally perpendicular} to the upper choroid boundary (see section \ref{subsec:ch1_INTRO_measure_bsan}). This was further exacerbated by unequal pixel length-scales, ultimately leading to a residual of 93 microns. For choroid area, this choroid was significantly thicker in the temporal region than it was in the nasal region. Thus, the fovea perturbation (6 pixels to the left) captured a significantly higher proportion of the choroid in its area calculation (red arrow), producing high error. Similarly, for \acrshort{CVI}, the fovea perturbation (6 pixels to the right) captured a region of the choroid not only vascular but also artificially darkened by superficial retinal vessel shadowing (red arrow), compared to the original \acrshort{ROI} which was not, causing high error. Nevertheless, on average the agreement is strong across the majority of cases. Thus, we propose that the fovea column quantitative error observed from Choroidalyzer does not significantly impact choroidal metrics.
                
            \end{mysubsubsection}

            \begin{mysubsubsection}[]{Comparison with manual segmentations} \label{sec:CHOROID_RESULTS_MANUAL}

                \begin{table}[tb]
                \centering
                \begin{tabular}{@{}rllll@{}}
\toprule
  \multicolumn{1}{c}{\multirow{2}{*}{Comparison}} &
  \multicolumn{2}{c}{Region} &
  \multicolumn{2}{c}{Vessel} \\
  \cmidrule(l){2-3}\cmidrule(l){4-5} 
  \multicolumn{1}{c}{} &
  \multicolumn{1}{c}{\acrshort{AUC}} &
  \multicolumn{1}{c}{Dice} &
  \multicolumn{1}{c}{\acrshort{AUC}} &
  \multicolumn{1}{c}{Dice} \\
  \midrule
M1 vs. M2 & 0.9639 & 0.9474 & 0.8891 & 0.7699 \\
Choroidalyzer vs. Manual & 0.9978 & 0.9375 & 0.9914 & 0.7669 \\
\acrshort{SOTA} vs. Manual & 0.9444 & 0.9333 & 0.9223 & 0.7742 \\ 
\bottomrule
\end{tabular}

                \caption[Choroidalyzer's segmentation performance metrics against manual graders and state-of-the-art methods.]{Segmentation metrics for the 20 images assessed manually and algorithmically from the external test set. M1 was grader supervisor Ian J.C. MacCormick and M2 was Jamie Burke (the author). \acrshort{SOTA}: state-of-the-art.}
                \label{tab:CHOROID_manual_segcomparison}
                \end{table}

                \begin{table}[tb]
                \begin{adjustwidth}{-1in}{-1in}  
                \centering
                
\scalebox{0.6}{\begin{tabular}{@{}rlllllllll@{}}
\toprule
  \multicolumn{1}{c}{\multirow{2}{*}{Comparison}} &
  \multicolumn{3}{c}{Thickness} &
  \multicolumn{3}{c}{Area} &
  \multicolumn{3}{c}{Vascular Index}\\ 
  \cmidrule(l){2-4}\cmidrule(l){5-7}\cmidrule(l){8-10} 
  \multicolumn{1}{c}{} &
  \multicolumn{1}{c}{Pearson} &
  \multicolumn{1}{c}{Spearman} &
  \multicolumn{1}{c}{MAE ($\mu$m)} &
  \multicolumn{1}{c}{Pearson} &
  \multicolumn{1}{c}{Spearman} &
  \multicolumn{1}{c}{MAE (mm$^2$)} &
  \multicolumn{1}{c}{Pearson} &
  \multicolumn{1}{c}{Spearman} &
  \multicolumn{1}{c}{MAE}\\ \midrule
M1 vs. M2  & 0.9503 & 0.9521 & 17.8833 & 0.9516 & 0.9248 & 0.1096 & 0.8074 & 0.6857 & 0.0618 \\
Choroidalyzer vs. Manual  & 0.9534 & 0.9663 & 20.9750 & 0.9554 & 0.9368 & 0.1150 & 0.6654 & 0.7383 & 0.0530 \\
\acrshort{SOTA} vs. Manual  & 0.9676 & 0.9636 & 19.9250 & 0.9548 & 0.9233 & 0.1202 & 0.6907 & 0.6105 & 0.1682 \\ \bottomrule
\end{tabular}}

                \end{adjustwidth}
                \caption[Choroidalyzer's derived-metric performance against manual graders and state-of-the-art methods.]{Derived metrics of thickness, area and \acrshort{CVI} for the 20 images assessed manually and algorithmically from the external test set. M1 was supervisor Ian J.C. MacCormick and M2 was Jamie Burke (the author). \acrshort{SOTA}: state-of-the-art.}
                \label{tab:CHOROID_manual_comparison}
                \end{table}
    
                \begin{table}[tb]
                \begin{adjustbox}{width=\textwidth}
                \centering
                \scalebox{0.99}{\begin{tabular}{@{}lccc@{}}
\toprule
\hspace{1em}Method\hspace{1.3em} & \hspace{1em}Region (s) \hspace{1.3em} & \hspace{1em} Vessel (s) \hspace{1.3em} & \hspace{1em} Total (s) \hspace{1.3em}\\ \midrule
M1 & 78.400 $\pm$ 12.261 & 1506.000 $\pm$ 744.073 & 1584.400 $\pm$ 771.284  \\
M2 & 165.000 $\pm$ 23.889 & 1176.700 $\pm$ 263.968 & 1341.700 $\pm$ 265.800   \\
\acrshort{SOTA} & 0.751 $\pm$ 0.081 & 0.370 $\pm$ 0.105 & 1.121 $\pm$ 0.140\\
Choroidalyzer & - & - & 0.299 $\pm$ 0.018 \\
\bottomrule
\end{tabular}}

                \end{adjustbox}
                \caption[Choroidalyzer's execution time compared with manual graders and state-of-the-art methods.]{Mean $\pm$ SD execution time in seconds (s) of the four different approaches to region and vessel segmentation for the 20 images assessed manually and algorithmically from the external test set. \acrshort{SOTA}: state-of-the-art.}
                \label{tab:CHOROID_manual_comparison_times}
                \end{table}
    
                Tables \ref{tab:CHOROID_manual_segcomparison} and \ref{tab:CHOROID_manual_comparison} show the results using segmentation metrics and choroid-derived metrics when comparing Choroidalyzer and the current state-of-the-art (\acrshort{SOTA}) (DeepGPET and Niblack for region and vessel analysis, respectively) to manual segmentations. Comparisons with Choroidalyzer/\acrshort{SOTA} against individual manual graders were averaged for brevity.
                
                Similar to the evaluation to ground truth labels, vessel Dice for Choroidalyzer (0.7410 vs M1 and 0.7927 vs M2; mean 0.7669) is worse than region Dice and even worse than the vessel Dice on both test sets. However, it is very similar to the manual grader's (inter-grader) agreement of 0.7699. More generally, the inter-grader agreements for all other metrics are similar to Choroidalyzer's agreement with the graders, with the notable exception of \acrshort{CVI}. Here, Choroidalyzer's \acrshort{MAE} is better (0.0555 vs M1 and 0.0506 vs M2; mean 0.0530) than the inter-grader agreement (0.0618), as is the Spearman correlation but Pearson correlation is worse. Compared to the respective \acrshort{SOTA}, Choroidalyzer has better agreement with the graders for most of the metrics, although methods are generally comparable.
    
                Table \ref{tab:CHOROID_manual_comparison_times} shows the execution per B-scan for all approaches. The manual graders on average needed more than 26 and 22 minutes (mean 24), with the vast majority of that time spent on the vessel segmentation. By contrast, the automatic methods on a standard laptop CPU needed about a second per B-scan and no human time at all. Thus, to get through a dataset of 100 scans, it would take manual graders about 40 hours of work, but with automatic methods it would be less than 2 minutes. With \acrshort{GPU}-acceleration, Choroidalyzer and DeepGPET could achieve throughputs of dozens or hundreds of B-scans per second even on consumer-grade hardware. Comparing the automatic methods with each other, Choroidalyzer took 73\% less time than DeepGPET and Niblack, while also detecting the fovea location. All three methods are very fast but for very large datasets or deployment on edge devices, Choroidalyzer's efficiency is an additional advantage over existing automatic methods.

            \end{mysubsubsection}
    
            \begin{mysubsubsection}[]{Detailed error analysis}

                \begin{table}[tb]
                \begin{adjustbox}{width=\textwidth}  
                \centering
                \begin{tabular}{@{}lccc@{}}
\toprule
 & Preferred Choroidalyzer & Preferred SOTA & Both equally good \\
\midrule
Region & 8/28 & 5/28 & 15/28 \\
Vessel & 13/29 & 4/29 & 12/29 \\
Fovea & 23/25 & 1/25 & 1/25\\ 
\midrule
& & & \\
\midrule
Method & Region: quality & Vessel: intravascular & Vessel: interstitial \\ \midrule
Choroidalyzer & VG: 3, G: 14, O: 9, B: 1, VB: 1 & VG: 0, G: 17, O: 12, B: 0, VB: 0 & VG: 0, G: 20, O: 9, B: 0, VB: 0 \\
SOTA & VG: 0, G: 17, O: 8, B: 1, VB: 2 & VG: 0, G: 5, O: 19, B: 3, VB: 2 & VG: 0, G: 17, O: 8, B: 2, VB: 2 \\ 
\bottomrule
\hspace{0.5em}
\end{tabular}%

                \end{adjustbox}
                \caption[Choroidalyzer's manual adjudication against state-of-the-art methods.]{Preference and segmentation scores from expert adjudicator (I.M.) comparing the highest region segmentation, vessel segmentation and fovea column errors between Choroidalyzer and the ground truth labels. VG: very good; G: good; O: okay; B: bad; VB: very bad.}
                \label{tab:CHOROID_adj}
                \end{table}
    
                \begin{figure}[tb]
                \centering
                \includegraphics[width=0.8\textwidth]{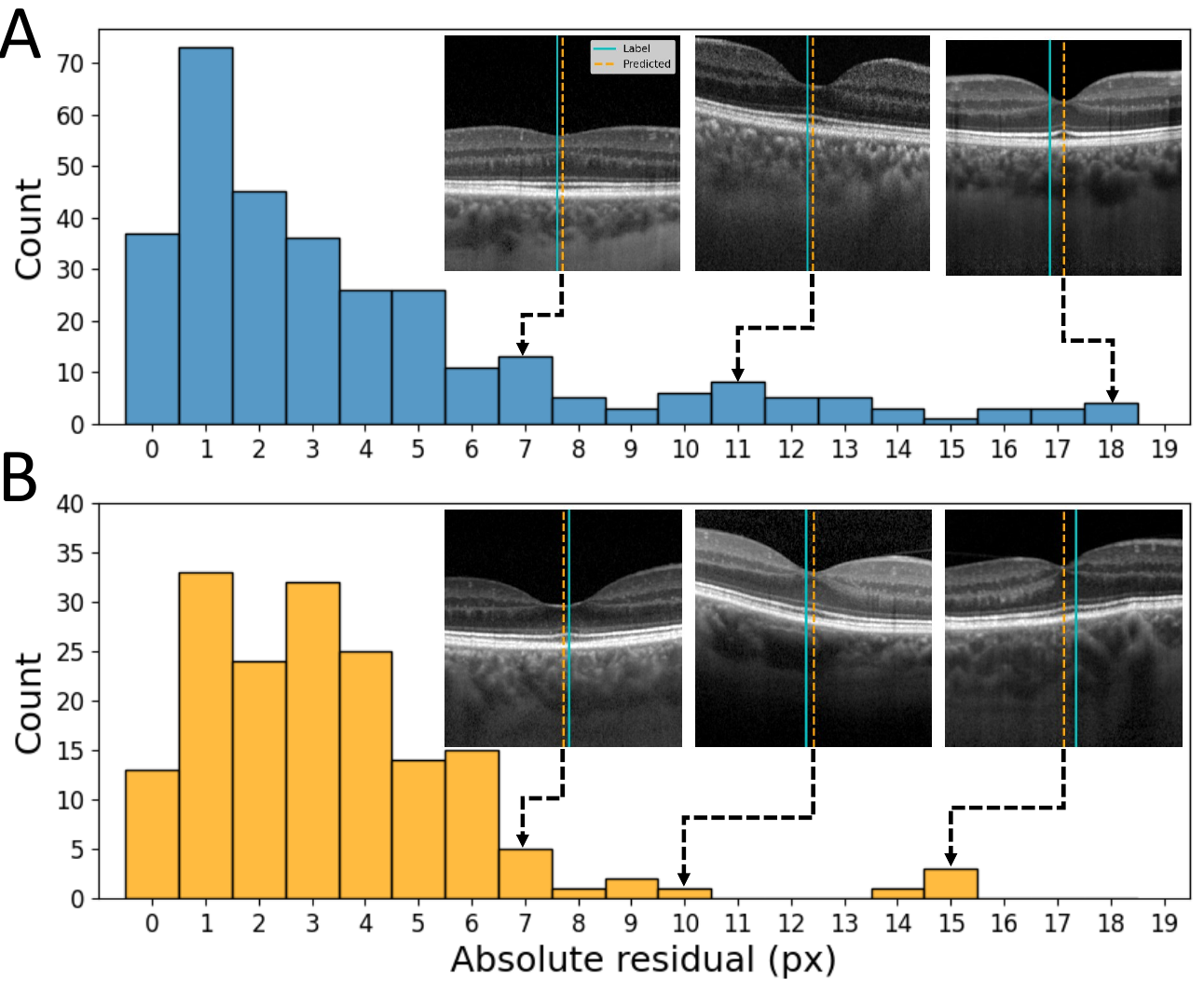}
                \caption[Choroidalyzer's fovea detection error distributions.]{Histogram of absolute errors for fovea column detection for the internal (A) and external (B) test sets. Examples for different levels of error are shown with dotted lines indicating which part of the distribution they come from. In the examples, the teal line indicates the ground truth label, the dashed orange line the prediction. px: pixels.}
                \label{fig:CHOROID_fovea}
                \end{figure}
    
                Table \ref{tab:CHOROID_adj} shows the results of manual inspection of scans where Choroidalyzer produced the highest error compared to the ground truth on the test sets. For region segmentation, Choroidalyzer was preferred in 8 cases, the ground truth in 5, and both methods were considered equally good in 15 cases. In terms of quality, Choroidalyzer was ``Very bad'' in only one case compared with 2 for the ground truth and ``Very good'' 3 times compared to none for the ground truth. For the vessels, Choroidalyzer was preferred in 13 cases, the ground truth in 4, and both were tied in 12 cases. Vessel segmentation is a harder task, with no methods achieving ``Very good''. However, the  intravascular scores for Choroidalyzer are substantially better, with no ``Bad'' or ``Very bad'' (vs. 3 and 2, respectively for ground truth) and far more ``Good'' (17 vs. 5), and the interstitial scores are similarly better. Finally, for the fovea, Choroidalyzer was preferred 23/25 times and the ground truth only once, indicating that large fovea errors are almost exclusively due to mistakes in the manual ground truth labels.
    
                Figure \ref{fig:CHOROID_fovea} shows the distributions of fovea errors for both test sets along with each example in both sets. For very large absolute error (>10 pixels), the ground truth labels are wrong and Choroidalyzer correctly identifies the fovea location. For errors around 7 pixels, still twice the \acrshort{MAE}, both methods are similar with either method sometimes being more correct. Further exploration revealed the majority of incorrectly labelled ground truths to be Topcon \acrshort{OCT} B-scans, as each radial scan did not have the foveal pit assigned to the same lateral position in each B-scan, unlike a Heidelberg Engineering OCT volume scan. This was only identified during error analysis, which was after the initial manual annotation which only identified the foveal pit for one representative B-scan per radial stack. Despite this oversight, Choroidalyzer learned to detect the fovea accurately.
                
            \end{mysubsubsection}

            \begin{mysubsubsection}[]{Potential generalisability}

                \begin{figure}[tb]
                    \centering
                    \includegraphics[width=\textwidth]{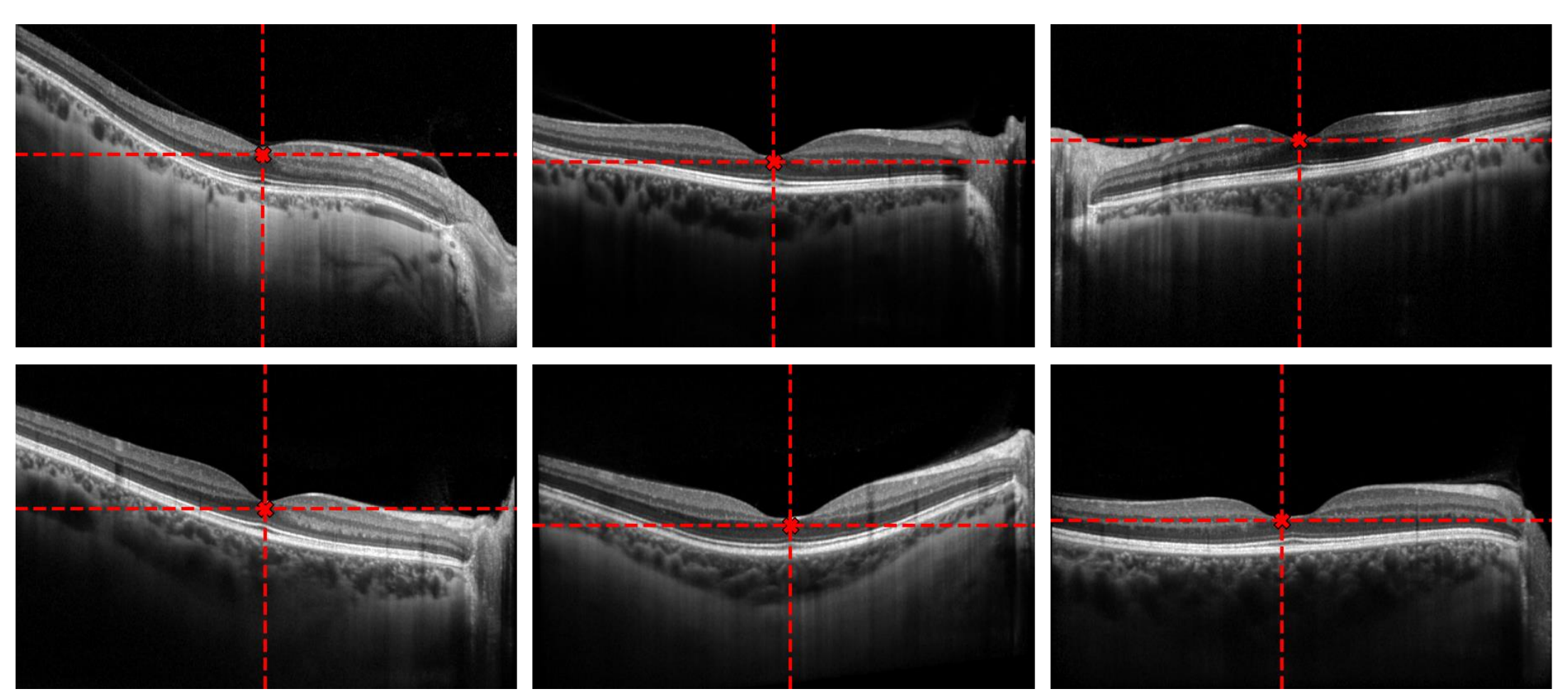}
                    \caption[Choroidalyzer's ability to locate the foveal pixel coordinate.]{\acrshort{OCT} B-scans from figure \ref{fig:INTRO_ROI_refr_skew_curve} with the detected axial and lateral foveal pit location overlaid as a red cross and the intersection of red vertical and horizontal lines.}
                    \label{fig:CHOROID_fovea_skew}
                \end{figure}

                \begin{figure}[tb]
                    \centering
                    \includegraphics[width=\textwidth]{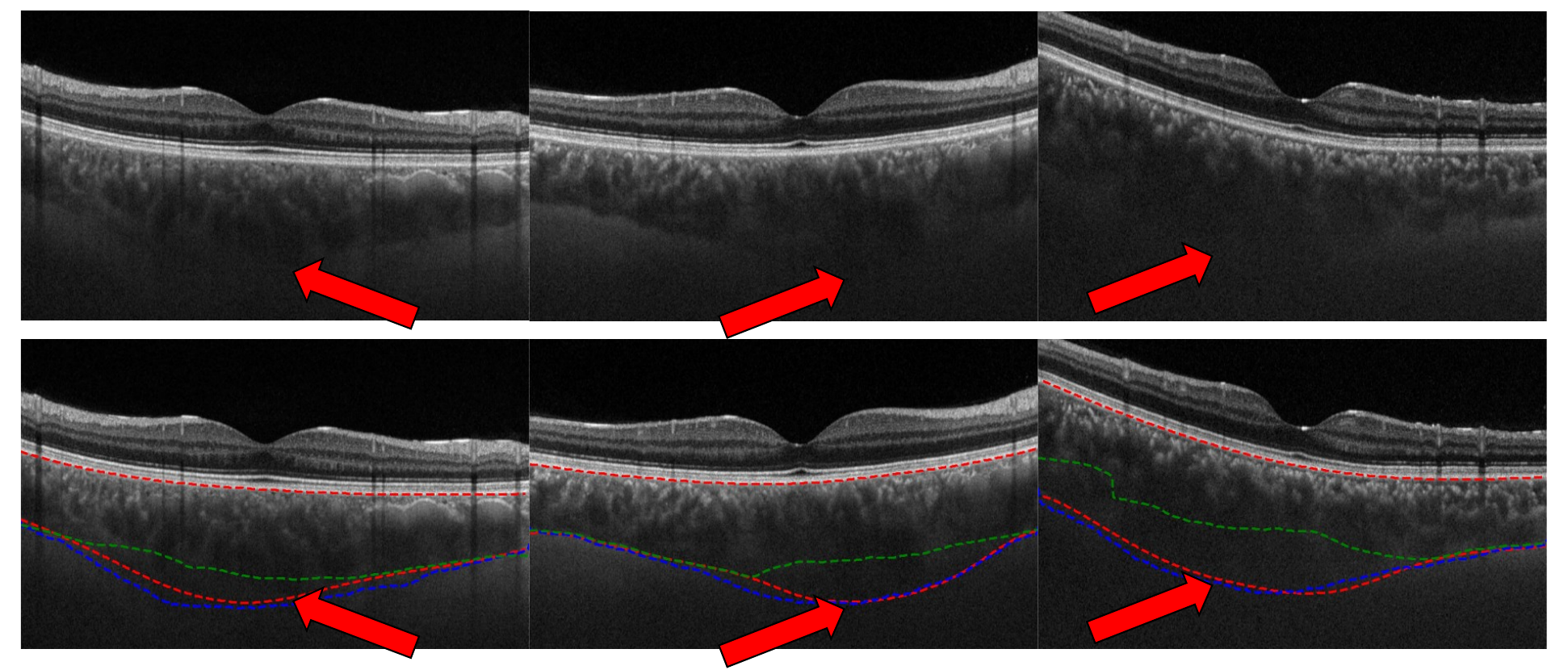}
                    \caption[Choroidalyzer's ability to segment challenging cases of poor Choroid-Sclera boundary visibility.]{Three example Topcon \acrshort{OCT} B-scans comparing performance of DeepGPET (green) and Choroidalyzer (blue) with manual labelling (red). Red arrows indicate poor, but visible Choroid-Sclera boundary.}
                    \label{fig:CHOROID_deepgpet}
                \end{figure}

                \begin{figure}[tb]
                    \centering
                    \includegraphics[width=\textwidth]{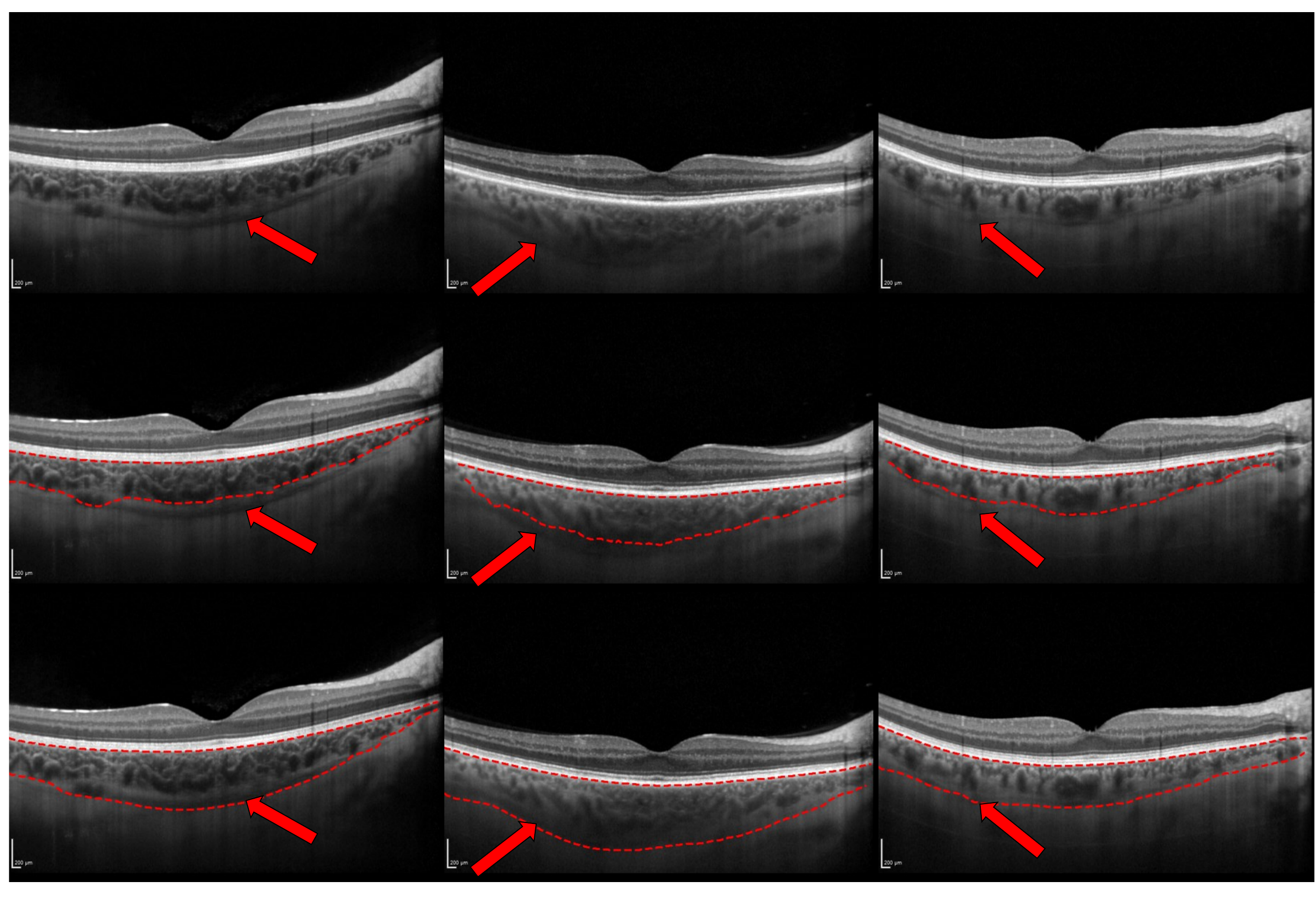}
                    \caption[Choroidalyzer's tunable region threshold enables different Choroid-Sclera boundary definitions to be applied.]{(Top) Selection of choroids from section \ref{sec:ch_app_sec_shock} with visible suprachoroidal spaces. Red boundaries delineate the choroid segmented by Choroidalyzer and according to definitions (1) and (3) from section \ref{subsec:ch1_INTRO_measure_choroid}, i.e. to exclude (Middle) / include (Bottom) the suprachoroid.}
                    \label{fig:CHOROID_SCS}
                \end{figure}
                
                While not explicitly part of Choroidalyzer's evaluation, we conducted some initial experiments on Choroidalyzer's potential to generalise to different presentations of the choroid in OCT B-scans. Note that while we highlight the following potential functionality of Choroidalyzer, these are yet to be officially validated.

                Firstly, our training procedure prioritised the detection of the lateral position over the axial position of the foveal pit. While this was a rational decision because the chorioretinal structures often appear predominantly aligned with the image-axis, it is often preferable to locate the axial position too. This is particularly important for choroids which exhibit skew (section \ref{subsec:INTRO_MEASURE_ROI}). In figure \ref{fig:CHOROID_fovea_skew}, we show the axial and lateral positions of the fovea detected through Choroidalyzer for choroids with a range of curvature and skew (selected from figure \ref{fig:INTRO_ROI_refr_skew_curve} in section \ref{subsec:INTRO_MEASURE_ROI}). 
                
                Note that across the board the general location of the fovea is correct, although the bottom-right prediction is off-centre. This particular choroid was from a highly hyperopic eye, which explains why the the intra-ocular structures are flattened and the foveola centralis is so large. Here, Choroidalyzer detected the fovea location as the point of highest illumination, but this was not centrally located, nor directly above the ridge which typically forms at the anterior point of outer retina. Nevertheless, even though Choroidalyzer's fovea detection map outputs a predominantly vertically-aligned probability map, our method of fovea pixel coordinate extraction (section \ref{subsubsec:CHOROID_model}) does reasonably well for these choroids which present skew and/or curvature. 

                Secondly, we tested Choroidalyzer's potential generalisability on a small sample of large and difficult choroids from the eyes of the \acrshort{GCU} Topcon cohort which were discarded during data collection (section \ref{subsec:CHOROID_sample_deriv}). In figure \ref{fig:CHOROID_deepgpet} we show three particularly challenging choroids which were part of the 94 images which were manually segmented and added to the dataset (in the internal test set) to help improve Choroidalyzer's generalisability to difficult choroids. In all cases, Choroidalyzer (red) lies closer to the manually detected Choroid-Sclera boundary (blue) than DeepGPET (green). This is an encouraging result, suggesting that Choroidalyzer is able to segment large choroids from \acrshort{SS-OCT} data with poor Choroid-Sclera boundary visibility better than DeepGPET.

                Finally, the majority of our dataset used to build Choroidalyzer were below the age of 50. Therefore, it's very likely that much of the data used to train and evaluate Choroidalyzer did not have a visible suprachoroidal space \cite{yiu2014characterization}. Thus, we could not strictly evaluate Choroidalyzer's ability to include/exclude the suprachoroid in it's region detection. While we anticipate Choroidalyzer to exclude the suprachoroidal space from the choroid (following definition (1) from section \ref{subsec:ch1_INTRO_measure_choroid}) end-users may wish to follow a different definition of the Choroid-Sclera boundary which does include the suprachoroid.
            
                Fortunately, Choroidalyzer's region detection outputs a raw probability map which provides an analogue scale and can be tuned by the end-user according to the particular definition of the Choroid-Sclera boundary they choose for their study. In figure \ref{fig:CHOROID_SCS}, the suprachoroid is visible for all four \acrshort{OCT} B-scans. Setting a very high threshold ($>$ 0.99) allows Choroidalyzer to segment only as far as the largest, posterior-most vessels of the choroid (following definition (1) from section \ref{subsec:ch1_INTRO_measure_choroid}), while setting a very low threshold ($<$ 0.05) allows Choroidalyzer to segment past Haller's layer to the inner scleral boundary (following definition (3) from section \ref{subsec:ch1_INTRO_measure_choroid}). Thus, Choroidalyzer appears to be reasonably flexible in permitting different definitions of the Choroid-Sclera boundary. However, if attempting this, we recommend the end-user inspects the resulting segmentations to ensure Choroidalyzer is not attempting to segment both the choroid and suprachoroid separately and simultaneously (resulting in a zig-zag like Choroid-Sclera boundary). 

            \end{mysubsubsection}
        
        \end{mysubsection}

    \end{mysection}

    \begin{mysection}[]{Choroidalyzer's Reproducibility}

        \begin{mysubsection}[]{Statistical analysis}

            Similar to \acrshort{MMCQ}'s reproducibility, we investigated Choroidalyzer's reproducibility using the i-Test and \acrshort{GCU} Topcon repeated samples. For all repeated pairs, Choroidalyzer segmented the choroidal space and vessels and detected the fovea to allow standardised, fovea-centred choroid-derived measurements.
            
            Briefly, for the i-Test sample, average choroid thickness and \acrshort{CVI} were measured for the nine sub-fields in the \acrshort{ETDRS} grid for each \acrshort{OCT} volume pair. Thus, there were 1080 comparisons made (9 values $\times$ 120 eyes). Choroidalyzer's fovea detection was used to centre each spatial map onto the macula. For the \acrshort{GCU} Topcon sample, 12-stack radial scans were collected throughout the day for all 33 eyes. To prevent sampling bias and remain objective in our analyses, we selected 13 repeated pairs from the morning, 12 from the afternoon and 8 from the evening. Thus, there were 396 comparisons made (33 eyes $\times$ 12 B-scans). Choroidalyzer was used to measure subfoveal choroidal thickness (\acrshort{SFCT}), choroidal area and \acrshort{CVI} in a 6 mm, fovea-centred \acrshort{ROI}. All derived metrics have been defined in section \ref{subsec:ch1_INTRO_measure_choroid}.

            We measure Choroidalyzer's reproducibility at the population-level, reporting population mean and standard deviation (\acrshort{SD}), as well as mean absolute error (\acrshort{MAE}), Pearson and Spearman correlation coefficients. Additionally, we show correlation plots and use Bland-Altman plots to assess agreement and error in the repeated samples. We also report Choroidalyzer's reproducibility at the eye-level using measurement noise $\lambda$ \cite{engelmann2024applicability}. All performance metrics have been described previously in section \ref{subsec:INTRO_metrics}.
            
        \end{mysubsection}

        \begin{mysubsection}[]{Results}

            \begin{mysubsubsection}[]{Population-level}
    
                Table \ref{tab:CHOROID_repr_results} presents the reproducibility performance of Choroidalyzer for macular \acrshort{OCT} volume and B-scan data. Choroidalyzer had excellent correlation across both cohorts (Pearson/Spearman for \acrshort{ETDRS} choroid thickness: 0.9933/0.9969, B-scan \acrshort{SFCT}: 0.9858/0.9889, B-scan choroid area: 0.9835/0.9870). Reproducibility for \acrshort{CVI} was slightly lower but still good (\acrshort{ETDRS} \acrshort{CVI}: 0.9669/0.9655, B-scan \acrshort{CVI}: 0.9090/0.9145).

                \begin{table}[tb]
                \begin{adjustbox}{width=\textwidth}  
                \centering
                
\begin{tabular}{rlllll}
\toprule
\multicolumn{1}{l}{Dataset} & Metric [unit] & Mean (\acrshort{SD}) & \acrshort{MAE} & Pearson & Spearman\\
\midrule
\multirow{1}{2.75cm}{i-Test (\acrshort{ETDRS})} & Choroid thickness [$\mu$m] & 274.09 (91.80) & 6.7342 & 0.9933 & 0.9969 \\
 & \acrshort{CVI} & 0.51 (0.03) & 0.0271 & 0.9669 & 0.9655 \\
 \cmidrule(l){2-6}
\multirow{1}{2.75cm}{\acrshort{GCU} Topcon (Fovea-centred)} & \acrshort{SFCT} {[}$\mu$m{]} & 392.63 (110.94) & 11.5535 & 0.9858 & 0.9889 \\
 & Choroid area {[}mm$^2${]} & 1.62 (0.46) & 0.0509 & 0.9835 & 0.9870 \\
 & \acrshort{CVI} & 0.53 (0.03) & 0.0130 & 0.9090 & 0.9145 \\
 \bottomrule
\end{tabular}

                \end{adjustbox}
                \caption[Choroidalyzer's population-level reproducibility agreement for derived metrics.]{Reproducibility performance of Choroidalyzer. All Pearson and Spearman correlations were statistically significant with P < 0.0001.}
                \label{tab:CHOROID_repr_results}
                \end{table}
    
                \begin{figure}[tb]
                \begin{adjustwidth}{-0.8in}{-0.8in}  
                \centering
                \includegraphics[width=\linewidth]{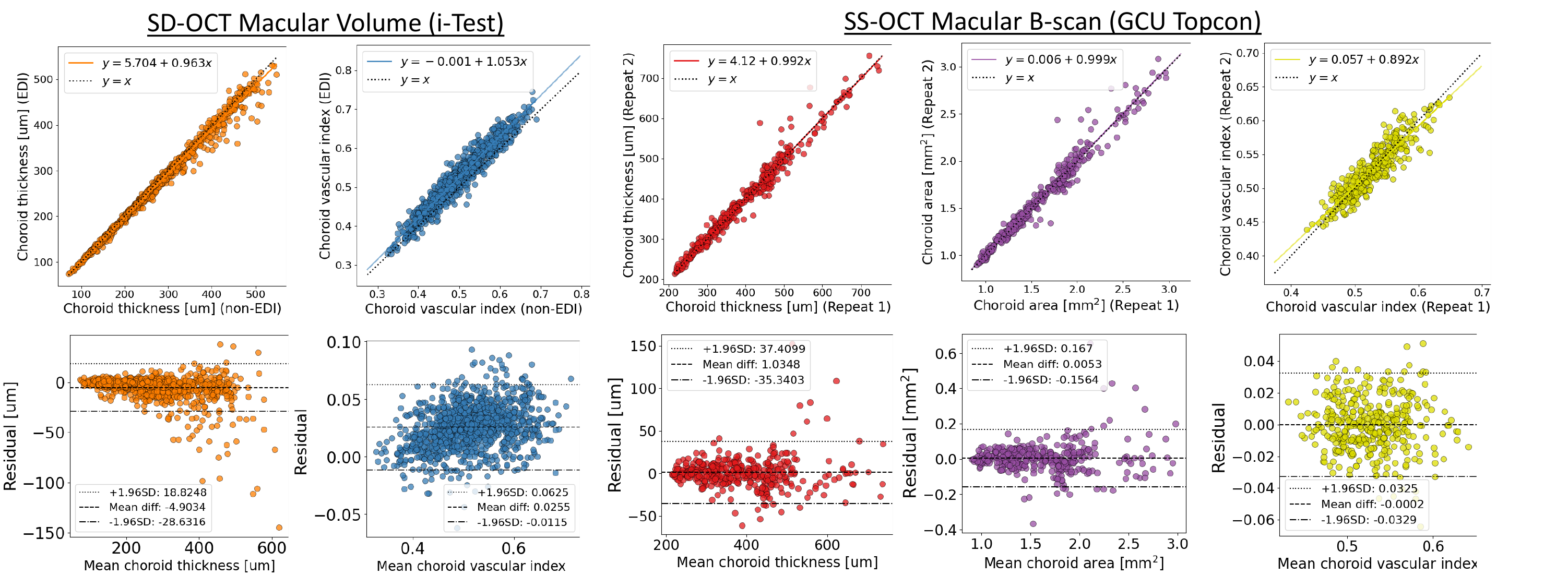}
                \end{adjustwidth}
                \caption[Choroidalyzer's population-level reproducibility agreement for derived metrics.]{Correlation and Bland-Altman plots for assessing the reproducibility of Choroidalyzer for macular \acrshort{OCT} data using Choroidalyzer.}
                \label{fig:CHOROID_pop_repr_macula}
                \end{figure}
    
                Measurements for the \acrshort{GCU} Topcon data suggested these individuals had much thicker subfoveal choroids (mean $\pm$ \acrshort{SD} of 386.9 $\pm$ 110.9 $\mu$m), compared to the i-Test measurements (mean $\pm$ \acrshort{SD} of 275.7 $\pm$ 91.8 $\mu$m). While the i-Test sample was older than the \acrshort{GCU} Topcon sample, the \acrshort{GCU} Topcon study did target hyperopes during recruitment, thus resulting in smaller eyes and thicker choroids. Moreover, the i-Test measurements represent \acrshort{ETDRS} sub-field averages allowing for extremal outliers, and potentially measurement error, to be averaged over unlike one-dimensional, point-source \acrshort{SFCT}.
                
                These observations explain why we see higher \acrshort{MAE} in the \acrshort{GCU} Topcon cohort (\acrshort{MAE}: 11.6 $\mu$m) compared with the \acrshort{ETDRS} data (\acrshort{MAE}: 6.7 $\mu$m). Additionally, this also highlights the importance of reporting values over an area through averaging, rather than taking a single point-source measurement. Nevertheless, these errors are still below any change expected due to diurnal variation \cite{chakraborty_diurnal_2011, tan2012diurnal, usui_circadian_2012, kinoshita_diurnal_2017, singh2019diurnal, ostrin_imi-dynamic_2023}, which changes by approximately 30 $\mu$m over the course of the day. These errors are also below the threshold expected to exceed manual rater agreement in \acrshort{SFCT} \cite{rahman2011repeatability}, and previously reported effect sizes in choroid thickness, such as in myopia progression which can be as low as 20 -- 30 $\mu$m \cite{breher2019metrological, flores2013relationship}.
                
                For \acrshort{CVI}, we report an \acrshort{MAE} of 0.0271 for \acrshort{SD-OCT} \acrshort{ETDRS} \acrshort{CVI} (\acrshort{EDI} and non-\acrshort{EDI} acquisition), and 0.0130 for \acrshort{SS-OCT} B-scan \acrshort{CVI}. The \acrshort{MAE} in the i-Test sample was double that in the \acrshort{GCU} Topcon sample. The \acrshort{MAE} was worse in the i-Test sample because of the additional speckle noise appearing in each non-\acrshort{EDI} member of each i-Test sample pair in the \acrshort{ETDRS} comparison, which can make vessel wall boundaries ambiguous. Nevertheless, it is unlikely that these \acrshort{MAE}s are clinically significant. Breher, et al. \cite{breher2020choroidal} tested the reproducibility of Sonoda's Niblack method \cite{sonoda2014choroidal} across different sub-fields of the \acrshort{ETDRS} macular region and reported a mean difference ranging from 3.9\% to 5.1\%. Additionally, Agrawal's major review on \acrshort{CVI} as a biomarker in retinal pathology reported changes between healthy and disease eyes between 2\% and 6\% \cite{agrawal2020exploring}.
                
                Figure \ref{fig:CHOROID_pop_repr_macula} show the correlation and Bland-Altman plots for the \acrshort{SD-OCT} i-Test and \acrshort{SS-OCT} \acrshort{GCU} Topcon macular \acrshort{OCT} samples. Generally we see that for larger choroids there is larger error in regional metrics. In the case for i-Test, this is entirely expected because the optical signal for non-\acrshort{EDI} \acrshort{OCT} acquisition degrades the deeper through the eye it penetrates, and the visibility of the deeper choroidal vasculature decays rapidly also. This is also the case for larger choroids in general, and thus why we see similarly larger error in the \acrshort{GCU} Topcon sample --- the largest choroid had an \acrshort{SFCT} of 756 $\mu$m (approximately 3.25 standard deviations away from the mean).
                
                The distributions in the Bland-Altman plots for \acrshort{SD-OCT} macular volume data in figure \ref{fig:CHOROID_pop_repr_macula} show under-prediction of choroid thickness and over-prediction of \acrshort{CVI} for the non-\acrshort{EDI} scans of each pair, while the distributions for macular \acrshort{SS-OCT} B-scan data are perfectly centred. The under-prediction of choroid thickness in the non-\acrshort{EDI} measurements was likely due to optical signal decay in non-\acrshort{EDI} scans, which had a direct consequence on \acrshort{CVI}, as this metric is normalised by the size of the choroid, leading to over-prediction of \acrshort{CVI} in non-\acrshort{EDI} measurements. However, this systematic bias of 0.026 (2.6\%) is unlikely to be clinically significant because it sits at the lower bound of previously reported effect sizes in retinal pathology \cite{agrawal2020exploring}. 
                            
            \end{mysubsubsection}
    
            \begin{mysubsubsection}[]{Eye-level}

                \begin{figure}[tb]
                \begin{adjustwidth}{-0.8in}{-0.8in}  
                \centering
                \includegraphics[width=\linewidth]{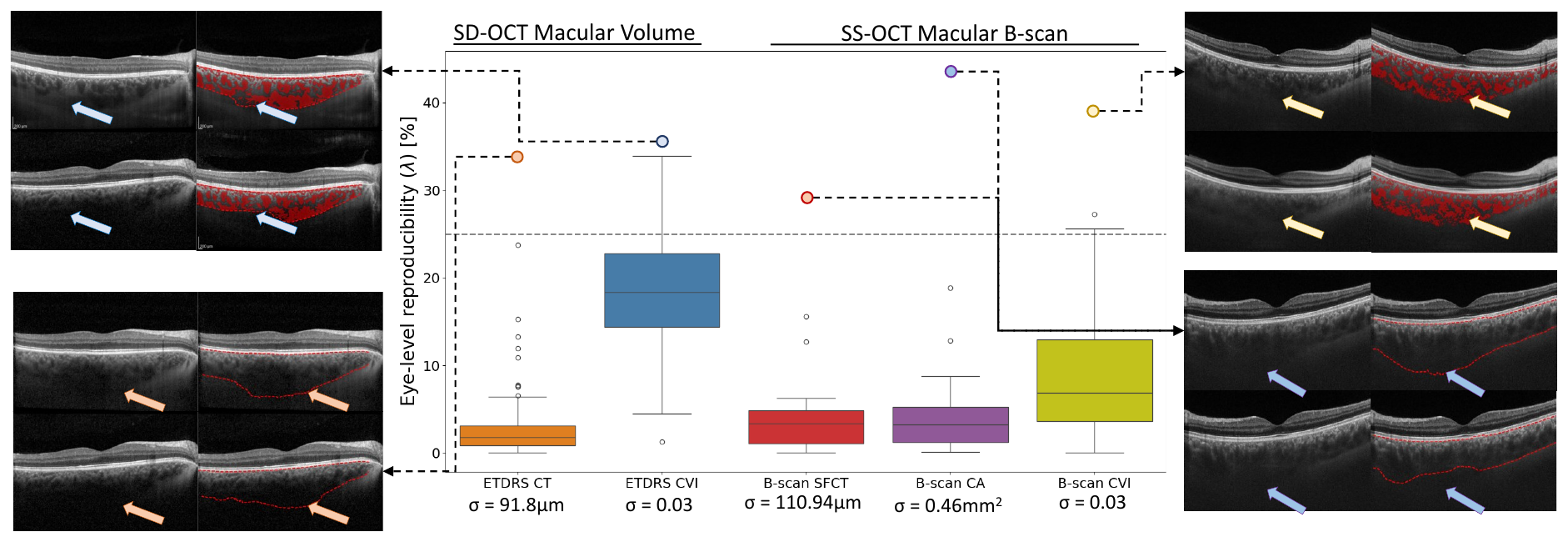}
                \end{adjustwidth}
                \caption[Choroidalyzer's eye-level reproducibility agreement for derived metrics.]{Choroidalyzer's reproducibility at the eye-level. We show distributions of eye-level measurement noise $\lambda$ \cite{engelmann2024applicability} for both the i-Test macular \acrshort{SD-OCT} volume data and \acrshort{GCU} Topcon macular \acrshort{SS-OCT} B-scan data. Representative B-scans, with segmentations overlaid in red, are selected and shown as major outliers with arrows indicating the source of the error. The between-eye standard deviations are shown below each box-plot.}
                \label{fig:CHOROID_ind_repr_macula}
                \end{figure}

                Figure \ref{fig:CHOROID_ind_repr_macula} presents the reproducibility of Choroidalyzer for macular \acrshort{OCT} data at the eye-level. We present box-plots to describe the distribution of measurement noise $\lambda$, expressed in terms of the overall population variability. 
                
                For region metrics, Choroidalyzer's measurement error was very low compared to the population's variability, with the upper quartile of the box-plot distributions for \acrshort{ETDRS} thickness/B-scan \acrshort{SFCT}/B-scan area sitting below 5\% of each metric's population variability. We observed higher variability in \acrshort{CVI} at the eye-level, with B-scan \acrshort{CVI} measurement error primarily sitting below 15\% of the population variability ($\sigma$=0.03) and 25\% for \acrshort{ETDRS} \acrshort{CVI} ($\sigma$=0.03). This suggests that \acrshort{CVI} is less reliable than regional metrics like area given the population's variability, a conclusion drawn also by Breher, et al. \cite{breher2020choroidal}. 
                
                \acrshort{ETDRS} \acrshort{CVI} could have higher measurement variability than B-scan \acrshort{CVI} because of the image quality difference between \acrshort{EDI} and non-\acrshort{EDI} volume pairs (ART 50 for \acrshort{EDI} volume data vs. \acrshort{ART} 12 for non-\acrshort{EDI} volume data), as well as the unregistered nature of the volume scans which have different B-scan sampling rates (31 B-scans per \acrshort{EDI} volume compared to 61 B-scans per non-\acrshort{EDI} volume) and slightly different regions of interest up to some rotation. These differences ultimately lead to different stacks of cross-sections of the dense, heterogeneous choroidal vasculature being analysed, although previous research suggests this should not make a difference \cite{agrawal2017influence,goud2019new, kim2022influence}.
                
                The major outliers for the macular \acrshort{OCT} B-scans shown in figure \ref{fig:CHOROID_ind_repr_macula} are mainly thick choroids whose lower boundaries (arrows) have poor visibility primarily due to decaying optical signal. This has a direct consequence for the \acrshort{CVI} outliers, as \acrshort{CVI} is normalised by the estimated size of the choroid. Moreover, image quality plays a clear role in poor quality segmentation, as the major outlier for B-scan \acrshort{CVI} shows one member of the repeated pair with significantly poorer contrast and thus poorer vessel wall definition. Overall, it is encouraging to see that major outliers are coming from challenging cases of thicker choroids with obscure Choroid-Sclera boundary and poor quality vessel wall definition.
                
            \end{mysubsubsection}

        \end{mysubsection}

    \end{mysection}

    \begin{mysection}[]{Discussion}

        \begin{mysubsection}[]{Evaluation}\label{subsec:ch5_CHOROID_discuss_eval}

            We developed Choroidalyzer, an end-to-end pipeline for choroidal analysis. Choroidalyzer shows excellent performance on the internal and external test sets. Where Choroidalyzer produced the highest errors were primarily cases of imperfect ground truth labels and Choroidalyzer was generally preferred by a blinded adjudicating ophthalmologist (supervisor Dr. Ian J.C. MacCormick), further indicating robustness and good performance. Its agreement with manual segmentations, which demand substantial time and attention from a human expert, is comparable to the inter-grader agreement. This suggests that Choroidalyzer performs well compared to laborious manual segmentation and also highlights the subjectivity introduced by manual graders. Choroidalyzer not only produces results similar to that of a skilled manual grader, but also does so fully-automatically without introducing subjectivity. Choroidalyzer is deterministic, meaning it is 100\% repeatable when applied to the same image more than once, unlike certain semi-automatic approaches like \acrshort{GPET} and especially manual approaches. Thus, researchers using Choroidalyzer would provide much more consistent, reliable and comparable results to other studies also using Choroidalyzer than if different manual graders were used in each case. Therefore, Choroidalyzer can help to improve the standardisation of choroid-derived metrics in \acrshort{OCT} image sequences.
    
            Additionally, Choroidalyzer saves a substantial amount of time per B-scan over manual segmentation, freeing up researcher time and enabling large scale analyses that otherwise would not be possible. Even compared to the current state-of-the-art for automatic methods, DeepGPET and Niblack, Choroidalyzer can do the analysis in roughly a quarter of the time, suggesting its feasibility for large-scale ophthalmic image analysis, even on a standard laptop CPU. 
            
            AlzEye \cite{wagner2022alzeye} is a recently collected, large-scale dataset of retinal images from the Moorefield's Eye Hospital NHS Foundation Trust. AlzEye contains 1,567,358 \acrshort{OCT} volume scans from 154,830 patients. This \acrshort{OCT} image set could be processed by Choroidalyzer in approximately 1 year using a low-grade CPU, compared to approximately 3.5 years using Niblack and DeepGPET combined. These estimates do not take into account the fact that Niblack requires parameter specification which may not be suitable across the entire \acrshort{OCT} image set, and that the inference time for Choroidalyzer (and DeepGPET, albeit) could be significantly improved with \acrshort{GPU} hardware.
            
            Importantly, Choroidalyzer also provides an end-to-end pipeline which makes it easier to implement and use than having to combine multiple methods like Niblack and DeepGPET. Ease-of-use is often underappreciated in the literature but key in saving researchers time and allowing them to focus on the science. Additionally, Choroidalyzer works similarly across different devices (Heidelberg and Topcon) and \acrshort{OCT} scan types (\acrshort{SD-OCT} and SS-OCT). Figure \ref{fig:CHOROID_examples} shows some notable examples of Choroidalyzer applied to two \acrshort{OCT} B-scans from each of the \acrshort{OCT} imaging devices used to develop it. This is a major advancement in the field as previous methods (tables \ref{tab:INTRO_region_methods} -- \ref{tab:INTRO_region_vessel_methods}) tend to only train/evaluate their methods on only on \acrshort{OCT} imaging device.

            \begin{figure}[tb]
            \begin{adjustwidth}{-0.75in}{-0.75in}
            \centering
            \includegraphics[width=\linewidth]{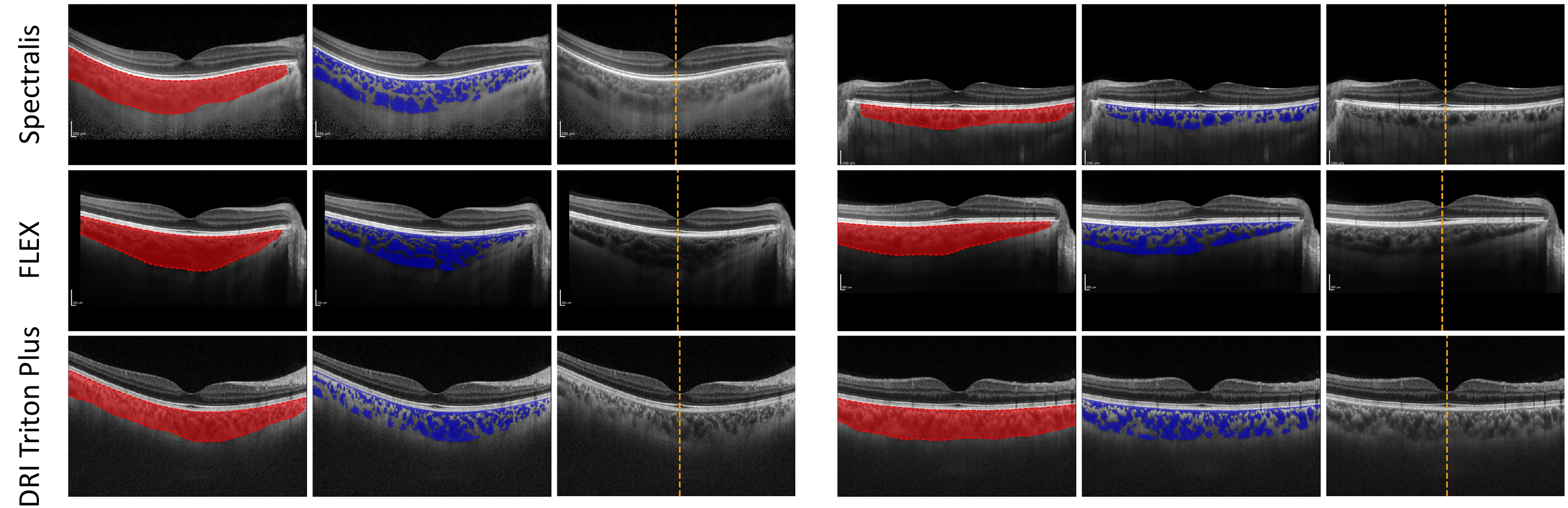}
            \end{adjustwidth}
            \caption[Examples of Choroidalyzer applied to \acrshort{OCT} data from different devices.]{Examples of Choroidalyzer being applied to B-scans from different imaging devices for region segmentations (left), vessel segmentations (middle) and fovea column location (right).}
            \label{fig:CHOROID_examples}
            \end{figure}
        
            Choroidalyzer performed well against manual graders relative to the state-of-the-art methods, reaching or surpassing the levels of agreement even between the two manual graders, particularly for \acrshort{CVI}, a far more difficult metric to calculate accurately than area and thickness. The inter-grader agreement between manual graders for these metrics indicate a potential lower bound of what effect sizes we might expect from these metrics. This has important downstream impact on the statistical confidence of results from cohort studies, particularly when assessing the choroidal vasculature.
    
            It is often difficult to visualise the choroid due to imaging noise, poor eye tracking and patient fixation, or operator inexperience. Thus, in some cases vessel boundaries can be hard to discern. This is why we proposed to use a soft version of \acrshort{CVI} where the probabilities that Choroidalyzer outputs are used instead of thresholded, binarised segmentations. The probabilities capture uncertainty about the precise location of the vessel wall and thus is more robust than using a single, somewhat arbitrary threshold. End-users could also tune the binarization threshold for their own images, if desired, which might help in instances of poor visibility of the choroidal vasculature.

            The dataset in the present work was substantially larger than the one used for DeepGPET and the majority of other methods from the literature (tables \ref{tab:INTRO_region_methods} -- \ref{tab:INTRO_region_vessel_methods}), allowing Choroidalyzer to learn from a vast number (5,600 B-scans from 385 eyes in 233 participants) of unique \acrshort{OCT} B-scans. The dataset also included manual ground truth labels of particularly challenging \acrshort{SS-OCT} choroids. As a result, we found that Choroidalyzer may have the ability to segment challenging choroids with poor quality Choroid-Sclera boundary, unlike DeepGPET (figure \ref{fig:CHOROID_deepgpet}). Moreover, Choroidalyzer was trained on region and vessel ground truths generated by fully automatic DeepGPET and semi-automatic \acrshort{MMCQ} --- the latter of which had fixed parameters to promote reproducibility and consistency in its segmentation procedure. Maloca, et al. \cite{maloca2023human} argues that such approaches to generating ground truth labels are preferable as they reduce subjectivity and thus bias and inconsistency, representing an additional benefit of Choroidalyzer over previous approaches in the literature which use manual graders for generating vessel segmentation ground truth labels \cite{khaing2021choroidnet, zhu2022synergistically, wang2023choroidal, wen2024transformer}. 
            
            Our model detected the fovea well, and the largest errors were cases where ground-truths were incorrectly labelled with the model correctly identifying the fovea location as confirmed by masked adjudication. Thus, the model performed even better than what the quantitative results suggest. Moreover, the error observed in both internal and external test sets did not result in significantly different choroid-derived metrics of thickness, area and \acrshort{CVI} even after simulating random, per-sample deviations of $\pm6$px, twice the Median absolute error on a subset of the dataset (figure \ref{fig:CHOROID_fovea_perturb}). 
            
            We did observe that small errors in the lateral fovea position can in rare cases lead to large error in choroid-derived metrics (figure \ref{fig:CHOROID_fovea_perturb}). These are where the local curvature of the segmented \acrshort{RPE}-Choroid boundary shifts significantly (figure \ref{fig:CHOROID_fovea_perturb}(C), left - red arrow), where the choroid is extremely heterogeneous in terms of size around the fovea (figure \ref{fig:CHOROID_fovea_perturb}(C), middle - red arrow), or the unfortunate combination of superficial retinal vessel shadowing and vessel heterogeneity (figure \ref{fig:CHOROID_fovea_perturb}(C), right - red arrow). Therefore, while on average small perturbations in the lateral position of the fovea does not impact choroid-derived metrics, this is not the case across all choroids.

            In the present work, we have focused on identifying the lateral position of the fovea which is needed to define the fovea-centred \acrshort{ROI}. The axial position is typically less important than the lateral position because the chorioretinal structures predominantly lie laterally. However, in cases related to poor image acquisition or high myopia, the choroid can exhibit skew or curvature relative to the image axis (figure \ref{fig:CHOROID_fovea_skew}). In these cases, it is not only important to make measurements which are choroid-aligned, i.e. locally perpendicular to the \acrshort{RPE}-Choroid boundary as in section \ref{subsec:ch1_INTRO_measure_bsan}, but it is also pertinent to know the axial location of the fovea so that the resulting \acrshort{ROI} is accurately accounting for this skew/curvature, which Choroidalyzer is able to do successfully already (figure \ref{fig:CHOROID_fovea_skew}).
    
            Segmentation performance for peripapillary scans was reasonable but much worse than for other scan patterns and locations. This could be due to those scans being relatively rare in our dataset and showing parts of the retina on the nasal side of the optic disc that are not captured in macular B-scans. Although, we found that Choroidalyzer was less robust than DeepGPET at segmenting the peripapillary choroidal space (figure \ref{fig:CHOROID_peri_deepgpet}). DeepGPET's model was previously trained on a large natural image dataset, ImageNet \cite{deng2009imagenet}, and only fine tuned for one task rather than training a network from scratch on three tasks like Choroidalyzer. Thus, DeepGPET's fine-tuning procedure may have instilled better generalisable features for choroid region segmentation. For example, DeepGPET's model weights were averaged over during training using exponential weight averaging which is known to improve generalisability \cite{morales2024exponential}. 
            
            Nevertheless, Choroidalyzer's lightweight model still outperforms DeepGPET on macular \acrshort{OCT} data in terms of execution time and functionality. More peripapillary training data would likely increase performance, as well as a more rigorous training procedure, such as use of exponential moving average of the model weights. At present, Choroidalyzer can be used for peripapillary B-scan but requires subsequent manual inspection and potential correction. Furthermore, adjusting the binarization threshold for the vessel predictions may improve results. Regardless, DeepGPET is likely to provide more reliable choroid region segmentations.

            The suprachoroidal space remains an elusive structure often not seen on \acrshort{OCT} images because of how thin it is \cite{moisseiev_suprachoroidal_2016}. As people age, their choroid thins and their sclera stiffens, so there is a higher chance of this opening being seen on \acrshort{OCT} \cite{yiu2014characterization}. However, there are clinical situations where it may be important to quantify this space, such as in post-operative care from surgery, where severely low eye pressure can sometimes cause the suprachoroidal space to bloat \cite{saidkasimova_suprachoroidal_2021}. For Choroidalyzer's model development, we imposed a strict definition of the Choroid-Sclera boundary for Choroidalyzer to learn which follows definition (1) from section \ref{subsec:ch1_INTRO_measure_choroid}, i.e. exclusion of the suprachoroidal space. However, we found that the end-user may fine-tune the threshold used to binarise the region segmentation map to detect the Choroid-Sclera boundary according to all three definitions from section \ref{subsec:ch1_INTRO_measure_choroid} (figure \ref{fig:CHOROID_SCS}). This may have important consequences in scenarios where the suprachoroidal space is clinically relevant.
            
        \end{mysubsection}

        \begin{mysubsection}[]{Reproducibility}

            Choroidalyzer had very high correlations across repeated measurements (Pearson and Spearman correlations for regional metrics were all greater than or equal to 0.98, and 0.91 for \acrshort{CVI} in the macular data). Encouragingly, \acrshort{MAE}s from all paired samples in the repeated macular \acrshort{OCT} data were below the values of previously reported reproducibility, agreement thresholds or effect sizes \cite{breher2019metrological, agrawal2020exploring, rahman2011repeatability, breher2020choroidal}. 
    
            The systematic bias observed in the macular \acrshort{OCT} volume data was a result of comparing repeated scans where \acrshort{EDI} mode was toggled on and off, increasing the potential for the non-\acrshort{EDI} scan of each pair having poor Choroid-Sclera boundary visibility. The residuals for \acrshort{CVI} were well balanced and appeared to show no apparent trend, apart from the minor systematic bias due to the aforementioned non-\acrshort{EDI} optical signal decay. Major outliers from eye-level reproducibility analysis for both regional and vessel metrics in macular \acrshort{OCT} data highlighted the challenge larger choroids pose for segmentation, given the propensity for optical signal degradation.
            
            Nevertheless, the majority of measurement error of regional metrics observed for macular \acrshort{OCT} B-scan data were within 5\% of the population's variability for regional metrics, 25\% for volumetric \acrshort{CVI} and 15\% for B-scan \acrshort{CVI}. However, the population variability is notably smaller for \acrshort{CVI} measurements, and the image quality, sampling rate and unregistered nature of the \acrshort{EDI} and non-\acrshort{EDI} volume pairs all likely played a role in producing greater variability in volumetric \acrshort{CVI}. It's also very possible that the high measurement variability in \acrshort{CVI}, relative to the regional metrics, is due to the exacerbation of measurement error in both vessel area and region area, as observed in chapter \ref{chp:chapter-mmcq}.
            
            The higher measurement variability in \acrshort{CVI} indicates the need to interpret this choroid-derived metric with care \cite{breher2020choroidal}. In chapter \ref{chp:chapter-mmcq} we considered the importance of reporting purely vascular metrics like vessel area and volume alongside \acrshort{CVI}, but this was not done here. It is possible that Choroidalyzer-derived metrics of vessel area and volume may be more reliable than \acrshort{CVI}, and thus should also be reported alongside. Ultimately, the macular regional metrics were more reliable than vessel metrics, but they were both still reasonable for the homogeneous (young and healthy) cohorts analysed.
    
            Thus, Choroidalyzer is able to provide fully automatic choroid-derived measurements of the choroidal space and vessels which are standardised according to a robust, fovea-detected \acrshort{ROI}. Importantly, the reproducibility analysis (both at the population- and eye-level) highlight the ability for Choroidalyzer to output reliable and clinically meaningful measurement of the choroid, and provide end-users with interpretable metrics to help differentiate measurement error from true biological change in future studies which adopt Choroidalyzer for \acrshort{OCT} image analysis.
      
        \end{mysubsection}

        \begin{mysubsection}[]{Limitations and future work} \label{subsec:ch5_CHOROID_limits}

            There were some limitations associated with Choroidalyzer. Choroidalyzer was built using only \acrshort{OCT} data related to systemic disease and has not been evaluated on data related to eye pathology. While we anticipate Choroidalyzer's value for ophthalmology in systemic health, it's performance in the context of ocular health is yet unknown. In particular, stress testing Choroidalyzer with examples of extreme choroid thinning or thickening such as in pathological myopia \cite{ohno2021imi} or central serous chorioretinopathy \cite{fung2023central}, respectively, or optical disturbance from photoreceptor degeneration in geographic atrophy \cite{vallino2024structural}, would be suitable next steps. Nevertheless, we anticipate Choroidalyzer will be a particularly useful tool in the field of oculomics, where it has demonstrated potential for generalisability.

            Regarding Choroidalyzer's reproducibility, and similar to DeepGPET's reproducibility limitations (section \ref{subsec:ch4_deepgpet_rpr_lims}), our paired macular \acrshort{OCT} volume data were not registered between acquisitions. Thus, the i-Test sample does not provide a perfect comparison to which segmentation could be the direct cause of measurement error. Moreover, there exists potential data leakage in our reproducibility analysis. Choroidalyzer's training and evaluation had a distinct overlap with the two reproducibility populations (using ground truth labels previously segmented by DeepGPET). 22\% of the i-Test sample (13 / 60) were used for training, but only the \acrshort{EDI} counterpart of their \acrshort{OCT} volume pair was used. Thus, while Choroidalyzer had potentially seen 26 eyes (of 120 used for reproducibility analysis) during training, the image artefacts introduced by deactivating \acrshort{EDI} mode still posed a significant challenge to Choroidalyzer's reliability as a method. For the \acrshort{GCU} Topcon sample, 20 eyes featured in at least one or more B-scans for training Choroidalyzer. However, the majority of B-scans from 62\% of participants in the reproducibility cohort (13 / 21) were excluded from Choroidalyzer's dataset because of DeepGPET's failure to generate ground-truth segmentation labels during data curation, with only a few of their B-scans being added back after manual annotation (figure \ref{fig:CHOROID_sample_deriv}).

            Choroidalyzer can process a single raw \acrshort{OCT} B-scan inputted as an image file into a set choroid-derived measurements. Thus, Choroidalyzer is set up to only process single B-scans at a time, and would require additional configuration to process and measure \acrshort{OCT} volumes (for example, to generate \acrshort{ETDRS} sub-field measurements as in section \ref{subsubsec:ch1_INTRO_oct_volume}). Moreover, it is common for researchers to collect \acrshort{OCT} image data saved using proprietary file formats which store the image and metadata direct from the imaging device --- a common proprietary file format for Heidelberg imaging devices are \verb|.vol| RAW files. To facilitate truly end-to-end and accessible choroidal analysis to the general researcher, it should be wrapped into a more automatic pipeline. 
            
            This is why we are working presently on such a toolkit, OCTolyzer, which folds Choroidalyzer into a segmentation module (as well as DeepGPET, as alluded to in section \ref{subsec:chp4_limitations}) with the aim of processing macular \acrshort{OCT} scan patterns (single-line, radial and volume) to provide reproducible and clinically meaningful measurements of the choroid in the macula. In this way, DeepGPET and Choroidalyzer provide fully automatic analysis of the posterior pole, further broadening the impact of these methods to the wider research community. While there are still hurdles preventing fully automatic large-scale ophthalmic image analysis due to the additional overhead of automating data extraction from the imaging device, OCTolyzer's ability to be run from the terminal eliminates certain pre-requisite knowledge in computer science or the Python programming language which Choroidalyzer requires to enable efficient and clinically meaningful analysis of the macular choroid.
            
        \end{mysubsection}

        \begin{mysubsection}[]{Outputs}

            In this chapter, there have been two publication outputs, one of which has been published and peer-reviewed (by December 2024) and another which is currently in peer-review, but has been uploaded as a pre-print on Arxiv. Author of thesis is in bold type, with any co-lead authors underlined.
    
            \begin{itemize}
    
                \item \underline{Engelmann, Justin}, \underline{\textbf{Jamie Burke}}, Charlene Hamid, Megan Reid-Schachter, Dan Pugh, Neeraj Dhaun, Diana Moukaddem et al. ``Choroidalyzer: An open-source, end-to-end pipeline for choroidal analysis in optical coherence tomography.'' Investigative Ophthalmology \& Visual Science 65, no. 6 (2024): 6-6. (Mathematical and Computational Ophthalmology Special Issue)
            
                \item \textbf{Burke, Jamie}, Justin Engelmann, Charlene Hamid, Diana Moukaddem, Dan Pugh, Neerah Dhaun, Niall Strang et al. ``OCTolyzer: Fully automatic analysis toolkit for segmentation and feature extracting in optical coherence tomography (OCT) and scanning laser ophthalmoscopy (SLO) data.'' arXiv preprint arXiv:2407.14128 (2024). Submitted and in peer-review at Elsevier's Medical Image Analysis.
                
            \end{itemize}
    
            The software associated with this method, Choroidalyzer, has been published as open-source and is freely available on GitHub \href{https://github.com/justinengelmann/choroidalyzer}{here}. Similarly, OCTolyzer is freely available on GitHub \href{https://github.com/jaburke166/OCTolyzer}{here}.

        \end{mysubsection}

        \begin{mysubsection}[]{Executive summary}

            In this chapter, we have developed a fully automatic, deep learning based method, Choroidalyzer, for generating automatic measurements of the choroid from raw \acrshort{SD-OCT} or \acrshort{SS-OCT} B-scans. Choroidalyzer segments the region and vessels of the choroid in an \acrshort{OCT} B-scan and detects the lateral position of the fovea, enabling standardised measurements such as thickness, area, vessel area and \acrshort{CVI} to be computed automatically in <0.3 seconds per B-scan on a standard laptop CPU. Choroidalyzer performed similarly well across internal and external test sets, suggesting Choroidalyzer's potential to generalise to new datasets in systemic disease. Choroidalyzer improved on current state-of-the-art approaches DeepGPET and Niblack for region and vessel detection --- according to comparison with manual ground truth labels --- and was preferred by a clinical ophthalmologist after blinded manual adjudication on examples exhibiting high error. We also found Choroidalyzer to output reproducible and clinically meaningful choroid-derived metrics at both the population- and eye-level, with regional metrics like thickness and area reporting lower measurement variability than \acrshort{CVI}. 
            
            Choroidalyzer's functionality potentially extends upon it's original design, such as being able to segment challenging \acrshort{SS-OCT} B-scans --- unlike DeepGPET --- and capturing the inherent uncertainty of the choroidal vessels in OCT B-scans by outputting a `soft' \acrshort{CVI}. Moreover, Choroidalyzer has the ability to consider different definitions of the Choroid-Sclera boundary (include/exclude suprachoroidal space). Moreover, while we anticipate Choroidalyzer to be particularly useful in datasets related to systemic health, Choroidalyzer has not been validated on examples related to retinal pathology. Furthermore, external validation and benchmarking of Choroidalyzer's deep learning model on publicly available datasets of abnormal retinochoroidal structures would further facilitate the model's applicability in wider ophthalmology, separate from its use in oculomics. As Choroidalyzer currently operates `per B-scan', future work would look to wrap Choroidalyzer within an automatic pipeline, OCTolyzer, capable of performing large-scale choroidal image analysis processing for batch-processing of different macular \acrshort{OCT} scan patterns.
            
        \end{mysubsection}
    
    \end{mysection}
	
\end{mychapter}

\begin{mychapter}[]{Applications of choroidal image analysis in OCT image sequences}\label{chp:chapter-applications}

    \begin{mysection}[]{The OCTANE Study: Chorioretinal changes in a longitudinal cohort of  kidney donor and recipients}\label{sec:ch_app_sec_ckd}

        \begin{mysubsection}[]{Introduction}
    
            Chronic kidney disease (\acrshort{CKD}) is a major cause of morbidity and mortality, affecting over 800 million people worldwide \cite{kovesdy2022epidemiology, macrae2021comorbidity, jager2019single}. \acrshort{CKD} is a gradual loss of kidney function, resulting in a failure to filter blood effectively and is associated with alterations in microvascular structure and function \cite{houben2017assessing}. Assessment of microvascular function is therefore helpful in the diagnosis, prognosis and treatment of renal disease. Current clinical evaluation is based on blood and urine testing and invasive kidney biopsy.
        
            The eye and the kidney show a striking resemblance in anatomy, physiology and response to disease \cite{wong2014kidney}, and in particular the choroidal microcirculation reflects the renal microcirculation \cite{farrah2020eye}. Developmentally, the choroid (specifically, the choriocapillaris) and cortical (glomerular) capillaries have similar vascular endothelium with similarly sized fenestrated vessel walls allowing subretinal fluid exchange in the choroid and blood filtration in the kidneys. Moreover, the circulatory systems in the choroid and renal cortex have similar proportions of blood flow in contrast to their retinal and renal medullary counterparts \cite{nickla2010multifunctional, satchell2009glomerular}. These similarities suggest the potential for choroidal microvasculature to reflect information about kidney function for the care of patients with \acrshort{CKD}.
        
            Choroid thickness has previously been shown to correlate strongly with systemic inflammation and renal dysfunction in patients with \acrshort{CKD}, compared with sex- and age-matched healthy and hypertensive individuals, suggesting that choroid thinning may reflect systemic microvascular injury in general and renal injury in particular \cite{balmforth2016chorioretinal}. An overactive sympathetic drive may contribute to disease progression in \acrshort{CKD} \cite{kaur2017sympathetic}, and could explain choroidal thinning since choroidal perfusion is influenced by its autonomic supply \cite{balmforth2016chorioretinal}. These observations suggest the potential clinical utility of choroidal biomarkers in renal disease. Moreover, a recent study reported that choroidal thinning significantly associates with decline in estimated glomerular filtration rate (eGFR), and is modifiable during treatment of \acrshort{CKD} \cite{farrah2023choroidal}.
        
            Microvascular dysfunction and systemic inflammation is observed in \acrshort{CKD}, and it is important to measure microvascular function through renal biopsies during therapy. With the striking similarities in circulation, structure and pathogenic pathways between the kidney and eye \cite{wong2014kidney, farrah2020eye}, and with the advent of \acrshort{OCT} imaging, examining the microvascular structures in the choroid may provide a more practical and less invasive way to monitor treatment for \acrshort{CKD}. We hypothesised that the observed effect of renal injury on the choroid at end-stage \acrshort{CKD} is also similarly reflected during treatment, whereby improved renal function corresponds to an increase in choroid thickness and area. Thus, we prospectively analysed a longitudinal cohort of patients with end-stage \acrshort{CKD} undergoing living donor kidney transplantation, to investigate the correspondence between the choroid and improving renal function/reduced systemic inflammation. Moreover, as a secondary analysis, we also explored whether a loss in renal function after unilateral nephrectomy results in thinning of the choroid in donors.
        
        \end{mysubsection}

        \begin{mysubsection}[]{Data}

            \begin{mysubsubsection}[]{Study population}

                Longitudinal data was prospectively collected on patients with end-stage \acrshort{CKD} undergoing living donor kidney transplantation and healthy kidney donors undergoing unilateral nephrectomy (NCT0213274) \cite{dhaun2014optical}. Data collection and subsequent analyses were conducted after ethical approval from the South East Scotland Research Ethics Committee, in accordance with the principles of the Declaration of Helsinki and all participants gave informed consent to recruitment. Eligibility criteria for recruitment were (1) donors must be living and healthy throughout the period of analysis, (2) recipients with end-stage \acrshort{CKD} have a functional kidney transplant and (3) must be aged 18 or over. To prevent ocular issues confounding our results, our exclusion criteria were (1) any ocular pathology pre-transplant, (2) any previous eye surgery, (3) a refractive error exceeding $\pm6$ dioptres or (4) a diagnosis of diabetes mellitus. We judged image quality by the OSCAR-IB criteria \cite{tewarie2012oscar} and excluded images with B-scan signal quality $\leq$ 15, using the imaging device's built-in \acrshort{SNR} (Heidelberg Engineering, Heidelberg, Germany), or whose Choroid-Sclera boundary was invisible due to speckle noise or partial image cropping.

                Data was sampled when possible at 8 time points: before transplant surgery (baseline), and seven time points post-transplant (1 week, 2 weeks, 4 weeks, 8 weeks, 16 weeks, 28 weeks and 52 weeks). Clinical data at each time point included standard biomarkers of renal function, including high sensitivity C-reactive protein (hsCRP), serum urea and creatinine, eGFR and urine protein to creatinine ratio (Urine P:Cr). At baseline, standard clinical measurements of systolic, diastolic and mean arterial blood pressure (BP) and body mass index (BMI) were also collected.
        
            \end{mysubsubsection}

             \begin{mysubsubsection}[]{Image collection and analysis}
        
                Horizontal-line, fovea-centred, \acrshort{EDI-OCT} B-scans of each participant's right eye was collected using the Heidelberg \acrshort{SD-OCT} Spectralis \acrshort{OCT}1 Standard Module. Each B-scan covered a 30$^{\circ}$ \acrshort{FOV} (approximately 9 mm laterally), and was extracted as a 768 $\times$ 768 (pixel height$\times$width) image for analysis. Patients were examined between 9am and 5pm and were endeavoured to be followed up at around the same time of day at each time point to limit the effects of diurnal variation on the choroid. Each scan was captured using active eye tracking with an \acrshort{ART} of 100. Although refractive error and axial length were not collected for this population, the built-in scan focus parameter (Heidelberg Engineering, Heidelberg, Germany) is used to compensate for the eye's refractive error \cite{engeineering_hardware_instructions}, and was extracted from the acquisition metadata and used as a surrogate value for refractive error.
        
                Semi-automatic \acrshort{GPET} was used to measure choroidal thickness and area using definitions described in section \ref{subsec:ch1_INTRO_measure_bsan}. We measured choroid thickness at three distinct locations: subfoveal and 2 mm temporally and nasally. Choroid area was measured in a 6 mm, fovea-centred \acrshort{ROI} centred. Note that DeepGPET, \acrshort{MMCQ} and Choroidalyzer were not used here as these approaches were developed after this work was conducted. 
                
            \end{mysubsubsection}

        \begin{mysubsubsection}[]{Statistical analysis}

            We investigated how choroid thickness and area measurements changed over time in donors and recipients. To improve the robustness of our thickness measurements and prevent spurious associations, we considered only a single measurement by averaging choroid thickness across all three macular locations \cite{xie2021evaluation}, resulting in a more representative value of thickness for each scan. We examined the difference in choroid thickness and area between measurements at baseline and one year post-transplant, testing for differences in their means longitudinally via the Student's t-test for dependent, paired samples. We uses a P-value of 0.05 as the threshold for statistical significance. 

            We also investigated the average longitudinal evolution of relative change that choroidal measures underwent over time, and tested for linear associations between choroid image measures and markers of renal function. Pairwise associations were estimated at the patient-level for each cohort using repeated measures Pearson correlation coefficient, $r_{rm}$ \cite{bland1994correlation, bakdash2017repeated}, which accounts for variability across and within patients.
            
            The nested structure within the data was also leveraged to fit linear mixed-effects models, setting an individual random effect to model the variation among individuals, accounting for age, sex, approximate refractive error (measured in dioptres, D) and the time of day the scan was taken, relative to when their baseline scan was taken (measured in decimal hours). Linear mixed-effect models were used to account for variation in measurements across time and across individuals. For each clinical measure of renal function, we fitted a linear mixed-effect model for each choroidal image measure in turn as one of the associated independent variables. Linear mixed-effect modelling was carried out using the \verb|lmer| package in base R (version 4.2.2) \cite{lmer_R}. We identified independent variables as statistically significant predictors if their associated 95\% confidence interval excluded 0.

        \end{mysubsubsection}

    \end{mysubsection}        

    \begin{mysubsection}[]{Results}

            \begin{figure}[!t]
            \centering
            \includegraphics[width=0.8\textwidth]{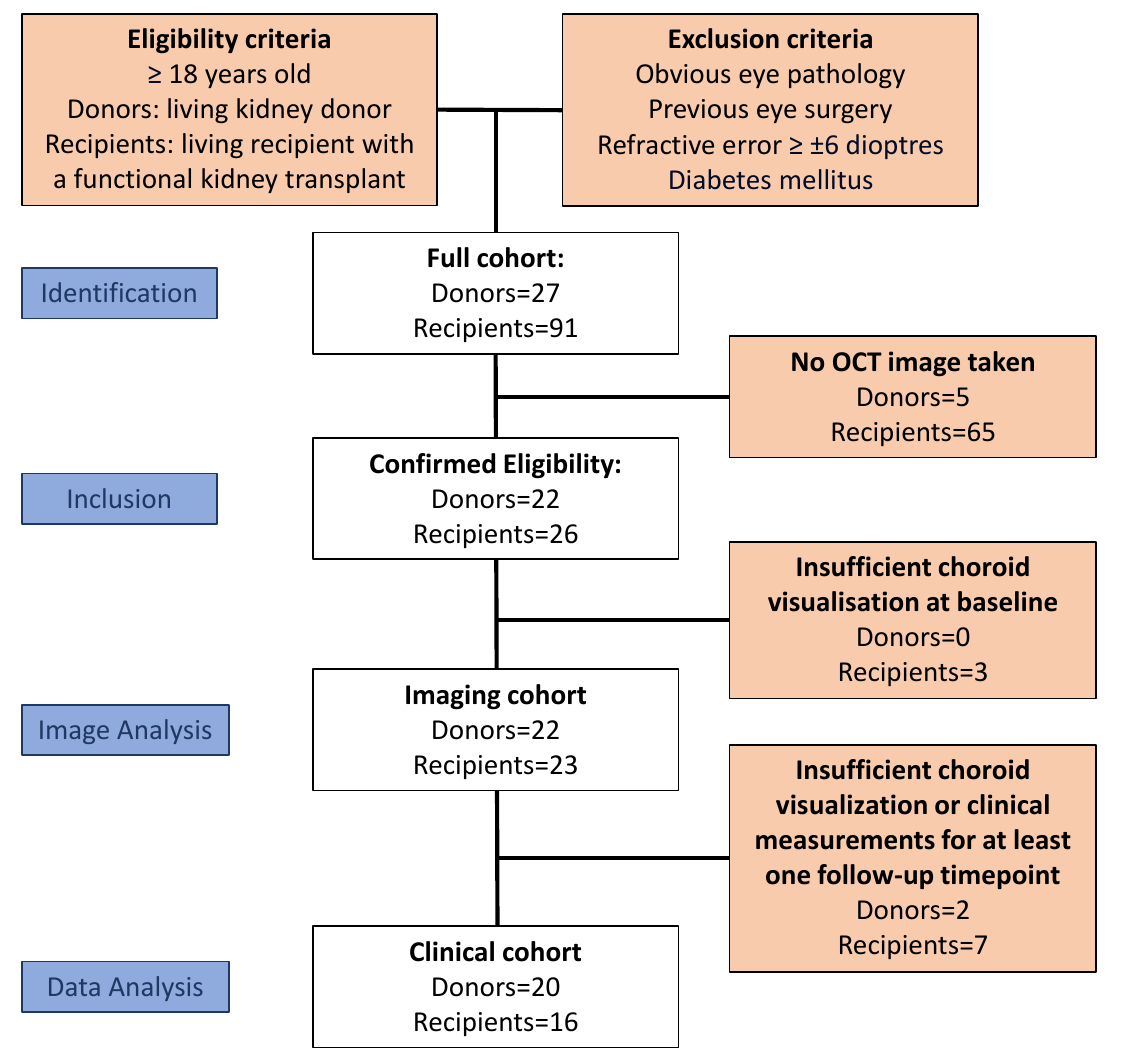}
            \caption[Sample derivation flowchart for the renal cohort.]{Population derivation for longitudinal renal cohort.}
            \label{fig:CKD_flowchart}
            \end{figure}
    
        \begin{mysubsubsection}[]{Participants}

            \begin{table}[tbp]
            \centering
            \scalebox{0.75}{\begin{tabular}{lll}
\toprule
 & \multicolumn{2}{c}{Cohort} \\
 \cmidrule(l){2-3}
 & Donors & Recipients \\
 \midrule
Sample, n (\%) & 20 & 16 \\
Age, years & 51 ± 10 & 45 ± 14 \\
Male sex, n (\%) & 8 (40) & 10 (63) \\
Approx. Refractive error (D) & 0.49 ± 1.17 & -0.33 ± 1.73 \\
Daytime (Hr:Min) & 13:50 ± 1:41 & 13:59 ± 2:19 \\
BMI, kg/m$^2$ (\%) & 25 ± 3 (95) & 27 ± 4 (100) \\
Systolic BP, mmHg (\%) & 113 ± 14 (85) & 136 ± 15 (88) \\
Diastolic BP, mmHg (\%) & 73 ± 18 (85) & 83 ± 8 (88) \\
MAP (\%) & 86 ± 16 (85) & 100 ± 9 (88) \\
hsCRP, mg/l (\%) & 1 ± 2 (60) & 3 ± 4 (44) \\
Serum urea, mmol/l (\%) & 5 ± 1 (95) & 18 ± 7 (100) \\
Creatinine, $\mu$mol/l (\%) & 66 ± 8 (95) & 678 ± 187 (100) \\
eGFR, ml/min/1.73m$^2$ (\%) & 91 ± 8 (95) & 8 ± 2 (100) \\
Urine P:Cr, mg/mmol (\%) & 1 ± 5 (38) & 403 ± 394 (35) \\
\bottomrule
\end{tabular}}

            \caption[Baseline demographics and clinical variables for the renal cohort.]{Baseline demographics collected for donors and recipients eligible. Values are shown as mean $\pm$ \acrshort{SD}, and those in parentheses represent proportion of data completeness relative to baseline populations. }
            \label{tab:CKD_base_pop}
            \end{table}

            \begin{table}[tbp]\footnotesize
            \begin{adjustwidth}{-5in}{-5in}  
            \centering
            \scalebox{0.7}{\begin{tabular}{p{2.5cm}p{1.5cm}p{2.5cm}p{2.25cm}p{2.25cm}p{2.25cm}p{2.25cm}p{2.25cm}p{2.25cm}p{2.25cm}}
\toprule
\multirow{2}{2.5cm}{} & \multirow{2}{*}{Cohort} & \multicolumn{8}{c}{Timepoint} \\
\cmidrule(l){3-10}
 &  & Baseline & 1 Week & 2 Weeks & 4 Weeks & 8 Weeks & 18 Weeks & 28 Weeks & 52 Weeks \\
 \midrule
\multirow{2}{2.5cm}{Sample, n (\%)} & Donors & 20 (100) & 10 (50) & 2 (10) & 8 (40) & 10 (50) & 8 (40) & 10 (50) & 11 (55) \\
 & Recipients & 16 (100) & 12 (75) & 4 (25) & 5 (31) & 9 (56) & 5 (31) & 7 (44) & 7 (44) \\
\multirow{2}{2.5cm}{Daytime (Hr:Min)} & Donors & 13:50 ± 1:41 & 13:48 ± 1:20 & 12:04 ± 1:34 & 12:48 ± 3:47 & 13:43 ± 2:12 & 13:54 ± 2:00 & 13:30 ± 2:03 & 13:48 ± 2:07 \\
 & Recipients & 13:59 ± 2:19 & 12:44 ± 1:42 & 11:45 ± 1:50 & 12:23 ± 2:01 & 11:31 ± 1:13 & 12:29 ± 1:55 & 11:47 ± 1:39 & 12:25 ± 1:31 \\
\multirow{2}{2.5cm}{Time from baseline (wks)} & Donors & 0 ± 0 & 1 ± 0 & 2 ± 0 & 4 ± 1 & 8 ± 2 & 16 ± 4 & 29 ± 3 & 53 ± 5 \\
 & Recipients & 0 ± 0 & 1 ± 0 & 2 ± 0 & 4 ± 1 & 8 ± 3 & 16 ± 4 & 28 ± 3 & 52 ± 4 \\
\multirow{2}{2.5cm}{hsCRP, mg/l (\%)} & Donors & 1 ± 2 (60) & 58 ± 28 (70) & 2 ± 1 (100) & 23 ± 34 (63) & 3 ± 3 (70) & 5 ± 9 (75) & 1 ± 1 (80) & 2 ± 4 (55) \\
 & Recipients & 3 ± 4 (44) & 17 ± 20 (42) & 3 ± 3 (50) & 7 ± 10 (100) & 3 ± 2 (33) & 2 ± 2 (60) & 2 ± 0 (29) & 25 ± 52 (86) \\
\multirow{2}{2.5cm}{Serum urea, mmol/l   (\%)} & Donors & 5 ± 1 (95) & 5 ± 2 (90) & 6 ± 1 (100) & 6 ± 1 (88) & 6 ± 1 (100) & 7 ± 2 (75) & 6 ± 1 (90) & 7 ± 1 (64) \\
 & Recipients & 18 ± 7 (100) & 6 ± 2 (75) & 6 ± 1 (100) & 6 ± 1 (100) & 7 ± 2 (89) & 7 ± 2 (100) & 8 ± 2 (86) & 6 ± 1 (100) \\
\multirow{2}{2.5cm}{Creatinine, $\mu$mol/l (\%)} & Donors & 66 ± 8 (95) & 102 ± 17 (90) & 102 ± 3 (100) & 89 ± 7 (88) & 102 ± 24 (100) & 101 ± 25 (75) & 100 ± 22 (90) & 91 ± 9 (64) \\
 & Recipients & 678 ± 187 (100) & 114 ± 36 (75) & 95 ± 26 (100) & 104 ± 25 (100) & 123 ± 20 (89) & 119 ± 27 (100) & 124 ± 37 (86) & 109 ± 27 (100) \\
\multirow{2}{3cm}{eGFR, ml/min/1.73m$^2$ (\%)} & Donors & 91 ± 8 (95) & 63 ± 7 (90) & 64 ± 9 (100) & 71 ± 10 (88) & 68 ± 18 (100) & 67 ± 14 (75) & 66 ± 12 (90) & 67 ± 9 (64) \\
 & Recipients & 8 ± 2 (100) & 62 ± 15 (65) & 77 ± 18 (100) & 71 ± 14 (100) & 61 ± 13 (89) & 63 ± 16 (100) & 58 ± 18 (86) & 69 ± 13 (100) \\
\multirow{2}{2.5cm}{Urine P:Cr, mg/mmol (\%)} & Donors & 1 ± 5 (38) & 165 ± 329 (50) & 1 ± 0 (50) & 3 ± 3 (25) & 1 ± 2 (60) & 1 ± 2 (50) & 1 ± 1 (60) & 1 ± 1 (27) \\
 & Recipients & 403 ± 394 (35) & 257 ± 28 (25) & 12 ± 0 (25) & 89 ± 57 (60) & 31 ± 19 (44) & 21 ± 8 (100) & 29 ± 20 (57) & 9 ± 13 (57)
 \\
 \bottomrule
\end{tabular}}

            \end{adjustwidth}
            \caption[Longitudinal demographics and clinical variables for the renal cohort.]{Population statistics for donors and recipients consistently collected over period of analysis. Where appropriate, values are shown as mean $\pm$ \acrshort{SD}, and values in parentheses represent proportion of data completeness relative to baseline populations.}
            \label{tab:CKD_long_pop}
            \end{table}

            Figure \ref{fig:CKD_flowchart} shows a flowchart on how the population was selected for this study. At the time of data extraction, 22 donors and 23 recipients were eligible for analysis, after excluding a total of 65 recipients and 5 donors. The majority of exclusions were because there were no eye scans performed for these individuals. This was because, once recruited, they were scheduled for transplantation and then immediately returned to their referring institution after transplant away from Edinburgh, thus making it impractical to retain in the study and re-scan. Each participant was assessed on the day of transplant and then attempts were made at 7 other time periods over the following year (8 time points in total). 20 donors and 16 recipients for eligible for association estimation with clinical variables related to renal function, after excluding 2 donors and 10 recipients due to missing clinical measurements/eye scans post-transplant, or poor quality \acrshort{EDI-OCT} images.

            Tables \ref{tab:CKD_base_pop} and \ref{tab:CKD_long_pop} present the baseline demographics and clinical variables collected longitudinally for both cohorts, respectively. There were a number of missing values for certain clinical variables in both cohorts, which rendered them too incomplete to be used for downstream modelling (BMI, BP, hsCRP, Urine P:Cr). Thus, based on the level of complete data (relative to baseline sample size) in table \ref{tab:CKD_long_pop}, only clinical variables eGFR, serum creatinine and serum urea were selected for downstream statistical analysis. 

        \end{mysubsubsection}

        \begin{mysubsubsection}[]{Choroidal variation}

            \begin{table}[tb]\footnotesize
                \begin{adjustwidth}{-1in}{-1in}  
                \centering
                \begin{tabular}{lllllll}
\toprule
 & \multicolumn{3}{c}{Donors} & \multicolumn{3}{c}{Recipients} \\ \cmidrule(l){2-4}\cmidrule(l){5-7}
\multirow{1}{*}{Measurement} & \multicolumn{1}{c}{Baseline} & \multicolumn{1}{c}{1 year PT} & \multicolumn{1}{c}{$\textit{P}$-value} & \multicolumn{1}{c}{Baseline} & \multicolumn{1}{c}{1 year PT} & \multicolumn{1}{c}{$\textit{P}$-value} \\ 
\midrule
Choroid area [mm$^2$] & 1.59 $\pm$ 0.41 & 1.49 $\pm$ 0.41 & \textbf{0.01} & 1.58 $\pm$ 0.41 & 1.74 $\pm$ 0.46 & \textbf{0.01} \\
Choroid thickness [$\mu$m] & 278 $\pm$ 69 & 259 $\pm$ 68 & \textbf{0.01} & 271 $\pm$ 70 & 301 $\pm$ 75 & \textbf{0.01} \\
\midrule
eGFR (ml/min/1.73m$^2$) & 91 $\pm$ 5 & 64 $\pm$ 7 & \textbf{<0.001} & 8 $\pm$ 3 & 69 $\pm$ 14 & \textbf{<0.001} \\
Creatinine ($\mu$mol/l) & 69 $\pm$ 9 & 92 $\pm$ 10 & \textbf{<0.001} & 659 $\pm$ 196 & 109 $\pm$ 29 & \textbf{<0.001} \\
Urea (mmol/l) & 5.0 $\pm$ 0.8 & 6.8 $\pm$ 1.3 & \textbf{0.002} & 18.5 $\pm$ 9.7 & 6.7 $\pm$ 1.7 &\textbf{0.005} \\
\bottomrule
\end{tabular}

                \end{adjustwidth}
                \caption[Choroidal variation pre-transplant and one year post-transplant in renal cohort.]{Choroidal measurements at baseline and one year post-transplant.}
                \label{tab:longitudinal_ttest}
            \end{table}

            \begin{figure}[tb]
                \centering
                \includegraphics[width=0.9\textwidth]{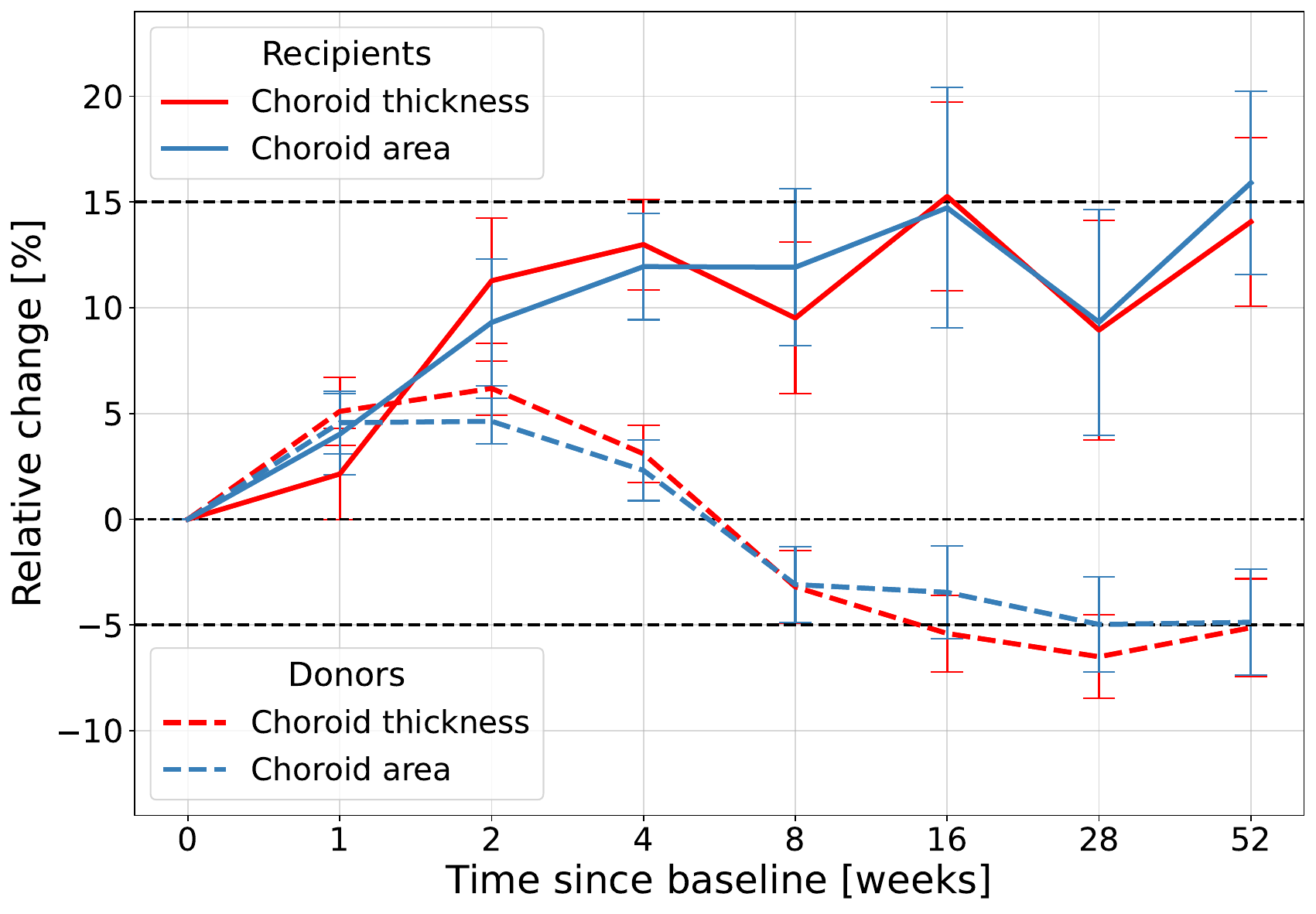}
                \caption[Longitudinal choroid variation in the renal cohort.]{Longitudinal change in choroid thickness and area over time, measured relative to baseline choroid measurements. Error bars refer to standard error measurements. Horizontal dashed lines at +15\% and -5\% are superimposed to aid readability.}
                \label{fig:CKD_longitudinal}
            \end{figure}

            \begin{table}[tb]\footnotesize
                \begin{adjustwidth}{-.5in}{-.5in}  
                \centering
                \begin{tabular}{lllllll}
\toprule
\multirow{2}{*}{} & \multicolumn{3}{c}{Donors} & \multicolumn{3}{c}{Recipients} \\ \cmidrule(l){2-4}\cmidrule(l){5-7} 
 & eGFR & Creatinine & Urea & eGFR & Creatinine & Urea \\ 
 \midrule
Choroid area & -0.01 & 0.04 & -0.14 & 0.68*** & -0.68*** & -0.42* \\
Choroid thickness & -0.09 & 0.13 & -0.27 & 0.51** & -0.51** & -0.26* \\
\bottomrule
\end{tabular}

                \end{adjustwidth}
                \caption[Association between choroidal measures and clinical variables in the renal cohort.]{Unadjusted, repeated Pearson correlation coefficient, $r_{rm}$ \cite{bakdash2017repeated} between choroidal thickness, area and markers of renal function eGFR, serum creatinine and serum urea. ***: P < 0.0001, **: P < 0.001, *: P < 0.05}
                \label{tab:CKD_clinical_correlation}
            \end{table}

            \begin{table}[tb]\footnotesize
                \centering
                \scalebox{0.7}{\begin{tabular}{llllllllll}
\toprule
\multicolumn{1}{c}{\multirow{2}{*}{\textbf{Transplant recipients}}} & \multicolumn{3}{c}{eGFR} & \multicolumn{3}{c}{Creatinine} & \multicolumn{3}{c}{Urea} \\
 \cmidrule(l){2-4}\cmidrule(l){5-7}\cmidrule(l){8-10}
 & \multicolumn{1}{c}{$\hat{\beta}$} & \multicolumn{1}{c}{95\% CI} & \multicolumn{1}{c}{$P$-value} & \multicolumn{1}{c}{$\hat{\beta}$} & \multicolumn{1}{c}{95\% CI} & \multicolumn{1}{c}{$P$-value} & \multicolumn{1}{c}{$\hat{\beta}$} & \multicolumn{1}{c}{95\% CI} & \multicolumn{1}{c}{$P$-value} \\
 \midrule
\underline{Choroid area} &  &  &  &  &  &  &  &  &  \\
Intercept & -0.10 & (-1.38, 1.17) & 0.84 & 0.09 & (-1.18, 1.36) & 0.86 & 0.03 & (-0.49, 0.55) & 0.90 \\
Age & 0.98 & (-0.47, 2.43) & 0.15 & -1.11 & (-2.54, 0.31) & 0.10 & -0.30 & (-0.95, 0.34) & 0.29 \\
Sex (Male) & -0.24 & (-1.55, 1.07) & 0.66 & 0.27 & (-1.03, 1.57) & 0.60 & 0.22 & (-0.31, 0.76) & 0.34 \\
\textbf{Daytime from Baseline} & -0.40 & (-0.59, -0.21) & \textbf{<0.001} & 0.59 & (0.39, 0.80) & \textbf{<0.001} & 0.57 & (0.32, 0.81) & \textbf{<0.001} \\
Approx. Refraction & -0.86 & (-2.17, 0.46) & 0.114 & 0.92 & (-0.37, 2.21) & 0.12 & 0.31 & (-0.28, 0.89) & 0.24 \\
\textbf{Choroid area} & 1.54 & (0.97, 2.10) & \textbf{<0.001} & -1.52 & (-2.12, -0.92) & \textbf{<0.001} & -0.60 & (-1.09, -0.11) & \textbf{0.02} \\
Patient Std. Dev. & 1.63 & \multicolumn{2}{l}{Cond. $R^2$: 0.93} & 0.60 & \multicolumn{2}{l}{Cond. $R^2$: 0.91} & 0.55 & \multicolumn{2}{l}{Cond. $R^2$: 0.57} \\
\midrule
\underline{Choroid thickness} & \multicolumn{3}{c}{} & \multicolumn{3}{c}{} & \multicolumn{3}{c}{} \\
Intercept & -0.10 & (-1.19, 0.99) & 0.83 & 0.07 & (-0.91, 1.05) & 0.84 & 0.03 & (-0.48, 0.53) & 0.90 \\
Age & 0.74 & (-0.51, 1.99) & 0.20 & -0.80 & (-1.92, 0.31) & 0.12 & -0.27 & (-0.90, 0.36) & 0.33 \\
Sex (Male) & -0.31 & (-1.43, 0.81) & 0.49 & 0.31 & (-0.69, 1.30) & 0.44 & 0.28 & (-0.25, 0.81) & 0.24 \\
\textbf{Daytime from Baseline} & -0.46 & (-0.67, -0.24) & \textbf{<0.001} & 0.65 & (0.42, 0.88) & \textbf{<0.001} & 0.55 & (0.30, 0.80) & \textbf{<0.001} \\
Approx. Refraction & -0.43 & (-1.55, 0.68) & 0.37 & 0.46 & (-0.54, 1.46) & 0.27 & 0.18 & (-0.35, 0.71) & 0.42 \\
\textbf{Choroid thickness} & 1.15 & (0.51, 1.78) & \textbf{<0.001} & -1.01 & (-1.62, -0.40) & \textbf{0.004} & -0.54 & (-1.06, -0.02) & \textbf{0.04} \\
Patient Std. Dev. & 1.37 & \multicolumn{2}{l}{Cond. $R^2$: 0.88} & 1.05 & \multicolumn{2}{l}{Cond. $R^2$: 0.81} & 0.50 & \multicolumn{2}{l}{Cond. $R^2$: 0.52}\\
\midrule
\multicolumn{1}{c}{\multirow{2}{*}{\textbf{Transplant donors}}} &  &  &  &  &  &  &  &  &  \\
 &  &  &  &  &  &  &  &  \\

 \midrule
\underline{Choroid area} &  &  &  &  &  &  &  &  &  \\
Intercept & 0.04 & (-0.28, 0.36) & 0.78 & -0.03 & (-0.34, 0.28) & 0.83 & -0.05 & (-0.36, 0.26) & 0.72 \\
Age & -0.34 & (-0.76, 0.08) & 0.10 & 0.07 & (-0.34, 0.47) & 0.73 & 0.08 & (-0.33, 0.48) & 0.69 \\
\textbf{Sex (Male)} & 0.06 & (-0.29, 0.40) & 0.73 & 0.36 & (0.02, 0.69) & \textbf{0.04} & 0.11 & (-0.22, 0.45) & 0.48 \\
Time from Baseline & 3.5e-3 & (-0.24, 0.25) & 0.98 & -0.02 & (-0.25, 0.21) & 0.86 & -0.06 & (0-0.31, 0.18) & 0.60 \\
Approx. Refraction & -0.05 & (-0.47, 0.36) & 0.79 & 0.13 & (-0.27, 0.53) & 0.51 & 0.22 & (-0.18, 0.62) & 0.26 \\
Choroid area & 6.7e-4 & (-0.33, 0.33) & 0.99 & -0.11 & (-0.42, 0.21) & 0.49 & -0.07 & (-0.39, 0.24) & 0.64 \\
Patient Std. Dev. & 0.41 & \multicolumn{2}{l}{Cond. $R^2$: 0.27} & 0.45 & \multicolumn{2}{l}{Cond. $R^2$: 0.38} & 0.35 & \multicolumn{2}{l}{Cond. $R^2$: 0.21} \\
\midrule
\underline{Choroid thickness}& \multicolumn{3}{c}{} & \multicolumn{3}{c}{} & \multicolumn{3}{c}{} \\
Intercept & 0.04 & (-0.28, 0.36) & 0.78 & -0.03 & (-0.34, 0.28) & 0.84 & -0.05 & (-0.35, 0.25) & 0.73 \\
Age & -0.34 & (-0.76, 0.36) & 0.10 & 0.06 & (-0.35, 0.46) & 0.78 & 0.07 & (-0.33, 0.47) & 0.72 \\
\textbf{Sex (Male)} & 0.05 & (-0.30, 0.40) & 0.75 & 0.36 & (0.02, 0.70) & \textbf{0.04} & 0.12 & (-0.22, 0.45) & 0.47 \\
Time from Baseline & 3.9e-3 & (-0.24, 0.25) & 0.98 & -0.02 & (-0.25, 0.21) & 0.86 & -0.06 & (-0.31, 0.18) & 0.60 \\
Approx. Refraction & -0.05 & (-0.47, 0.36) & 0.80 & 0.13 & (-0.27, 0.54) & 0.49 & 0.22 & (-0.17, 0.62) & 0.24 \\
Choroid thickness & -0.01 & (-0.34, 0.31) & 0.95 & -0.08 & (-0.39, 0.23) & 0.60 & -0.07 & (-0.38, 0.25) & 0.66 \\
Patient Std. Dev. & 0.41 & \multicolumn{2}{l}{Cond. $R^2$: 0.27} & 0.45 & \multicolumn{2}{l}{Cond. $R^2$: 0.37} & 0.34 & \multicolumn{2}{l}{Cond. $R^2$: 0.21} \\
\bottomrule

\end{tabular}}

                \caption[Linear mixed-effects modelling between the choroid and clinical variables in the renal cohort.]{Linear mixed-effects modelling summary tables for predicting eGFR, serum creatinine and serum urea in transplant recipients (top) and donors (bottom) for choroid thickness and choroid area. Statistically significant covariates are shown in bold.}
                \label{tab:CKD_mixed_lm}
            \end{table}
            
            Table \ref{tab:longitudinal_ttest} shows the average choroidal measurements and markers of renal function at baseline and one year post-transplant. All measurements were significantly different one year post-transplant, relative to baseline measurements in both cohorts. Figure \ref{fig:CKD_longitudinal} shows the average percentage change in choroidal measurements (with standard errors annotated as vertical lines at each time point) from baseline measurements in both cohorts. In transplant recipients, there was a significant increase in choroidal thickness of 12.8 ± 4.8\% (median: 11.4\%) from baseline after 4 weeks post-transplant (P=0.02). After 1 year post-transplant, we observed a significant increase of 14.1 ± 11.4\% (median: 10.7\%) from baseline (P=0.01). Interestingly, we observed a significant increase of 5.1 ± 5.6\% (median: 5.4\%) 1 week post-transplant in the choroid of donors (P=0.01). The average choroid remained enlarged relative to baseline measurements and at some point between 4 and 8 weeks post-transplant the choroid deflated. After 1 year post-transplant, we observed a significant decrease of -4.9 ± 7.9\% (median: -3.5\%) in automatic choroid thickness (P=0.01).

            Table \ref{tab:CKD_clinical_correlation} presents the repeated Pearson correlation coefficients between choroidal measurements and clinical variables. In the recipient cohort, all choroidal measurements linearly correlated with eGFR, serum creatinine and serum urea with statistical significance (P < 0.05). However, there were no significant, linear associations found between choroidal measurements and renal function in the donor cohort. Choroid area appeared to have stronger linear correspondence with markers of renal function in recipients compared to choroid thickness ($r_{rm}$ between eGFR and choroid area, 0.68; choroid thickness, 0.51).

            Table \ref{tab:CKD_mixed_lm} shows the model summaries for predicting eGFR, serum creatinine and serum urea using each choroidal measurement in turn for transplant recipients (top) and donors (bottom). In transplant recipients linear mixed-effects modelling showed evidence of statistically significant positive linear association with eGFR, and negative associations with serum creatinine and urea levels. 
            
            Interestingly, the time difference (`daytime from baseline' in table \ref{tab:CKD_mixed_lm}) between follow-up and baseline image acquisition was a statistically significant predictor of all markers of renal functions for models for transplant recipients. This covariate had the opposite effect that choroid thickness and area had in each model, i.e. $\hat{\beta}$ for the model using choroid area in transplant recipients was +1.54, and for daytime from baseline this was -0.40. There were no statistically significant associations found between any choroidal measurement and marker of renal function in the donor cohort. However, sex was associated with serum creatinine, albeit with weak statistical significance.

        \end{mysubsubsection}

    \end{mysubsection}

    \begin{mysubsection}[]{Discussion}
        Correlation analyses and linear mixed-effects modelling reported significant, linear correspondence between choroidal image measures and markers of renal function over time. All choroidal measurements changed substantially over time for all study participants (figure \ref{fig:CKD_longitudinal}), as did eGFR, serum creatinine and serum urea (table \ref{tab:CKD_long_pop}). This is consistent with the choroid reflecting renal function during treatment of \acrshort{CKD}. The choroid does vary naturally during the day due to diurnal variation, and Tan et al. \cite{tan2012diurnal} described the average change in choroid thickness across daytime hours to be approximately 8.5 ± 5.2\%. The change reported here in choroid thickness in transplant recipients one year post-transplant was 14.1 ± 11.4\% (14.7 ± 11.2\% for choroid area), suggesting that choroidal inflation in transplant recipients cannot be explained fully through diurnal variation, but potentially through improved renal function \cite{choi2020strong}. For donors, saw an average decrease in choroid thickness of -4.9 ± 7.9\% (-5.3 ± 7.7\% for area) one year post-transplant (figure \ref{fig:CKD_longitudinal}). While we cannot rule out diurnal variation contributing to the reduced choroidal space, from a sub-clinical perspective the reduced choroidal measurements may reflect relative reduction in renal function, as observed in the eGFR levels in table \ref{tab:CKD_long_pop}.

        In our linear mixed-effects models for transplant recipients we accounted for covariates to quantify the impact of myopia and diurnal variation (`approximate refraction' and `daytime from baseline' in table \ref{tab:CKD_mixed_lm}, respectively). Interestingly, relative daytime proved to be a significant predictor variable for markers of renal function alongside the choroid in transplant recipients, while approximate refractive error did not. In every transplant recipient linear mixed-effects model, there was an opposite effect of relative daytime compared with the choroid --- model coefficients for these covariates had opposite signs. This suggests that the size of the choroid is negatively offset such that the choroid thickness and choroid area in a follow-up scan taken later in the day (relative to the baseline acquisition time) is decreased in absolute value --- and vice versa for follow-up scans taken earlier than the original baseline scan. This makes sense given we know that the choroid naturally decreases throughout the course of the day \cite{tan2012diurnal} from diurnal variation. Note however that the magnitude of effect (standardised model coefficient $\hat{\beta}$) for `daytime from baseline' is consistently lower in absolute value, relative to the model coefficient for choroid thickness and area in each transplant recipient linear mixed-effects model (table \ref{tab:CKD_mixed_lm}). Therefore, it remains important that daytime be recorded to account for within-patient choroidal fluctuation in downstream longitudinal analysis, or image acquisition should be performed around the same time of day so as not to diminish any longitudinal signal present in the data. 

        Depending on whether dialysis had been performed prior to transplant, and whether the recipient's new kidney is able to start making urine immediately, patients may need to undergo (haemo)dialysis post-transplant \cite{jain2019choice}, which can contribute to dynamic fluid shifts during and after transplantation \cite{canaud2019fluid}. However, we have not been able to account for any change in body fluid volume post-transplant and what impact that may have had on the choroid. Cho, et al. \cite{cho2019evaluation} found that the choroid thinned after haemodialysis, and that this was more pronounced in those patients with diabetes mellitus (exclusion criteria for transplant recipients in this analysis). However, eligibility criteria assumed that the renal transplant was functional, so temporary dialysis may only have lasted a few of days. Thus, while we cannot account for any potential choroidal fluctuation in response to post-operative dialysis in the short-term, it is unlikely these fluid shifts would have any long-term effects on body fluid volume and the choroid.
         
        eGFR is a high-level aggregate measurement of kidney function, derived from taking into account demographic factors such as age, sex and how much waste is in the blood via creatinine. Serum creatinine and urea are excreted by the kidney and are natural waste products. Thus, in the case for eGFR, higher values indicate good kidney function, and vice versa for serum creatinine and urea. Therefore, the significant correspondence observed between improved kidney function (increased eGFR, lowered creatinine and urea) and increased choroidal measures over time further indicates the potential for the choroid to reflect systemic microvascular injury in the context of kidney health. The progressively increasing choroidal area and thickness over time may indicate the progressive return to normal of it's perfusion, which is influenced by it's autonomic supply, a regulatory system which was overactive during end-stage \acrshort{CKD} \cite{balmforth2016chorioretinal}. 
        
        Our results are consistent with the literature. Aksoy, et al. \cite{aksoy2023choroidal} similarly found progressively increased choroidal thickness one week and one month post-transplant in recipients. Moreover, Farrah, et al. \cite{farrah2023choroidal} recently conducted a study with a larger cohort (of which this dataset is a subsample of) in \acrshort{OCT} chorioretinal image analysis in \acrshort{CKD} patients and similarly observed significant choroidal thinning in the presence of CKD, which was indeed reversible after kidney transplantation. Moreover, they also reported long-term, gradual choroidal thinning in healthy kidney donors, also consistent with this work. Thus, high-resolution and non-invasive \acrshort{OCT} scans of the choroidal microvasculature may be informative in the monitoring and prediction of renal injury \cite{farrah2023choroidal}, but a larger and more diverse cohort is needed to further statistically validate these initial findings. 

        In the donor cohort, we observed eGFR to be at approximately 66\% of baseline function, which increased very marginally to approximately 75\% one year post-transplant (table \ref{tab:CKD_long_pop}). This is mostly consistent with current literature on kidney function after unilateral nephrectomy \cite{krohn1966renal}, which report immediate post-nephrectomy kidney function to be 50\% of baseline, which marginally rises to 60\% as a result of compensatory renal hypertrophy. Moody, et al. \cite{moody2016cardiovascular} reported that healthy kidney donors thus had a greater risk of cardiovascular mortality, observing higher levels of microalbuminuria. While Mule, et al. \cite{mule2019association} reported choroidal thinning with lower levels of eGFR and microalbuminuria, the longitudinal change in an average donor's choroid from our sample did not reflect the immediate decline and subsequent plateau consistent with their eGFR levels, and neither through linear mixed-effects modelling, ultimately suggesting a more complex and very likely non-linear relationship between these variables, if any at all.

        It may be that the immediate inflation of the choroid in the first 4 weeks after kidney donation could be a result of a prompt systemic, cardiovascular response to the unilateral nephrectomy, as it is within the first month in a donor's recovery period that compensatory renal hypertrophy takes place \cite{anderson1991short, rojas2019compensatory}. The relative change after 8 weeks post-transplant showed gradual choroidal thinning, which was consistent with Mule at al. \cite{mule2019association}, and indeed with Choi and Kim \cite{choi2020strong}, who cautiously suggest that a decrease in choroid thickness is associated with a decline in renal function but its increase is associated with renal hypertrophy, albeit in patients with diabetic retinopathy.

        There were a number of limitations in this study. Firstly, our two cohorts were limited in sample size and not all participants or clinical variables were collected at every time point. Moreover, of the 20 donors and 16 recipients included for clinical evaluation, only 11 donors and 8 recipients were followed up 1 year post-transplant, which potentially introduced bias into our clinical evaluation, and contributed to a reduction of statistical power. Furthermore, using a sub-cohort for clinical evaluation could potentially introduce selection bias when estimating associations between the choroid and renal function. Therefore, further studies with a larger and more complete dataset would permit a higher confidence regarding the significance of our analyses. While Farrah, et al. \cite{farrah2020eye} has been able to do so at a larger-scale and confirm these initial findings, image analysis was conducted manually. Thus, reproducing these findings using automated systems which are now readily available would be an obvious next step.

        Additionally, from an image analysis perspective, we have only provided an analysis on regional quantities of the choroid and their links to \acrshort{CKD} treatment and kidney transplantation. With the introduction of improved automated systems for characterising the choroid, such as Choroidalyzer, further work would aim to quantify any relationships with the choroidal vasculature. This would enable measurement of choroid vessel area and \acrshort{CVI}, which would provide a more representative measurement of the choroid. However, Aksoy, et al. \cite{aksoy2023choroidal} did not observe \acrshort{CVI} to change significantly pre- and post-transplant, which may be due to improper use of Niblack's method, or simply that the increase in choroidal area was of a similar magnitude of increase as choroidal vessel area, thus stabilising reported \acrshort{CVI} values pre- and post-transplant. This necessitates the need to also report choroid vessel area alongside \acrshort{CVI}, as highlighted in chapters \ref{chp:chapter-mmcq} and \ref{chp:chapter-choroidalyzer}.

        \begin{mysubsubsection}[]{Outputs}

            In this section, there has been one publication output:
            \begin{itemize}\setlength\itemsep{0em}
                \item \underline{\textbf{Burke, Jamie}}, Dan Pugh, Tariq Farrah, Charlene Hamid, Emily Godden, Thomas J. MacGillivray, Neeraj Dhaun, J. Kenneth Baillie, Stuart King, and Ian J.C. MacCormick. ``\textit{Evaluation of an automated choroid segmentation algorithm in a longitudinal kidney donor and recipient cohort.}'' Translational Vision Science \& Technology 12, no. 11 (2023): 19-19.
            \end{itemize}
  
        \end{mysubsubsection}

        \begin{mysubsubsection}[]{Executive summary}

            In this section, we assessed 22 healthy kidney donors and 23 patients requiring renal transplantation up to 1 year post-transplant. We measured choroidal thickness and area and estimated associations between choroidal measurements and markers of renal function (estimated glomerular filtration rate (eGFR), serum creatinine and urea). We observed significant choroidal thickening in recipients 1-year post-transplant and the choroid corresponded to improved eGFR, serum creatinine and urea levels. In donors, we observed choroidal inflation in the first 4 weeks after the unilateral nephrectomy, followed by significant thinning 1-year post-transplant. 
            
            In recipients, the relationship between choroid and kidney reported here adds further evidence to the growing literature around the choroid reflecting microvascular injury in \acrshort{CKD}. In donors, initial inflation of the choroid could possibly be related to compensatory renal hypertrophy, while long-term thinning may be linked to long-term impairment of renal function. Future work should aim at a more complete analysis with a larger and more diverse sample in order to increase statistical power.
            
        \end{mysubsubsection}

        \vfill
        
    \end{mysubsection}

    \end{mysection}

    \begin{mysection}[sec:]{The PREVENT Dementia Study: Association between choroidal microvasculature and Alzheimer’s disease risk}\label{sec:ch_app_sec_prevent}

        \begin{mysubsection}[]{Introduction}
    
            Dementia is a multi-factorial neurological disorder characterised by progressive cognitive decline that significantly impacts daily functioning and quality of life. As the global population ages, the prevalence of dementia continues to rise, posing a significant public health challenge \cite{winblad2016defeating, livingston2020dementia}. While the pathophysiology of dementia is complex, accumulating evidence suggests that genetic and vascular factors play a crucial role in its aetiology and progression \cite{livingston2020dementia}. Genetic factors such as the apolipoprotein E4 (\acrshort{APOE4}) allele \cite{stocker2021prediction}, and family history of dementia \cite{scarabino2016influence, mak2021proximity}, have been identified as important determinants of dementia risk. 
            
            The retina is part of the central nervous system \cite{purves2001neuroscience}, which means the retinal microvasculature might reflect changes to similar vessels in the brain. Their similar embryonic origin and cellular composition has been leveraged previously to understand brain health through the lens of the ocular microvasculature \cite{wagner_retinal_2023, chiquita2019retina}. While the primary goal of \acrshort{OCT} has often been to image and analyse the cross-sectional retinal layers (a \textit{neuronal} measure and marker of neurodegeneration), advances in \acrshort{EDI-OCT} \cite{spaide2008enhanced} now permit exploration of possible associations between the choroid and brain diseases. In comparison to the retinal circulation, the choroidal circulation has significantly greater blood flow and perfusion pressure \cite{villa2021ocular, martin2014glia} and is innervated by the central autonomic network \cite{reiner2018neural}.
        
            However, there have been conflicting reports of choroidal changes in established Alzheimer’s disease in older populations \cite{gharbiya2014choroidal, bayhan2015evaluation, bulut2016choroidal, cunha2017choroidal, asanad2019retinal, robbins_choroidal_2021} raising questions about the underlying mechanisms and stages of the disease that might influence choroidal vasculature. One possibility is that genetic factors like \acrshort{APOE4} and family history of dementia (\acrshort{FH}) may play a role. To our knowledge, only one study \cite{ma2022longitudinal} directly assessed the relationship between choroidal measures and genetic risk (\acrshort{APOE4}) in cognitively normal, old-age individuals but found no significant evidence of differences between carriers and non-carriers. Choroidal changes at the very beginning of the Alzheimer’s disease trajectory remain unexplored. This could provide valuable insights into the prognostic value of the eye in reflecting the underlying processes linking vascular pathology to prospective cognitive decline. 
            
            The \textit{PREVENT Dementia study} \cite{ritchie_prevent_2012, ritchie_prevent_2024} investigates concurrent cerebral and retinal vascular changes in a cohort of mid-life individuals, half of whom are at increased risk of Alzheimer’s disease. While most research on ocular microvascular changes in Alzheimer’s disease focuses on older age groups \cite{gharbiya2014choroidal, bayhan2015evaluation, bulut2016choroidal, cunha2017choroidal, asanad2019retinal, robbins_choroidal_2021, ma2022longitudinal}, the PREVENT cohort presents a unique opportunity to examine the microvasculature in individuals who may be in the very early prodromal stage of disease. Accordingly, the study presented here aimed to explore the associations between the choroidal vasculature and dementia risk in cognitively healthy midlife adults at baseline.
            
        \end{mysubsection}

        \begin{mysubsection}[]{Data}
    
            \begin{mysubsubsection}[]{Study population and image capture}
        
                The protocol for the PREVENT Dementia programme has been outlined previously \cite{ritchie_prevent_2012, ritchie_prevent_2024}. The study aimed to recruit participants aged between 40 and 59 from five different research institutes, and to include a significant portion with a family history of dementia (approximately 50\%). The Edinburgh site recruited 224 participants, and 132 participants provided written informed consent to the retinal sub-study at baseline. The study adhered to the principles of the Declaration of Helsinki.
            
                Eligibility criteria for the Edinburgh arm was competent and consenting adults recruited from the wider PREVENT study who had the capacity to manoeuvre themselves to the retinal imaging machines unaided and follow instructions to facilitate patient fixation during imaging at baseline. Exclusion criteria were those with a clinical diagnosis of dementia, those without capacity to consent at baseline or individuals not able to fully understand written and verbal English. Exclusion criteria also included participants with current or previous ocular disease such as glaucoma, macular degeneration, diabetic retinopathy, uveitis, vitreous haemorrhage, ischaemic optic neuropathy, optic neuritis or other optic nerve diseases, or those who had undergone previous ocular surgery such as cataract surgery or retinal surgery.
                
                \acrshort{OCT} capture was performed with the \acrshort{SD-OCT} Heidelberg Engineering Spectralis \acrshort{OCT}1 Standard Module (Heidelberg Engineering, Heidelberg, Germany) for both eyes. Single, horizontal- and vertical-line scans were taken using active eye tracking with an \acrshort{ART} of 100 to reduce speckle noise and covered a 30$^\circ$ angle, resulting in approximately a 9 mm \acrshort{FOV}. Posterior pole volumetric scans were also collected, with 61 parallel line-scans covering the macula and spaced approximately 125 microns apart without \acrshort{EDI} mode activated, also using active eye tracking but with an \acrshort{ART} of 9. B-scans with an \acrshort{SNR} index <15 were excluded, according to the OSCAR-IB criteria \cite{tewarie2012oscar} for retinal \acrshort{OCT} quality assessment.
                
            \end{mysubsubsection}
        
            \begin{mysubsubsection}[]{Image analysis}
            
                \begin{figure}[!t]
                    \centering
                    \includegraphics[width=\textwidth]{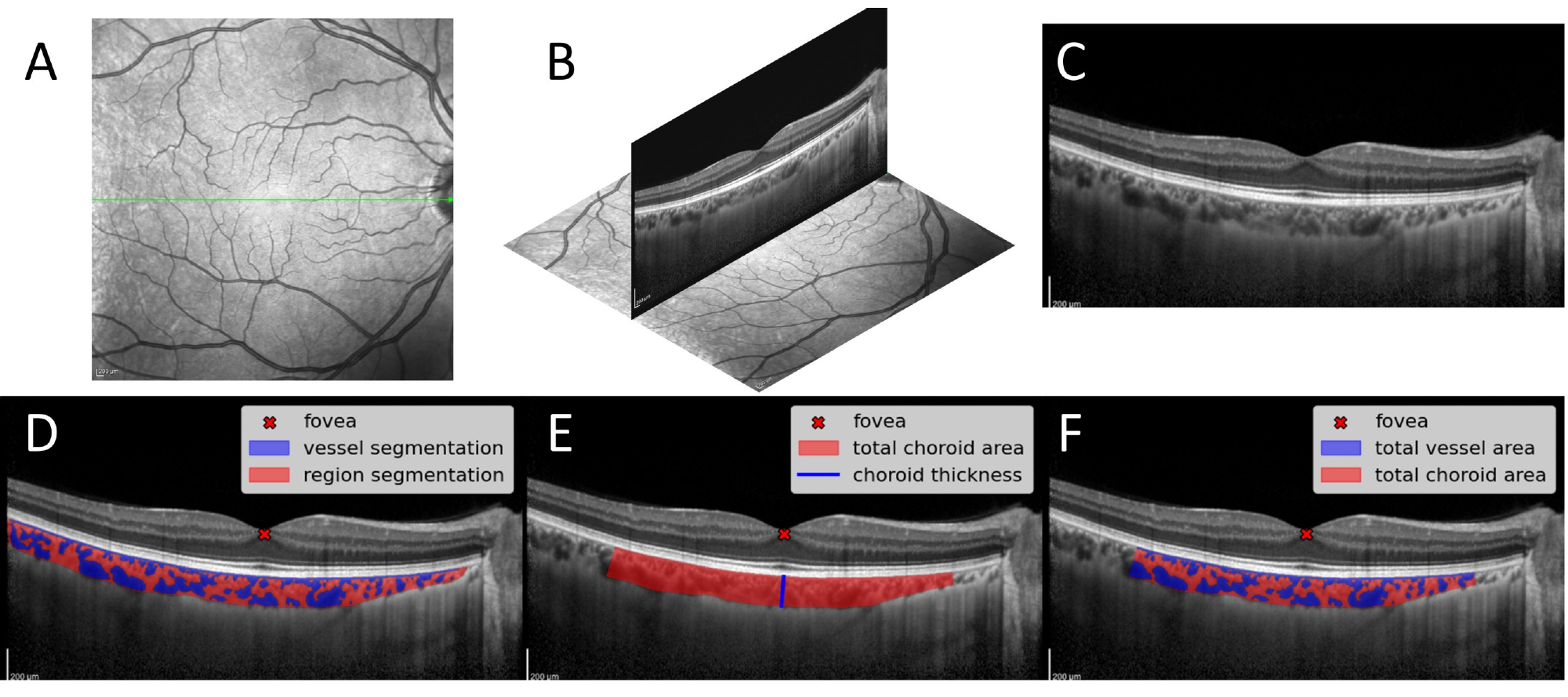}
                    \caption[Derivation of choroidal measurements in PREVENT cohort.]{Example derivation of choroidal measurements. (A) En face \acrshort{SLO} image of the retina with location of B-scan shown in green. (B) \acrshort{OCT} B-scan cross-section overlaid onto the en face \acrshort{SLO} image. (C) \acrshort{OCT} B-scan. (D) Automatic segmentation of region, vessels and foveal pit detection using Choroidalyzer. (E) Measurements of choroid thickness and choroid area and (F) vessel area and \acrshort{CVI}.}
                    \label{fig:PREVENT_oct_fig}
                \end{figure}
            
                Measurements of fovea-centred subfoveal choroid thickness, region area, vessel area and \acrshort{CVI} were computed using Choroidalyzer (chapter \ref{chp:chapter-choroidalyzer}), which detected the fovea, and segmented the choroid region and vessels in every \acrshort{OCT} B-scan. Figure \ref{fig:PREVENT_oct_fig} shows an \acrshort{OCT} B-scan with the corresponding localiser \acrshort{SLO} (figure \ref{fig:PREVENT_oct_fig}(A -- C)) and segmentations of the choroidal space, vessels and fovea using Choroidalyzer (figure \ref{fig:PREVENT_oct_fig}(D -- F)). Details on how these metrics were computed are described in section \ref{subsec:ch1_INTRO_measure_bsan}. Choroid measurements were computed for three distinct \acrshort{ROI}s defined according to the different \acrshort{ETDRS} circular grids (diameters 1 mm, 3mm and 6 mm)
                \cite{early1991grading}.
                
            \end{mysubsubsection}
        
            \begin{mysubsubsection}[]{Statistical analysis}
    
                Participants in the PREVENT cohort \cite{ritchie_prevent_2012} were prospectively stratified into three risk groups based on \acrshort{APOE4} presence and family history (\acrshort{FH}) by design: `high' risk was defined where \acrshort{APOE4} and \acrshort{FH} were both present, `medium' risk where either \acrshort{APOE4} or \acrshort{FH} were present and `low' risk where neither were present. 
                
                As this was an exploratory study with a relatively small sample size, our primary objective was to report descriptive statistics and investigate potential trends and associations. We first graphed choroidal measurements, testing for normality using the Shapiro-Wilks test, and performed univariate hypothesis testing on the mean differences between choroidal measures stratified for each risk for dementia. Alongside univariate statistical tests with individual risk markers, we assumed a biological ordering between risk groups and used ordinal logistic regression with each choroidal measure in turn as a single covariate in an unadjusted model, as well as in adjusted models controlling for age, sex, and mean arterial blood pressure \cite{polak2003choroidal, sansom2016association, waghamare2021comparison}. Unadjusted models were estimated so as to retain maximum statistical power in a relatively low sample size. 

                Because much of the posterior pole data was unusable due to the combined effect of acquisition without \acrshort{EDI} mode and low \acrshort{ART}, and to minimise the risk of spurious associations from analyses of multiple \acrshort{ROI}s per eye, we only selected choroidal measurements from horizontal-line scans with a 1.5mm fovea-centred \acrshort{ROI} with a preference for the left eye. The left eye tends to have a thinner choroid than the right \cite{lu2022interocular}, helping overcome issues with image quality from optical signal degradation. Any missing values were removed at the point of analysis. Analyses were performed using Python's statsmodels (version 0.14.0).
            
            \end{mysubsubsection}
        
        \end{mysubsection}

        \begin{mysubsection}[]{Results}

            \begin{mysubsubsection}[]{Participants}
    
                \begin{figure}[tb]
                    \centering
                    \includegraphics[width=0.75\textwidth]{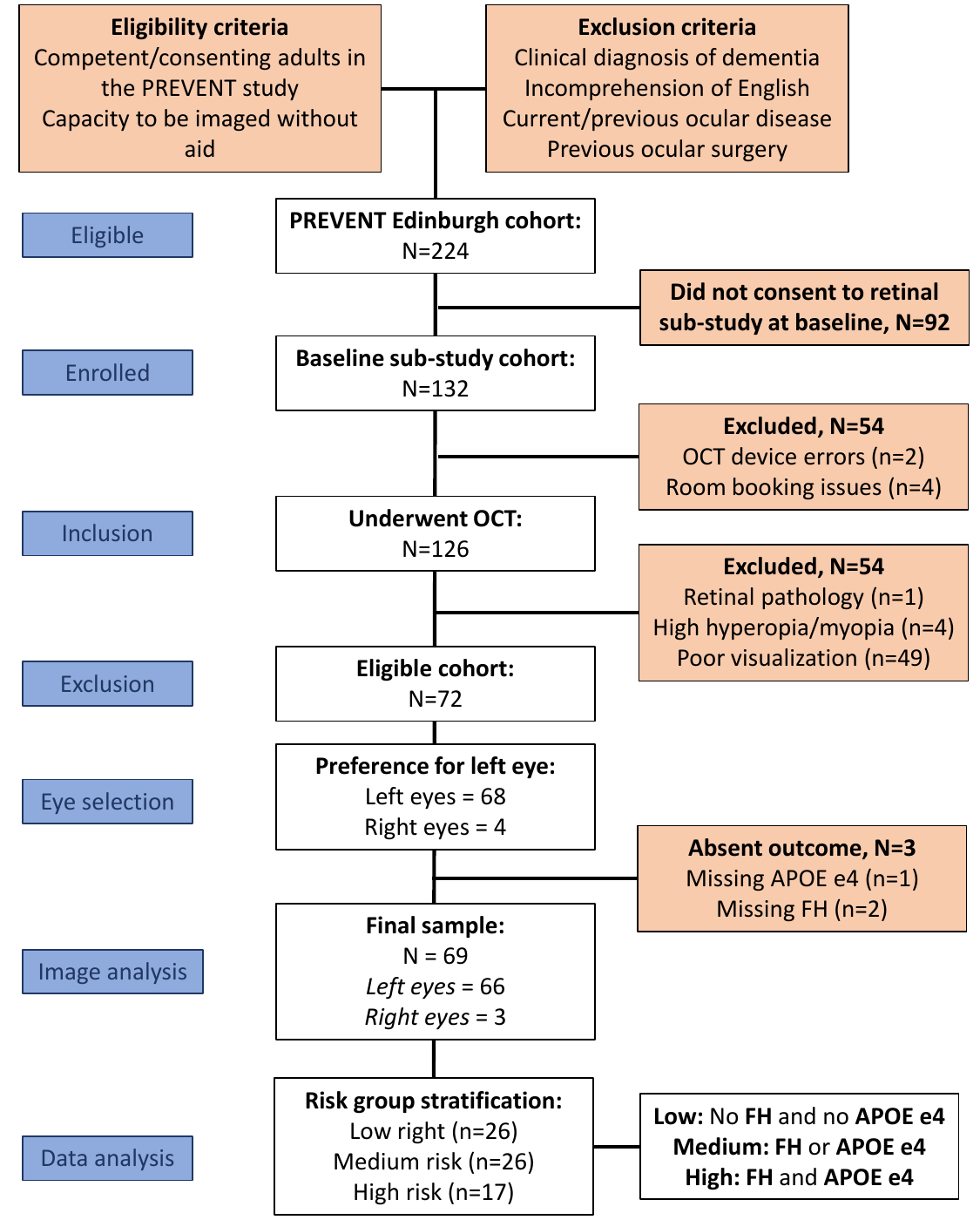}
                    \caption[Sample derivation flowchart for the PREVENT cohort.]{Sample derivation flowchart.}
                    \label{fig:PREVENT_flowchart}
                \end{figure}
            
                \begin{figure}[tb]
                    \centering
                    \includegraphics[width=\textwidth]{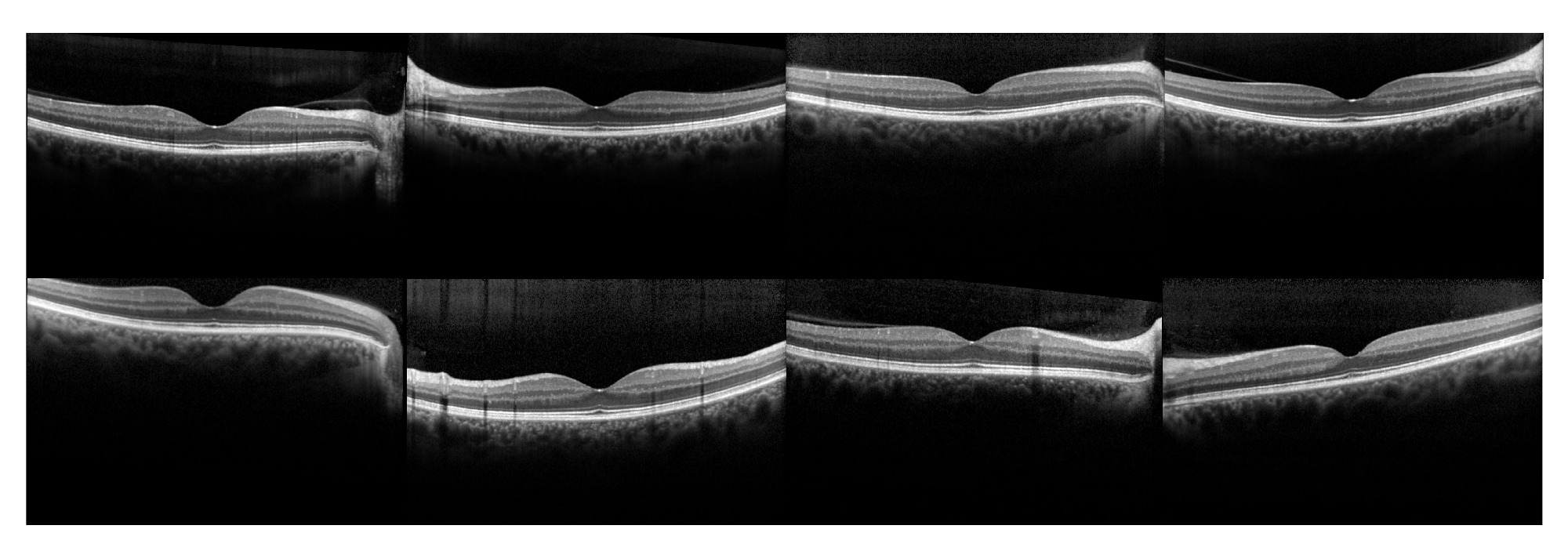}
                    \caption[Examples demonstrating poor choroid visualisation using conventional \acrshort{SD-OCT}.]{Example \acrshort{OCT} B-scans from the PREVENT retinal sub-study which were captured without \acrshort{EDI} mode activated.}
                    \label{fig:PREVENT_nEDI_quality}
                \end{figure}
            
                \begin{figure}[tb]
                    \centering
                    \includegraphics[width=\textwidth]{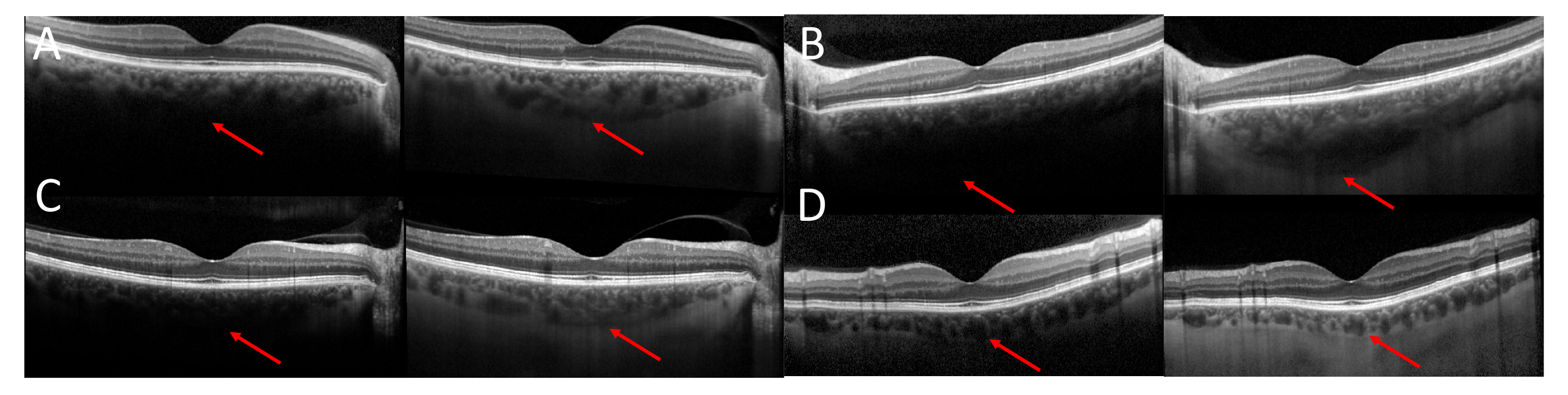}
                    \caption[Examples demonstrating the added benefit of \acrshort{EDI-OCT} over conventional \acrshort{SD-OCT}.]{Paired examples of \acrshort{OCT} B-scans where \acrshort{EDI} mode was toggled off and on, with red arrows indicating the lowest point of the Choroid-Sclera junction in each instance.}
                    \label{fig:PREVENT_nEDI_EDI}
                \end{figure}
    
                 \begin{table}[tb]\footnotesize
                    \centering \begin{tabular}{lllll}
\toprule
 & \multicolumn{1}{c}{\multirow{2}{*}{Overall}} & \multicolumn{3}{c}{Included in final sample} \\
 \cmidrule(l){3-5}
 & \multicolumn{1}{c}{} & No & Yes & P-Value \\
 \midrule
\textbf{Participants} & 224 & 155 & 69 &  \\
\textbf{Age} (years) & 51.347 (5.532) & 51.255 (5.467) & 51.551 (5.707) & 0.718* \\
\textbf{Sex} (female) & 128 (57.658) & 87 (56.863) & 41 (59.420) & 0.834\textsuperscript{\textdagger} \\
\textbf{Hypertension} & 23 (10.407) & 15 (9.804) & 8 (11.765) & 0.840\textsuperscript{\textdagger} \\
\textbf{Blood pressure} (mmHg) & 128.205 (14.733) & 127.121 (14.204) & 130.594 (15.678) & 0.119* \\
\textbf{BMI} (kg/m$^2$) & 28.892 (6.207) & 28.782 (6.415) & 29.143 (5.741) & 0.680* \\
\textbf{Smoking} &  &  &  & 0.704\textsuperscript{\textdagger} \\
\multicolumn{1}{r}{Current} & 10 (4.545) & 8 (5.298) & 2 (2.899) &  \\
\multicolumn{1}{r}{Ex} & 78 (35.455) & 54 (35.762) & 24 (34.783) &  \\
\multicolumn{1}{r}{None} & 132 (60.000) & 89 (58.940) & 43 (62.319) &  \\
\textbf{Diabetes} & 8 (3.620) & 5 (3.289) & 3 (4.348) & 0.707\textsuperscript{\textdagger} \\
\textbf{Risk factors} &  &  &  &  \\
\multicolumn{1}{r}{APOE4 status} & 89 (40.455) & 64 (42.384) & 25 (36.232) & 0.475\textsuperscript{\textdagger} \\
\multicolumn{1}{r}{Family History} & 104 (47.489) & 69 (46.000) & 35 (50.725) & 0.614\textsuperscript{\textdagger} \\
\bottomrule
\end{tabular}

                    \caption[Population demographics and clinical variables in the entire PREVENT cohort.]{Demographics and study variables of the entire PREVENT Dementia study cohort at the Edinburgh site, stratified by participant inclusion in this study's final sample. BMI (body mass index). Notes: All values are N (\%) or Mean (standard deviation). Missing data (number of eyes, percentage): Sex (2, 0.01\%), Age (2, 0.01\%), Hypertension (3, 0.01\%); Blood Pressure (3, 0.01\%), BMI (4, 0.02\%); Smoker (4, 0.02\%), Diabetes (3, 0.01\%), \acrshort{APOE4} (4, 0.02\%) and Family History (5, 0.02\%). *, Two-sample $t$-test; \textsuperscript{\textdagger}, Chi-squared test.}
                    \label{tab:PREVET_samp_vs_cohort}
                \end{table}

                \begin{table}[tb]\footnotesize
                    \centering    
                    \scalebox{0.95}{\begin{tabular}{llllll}
\toprule
 & \multicolumn{3}{c}{Risk group} &  & \multirow{2}{*}{Overall} \\
 \cmidrule(l){2-4}
 & Low & Medium & High &  \\
 \midrule
\textbf{Participants} & 26 & 26 & \multicolumn{2}{l}{17} & 69 \\
\textbf{Age} (years) & 50.615 (6.32) & 52.423 (4.54) & \multicolumn{2}{l}{51.647 (6.42)} & 51.551 (5.71) \\
\textbf{Sex} (female) & 14 (53.9\%) & 17 (65.4\%) & \multicolumn{2}{l}{10 (58.8\%)} & 41 (59.4\%) \\
\textbf{Hypertension} & 4 (15.4\%) & 3 (11.5\%) & \multicolumn{2}{l}{1 (6.3\%)} & 8 (11.8\%) \\
\textbf{Blood Pressure} (mmHg) & 138.603 (13.02) & 124.654 (16.37) & \multicolumn{2}{l}{127.431 (13.73)} & 130.594 (15.68) \\
\textbf{BMI} (kg/m$^2$) & 30.362 (6.87) & 28.103 (5.27) & \multicolumn{2}{l}{28.748 (4.23)} & 29.143 (5.74) \\
\textbf{Smoking} &  &  & \multicolumn{2}{l}{} &  \\
\multicolumn{1}{r}{Current} & 1 (3.9\%) & 1 (3.8\%) & \multicolumn{2}{l}{0 (0\%)} & 2 (2.9\%) \\
\multicolumn{1}{r}{Ex} & 8 (30.8\%) & 7 (26.9\%) & \multicolumn{2}{l}{9 (52.9\%)} & 24 (34.8\%) \\
\multicolumn{1}{r}{Non} & 17 (65.4\%) & 18 (69.2\%) & \multicolumn{2}{l}{8 (47.1\%)} & 43 (62.3\%) \\
\textbf{Diabetes} & 1 (3.9\%) & 2 (7.7\%) & \multicolumn{2}{l}{0 (0\%)} & 3 (4.3\%) \\
\textbf{Risk factors} &  &  & \multicolumn{2}{l}{} &  \\
\multicolumn{1}{r}{APOE4 status} & 0 (0\%) & 8 (30.8\%) & \multicolumn{2}{l}{17 (100.0\%)} & 25 (36.2\%) \\
\multicolumn{1}{r}{Family History} & 0 (0\%) & 18 (69.2\%) & \multicolumn{2}{l}{17 (100.0\%)} & 35 (50.7\%) \\
\textbf{Choroidal measures} &  &  & \multicolumn{2}{l}{} &  \\
\multicolumn{1}{r}{Area (mm$^2$)} & 0.68 (0.25) & 0.71 (0.194) & \multicolumn{2}{l}{0.87 (0.30)} & 0.74 (0.26) \\
\multicolumn{1}{r}{Thickness ($\mu$m)} & 250.39 (84.48) & 257.00 (78.37) & \multicolumn{2}{l}{303.35 (105.56)} & 265.93 (89.32) \\
\multicolumn{1}{r}{\acrshort{CVI}} & 0.477 (0.06) & 0.533 (0.07) & \multicolumn{2}{l}{0.528 (0.08)} & 0.511 (0.07) \\
\multicolumn{1}{r}{Vessel area (mm$^2$)} & 0.307 (0.14) & 0.364 (0.14) & \multicolumn{2}{l}{0.459 (0.18)} & 0.366 (0.16) \\
\bottomrule
\end{tabular}}

                    \caption[Population demographics and clinical variables in the PREVENT cohort eligible for analysis.]{Demographics and study variables, stratified by risk group. BMI (body mass index); \acrshort{CVI} (choroidal vascular index). All values are N (\%) or Mean (standard deviation). Missing data (number of eyes, percentage): Hypertension (1, 1.5\%); BMI (2, 2.9\%); \acrshort{CVI} (2, 2.9\%); vessel area (2, 2.9\%).}
                    \label{tab:PREVENT_pop}
                \end{table}
    
                Figure \ref{fig:PREVENT_flowchart} describes the derivation of our final sample (N=69). Of the 224 participants enrolled at the Edinburgh arm of the PREVENT dementia study, 132 consented to the Edinburgh site's retinal sub-study. Of 132 consented, 126 participants underwent \acrshort{OCT} capture (OCT device/computer error, N=3; room booking issues, N=4). Furthermore, as the primary aim of \acrshort{OCT} in the study protocol was to only image the retina (which does not require \acrshort{EDI} mode activated), \acrshort{EDI-OCT} capture was not collected for many of the earlier participants, which rendered the choroids of some eyes too poor quality for analysis (N=49).

                As described in section \ref{sec:INTRO_OCT_EDI}, \acrshort{EDI-OCT} is an important functionality in \acrshort{SD-OCT} image capture for improved visualisation of the choroid. Without it, particularly for older devices and hardware, can render the choroids of some eyes too poor quality for analysis. Figure \ref{fig:PREVENT_nEDI_quality} shows a selection of 8 eyes which were excluded from the analysis in this cohort due to imaging without \acrshort{EDI} mode, which ultimately led to only the retinal layers being visible. In figure \ref{fig:PREVENT_nEDI_EDI} we show paired examples where \acrshort{EDI} mode was toggled on and off. 
    
                Alongside the 49 participants excluded due to poor choroid visualisation, participants with retinal pathology (N=1) were excluded from our analysis to prevent abnormal retinae from confounding our results. While ocular pathology was part of the exclusion criteria, the image technician identified one participant with macular degeneration in both eyes during acquisition and after recruitment. Moreover, participants who were highly myopic/hyperopic ($\geq$ $\pm$ 6 dioptres, measured using the scan focus on the Heidelberg imaging device \cite{engineering_Spectralis_2022}) were also excluded (N=4), since extreme myopia/hyperopia is known to shrink/enlarge the choroidal space due to the elongation/compression that the eye undergoes \cite{liu2021influence}. 
            
                After selecting predominantly left eyes (68 left eyes, 4 right eyes where left was not available), and excluding participants with missing \acrshort{APOE4} and \acrshort{FH} (N=3), this left a final sample of 69 participants (66 left eyes, 3 right eyes), stratified into low (N=26), medium (N=26) and high (N=17) risk groups. 
            
                Table \ref{tab:PREVET_samp_vs_cohort} presents the demographics of the entire PREVENT Dementia Study cohort, stratified into whether a participant was included into our final sample or not. We ran standard statistical tests to compare the distributions between those participants included and excluded from the final sample and found no evidence of significant differences between baseline demographics and clinical variables. We did not run any statistical tests between the entire cohort and the final sample as this would violate the independence assumption underlying most statistical tests, due to overlapping samples. 
                 
                Demographics and study variables are summarised in table \ref{tab:PREVENT_pop}. The overall population had a mean age of 51.6 years (standard deviation, \acrshort{SD} = 5.71) and a slight female majority (N=41, 59.4\%). Mean arterial blood pressure was slightly higher in the low-risk group compared to the medium and high-risk groups which was statistically significant (low vs. medium, P=0.001; low vs. high, P=0.01). No other evidence of significant differences in demographic or cardiovascular variables were observed between risk groups. 
    
            \end{mysubsubsection}
    
            \begin{mysubsubsection}[]{Choroidal variation}
        
                \begin{figure}[tb]
                    \centering
                    \includegraphics[width=\textwidth]{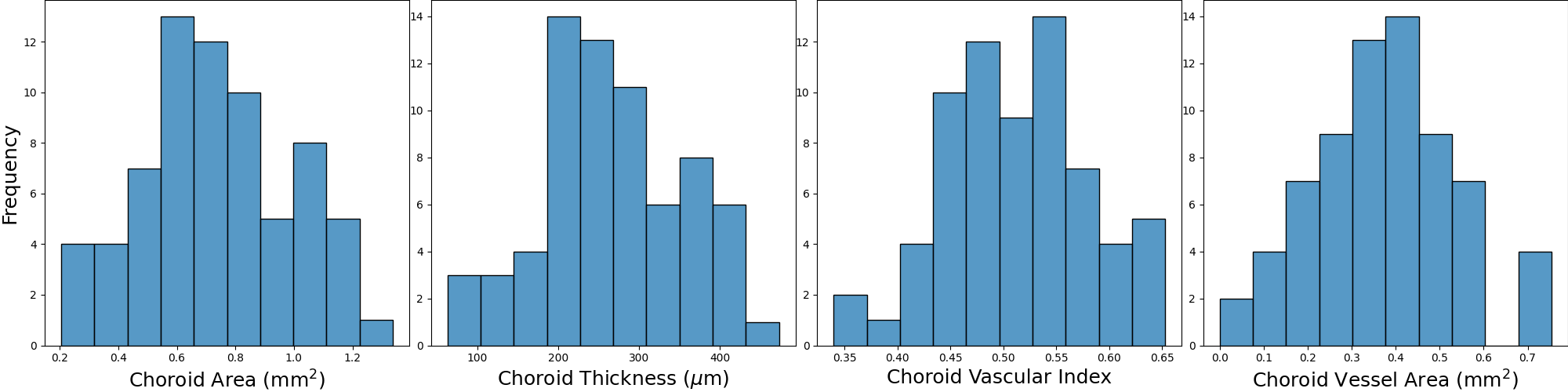}
                    \caption[Distribution histograms of choroidal measurements in the PREVENT cohort.]{Histogram distributions of each choroidal measure.}
                    \label{fig:PREVENT_Chor_Dists}
                \end{figure}
            
                \begin{figure}[tb]
                \centering
                \includegraphics[width=\textwidth]{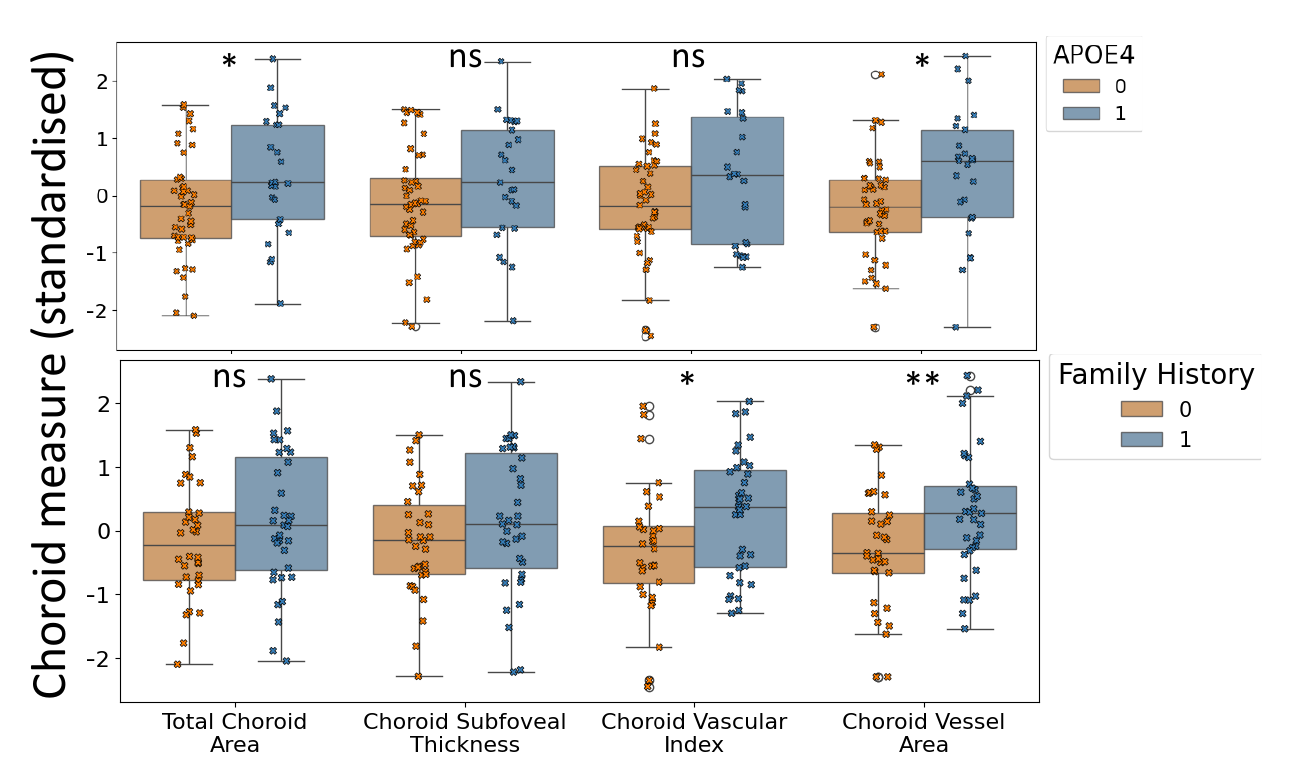}
                \caption[Distribution of choroidal measurements for \acrshort{APOE4}/\acrshort{FH} carriers and non-carriers in the PREVENT cohort.]{Box-plots showing the relationship between (standardised) choroidal measures and \acrshort{APOE4} (top) and \acrshort{FH} presence (bottom). **: P < 0.01; *: P < 0.05; ns: P > 0.05.}
                \label{fig:PREVENT_ind_risk}
                \end{figure}
            
                \begin{table}[tb]\footnotesize
                \centering    \begin{tabular}{llll}
\toprule
 & \multicolumn{2}{c}{APOE4 carrier status} &  \\
 \cmidrule(l){2-3}
 & Non-carrier & Carrier & P-Value \\
 \midrule
N & 47 & 25 &  \\
Area [mm$^2$] & 0.67 (0.23) & 0.83 (0.27) & \textbf{0.020} \\
Thickness [$\mu$μm] & 248.40 (85.36) & 288.64 (94.13) & 0.081 \\
\acrshort{CVI} & 0.50 (0.06) & 0.53 (0.08) & 0.104 \\
Vessel area [mm$^2$] & 0.33 (0.14) & 0.43 (0.18) & \textbf{0.022} \\
\midrule
 & \multicolumn{2}{c}{Family History of Dementia} &  \\
 \cmidrule(l){2-3}
 & None & Yes & P-Value \\
 \midrule
N & 34 & 36 &  \\
Area [mm$^2$] & 0.69 (0.23) & {0.79 (0.28)} & 0.122 \\
Thickness [$\mu$m] & 252.03 (78.16) & 283.50 (99.83) & 0.146 \\
\acrshort{CVI} & 0.49 (0.07) & 0.53 (0.07) & \textbf{0.022} \\
Vessel area [mm$^2$] & 0.32 (0.15) & 0.42 (0.16) & \textbf{0.009} \\
\bottomrule
\end{tabular}

                \caption[Distribution of choroidal measurements between ordinal Dementia risk groups in the PREVENT cohort.]{Distribution of choroidal measures by dementia risk with P-values for $t$-tests. Values are reported as mean (\acrshort{SD}).}
                \label{tab:PREVENT_ind_risk}
                \end{table}
            
                Figure \ref{fig:PREVENT_Chor_Dists} shows the histogram distributions for each choroidal measurement. Using the Shapiro-Wilks normality test, we found that each choroid measurement was approximately normal (test statistic (P-value) for choroid thickness, 0.984 (0.50); choroid area: 0.979 (0.31); \acrshort{CVI}: 0.984 (0.52); choroid vessel area: 0.987 (0.72)). 
                
                Figure \ref{fig:PREVENT_ind_risk} shows box-plots and distributions of each choroidal measurement stratified by the presence of each individual risk marker for Alzheimer's disease. We observed that choroids had larger vascular tissue and appeared larger in area for \acrshort{APOE4} carriers and \acrshort{FH} individuals, compared to non-carriers and those with no \acrshort{FH}. These differences were supported by independent $t$-tests for choroid vessel area (P=0.022, P=0.009). This was also observed for choroid area (P=0.020) and \acrshort{CVI} (P=0.022) for \acrshort{APOE4} carriers and \acrshort{FH} individuals, respectively (Table \ref{tab:PREVENT_ind_risk}).
            
                Figure \ref{fig:PREVENT_covariates} presents the choroidal measurements plotted against the covariates chosen for ordinal logistic modelling (age, sex and mean arterial blood pressure). Choroidal measurements appeared to decrease weakly with age (figure \ref{fig:PREVENT_covariates}, top) but the observed trend was much clearer for the low-risk group. Choroid vessel area and \acrshort{CVI} appeared lower in males (figure \ref{fig:PREVENT_covariates}, middle), but had very similar distributions for choroid area and thickness. For low-/medium-risk groups, we observed weak trends of smaller choroidal measures with increasing blood pressure (figure \ref{fig:PREVENT_covariates}, bottom). However, in the high-risk group, these trends appeared much more pronounced, and blood pressure was lower in higher-risk groups in general as seen from table \ref{tab:PREVENT_pop}. These observed trends in our sample justified our selected covariates when reporting adjusted ordinal logit models for regressing choroidal measurements on Dementia risk-group classification.
            
                \begin{figure}[tb]
                    \centering
                    \includegraphics[width=0.95\textwidth]{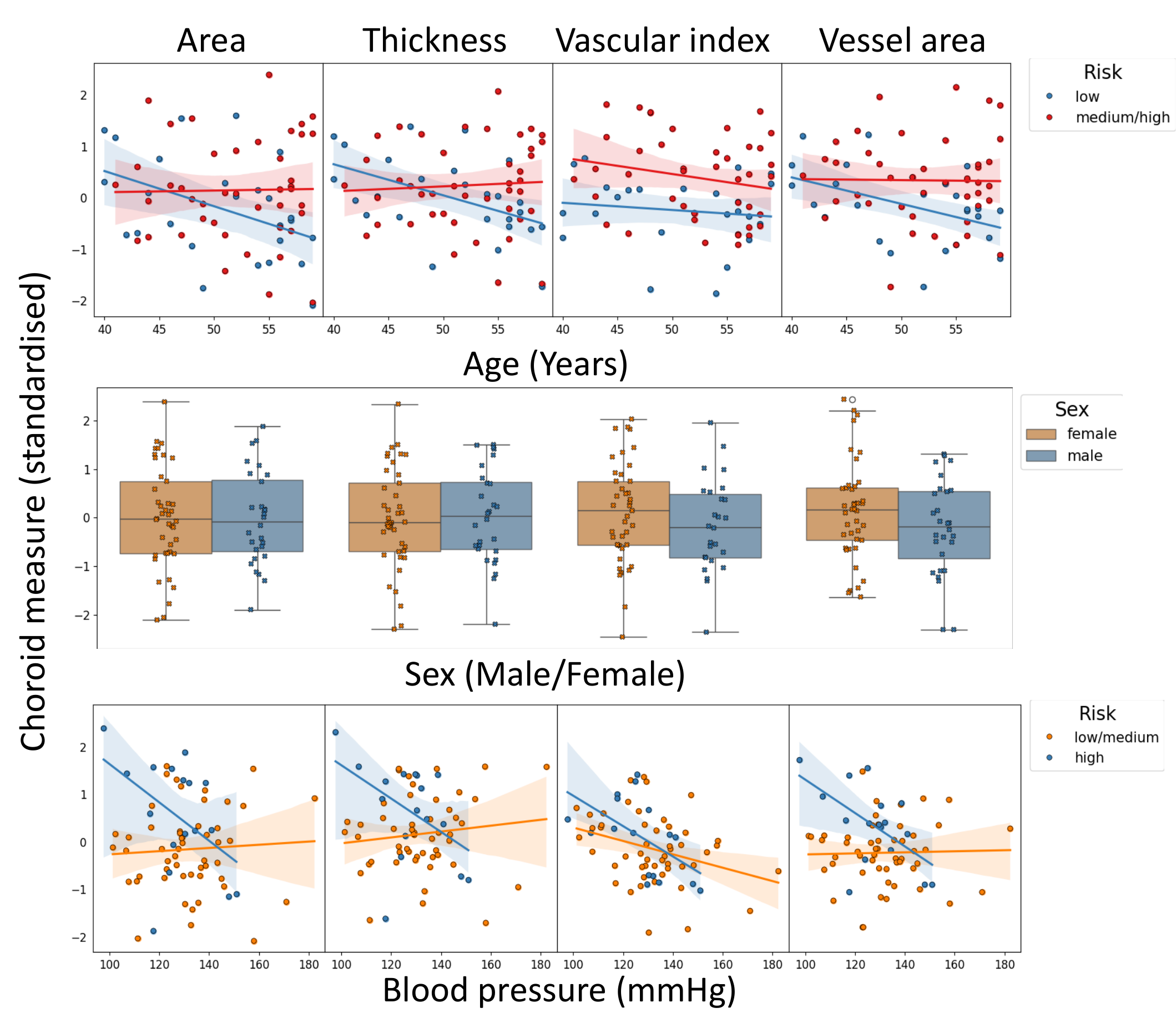}
                    \caption[Distribution of choroidal measurements against age, sex and blood pressure in the PREVENT cohort.]{Observed trends between choroidal measures and chosen covariates for ordinal logistic modelling, with each choroidal measure listed column-wise, and each covariate row-wise. Age (top), sex (middle) and mean arterial blood pressure (bottom).}
                    \label{fig:PREVENT_covariates}
                \end{figure}
            
                \begin{figure}[tb]
                    \centering
                    \includegraphics[width=\textwidth]{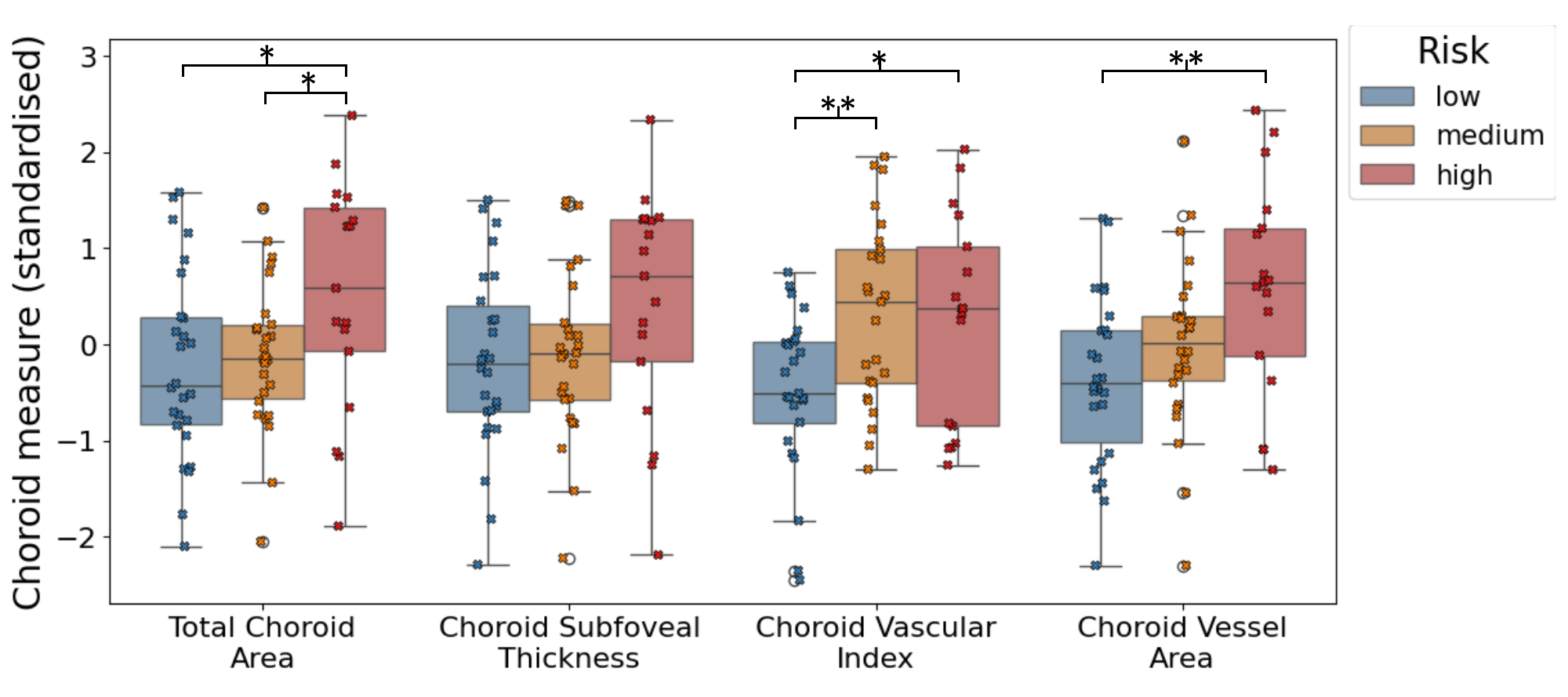}
                    \caption[Distribution of choroidal measurements between ordinal Dementia risk groups in the PREVENT cohort.]{Boxplots showing the relationship between choroidal measures and risk groups, with individual data points overlaid in a swarm-plot. **: P < 0.01; *: P < 0.05.}
                    \label{fig:PREVENT_comb_risk_chor}
                \end{figure}
            
                \begin{table}[tb]\footnotesize
                    \begin{adjustwidth}{-1in}{-1in}  
                    \centering    
                    \scalebox{0.8}{\begin{tabular}{lllllllll}
\toprule
 & \multicolumn{2}{c}{Choroid area} & \multicolumn{2}{c}{Choroid thickness} & \multicolumn{2}{c}{\acrshort{CVI}} & \multicolumn{2}{c}{Choroid vessel area} \\
 \cmidrule(l){2-3}\cmidrule(l){4-5}\cmidrule(l){6-7}\cmidrule(l){8-9}
 & \acrshort{OR}   (95\% \acrshort{CI}) & P-value & \acrshort{OR}   (95\% \acrshort{CI}) & P-value & \acrshort{OR}   (95\% \acrshort{CI}) & P-value & \acrshort{OR}   (95\% \acrshort{CI}) & P-value \\
 \midrule
\textbf{Covariates} &  &  & &  &  &  &  &  \\
\multicolumn{1}{r}{Age} & 1.33   (0.83, 2.15) & 0.2395 & 1.31   (0.81, 2.11) & 0.2664 & 1.34   (0.83, 2.18) & 0.2288 & 1.36   (0.84, 2.21) & 0.2160 \\
\multicolumn{1}{r}{Blood Pressure} & 0.5 (0.3,   0.85) & \textbf{0.0095} & 0.49   (0.29, 0.83) & \textbf{0.0076} & 0.57   (0.34, 0.98) & \textbf{0.0415} & 0.5   (0.29, 0.85) & \textbf{0.0100} \\
\multicolumn{1}{r}{Sex (Male)} & 1.13   (0.42, 3.01) & 0.8140 & 1.09   (0.41, 2.9) & 0.8687 & 1.35   (0.49, 3.7) & 0.5662 & 1.43   (0.52, 3.91) & 0.4907 \\
\multicolumn{1}{r}{Choroid measure} & 1.87   (1.12, 3.11) & \textbf{0.0162} & 1.69   (1.03, 2.79) & \textbf{0.0390} & 1.7   (1.02, 2.82) & \textbf{0.0408} & 2.25   (1.33, 3.8) & \textbf{0.0025} \\
\textbf{Intercept} &  &  &  &  &  &  &  &  \\
\multicolumn{1}{r}{Low/Medium} & 0.6   (0.31, 1.16) & 0.1317 & 0.61   (0.32, 1.17) & 0.1344 & 0.67   (0.35, 1.27) & 0.2194 & 0.65 (0.33,   1.26) & 0.2031 \\
\multicolumn{1}{r}{Medium/High} & 1.91   (1.35, 2.7) & \textbf{0.0002} & 1.88   (1.33, 2.65) & \textbf{0.0003} & 1.87   (1.31, 2.66) & \textbf{0.0005} & 2.0   (1.42, 2.83) & \textbf{0.0001}\\
\bottomrule
\end{tabular}}

                    \end{adjustwidth}
                    \caption[Ordinal logistic odds ratios from regressing choroidal measurements on Dementia risk in the PREVENT cohort.]{Odds ratios (with 95\% confidence intervals and corresponding P-values) for each ordinal logistic regression model, predicting risk group with each choroidal measure in turn. All models were adjusted for age (standardised), blood pressure (standardised) and sex. To provide stable confidence intervals in logistic models, all choroidal measures were standardised. P-values lower than 0.05 are in bold type.}
                    \label{tab:PREVENT_comb_risk_chor}
                \end{table}

                Figure \ref{fig:PREVENT_comb_risk_chor} presents grouped box-plots of choroidal measurements and prospectively defined Alzheimer’s disease risk groups, showing a biological gradient between increasing risk and increasing choroidal measurements, particularly for choroid vessel area.
                
                After adjusting for age, sex, and blood pressure, all choroidal measurements were significantly associated with increased combined risk for Alzheimer’s disease. Table \ref{tab:PREVENT_comb_risk_chor} presents the odds ratios (\acrshort{OR}) from the ordinal logistic models. Specifically, the difference between having no markers and at least one or both was significantly associated with increased choroidal area (\acrshort{OR} per \acrshort{SD} increase = 1.87, 95\% \acrshort{CI}:[1.12 – 3.11], P = 0.016), choroidal thickness (\acrshort{OR} per \acrshort{SD} increase = 1.69, 95\% CI:[1.03 – 2.79], P = 0.039), \acrshort{CVI} (\acrshort{OR} per \acrshort{SD} increase = 1.70, 95\% \acrshort{CI}: [1.02 – 2.82], P = 0.041), and vessel area (\acrshort{OR} per \acrshort{SD} increase = 2.25, 95\% \acrshort{CI}:[1.33 – 3.80], P = 0.003). The magnitude of effect for choroidal vessel area was the largest of all measures. Put another way, for every unit standard deviation increase in choroidal vessel area (0.16 mm$^2$), on average it is 2.25 times more likely that this individual has at least one marker of risk for Alzheimer’s disease (\acrshort{APOE4}, \acrshort{FH}) rather than none, after holding all other covariates fixed. 
                    
            \end{mysubsubsection}
            
        \end{mysubsection}
    
        \begin{mysubsection}[]{Discussion}
        
            We explored the association between the choroidal vasculature and risk factors for Alzheimer’s disease and observed a significant, positive association between each choroidal measure and degree of combined risk (\acrshort{APOE4}, \acrshort{FH}). To our best knowledge, this is the first exploration of the choroidal vasculature in cognitively healthy midlife adults at risk of Alzheimer’s disease. Previous, related studies have focused only on retinal difference between groups \cite{lopez2023exploratory}.

            In this exploratory study of asymptomatic people at risk of Alzheimer’s disease, we observed trends which, independent of covariates, indicate an increased retinal choroid vasculature in participants who carry the \acrshort{APOE4} genotype or have a family history of dementia, relative to those who do not. This reached statistical significance in choroid vessel area and choroid area. Subfoveal choroid thickness is a one-dimensional point-source measurement, and is less robust in measuring the choroid relative to two-dimensional measurements such as area and \acrshort{CVI}.
            
            The relationship between \acrshort{CVI} and vessel area/choroid area likely played a role in any significant, distributional differences observed between groups. Both choroidal area and vessel area were significantly higher in \acrshort{APOE4} carriers relative to non-carriers. Area and vessel area are directly and inversely proportional to \acrshort{CVI}, respectively, and so these two measures directly contribute toward estimated associations with \acrshort{CVI}. Thus, the increase in \acrshort{CVI} (mean increase of 0.03) did not reach significance likely because of the greater mean increase in region area (mean increase of 0.16 mm$^2$) relative to vessel area (mean increase of 0.10 mm$^2$). For individuals with a family history of dementia, we observed a smaller magnitude of increase in total choroid area relative to \acrshort{APOE4} carriers (mean increase of 0.10 mm$^2$). Put in the context of the significant increase observed in choroidal vessel area (mean increase of 0.10 mm$^2$), this very likely contributed to the significant increase observed in \acrshort{CVI} (mean increase of 0.04). 
            
            While these observations highlight the interplay between these representative choroidal measurements, they ultimately suggest that choroidal vessel area had the strongest signal among measurements, and that the vasculature itself might be the driving factor for increased choroidal area and thickness, rather than surrounding extravascular tissue (for our sample). Yet again, like in previous chapters, we highlight the importance of reporting choroidal vascular measures like area rather than only \acrshort{CVI}.

            Encouragingly, measurement error and diurnal variation are unlikely to have had a major effect on our results. The differences in mean choroid measurements between \acrshort{APOE4} carriers and non-carriers (thickness, 53 microns; area, 0.19 mm$^2$; \acrshort{CVI}, 0.05; and vessel area, 0.15 mm$^2$), were greater than expected from approximate choroidal fluctuation due to diurnal variation \cite{chakraborty_diurnal_2011, usui_circadian_2012, tan_diurnal_2012, kinoshita_diurnal_2017, singh2019diurnal, ostrin_imi-dynamic_2023}, (thickness, 30 microns; area, 0.035 mm$^2$; vascular index, 0.015; vessel area, 0.02 mm$^2$). Moreover, our results are greater than the threshold expected to exceed Choroidalyzer's estimated measurement error for single-line \acrshort{OCT} B-scans (thickness, 11.5 microns; area, 0.05 mm$^2$; \acrshort{CVI}, 0.013). This was also the case for individuals with and without a family history of dementia (table \ref{tab:PREVENT_ind_risk}).
            
            The ordinal models give us an indication of the size of the observed associations found in our sample and can be easily interpreted. For example, consider two males of equal age, measuring the same blood pressure. If the first male's choroidal vascular tissue was greater in area by at least 0.16 mm$^2$, they are on average over two times more likely to have at least one (or both) of the risk markers, assuming the comparative individual had no risk markers, thus placing them in a higher risk group for developing Alzheimer's disease in later life. Whether this effect exists within the general population should be investigated in cohorts designed to test this hypothesis.

            The \acrshort{APOE4} variant is a major risk factor for Alzheimer’s disease \cite{stocker2021prediction} and has been associated with a wide range of negative health-related outcomes or features including cardiovascular disease \cite{lahoz2001apolipoprotein}, blood-brain barrier dysfunction \cite{montagne2020apoe4} and inflammation \cite{tao2018association}. However, there is also some debate around its potential role in reducing/increasing risk for ocular disorders like age-related macular degeneration \cite{rasmussen2023associations, liuska2023association}, a disease of the choroid. Thus, it could be that these observed associations are simply reflecting an effect of the \acrshort{APOE4} allele (or related genotypes) on the choroid itself, rather than a direct association with brain health.
            
            An alternative may be related to autonomic dysfunction, of which the \acrshort{APOE4} variant has been associated with \cite{lohman2024central}. Lohman, et al. \cite{lohman2024central} reports that \acrshort{APOE4} carriers exhibit central autonomic dysfunction in early-stage Alzheimer’s disease, including the parasympathetic control of cardiovascular functions \cite{lohman2024blood}. Dillon et al. \cite{collins2012parasympathetic} found that mildly cognitively impaired patients were on average 5.6 times more likely to have autonomic dysfunction, showing significant parasympathetic deficits, which may be involved in the pathogenesis of hypotension in dementia. Given that choroidal blood flow is regulated to some extent by autonomic input, particularly parasympathetic-mediated vasodilation, and systemic hypotension does not cause sympathetic input to the choroid \cite{reiner2018neural}, dysregulation of this autonomic supply could have played a role in the increased choroidal measurements observed in our sample. While this would correspond with larger calibre, the choroid is a highly heterogeneous vascular compartment, and the current device resolution and cross-sectional nature of \acrshort{OCT} capture prevents us from definitively concluding whether increased vasculature corresponds to larger calibre, or simply more vessels.

            Despite its promise as a marker of disease risk for Alzheimer's disease \cite{safieh2019apoe4}, we found just one other study that used \acrshort{EDI-OCT} choroid measures to investigate asymptomatic individuals with known \acrshort{APOE4} status and family history \cite{ma2022longitudinal}. However, in contrast with our findings, Ma et al. \cite{ma2022longitudinal} found no evidence of significant differences in choroidal measures between carriers and non-carriers. One reason for the conflicting findings could be that Ma et al. \cite{ma2022longitudinal} reported choroidal measurements in pixel units. By contrast, we converted choroidal measurements from pixel units into physical units (i.e. microns) according to the unique axial and lateral pixel length-scales corresponding to each B-scan, so that measurements across the population could be compared more appropriately. Moreover, the average participant age in Ma et al. \cite{ma2022longitudinal} was around 20 years older than ours, suggesting that age differences may also play a role in the contrasting findings. 
            
            \acrshort{EDI-OCT} has previously been leveraged to investigate the link between diagnosed dementia and the choroid, with mixed results. Some have found \textit{reduced} choroidal thickness in Alzheimer's disease and mild cognitive impairment compared to controls \cite{bulut2016choroidal, gharbiya2014choroidal, bayhan2015evaluation, kwapong2024choriocapillaris}. Bulut et al. \cite{bulut2016choroidal} and Gharbiya et al. \cite{gharbiya2014choroidal} showed significant choroidal thinning in Alzheimer's disease, but selected both eyes per participant for their statistical analyses without accounting for intra-participant correlation, violating a core statistical principle. Gharbiya et al. \cite{gharbiya2014choroidal} and Bayhan et al. \cite{bayhan2015evaluation} analysed smaller cohorts than ours and performed measurement of the choroid manually, which has the potential for significant measurement error \cite{rahman2011repeatability}. Recently, Kwapong et al. \cite{kwapong2024choriocapillaris} found reduced choriocapillaris density in OCT-Angiography imaging of early-age onset Alzheimer’s disease patients compared with healthy controls, and was able to discriminate between such groups. However, compared to the PREVENT Dementia study, the Kwapong et al. \cite{kwapong2024choriocapillaris} cohort was older and already presenting with symptoms sufficient for clinical diagnosis– whether the same approach can be leveraged at the preclinical stage of Alzheimer’s disease still requires further validation.
            
            On the other hand, other works have found \textit{increased} choroidal measurements in Alzheimer's disease \cite{robbins_choroidal_2021, asanad2019retinal}. Asanad et al. \cite{asanad2019retinal} found an increased choroidal thickness in post mortem human tissue in Alzheimer's disease patients relative to controls using histopathology, while Robbins et al. \cite{robbins_choroidal_2021} found increased choroidal area and vessel area in Alzheimer's disease compared with controls in \acrshort{EDI-OCT}. However, these measurements were also reported in pixel space. Moreover, the authors paradoxically observed significantly decreased choroidal thickness in their sample, but these distances were measured as vertical lines without accounting for choroidal skew or curvature. Thus, where our results fit into the disease trajectory of Alzheimer’s disease is yet to be realised. 
            
            However, the Dementia study is ongoing, collecting longitudinal data on this mid-life cohort. This allows future temporal modelling that may further validate our findings. Nevertheless, our preliminary results in this relatively small sample suggest choroidal differences may exist between prospective risk groups for Alzheimer's disease, and may potentially take place long before the onset of disease. 
            
            Strengths of the current study include the use of Choroidalyzer as a tool to extract measurements from the choroidal vasculature using a standardised, fovea-centred \acrshort{ROI} across the whole cohort, reporting values in physical units rather than pixel units \cite{ma2022longitudinal}, aiding interpretability and helping mitigate spurious associations. Another strength is the uniqueness of the PREVENT cohort, permitting characterisation of the cerebral and ocular microvasculature in a cohort of asymptomatic individuals, at increased risk and 20 years younger than most individuals in similar studies.
            
            There were some limitations to this study. Firstly, of 132 consented at baseline, 49 participants had too poor choroid visualisation for image analysis and all OCT volume data was unusable because the \acrshort{OCT} acquisition protocol was originally designed for retinal \acrshort{OCT} and hence did not prioritise \acrshort{EDI} imaging. Therefore, much of the data collected was unsuitable for analysis due to poor choroid visualisation, resulting in a low sample size. Thus, statistical analysis was limited by low power due to small sample size, and we cannot be certain that selection bias did not play a role in our results. Additionally, participants were primarily white in ethnicity, therefore, generalisability of our findings to other ethnic groups or the wider population may be limited. Thus, while our associations could relate to mechanistic links between the choroid microcirculation and pathogenesis of Alzheimer’s disease, more research is needed. However, future temporal modelling of this ongoing, longitudinal study may enable more impactful results by tracking evolution from mere risk, to cognitive impairment and potential clinical diagnosis of Alzhemer's disease.
        
            \begin{mysubsubsection}[]{Outputs}

                In this section, there has been one publication output which has been published and peer-reviewed (as of February 2025). The author's name is in bold type, with co-load authors underlined:
                \begin{itemize}\setlength\itemsep{0em}
                    \item \underline{\textbf{Burke, Jamie}}, \underline{Samuel Gibbon}, Audrey Low, Charlene Hamid, Megan Reid‐Schachter, Graciela Muniz‐Terrera, Craig W. Ritchie et al. ``\textit{Association between choroidal microvasculature in the eye and Alzheimer's disease risk in cognitively healthy mid‐life adults: A pilot study.}'' Alzheimer's \& Dementia: Diagnosis, Assessment \& Disease Monitoring 17, no. 1 (2025): e70075. 
                \end{itemize}
                
            \end{mysubsubsection}

            \begin{mysubsubsection}[]{Executive summary}
            
                We investigated associations between choroidal measurements obtained with \acrshort{EDI-OCT}, with risk factors (\acrshort{APOE4} and parental family history) for Alzheimer's disease in 69 healthy, mid-life individuals from the PREVENT Dementia study. We found that the choroidal space and vascular tissue was significantly larger in participants who were both \acrshort{APOE4} carriers and had a positive family history of dementia, and this increase was consistent and ordinal between risk groups (vascular tissue in those with neither risk factor < only one risk factor < both risk factors). In our sample, a 0.16 mm$^2$ increase in choroidal vascular area was associated with a two-fold increase in the chance of having one or more markers of Alzheimer’s disease risk, compared with none. Thus, our results suggest a potential link between the choroidal vasculature and risk of Alzheimer’s disease. However, these findings are exploratory and should be replicated in a larger and more diverse sample.
            
            \end{mysubsubsection}
        
            \vfill
        \end{mysubsection}
    
    \end{mysection}

    \begin{mysection}[]{D-RISC II: A phase two evaluation of retinal choroid biomarkers for shock resuscitation in critically Ill adults}\label{sec:ch_app_sec_shock}

        \begin{mysubsection}[]{Introduction}
    
            Circulatory shock is a non-specific pathophysiological state constituting central organ hypoperfusion and can result from hypovolaemic, cardiogenic, obstructive, distributive, or endocrine aetiologies. Shock represents a large proportion of morbidity and mortality in acute care settings, with Sakr, et al. \cite{sakr_characterization_2010} estimating in 2010 that 1 in 3 patients experience some kind of circulatory shock throughout the course of intensive care. More recently, circulatory shock has been a key driver of multi-organ failure and cause of COVID-19 mortality within intensive care \cite{elezkurtaj2021causes}.
            
            Current management of circulatory shock targets improved vascular perfusion through inotropes and vasopressors, with fluid resuscitation to replete intravascular volume. These early goal-directed therapies are effective at restoring measurements of large vessel function, such as systemic blood pressure, but assume that measurements of macrovascular function reflect the microvascular environment and thus informs vital organ perfusion and clinical outcomes. 
            
            However, macrovascular function does not necessarily correspond to microvascular perfusion \cite{gazmuri_pressure-guided_2020}. This macrophysical perspective of treatment may explain why early goal-directed therapies to restore large vessel parameters appear to be ineffective \cite{null_null_randomized_2014, mouncey_paul_r_trial_2015, null_null_early_2017, meyhoff2022restriction} and have the potential to mislead clinicians on the state of the patient and increase risk of mortality. Byrne, et al. \cite{byrne2018unintended} showed that fluid resuscitation in an ovine model increased morbidity through altering the function and permeability of the vascular endothelium. Furthermore, Maitland, et al. \cite{maitland_kathryn_mortality_2011} highlighted the potential for overly aggressive fluid resuscitation to lead to negative patient outcomes after reporting the increased 48-hour mortality in critically ill children with impaired perfusion after fluid boluses.
        
            While macrovascular function does not necessarily correspond to microvascular perfusion, microvascular function within central organs \textit{is} directly relevant to perfusion. It's been observed that microvascular perfusion of sublingual capillaries appears to be a better predictor of multi-organ failure in haemorrhagic shock than serum lactate or systolic blood pressure \cite{hutchings_microcirculatory_2018}. Moreover, strategies targeting microcirculatory dysfunction by protecting the endothelium and preventing vascular leakage may prove more effective therapeutic pathways for patients \cite{iba2024managing, mcmullan2024vascular, levy2024endothelial}. With fluid resuscitation playing a key role in sepsis care, being able to monitor potential microvascular injury by consistently reassessing patient response could help guide treatment at the patient-level, reduce mortality and prevent unnecessary deaths from fluid overload.
        
            However, the microvascular circulation is a highly heterogeneous environment and it is generally difficult to measure the microvascular perfusion of central organs consistently in humans, making interpretation in the clinical setting challenging. The retinal and choroidal circulations are an exception, since they are both directly accessible with non-invasive optical imaging. Recent examination of the retina using \acrshort{OCT} highlight its potential as a non-invasive, safe, feasible and valid approach to observe and quantify intraocular structures in supine, critically ill patients \cite{courtie_retinal_2020, liu_optical_2019}. 
            
            Moreover, the choroidal vascular bed represents a novel microvascular sample site which is preserved as part of the physiological response to shock. This is because it is part of the central nervous system and adapts dynamically to systemic physiology \cite{reiner2018neural}. Recent pre-clinical evidence \cite{park2021visualization} in a septic rat model reported reduced choroidal blood flow during shock, and we already observed it's potential to reflect the perfusion of other central organs, such as the kidney (section \ref{sec:ch_app_sec_ckd}).
            
            Thus, we hypothesised that choroidal microvascular blood flow may represent the microvascular environment in vital organs such as the brain, heart and lungs in patients experiencing circulatory shock. Accordingly, we aimed to establish the feasibility of bedside \acrshort{OCT} within a heterogeneous group of critically ill intensive care and high dependency unit patients, assess whether the intra-ocular vasculature is responsive to progression within a patient's disease course, and identify any physiological correlates with this change.
    
        \end{mysubsection}

        \begin{mysubsection}[]{Data}

            \begin{mysubsubsection}[]{Study population}

                \begin{table}[tb]\footnotesize
                \begin{adjustwidth}{-2in}{-2in}  
                \centering    
                \scalebox{0.95}{\begin{tabular}{p{5cm}p{5cm}p{5cm}}
\toprule
Age (Years)   {[}BL:100\%, FU:100\%{]} & Fluid 24 Hour Output (ml) {[}BL:83\%, FU:100\%{]} & Neutrophils (x10\textasciicircum{}3/mL) {[}BL:100\%, FU:100\%{]} \\
\midrule
Albumin (g/L) {[}BL:83\%, FU:100\%{]} & Fraction Inspired Oxygen (\%) {[}BL:92\%, FU:83\%{]} & Oxygen saturation (\%) {[}BL:100\%, FU:100\%{]} \\
\midrule
APACHE2 (N) {[}BL:75\%, FU:67\%{]} & Glasgow Coma Score (N) {[}BL:100\%, FU:100\%{]} & Peripheral Capillary Refill Time (seconds) {[}BL:100\%, FU:100\%{]} \\
\midrule
Bicarbonate (mmol/L) {[}BL:92\%, FU:83\%{]} & Glucose (mmol/L) {[}BL:92\%, FU:83\%{]} & Peripheral Temperature (Celsius) {[}BL:100\%, FU:100\%{]} \\
\midrule
Calcium (mmol/L) {[}BL:75\%, FU:50\%{]} & Haematocrit (\%) {[}BL:100\%, FU:100\%{]} & Phosphate (mg/dL) {[}BL:92\%, FU:100\%{]} \\
\midrule
Creatinine (mmol/L) {[}BL:100\%, FU:100\%{]} & Haemoglobin (g/L) {[}BL:100\%, FU:100\%{]} & Platelets (x10\textasciicircum{}3/mL) {[}BL:100\%, FU:100\%{]} \\
\midrule
Cumulative Fluid Balance (ml) {[}BL:100\%,   FU:100\%{]} & Heart Rate (bpm) {[}BL:100\%, FU:100\%{]} & Potassium (mmol/kg) {[}BL:100\%, FU:100\%{]} \\
\midrule
Cumulative Fluid Input (ml) {[}BL:100\%,   FU:100\%{]} & Highest Lactate (24 Hour) (mmol/L) {[}BL:92\%, FU:83\%{]} & Respiratory Rate (bpm) {[}BL:100\%, FU:100\%{]} \\
\midrule
Cumulative Fluid Output (ml) {[}BL:100\%,   FU:100\%{]} & Lowest Perfusion Pressure of Oxygen (kPa) {[}BL:92\%, FU:0\%{]} & Sodium (mmol/L) {[}BL:100\%, FU:100\%{]} \\
\midrule
Diastolic Blood Pressure (mmHg) {[}BL:100\%,   FU:100\%{]} & Lymphocytes (x10\textasciicircum{}3/mL) {[}BL:100\%, FU:100\%{]} & Systolic Blood Pressure (mmHg) {[}BL:100\%, FU:100\%{]} \\
\midrule
Fluid 24 Hour Balance (ml) {[}BL:83\%,   FU:100\%{]} & Magnesium (mmol/L) {[}BL:75\%, FU:67\%{]} & Urea (mmol/L) {[}BL:100\%, FU:100\%{]} \\
\midrule
Fluid 24 Hour Input (ml) {[}BL:83\%,   FU:100\%{]} & Mean Arterial Blood Pressure (mmHg) {[}BL:100\%, FU:100\%{]} & White Cell Count (x10\textasciicircum{}3/mL) {[}BL:100\%, FU:100\%{]} \\
\bottomrule
\end{tabular}}

                \end{adjustwidth}
                \caption[Clinical variables collected for the \acrshort{D-RISC}-ii study.]{Clinical variables collected and used in downstream statistical analysis. Individual elements are listed as `variable name (units) [completeness at baseline (BL), completeness at follow-up (FU)]'. APACHE2, Acute Physiological and Chronic Health Evaluation.}
                \label{tab:DRISC_clinvar}
                \end{table}

                Between March 2023 and July 2023, patients were recruited to the Direct Retinal Imaging for Shock Resuscitation in Critically Ill Adults- II (\acrshort{D-RISC}-ii) study --- a single-centre cohort study within tertiary intensive (\acrshort{ITU}) and high-dependency care units (\acrshort{HDU}) at the Royal Infirmary of Edinburgh. Local research and developmental approval (2022/0190) and national ethical approval (22/SS/0055) were obtained and tenets of the Declaration of Helsinki followed. Patients were screened for eligibility by the departmental research nurses. Eligible patients were aged 16 years or older and receiving care corresponding to Intensive Care Society Level 2 or 3 \cite{ics_levels_nodate}. Exclusion criteria included anticipated survival time < 24 hours, pregnancy, obvious ocular pathology or facial trauma, previous clinical exclusion for pupillary dilatation, if \acrshort{OCT} equipment was not available, if study enrolment would be disruptive to care provision, or if consent was not given. Due to ethical approval, follow-up imaging was only done within \acrshort{ITU}/\acrshort{HDU}. We aimed to recruit 15 participants to assess feasibility of choroidal imaging. 
    
                We recorded several clinical parameters prospectively from clinical records. Only clinical variables which had at most 25\% missing values at baseline were used for downstream analysis and are listed in table \ref{tab:DRISC_clinvar}, with units in square brackets and completeness at baseline and follow-up in round brackets. 

            \end{mysubsubsection}

            \begin{mysubsubsection}[]{Image acquisition}

                \begin{figure}[tb]
                    \centering
                    \includegraphics[width=\textwidth]{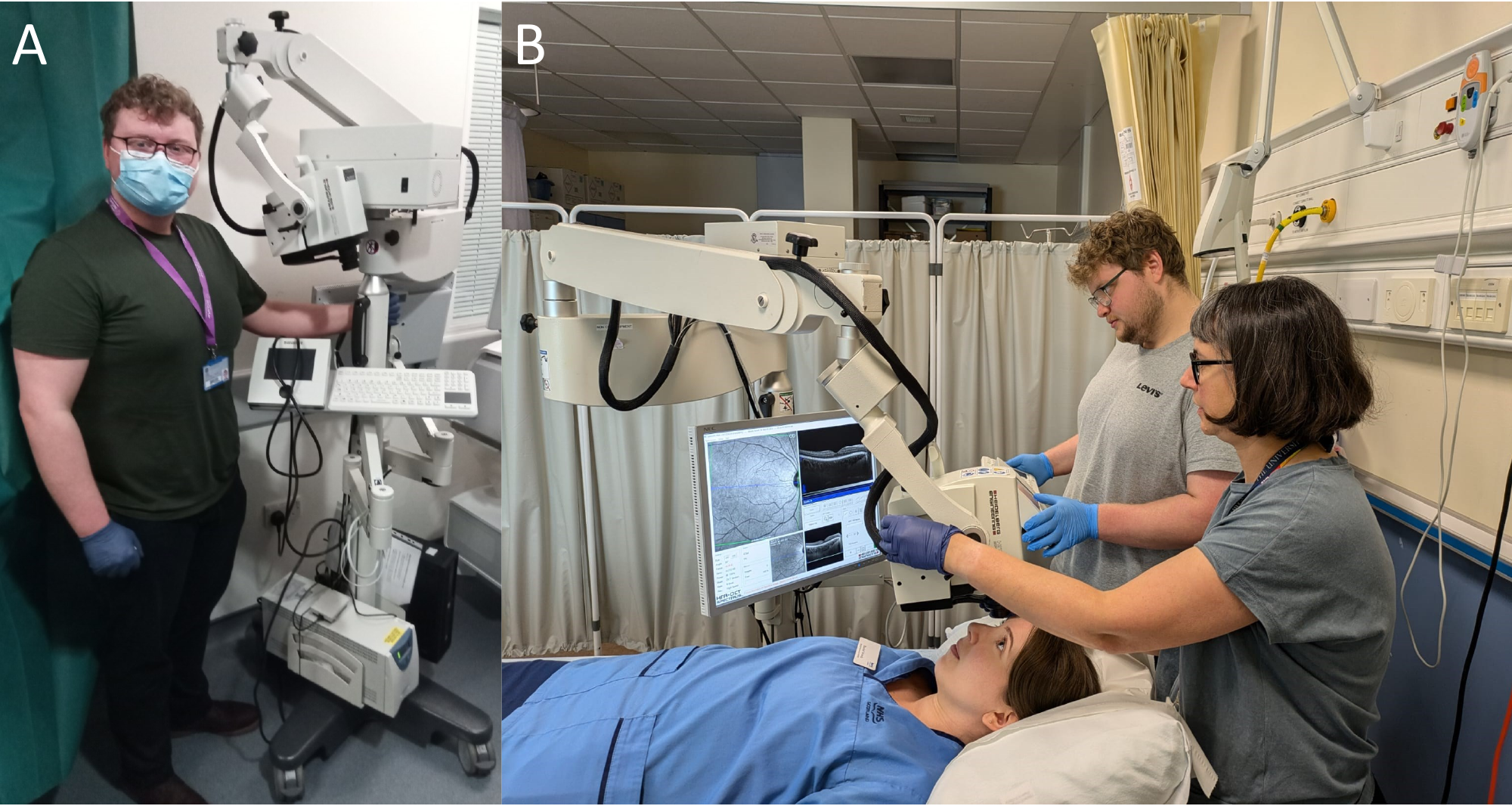}
                    \caption[\acrshort{OCT} FLEX Module in \acrshort{ITU} and demonstrative acquisition.]{(A) Image of the author (Jamie Burke) with the \acrshort{OCT} Spectralis FLEX module in tertiary care \acrshort{ITU}/\acrshort{HDU} unit at the Royal Infirmary of Edinburgh. Image taken by Dr Ian J.C. MacCormick. (B) Demonstration of the \acrshort{OCT} Spectralis FLEX operated by two trained technicians, the author and Charlene Hamid, on a healthy volunteer, Rachel Woods. One technician operated the camera head, while the other operated the computer. Image taken by Rosie Keane.}
                    \label{fig:DRISC_flex}
                \end{figure}
                 
                We sampled patients at two time points, with follow-up examination occurring 12 – 72 hours post-baseline. We performed retinal imaging as soon as feasibly possible once recruited. In each instance, and with preference to the right eye, retinal imaging was done after pharmacological pupil dilation (Tropicamide 1\% weight-by-volume) 30 minutes before examination to facilitate visualisation \cite{kuriakose_slit_2020}. Pupil dilation within a shallow ocular anterior chamber can cause acute angle closure, so anterior chamber depth was screened before dilation using the oblique illumination test (\acrshort{OIT}). 
                 
                All scans collected were macula-centred with a 30-degree \acrshort{FOV} with B-scans covering approximately 9 mm$^2$. All image capture was collected with \acrshort{EDI} mode toggled on to improve visualisation of the Choroid-Sclera boundary and vasculature. Active eye tracking with \acrshort{ART} was also used to help improve image quality, longitudinal registration and reduce speckle noise. 
                 
                Our imaging protocol is enumerated below in order of image capture: A fovea-centred, horizontal-line B-scan with an \acrshort{ART} of 100 followed by a vertical-line B-scan with similar parameters. A fovea-centred posterior pole volume scan, consisting of equally spaced B-scans approximately 240 microns apart using an \acrshort{ART} of 9. Dependent upon patient cooperation, the volume scan either contained 31 B-scans covering a 9.0 $\times$ 6.6 mm \acrshort{FOV}, 25 B-scans covering a 9.0 $\times$ 5.3 mm \acrshort{FOV} or 25 B-scans covering a 5.3 $\times$ 5.3 mm \acrshort{FOV}. 
            
                \acrshort{OCT} scans were collected using the \acrshort{SD-OCT} Heidelberg Spectralis FLEX \acrshort{OCT}1 module (Heidelberg Engineering, Heidelberg, Germany). The Spectralis FLEX module is a mobile version of the table-mounted Spectralis Standard module, with the camera head mounted on a boom to facilitate retinal imaging of supine patients. This was operated by two trained technicians: one to position the camera, and one to operate the computer. In some cases a third person helped to stabilise the head and eyelids. \acrshort{OCT} imaging was performed by three trained operators, Charlene Hamid, (C.H.), Jamie Burke (the author) (J.B.) and Emily Godden (E.G.).
                
                Figure \ref{fig:DRISC_flex}(A) shows the FLEX folded into resting position next to the author after successful \acrshort{OCT} capture of the first patient recruited to \acrshort{D-RISC}-ii. In panel (B) we show a demonstration of the FLEX in use on a healthy volunteer being handled by two persons: the author operating the computer during acquisition, and Charlene Hamid operating the camera head.

            \end{mysubsubsection}
    
             \begin{mysubsubsection}[]{Image analysis}
    
                \begin{figure}[tb]
                    \centering
                    \includegraphics[width=\textwidth]{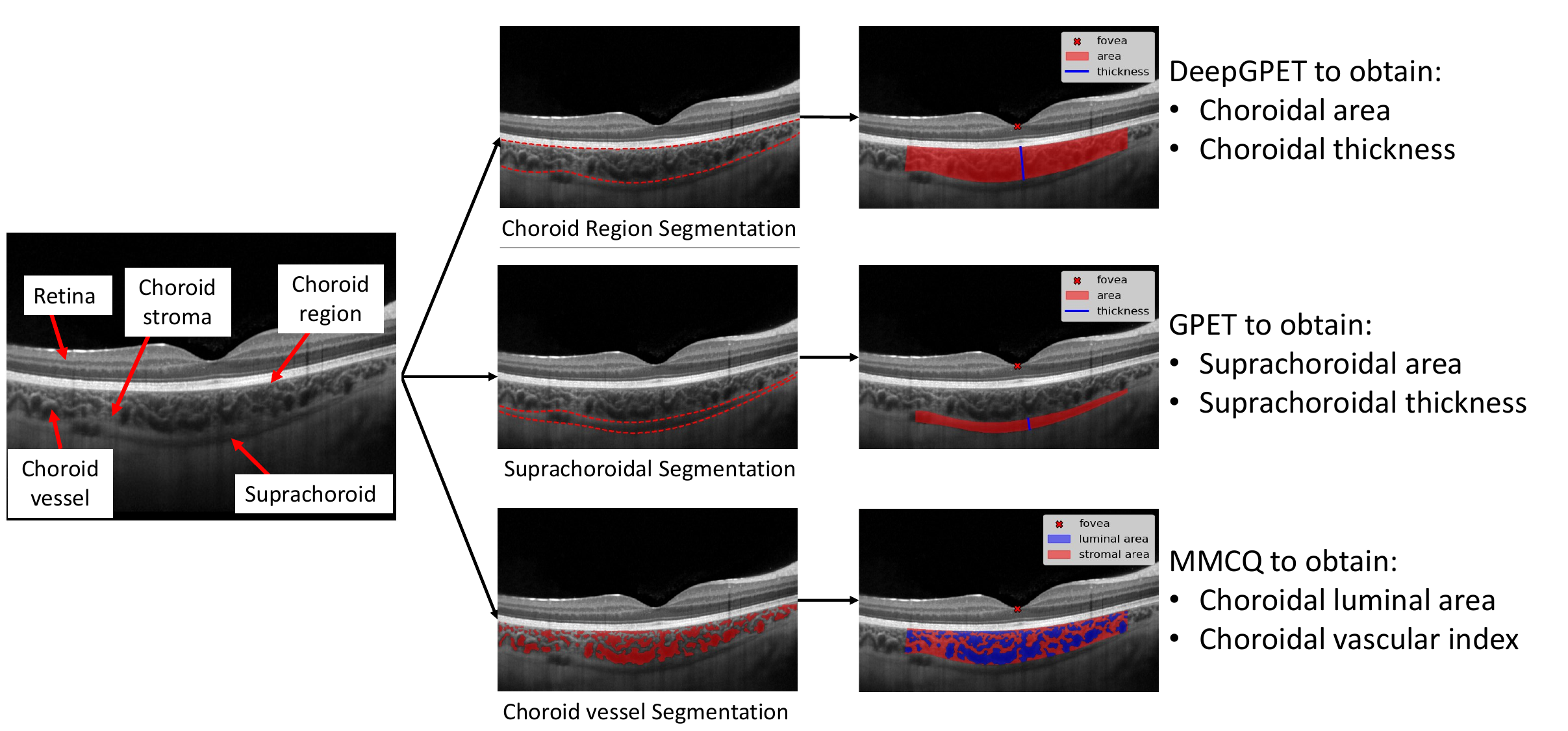}
                    \caption[Description of choroidal measurements collected for the \acrshort{D-RISC}-ii study.]{An \acrshort{OCT} B-scan with landmarks of interest annotated, and pixel-level anatomical segmentations and measurements of the choroidal and suprachoroidal spaces, and choroidal vessels overlaid.}
                    \label{fig:DRISC_ChorMeas}
                \end{figure}
            
                Figure \ref{fig:DRISC_ChorMeas} shows the anatomical layers of interest and the methods used and measurements taken for this study. Delineation of the choroidal space in \acrshort{OCT} B-scans was performed using DeepGPET (chapter \ref{chp:chapter-deepgpet}) and the suprachoroid, if observed, was delineated using \acrshort{GPET} (chapter \ref{chp:chapter-GPET}) which allowed human-assisted pixel-level segmentation (due to its size and co-location to the Choroid-Sclera boundary). Choroidal vessels were segmented using \acrshort{MMCQ} (chapter \ref{chp:chapter-mmcq}) --- Choroidalyzer (chapter \ref{chp:chapter-choroidalyzer}) was not developed until after data collection and image analysis. For each \acrshort{OCT} B-scan, we measure subfoveal choroid thickness, area, vessel area and \acrshort{CVI} using a fovea-centred \acrshort{ROI}. The definition of the choroid, suprachoroid and choroidal measurements have been described previously in section \ref{subsec:ch1_INTRO_measure_bsan}.

            \end{mysubsubsection}
    
            \begin{mysubsubsection}[]{Statistical analysis}
    
                Qualitative feasibility was assessed by collecting several process and safety outcomes to help guide future protocols. These include anterior chamber depth estimated using the \acrshort{OIT} \cite{kuriakose_slit_2020}, application of topical mydriatic agent for pupil dilation, adverse event reporting, patient positions, and case-specific feedback from the image team for each patient imaged. We also leveraged metadata from the image data to quantitatively assess feasibility in terms of acquisition time, image quality index (reported from the imaging device’s built-in \acrshort{SNR} estimate as an unbounded value starting at 0 and plateauing around 50 (Heidelberg Engineering, Heidelberg, Germany) \cite{engineering_Spectralis_2022}) and the number of repeated scans needed for protocol completion.
            
                We sought to generate hypotheses around how patients' physiological and clinical status might correlate with choroidal measurements. For this analysis, we selected only horizontal-line, fovea-centred \acrshort{OCT} B-scans. A 4 mm, fovea-centred \acrshort{ROI} was chosen to include all patients, as larger \acrshort{ROI}s precluded challenging patients from analysis.
                
                We reviewed the distributions of choroidal and clinical data graphically both at baseline and across time points and explored potential pairwise associations. Since these associations are exploratory and hypothesis-generating we did not report P-values associated with the correlation coefficients.  
                
                At baseline, we graphed choroidal metrics against clinical variables using scatter plots and fitted curves to plots that suggested linear or non-linear trends. After graphing, we estimated associations using Pearson's linear correlation coefficient ($r$), or using Spearman's monotonic correlation coefficient ($s$) to assess non-linear association. 
            
                For those successfully imaged at follow-up, we estimated the overall pairwise associations at the patient-level using repeated measures Pearson correlation coefficient ($r_{rm}$) \cite{bland1994correlation, bakdash2017repeated}, accounting for variability within and across patients.
                    
            \end{mysubsubsection}
        
        \end{mysubsection}

        \begin{mysubsection}[]{Results}

            \begin{mysubsubsection}[]{Participants}
        
                \begin{figure}[tb]
                    \centering
                    \includegraphics[width=0.75\textwidth]{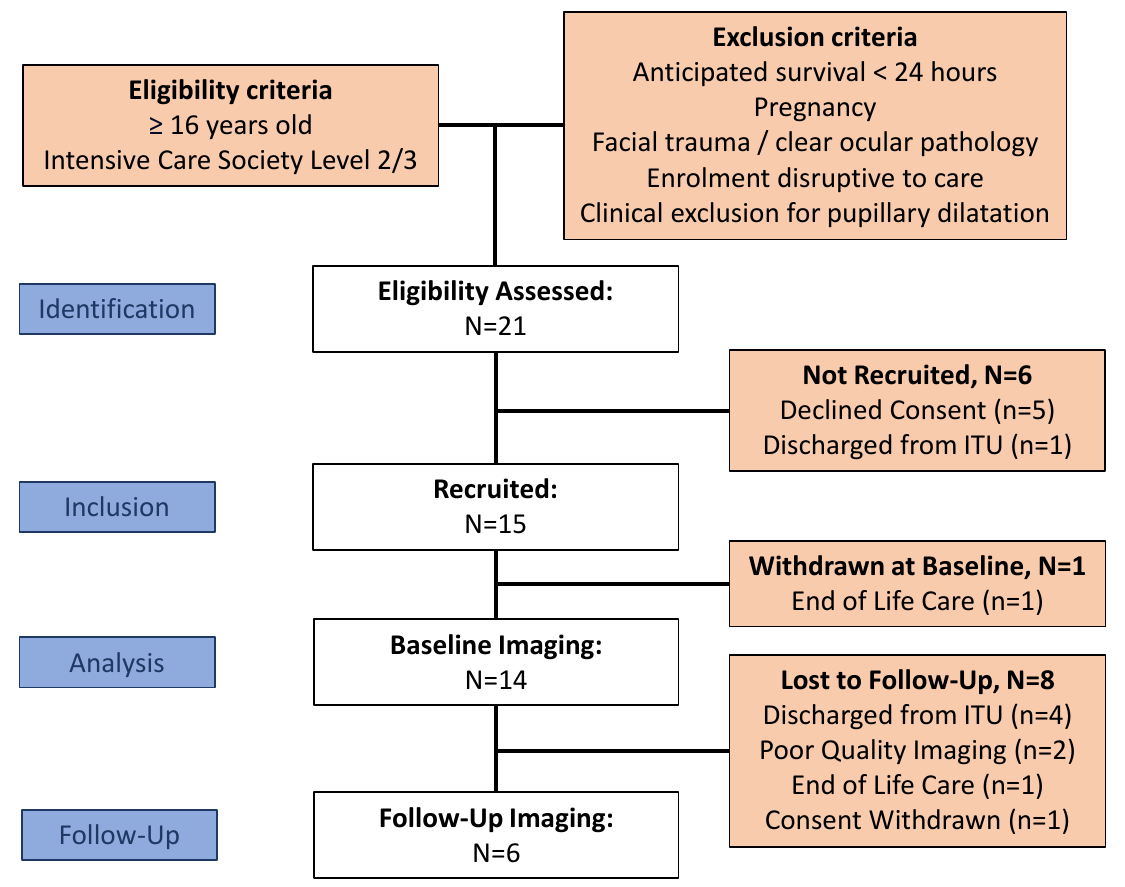}
                    \caption[Sample derivation flowchart for the \acrshort{D-RISC}-ii study.]{Flowchart summarising patient recruitment. From twenty-one eligible patients, fourteen patients were imaged at baseline and six at follow-up.}
                    \label{fig:DRISC_flowchart}
                \end{figure}

                \begin{table}[tbp]\footnotesize
                    \begin{adjustwidth}{-1in}{-1in}  
                    \centering    
                    \scalebox{0.75}{
\begin{tabular}{rlll}
\toprule
\multicolumn{1}{l}{\multirow{2}{*}{Patient Characteristics}} & \multicolumn{2}{c}{Time point} & \multirow{2}{*}{Laboratory Reference¶} \\ 
\cmidrule(l){2-3}
\multicolumn{1}{l}{} & Baseline & Follow-Up &  \\ 
\midrule
\multicolumn{1}{l}{N (Patients)} & 14 & 6 &  \\
\multicolumn{1}{l}{Age (Mean, Years):} & 56.2±15.3 & 47.0±19.0 & - \\
\multicolumn{1}{l}{Sex (male)} & 10 (71) & 3 (50) & - \\
\multicolumn{1}{l}{Body Mass Index (kg/m2):} & 28.4±5.9 & 27.3±7.3 & - \\
\multicolumn{1}{l}{APACHE2* Score at Critical Care Admission} & 20.4±5.9 & 19±6.6 & - \\
\multicolumn{1}{l}{Days Since Escalation to Level 2 or Level 3 Care**} & 9.1±8.8 & 11.7±11.0 & - \\
\multicolumn{1}{l}{Proportion of Patients Receiving Level 3 Care**} & 14 (100) & 6 (100) & - \\
\multicolumn{1}{l}{Primary Diagnosis on Admission} &  &  &  \\
Cardiac Arrest: & 1 (7) & 0 (0) & - \\
Type 1 Respiratory Failure† & 2 (14) & 2 (34) & - \\
Overdose & 1 (7) & 1 (17) & - \\
Toxic Shock Syndrome & 1 (7) & 1 (17) & - \\
Acute Biliary Disease‡ & 2 (14) & 1 (17) & - \\
Elective Igor-Lewis Oesophagectomy & 2 (14) & 0 (0) & - \\
Elective Whipple's Procedure & 1 (7) & 1 (17) & - \\
Elective Liver Transplantation & 3 (21) & 0 (0) & - \\
Elective Thoraco-aortic Aneurysm Repair & 1 (7) & 0 (0) & - \\
\multicolumn{1}{l}{\textbf{Indicators of Circulatory Function}} &  &  & - \\
Heart Rate (min-1) & 87.9±16.9 & 87.8±18.2 & - \\
Mean Arterial Pressure (mmHg) & 84.0±15.3 & 87.6±14.3 & - \\
24 Hour Fluid Balance (mL) & 460.4±1931.2 & -366.5±560.1 &  \\
Cumulative Fluid Balance Since Treatment Escalation (mL) & 5376.8±8461.5 & 6423±11896.1 & - \\
\multicolumn{1}{l}{\textbf{Predictors of Circulatory Function}} &  &  &  \\
Highest Lactate 24-Hours Preceding Sampling (mmol/L) & 1.54±1.00 & 0.98±0.30 & 0.5-1.6 \\
Creatinine Preceding Sampling (mmol/L) & 123.6±83.3 & 117.2±136.7 & 50-111 \\
Haemoglobin Preceding Sampling (g/L) & 95.2±20.1 & 92.7±21.3 & 115-180 \\
Haematocrit Preceding Sampling & 0.28±0.06 & 0.28±0.06 & 0.36-0.52 \\
Albumin Preceding Sampling (g/L) & 18.3±4.2 & 17.5±3.8 & 36-47 \\
\multicolumn{1}{l}{\textbf{Receipt of Organ Supportive Therapies}} &  &  &  \\
Receipt of Invasive Mechanical Ventilation & 4 (29) & 0 (0) & - \\
FiO2 Preceding Sampling & 31.0±19.5 & 23.8±5.6 & - \\
SaO2 Preceding Sampling & 96.7±2.2 & 98.0±1.4 & - \\
Receipt of Vasopressor Therapy & 3 (21) & 1 (17) & - \\
Receipt of Intra-Aortic Balloon Pump & 6 (43) & 2 (33) & - \\
Receipt of Renal Replacement Therapy & 0 (0) & 0 (0) & - \\
Receipt of Intravenous Sedation§ & 3 (21) & 1 (17) & - \\
Glasgow Coma Scale Preceding Sampling (Median, IQR) & 15, 3 & 15, 0 & - \\
\multicolumn{1}{l}{\textbf{Choroid measurements (Median, IQR)}} &  &  & - \\
Subfoveal choroid thickness ($\mu$m) & 249 (134) & 309 (150) & - \\
Area (mm$^2$) & 0.82 (0.62) & 1.15 (0.68) & - \\
Luminal area (mm$^2$) & 0.36 (0.35) & 0.60 (0.39) & - \\
\acrshort{CVI} & 0.44 (0.05) & 0.52 (0.07) & - \\
\multicolumn{1}{l}{\textbf{Suprachoroidal measurements (Median, IQR)}} &  &  &  \\
Subfoveal suprachoroidal thickness ($\mu$m) & 64 (34) & 48 (26) & - \\
Suprachoroidal area (mm$^2$) & 0.22 (0.16) & 0.21 (0.13) & - \\ 
\bottomrule
\end{tabular}
}

                    \end{adjustwidth}
                    \caption[Patient summary statistics for the \acrshort{D-RISC}-ii study.]{Patient summary characteristics at baseline and follow-up. All values are N (\%) or Mean (standard deviation), unless specified otherwise. *, Acute Physiology and Chronic Health Evaluation Score 2 (APACHE2); **, Intensive Care Society Level 2 (\acrshort{HDU}) or Level 3 (\acrshort{ITU}) care; †, Type 1 Respiratory Failure: one case secondary to pneumonia (concomitant bacterial and viral); ‡, Biliary Disease: one case of sepsis secondary to cholangitis, one patient with acute kidney injury following cholecystitis; ¶, Biochemical reference ranges as indicated by local health board (NHS Lothian); §, Sedation including either intravenous propofol or clonidine infusion.}
                    \label{tab:DRISC_pop}
                \end{table}
                
                Figure \ref{fig:DRISC_flowchart} summarises patient recruitment. 21 patients were screened and found eligible for the study and 15 were recruited (declined consent, N=5; \acrshort{ITU}/\acrshort{HDU} discharged, N=1). Of 15 recruited, 14 underwent \acrshort{OCT} capture (end-of-life care, N=1). Follow-up capture was attempted on 6 patients, with 8 patients unavailable (\acrshort{ITU}/\acrshort{HDU} discharged, N=4; end-of-life care, N=1; withdrew consent, N=1; poor baseline imaging, N=2).
            
                Demographics of participants who had retinal imaging are reported in Table \ref{tab:DRISC_pop}. At baseline, the cohort had a mean age of 56.2 years, with 71\% male. All patients were receiving Level 3 care for a mean of 9.1 days. Elective surgical patients contributed approximately half of the cohort (7/15). Mean 24-hour fluid balance was +460ml with a haemoglobin of 95.2g/L (115 -- 180), albumin of 18.3 g/L (normal range 36 -- 47) and APACHE2 score 20.4 (5.9). Mean time to follow-up was 46.3 $\pm$ 31.7 hours. 
                
            \end{mysubsubsection}
    
            \begin{mysubsubsection}[]{Feasibility}
    
                \begin{table}[tbp]\footnotesize
                    \begin{adjustwidth}{-1in}{-1in}  
                    \centering    
                    \scalebox{0.75}{
\begin{tabular}{lllllllllp{6cm}}
\toprule
\begin{tabular}[c]{@{}l@{}}Patient\\ ID**\end{tabular} & Captured & GCS & Tropicamide & Laterality & Position & Time (m:s)* & Operator\$ & \begin{tabular}[c]{@{}l@{}}SCS \\ \\ visible\end{tabular} & Case-Specific Notes \\
\midrule
\multicolumn{10}{l}{Patients Imaged at Baseline: N=14} \\
\midrule
1 & Yes & 4 & Applied & Right & Supine & 20:27 & A & Yes & Sedated, unable to fixate, large suprachoroid, image artefacts. Camera position affected by ventilator in-situ. \\
2 & No & 11 & Applied & Right & Upright & N/A¶ & A, B & No & Unable to fixate, cooperate, or follow instructions. Camera position affected by ventilator in-situ. \\
3 & Yes & 15 & Applied & Right & Upright & 04:32 & A & Yes & Required suctioning of airway secretions during protocol. Camera position affected by ventilator in-situ. \\
4 & Yes & 15 & Not applied§ & Right & Supine & 01:59 & B & Yes &  \\
5 & No & 3 & Applied & Both & Supine & 13:32 & A & No & Sedated, required saline eye drops for dry eyes. Difficulty locating macula due to previous ophthalmic conditions (\acrshort{AMD}). Camera position affected by ventilator in-situ. \\
7 & Yes & 15 & Applied & Right & Supine & 02:53 & A & Yes & Oxygen face mask in-situ. \\
8 & Yes & 15 & Applied & Right & Upright & 05:44 & A & Yes & Camera position affected by nebuliser in-situ. Required suctioning mid-protocol. Bedside access blocked ipsilaterally. \\
9 & Yes & 15 & Applied & Right & Upright & 03:26 & A & Yes & Lack of head support due to upright positioning, lead to difficulty fixating. \\
10 & Yes & 15 & Applied & Right & Upright & 02:19 & A & Yes & Patient hearing impaired with difficulty fixating. \\
11 & Yes & 10 & Applied & Right & Supine & 07:15 & A & No & Required saline eye drops for dry eyes. Partially sedated. Highly myopic eyes, large area of peripapillary atrophy. Difficultly fixating. Camera position affected by ventilator in-situ. \\
12 & Yes & 15 & Applied & Right & Upright & 01:11 & B & No & Poor cooperating and fixation due to patient discomfort. \\
13 & Yes & 15 & Applied & Right & Upright & 04:09 & A & No & Patient drowsy with difficulty fixating. \\
14 & Yes & 15 & Applied & Right & Supine & 01:30 & A & Yes &  \\
15 & Yes & 15 & Applied & Right & Supine & 03:04 & A & Yes & Difficulty imaging deep choroid with enhanced depth OCT. \\
\midrule
\multicolumn{10}{l}{Patients Imaged at Follow-Up: N=6} \\
\midrule
1 & Yes & 11 & Applied & Right & Supine & 28:40 & A & Yes & Sedated, unable to fixate, Camera position affected by ventilator in-situ \\
3 & Yes & 15 & Applied & Right & Supine & 05:27 & A & Yes &  \\
4 & Yes & 15 & Applied & Right & Supine & 01:52 & A & Yes & Protocol completed despite \acrshort{OCT} power supply issues. \\
8 & Yes & 15 & Applied & Right & Supine & 04:14 & B & Yes & Horizontal imaging repeated to improve capture quality. \\
9 & Yes & 15 & Not applied§ & Right & Supine & 01:32 & A & Yes & \acrshort{OCT} imaging facilitated by supine position. \\
14 & Yes & 15 & Applied & Right & Upright & 01:04 & A & Yes & \\
\bottomrule
\end{tabular}
}

                    \end{adjustwidth}
                    \caption[Patient-level feasibility data for the \acrshort{D-RISC}-ii study.]{Feasibility data for \acrshort{OCT} in critical care settings. SCS: suprachoroidal space, m:s, minutes:seconds. GCS, Glasgow Coma Scale. ** : Patient 6 was excluded before baseline imaging due to clinical deterioration and end of life care; * : Time was the intra-session difference in acquisition time between first and last scan acquired; § : Patient right eye was clinically dilated, with no requirement for mydriasis. ¶ : Only one scan was collected so timing could not be estimated; \$ : Operators A were Charlene Hamid and Jamie Burke (the author), and B were Emily Godden and Jamie Burke (the author).}
                    \label{tab:DRISC_feasibility}
                \end{table}

                \begin{figure}[tb]
                    \centering
                    \includegraphics[width=\textwidth]{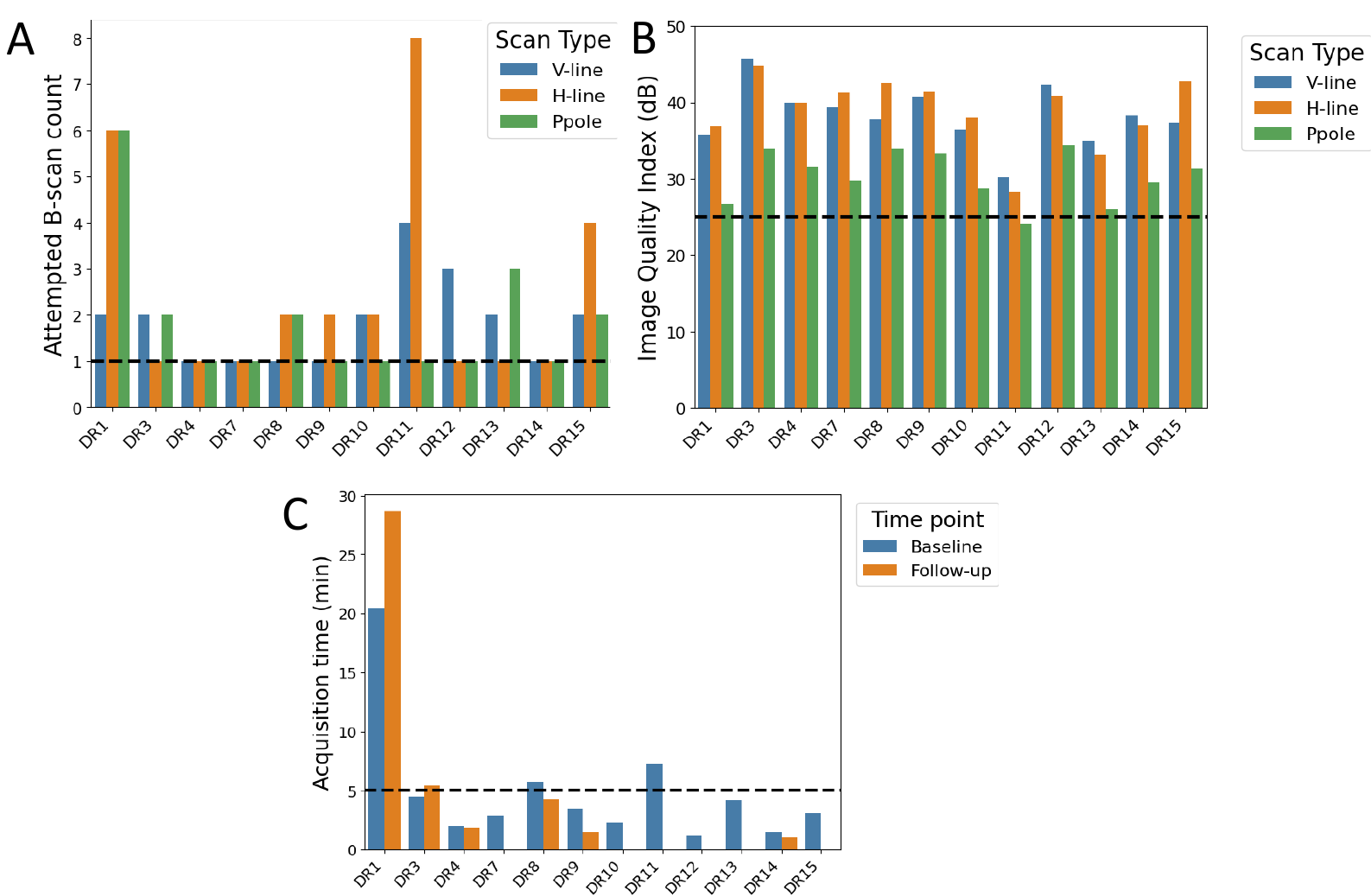}
                    \caption[Patient-level \acrshort{OCT} feasibility data for the \acrshort{D-RISC}-ii study.]{OCT acquisition information for each successfully imaged patient. (A) Acquisition time, stratified by scan type, with a black dashed line at 1. (B) \acrshort{OCT} image quality index, stratified by scan type, with a black dashed line at 25dB. (C) Acquisition time between first and last B-scan for each time point, with a black dashed line at 5 minutes. Dashed lines are used as a visual aid.}
                    \label{fig:DRISC_feasibility}
                \end{figure}
            
                Table \ref{tab:DRISC_feasibility} presents case-specific feasibility data. Baseline imaging was attempted in 14/15 recruited participants (the exception was owing to clinical deterioration and end-of-life care) and follow-up imaging attempted in six of these. Cumulatively, the imaging protocol was attempted twenty times, with eighteen of these (18/20, 90\%) producing satisfactory outputs for analysis. In particular, baseline imaging was successful in 12 of 14 (86\%) participants and follow-up imaging was successful in all six attempts (100\%). All recruitment and imaging took place in \acrshort{ITU}. There were no adverse events from imaging, and case-specific comments largely cite patient difficulty fixating or camera head-related issues due to in-situ ventilators. However, these did not negatively impact image acquisition, implying there was no significant issue for those whose protocol was successfully completed.
            
                Figure \ref{fig:DRISC_feasibility} presents bar plots on acquisition metadata related to each successfully imaged patient. Median (IQR) acquisition time per visit was 02:09 (3:30) minutes:seconds, with thirteen out of eighteen attempted imaging sessions completed within 5 minutes. Average (\acrshort{SD}) scores for \acrshort{SNR} across the cohort was 38.9 $\pm$ 4.6, 38.3 $\pm$ 3.9 and 30.3 $\pm$ 3.3 for horizontal-line, vertical-line and posterior pole volume scans, respectively, with all image quality strengths exceeding the minimum score outlined by the manufacturer (25dB) \cite{engineering_Spectralis_2022}. Median (IQR) number of scan attempts needed per visit was 1.5 (1).
            
                Two patients (ID2, ID5) failed the imaging protocol. For Patient ID2, who developed respiratory failure after a planned surgical admission, imaging was attempted in an upright position with topical mydriasis applied to the right eye. The patient, partially sedated, became agitated and was unable to fixate or follow instructions. A second imaging attempt on the fellow eye was unsuccessful, leading to their exclusion from the study with only one scan collected overall. Patient ID5 was highly myopic with advanced \acrshort{AMD} in the right eye and vitreo-macular traction in the fellow eye, which made locating the macula centre a significant challenge. The patient's sedation further hindered cooperation, and after 13.5 minutes of attempted acquisition, the protocol was abandoned. Any collected scans were excluded due to insufficient choroidal visualisation and poor macula-centring.
            
            \end{mysubsubsection}

            \begin{figure}[!t]
                    \centering
                    \includegraphics[width=\textwidth]{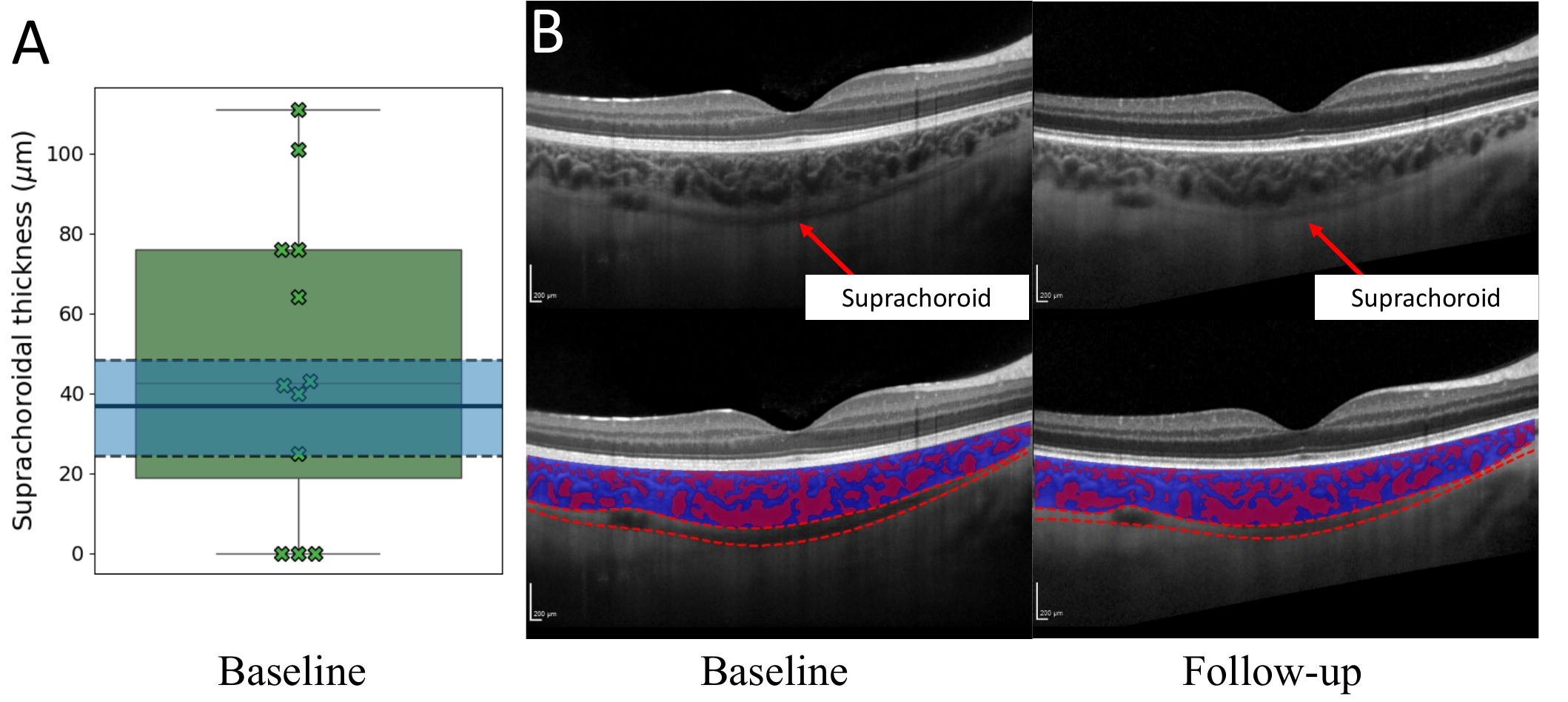}
                    \caption[Representative suprachoroidal space in patients in the \acrshort{D-RISC}-ii study.]{(A) Choroidal variation at baseline for suprachoroidal thickness, with shaded blue region representing median and IQR estimated from Yiu, et al. \cite{yiu2014characterization}. (B) Visible suprachoroidal space in youngest patient (age: 16) at baseline and at follow-up 22 hours later. (Top) suprachoroidal space highlighted by the red arrows. (Bottom) suprachoroidal space outlined in dotted red.}
                    \label{fig:DRISC_suprachoroid}
                \end{figure}
    
            \begin{mysubsubsection}[]{Choroidal variation}

                \begin{figure}[tbp]
                    \centering
                    \includegraphics[width=0.75\textwidth]{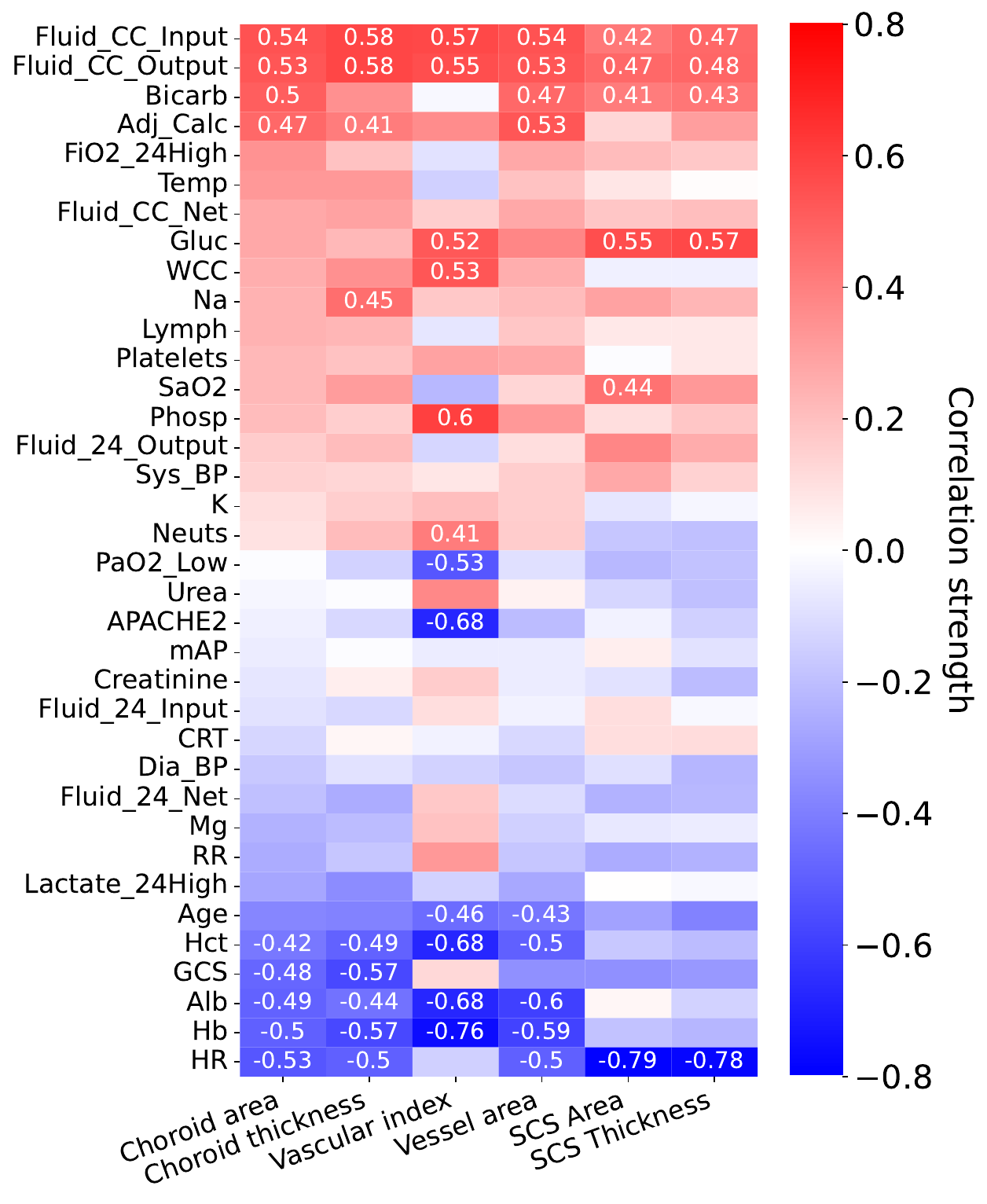}
                    \caption[Pairwise correlation analysis between choroidal measurements and clinical variables in the \acrshort{D-RISC}-ii study.]{Correlation heat map between choroidal measurements (horizontal axis) and clinical measurements (vertical axis) at baseline, with correlation coefficients superimposed and ordered by choroid area (far left). All correlations are Pearson correlation coefficients, with exception of cumulative fluid input/output/net which use the Spearman correlation coefficient. Only correlation coefficients $\geq$ 0.4 in absolute value are annotated. CC, cumulative; Bicarb, bicarbonate; Adj\_Calc: Calcium; FiO2, fraction inspired oxygen; Temp, peripheral temperature; Gluc, glucose; WCC, white cell count; Na, sodium; Lymph, lymphocites; Sa02, oxygen saturation; Phosp, phosphate; Sys\_BP, systolic blood pressure; K, potassium; Neuts, Neutrophils; PaO2\_Low, lowest partial pressure of oxygen; APACHE2, Acute Physiology and Chronic Health Evaluation Score 2; mAP, mean arterial pressure; CRT, capillary refill time; Dia\_BP, diastolic blood pressure; Mg, magnesium; RR, respiratory rate; CRP, C-reactive protein; Hct, haematocrit; GCS, Glasgow coma scale; Alb, albuminuria; Hb, haemoglobin; HR, heart rate; Hb, haemoglobin; SCS, suprachoroidal space.}
                    \label{fig:DRISC_corrplot}
                \end{figure}
             
                \begin{figure}[!t]
                    \centering
                    \includegraphics[width=\textwidth]{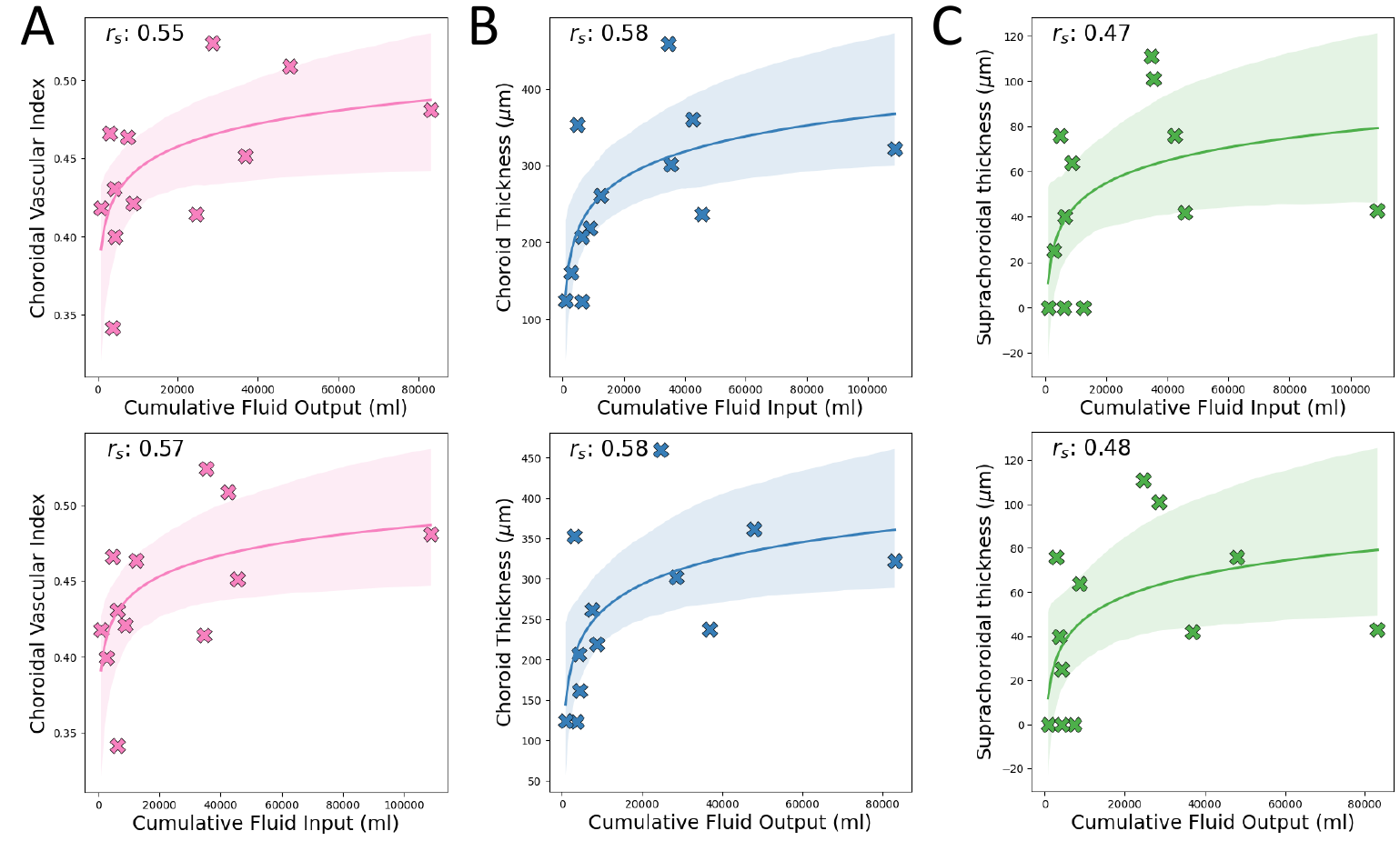}
                    \caption[Potential non-linear and monotonic association between choroidal measurements and fluid measurements in the \acrshort{D-RISC}-ii study.]{Observed non-linear and monotonic trends between choroidal measurements and cumulative fluid input (top) and output (bottom). (A) \acrshort{CVI}, (B) subfoveal choroid thickness and (C) subfoveal suprachoroidal thickness.}
                    \label{fig:DRISC_ChorFluid}
                \end{figure}

                 \begin{figure}[!t]
                    \centering
                    \includegraphics[width=\textwidth]{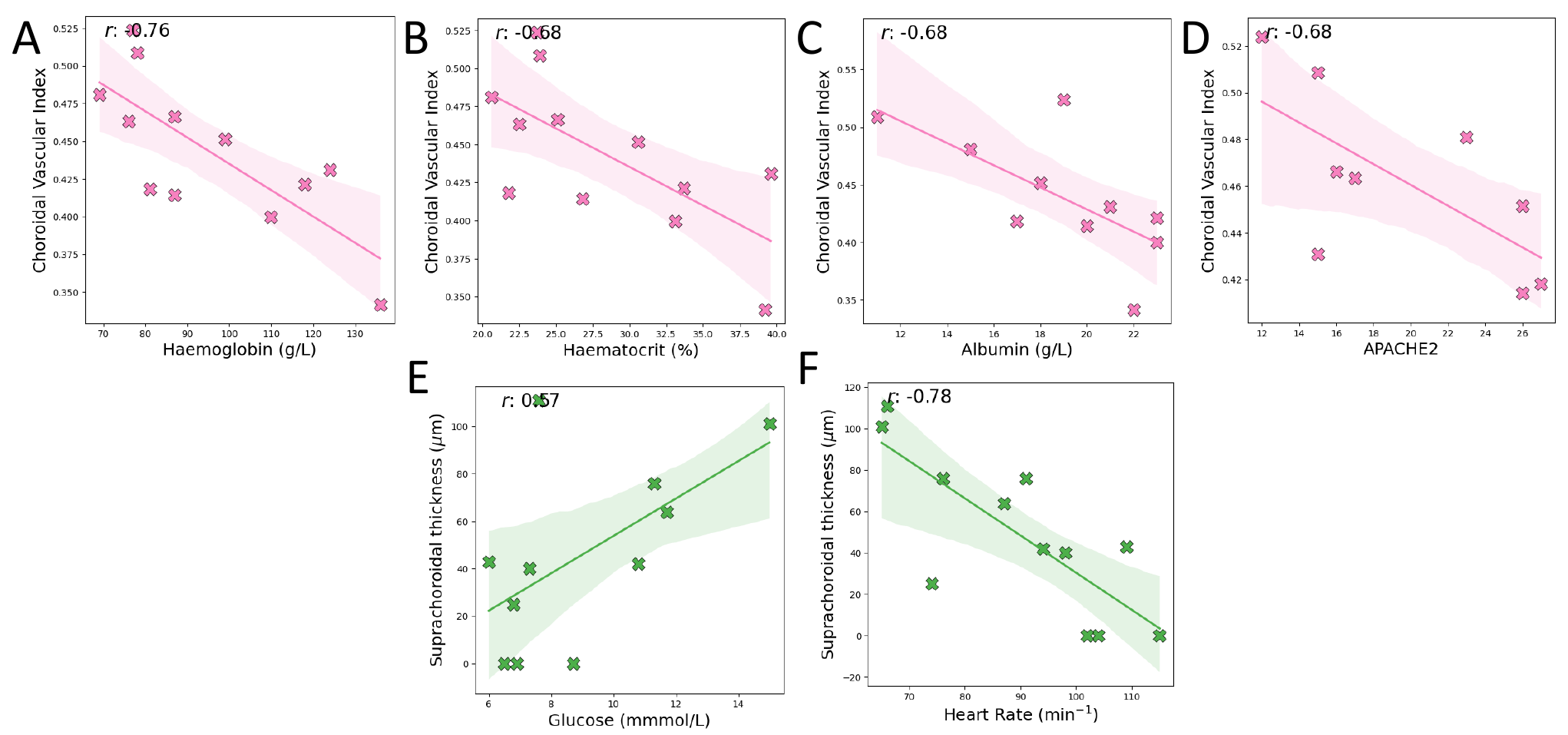}
                    \caption[Potential linear association between choroidal measurements and various clinical variables measurements in the \acrshort{D-RISC}-ii study.]{(A -- D) Observed linear trends between \acrshort{CVI} and (A) haemoglobin, (B) haematocrit, (C) albumin and (D) APACHE2 score. (E -- F) Observed linear trends with subfoveal suprachoroidal space thickness and (E) glucose and (F) heart rate. APACHE2, Acute Physiology and Chronic Health Evaluation Score 2;}
                    \label{fig:DRISC_CVI_association}
                \end{figure}
            
                \begin{figure}[!t]
                    \centering
                    \includegraphics[width=\textwidth]{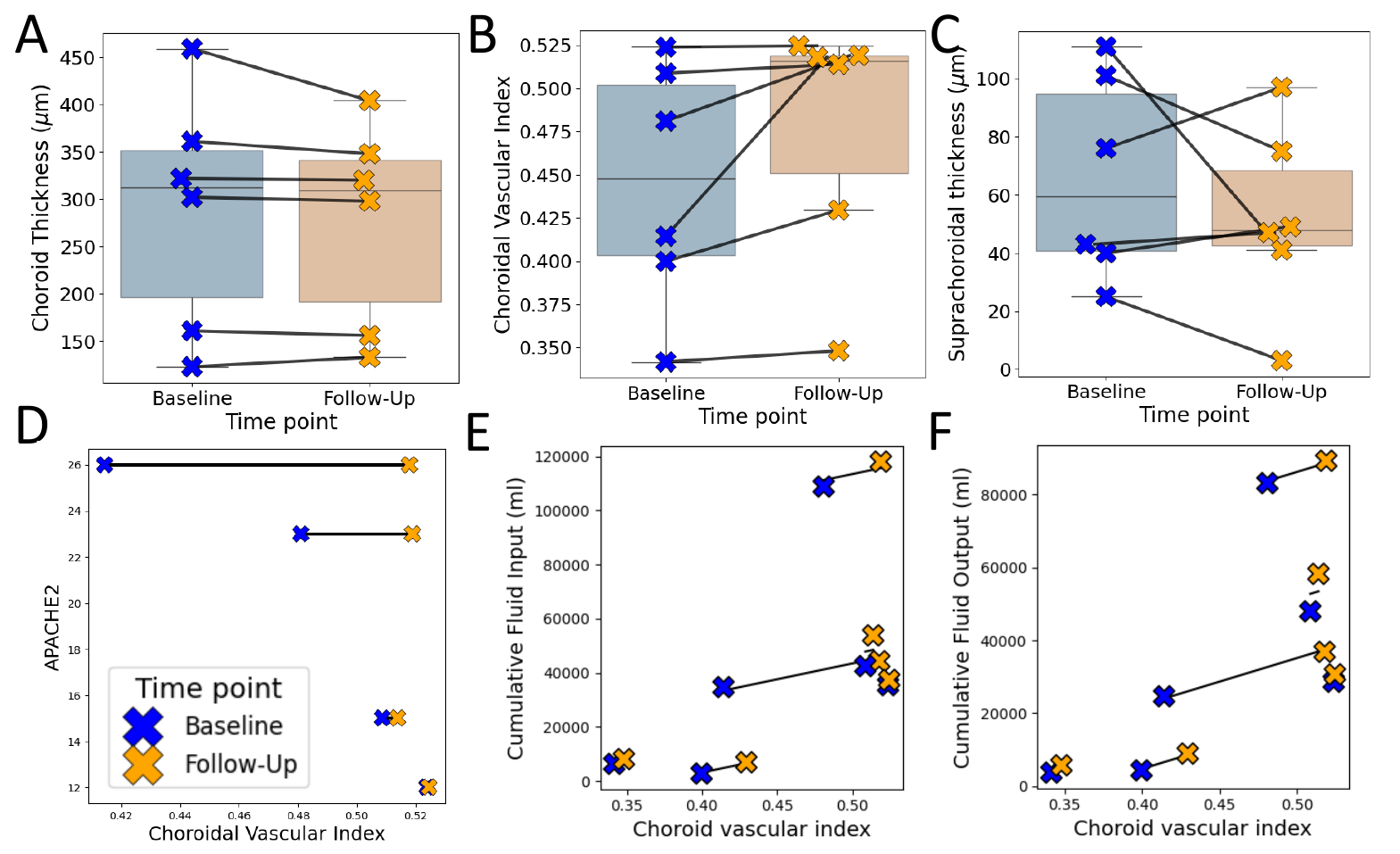}
                    \caption[Observed longitudinal change in choroidal measurements and fluid in the \acrshort{D-RISC}-ii study.]{(A -- C) Observed change in measurements between time points for (A) \acrshort{CVI}, (B) subfoveal choroid thickness and (C) subfoveal suprachoroidal thickness. (D -- F) Longitudinal change, with common slope in black, between \acrshort{CVI} and clinical markers across time points for (D) APACHE2, (E) cumulative input fluid and (F) cumulative output fluid. 2 patients of the 6 patients who were followed-up did not have an APACHE2 score. APACHE2, Acute Physiology and Chronic Health Evaluation Score 2.}
                    \label{fig:DRISC_longitudinal}
                \end{figure}
                
                Baseline suprachoroidal space thickness is shown in figure \ref{fig:DRISC_suprachoroid}(A). The distribution of baseline suprachoroidal thickness had an interquartile range notably larger than previously reported in a healthy, elderly cohort aged 55 -- 85 \cite{yiu2014characterization}. In nine of the twelve (75\%) successfully imaged patients at baseline, the suprachoroidal space was visible and markedly enlarged, and was largest in the younger patients of our cohort --- the suprachoroid of the youngest patient (aged 16) decreased in subfoveal thickness by 26 microns between baseline and follow-up after 22 hours (figure \ref{fig:DRISC_suprachoroid}(B)). Of these nine, six had follow-up scans with the suprachoroid still visible which decreased from baseline median thickness of 64 microns to 48 microns (table \ref{tab:DRISC_pop}).

                Figure \ref{fig:DRISC_corrplot} presents a correlation heat map between choroid and clinical measurements at baseline, and is colour coordinated according to the sign and strength of the correlation coefficient. All correlations represent Pearson's linear correlation coefficient, $r$, with exception of cumulative fluid input/output/net which was estimated using Spearman's monotonic correlation coefficient, $s$, due to the observed non-linear trend after graphing. The correlation matrix is ordered according to choroid area, and only correlation coefficients $\geq$ 0.4 in absolute value are annotated. 
            
                All baseline choroidal measurements appeared to have positive non-linear, monotonic (Spearman $s$) associations with cumulative fluid input and output (input fluid: choroidal thickness, $s$=+0.58; suprachoroidal space thickness, $s$=+0.47; \acrshort{CVI} $s$=+0.57; output fluid: choroidal thickness, $s$=+0.58; suprachoroidal space thickness, $s$=+0.48; \acrshort{CVI} $s$=+0.55). These non-linear trends are shown in figure \ref{fig:DRISC_ChorFluid}.
            
                Baseline \acrshort{CVI} appeared to have many strong, linear associations with the clinical variables collected, in particular blood markers. Figure \ref{fig:DRISC_CVI_association}(A -- D) presents the strongest linear associations estimated between baseline \acrshort{CVI} and the clinical variables selected for analysis. Baseline \acrshort{CVI} appeared negatively associated with Acute Physiology and Chronic Health Evaluation (APACHE) II score ($r$= -0.68), albumin ($r$= -0.68), haematocrit ($r$= -0.68), haemoglobin ($r$= -0.76). A weaker, negative association was observed with partial pressure oxygen ($r$= -0.53) and positively with phosphate ($r$=+0.60), glucose ($r$= +0.52) and white cell count ($r$= +0.53).
                
                For those blood markers, haemoglobin, haematocrit and albumin, we observed relatively strong associations with measurements of thickness, area and vessel area. Suprachoroidal thickness and area appeared to be positively associated with glucose (thickness $r$= +0.57, area $r$= +0.55) and negatively associated with heart rate (thickness $r$= -0.78, area $r$= -0.79), as shown in figure \ref{fig:DRISC_CVI_association}(E -- F). There were no obvious associations at baseline with systemic blood pressure, nor was there an obvious association between these rheological markers and systemic blood pressure (mean arterial blood pressure and haemoglobin $r$= +0.02, haematocrit $r$= -0.08 and albumin $r$= -0.03).
            
                Figure \ref{fig:DRISC_longitudinal} shows longitudinal change in choroidal measurements. Subfoveal choroidal thickness did not vary substantially, but \acrshort{CVI} and suprachoroidal thickness did in some individuals (figure \ref{fig:DRISC_longitudinal}(A -- C)). The largest increases in \acrshort{CVI} appeared to be in patients with the highest APACHE II scores (Figure \ref{fig:DRISC_longitudinal}(D)), however two of the six patients who were followed-up did not have an APACHE2 score. Finally, \acrshort{CVI} appeared to linearly correlate with cumulative fluid markers over time and within patients (repeated measures Pearson correlation for \acrshort{CVI} with input fluid, $r_{rm}$= +0.74; and output fluid $r_{rm}$= +0.83). 
            
            \end{mysubsubsection}
            
        \end{mysubsection}

        \begin{mysubsection}[]{Discussion}

            This study has demonstrated that bedside \acrshort{OCT} is feasible and effective in diverse clinical presentations from a single-centre, tertiary critical care unit. The majority (60\%) of encounters were completed with patients positioned supine, supporting feasible imaging of hospitalised patients. The need to perform multiple, repeated \acrshort{OCT} scans of the same location per visit were primarily for patients who were sedated, had thick choroids, or had poor fixation. Nevertheless, 50\% of all examinations (and macular \acrshort{OCT} scan types) only required one attempt, and 81\% within two attempts. Repeated attempts at scanning were also associated with elongated acquisition times, but the majority (72\%) of examinations were finished within 5 minutes which, considering the time configuring the device at bedside ($\sim$ 5 minutes), is a reasonable length of time to ensure any disruption to provisional care is minimised.
        
            Our exploratory results in this small sample suggest that choroidal measurements may reflect systemic fluid status and rheology, and that \acrshort{EDI-OCT} is sensitive to changes within individuals over time. In particular, we did observe a potential monotonic relationship between choroidal measures and cumulative input/output fluid, both at baseline and over time, while \acrshort{CVI} appeared to associate negatively with blood markers haemaglobin, haematocrit and albumin. These observations are particularly interesting given the apparent absence of relationship with macrovascular markers of systemic blood pressure. Moreover, there was no obvious association with systemic blood pressure and these blood markers. This may suggest the choroidal vascular bed is more informative at reflecting the microvascular environment and could be a potentially useful sampling site for assessing organ perfusion. 
            
            This study is also the first to report the visibility of the suprachoroid in critically ill patients, which was present in 75\% of successfully imaged patients. This frequency and magnitude of suprachoroidal space we found is unusual. The suprachoroid is a potential space between the choroid and sclera traversed by vessels and nerves, and ranges in thickness from 10 to 35 microns \cite{saidkasimova_suprachoroidal_2021}, and is often not visible due to limited \acrshort{OCT} device resolution. Although we observed this space as open for younger patients, a thin suprachoroid may be visible in up to 44\% of healthy people aged 55 -- 85 \cite{yiu2014characterization}, but an obviously visible space is usually associated with ocular disease such as severely low eye pressure or retinochoroidal inflammation \cite{saidkasimova_suprachoroidal_2021, moisseiev_suprachoroidal_2016, sadda_ryans_2022}. Whilst we did not measure eye pressure, the presence of this space in these \acrshort{ITU} patients, and its ability to change during admission, might indicate a response to fluid build-up and choroidal effusion \cite{waheed_choroidal_2022}. Further work in this space is needed to elucidate the potential role of the suprachoroid during fluid therapy, while considering eye pressure.
        
            Additionally, to the authors knowledge, this study is the first to successfully quantify choroidal variation in both cross-sectional and longitudinal settings using \acrshort{EDI-OCT} in a heterogeneous cohort of critically ill patients across multiple diagnoses. Previous studies have shown \acrshort{OCT} feasibility in tertiary critical care units \cite{liu_optical_2019, courtie2021stability}, but this study is the first to correlate intra-ocular changes with patients' acute physiology across heterogeneous aetiologies. Liu et al. \cite{liu_optical_2019} demonstrated \acrshort{OCT} and OCT-Angiography (OCT-A) feasibility in a similar setting, and our study extends this by investigating a marginally larger cohort and suggesting the feasibility of \acrshort{EDI-OCT} of the choroid. Additionally, Courtie et al. \cite{courtie2021stability} found changes in retinal blood flow using OCT-A in patients undergoing elective oesophagostomy, supporting our findings of abnormal intra-ocular anatomy in critical illness. We hypothesise that acute changes in patients' intravascular compartment and physiological status are linked to microcirculatory adaptations in intra-ocular structures. Our study is a first step towards understanding how choroidal responses to systemic stress might be used to monitor circulatory shock progression in critically ill adults.
            
            The main limitation of this study was the small and heterogeneous study population. While the wide inclusion criteria are \textit{advantageous} to assess bedside \acrshort{OCT} feasibility, they also introduce \textit{confounding} factors into the statistical analysis. The limited number of participants, due to the exploratory nature and resource constraints of this preliminary study, prevented application of more comprehensive statistical analysis methods or reporting of statistical significance of estimated associations. A larger cohort may help further statistically validate our exploratory findings.
            
            Moreover, whilst 14 patients were imaged at baseline, only 6 (43\%) were followed-up per-protocol. Of the 8 patients lost, 4 (29\%) were discharged from critical care before follow-up imaging could be arranged --- a result of limited resources to implement and continue the study protocol outside of the tertiary critical care unit. Thus, future studies designed to test hypotheses about associations between choroidal metrics and systemic physiology should follow patients up after discharge. 
        
            Finally, we also note that while it is feasible, effective imaging with the Heidelberg Engineering \acrshort{OCT} Spectralis FLEX required 2 -- 3 operators and co-operation of nurses. However, the size, cost, and ease of use of portable \acrshort{OCT} equipment is likely to improve, creating the potential for investigating retinal microcirculatory biomarkers of multi-organ failure and shock further \cite{chopra_optical_2021, song_review_2021}, especially since retinochoroidal \acrshort{OCT} images can now be analysed with open-source tools like Choroidalyzer.
        
            Future work will seek to perform \acrshort{OCT} and \acrshort{OCT-A} imaging of heterogeneous intensive care and high dependency unit patients, to validate the initial findings presented in this study, and explore the utility of the suprachoroid and choroidal vascular bed as biomarkers for circulatory shock. Ultimately, it is hoped that this research will enable patient-level physiological optimisation of fluid resuscitation and to prevent excess iatrogenic morbidity from fluid resuscitation within critically ill patients.

            \begin{mysubsubsection}[]{Outputs}

                In this section, there has been one publication output which is currently in submission and under peer-review at BMC Critical Care as of February 2025. The manuscript has been uploaded to medRxiv, with the author of this thesis in bold and any co-lead author contributions in italic:

                \begin{itemize}\setlength\itemsep{0em}
                    \item \underline{George Cooper}, \underline{\textbf{Burke, Jamie}}, Charlene Hamid, Emily Godden, Neeraj Dhaun, Stuart King, Tom MacGillivray, J. Kenneth Baillie, David M. Griffith and Ian J.C. MacCormick. ``\textit{D-RISC II: Direct Retinal Imaging for Shock Resuscitation in Critical Ill Adults II.}'' Submitted and in peer-review at BMC Critical Care as of February 2025. 
                \end{itemize}
                
            \end{mysubsubsection}

            \begin{mysubsubsection}[]{Executive summary}
            
                We tested the feasibility of measuring the choroid in intensive care and explored associations between choroidal measurements and severity of disease. We performed \acrshort{OCT} capture on patients admitted to \acrshort{ITU}/\acrshort{HDU} and repeated imaging once 12 -- 72 hours later. We measured the choroid and compared this to prospectively collected clinical variables. Attempted \acrshort{OCT} imaging on fourteen patients was successful in twelve at baseline, and all six attempts at follow-up was successful. At baseline, choroidal measurements were positively associated with fluid balance, and negatively with Acute Physiology and Chronic Health Evaluation (APACHE) II score, haematocrit, albumin and haemoglobin. A measurable suprachoroidal space was seen in 75\% (9) patients and was inversely associated with heart rate. There was substantial intra-individual variation in choroidal measurements over time. We found that measuring the choroid was feasible in patients with critical illness. Exploratory associations with systemic variables suggest that the choroid may provide information about the microvascular function of other major organs. Size and change of choroidal measurements may reflect perfusion pressure or inflammation. To the best of our knowledge, this is the first report of choroidal imaging in intensive care.
            
            \end{mysubsubsection}
    
        \end{mysubsection}
    
    \end{mysection}

\end{mychapter}

\begin{mychapter}[]{Discussion} \label{chp:chapter-discussion}

    \begin{mysection}[]{Core contributions to the field}

        Figure \ref{fig:DISCUSS_PhD} presents the diagram of the PhD research conducted during the study period, highlighting the accomplishments of this thesis in addressing the lack of standardised tools for measuring the choroid in \acrshort{OCT} image sequences.

        \begin{figure}[tb]
            \centering
            \includegraphics[width=0.75\linewidth]{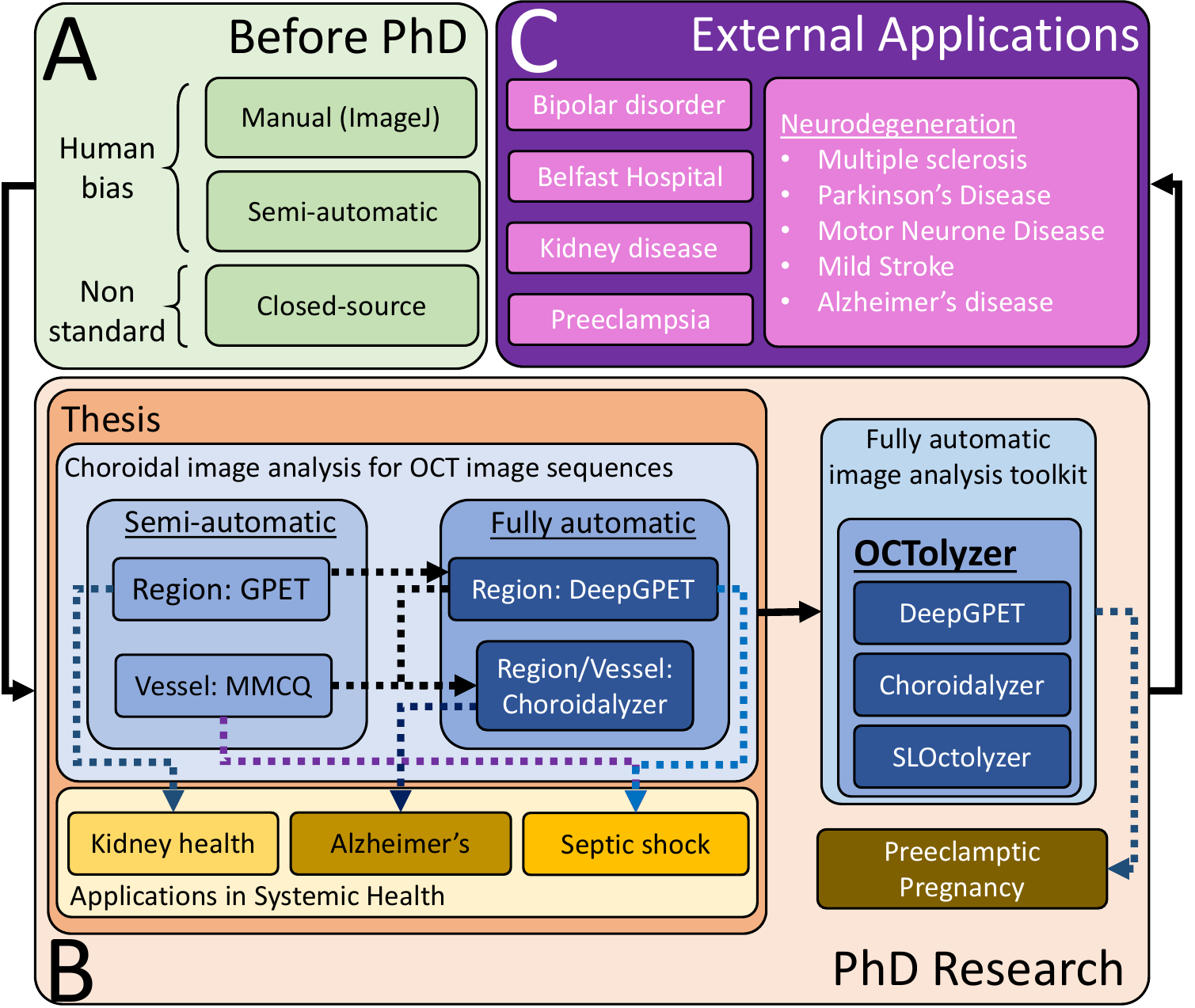}
            \caption[Schematic of work conducted during research period.]{(A) Previous approaches to choroidal image analysis. (B) Work conducted during research period. (C) Past/current/future applications of developed approaches.}
            \label{fig:DISCUSS_PhD}
        \end{figure}

        In this thesis, we described and evaluated the accuracy and reproducibility of several new approaches for measuring the choroid in \acrshort{OCT} image sequences. These approaches address the issue of reproducibility in choroidal image analysis, enabling the research community to better interpret their results when using one of these methods.

        The approaches are freely available online in user-friendly formats, allowing the research community to adopt them for their own research studies, thereby tackling the significant lack of open-source tools for choroid measurement. Fully automatic segmentation methods, such as DeepGPET and Choroidalyzer, eliminate the need for manual intervention or semi-automatic alternatives, do not require domain-specific knowledge or specialised training, and provide consistent definitions and protocols for the segmentation and measurement of the choroidal space. This, in turn, contributes to the standardisation of choroidal measurements in \acrshort{OCT} image sequences.

        Moreover, the quantitative methods developed in this thesis evolved in a series, reflecting the iterative nature of the research process. Specifically, the discussion and recognition of the biases and limitations inherent in each method led to the development of the next method as an improvement, addressing some of the shortcomings of its predecessors. This iterative research process promoted transparency, critical self-reflection, and scientific rigour, ensuring the development of more accurate and reliable choroidal measurements for the research community.

        Finally, we demonstrated the use of these methods in three specific applications of choroidal image analysis in systemic disease: renal health, Alzheimer’s disease risk, and circulatory shock. These applications not only illustrate the potential role of the choroid as a biomarker for systemic disease, but also highlight the importance of accurate measurements of the choroid in understanding this role better. 

    \end{mysection}

    \begin{mysection}[]{The technical work accomplished in this thesis}

        The quantitative methods developed in this work were tested rigorously in our evaluation. We performed population-level comparisons against ground truth labels generated using deterministic, semi-automatic approaches. We also provided transparent and detailed analyses of errors, both quantitatively and qualitatively. In the latter, strict protocols were used to educate manual adjudicators and image graders on image quality, segmentation quality and manual segmentations to ensure consistency (appendix \ref{apdx:protocol_appendix}).
        
        Moreover, we assessed the reproducibility of our methods to provide researchers with the ability to differentiate measurement error from true biological change. For choroidal metrics or applications where effect sizes are likely to be small, such as measuring \acrshort{CVI} between healthy and diseased eyes (2\% -- 6\%) \cite{agrawal2020exploring} or tracking myopia progression with choroidal thickness (20 -- 30 $\mu$m) \cite{breher2019metrological, flores2013relationship}, respectively, it's imperative that clinical measurements are interpreted by accounting for the expected effect size and potential measurement variability \cite{breher2020choroidal}. Fortunately, we have assessed each method's measurement error at the eye-level \cite{engelmann2024applicability}, which allows end-users to better interpret their study results if using these new approaches to measure the choroid.
        
        This thesis has contributed new approaches to the field which are easier to interpret than predictive models which output risk or classification of disease. The approaches introduced and validated in chapters \ref{chp:chapter-GPET} -- \ref{chp:chapter-choroidalyzer} output visual delineation of a particular anatomical tissue, which is something that can be easily assessed for accuracy, and excluded or corrected if desired. Conversely, deep learning-based disease classification models encode any tissue quantification step within their `black-box' architecture as a latent feature which is then subsequently used for prediction. This does not allow for any assessment of the working which contributes to the overall low explainability of any decision made from these models. While ultimately the output segmentations are only one step of the pipeline from bedside to biological interpretation, the visual output of such methods are easier to verify than a continuous risk score of disease.

        \begin{mysubsection}[]{Outline of thesis progression}

            We first developed \acrshort{GPET} (chapter \ref{chp:chapter-GPET}), a semi-automatic approach to general purpose edge tracing which was designed around choroid region segmentation in \acrshort{OCT} B-scans. This allowed flexible edge modelling with human intervention, and removed measurement error present in manual measurements. However, it was dependent on critical pre-processing and manual initialisation which resulted in poor execution time ($\approx$ 35 seconds/B-scan), thus making it unfeasible for large-scale \acrshort{OCT} image analysis. While the underlying theory of the approach was sound, \acrshort{GPET} was not able to be used routinely in research because its semi-automatic pipeline required domain-specific knowledge. This precluded \acrshort{GPET}'s accessibility for those unequipped with the pre-requisite knowledge in image analysis or mathematics.
           
            The human bias, poor execution time and challenging end-user accessibility motivated a more fully automatic approach, DeepGPET (chapter \ref{chp:chapter-deepgpet}), which was developed using deep learning as a fast and significantly more accessible approach to choroid region segmentation. We found DeepGPET to be highly reproducible for \acrshort{SD-OCT} scans both across the macula and also in the peripapillary at the population- and eye-level. Additionally, DeepGPET reduced the execution time for region segmentation by 96\% relative to \acrshort{GPET} and, qualitatively, it was preferred in cases of large disagreement with ground truth labels generated by \acrshort{GPET}. However, using DeepGPET to generate standardised, fovea-centred \acrshort{ROI} measurements of the choroid still required an element of manual intervention to detect the fovea. DeepGPET was also not robust to thick choroids in \acrshort{SS-OCT} data, preventing it's wider use in other OCT data types. Importantly, DeepGPET was unable to segment the choroidal vessels.
    
            For choroid vessel segmentation we developed a semi-automatic procedure \acrshort{MMCQ} for multi-scale vessel enhancement and segmentation (chapter \ref{chp:chapter-mmcq}). \acrshort{MMCQ} was developed as a domain-specific solution to address the heterogeneity of the choroidal vasculature in terms of size and shape, and the poor contrast of which the choroid often suffers. Developing this method also prevented the need to have large amounts of manually labelled data for downstream deep learning model development, given that grading is hard to do unambiguously for a human and does not represent a gold standard. 
            
            We showed that the multi-scale nature of \acrshort{MMCQ}'s enhancement procedure addressed the heterogeneity and contrast issues related to the choroidal vasculature. We also showed that it was reproducible across \acrshort{SD-OCT} and \acrshort{SS-OCT} data, and was able to better preserve the fidelity of the choroidal vasculature than local thresholding approaches such as the Niblack method outlined by Agrawal, et al. \cite{agrawal2016choroidal}. \acrshort{MMCQ} is a semi-automatic, open-source tool  which should require no parameter tuning or manual intervention, and only takes approximately 2 seconds/B-scan. Thus, MMCQ is a suitable replacement for choroid vessel segmentation to popularised approaches like Niblack \cite{sonoda2014choroidal, agrawal2016choroidal}.
            
            At the time of writing, local thresholding of choroidal vessels using the Niblack method is still at the forefront of choroid vessel segmentation. This is despite issues related to agreement among different versions of Niblack's method \cite{wei2018comparison}, improper usage across various themes such as reproducibility \cite{breher2020choroidal}, disease \cite{liu2018choroidal, agrawal2016Koyanagi}, robustness across devices \cite{ma_validation_2024, agrawal2019choroidal}, brightness sensitivity \cite{rosa2024impact}, scale sensitivity \cite{elbay2022comparison} and parameter sensitivity \cite{muller2022application}. All of these contribute to Niblack's variability in providing optimal vessel segmentations of the choroid. 
            
            While we showed that the Niblack method had similar levels of reproducibility to \acrshort{MMCQ} for macular \acrshort{OCT} data on a fixed setting of parameters, Niblack's segmentations were of poorer quality across choroids of varying size. This is a problem which could be solved by tuning Nilack's parameters according to the size of choroidal vasculature, thus injecting subjectivity into the vessel segmentation procedure and making Niblack a human-driven methodology. We demonstrated earlier that the range of acceptable values for Niblack's internal parameters can vary significantly when tuned for optimal segmentation agreement to different manual graders (section \ref{subsec:MMCQ_intro_niblack}).
    
            Regardless of the ability for \acrshort{MMCQ} to provide more precise vessel segmentation than the current reference standard, the method still fell short of providing automated software to the research community. This was because of its multi-stage pipeline which has the ability to be further tuned by the end-user and requiring manual fovea detection to generate standardised measurements of the choroidal vasculature. This unfortunately injects human bias into the segmentation procedure. The ultimate goal was to develop an algorithm capable of producing high quality choroid vessel segmentations, which was parameter-free to the end-user, and worked on a variety of choroids. 
     
            In realising the bias and limitations of GPET, DeepGPET and MMCQ, and identifying a need for a standardised and fully automatic approach to segment and measure the choroid, we developed Choroidalyzer. 
            
            Choroidalyzer has several benefits to the research community, satisfying many of the key element missing from much of the current literature (tables \ref{tab:INTRO_region_methods} -- \ref{tab:INTRO_region_vessel_methods}). Namely, it was developed using a substantially larger dataset than most previous methods, and whose ground-truth labels were generated by semi-automatic methods rather than manual ones, it permits fully-automatic measurement of the choroid for a single B-scan, it is generalisable across different OCT imaging devices and data types, and has been rigorously evaluated to provide interpretable reproducibility metrics for downstream choroidal image analysis in \acrshort{OCT} image sequences.

        \end{mysubsection}

        \begin{mysubsection}[]{Choroidalyzer's key contributions}
            
            Choroidalyzer combined all four stages of choroidal image analysis into a single procedure (figure \ref{fig:CHOROID_schematic}) and was trained on 5,600 OCT B-scans from two market-leading \acrshort{OCT} device manufacturers (Heidelberg Engineering and Topcon) and \acrshort{OCT} data types (\acrshort{SD-OCT} and \acrshort{SS-OCT}). Additionally, the ground truth labels generated from these B-scans used semi-automatic methods rather than manual grading, mitigating any potential human error \cite{maloca2023human}. Besides, manual ground truth labelling was simply not practical given the amount of time it takes for a human grader to annotate choroidal vessels (section \ref{sec:CHOROID_RESULTS_MANUAL}).
    
            Choroidalyzer's functionality suggests its feasibility for large-scale ophthalmic image analysis, given its low CPU execution time (0.3 seconds/B-scan) and demonstrated successful segmentation of large and poor quality choroids (figure \ref{fig:CHOROID_deepgpet}) from \acrshort{SS-OCT}. Choroidalyzer also demonstrated generalisability on an external test set in systemic disease, which performed in parallel with the internal test set, and detailed error analyses showed errors within ground truth labels rather than error on Choroidalyzer alone. The model also outperformed the current reference standard (Niblack for vessels and DeepGPET for region segmentation) used by most in the research community, and was preferred in cases of high error by a qualitative adjudicator.
            
            Choroidalyzer also reported strong levels of reproducibility across \acrshort{SD-OCT} and \acrshort{SS-OCT} data, with examples of major outliers representing challenging cases of poor vessel visualisation and invisible Choroid-Sclera boundary. In particular, it's eye-level reproducibility enables the research community to differentiate measurement error from true biological effect. Additionally, this reproducibility analysis demonstrated Choroidalyzer's ability to process volumetric data and successfully measure the choroid when \acrshort{EDI} mode is toggled on and off, making it an attractive approach for fully automatic analysis of the macular choroid in \acrshort{OCT} image sequences. 

            Choroidalyzer's deterministic output ensures a strict segmentation and measurement protocol is followed, preventing any subjectivity around the definition of the Choroid-Sclera boundary. Although, it may also be possible to tune Choroidalyzer's probabilistic output according to the research problem which may consider different definitions of the Choroid-Sclera boundary, permitting segmentation of the choroid and/or suprachoroidal space (figure \ref{fig:CHOROID_SCS}, further exploration required). Choroidalyzer also outputs a soft \acrshort{CVI} to account for the inherent uncertainty around choroidal vessels in \acrshort{OCT}. This is particularly advantageous over semi-automatic approaches which consider the problem of vessel segmentation a binary one, or still have human bias injected into the segmentation and measurement procedures from end-users performing manual pre-processing or selecting optimal parameters per B-scan \cite{sonoda2014choroidal, agrawal2016choroidal}.
      
            While these attributes demonstrate the rigour in Choroidalyzer's model development and performance, it is not without it's limitations. The model struggled with peripapillary \acrshort{OCT} B-scans (figure \ref{fig:CHOROID_peri_deepgpet}) and can presently only process single B-scans inputted as an image, requiring minimal interfacing with Python programming for the end-user. Additionally, it's lack of validation across examples related to retinal pathology highlight the need for external validation. Nevertheless, we anticipate the significant benefit this toolkit will have for choroidal image analysis in \acrshort{OCT} image sequences in systemic disease.
            
        \end{mysubsection}
        
        \begin{mysubsection}[]{From imaging device to choroid measurement}
        
            \begin{figure}[tbp]
                \begin{adjustwidth}{-1in}{-1in}
                \centering
                \includegraphics[width=0.75\linewidth]{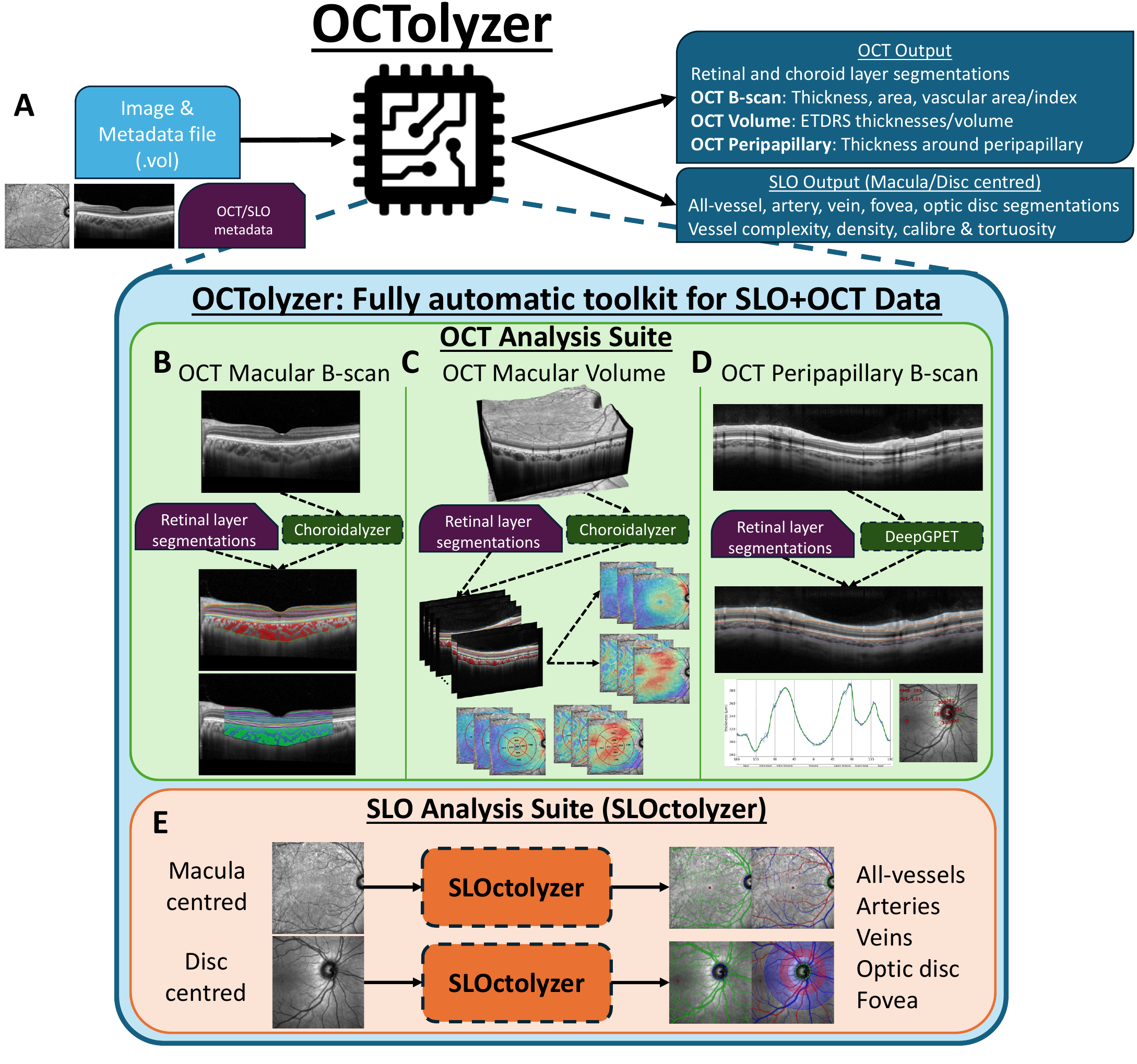}
                \end{adjustwidth}
                \caption[OCTolyzer's toolkit.]{OCTolyzer's pipeline \cite{burke2024octolyzer}. (A) OCTolyzer expects a file containing the \acrshort{OCT} (and \acrshort{SLO}) image data, and necessary \acrshort{OCT} file metadata. (B -- E) A deeper dive into OCTolyzer's internal structure. (B -- D) \acrshort{OCT} analysis suite: analysis pipelines for when OCTolyzer detects a single/radial macular B-scan (B), macular volume scan (C) or peripapillary B-scan (D). (E) \acrshort{SLO} analysis suite for macula- or disc-centred \acrshort{SLO} localiser images, which accompany \acrshort{OCT} capture, using SLOctolyzer \cite{burke2024sloctolyzer}.}
                \label{fig:DISCUSS_OCTolyzer}
            \end{figure}
    
            A significant limitation of Choroidalyzer was that it worked on a single B-scan presented as an image file, and was not equipped to handle proprietary file formats or different scan patterns. To ensure methods like Choroidalyzer and DeepGPET are accessible to the general researcher and require no technical background in image analysis or indeed computer science, Choroidalyzer and DeepGPET have been wrapped into a fully automatic analysis toolkit for \acrshort{OCT} data, OCTolyzer \cite{burke2024octolyzer} --- the first of its kind. Figure \ref{fig:DISCUSS_OCTolyzer} shows a schematic diagram of this toolkit.
            
            OCTolyzer is a Python-based toolkit which (currently) supports processing $\verb|.vol|$ RAW files (native to Heidelberg Engineering) which contain the \acrshort{OCT} (and localiser \acrshort{SLO}) image data and necessary file metadata. Due to lack of data availability at the time of development, other file formats, such as \verb|.fda|, \verb|.fds|, \verb|.img| and \verb|.oct| are not supported. Additionally, \verb|.e2e| (also native to Heidelberg Engineering) is not supported because current python-based file readers \cite{morelle2023eyepy, graham2024octconverter} are unable to locate the necessary pixel length-scales for converting from pixel space into physical space. However, it would be desirable for OCTolyzer to be fully compatible with DICOM (\verb|.dcm|) format for purposes of vendor neutrality. It is hoped that continued development from the author or future collaborators may facilitate this. 
    
            OCTolyzer has an \acrshort{OCT} analysis suite which supports processing of macular single-line/radial-line B-scans, macular volume scans and peripapillary B-scans. OCTolyzer can automatically process and measure the choroid using the methods discussed in section \ref{subsec:INTRO_MEASURE_ROI}. As previously discussed in sections \ref{subsec:chp4_limitations} and \ref{subsec:ch5_CHOROID_limits}, Choroidalyzer is used for processing macular \acrshort{OCT} data, while DeepGPET is used for processing peripapillary \acrshort{OCT} data.
            
            Although outside the scope of this thesis (figure \ref{fig:DISCUSS_PhD}), OCTolyzer also has the ability to measure the retina through layer segmentations saved within the input file metadata, since most \acrshort{OCT} imaging devices have built-in software to process retinal layer segmentations. Moreover, OCTolyzer is equipped to segment and measure the retinal vessels on the en face localiser \acrshort{SLO} using OCTolyzer's \acrshort{SLO} analysis suite, SLOctolyzer \cite{burke2024sloctolyzer}. Thus, OCTolyzer is capable of converting a raw proprietary \acrshort{OCT} file into a set of clinically meaningful and reproducible measurements of the choroid, alongside cross-sectional and en face measurements of the retina. The work on developing OCTolyzer and SLOctolyzer has resulted in two articles \cite{burke2024octolyzer, burke2024sloctolyzer}. At the time of writing (December 2024), one has been published \cite{burke2024sloctolyzer} and the other is in peer review \cite{burke2024octolyzer}. The codebases for OCTolyzer and SLOctolyzer have been released as open-source and can be found \href{https://www.github.com/jaburke166/OCTolyzer}{here} and \href{https://www.github.com/jaburke166/SLOctolyzer}{here}, respectively.

            The development and subsequent release of toolkits like OCTolyzer and SLOctolyzer has started to streamline much of the \acrshort{OCT} image analysis conducted at the author's home research institution (University of Edinburgh). Researchers (Miracle Ozzoude, Borja Marin, Francesca Pentimalli) and technicians (Charlene Hamid, Rosie Keane) are currently using these tools for their own applications, such as in neurodegeneration, bipolar disorder, mild stroke, Alzheimer's disease and preeclampsia. There has also been interest outside of the research institute in using these tools, such as with fellow collaborators at University College London and at Queen's University Belfast. This not only highlights the continued interest for the choroid (and retina) in systemic disease, but also demonstrates the potential utility that open-source methods OCTolyzer (and SLOctolyzer) can and will have to the research community.

        \end{mysubsection}

        \begin{mysubsection}[]{Closing the gap between research and clinical practice}

            There is a growing recognition in healthcare-oriented artificial intelligence (AI) of the need for publicly available methods that have undergone rigorous evaluation \cite{ciobanu2024critical}. The distinct lack of such methods contributes to the significant gap between academic research and clinical AI applications \cite{weissler2021role}. 

            Despite the abundance of deep learning and AI methods applied to healthcare problems, there is a distinct lack of transparency in the field surrounding their reproducibility, validation strategies and publicly available source code. McDermott, et al. \cite{mcdermott2021reproducibility} surveyed these themes for 511 machine learning studies presented at machine learning conferences between 2017 and 2019, of which 211 were in the context of healthcare. Of these 211, they found that only 44\% reported their measurement variability and 21\% made their code and model available. In addition to model availability and reproducibility, external validation has also become a key element of algorithm validation \cite{cabitza2021importance}.

            Model and measurement reproducibility are key to understanding measurement variability and model reliability, and are core elements of a method's validation. The work in this thesis endeavoured to follow current guidelines and reporting standards which exist around machine learning reproducibility \cite{pineau2021improving}. We provided sufficient model and hyperparameter specification for model reproduction, as well as detailed description of our training procedures to help facilitate reproducibility and transparency for both DeepGPET and Choroidalyzer. Additionally, we assessed the measurement reproducibility of these tools using \acrshort{SD-OCT} and \acrshort{SS-OCT} at both the population-level and eye-level. However, we were unable to share the datasets used for model development, and nor were there publicly available datasets to benchmark model performance with.

            Open-source tools can play a critical role in helping bridge the gap between research-only software and their use-case in research studies by making models more accessible to clinicians and researchers \cite{harish2022open}. This is why we have made a concerted effort to make the tools developed in this PhD publicly available, which we hope will facilitate future collaboration to help externally validate, extend and maintain. Importantly, without proper long-term management and maintenance, their impact may be ephemeral at most \cite{harish2022open}.
            
            While the approaches introduced during this research period will make broader impact to the research community by helping improve our understanding of the choroid in disease, their entry into clinical practice would reach the clinician and patient more directly. However, introducing automated tools into clinical practice is no easy feat, particularly those using AI given their black-box nature resulting in poor interpretability \cite{reddy2022explainability, weissler2021role}. Nevertheless, Choroidalyzer and DeepGPET advantageously only provide visual, anatomical labelling rather than disease detection which removes a level of abstraction that the latter models encode in their decision making.

            The ability for a method to generalise across out-of-sample distributions is a key concern among the healthcare community \cite{liu2019comparison}, contributing to the difficulty of adopting and integrating these technologies within the medical community. The current landscape is dominated by academic research where methods are often developed and tailored to specific in-house datasets, are closed-source or held behind a membership-service, and thus may not generalise well to broader clinical settings \cite{ciobanu2024critical}. For example, Liu, et al. \cite{liu2019comparison} found a distinct lack of external validation of deep learning models for disease detection, despite finding their performance equivalent to that of healthcare professionals. While we acknowledge the importance of external validation, this was only performed with Choroidalyzer on data related to one systemic disease --- containing examples with relatively normal retinae --- on prospectively collected research data. Thus, further out-of-sample validation is an important next step, particularly from external datasets which were not collected in-house and which also reflect image acquisition of real-world clinical care.

            Overcoming these barriers around reproducibility, validation and model availability requires significant interdisciplinary collaboration from researchers, clinicians, regulators, and technicians. However, the research community has acknowledged this and attempted to report guidelines for the use of AI in healthcare \cite{liu2020reporting}. For example, the CONSORT-AI \cite{liu2020reporting} guidelines aim to improve the reporting standards, explainability and reproducibility of AI in healthcare and hopes that these will facilitate increased trust in clinical practice. A clear framework for AI in clinical trial reports may also help regulatory bodies better evaluate these models as they attempt to keep pace with the ever-evolving nature of AI \cite{van2022clinical}. 

            The development of these methods in a healthcare environment is not just about creating tools but also about fostering a collaborative environment where researchers, clinicians, and technicians can work together to conceptualise, develop, refine and validate models \cite{boag2024algorithm}. This is a sentiment which the work during this research period aimed to align with. The methods developed in this thesis were developed not only in consultation with \acrshort{OCT} imaging technicians, but also clinicians who were available for crucial image annotation and model adjudication. The use of AI in healthcare must leverage multi-disciplinary teams and collaboration throughout the developmental process \cite{bertagnolli2023advancing}. This cross-functional effort is what drives continuous improvement and adaptation to the evolving needs of healthcare systems.

            Choroidal image analysis in \acrshort{OCT} is still an evolving field, and we hope that tools like Choroidalyzer will motivate other researchers to consider the importance of distributing, evaluating and validating transparent, open-source methods. In the field of choroid analysis in \acrshort{OCT} --- one which has grown substantially with the advancements in AI, \acrshort{OCT} technology and the observed associations between the choroid and systemic disease --- standardisation of choroidal measurements has become a necessity. Fully automatic, deep learning-based tools like Choroidalyzer may help mitigate measurement error introduced from manual or semi-automatic methods, and make a positive step toward standardisation of ocular parameters associated with the choroid. With additional validation from external sources, we anticipate Choroidalyzer to make further impact to the field.

        \end{mysubsection}

    \end{mysection}

    \begin{mysection}{Current limitations}

        The primary limitation of this work surround the datasets used for model development which were all related to systemic health from prospective studies. Thus, not only did the tools developed during the research period did not undergo any real internal or external validation on ocular pathology, but the development of all models was conducted on high-quality, idealised datasets whose image acquisition was conducted by a trained and experienced image technician. 
        
        Real-world data from routine clinical care is often marred by virtue of unpredictable or uncontrollable elements such as patient cooperation, camera operator experience and incidental findings. In contrast, research data from prospective studies is collected within a controlled clinical environment, which can often correct for the aforementioned elements impacting real-world data. Thus, it would be a natural next step to quantify the performance of Choroidalyzer on real-world data from clinical care.
        
        Moreover, the majority of researchers in ophthalmology will be investigating the choroid from \acrshort{OCT} in chorioretinopathy, and the generalisability of these models on these datasets are currently unknown, which represents a significant limitation. This is especially pertinent to extreme cases of choroidal thinning due to pathological myopia \cite{ohno2021imi}, extreme choroidal enlargement such as central serous chorioretinopathy \cite{fung2023central}, or conditions affecting the photoreceptors and \acrshort{RPE} leading to signal hypertransmission into the choroidal space, such as geographic atrophy \cite{vallino2024structural}.

        However, organs are not closed systems, and the emerging evidence of the choroid's role in systemic disease motivated much of the clinical applications (chapter \ref{chp:chapter-applications}). Thus, the rigorous evaluation on the research data related to systemic health presented in this thesis highlight the feasibility of Choroidalyzer and DeepGPET being used for such applications by the research community. Thus, we anticipate them to still be of significant value to the nascent field of ophthalmology for systemic disease, or ``oculomics'' \cite{wagner2020insights, wagner_retinal_2023, wagner2022alzeye}.

        A significant limitation associated with our clinical applications is their relatively small sample sizes. In some cases, larger-scale image analysis was precluded due to historical imaging protocols capturing \acrshort{OCT} images without \acrshort{EDI} mode resulting in poor choroid visualisation. This had a direct impact on the amount of \acrshort{OCT} data that was available for image analysis and downstream statistical analysis, and thus impacted the statistical power to which we could draw our conclusions with. This was particularly relevant to the applications in \acrshort{CKD} (section \ref{sec:ch_app_sec_ckd}) and Alzheimer's disease (section \ref{sec:ch_app_sec_prevent}), as much of the \acrshort{OCT} volume data was not usable because \acrshort{OCT} volume scans were collected without \acrshort{EDI} mode activated. Consequently, statistical analysis was simplified to fovea-centred single line-scans.

        This ultimately introduced an element of selection bias, since the usability of non-\acrshort{EDI} \acrshort{OCT} for choroid analysis is correlated with the size of the choroid, i.e. thicker choroids imaged with non-\acrshort{EDI} \acrshort{OCT} will be at higher risk of poor Choroid-Sclera boundary visualisation. Moreover, it's likely that poor \acrshort{OCT} acquisition is associated with degree of sickness (as shown by the feasibility of \acrshort{OCT} capture in \acrshort{ITU} from section \ref{sec:ch_app_sec_shock}). However, while this thesis has been impacted by \acrshort{OCT} protocols neglecting the choroid during \acrshort{OCT} capture, the work has motivated a change of protocols at the Edinburgh Imaging Facility (Royal Infirmary of Edinburgh) to consider \acrshort{EDI} mode as an important attribute in many of the ongoing research studies related to preeclampsia, Alzheimer's disease, stroke and kidney disease. Moreover, better training is being provided so that imaging technicians are well equipped with the ability to image deep into the eye and visualise the choroid.
    
        Additionally, at the time of model development there were no publicly available datasets for choroid segmentation in either systemic disease or retinal pathology, which precluded benchmarking performance. Benchmarking tools are essential in research as they enable objective performance comparison between different methods, ensuring transparency and reproducibility by providing a standardised framework for evaluating performance and replicating results. While this does not yet exist in \acrshort{OCT} choroid analysis, it does for other popular modalities like retinal \acrshort{CFP} which have many publicly available datasets such as the DRIVE \cite{staal2004ridge}, ORIGA \cite{zhang2010origa}, STARE \cite{hoover2000locating}, and RIM-ONE \cite{fumero2011rim}.
    
        While Choroidalyzer was developed to be robust to both \acrshort{SD-OCT} and \acrshort{SS-OCT} data, older methods were developed for \acrshort{SD-OCT} data which resulted in DeepGPET's poor generalisation to large choroids in \acrshort{SS-OCT} data (figure \ref{fig:DEEPGPET_topcon_large_small}). Even with Choroidalyzer's robustness to both OCT data types, only two \acrshort{OCT} imaging device manufacturers were used (Heidelberg Engineering, Topcon), and thus the performance of Choroidalyzer on data from other device manufacturers, such as Zeiss, Optovue and Nidek is unknown. Nevertheless, Topcon and Heidelberg are market-leading \acrshort{OCT} device manufacturers and providing tools which are fully compatible with these devices is an important advance.
            
    \end{mysection}
        
    \begin{mysection}[]{Future work}

        \begin{mysubsection}[]{Large-scale analysis, external validation and benchmarking}

            Choroidalyzer's accessibility can enable reproducible and clinically meaningful analysis of the choroid which, coupled with its strict measurement protocol, permits standardisation of choroidal measurements. This has potential to improve consistency in and impact of future research studies which adopt Choroidalyzer. Additionally, Choroidalyzer presents massive opportunities for large-scale automation of choroidal image analysis. Applications presented in this thesis were on relatively small datasets, but future work would seek to apply Choroidalyzer to large-scale, real-world datasets from clinical care such as AlzEye \cite{wagner2022alzeye} or UK Biobank \cite{sudlow2015uk}. However, while the development of OCTolyzer mitigates any requirement of Python programming and computer science, challenges remain surrounding efficient, large-scale data extract protocols from different machine, as well as Choroidalyzer's lack of validation on these kinds of datasets.

            Future work would also seek external validation of Choroidalyzer on datasets related to retinochoroidal pathology to evaluate its generalisability to abnormal retinae. A recent dataset of publicly available \acrshort{OCT} B-scans and corresponding segmentations related to macular holes was published, which included choroid region segmentation labels \cite{ye2023oimhs}. Not only could this be used to measure Choroidalyzer's generalisability, but could also enable transparent benchmarking of its performance for future iterations (or different models entirely) to be evaluated against in the public domain. Additionally, Choroidalyzer should be re-trained with the original dataset plus these additional 3,859 \acrshort{OCT} images, with more superior, transformer-based architectures \cite{wang2024pgkd, wen2024transformer} to leverage more state-of-the-art AI techniques in the field of computer vision. 
            
            In time, the availability of public datasets of \acrshort{OCT} choroid data will improve, as exemplified by Arian, et al. \cite{arian2023automatic}, who have released their \acrshort{OCT} data (albeit without annotations) related to diabetic retinopathy and pachychoroid spectrum disorder. This dataset, and others in the future, will permit external validation and benchmarking of open-source algorithms such as the ones developed in this thesis. With external validation comes re-evaluation of the model's limitations and identification of where improvements should be made. This cyclical process will hopefully continue to facilitate accurate and reliable measurements of the choroid, thus continuing to improve the standardisation of ocular parameters for choroidal image analysis in \acrshort{OCT} image sequences.
            
        \end{mysubsection}

        \begin{mysubsection}[]{Towards choroidal vessel standardisation}

            As discussed in chapter \ref{chp:chapter-mmcq}, choroid vessel segmentation is a challenging problem, and there is currently no gold standard. Choroidalyzer outputs a soft \acrshort{CVI} to account for the oblique presentation of vessels and ambiguity of their vessel walls seen in \acrshort{OCT} data. To begin to address this ambiguity it would be pertinent to quantify it, agreed upon by experts in the field (from both those who acquire the images, i.e. technicians, and those who interpret them, i.e. clinicians). We showed in chapter \ref{chp:chapter-mmcq} that choroid-derived vessel measurements may have the potential to vary significantly across raters using a small sample of B-scans, but a more systematic approach would be to have several experienced raters tackle a larger and more diverse dataset. 
        
            Fortunately, Choroidalyzer outputs vessel segmentation masks as probabilistic maps which can be binarised given a threshold, if desired. Thus, future work would consider asking multiple experts to apply Choroidalyzer in a large, varied \acrshort{OCT} image set and select the optimal threshold (using an analogue scale from 0 to 1). We hypothesise that the distribution of selected thresholds across raters would differ significantly, indicating that human ground truth labels are poor reference standards to benchmark automatic models for choroid vessel segmentation against, or train with. Laurik-Feuerstein, et al. \cite{laurik2022assessment} did a similar procedure with colour fundus photography in the context of ground-truth label generation for image quality labels. 
            
        \end{mysubsection}

        \begin{mysubsection}[]{Context-dependent image quality}

            As with any large-scale image analysis pipeline, the ability to reject images based on image quality or segmentation accuracy is crucial for forming a meaningful dataset of clinical measurements. \acrshort{SNR} estimates built into commercial \acrshort{OCT} imaging devices do not characterise the quality of the image in the context to which it will be analysed. Such context-dependent image quality methods are essential in large-scale ophthalmic image analysis, such that automatic rejection plays a precursory role in the image analysis pipeline. 
        
            However, definitions for image quality and taxonomy generation require effective collaboration from clinicians, technicians and researchers because of the different, but plausible, definitions surrounding quality. For example, quality with respect to choroid image segmentation may be oriented around clear visualisation of the Choroid-Sclera boundary and vessels, while quality with respect to chorioretinopathy may simply be the ability to observe the disease manifested on the \acrshort{OCT} scan. Therein lies the significant challenge which requires technicians, researchers and clinicians collaborating on a taxonomy that is context-dependent. Encouragingly, such context-dependent quality methods are currently being developed for \acrshort{CFP}s \cite{hamid2024finequal}.
            
        \end{mysubsection}

        \begin{mysubsection}[]{Choroid depth inference from en face imaging}
        
            With the availability of methods like Choroidalyzer enabling large-scale \acrshort{OCT} choroid image analysis, choroid quantification across the macula can be done reasonably fast across large volumes of data. Advancements in deep learning have enabled the inference of three dimensional information from two dimensional input, such as depth-maps for natural images \cite{ming2021deep}. In ophthalmic images, this has also been attempted \cite{sun2023retinal}. Sun, et al. \cite{sun2023retinal} was able to use en face \acrshort{CFP} and \acrshort{SLO} images to estimate cross-sectional retinal thickness maps across the macula, typically measured from \acrshort{OCT}. This could enable more scalable image capture to take place where less focus is needed on \acrshort{OCT} capture which is typically more challenging and time-intensive than \acrshort{CFP}. 
        
            Translating this task to estimating choroid thickness maps across the macula using a combination of \acrshort{CFP} and \acrshort{SLO} images would also have significant benefit. Importantly, the choroid has historically been poorly imaged, and it is likely that vast volumes of historical \acrshort{OCT} data have been captured without \acrshort{EDI} mode activated, or before the advent of \acrshort{SS-OCT}. Thus, there is potential to recover lost information on the choroid from large \acrshort{OCT} image sets which have been plagued from historic data collection, such as the UK Biobank's focus on conventional \acrshort{SD-OCT} \cite{sudlow2015uk}. This is also applicable in research institutions where protocols for \acrshort{OCT} capture neglected the choroid, such as in the initial PREVENT dementia and \acrshort{OCTANE} \acrshort{OCT} imaging protocols.
            
        \end{mysubsection}
        
    \end{mysection}

    \begin{mysection}[]{Concluding remarks}

        This thesis has addressed several key issues within the field of choroidal image analysis for \acrshort{OCT} image sequences, namely around accuracy, reproducibility and accessibility. We developed several tools of which the latest, Choroidalyzer, now enables end-to-end choroidal analysis in \acrshort{OCT} images. Ultimately, Choroidalyzer is the culmination of the previous methods introduced in this research period and was developed to equip the general researcher with the ability to measure the choroid more accurately, consistently and reliably. Ultimately, we hope the dissemination of tools like Choroidalyzer into the research community will facilitate more accurate and reliable measurements of the choroid than existing tools, helping aid interpretability in future research studies. 

        A major deliverable of these tools was their public release in user-friendly frameworks for end-to-end choroidal analysis. This was to promote accessibility, ensuring that researchers and clinicians alike can adopt these off-the-shelf approaches to measure the choroid in their own studies. 
        
        Importantly, these tools remove the need for manual and semi-automatic measurement of the choroid which, due to the distinct lack of available and fully automatic methods, are still the current reference standard despite their subjective approach to segmentation. This open-source approach is key to standardising choroidal measurements.

        This research has also demonstrated the application of choroidal analysis in systemic disease, continuing to provide evidence of the links between the microvasculature of the eye and renal health, brain health and critical illness. This illustrates the potential importance of \acrshort{SS-OCT} and \acrshort{EDI-OCT} imaging of the choroid outside ocular health. 

        Although current limitations remain in their generalisability to retinal pathologies and validation on real-world data from clinical care, future work aims to address this and benchmark their performance. Regardless, these contributions are anticipated to play a crucial role in advancing large-scale choroidal image analysis and accelerate progress in the nascent field of oculomics.

    \end{mysection}
        
\end{mychapter}

\bibliography{biblio.bib}

\myappendix

\setcounter{figure}{0}
\renewcommand{\thefigure}{S\arabic{figure}}
\setcounter{table}{0}
\renewcommand{\thetable}{S\arabic{table}}

\begin{mychapter}[]{Protocol Documentation} \label{apdx:protocol_appendix}

\begin{mysection}[]{OCT B-scan definitions} \label{apdx:definitions}

    Below are a list of definitions of various landmarks related to the choroid, as seen on an OCT B-scan. The definitions relate to the features on the B-scan, rather than the biology. See figure \ref{fig:APPDX_annot_bscan} for an exemplar OCT B-scan with regions of the choroid labelled.
    
    \begin{figure}[tb]
        \centering
        \includegraphics[width=0.8\textwidth]{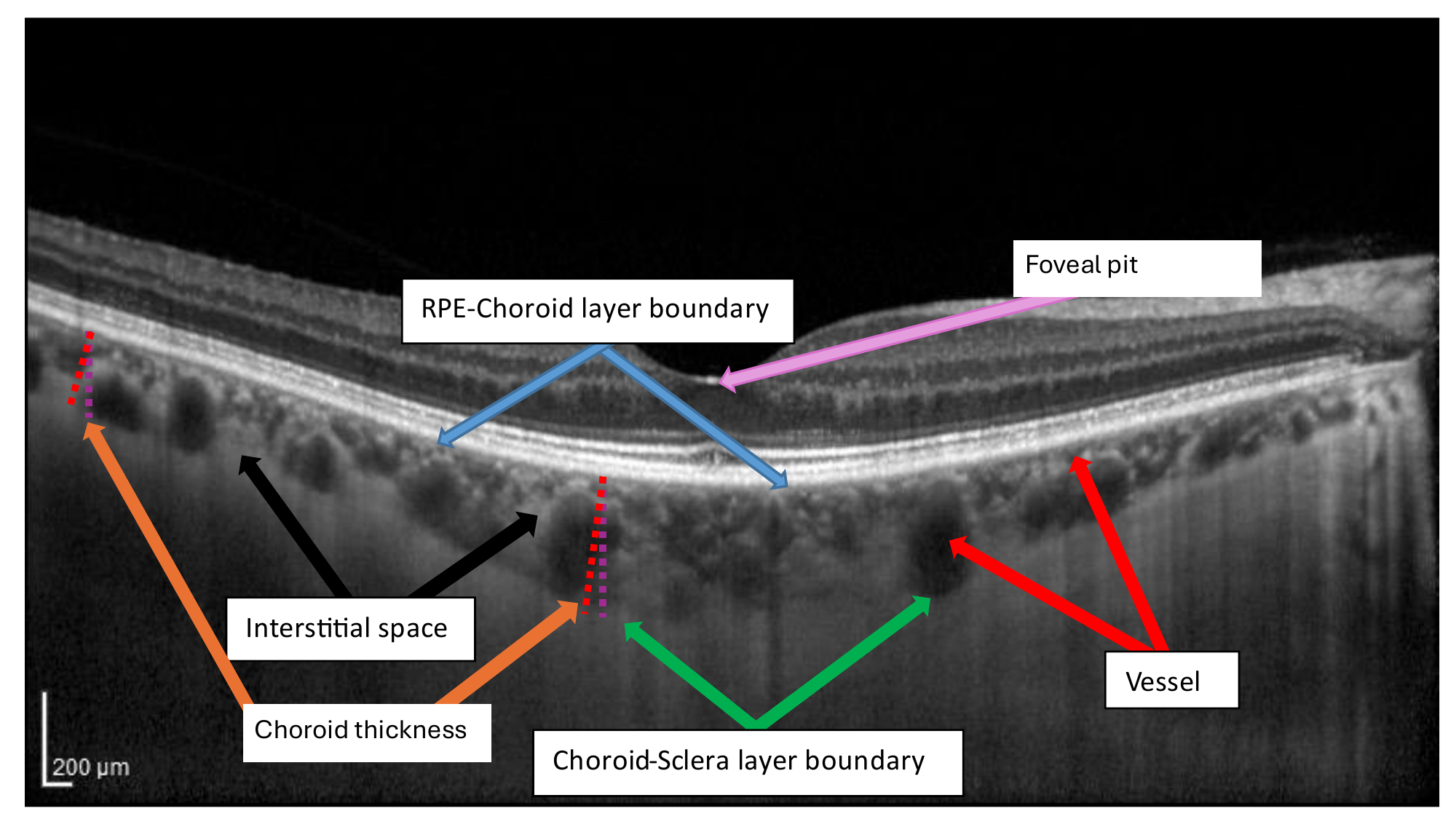}
        \caption[Annotated \acrshort{OCT} B-scan.]{OCT B-scan with arrows directed at various features of the choroid.}
        \label{fig:APPDX_annot_bscan}
    \end{figure}
    
    \begin{itemize}\setlength\itemsep{0em}
        \item \textbf{\acrshort{RPE}-Choroid boundary}: The lower surface of the junction between the hyperreflective \acrshort{RPE} layer and Bruch's membrane complex (figure \ref{fig:APPDX_annot_bscan}, blue).
        
        \item \textbf{Choroid-sclera boundary}: The upper surface of the junction between the choroid and sclera. A predominantly smooth and concave line posterior to the posterior-most choroidal vessels (figure \ref{fig:APPDX_annot_bscan}, green).

        \item \textbf{Choroid thickness}: This measures a straight-line extending from the \acrshort{RPE}-Choroid boundary to the Choroid-Sclera boundary, measured locally perpendicular to the \acrshort{RPE}-Choroid boundary (figure \ref{fig:APPDX_annot_bscan}, red-dashed line), not vertically (figure \ref{fig:APPDX_annot_bscan}, purple-dashed line). 
        
        \item \textbf{Choroidal space}: Region between the \acrshort{RPE}-Choroid and Choroid-Sclera boundaries. 
        
        \item \textbf{Choriocapillaris}: A small strip below and parallel to the \acrshort{RPE}-Choroid boundary, which extends across the choroid seen in OCT B-scans. Typically $\approx$3 pixels deep for Heidelberg Engineering Spectralis \acrshort{OCT}1 devices. This is typically included in segmentations as a thin strip below the \acrshort{RPE}-Choroid boundary (figure \ref{fig:APPDX_chor_defn}(D)).
        
        \item \textbf{Choroidal medium/large vessel}: Large, dark and irregularly shaped and sized blob seen with lower intensity relative to locally adjacent and higher intensity strips in the choroid. This is located within the Sattler and Haller’s layers (figure \ref{fig:APPDX_chor_defn}(A)).
        
        \item \textbf{Choroidal small vessel}: Smaller, but still irregularly shaped and sized blobs than larger choroidal vessels and appear below the choriocapillaris (figure \ref{fig:APPDX_chor_defn}(A)). These smaller vessels are darker in intensity than adjacent regions, but their contrast is much poorer than larger choroidal vessels.
        
        \item \textbf{Choroid vessel wall/boundary}: Edges of choroidal vasculature are inherently more difficult to define as they blend into adjacent interstitial fluid – their exact definition is image-dependent based on local contrast, but is generally brighter than the centre of the vessel itself (figure \ref{fig:APPDX_chor_defn}(C)). 

        \begin{figure}[tb]
            \centering
            \includegraphics[width=\textwidth]{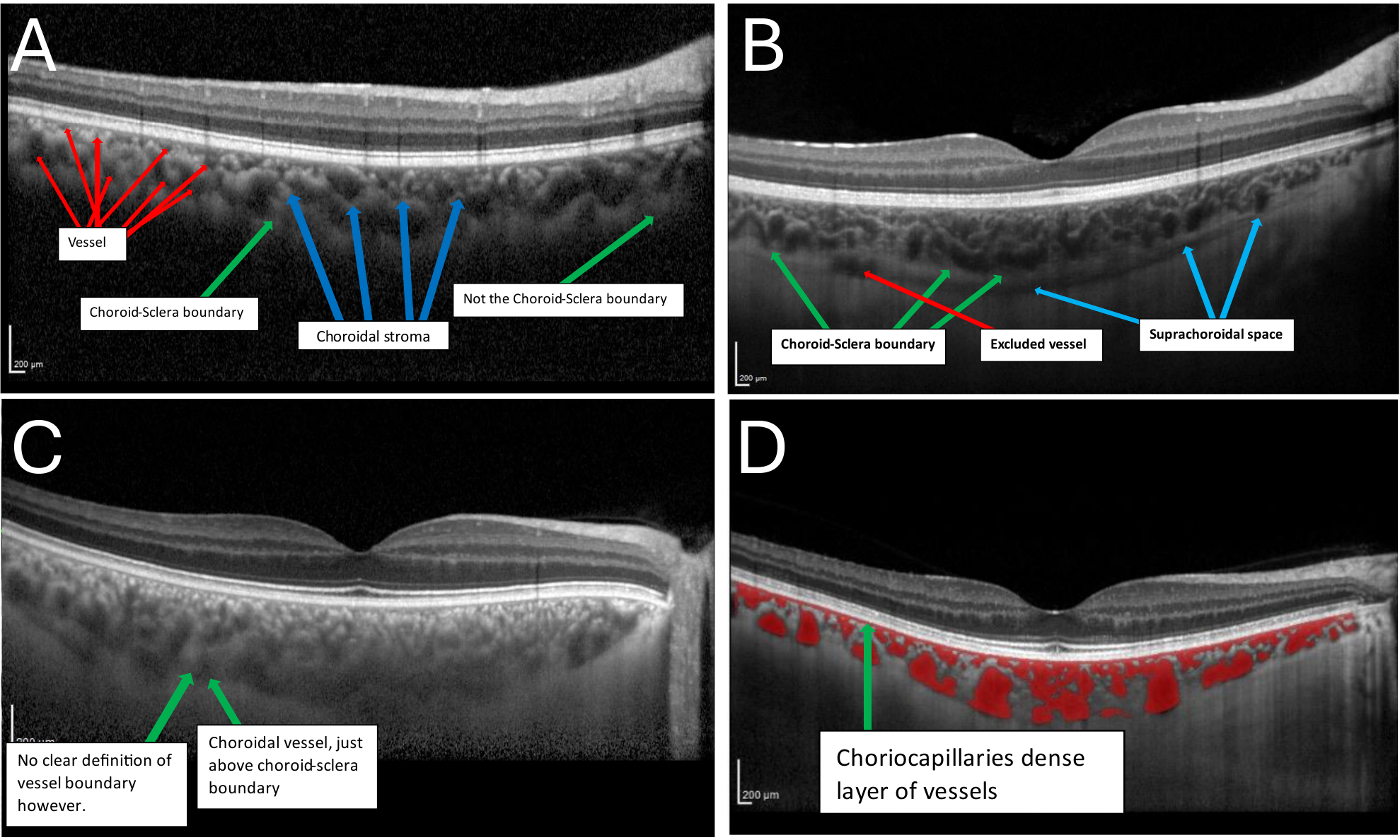}
            \caption[Collection of annotated \acrshort{OCT} B-scans outlining region and vessel landmarks.]{(A) OCT B-scan with arrows indicating Choroid-Sclera boundary (green), vessels (red) and stroma (blue). (B) OCT B-scan with visible suprachoroidal space (blue) and specification of Choroid-Sclera boundary (green) and excluded vessel (red). (C) Large choroid with poor contrast vessel visualisation (green arrow). (D) OCT B-scan with vessel segmentation overlaid with choriocapillaris band (green arrow).}
            \label{fig:APPDX_chor_defn}
        \end{figure}
        
        \item \textbf{Choroidal vasculature}: Non-uniform distribution of irregularly sized and shaped blobs constituting the choriocapillaris, medium/large and small vessels. The vasculature can suffer from poor contrast and visualisation, making vessel walls difficult to define, even for larger vessels (figure \ref{fig:APPDX_chor_defn}(C)).
        
        \item \textbf{Choroidal stroma/interstitial space}: Brightly illuminated regions distributed around choroidal vasculature in the choroidal space (figure \ref{fig:APPDX_chor_defn}(A)).
        
        \item \textbf{Suprachoroid}: A potential space posterior to the Choroid-Sclera boundary, and is visualised as a faint, homogeneous strip parallel with the Choroid-Sclera boundary between the brighter sclera and heterogeneous (in intensity) choroidal space. This, and any vessels traversing it, is \textbf{not} considered part of the choroidal space and should be avoided during segmentation or adjudication (figure \ref{fig:APPDX_chor_defn}(B)).

        \item \textbf{Foveal pit}: Pixel of highest illumination at the deepest point of the foveola depression in the B-scan (figure \ref{fig:APPDX_annot_bscan}, pink), typically vertically-aligned with a ridge forming at the anterior point of the outer retina.
    \end{itemize}
  
\end{mysection}

\begin{mysection}[]{Equipment check} \label{apdx:equipment_protocol}

    Be it adjudication or segmentation, the following list describes essential equipment. Figure \ref{fig:APPDX_setup_screen} shows the ideal monitor and zoom setting for viewing an OCT B-scan for adjudication or segmentation.
    \begin{figure}[!b]
        \centering
        \includegraphics[width=0.8\textwidth]{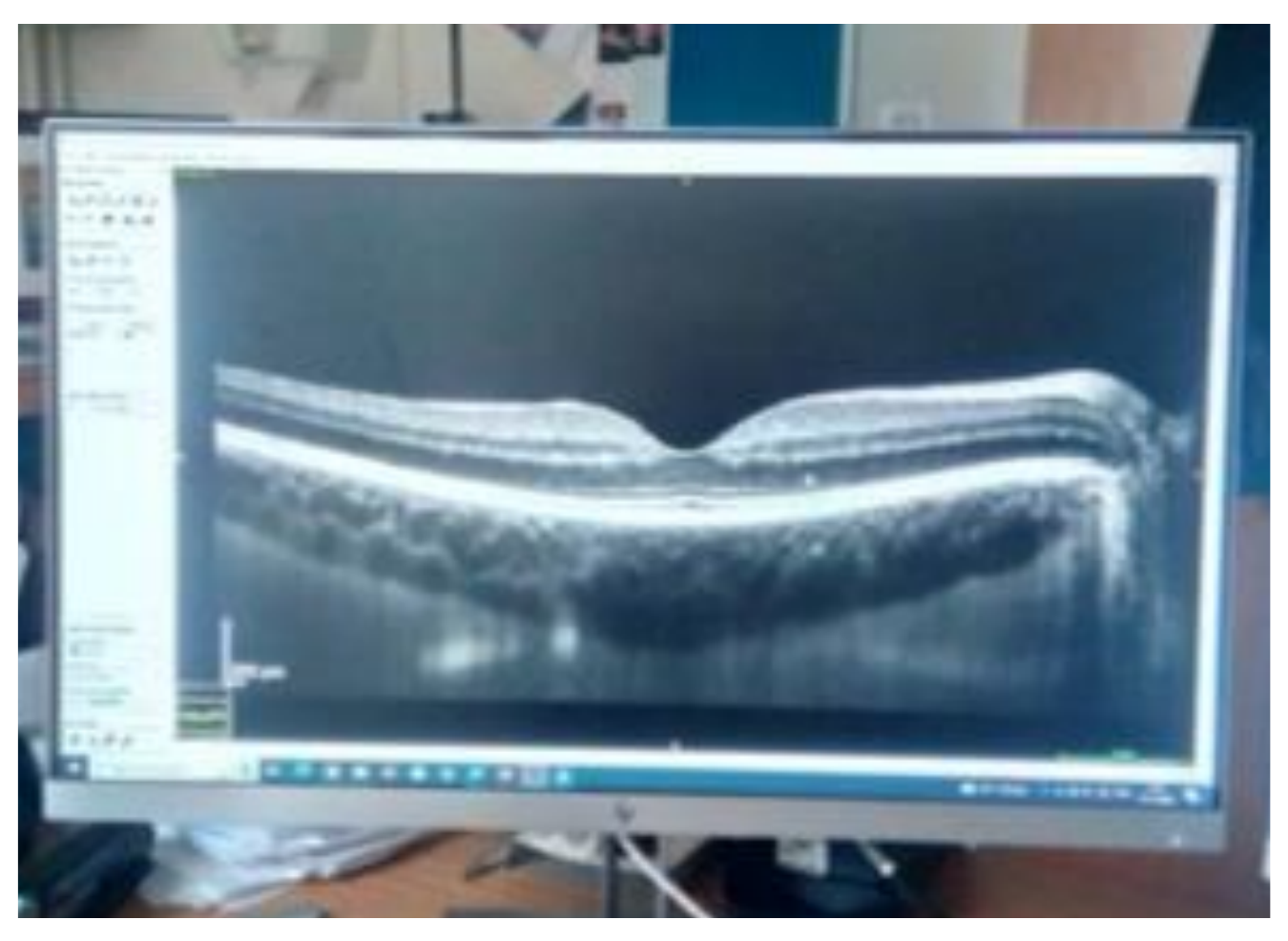}
        \caption[Example set-up for manual adjudication/segmentation.]{Example setup with ITK-Snap open covering the entire screen of a standard HP monitor. Image courtesy of Dr. Ian J.C. MacCormick.}
        \label{fig:APPDX_setup_screen}
    \end{figure}    
    \begin{itemize}\setlength\itemsep{0em}
        \item You will need a separate mouse and mouse pad, not a laptop trackpad.
        \item You will need a large monitor with high resolution (greater than the standard 1366 $\times$ 768 pixel resolution, i.e. 1920 $\times$ 1080 pixels) so that you can ideally see the whole B-scan and detailed segmentation together (figure \ref{fig:APPDX_setup_screen}).
        \item Two monitors together are also ideal for pairwise, visual comparisons, but not strictly necessary.
        \item Use the zoom function at your discretion for assessing smaller-scale regions.
        \item Perform adjudication in a darkly lit room, for example, the lights off on a dull day without much sunlight coming in.
    \end{itemize}
    
\end{mysection}

\begin{mysection}[]{OCT B-scan quality guide} \label{apdx:quality_protocol}

    Below is a list of definitions for each scale when assessing the quality of the choroidal region in OCT B-scans. Accompanying each scale is a representative OCT B-scan, shown in figure \ref{fig:APPDX_reg_scale}.

    \begin{figure}[tb]
        \centering
        \includegraphics[width=\textwidth]{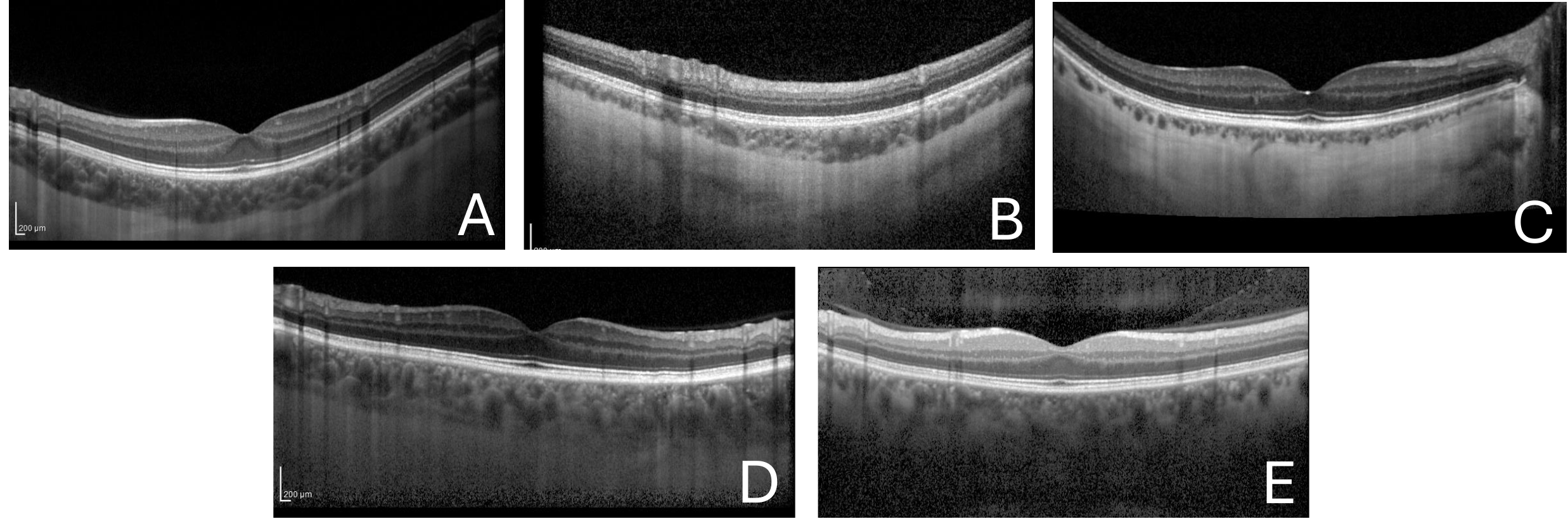}
        \caption[Exemplar \acrshort{OCT} B-scans with varying image quality in choroidal space.]{Exemplar OCT B-scans to accompany each scale to assess image quality for the choroidal space.}
        \label{fig:APPDX_reg_scale}
    \end{figure}

    \begin{mysubsection}[]{Choroidal space OCT image quality}
        \begin{itemize}\setlength\itemsep{0em}
        
            \item \textbf{2 Very good}: The definition of the Choroid-Sclera boundary is excellent, and is particularly enhanced in regions by darker, posterior-most vessels above it. Its definition is clear across the whole OCT B-scan (figure \ref{fig:APPDX_ves_scale}(A)).
            
            \item \textbf{1 Good}: The contrast of the Choroid-Sclera boundary is still greater than Haller's layer of the choroid above it, but elements of poor signal (i.e. speckle noise), or interstitial fluid is obscuring parts across the OCT B-scan (relative to Very good) (figure \ref{fig:APPDX_ves_scale}(B)).
            
            \item \textbf{0 Okay}: Choroid-Sclera boundary is still visible enough to be traced, but the contrast between it and the choroidal space is poor (relative to Good) (figure \ref{fig:APPDX_ves_scale}(C)).
            
            \item \textbf{-1 Bad}: Exact definition of Choroid-Sclera boundary is uncertain and made ``fuzzy'' by poor signal penetration and speckle noise --- posterior-most vessels unable to be used as a reference/guide to trace boundary (figure \ref{fig:APPDX_ves_scale}(D)).
            
            \item \textbf{-2 Very bad}: The choroid region is cropped in parts, or there is little signal penetrating deep enough to make the Choroid-Sclera boundary visible in most locations along the OCT B-scan (figure \ref{fig:APPDX_ves_scale}(E)).

        \end{itemize}
    \end{mysubsection}
    
    \begin{mysubsection}[]{Choroidal vasculature OCT image quality}
        Below is a list of definitions for each scale when assessing the quality of the choroidal vasculature in OCT B-scans. Accompanying each scale is a representative OCT B-scan, shown in figure \ref{fig:APPDX_ves_scale}.

        \begin{figure}[!t]
            \centering
            \includegraphics[width=\textwidth]{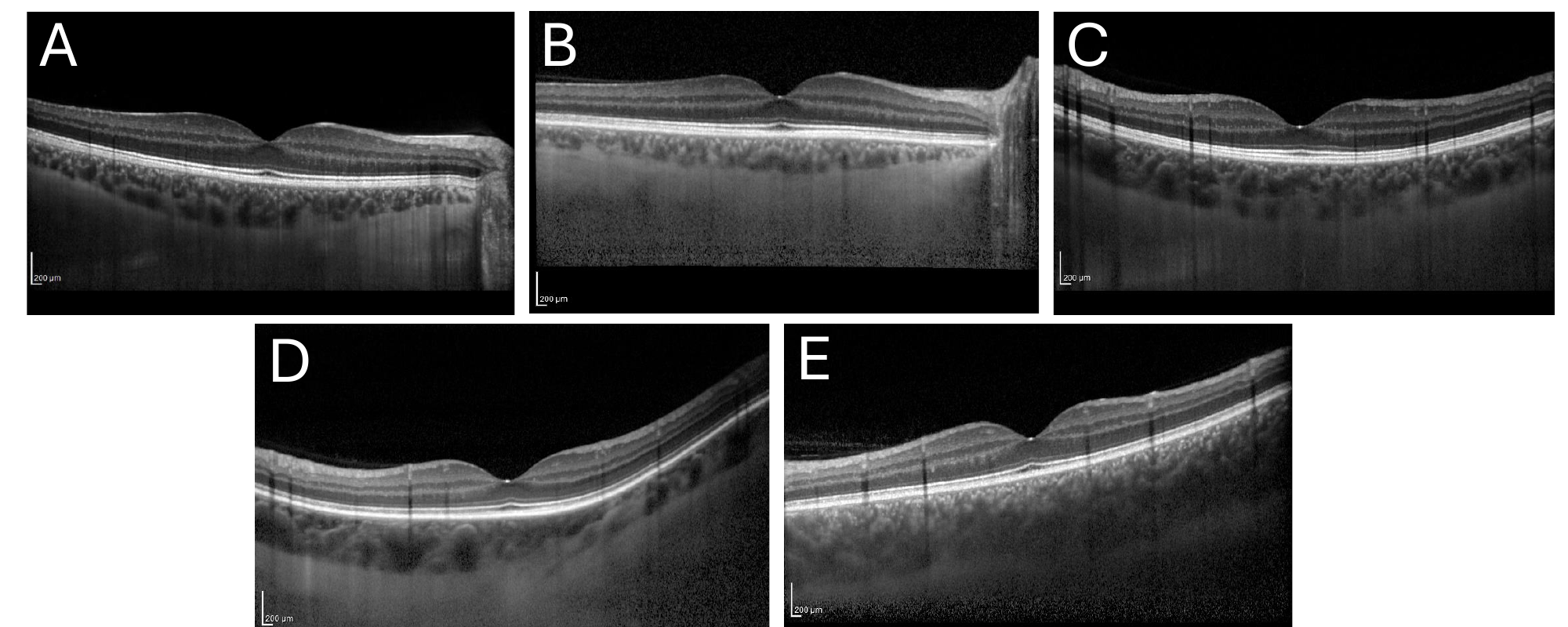}
            \caption[Exemplar \acrshort{OCT} B-scans with varying image quality in choroidal vasculature.]{Exemplar OCT B-scans to accompany each scale to assess image quality for the choroidal vasculature.}
            \label{fig:APPDX_ves_scale}
        \end{figure}

        \begin{itemize}\setlength\itemsep{0em}
            \item \textbf{2 Very good}: Large contrast between large vessels and interstitial fluid, with clear vessel wall definition. Smaller vessels are easily identifiable relative to interstitial fluid (figure \ref{fig:APPDX_ves_scale}(A)).
            
            \item \textbf{1 Good}: Large vessels still significantly darker in intensity than adjacent interstitial fluid, but vessel wall definition has poorer contrast. Small vessels are still easily identifiable. Some medium vessels aren't as dark as larger ones, leading to some uncertainty in their vessel wall definition (figure \ref{fig:APPDX_ves_scale}(B)).

            \item \textbf{0 Okay}: Good definition across all vessels, but many vessels clumped together so vessel wall separation challenging, with speckle noise corrupting vessels at all scales (figure \ref{fig:APPDX_ves_scale}(C)).
            
            \item \textbf{-1 Bad}: Low contrast choroid and speckle noise corrupting vessel wall definition, the centre of larger vessels still visible, but smaller vasculature difficult to confidently define (figure \ref{fig:APPDX_ves_scale}(D)).
            
            \item \textbf{-2 Very bad}: Choroidal region is cropped and/or speckle noise destroys any clear vessel wall definition for larger vessels, and smaller vessels are indescribable relative to adjacent interstitial fluid (figure \ref{fig:APPDX_ves_scale}(E)).
        \end{itemize}

    \end{mysubsection}

\end{mysection}

\begin{mysection}[]{OCT B-scan vessel guide} \label{apdx:vessel_protocol}

    \begin{figure}[tb]
        \centering
        \includegraphics[width=\textwidth]{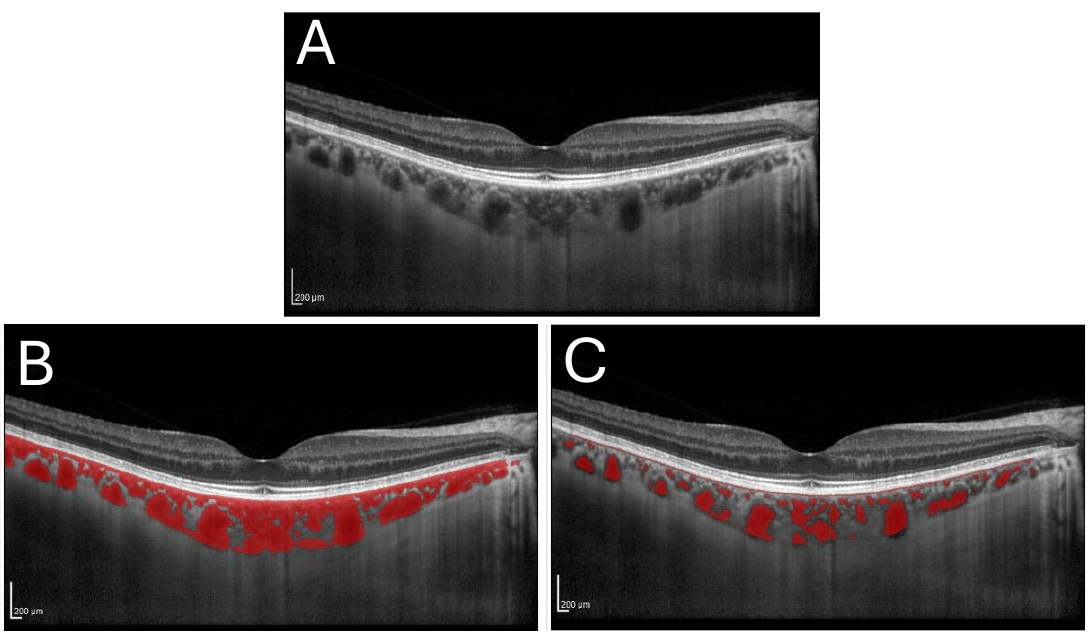}
        \caption[Example of over- and under-segmentation of choroidal vasculature.]{(A) OCT B-scan. (B) Over-segmentation and (C) under-segmentation of the choroidal vasculature.}
        \label{fig:APPDX_vess_seg}
    \end{figure}
    
    The example in figure \ref{fig:APPDX_vess_seg} highlights the effect of over- and under-segmentation of the choroidal vasculature. In figure \ref{fig:APPDX_vess_seg}(B), over-segmentation leads to all vessel pixels correctly segmented, as well as much of the interstitial space. This represents good \textit{completeness of vessels} but poor \textit{interstitial segmentation}. Conversely, in figure \ref{fig:APPDX_vess_seg}(C), under-segmentation leads to only vessel pixels segmented, but many pixels representing vessels were miss-classified as interstitial space (not coloured red). This represents poor \textit{completeness of vessels} but good \textit{interstitial segmentation}. Figure \ref{fig:APPDX_chor_defn}(D) shows the ideal segmentation for this OCT B-scan.

    The two examples above highlight an inherent problem with choroidal vessel segmentation. There are two goals in mind: successfully segmenting every vessel in the intravascular space, and successfully excluding all pixels representing interstitial fluid/choroidal stroma. In reality, contrast issues and noise corruption can make this problem a very difficult one, leading to uncertainty in what constitutes a vessel boundary or not. This results in two design choices made on the objective behind choroidal vessel segmentation: 
    \begin{enumerate}\setlength\itemsep{0em}
        \item Segment all vessels at the cost of segmenting interstitial fluid as well (over-segmentation)
        \item The segmentation only identifies vessel in order to exclude the interstitial space, at the cost of potentially missing pixels representing vessel boundaries/low contrast vessels.
    \end{enumerate}
    We believe (2) is the preferred approach, so that the fidelity of the vessels are preserved. Nevertheless, it’s important a segmentation technique be consistent and repeatable, using the same definition/design choice across all images (which is something quite difficult for a human grader to apply).

\end{mysection}

\begin{mysection}[]{OCT B-scan manual segmentation protocol} \label{apdx:seg_protocol}

    \begin{mysubsection}[]{ITK-SNAP setup}
        \begin{enumerate}\setlength\itemsep{0em}
            \item Go to ITK-SNAP Version 3.x Downloads and click the link for ITK-SNAP 3.8.0 (stable version) windows Self-Install Package (64-bit).
            \item Fill in the form and select next. On the next page and select the download. It will take you to another page and the download should start.
            \item Navigate to the downloaded \verb|.exe| file and double click to begin the install process.
            \item Select next, agree to the terms and conditions and select next. Tick the box next to `Do not add SNAP to the system PATH', keep selecting next until the setup is complete, press finish.
            \item Navigate to the SNAP program and open, it will prompt you if you would like to enable automatic updates, select yes.
            \item (\textbf{Optional}) Go to \href{https://www.youtube.com/watch?v=-tjVN5GwjKg&feature=youtu.be}{RSNA 2016 ITK-SNAP Training - Intro \& Manual Segmentation - YouTube} and watch the ~15min video. This is a training video to learn how to use SNAP, they give an example in 3D but we will be annotating in 2D. The slides for the training video can be found \href{http://www.itksnap.org/pmwiki/uploads/Train/RSNA2016-Manual-Guido-Final.pdf}{here}. 
        \end{enumerate}
    \end{mysubsection}

    \begin{mysubsection}[]{Load in image, label definition, main tools} \label{sec:APPDX_load_bscan}
    
        \begin{enumerate}\setlength\itemsep{0em}
            \item Go to File > Open Main Image. For image file name use the browse button to navigate to the image. Under File Format select `Generic ITK image'. Select Next and Finish.
            \item Click the `A' in the top right corner of the top left window to expand the view to full size and set the window to full screen (figure \ref{fig:APPDX_tik_window}).
            \item Next, we need to set up some labels. Click on the label editor which allows you to define separate segmentation labels (figure \ref{fig:APPDX_tik_window}).
            \item We will have three separate labels (figure \ref{fig:APPDX_tik_window}):
            \begin{enumerate}\setlength\itemsep{0em}
                \item Label 1, called `Choroid' represents segmenting the choroidal region, in blue.
                \item Label 2, called `Vessel' represents labelling vessels within the choroidal space, in red. Here, this will have an opacity of 113 to aid segmentation.
                \item Label 0 is the `Clear' label which is your `rubber'.
            \end{enumerate}

            \begin{figure}[tb]
                \centering
                \includegraphics[width=0.8\textwidth]{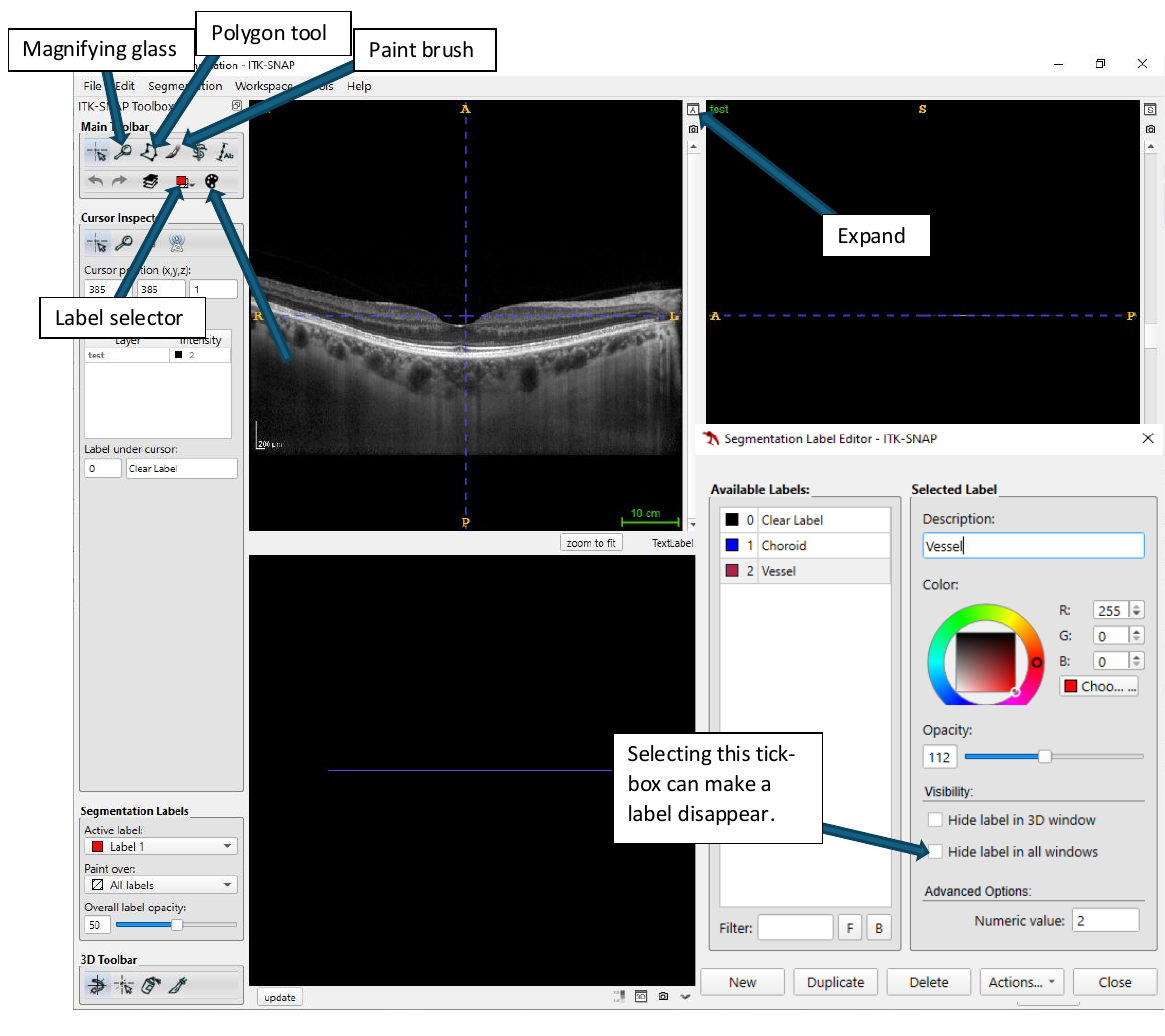}
                \caption[Screenshot of ITK-Snap for manual choroid segmentation.]{Screenshot of computer screen of ITK-Snap after loading an \acrshort{OCT} B-scan as a generic ITK image, with various functions annotated. The label editor also shows the creation of labels for the choroid region (blue) and vessels (red).}
                \label{fig:APPDX_tik_window}
            \end{figure}
            
            \item On the Main Toolbar, the label selector allows you to toggle between different segmentation labels.
            \begin{enumerate}\setlength\itemsep{0em}
                \item Alternatively, you can do the same on the `Segmentation Labels' toolbar, and there is also an option to change the colour opacity. There is a drop-down list of all the labels in the `Active label' section.
            \end{enumerate}
            \item On the Main Toolbar, the magnifying glass can be used (up to 4x) to zoom into the portion of the image you would like to annotate. If you hover your mouse over this icon it provides instructions on how to do that
            \item Label 1 segmentation, defining the choroidal region segmentation, are done using the Polygon tool seen on the Main Toolbar.
            \item Label 2 segmentations, defining the choroidal vessel segmentation, are done using the Paintbrush tool seen on the Main Toolbar.
            \item Image enhancement and brightness tool features will not be used in this protocol. 
        
        \end{enumerate}
    
    \end{mysubsection}

    \begin{mysubsection}[]{Segmentation}

        \begin{mysubsubsection}[]{Region segmentation}

            \begin{figure}[tb]
                \centering
                \includegraphics[width=\textwidth]{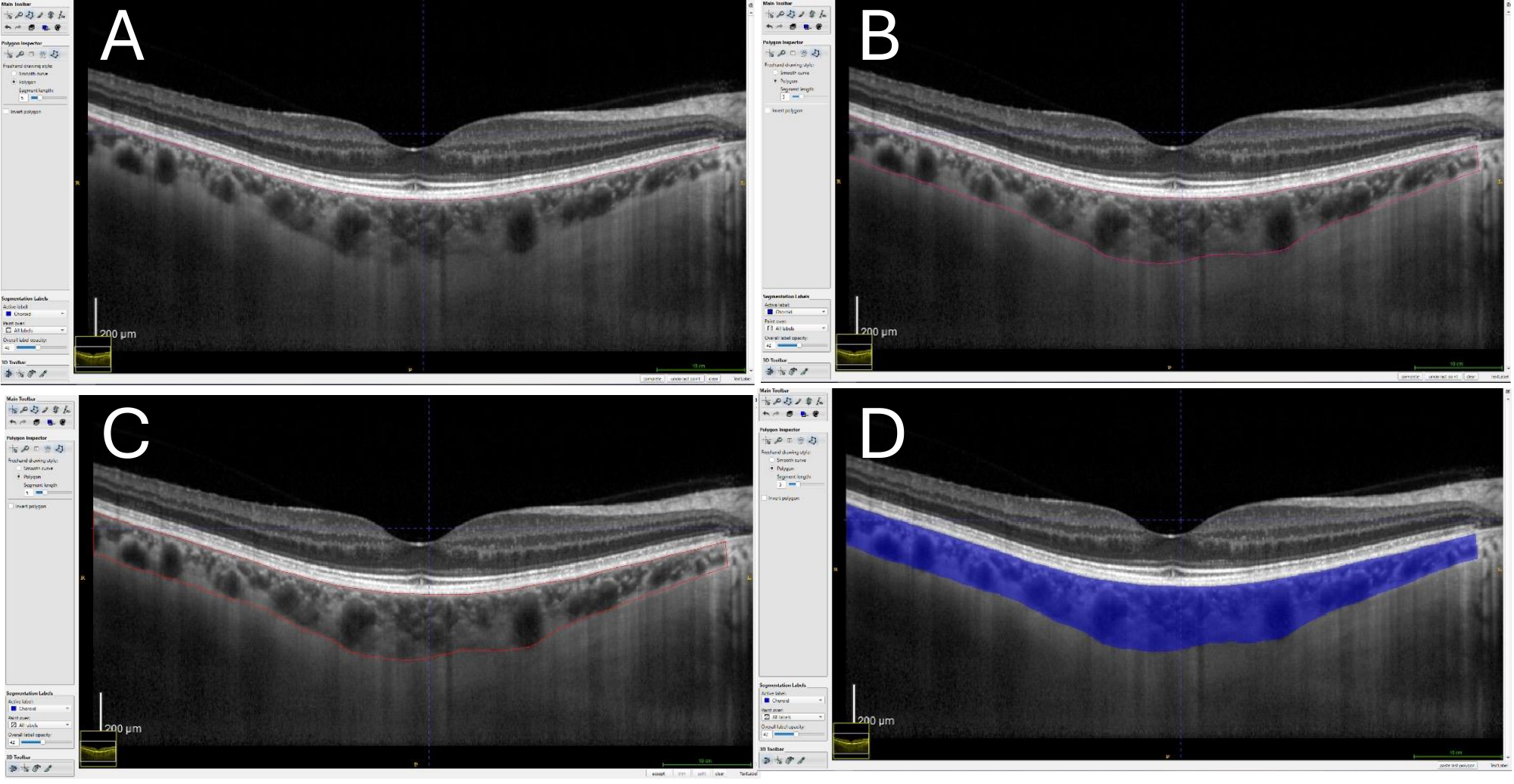}
                \caption[Demonstration of manual choroid region segmentation.]{(A) Manual selection of pixels from left--right along \acrshort{RPE}-Choroid boundary using Polygon tool. (B) Within the same Polygon tool selection, select pixels from right--left along Choroid-Sclera boundary. (C) Clicking `Complete' forms polygonal shape representing choroidal space, linking first and last selected pixels. (D) Clicking `Accept' converts polygonal shape into segmentation object.}
                \label{fig:APPDX_annot_regchor}
            \end{figure}
    
            \begin{enumerate}\setlength\itemsep{0em}
                \item Select the Polygon tool. This allows you to draw lines by either selecting pixels to connect linearly, or by holding the left-mouse clicker to draw a smooth line.
                \item Select the “Choroid” segmentation label and make sure you are painting over All Labels.
                \item Select the Polygon style button within the Freehand drawing style window. Choose a segment length of 5.
                \item You may use the magnifying glass freely. Holding the right mouse-click will zoom in and out for you
                \item There are buttons on the bottom-right of the window to allow you to undo the most recently selected pixel, or to clear all selected pixels entirely.
                
                \item Core instructions:
                \begin{enumerate}\setlength\itemsep{0em}
                    \item Start at the top-left most pixel of the \acrshort{RPE}-Choroid layer boundary and move left-to-right until all of the visible layer boundary has been traced (figure \ref{fig:APPDX_annot_regchor}(A)).
                    \item On the same polygon segmentation, select the bottom-right most pixel of the Choroid-Sclera layer boundary and trace right-left ((figure \ref{fig:APPDX_annot_regchor}(B)). You should see a polygonal shape begin to form, defining the whole choroidal space.
                    \item When you reach the left-most pixel of the Choroid-Sclera layer boundary, click Complete, this will join the first and last pixels you have selected (figure \ref{fig:APPDX_annot_regchor}(C)).
                    \item Here, you can move (or remove) individually selected pixels around to better fit the polygonal shape to the choroidal region.
                    \item When you are finished, click Accept, this will turn your polygonal shape into a Blue segmented region (figure \ref{fig:APPDX_annot_regchor}(D)).
                \end{enumerate}
                \item If you wish, you can go straight to section \ref{sec:APPDX_save_segs} to save your region segmentation, please save as `{filename}\_RegionAnnotations'.
            \end{enumerate}
        
        \end{mysubsubsection}
        
        \begin{mysubsubsection}[]{Vessel segmentation}

            \begin{figure}[tb]
                \centering
                \includegraphics[width=\textwidth]{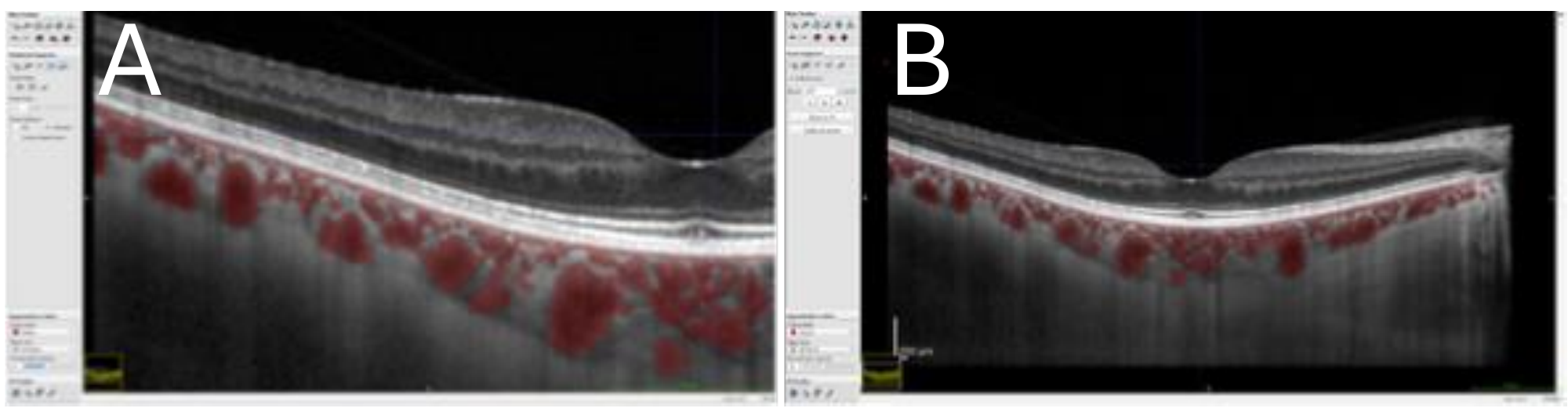}
                \caption[Demonstration of manual choroid vessel segmentation.]{(A) Zoomed in temporal side of choroid, as we progress through segmenting with half-opacity using the Paintbrush. (B) completed vessel segmentation. Note the choriocapillaries (approximately 3 pixels deep below the \acrshort{RPE}-Choroid layer boundary on Heidelberg Engineering Spectralis OCT1 data) is always assumed to be vessel due to its dense capillary structure, hence why it is highlighted also.}
                \label{fig:APPDX_annot_veschor}
            \end{figure}

            \begin{enumerate}\setlength\itemsep{0em}
                \item In order for the region segmentation above to not block your view, click the Label Editor and select the box `Hide label in all windows' for Choroid Label 1. This removes the blue, choroidal space segmentation, giving you a clearer view of the vessels.
                \item Select the Paintbrush tool. This is a paint brush that allows you to ‘colour in’ the vessels. You can change the size and shape of the brush.
                \item Change the `Active label' to `Vessel' or the `Segmentation Labels' toolbar and make sure you select the `All labels' layer in the `Paint over' dropdown list.
                \item You may alter the opacity to aid segmentation.
                \item You may use up to a brush size between 2 and 12 (for small and large vessel labelling), selecting the circular shape and ticking the isotropic box (figure \ref{fig:APPDX_annot_veschor}).
                \item You may use the magnifying glass to zoom freely for small regions.
                \item Some tips:
                \begin{enumerate}\setlength\itemsep{0em}
                    \item You can select the `Clear label' as a `rubber', in case you label parts of the image mistakenly.
                    This can also be done with the right mouse-click.
                    \item Ctrl+Z will undo the most recent Paintbrush annotation.
                    \item Press 2 or 4 to toggle between pan/zoom (2) and brush tool (4)
                    \item Regularly reduce the `Overall label opacity' on the `Segmentation Labels' toolbar to sanity check any selected regions to make sure you are happy with your selections.
                \end{enumerate}
                \item If you wish, you can go straight to (5) to save your vessel segmentation, please save as `{filename}\_VesselAnnotations'.
            \end{enumerate}
            
        \end{mysubsubsection}

    \end{mysubsection}

    \begin{mysubsection}[]{Saving segmentations} \label{sec:APPDX_save_segs}

        \begin{enumerate}\setlength\itemsep{0em}
            \item Once completed the region and vessel segmentations, you can view your final segmentation and make any amendments by using the right-click `rubber' i.e. use the default Clear label (below).
            \begin{enumerate}\setlength\itemsep{0em}
                \item Unselect the `Hide label in all windows' for Choroid Label 1 on the Label Editor to see the choroid region segmentation, alongside the choroid vessel segmentation.
                \item Note that you must select the `Choroid' label to make any amendments to the choroid region segmentation, and likewise for amendments to the choroid vessel segmentation.
            \end{enumerate}
            \item Once happy, to save the segmentation go to Segmentation > Save Segmentation Image and call the filename the same as the beginning of the image’s filename and add `\_Annotations' to the end of it. For example, `test\_Annotations'. Make sure the format is NiFTI and select Finish.
            \item Save the labels to a \verb|.txt| file by selecting Segmentation > Export Label Descriptions and save the name with the beginning of the image file name and add `labels.txt' – e.g. `test\_Labels'.
            \item You can do step 2 \& 3 as you go along. If you need to finish an annotation later follow step 2 and 3 and close SNAP. When you want to come back to it, load the image as in section \ref{sec:APPDX_load_bscan} and load the segmentation by going to Segmentation > Open Segmentation and select the file to open and go to Segmentation > Import Label Descriptions and select the  \verb|.txt| file to open and continue annotating where you left off.
        \end{enumerate}
        
    \end{mysubsection}

\end{mysection}

\end{mychapter}

\end{document}